# HERA AND THE LHC

A workshop on the implications of HERA for LHC physics

March 2004 – March 2005

Parton density functions
Multijet final states
and energy flow
Heavy quarks
Diffraction
Monte Carlo tools

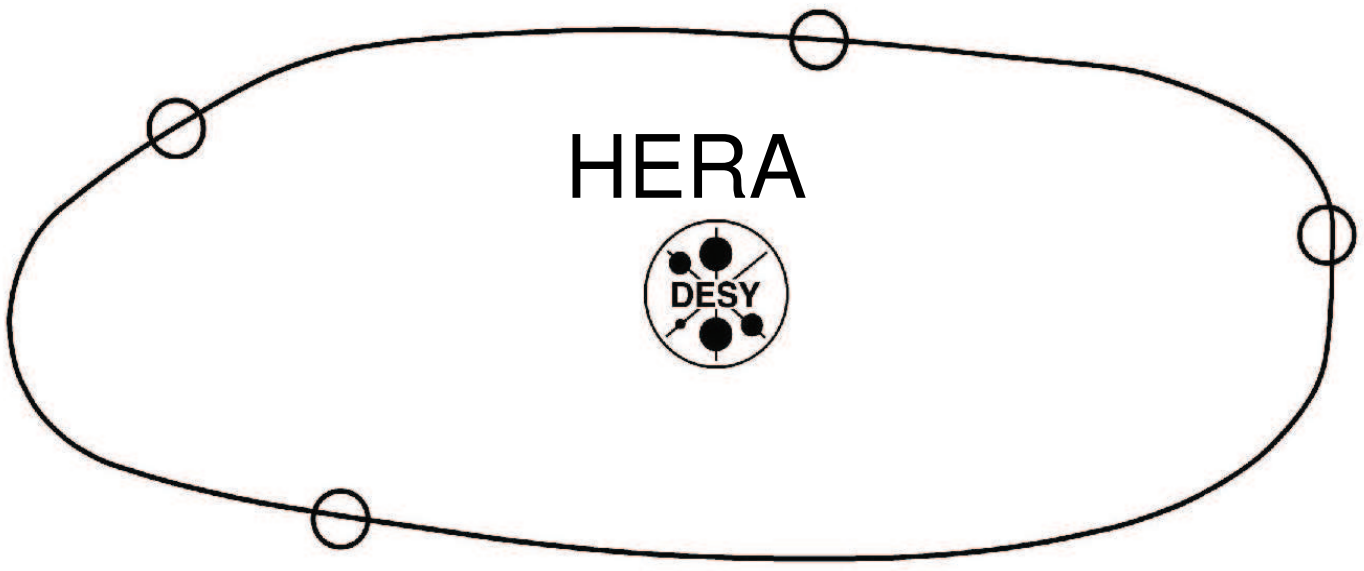

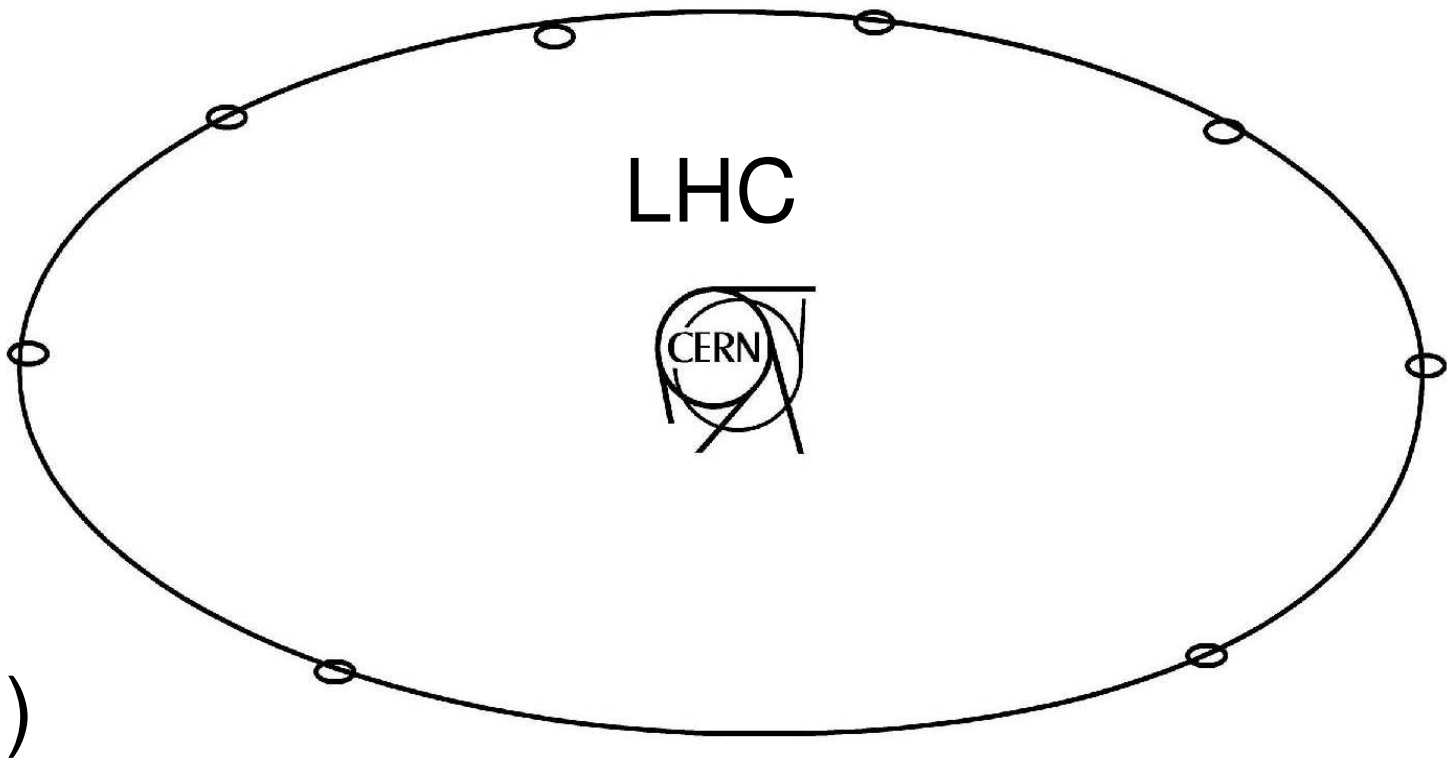

Proceedings
Editors:
A. De Roeck (CERN)
H. Jung (DESY)

**Part A**

**Organizing Committee:**

G. Altarelli (CERN), J. Blümlein (DESY), M. Botje (NIKHEF),
J. Butterworth (UCL), A. De Roeck (CERN) (chair), K. Eggert (CERN),
H. Jung (DESY) (chair), M. Mangano (CERN), A. Morsch (CERN),
P. Newman (Birmingham), G. Polesello (INFN), O. Schneider (EPFL),
R. Yoshida (ANL)

**Advisory Committee:**

J. Bartels (Hamburg), M. Della Negra (CERN), J. Ellis (CERN),
J. Engelen (CERN), G. Gustafson (Lund), G. Ingelman (Uppsala), P. Jenni (CERN),
R. Klanner (DESY), M. Klein (DESY), L. McLerran (BNL), T. Nakada (CERN),
D. Schlatter (CERN), F. Schrempp (DESY), J. Schukraft (CERN),
J. Stirling (Durham), W.K. Tung (Michigan State), A. Wagner (DESY),
R. Yoshida (ANL)

# Abstract

The HERA electron–proton collider has collected 100 $pb^{-1}$ of data since its start-up in 1992, and recently moved into a high-luminosity operation mode, with upgraded detectors, aiming to increase the total integrated luminosity per experiment to more than 500 $pb^{-1}$. HERA has been a machine of excellence for the study of QCD and the structure of the proton. The Large Hadron Collider (LHC), which will collide protons with a centre-of-mass energy of 14 TeV, will be completed at CERN in 2007. The main mission of the LHC is to discover and study the mechanisms of electroweak symmetry breaking, possibly via the discovery of the Higgs particle, and search for new physics in the TeV energy scale, such as supersymmetry or extra dimensions. Besides these goals, the LHC will also make a substantial number of precision measurements and will offer a new regime to study the strong force via perturbative QCD processes and diffraction. For the full LHC physics programme a good understanding of QCD phenomena and the structure function of the proton is essential. Therefore, in March 2004, a one-year-long workshop started to study the implications of HERA on LHC physics. This included proposing new measurements to be made at HERA, extracting the maximum information from the available data, and developing/improving the theoretical and experimental tools. This report summarizes the results achieved during this workshop.

# Preface

The workshop on 'HERA and the LHC' successfully brought together experimental and theory experts working on electron–proton and proton–proton collider physics. It offered a forum to discuss the impact of present and future measurements at HERA on the physics programme of the LHC. The workshop was launched with a meeting at CERN in March 2004 and its first phase was terminated with a summary meeting in April 2005 at DESY. The workshop was very timely with on the one hand HERA-II, expected to deliver more than 500 $pb^{-1}$ per experiment by 2007, ramping up to full strength, and on the other hand three years before the first collisions at the LHC.

The following aims were defined as the charge to the workshop:

– To identify and prioritize those measurements to be made at HERA which have an impact on the physics reach of the LHC.

– To encourage and stimulate transfer of knowledge between the HERA and LHC communities and establish an ongoing interaction.

– To encourage and stimulate theory and phenomenological efforts related to the above goals.

– To examine and improve theoretical and experimental tools related to the above goals.

– To increase the quantitative understanding of the implication of HERA measurements on LHC physics.

Five working groups were formed to tackle the workshop charge. Results and progress were presented and discussed at six major meetings, held alternately at CERN and at DESY.

Working group one had a close look at the parton distribution functions (PDFs), their uncertainties and their impact on the LHC measurements. The potential experimental and theoretical accuracy with which various LHC processes such as Drell–Yan, the production of W's, Z's and dibosons, etc. can be predicted was studied. Cross-section calculations and differential distributions were documented and some of these processes are used as benchmark processes for PDF and other QCD uncertainty studies. In particular W and Z production at the LHC has been scrutinized in detail, since these processes will be important standard candles. It is even planned to use these for the luminosity determination at the LHC. The impact of PDFs on LHC measurements and the accuracy with which the PDFs can be extracted from current and forthcoming data, particular the HERA-II data, have been investigated, as well as the impact of higher order corrections, small-$x$ and large-$x$ resummations. Initial studies have been started to provide a combined data set on structure function measurements from the two experiments H1 and ZEUS. Arguments for running HERA at lower energies, to allow for the measurement of the longitudinal structure function, and with deuterons, have been brought forward.

The working group on multi-jet final states and energy flows studied processes in the perturbative and non-perturbative QCD region. One of the main issues of discussion during the workshop was the structure of the underlying event and of minimum-bias events. New models were completed and presented during the workshop, and new tunes on p–p data were discussed. A crucial test will be to check these generator tunes with e–p and $\gamma$–p data from HERA, and thus check their universality. Other important topics tackled by this working group concern the study of rapidity-gap events, multi-jet topologies and matrix-element parton-shower matching questions. The understanding of rapidity gaps and in particular their survival probability is of crucial importance to make reliable predictions for central exclusive processes at the LHC. HERA can make use of the virtuality of the photon to study in detail the onset of multiple interactions. Similarly HERA data, because of its handles on the event kinematics via the scattered electron, is an ideal laboratory to study multiple-scale QCD problems and improve our understanding in that area such that it can be applied with confidence to the LHC data. For example, the HERA data give strong indications that in order to get reliable and precise predictions, the use of unintegrated parton distributions will be necessary. The HERA data should be maximally exploited to extract those distributions.

The third group studied heavy flavours at HERA and the LHC. Heavy quark production, in particular at small momenta at the LHC, is likely to give new insight into low-$x$ phenomena in general and saturation in particular. The possibilities for heavy quark measurements at LHC were investigated. The charm and bottom content of the proton are key measurements, and the anticipated precision achievable with HERA-II is very promising. Furthermore, heavy quark production in standard QCD processes may form an important background in searches for new physics at the LHC and has therefore to be kept as much as possible under control. Again, heavy quark production results from mostly multi-scale processes where topics similar to those discussed in working group two can be studied and tested. Important steps were taken for a better understanding of the heavy quark fragmentation functions, which are and will be measured at HERA. The uncertainties of the predicted heavy quark cross-section were studied systematically and benchmark cross-sections were presented, allowing a detailed comparison of different calculations.

Diffraction was the topic of working group four. A good fraction of the work in this group went into the understanding of the possibility of the exclusive central production of new particles such as the Higgs pp→p+H+p at the LHC. With measurable cross-sections, these events can then be used to pin down the CP properties of these new particles, via the azimuthal correlation of the two protons, and thus deliver an important added value to the LHC physics programme. The different theoretical approaches to calculate cross-sections for this channel have been confronted, and scrutinized. The Durham approach, though the one that gives the most conservative estimate of the event cross-section, namely in the order of a few femtobarns, has now been verified by independent groups. In this approach the generalized parton distributions play a key role. HERA can determine generalized parton distributions, especially via exclusive meson production. Other topics discussed in this group were the factorization breaking mechanisms and parton saturation. It appears that the present diffractive dijet production at HERA does not agree with a universal description of the factorization breaking, which is one of the mysteries in the present HERA data. Parton saturation is important for event rates and event shapes at the LHC, which will get large contributions of events at very low-$x$. Furthermore, the precise measurement of the diffractive structure functions is important for any calculation of the cross-section for inclusive diffractive reactions at the LHC. Additionally, this working group has really acted as a very useful forum to discuss the challenges of building and operating beam-line integrated detectors, such as Roman Pots, in a hadron storage ring. The experience gained at HERA was transferred in detail to the LHC groups which are planning for such detectors.

Finally, working group five on the Monte Carlo tools had very productive meetings on discussing and organizing the developments and tunings of Monte Carlo programs and tools in the light of the HERA–LHC connection. The group discussed the developments of the existing generators (e.g., PYTHIA, HERWIG) and new generators (e.g., SHERPA), or modifications of existing ones to include p–p scattering (e.g., RAPGAP, CASCADE). Many of the other studies like tuning to data, matrix-element and parton shower matching, etc., were done in common discussions with the other working groups. Validation frameworks have been compared and further developed, and should allow future comparisons with new and existing data to be facilitated.

In all it has been a very productive workshop, demonstrated by the content of these proceedings. Yet the ambitious programme set out from the start has not been fully completed: new questions and ideas arose in the course of this workshop, and the participants are eager to pursue these ideas. Also the synergy between the HERA and LHC communities, which has been built up during this workshop, should not evaporate. Therefore this initiative will continue and we look forward to further and new studies in the coming years, and the plan to hold a workshop once a year to provide the forum for communicating and discussion the new results.

We thank all the convenors for the excellent organization of their working groups and all participants for their work and enthusiasm and contribution to these proceedings.

We are grateful to the CERN and DESY directorates for the financial support of this workshop and for the hospitality which they extended to all the participants. We are grateful to D. Denise, A. Grabowksi and S. Platz for their continuous help and support during all the meeting weeks. We would like to thank also B. Liebaug for the design of the poster for this first HERA–LHC workshop.

Hannes Jung and Albert De Roeck

# List of Authors

S. Alekhin [1], G. Altarelli [2,3], N. Amapane [4], J. Andersen [5], V. Andreev [6], M. Arneodo [7], V. Avati [8], J. Baines [9], R.D. Ball [10], A. Banfi [5], S.P. Baranov [6], J. Bartels [11], O. Behnke [12], R. Bellan [4], J. Blümlein [13], H. Böttcher [13], S. Bolognesi [4], M. Boonekamp [14], D. Bourilkov [15], J. Braciník [16], A. Bruni [17], G. Bruni[18], A. Buckley [19], A. Bunyatyan [20], C.M. Buttar [21], J.M. Butterworth [22], S. Butterworth [22], M. Cacciari [23], T. Carli [24], G. Cerminara [4], S. Chekanov [25], M. Ciafaloni [26], D. Colferai [26], J. Collins [27], A. Cooper-Sarkar [28], G. Corcella [2], M. Corradi [29], B.E. Cox [30], R. Croft [31], Z. Czyczula [32], A. Dainese [33], M. Dasgupta [2], G. Davatz [34], L. Del Debbio [2,10], Y. Delenda [30], A. De Roeck [24], M. Diehl [35], S. Diglio [3], G. Dissertori [34], M. Dittmar [34], J. Ellis [2], K.J. Eskola [36], T.O. Eynck [37], J. Feltesse [38], F. Ferro [39], R.D. Field [40], J. Forshaw [30], S. Forte [41], A. Geiser [35], S. Gieseke [42], A. Glazov [35], T. Gleisberg [43], P. Golonka [44], E. Gotsman [45], G. Grindhammer [16], M. Grothe [46], C. Group [40], M. Groys [45], A. Guffanti [13], G. Gustafson [47], C. Gwenlan [28], S. Höche [43], C. Hogg [48], J. Huston [49], G. Iacobucci [18], G. Ingelman [50], S. Jadach [51], H. Jung [35], J. Kalliopuska [52], M. Kapishin [53], B. Kersevan [54], V. Khoze [19], M. Klasen [11,55], M. Klein [13], B.A. Kniehl [11], V.J. Kolhinen [36], H. Kowalski [35], G. Kramer [13], F. Krauss [43], S. Kretzer [56], K. Kutak [11], J.W. Lämsä [52], L. Lönnblad [47], T. Laštovička [24], G. Laštovička-Medin [57], E. Laenen [37], Th. Lagouri [58], J.I. Latorre [59], N. Lavesson [47], V. Lendermann [60], E. Levin [45], A. Levy [45], A.V. Lipatov [61], M. Lublinsky [62], L. Lytkin [63], T. Mäki [52], L. Magnea [64], F. Maltoni [65], M. Mangano [2], U. Maor [45], C. Mariotti [66], N. Marola [52], A.D. Martin [19], A. Meyer [35], S. Moch [13], J. Monk [30], A. Moraes [21], A. Morsch [24], L. Motyka [11], E. Naftali [45], P. Newman [67], A. Nikitenko [68], F. Oljemark [52], R. Orava [52], M. Ottela [52], K. Österberg [52], K. Peters [30,35], F. Petrucci [44], A. Piccione [64], A. Pilkington [30], K. Piotrzkowski [69], O.I. Piskounova [6], A. Proskuryakov [45], A. Prygarin [45], J. Pumplin [70], K. Rabbertz [71], R. Ranieri [26], V. Ravindran [72], B. Reisert [73], E. Richter-Was [74], L. Rinaldi [18], P. Robbe [75], E. Rodrigues [76], J. Rojo [60], H. Ruiz [24], M. Ruspa [18], M.G. Ryskin [77], A. Sabio Vera [11], G.P. Salam [78], A. Schälicke [43], S. Schätzel [35], T. Schörner-Sadenius [79], I. Schienbein [80], F-P. Schilling [24], F. Schrempp [35], S. Schumann [43], M.H. Seymour [81], F. Siegert [24], T. Sjöstrand [2,47], M. Skrzypek [51], J. Smith [37,82], M. Smizanska [83], H. Spiesberger [84], A. Staśto [85], H. Stenzel [86], W.J. Stirling [19], P. Szczypka [31], S. Tapprogge [52,84], C. Targett-Adams [22], M. Tasevsky [87], T. Teubner [88], R.S. Thorne [5], A. Tonazzo [3], A. Tricoli [28], N. Tuning [37], J. Turnau [85], U. Uwer [12], P. Van Mechelen [87], R. Venugopalan [56], M. Verducci [24], J.A.M. Vermaseren [37], A. Vogt [19], R. Vogt [89], B.F.L. Ward [90], Z. Was [44], G. Watt [35], B.M. Waugh [22], C. Weiser [91], M.R. Whalley [19], M. Wing [22], J. Winter [43], S.A. Yost [90], G. Zanderighi [73], N.P. Zotov [61]

[1] Institute for High Energy Physics, 142284 Protvino, Russia
[2] CERN, Department of Physics, Theory Unit, CH 1211 Geneva 23, Switzerland
[3] Dipartimento di Fisica “E.Amaldi”, Università Roma Tre and INFN, Sezione di Roma Tre, via della Vasca Navale 84, I 00146 Roma, Italy
[4] Torino University and INFN Torino, Italy
[5] Cavendish Laboratory, University of Cambridge, Madingley Road, Cambridge, CB3 0HE, UK
[6] P.N. Lebedev Physical Institute of the Russian Academy of Science, Moscow, Russia
[7] Università del Piemonte Orientale, Novara, and INFN-Torino, Italy
[8] CERN, Geneva, Switzerland, and Case Western Reserve University, Cleveland, OH, USA
[9] Rutherford Laboratory, UK
[10] School of Physics, University of Edinburgh, Edinburgh EH9 3JZ, UK
[11] II. Institut für Theoretische Physik, Universität Hamburg Luruper Chaussee 149, D-22761 Hamburg, Germany
[12] Universität Heidelberg, Philosophenweg 12 69120 Heidelberg, Germany
[13] DESY, Platanenallee 6, D 15738 Zeuthen, Germany
[14] Service de physique des particules, CEA/Saclay, 91191 Gif-sur-Yvette CEDEX, France
[15] University of Florida, Gainesville, FL 32611, USA
[16] Max-Planck-Institut für Physik, München, Germany
[17] INFN Bologna, Via Irnerio 46, 40156 Bologna, Italy
[18] INFN Bologna and University of Eastern Piedmont, Italy
[19] Institute for Particle Physics Phenomenology, University of Durham, DH1 3LE, UK
[20] Yerevan Physics Institute, Armenia and MPI-K Heidelberg, Germany
[21] Dept. of Physics and Astronomy, University of Glasgow, UK
[22] Dept. of Physics and Astronomy, University College London, UK
[23] LPTHE - Université P. et M. Curie (Paris 6), Paris, France
[24] CERN, Department of Physics, CH 1211 Geneva 23, Switzerland
[25] HEP Division, Argonne National Laboratory, 9700 S. Cass Avenue, Argonne, IL 60439, USA
[26] Dipartimento di Fisica, Università di Firenze and INFN, Sezione di Firenze, I 50019 Sesto Fiorentino, Italy
[27] Physics Department, Penn State University, USA
[28] Department of Physics, Nuclear and Astrophysics Lab., Keble Road, Oxford, OX1 3RH, UK
[29] INFN Bologna, via Irnerio 46, Bologna, Italy
[30] School of Physics and Astronomy, The University of Manchester, Manchester M13 9PL, UK
[31] University of Bristol, Bristol, UK
[32] Institute of Physics, Jagiellonian University, Krakow, Poland and Niels Bohr Institute, University of Copenhagen, Copenhagen, Denmark
[33] University and INFN, Padova, Italy
[34] Institute for Particle Physics, ETH-Zürich Hönggerberg, CH 8093 Zürich, Switzerland
[35] DESY, Notkestrasse 85, D 22603 Hamburg, Germany
[36] Department of Physics, University of Jyväskylä, Jyväskylä, Finland
[37] NIKHEF Theory Group, Kruislaan 409, 1098 SJ Amsterdam, The Netherlands
[38] DSM/DAPNIA, CEA, Centre d’Etudes de Saclay, F 91191 Gif-sur-Yvette, France
[39] University of Genova and INFN-Genova, Italy
[40] Dept. of Physics, University of Florida, USA
[41] Dipartimento di Fisica, Universitá di Milano, INFN Sezione di Milano, Via Celoria 16, I 20133 Milan, Italy
[42] Institut für Theoretische Physik, Universität Karlsruhe, 76128 Karlsruhe, Germany
[43] Institut für Theoretische Physik, TU Dresden, D-01062 Dresden, Germany
[44] CERN, 1211 Geneva 23, Switzerland, and Institute of Nuclear Physics, ul. Radzikowskiego 152, 31-342 Kraków, Poland
[45] HEP Department, School of Physics and Astronomy, Raymond and Beverly Sackler Faculty of Exact Science, Tel Aviv University, Tel Aviv, 69978, Israel

[46] University of Torino and INFN-Torino, Italy; also at University of Wisconsin, Madison, WI, USA
[47] Dept. of Theoretical Physics, Lund University, Sweden
[48] University of Wisconsin, Madison, WI, USA
[49] Department of Physics and Astronomy, Michigan State University, E. Lansing, MI 48824, USA
[50] High Energy Physics, Uppsala University, Box 535, SE 75121 Uppsala, Sweden
[51] Institute of Nuclear Physics, Academy of Sciences, ul. Radzikowskiego 152, 31-342 Cracow, Poland and CERN, Department of Physics, Theory Unit, CH 1211 Geneva 23, Switzerland
[52] High Energy Physics Division, Department of Physical Sciences, University of Helsinki and Helsinki Institute of Physics, P.O. Box 64, FIN-00014, Finland
[53] JINR Dubna, Russia
[54] Faculty of Mathematics and Physics, University of Ljubljana, Jadranska 19, SI-1000, Slovenia Experimental Particle Physics Department, Jozef Stefan Institute, P.P. 3000, Jamova 39, SI-1000 Ljubljana, Slovenia
[55] Laboratoire de Physique Subatomique et de Cosmologie, Université Joseph Fourier/CNRS-IN2P3, 53 Avenue des Martyrs, 38026 Grenoble, France
[56] Brookhaven National Laboratory, Upton, NY 11973, USA
[57] University of Podgorica, Cetinjski put bb, CNG 81000 Podgorica, Serbia and Montenegro
[58] Institute of Nuclear and Particle Physics, Charles University, Prague, Czech Republic
[59] Departament d'Estructura i Constituents de la Matèria, Universitat de Barcelona, Diagonal 647, E 08028 Barcelona, Spain
[60] Kirchhoff-Institut für Physik, Universität Heidelberg, Im Neuenheimer Feld 227, 69120 Heidelberg, Germany
[61] D.V. Skobeltsyn Institute of Nuclear Physics, Moscow, Russia
[62] University of Connecticut, USA
[63] MPI-K Heidelberg, Germany and JINR Dubna, Russia
[64] Dipartimento di Fisica Teorica, Università di Torino and INFN Sezione di Torino, Via P. Giuria 1, I 10125 Torino, Italy
[65] Institut de Physique Théorique, Université Catholique de Louvain, Chemin du Cyclotron, 2, B-1348, Louvain-la-Neuve, Belgium
[66] INFN Torino, Italy
[67] School of Physics and Astronomy, University of Birmingham, B15 2TT, UK
[68] Imperial College, London, UK
[69] Institut de Physique Nucléaire, Université Catholique de Louvain, Louvain-la-Neuve, Belgium
[70] Department of Physics and Astronomy, Michigan State University, E. Lansing, MI 48824, USA
[71] University of Karlsruhe, EKP, Postfach 6980, D-76128 Karlsruhe, Germany
[72] Harish-Chandra Research Institute, Chhatnag Road, Jhunsi, Allahabad, India
[73] Fermi National Accelerator Laboratory, Batavia, IL 60126, USA
[74] Institute of Physics, Jagiellonian University, 30-059 Krakow, ul. Reymonta 4, Poland. Institute of Nuclear Physics PAN, 31-342 Krakow, ul. Radzikowskiego 152, Poland.
[75] Laboratoire de l'Accélérateur Linéaire, Université Paris-Sud, 91898 Orsay, France
[76] NIKHEF, Amsterdam, The Netherlands
[77] Petersburg Nuclear Physics Institute, Gatchina, St. Petersburg, Russia
[78] LPTHE, Universities of Paris VI and VII and CNRS, F 75005, Paris, France
[79] University of Hamburg, IExpPh, Luruper Chaussee 149, D-22761 Hamburg, Germany
[80] Southern Methodist University Dallas, 3215 Daniel Avenue, Dallas, TX 75275-0175, USA
[81] School of Physics & Astronomy, University of Manchester, UK and Theoretical Physics Group, CERN, Geneva, Switzerland
[82] C.N. Yang Institute for Theoretical Physics, Stony Brook University, Stony Brook, NY 11794, USA
[83] Lancaster University, Lancaster, UK
[84] Johannes-Gutenberg-Universität Mainz, D-55099 Mainz, Germany
[85] H. Niewodniczański Institute of Nuclear Physics, PL 31-342 Kraków, Poland

[86] II. Physikalisches Institut, Universität Giessen, Heinrich-Buff-Ring 16, D 35392 Giessen, Germany
[87] Universiteit Antwerpen, Antwerpen, Belgium
[88] University of Liverpool, UK
[89] Lawrence Berkeley National Laboratory, Berkeley, CA, USA and Physics Department, University of California, Davis, CA, USA
[90] Department of Physics, Baylor University, Waco, TX, USA
[91] Institut für Experimentelle Kernphysik, Universität Karlsruhe, Karlsruhe, Germany

# Contents

# Part I

# Plenary Presentations

# Instanton-Induced Processes
# An Overview

*F. Schrempp*
Deutsches Elektronen-Synchrotron DESY, Hamburg, Germany

**Abstract**

A first part of this review is devoted to a summary of our extensive studies of the discovery potential for instanton ($I$)-induced, deep-inelastic processes at HERA. Included are some key issues about $I$-perturbation theory, an exploitation of crucial lattice constraints and a status report about the recent $I$-search results by the HERA collaborations H1 and ZEUS in relation to our predictions. Next follows a brief outline of an ongoing project concerning a broad exploration of the discovery potential for hard instanton processes at the LHC. I then turn to an overview of our work on high-energy processes, involving larger-sized instantons. I shall mainly focus on the phenomenon of saturation at small Bjorken-$x$ from an instanton perspective. In such a framework, the saturation scale is associated with the conspicuous average instanton size, $\langle\rho\rangle \sim 0.5$ fm, as known from lattice simulations. A further main result is the intriguing identification of the "Colour Glass Condensate" with the QCD *sphaleron* state.

## 1 Setting the stage

*Instantons* represent a basic non-perturbative aspect of non-abelian gauge theories like QCD. They were theoretically discovered and first studied by Belavin *et al.* [1] and 't Hooft [2], about 30 years ago.

Due to their rich vacuum structure, QCD and similar theories include topologically non-trivial fluctuations of the gauge fields, which in general carry a conserved, integer topological charge $Q$. Instantons ($Q = +1$) and anti-instantons ($Q = -1$) represent the simplest building blocks of topologically non-trivial vacuum structure. They are explicit solutions of the euclidean field equations in four dimensions [1]. They are known to play an important rôle in the transition region between a partonic and a hadronic description of strong interactions [3]. Yet, despite substantial theoretical evidence for the importance of instantons in chiral symmetry breaking and hadron spectroscopy, their direct experimental verification is lacking until now.

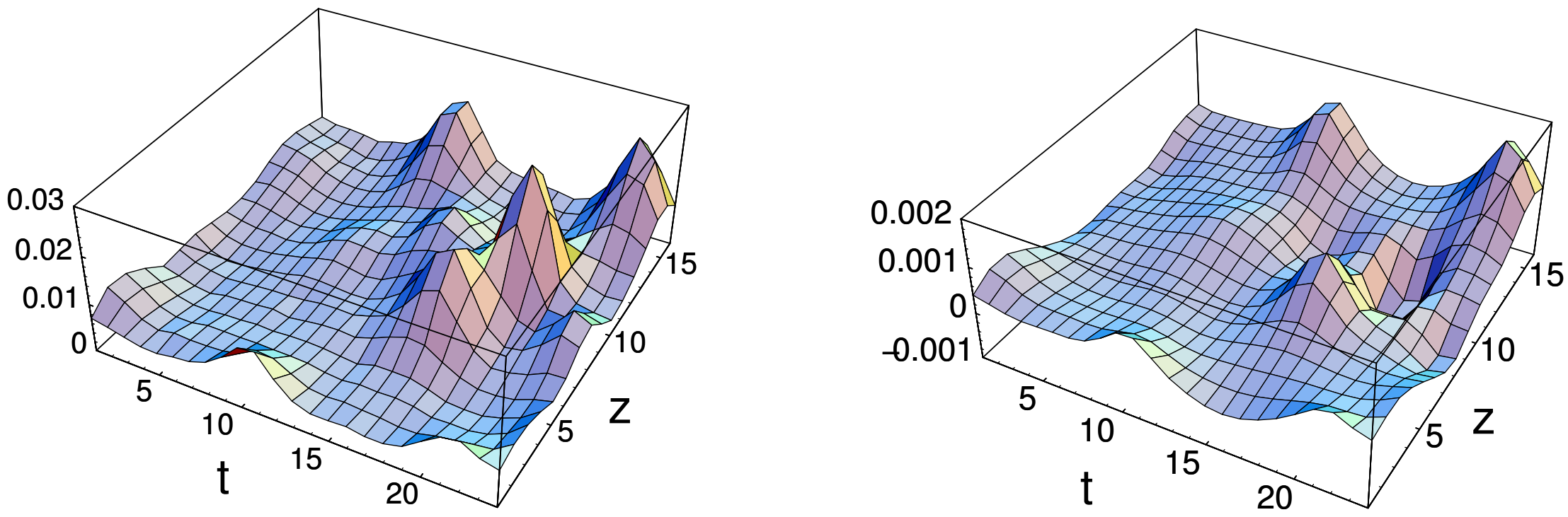

**Fig. 1:** Contribution from three instantons ($Q = +1$) and two anti-instantons ($Q = -1$) to the Lagrangian (left)) and the topological charge density (right) in a lattice simulation [4] (after cooling). The euclidean coordinates x and y are kept fixed while the dependence on z and t is displayed.

It turns out, however, that a characteristic *short distance* manifestation of instantons can be exploited [5] for an experimental search: Instantons induce certain (hard) processes that are forbidden in usual perturbative QCD. These involve all (light) quark flavours democratically along with a violation of chirality, in accord with the general chiral anomaly relation [2]. Based on this crucial observation, deep-inelastic scattering (DIS) at HERA has been shown to offer a unique opportunity [5] to discover such instanton-induced processes. It is of particular importance that a theoretical prediction of both the corresponding rate [6–8] and the characteristic event signature [5, 10–12] is possible in this hard scattering regime[1]. The instanton-induced cross section turns out to be in a measurable range [7, 10]. Crucial information on the region of validity for this important result, based on instanton-perturbation theory, comes from a high-quality lattice simulation [8, 13]. Another interesting possible spin-dependent signature of instantons in DIS, in form of a characteristic azimuthal spin asymmetry, has recently been discussed in Ref. [14].

In a first part (Sect. 2), I shall review our extensive investigations of deep-inelastic processes induced by small instantons. This includes a "flow-chart" of our calculations based on $I$-perturbation theory [6, 7], an exploitation of crucial lattice constraints [8, 13] and a confrontation [12] of the recent $I$-search results by the HERA collaborations H1 and ZEUS [15, 16] with our predictions. Next I shall briefly outline in Sect. 3 an ongoing project [17] to investigate theoretically and phenomenologically the discovery potential of hard instanton processes at the LHC. In Sect. 4, I then turn to an overview of our work [18–21] on high-energy processes involving larger-sized instantons. I shall focus mainly on the important theoretical challenge of the phenomenon of saturation at small Bjorken-$x$ from an instanton perspective. In such a framework we found [18–21] that the conspicuous average instanton size scale, $\langle\rho\rangle \sim 0.5$ fm, as known from lattice simulations [13], plays the rôle of the saturation scale. As a further main and intriguing result, we were led to associate the "Colour Glass Condensate" [22] with the QCD *sphaleron* state [23]. For another more recent approach to small-$x$ saturation in an instanton background with main emphasis on Wilson loop scattering and lacking direct lattice input, see Ref. [24]. The conclusions of this overview may be found in Sect. 5.

## 2 Small instantons in deep-inelastic scattering

### 2.1 Instanton-perturbation theory

Let us start by briefly summarizing the essence of our theoretical calculations [6, 7] based on so-called $I$-perturbation theory. As we shall see below, in an appropriate phase-space region of deep-inelastic scattering with generic hard scale $\mathcal{Q}$, the contributing $I$'s and $\overline{I}$'s have *small size* $\rho \lesssim \mathcal{O}(\frac{1}{\alpha_s(\mathcal{Q})\mathcal{Q}})$ and may be self-consistently considered as a *dilute* gas, with the small QCD coupling $\alpha_s(\mathcal{Q})$ being the expansion parameter like in usual perturbative QCD (pQCD). Unlike the familiar expansion about the trivial vacuum $A_\mu^{(0)} = 0$ in pQCD, in $I$-perturbation theory the path integral for the generating functional of the Green's functions in Euclidean position space is then expanded about the known, classical one-instanton solution, $A_\mu = A_\mu^{(I)}(x) + \ldots$. After Fourier transformation to momentum space, LSZ amputation and careful analytic continuation to Minkowski space (where the actual on-shell limits are taken), one obtains a corresponding set of modified Feynman rules for calculating $I$-induced scattering amplitudes. As a further prerequisite, the masses $m_q$ of the active quark flavours must be light on the scale of the inverse effective $I$-size $1/\rho_{\rm eff}$, i.e. $\mathrm{m}_q \cdot \rho_{\rm eff} \ll 1$. The leading, $I$-induced, chirality-violating process in the deep-inelastic regime of $e^{\pm}\mathrm{P}$ scattering is displayed in Fig. 2 (left) for $n_f = 3$ massless flavors. In the background of an $I$ ($\overline{I}$) (of topological charge $Q = +1$ $(-1)$), all $n_f$ massless quarks and anti-quarks are right (left)-handed such that the $I$-induced subprocess emphasized in the dotted box of Fig. 2 (left) involves a violation of chirality $Q_5 = \#\,(q_{\rm R} + \overline{q}_{\rm R}) - \#\,(q_{\rm L} + \overline{q}_{\rm L})$ by an amount,

$$\Delta Q_5 = 2\,n_f\,Q, \tag{1}$$

[1]For an exploratory calculation of the instanton contribution to the gluon-structure function, see Ref. [9].

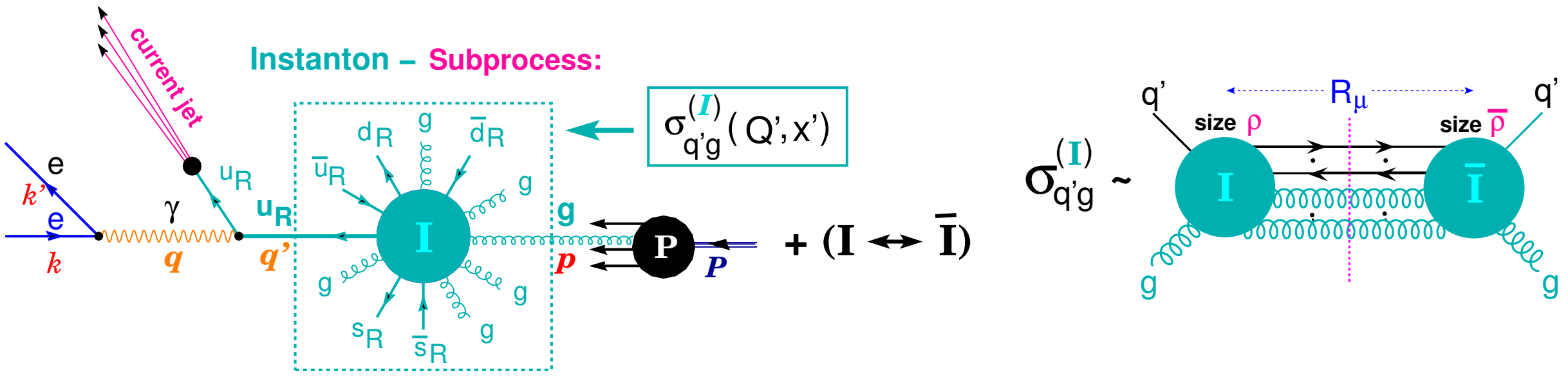

**Fig. 2:** (left): Leading, instanton-induced process in deep-inelastic $e^{\pm}$P scattering for $n_f = 3$ massless flavours. (right): Structure of the total cross section $\sigma^{(I)}_{q'\,g}$ for the chirality-violating "instanton-subprocess" $q'\,g \overset{(I)}{\Rightarrow} X$ according to the optical theorem. Note the illustration of the collective coordinates $\rho, \overline{\rho}$ and $R_\mu$.

in accord with the general chiral anomaly relation [2]. Within $I$-perturbation theory, one first of all derives the following factorized expression in the Bjorken limit of the $I$-subprocess variables $Q'^2$ and $x'$ (c. f. Fig. 2 (left)):

$$\frac{\mathrm{d}\sigma^{(I)}_{\mathrm{HERA}}}{\mathrm{d}x'\mathrm{d}Q'^2} \simeq \frac{\mathrm{d}\mathcal{L}^{(I)}_{q'g}}{\mathrm{d}x'\mathrm{d}Q'^2}\cdot\sigma^{(I)}_{q'g}(Q',x') \quad \text{for} \quad \begin{cases} Q'^2 = -q'^2 > 0 \text{ large}, \\ 0 \le x' = \frac{Q'^2}{2p\cdot q'} \le 1 \text{ fixed}. \end{cases} \tag{2}$$

In Eq. (2), the differential luminosity, $\mathrm{d}\mathcal{L}^{(I)}_{q'\,g}$ counts the number of $q'\,g$ collisions per $eP$ collisions. It is given in terms of integrals over the gluon density, the virtual photon flux, and the (known) flux of the virtual quark $q'$ in the instanton background [7].

The essential instanton dynamics resides, however, in the total cross-section of the $I$-subprocess $q'\,g \overset{I}{\Rightarrow} X$ (dotted box of Fig. 2 (left) and Fig. 2 (right)). Being an observable, $\sigma^{(I)}_{q'g}(Q',x')$ involves integrations over all $I$ and $\overline{I}$-"collective coordinates", i. e. the $I$ ($\overline{I}$) sizes $\rho$ ($\overline{\rho}$), the $I\overline{I}$ distance four-vector $R_\mu$ and the relative $I\overline{I}$ color orientation matrix $U$.

$$\sigma^{(I)}_{q'\,g} = \int d^4R\, \mathrm{e}^{\mathrm{i}\,(p+q')\cdot R} \int_0^\infty d\rho \int_0^\infty d\overline{\rho}\; \mathrm{e}^{-(\rho+\overline{\rho})Q'}\; D(\rho)\, D(\overline{\rho}) \int dU\, \mathrm{e}^{-\frac{4\pi}{\alpha_s}\Omega\left(U,\frac{R^2}{\rho\overline{\rho}},\frac{\overline{\rho}}{\rho}\right)} \{\ldots\} \tag{3}$$

Both instanton and anti-instanton degrees of freedom enter here, since the I-induced cross-section results from taking the modulus squared of an amplitude in the single $I$-background. Alternatively and more conveniently (c. f. Fig. 2 (right)), one may invoke the optical theorem to obtain the cross-section (3) in Minkowski space as a discontinuity of the $q'\,g$ forward elastic scattering amplitude in the $I\overline{I}$-background [7]. The $\{\ldots\}$ in Eq. (3) abreviates smooth contributions associated with the external partons etc. Let us concentrate on two crucial and strongly varying quantities of the $I$-calculus appearing in Eq. (3): $D(\rho)$, the (reduced) $I$-size distribution [2,28], and $\Omega\left(U, \frac{R^2}{\rho\overline{\rho}}, \frac{\overline{\rho}}{\rho}\right)$, the $I\overline{I}$ interaction, associated with a resummation of final-state gluons. Both objects are *known* within $I$-perturbation theory, formally for $\alpha_s(\mu_r)\ln(\mu_r\,\rho) \ll 1$ and $\frac{R^2}{\rho\overline{\rho}} \gg 1$ (diluteness), respectively, with $\mu_r$ being the renormalization scale. In the $I\overline{I}$-valley approach [25], the functional form of $\Omega^{I\overline{I}}_{\mathrm{valley}}$ is analytically known [26, 27] (formally) for *all* values of $R^2/(\rho\overline{\rho})$. The *actual* region of validity of the valley approach is an important issue to be addressed again later.

Most importantly, the resulting power-law behaviour for the $I$-size distribution,

$$D(\rho) \propto \rho^{\beta_0 - 5 + \mathcal{O}(\alpha_s)}, \tag{4}$$

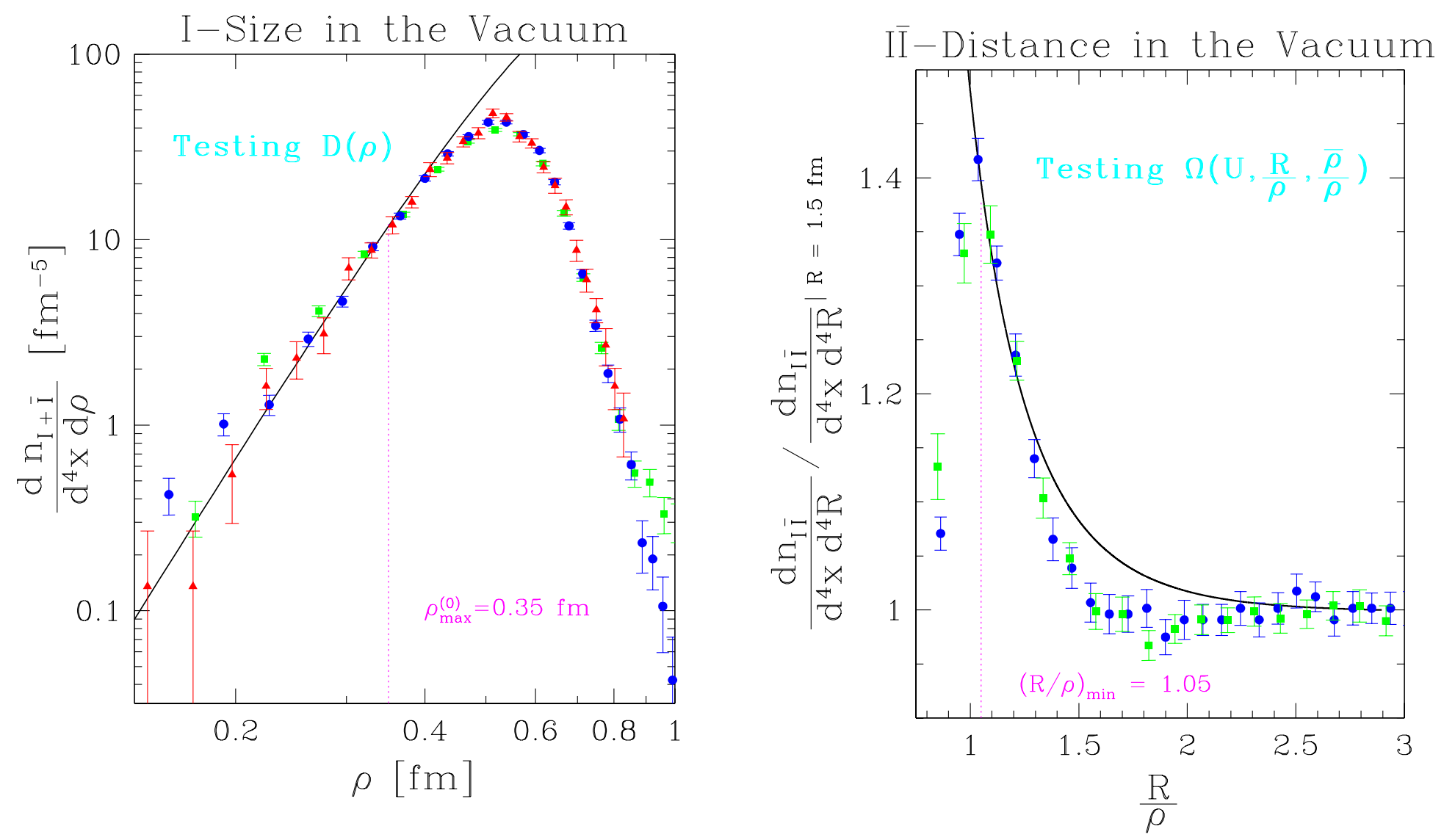

**Fig. 3:** Illustration of the agreement of recent high-quality lattice data [8,13] for the instanton-size distribution (left) and the normalized $I\overline{I}$-distance distribution (right) with the predictions from instanton-perturbation theory [8] for $\rho \lesssim 0.35$ fm and $R/\rho \gtrsim 1.05$, respectively. $\alpha_{\overline{\rm MS}}^{3-{\rm loop}}$ with $\Lambda_{\overline{\rm MS}}^{(n_f=0)}$ from the ALPHA collaboration [29] was used.

involving the leading QCD $\beta$-function coefficient, $\beta_0 = \frac{11}{3}\,N_c - \frac{2}{3}\,n_f$, $(N_c = 3)$, generically spoils the calculability of $I$-observables due to the bad IR-divergence of the integrations over the $I$ ($\overline{I}$)-sizes for large $\rho$ $(\overline{\rho})$. Deep-inelastic scattering represents, however, a crucial exception: The *exponential* "form factor" $\exp(-Q'(\rho+\overline{\rho}))$ that was shown [6] to arise in Eq. (3), insures convergence and *small* instantons for large enough $Q'$, despite the strong power-law growth of $D(\rho)$. This is the key feature, warranting the calculability of $I$-predictions for DIS.

It turns out that for (large) $Q' \neq 0$, all collective coordinate integrations in $\sigma_{q'g}^{(I)}$ of Eq. (3) may be performed in terms of a *unique saddle point*:

$$U^* \quad \Leftrightarrow \quad \text{most attractive relative } I\overline{I} \text{ orientation in color space,}$$

$$\rho^* \quad = \quad \overline{\rho}^* \sim \frac{4\pi}{\alpha_s(\frac{1}{\rho^*})}\,\frac{1}{Q'}; \quad \frac{R^{*\,2}}{\rho^{*\,2}} \overset{Q'\ \text{large}}{\sim} 4\frac{x'}{1-x'} \tag{5}$$

This result underligns the self-consistency of the approach, since for large $Q'$ and small $(1-x')$ the saddle point (5), indeed, corresponds to widely separated, small $I$'s and $\overline{I}$'s.

## 2.2 Crucial impact of lattice results

The $I$-size distribution $D(\rho)$ and the $I\overline{I}$ interaction $\Omega\left(U, \frac{R^2}{\rho\overline{\rho}}, \frac{\overline{\rho}}{\rho}\right)$ form a crucial link between deep-inelastic scattering and lattice observables in the QCD vacuum [8].

Lattice simulations, on the other hand, provide independent, non-perturbative information on the *actual* range of validity of the form predicted from $I$-perturbation theory for these important functions of $\rho$ and $R/\rho$, respectively. The one-to-one saddle-point correspondence (5) of the (effective) collective $I$-coordinates $(\rho^*, R^*/\rho^*)$ to $(Q', x')$ may then be exploited to arrive at a "fiducial" $(Q', x')$ region for our predictions in DIS. Let us briefly summarize the results of this strategy [8].

We have used the high-quality lattice data [8, 13] for quenched QCD ($n_f = 0$) by the UKQCD collaboration together with the careful, non-perturbative lattice determination of the respective QCD $\Lambda$-parameter, $\Lambda_{\overline{\rm MS}}^{(n_f=0)} = (238\pm19)$ MeV, by the ALPHA collaboration [29]. The results of an essentially parameter-free comparison of the continuum limit [8] for the simulated $(I+\overline{I})$-size and the $I\overline{I}$-distance distributions with $I$-perturbation theory versus $\rho$ and $R/\rho$, respectively, is displayed in Fig. 3. The UKQCD data for the $I\overline{I}$-distance distribution provide the first direct test of the $I\overline{I}$ interaction $\Omega\left(U, \frac{R^2}{\rho\overline{\rho}}, \frac{\overline{\rho}}{\rho}\right)$ from the $I\overline{I}$-valley approach via [8]

$$\frac{\mathrm{d}\, n_{I\overline{I}}}{\mathrm{d}^4x\, \mathrm{d}^4R}_{|\mathrm{UKQCD}} \overset{?}{\simeq} \int_0^\infty d\rho \int_0^\infty d\overline{\rho}\, D(\rho)\, D(\overline{\rho}) \int dU\, \mathrm{e}^{-\frac{4\pi}{\alpha_s}\Omega\left(U, \frac{R^2}{\rho\overline{\rho}}, \frac{\overline{\rho}}{\rho}\right)}, \tag{6}$$

and the lattice measurements of $D(\rho)$.

From Fig. 3, $I$-perturbation theory appears to be quantitatively valid for

$$\left.\begin{array}{lll} \rho \cdot \Lambda_{\overline{\mathrm{MS}}}^{(n_f=0)} & \lesssim & 0.42 \\ R/\rho & \gtrsim & 1.05 \end{array}\right\} \overset{\textbf{saddle point}}{\Rightarrow} \left\{\begin{array}{lll} Q'/\Lambda_{\overline{\mathrm{MS}}}^{(n_f)} & \gtrsim & 30.8, \\ x' & \gtrsim & 0.35, \end{array}\right. \tag{7}$$

Beyond providing a quantitative estimate for the "fiducial" momentum space region in DIS, the good, parameter-free agreement of the lattice data with $I$-perturbation theory is very interesting in its own right. Uncertainties associated with the inequalities (7) are studied in detail in Ref. [12].

### 2.3 Characteristic final-state signature

The qualitative origin of the characteristic final-state signature of $I$-induced events is intuitively explained and illustrated in Fig. 4. An indispensable tool for a quantitative investigation of the characteristic final-state signature and notably for actual experimental searches of $I$-induced events at HERA is our Monte-Carlo generator package QCDINS [10].

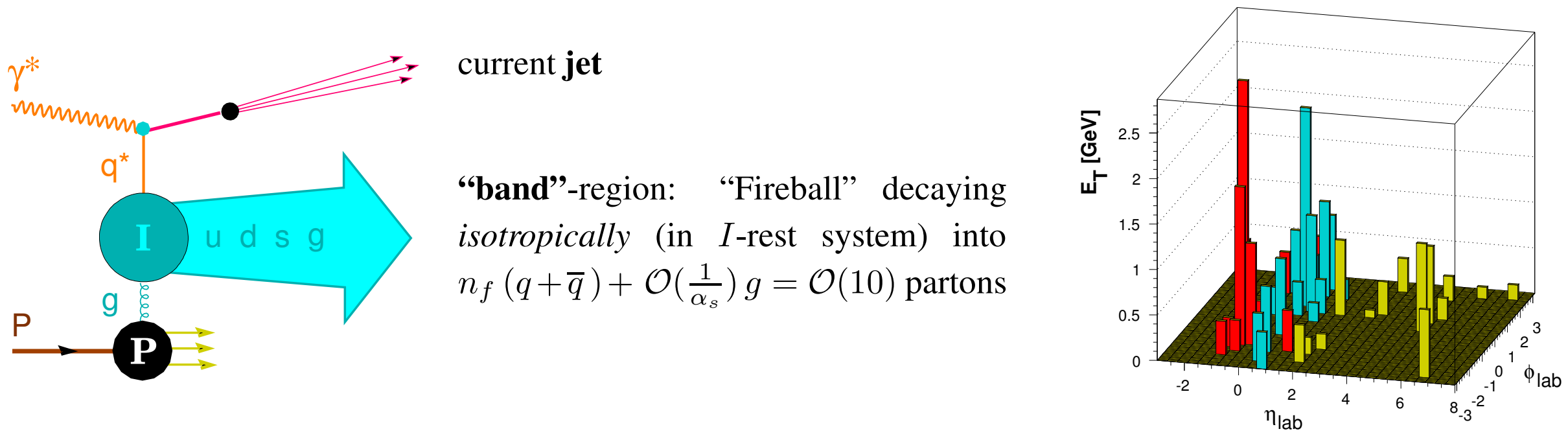

**Fig. 4:** Characteristic signature of $I$-induced events: *One* (current) *jet* along with a densely filled *band* of hadrons in the $(\eta, \phi)$ plane. Each event has large hadron multiplicity, large total $E_t$, u-d-s flavor democracy with 1 $s\overline{s}$-pair/event leading to $K's, \Lambda's\ldots$. An event from our QCDINS [10] generator (right) illustrates these features.

### 2.4 Status of searches at HERA

The results of dedicated searches for instanton-induced events by the H1 and ZEUS collaborations [15, 16], based on our theoretical work, have been finalized meanwhile. The H1 analysis was based on $\int \mathcal{L}dt \approx 21$ pb$^{-1}$, while ZEUS used $\int \mathcal{L}dt \approx 38$ pb$^{-1}$, with somewhat differing kinematical cuts. Since the upgraded HERA II machine is now performing very well, forthcoming searches based on a several times higher luminosity might turn out most interesting. Let me briefly summarize the present status from a theorist's perspective.

While H1 indeed observed a statistically significant excess of events with instanton-like topology and in good agreement with the theoretical predictions, *physical* significance could not be claimed, due to remaining uncertainties in the standard DIS (sDIS) background simulation. The ZEUS collaboration obtained a conservative, background-independent upper limit on the instanton-induced HERA cross section of 26 $\mathrm{pb@95\%~CL}$, to be compared to our theoretical prediction of 8.9 pb for the given cuts. In both experiments it was demonstrated that a decisive experimental test of the theoretical predictions based on I-perturbation theory is well within reach in the near-future. In view of the present situation and the interesting prospects for HERA II, let me proceed with a number of comments.

**Table 1:** Comparison of implemented fiducial cuts that are required in principle to warrant the validity of I-perturbation theory.

| Fiducial | Cuts | | **H1** | **ZEUS** |
|---|---|---|---|---|
| $Q^2$ | $\gtrsim$ | 113 GeV$^2$ ? | **no** | **yes** |
| $Q'^2$ | $\gtrsim$ | 113 GeV$^2$ ? | **yes** | **yes** |
| $x'$ | $\gtrsim$ | 0.35 ? | **no** | **no** |

A first important task consists in reconstructing the instanton-subprocess variables $(Q'^2, x')$ from Eq. (2) and in implementing the theoretically required fiducial cuts (cf. Eq. (7)). The actual status is displayed in Table 1 for comparison. The implications of the lacking $x'$-cut both in the H1 and ZEUS data are presumably not too serious, since QCDINS — *with* its default $x'$-cut — models to some extent the sharp suppression of $I$-effects, apparent in the lattice data (cf. Fig. 3 (right)) for $R/\rho \lesssim 1.0 - 1.05$, i.e. $x' \lesssim 0.3 - 0.35$. Yet, this lacking, experimental cut introduces a substantial uncertainty in the predicted magnitude of the $I$-signal that hopefully may be eliminated soon. The lacking $Q^2$-cut in the H1 data is potentially more serious. As a brief reminder [6, 10], this cut assures in particular the dominance of "planar" handbag-type graphs in $\sigma^{(I)}_{\rm HERA}$ and all final-state observables. Because of computational complications, the non-planar contributions are *not* implemented in the QCDINS event generator, corresponding to unreliable QCDINS results for small $Q^2$.

The main remaining challenge resides in the fairly large sDIS background uncertainties. The essential reason is that the existing Monte Carlo generators have been typically designed and tested for kinematical regions different from where the instanton signal is expected! Although the residual problematics is not primarily related to lacking statistics, the near-future availability of many more events will allow to strengthen the cuts and thus hopefully to increase the gap between signal and background. A common search strategy consists in producing I-enriched data samples by cutting on several discriminating observables, each one being sensitive to *different* basic instanton characteristics. An optimized set may be found according to the highest possible

$$\text{instanton separation power} = \frac{\epsilon_I}{\epsilon_{\rm sDIS}}, \tag{8}$$

in terms of the sDIS and instanton efficiencies, with $\epsilon_I \gtrsim 5 - 10\%$. Substantial enhancements of the instanton sensitivity were obtained, by means of various multivariate discrimination methods, involving only a single cut on a suitable discriminant variable. In case of ZEUS, cuts on the Fisher discriminant have been used to obtain instanton-enhanced subsamples.

Let me summarize the results obtained so far in form of a theorist's "unified plot" of the H1 and ZEUS "excess" versus the $I$-separation power. Any visible correlation of a rising *experimental* "excess" with the (Monte-Carlo) *theoretical* I-separation power in Fig. 5 would be an intriguing first signature for a signal. The behaviour seen from the end of the ZEUS range into the H1 domain, might indeed suggest some increase of the excess towards rising I-sparation power. The comparatively low $I$-separation power of the ZEUS data (and thus perhaps also their *negative* excess?) is mainly due to the implementation of the fiducial cut in $Q^2$ that is lacking in case of H1.

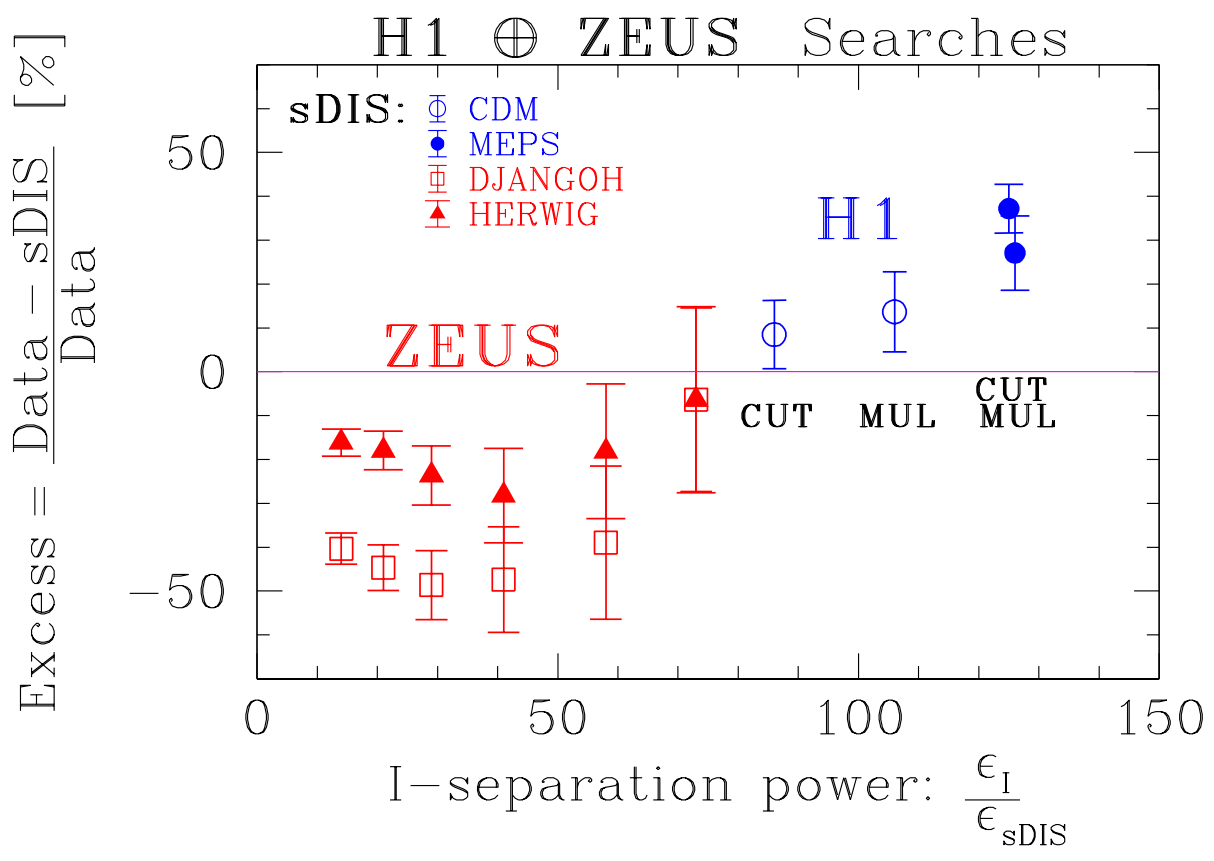

**Fig. 5:** A theorist's "unified plot" of the H1 and ZEUS "excess" versus $I$-separation power. The H1 and ZEUS data are seen to join smoothly. A first sign of a rising excess towards higher separation powers might be suspected.

## 3 Study of the discovery potential at the LHC

Given our extended experience with instanton physics both theoretically and experimentally at DESY, it is natural to ask about the discovery potential for instanton-induced processes at the forthcoming LHC. Indeed, a respective project has been set up around a theoretical PhD Thesis [17], but is still in a relatively early stage.

### 3.1 Outline of the project

We attempt to do a broad study, focussing both on theoretical and phenomenological issues. Let me just enumerate some interesting aspects that differ essentially from the familiar situation for spacelike hard scattering in DIS at HERA.

*Theoretically*: The first and foremost task is to identify and calculate the leading $I$-subprocess at the LHC within $I$-perturbation theory. Unlike HERA (Fig. 2 (left)), one starts from a $g\,g$-initial state at the LHC. Hence, the rate will be enhanced by a factor $\propto \frac{1}{\alpha_{\rm e.m.}\,\alpha_s}$ compared to $\gamma^*\,g$ scattering at HERA. Then, the next crucial question is how to enforce some parton virtuality in the respective instanton-induced $g\,g$-subprocess, such as to retain the applicability of $I$-perturbation theory.

An interesting possibility we are exploring is to enter the required virtuality through the *final state* in case of the LHC! One may consider the fragmentation of one or even two outgoing quarks from the $g\,g$-initiated $I$-instanton subprocess into a *large* $E_\perp$ photon or $W$-boson and other particles. The requirement of large $E_\perp$ then enforces a *timelike* virtuality onto the outgoing parent quark.

*Experimentally*: Crucial criteria will be a good signature paired with the lowest possible background, as well as a good trigger. At the experimental front we forsee the collaboration of T. Carli/CERN, who will be able to merge his actual knowledge of the LHC with many years of experience from searches for instantons at HERA. After the theoretical calculations are under control, the next task is to adapt our QCDINS event generator to the LHC, to work out characteristic event signatures, optimal observables, fiducial cuts etc.

## 4 Instanton-driven saturation at small $x$

One of the most important observations from HERA is the strong rise of the gluon distribution at small Bjorken-$x$ [30]. On the one hand, this rise is predicted by the DGLAP evolution equations [31] at high $Q^2$ and thus supports QCD [32]. On the other hand, an undamped rise will eventually violate unitarity. The reason for the latter problem is known to be buried in the linear nature of the DGLAP- and the BFKL-equations [33]: For decreasing Bjorken-$x$, the number of partons in the proton rises, while their effective size $\sim 1/Q$ increases with decreasing $Q^2$. At some characteristic scale $Q^2 \approx Q_s^2(x)$, the gluons in the proton start to overlap and so the linear approximation is no longer applicable; non-linear corrections to the linear evolution equations [34] arise and become significant, potentially taming the growth of the gluon distribution towards a "saturating" behaviour.

From a theoretical perspective, $eP$-scattering at small Bjorken-$x$ and decreasing $Q^2$ uncovers a novel regime of QCD, where the coupling $\alpha_s$ is (still) small, but the parton densities are so large that conventional perturbation theory ceases to be applicable, eventually. Much interest has recently been generated through association of the saturation phenomenon with a multiparticle quantum state of high occupation numbers, the "Colour Glass Condensate" that correspondingly, can be viewed [22] as a strong *classical* colour field $\propto 1/\sqrt{\alpha_s}$.

### 4.1 Why instantons?

Being extended non-perturbative fluctuations of the gluon field, instantons come to mind naturally in the context of saturation, since

- classical *non-perturbative* colour fields are physically appropriate in this regime; $I$-interactions always involve many non-perturbative gluons with multiplicity $\langle n_g \rangle \propto \frac{1}{\alpha_s}$!
- the functional form of the instanton gauge field is explicitely known and its strength is $A_\mu^{(I)} \propto \frac{1}{\sqrt{\alpha_s}}$ as needed;
- an identification of the "Colour Glass Condensate" with the QCD-sphaleron state appears very suggestive [20, 21] (cf. below and Sec 4.4).
- At high energies ($x \to 0$), larger $I$-sizes ($\rho \gtrsim 0.35$ fm) are probed! Unlike DIS, now the sharply defined average $I$-size $\langle \rho \rangle \approx 0.5$ fm (known from lattice simulations [13]) comes into play and becomes a relevant and conspicuous length scale in this regime (cf. Fig. 6 (left)).
- An intriguing observation is that the $I$-size scale $\langle \rho \rangle$ coincides surprisingly well with the transverse resolution $\Delta x_\perp \sim 1/Q$, where the small-$x$ rise of the structure function $F_2(x, Q^2)$ *abruptly* starts to *increase* with falling $\Delta x_\perp$! This striking feature[2] is illustrated in Fig. 6 (right), with the power $\lambda(Q)$ being defined via the ansatz $F_2(x, Q^2) = c(Q)\, x^{-\lambda(Q)}$ at small $x$. A suggestive interpretation is that instantons are getting resolved for $\Delta x_\perp \lesssim \langle \rho \rangle$.
- We know already from $I$-perturbation theory that the instanton contribution tends to strongly increase towards the softer regime [5,7,10]. The mechanism for the decreasing instanton suppression with increasing energy is known since a long time [35,36]: Feeding increasing energy into the scattering process makes the picture shift from one of tunneling between adjacent vacua ($E \approx 0$) to that of the actual creation of the sphaleron-like, coherent multi-gluon configuration [23] on top of the potential barrier of height [5, 37] $E = m_{\rm sph} \propto \frac{1}{\alpha_s \rho_{\rm eff.}}$.

### 4.2 From instanton-perturbation theory to saturation

The investigation of saturation becomes most transparent in the familiar colour-dipole picture [38] (cf. Fig. 7 (left)), notably if analyzed in the so-called dipole frame [39]. In this frame, most of the energy is still carried by the hadron, but the virtual photon is sufficiently energetic, to dissociate before scattering into a $q\overline{q}$-pair (a *colour dipole*), which then scatters off the hadron. Since the latter is Lorentz-contracted, the dipole sees it as a colour source of transverse extent, living (essentially) on the light cone. This colour field is created by the constituents of the well developed hadron wave function and – in view of its high intensity, i.e. large occupation numbers – can be considered as classical. Its strength near saturation is $\mathcal{O}(1/\sqrt{\alpha_s})$. At high energies, the lifetime of the $q\overline{q}$-dipole is much larger than the interaction time between this $q\overline{q}$-pair and the hadron and hence, at small $x_{\rm Bj}$, this gives rise to the familiar factorized expression of the inclusive photon-proton cross sections,

$$\sigma_{L,T}(x_{\rm Bj}, Q^2) = \int_0^1 dz \int d^2 r\, |\Psi_{L,T}(z,r)|^2\, \sigma_{\rm DP}(r,\ldots). \tag{9}$$

[2] I wish to thank A. Levy for the experimental data in Fig. 6 (right)

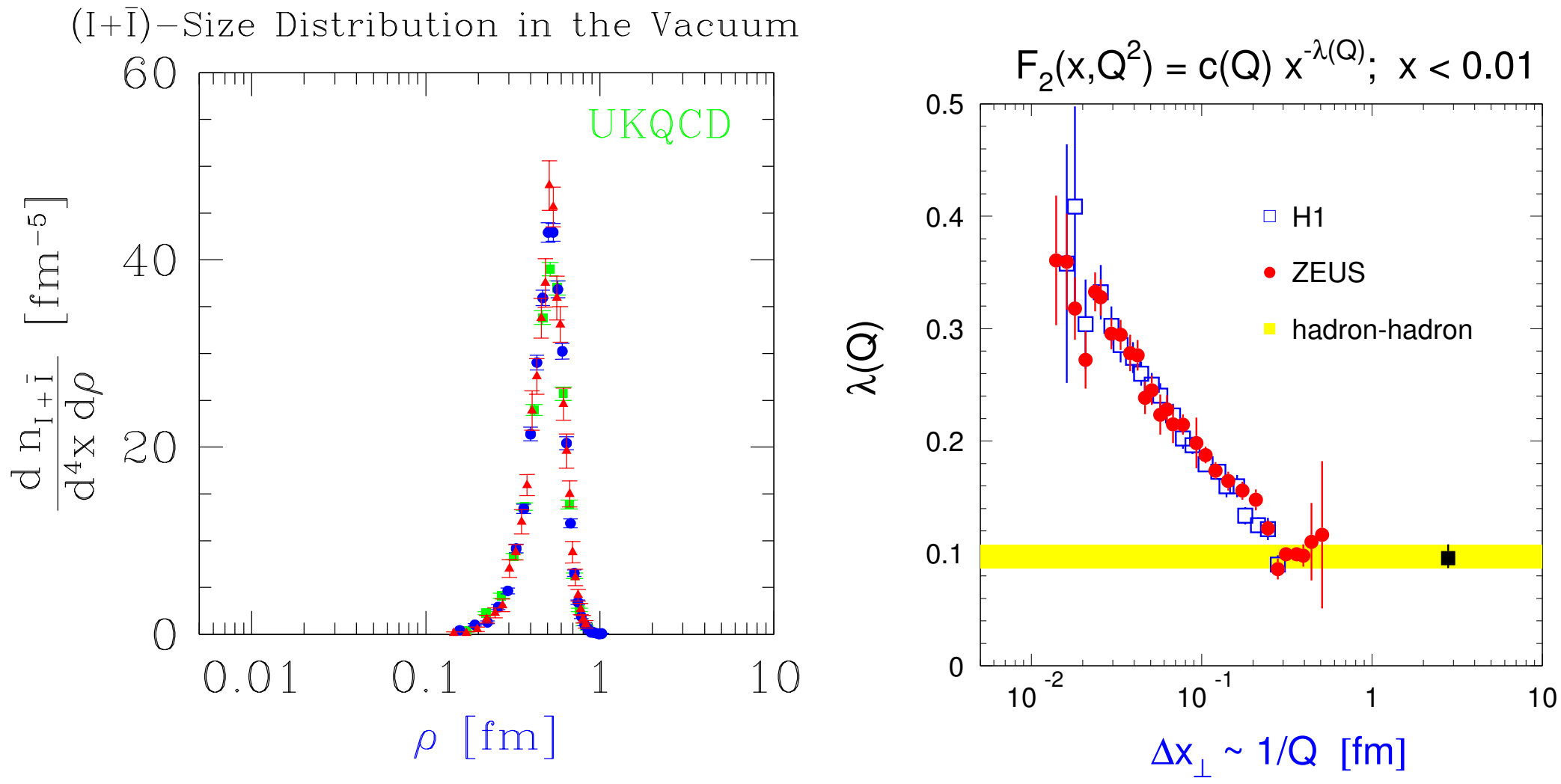

**Fig. 6:** The $I$-size scale $\langle\rho\rangle$ from lattice data [8,13] (left) coincides surprisingly well with the transverse resolution $\Delta x_\perp \sim 1/Q$, where the small-$x$ rise of the structure function $F_2(x, Q^2)$ *abruptly* starts to *increase* with falling $\Delta x_\perp$ (right).

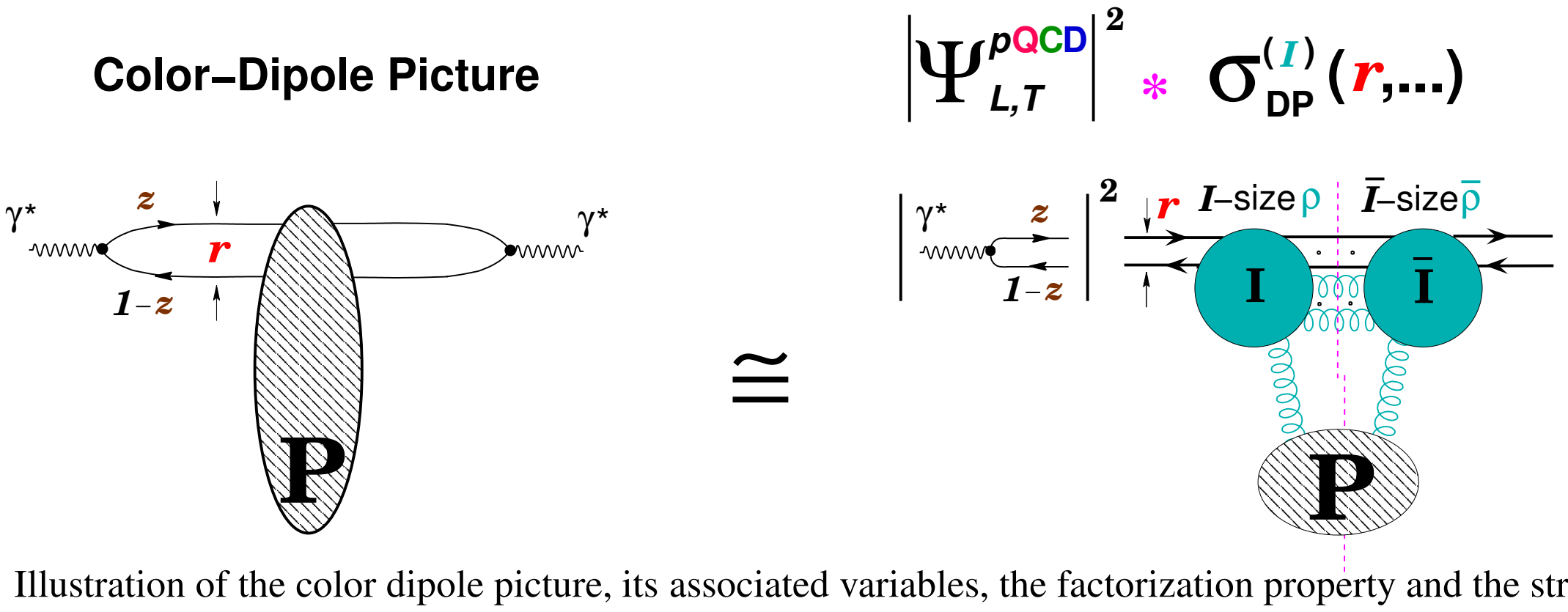

**Fig. 7:** Illustration of the color dipole picture, its associated variables, the factorization property and the structure of the dipole cross section in an instanton approach.

Here, $|\Psi_{L,T}(z,r)|^2$ denotes the modulus squared of the (light-cone) wave function of the virtual photon, calculable in pQCD, and $\sigma_{\rm DP}(r,\ldots)$ is the $q\overline{q}$-dipole - nucleon cross section. The variables in Eq. (9) are the transverse $(q\overline{q})$-size $\mathbf{r}$ and the photon's longitudinal momentum fraction $z$ carried by the quark. The dipole cross section is expected to include in general the main non-perturbative contributions. For small $r$, one finds within pQCD [38, 40] that $\sigma_{\rm DP}$ vanishes with the area $\pi r^2$ of the $q\overline{q}$-dipole. Besides this phenomenon of "colour transparency" for small $r = |r|$, the dipole cross section is expected to saturate towards a constant, once the $q\overline{q}$-separation $r$ exceeds a certain saturation scale $r_s$ (cf. Fig. 7 (right)). While there is no direct proof of the saturation phenomenon, successful models incorporating saturation do exist [41] and describe the data efficiently.

Let us outline more precisely our underlying strategy:

– We start from the large $Q^2$ regime and appropriate cuts such that $I$-perturbation theory is strictly

valid. The corresponding, known results on $I$-induced DIS processes [6] are then transformed into the colour-dipole picture.

– The guiding question is: Can background instantons of size $\sim \langle\rho\rangle$ give rise to a saturating, geometrical form for the dipole cross section,

$$\sigma_{\rm DP}^{(I)}(r,\ldots) \overset{r \gtrsim \langle\rho\rangle}{\sim} \pi\langle\rho\rangle^2. \tag{10}$$

– With the crucial help of lattice results, the $q\bar{q}$-dipole size $r$ is next carefully increased towards hadronic dimensions. Thanks to the lattice input, IR divergencies are removed and the original cuts are no longer necessary.

### 4.3 The simplest process: $\gamma^* + g \overset{(I)}{\rightarrow} q_{\rm R} + \overline{q}_{\rm R}$

Let us briefly consider first the simplest $I$-induced process, $\gamma^*\, g \Rightarrow q_{\rm R}\overline{q}_{\rm R}$, with one flavour and no final-state gluons. More details may be found in Ref. [20]. Already this simplest case illustrates transparently that in the presence of a background instanton, the dipole cross section indeed saturates with a saturation scale of the order of the average $I$-size $\langle\rho\rangle$.

We start by recalling the results for the total $\gamma^* N$ cross section within $I$-perturbation theory from Ref. [6],

$$\sigma_{L,T}(x_{\rm Bj}, Q^2) = \int\limits_{x_{\rm Bj}}^{1} \frac{dx}{x}\left(\frac{x_{\rm Bj}}{x}\right) G\left(\frac{x_{\rm Bj}}{x},\mu^2\right) \int dt \frac{d\hat{\sigma}_{L,T}^{\gamma^* g}(x,t,Q^2)}{dt}; \tag{11}$$

$$\frac{d\hat{\sigma}_L^{\gamma^* g}}{dt} = \frac{\pi^7}{2}\frac{e_q^2}{Q^2}\frac{\alpha_{\rm em}}{\alpha_{\rm s}}\left[x(1-x)\sqrt{tu}\,\frac{\mathcal{R}(\sqrt{-t})-\mathcal{R}(Q)}{t+Q^2} - (t\leftrightarrow u)\right]^2, \tag{12}$$

with a similar expression for $d\hat{\sigma}_T^{\gamma^*\, g}/dt$. Here, $G\left(x_{\rm Bj},\mu^2\right)$ denotes the gluon density and $L,T$ refers to longitudinal and transverse photons, respectively.

Note that Eqs. (11), (12) involve the resolution dependent length scale

$$\mathcal{R}(\mathcal{Q}) = \int_0^\infty d\rho\; D(\rho)\rho^5(\mathcal{Q}\rho){\rm K}_1(\mathcal{Q}\rho). \tag{13}$$

which is of key importance for continuing towards $\mathcal{Q}\langle\rho\rangle \Rightarrow 0$! For sufficiently large $\mathcal{Q}\langle\rho\rangle$, the crucial factor $(\mathcal{Q}\rho)\,K_1(\mathcal{Q}\rho) \sim e^{-\mathcal{Q}\rho}$ in Eq.(13) exponentially suppresses large size instantons and $I$-perturbation theory holds, as shown first in Ref. [6]. In our continuation task towards smaller $\mathcal{Q}\langle\rho\rangle$, crucial lattice information enters. We recall that the $I$-size distribution $D_{\rm lattice}(\rho)$, as *measured* on the lattice [8, 12, 13], is strongly peaked around an average $I$-size $\langle\rho\rangle \approx 0.5$ fm, while being in excellent agreement with $I$-perturbation theory for $\rho \lesssim 0.35$ fm (cf. Sect. 2.2 and Fig. 3(left)). Our strategy is thus to generally identify $D(\rho) = D_{\rm lattice}(\rho)$ in Eq.(13), whence

$$\mathcal{R}(0) = \int_0^\infty d\rho\; D_{\rm lattice}(\rho)\rho^5 \approx 0.3\ {\rm fm} \tag{14}$$

becomes finite and a $\mathcal{Q}^2$ cut is no longer necessary.

By means of an appropriate change of variables and a subsequent $2d$-Fourier transformation, Eqs. (11), (12) may indeed be cast [20] into a colour-dipole form (9), e.g. (with $\hat{Q} = \sqrt{z\,(1-z)}\,Q$)

$$\left(|\Psi_L|^2\,\sigma_{\rm DP}\right)^{(I)} \approx |\,\Psi_L^{\rm pQCD}(z,r)\,|^2\;\frac{1}{\alpha_{\rm s}}\,x_{\rm Bj}\,G(x_{\rm Bj},\mu^2)\,\frac{\pi^8}{12} \tag{15}$$

$$\times \left\{ \int_0^\infty d\rho D(\rho)\,\rho^5 \left( \frac{-\frac{d}{dr^2}\left(2r^2 \frac{\mathrm{K}_1(\hat{Q}\sqrt{r^2+\rho^2/z})}{\hat{Q}\sqrt{r^2+\rho^2/z}}\right)}{\mathrm{K}_0(\hat{Q}r)} - (z \leftrightarrow 1-z) \right) \right\}^2 .$$

The strong peaking of $D_{\rm lattice}(\rho)$ around $\rho \approx \langle\rho\rangle$, implies

$$\left(|\Psi_{L,T}|^2 \sigma_{\rm DP}\right)^{(I)} \Rightarrow \begin{cases} \mathcal{O}(1) \text{ but exponentially small}; & r \to 0, \\ \mid \Psi_{L,T}^{\rm pQCD} \mid^2 \frac{1}{\alpha_{\rm s}}\, x_{\rm Bj}\, G(x_{\rm Bj}, \mu^2)\, \frac{\pi^8}{12}\, \mathcal{R}(0)^2; & \frac{r}{\langle\rho\rangle} \gtrsim 1. \end{cases} \tag{16}$$

Hence, the association of the intrinsic instanton scale $\langle\rho\rangle$ with the saturation scale $r_s$ becomes apparent from Eqs. (15), (16): $\sigma_{\rm DP}^{(I)}(r,\ldots)$ rises strongly as function of $r$ around $r_s \approx \langle\rho\rangle$, and indeed *saturates* for $r/\langle\rho\rangle > 1$ towards a *constant geometrical limit*, proportional to the area $\pi\,\mathcal{R}(0)^2 = \pi\left(\int_0^\infty d\rho\, D_{\rm lattice}(\rho)\,\rho^5\right)^2$, subtended by the instanton. Since $\mathcal{R}(0)$ is divergent within $I$-perturbation theory, the information about $D(\rho)$ from the lattice (Fig. 6 (left)) is crucial for the finiteness of the result.

### 4.4 Identification of the color glass condensate with the QCD-sphaleron state

Next, let us consider the realistic process, $\gamma^* + g \stackrel{(I)}{\to} n_f\,(q_{\rm R} + \overline{q}_{\rm R}) +$ gluons. On the one hand, the inclusion of final-state gluons and $n_f > 1$ causes a significant complication. On the other hand, it is due to the effect of those gluons that the identification of the QCD-sphaleron state with the colour glass condensate has emerged [20, 21], while the qualitative "saturation" features remain unchanged. Most of the $I$-dynamics resides in the $I$-induced $q^*\,g$-subprocess with an incoming off-mass-shell quark $q^*$ originating from photon dissociation. The important kinematical variables are the $I$-subprocess energy $E = \sqrt{(q'+p)^2}$ and the quark virtuality $Q'^2 = -q'^2$, with the gluon 4-momentum denoted by $p_\mu$.

It is most convenient to account for the final-state gluons by means of the $I\bar{I}$-valley method [25] (cf. also Sect. 2.1). It allows to achieve via the optical theorem, an elegant summation over the gluons. The result leads to an exponentiation of the final-state gluon effects, residing entirely in the $I\bar{I}$-valley interaction $-1 \le \Omega^{I\bar{I}}_{\rm valley}(\frac{R^2}{\rho\bar{\rho}} + \frac{\rho}{\bar{\rho}} + \frac{\bar{\rho}}{\rho}; U) \le 0$, introduced in Eq. (3) of Sect. 2.1. Due to the new gluon degrees of freedom, the additional integrations over the $I\bar{I}$-distance $R_\mu$ appear (cf. Fig. 2 (right)), while the matrix $U$ characterizes the relative $I\bar{I}$ orientation in colour space. We remember from Sect. 2.1 that the functional form of $\Omega^{I\bar{I}}_{\rm valley}$ is analytically known [26, 27] (formally) for *all* values of $R^2/(\rho\bar{\rho})$. Our strategy here is identical to the one for the "simplest process" above: Starting point is the $\gamma^* N$ cross section, this time obtained by means of the $I\bar{I}$-valley method [7]. The next step is a variable and Fourier transformation into the colour-dipole picture. The dipole cross section $\tilde{\sigma}_{\rm DP}^{(I),\rm gluons}(\mathbf{l}^2, x_{\rm Bj}, \ldots)$ before the final 2d-Fourier transformation of the quark transverse momentum $\mathbf{l}$ to the conjugate dipole size $\mathbf{r}$, arises simply as an energy integral over the $I$-induced total $q^*g$ cross section in Eq. (3) from Ref. [7],

$$\tilde{\sigma}_{\rm DP}^{(I),\rm gluons} \approx \frac{x_{\rm Bj}}{2}\, G(x_{\rm Bj}, \mu^2) \int_0^{E_{\rm max}} \frac{d\,E}{E} \left[ \frac{E^4}{(E^2 + Q'^2)\,Q'^2}\, \sigma^{(I)}_{q^*g}\left(E, \mathbf{l}^2, \ldots\right) \right], \tag{17}$$

involving in turn integrations over the $I\bar{I}$-collective coordinates $\rho, \bar{\rho}, U$ and $R_\mu$.

In the softer regime of interest for saturation, we again substitute $D(\rho) = D_{\rm lattice}(\rho)$, which enforces $\rho \approx \bar{\rho} \approx \langle\rho\rangle$ in the respective $\rho, \bar{\rho}$-integrals, while the integral over the $I\bar{I}$-distance $R$ is dominated by a *saddle point*,

$$\frac{R}{\langle\rho\rangle} \approx \text{function}\left(\frac{E}{m_{\rm sph}}\right); \quad m_{\rm sph} \approx \frac{3\pi}{4} \frac{1}{\alpha_{\rm s}\,\langle\rho\rangle} = \mathcal{O}(\text{ few GeV }). \tag{18}$$

At this point, the mass $m_{\rm sph}$ of the QCD-sphaleron [5, 37], i.e the barrier height separating neighbouring topologically inequivalent vacua, enters as the scale for the energy $E$. The saddle-point dominance

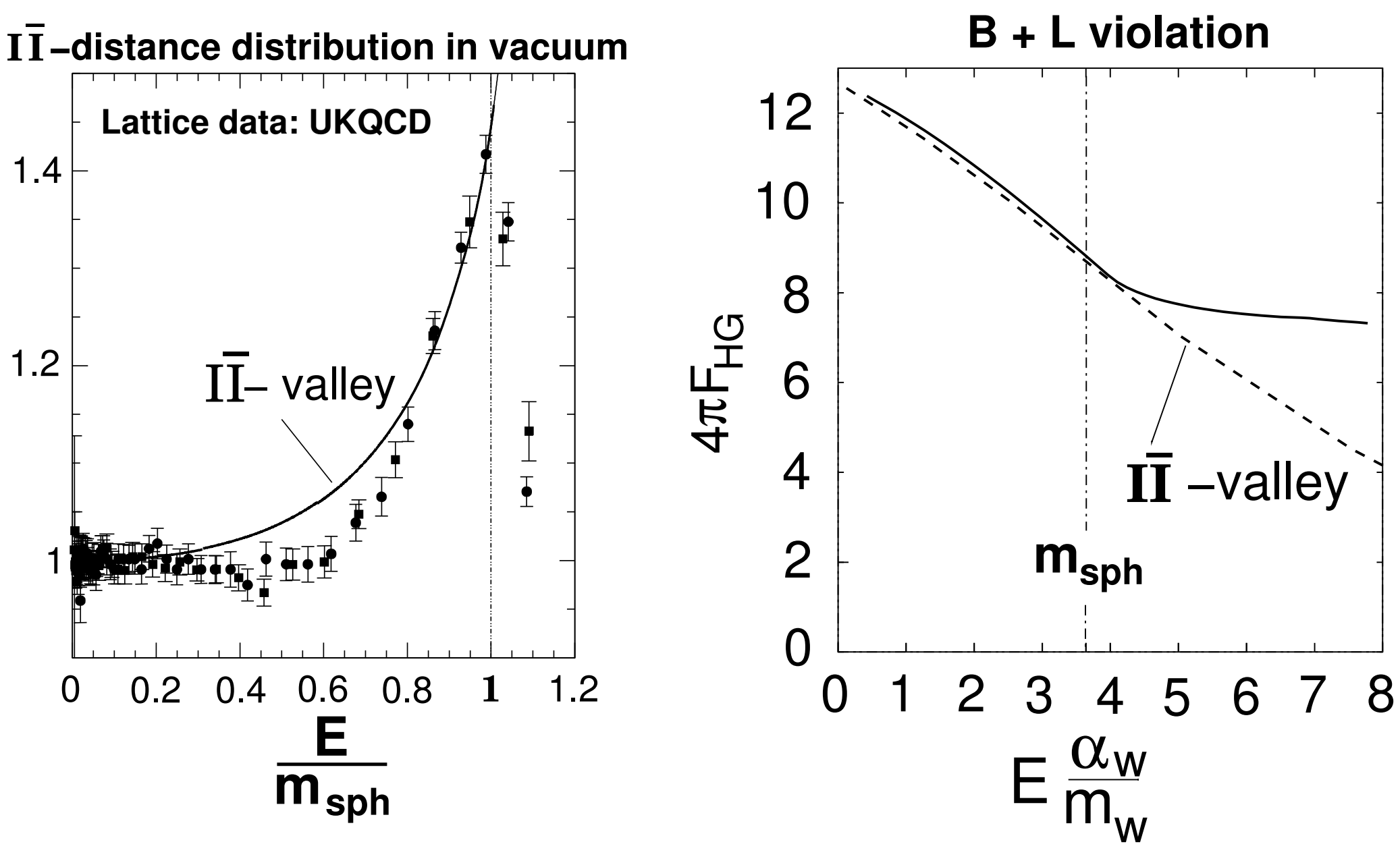

**Fig. 8:** (left) The UKQCD lattice data [8, 13] of the (normalized) $I\bar{I}$-distance distribution together with the corresponding $I\bar{I}$-valley prediction [20] from Fig. 3 (right) are re-displayed versus energy in units of the QCD sphaleron mass $m_{\rm sph}$.This illustrates the validity of the valley approach right until the sphaleron peak! (right) The same trend for electroweak $B+L$ -violation is apparent from an independent numerical simulation of the suppression exponent for two-particle collisions ('Holy Grail' function) $F_{\rm HG}(E)$ [42, 43]

implies a one-to-one relation,

$$\frac{R}{\langle\rho\rangle} \Leftrightarrow \frac{E}{m_{\rm sph}}; \quad \text{with } R = \langle\rho\rangle \Leftrightarrow E \approx m_{\rm sph}. \tag{19}$$

Our continuation to the saturation regime now involves crucial lattice information about $\Omega^{I\bar{I}}$. The relevant lattice observable is the distribution of the $I\bar{I}$-distance [8, 20] $R$, providing information on $\left\langle \exp[-\frac{4\pi}{\alpha_s}\Omega^{I\bar{I}}]\right\rangle_{U,\rho,\bar{\rho}}$ in euclidean space (cf. Fig. 3 (right)). Due to the crucial saddle-point relation Eqs. (18, 19), we may replace the original variable $R/\langle\rho\rangle$ by $E/m_{\rm sph}$. A comparison of the respective $I\bar{I}$-valley predictions with the UKQCD lattice data [8, 13, 20] versus $E/m_{\rm sph}$ is displayed in Fig. 8 (left). It reveals the important result that the $I\bar{I}$-valley approximation is quite reliable up to $E \approx m_{\rm sph}$. Beyond this point a marked disagreement rapidly develops: While the lattice data show a *sharp peak* at $E \approx m_{\rm sph}$, the valley prediction continues to rise indefinitely for $E \gtrsim m_{\rm sph}$! It is remarkable that an extensive recent and completely independent semiclassical numerical simulation [42] shows precisely the same trend for electroweak $B + L$-violation, as displayed in Fig. 8 (right).

It is again at hand to identify $\Omega^{I\bar{I}} = \Omega^{I\bar{I}}_{\rm lattice}$ for $E \gtrsim m_{\rm sph}$. Then the integral over the $I$-subprocess energy spectrum (17) in the dipole cross section appears to be dominated by the sphaleron configuration at $E \approx m_{\rm sph}$. The feature of saturation analogously to the "simplest process" in Sect. 4.3 then implies the announced identification of the colour glass condensate with the QCD-sphaleron state.

## 5 Conclusions

As non-perturbative, topological fluctuations of the gluon fields, *instantons* are a basic aspect of QCD. Hence their experimental discovery through hard instanton-induced processes would be of fundamental significance. A first purpose of this overview was to present a summary of our systematic theoretical and phenomenological investigations of the discovery potential in DIS at HERA, based on a calculable rate of measurable range and a characteristic "fireball"-like event signature. In a summary of the present status of experimental searches by H1 and ZEUS, the typical remaining challenges were particularly emphasized. In view of the good performance of the upgraded HERA II machine, one may expect further possibly decisive instanton search results in the near future.The existing H1 and ZEUS results have demonstrated already that the required sensitivity according to our theoretical predictions is within reach. Looking ahead, I have briefly discussed an ongoing project concerning a broad investigation of the discovery potential of instanton processes at the LHC. A final part of this review was devoted to our work on small-$x$ saturation from an instanton perspective. After summarizing the considerable motivation for the relevance of instantons in this regime, the emerging intuitive, geometrical picture was illustrated with the simplest example, where indeed, saturation does occur. The form of the dipole cross section depends on the relation of two competing areas: the area $\pi\, r^2$, subtended by the $\bar{q}q$-dipole, and the area $\pi\, \langle\rho\rangle^2$ associated with the average size, $\langle\rho\rangle \approx 0.5$ fm, of the background instanton. For $r/\langle\rho\rangle \ll 1$, the dipole cross section is dominated by the dipole area, corresponding to 'color transparency'. For $r/\langle\rho\rangle \gtrsim 1$ it saturates towards a constant proportional to the background instanton area. Correspondingly, the average $I$-size scale $\langle\rho\rangle$ is associated with the saturation scale. A further central and intriguing result concerned the identification of the Color Glass Condensate with the QCD-sphaleron state. Throughout, the non-perturbative information from lattice simulations was instrumental.

# Heavy quark production at HERA and the LHC

*Matthew Wing*
University College London and DESY

**Abstract**
Measurements of heavy quark production, particularly from HERA, their theoretical understanding and their relevance for the LHC are reviewed[1]. The status of beauty and charm production is discussed in the context of the different components of the production process: the parton density function of the colliding hadrons; the hard scatter; and the fragmentation of the quarks into hadrons. The theory of QCD at next-to-leading order generally describes well the hadronic structure and the production of heavy quarks although sometimes fails in details which are highlighted. The fragmentation of heavy quarks measured at HERA is consistent with that at LEP and hence supports the notion of universality.

## 1 Why study heavy quark production?

The measurement of heavy quarks can give insights into many physical phenomena such as: new particles which are expected to decay predominantly to beauty (and charm); precise measurements of electroweak parameters; and, the subject of this paper, a deeper understanding of the strong force of nature. The strong force as described within perturbative Quantum Chromodynamics (QCD) should be able to give a precise description of heavy quark production. This postulate is described and tested here. The measurement of heavy quark production also yields valuable information on the structure of colliding hadrons. The production of a pair of heavy quarks in a generic hadron collision is shown in Fig. 1 where it can be seen that the process is directly sensitive to the gluon content of the hadron. Most information on the structure of a hadron comes from inclusive deep inelastic scattering where the gluon content is determined in the evolution of the QCD equations. Therefore measurement of such a process in Fig. 1 provides complimentary information to that from inclusive measurements.

As well as understanding for its own sake, knowledge of the structure of hadrons will be important at future colliders such as the LHC and International Linear Collider where hadronic photons will have large cross section in both $e^+e^-$ and $\gamma\gamma$ modes. Heavy quarks will be copiously produced at future colliders as a background to the more exotic processes expected. Therefore a precise description of their production properties within QCD will aid in the discovery of physics beyond the Standard Model. An example of this was studied by the ATLAS collaboration using Monte Carlo to simulate the production at the LHC of a $b\bar{b}$ pair along with a supersymmetric Higgs particle ($H/A$) which subsequently decays to a $b\bar{b}$ pair [1]. For an assumed mass $m_A = 500$ GeV, even requiring four beauty jets, a signal-to-background ratio of only a few percent would be achieved. The irreducible background arises from QCD processes where the dominant processes are $gg$ and $gb$ with a gluon splitting to a $b\bar{b}$ pair. A discovery in this channel would therefore only be possible with precise knowledge these QCD background processes.

## 2 Theoretical and phenomenological overview

For a generic collision between two hadrons, $H_a$ and $H_b$, in which a heavy quark pair is produced (see Fig. 1),

$$H_a + H_b \rightarrow Q\bar{Q} + X,$$

[1]Since the presentation, some results have been updated; these are used in what follows.

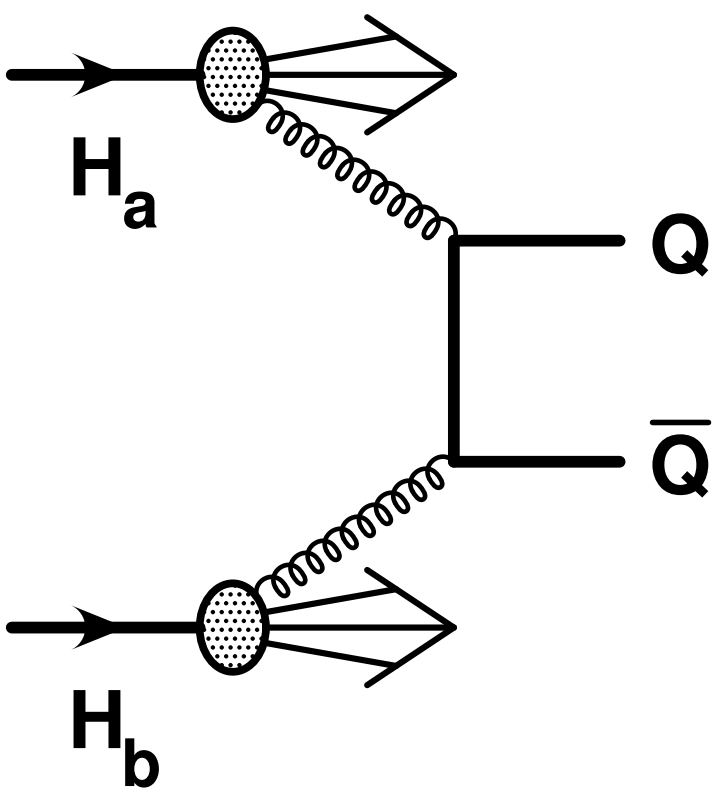

**Fig. 1:** Example of the production of a heavy quark pair in the collision of two hadrons.

the production cross section, $\sigma(S)$, for such a reaction at a centre-of-mass energy, $S$, can be written as:

$$\sigma(S) = \sum_{i,j} \int dx_1 \int dx_2 \; \hat{\sigma}_{ij}(x_1 x_2 S, m^2, \mu^2) f_i^{H_a}(x_1, \mu) f_j^{H_b}(x_2, \mu),$$

where the right-hand side is a convolution of the parton densities in the colliding hadrons, $f_i^{H_a}$ and $f_j^{H_b}$, and the short-distance cross section, $\hat{\sigma}_{ij}$. These are evaluated at a renormalisation and factorisation scale, $\mu$, and momentum fractions of the colliding partons, $x_1$ and $x_2$. The parton densities are extracted from QCD fits to inclusive deep inelastic scattering and other data. The short-distance cross section is calculable in QCD and is a perturbative expansion in the mass of the heavy quark, $m$:

$$\hat{\sigma}_{ij}(s, m^2, \mu^2) = \frac{\alpha_s^2(\mu^2)}{m^2} \left[ f_{ij}^{(0)}(\rho) + 4\pi\alpha_s(\mu^2) \left[ f_{ij}^{(1)}(\rho) + \bar{f}_{ij}^{(1)}(\rho) \log(\mu^2/m^2) \right] + \mathcal{O}(\alpha_s^2) \right], \quad \rho = 4m^2/s.$$

The expansion demonstrates that the larger the mass the faster the convergence. Hence predictions for beauty production should be more accurate than those for charm.

The treatment of the mass of the heavy quark is an important consideration for the implementation of the perturbative formalism in calculations. There are three schemes used: the fixed-order (FO) or "massive" scheme, the resummed to next-to-leading logarithms (NLL), or "massless" scheme and more recently a scheme matching the two, known as FONLL [2]. In the FO scheme, the predictions should be valid for transverse momenta of the order of the mass of the heavy quark. In this scheme, the heavy quarks are not active flavours in the parton distributions of the incoming hadron(s); they are produced in the hard scatter through processes such as $gg \to Q\bar{Q}$ shown in Fig. 1. The resummed scheme is valid for transverse momenta much larger than the heavy quark mass. The heavy quarks are active flavours in the parton distributions of the incoming hadron(s), so can be produced by reactions such as $gQ \to gQ$. The FONLL calculations match the two schemes and are valid for all transverse momenta. The validity of the different calculations is investigated in comparison with data, particularly as a function the energy scale.

The fixed-order calculations used are from Frixione et al. (FMNR) [3] for photoproduction processes and HVQDIS from Harris and Smith [4] for deep inelastic scattering. Resummed calculations are only available for photoproduction at HERA from two groups of authors, Cacciari et al. [5] and Kniehl et al. [6]. The FONLL calculation is also only available in photoproduction. A calculation which is already available for some processes in $pp$ collisions, MC@NLO [7], combines a fixed-order calculation with the parton showering and hadronisation from the HERWIG Monte Carlo generator [8]. Processes at HERA are not yet included, but it is hoped they will be done in the future and thereby provide a new level of detail in comparison with experimental data.

The advantages of a programme such as MC@NLO are its simulation of higher orders and also its sophisticated approach to hadronisation which attempts to describe the whole of the final state. The other programmes produce partons in the final state and fragment the outgoing quark to a hadron usually via the Peterson function [9]. Therefore these calculations may not be able to describe the full hadronic final state of an event. The validity of the fragmentation functions used also needs to be tested; they are usually extracted from fits to $e^+e^-$ data and their applicability to $ep$ or $pp$ needs to be demonstrated. Therefore the fragmentation function should be measured at HERA, and is discussed later, or measurements need to be made at high transverse energy or using jets where the effects of fragmentation are reduced.

Hadron-hadron collisions producing heavy quarks pairs can be simplified to and provide information on: the parton densities and in particular the gluon and heavy quark content of the hadron; the hard scatter and the dynamics of QCD as implemented into programmes; fragmentation or description of the parton to hadron transition. All of these aspects are discussed in this write-up.

## 3 Information needed by the LHC experiments

The information needed by the LHC which can be provided by the HERA experiments is the following:

- the state of the description of heavy quark production data by theoretical predictions. The production of heavy quarks in the hard scattering process is discussed here in detail. Information on heavy quarks produced in the splitting of a gluon outgoing from the hard sub-process is also important for the LHC, but the information from HERA is currently limited;
- the gluon and heavy quark content of the proton parton density functions;
- details of fragmentation in a hadronic environment;
- the effect of the underlying event in heavy quark processes. This information is limited at HERA but may be studied in the future;
- HERA results can provide general information on event and jet topologies which will be useful for designing algorithms or triggers at the LHC experiments.

The designing of effective triggers for $b$ physics is particularly acute for the LHCb experiment [10]. Large backgrounds are expected although event topologies should be different to the signal $b$ physics. For example minimum bias events will have a smaller track multiplicity and a lower transverse momentum for the highest $p_T$ track. Therefore using Monte Carlo simulation, cuts can be found to be able to reduce the rate of minimum bias whilst triggering efficiently on $b$ events. Such simulations require reliable Monte Carlo simulation of the event topologies of both classes of events.

Measurements of the proton structure function at HERA will constrain the parton densities in a large region of the kinematic plane where $B$ mesons will be produced within the acceptance of the LHCb detector. According to Monte Carlo simulations, these events are produced predominantly with a $b$ quark in the proton. However, this is just a model (PYTHIA [11]) and at NLO some of the events will be summed into the gluon distribution of the proton. Nevertheless, measuring all flavours in the proton at HERA is one of the goals of the experiments and recent results on the beauty contribution to the proton structure function [12] shed some light on the issue.

## 4 Open beauty production

The production of open beauty and its description by QCD has been of great interest in the last 10–15 years.The difference between the rates observed by the Tevatron experiments [13] and NLO QCD predictions led to a mini crisis with many explanations put forward. Several measurements were performed in different decay channels and then extrapolated to the quark level to facilitate a comparison with QCD and between themselves. The NLO QCD prediction was found to be a factor of 2–3 below the data for all measurements as shown in Fig. 2a. As mentioned, these results were extrapolated to the $b$-quark level using Monte Carlo models which may or may not give a good estimate of this extrapolation. To facilitate a particular comparison, an extrapolation can be useful, but should always be treated with caution and the procedure clearly stated and values of extrapolation factors given. Initial measurements in terms of measured quantities should also always be given.

The CDF collaboration also published measurements of $B$ meson cross sections. They were also found to be significantly above NLO calculations, but allowed for phenomenological study. Work on the fragmentation function was performed by Cacciari and Nason [14] which in combination with updated parton density functions and the FONLL calculation gave an increased prediction. New measurements at Run II have also been made by the CDF collaboration which probe down to very low transverse momenta. In combination with a measured cross section lower (but consistent) than the Run I data, and the above theoretical improvements, the data and theory now agree very well as shown in Fig. 2b. The programme MC@NLO also gives a good description of the data.

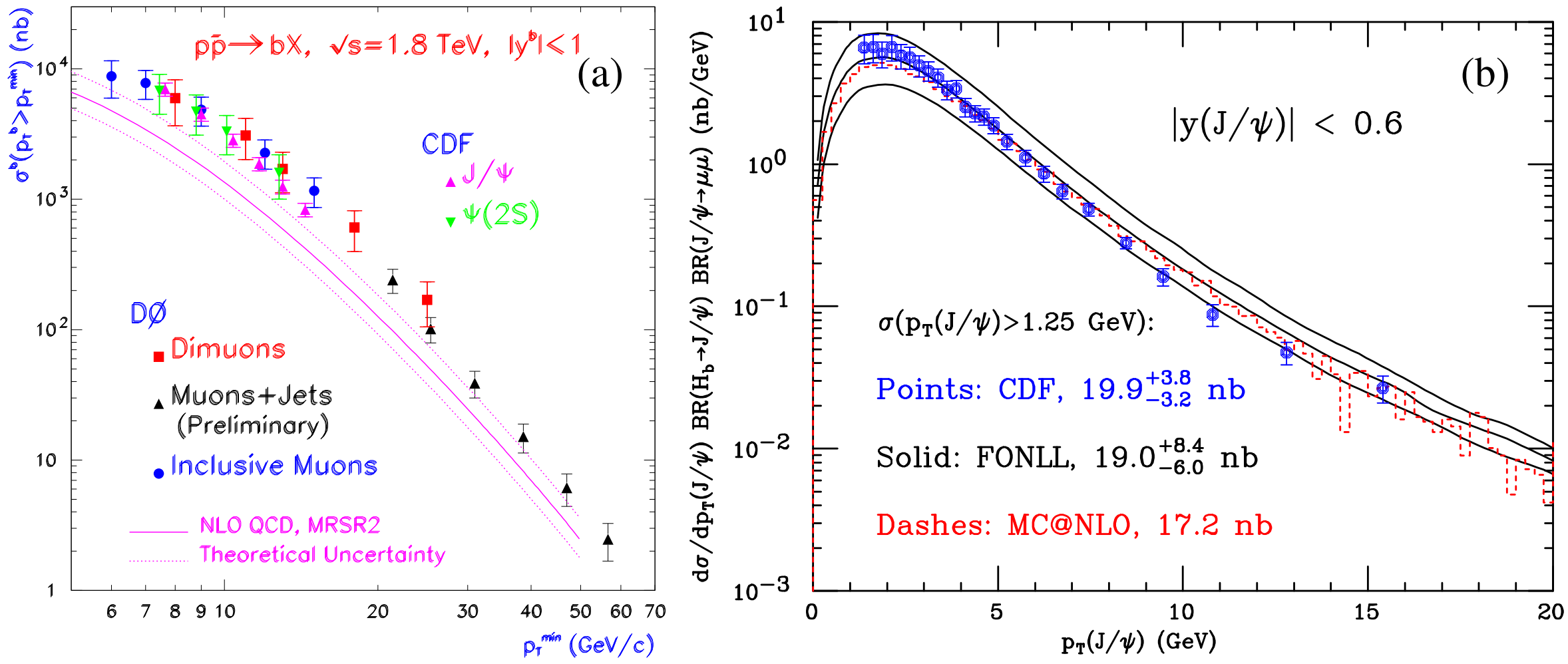

**Fig. 2:** (a) Tevatron Run I data extrapolated and compared to NLO QCD predictions and (b) Run II data presented in terms of the measured quantities and compared to improved QCD theory.

The first result from HERA [15] also revealed a large discrepancy with NLO QCD predictions. This analysis also presented an extrapolated quantity, whereas later measurements [16–18] also presented measured quantities. The most recent and precise measurements [17] of beauty production with accompanying jets are shown in Fig. 3 compared with predictions from NLO QCD. The measurements in photoproduction (Fig. 3a) are shown to be very well described by the prediction and the data from the two collaborations also agree well. The H1 data is somewhat higher than that from ZEUS; the difference is concentrated at low $p_T^\mu$ where the H1 data is also above the NLO calculation. The measurements in deep inelastic scattering are also generally described by NLO QCD although some differences at forward $\eta^\mu$ and low $p_T^\mu$ are observed by both collaborations. However, inclusive measurements which lead to a measurement of the beauty contribution to the proton structure function [12] are well described by QCD (see next Section).

The situation for the QCD description of $b$ production has recently changed significantly. In general, QCD provides a good description of the data with some hints at differences in specific regions. Certainly, there is no longer a difference of a factor of 2–3 independent of $p_T$. The HERA experiments will produce several new measurements in the next few years of higher precision and covering a larger kinematic region at both low and high $p_T$ and forward $\eta$. Allied with expected calculational and phenomenological improvements, a deep understanding of beauty production should be achieved by the turn-on of the LHC.

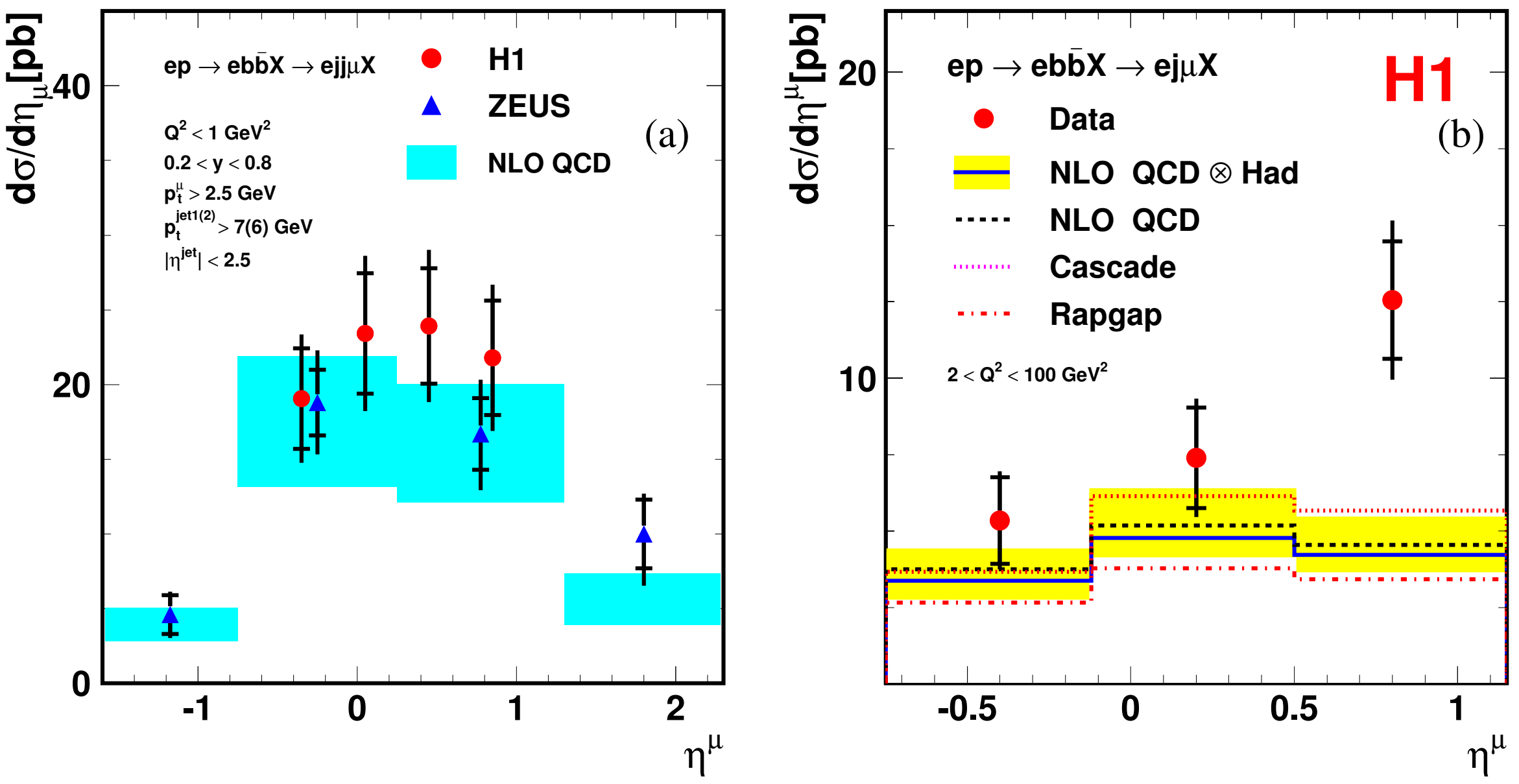

**Fig. 3:** Measurement of open beauty production as a function of the pseudorapidity of the decay muon for (a) dijet photoproduction from the H1 and ZEUS experiments and (b) inclusive jet deep inelastic scattering from the H1 experiment. (The measurement from ZEUS experiment for (b) is in a different kinematic region but reveals the same physics message and so for brevity, is not shown)

## 5 Open charm production

Due to its smaller mass, predictions for charm production are less accurate than for beauty. However large data samples allow detailed comparisons with theory. An example of a measured $D^*$ cross section in deep inelastic scattering is shown in Fig. 4a; data from the two experiments agree with each other and are well described by the prediction of QCD. Similar measurements have been made in photoproduction in which the data is less well described. Due to the larger cross section, the photoproduction data could prove valuable in constraining the photon as well as the proton structure. However, as can be seen from Fig. 4b, the theoretical precision is lagging well behind that of the data. Therefore more exclusive quantities and regions, with smaller theoretical uncertainties, are measured.

Measurements of charm photoproduction accompanied with jets pose a challenge for theory due the extra scale of the jet transverse energy. Such complicated final states will be copious at the LHC, so the verification of theory to HERA data will aid in the understanding of these high-rate QCD events. Dijet correlations in photoproduction have recently been measured [19] and compared with available calculations. Events were selected in two regions: one enriched in direct photon events where the photon acts as a pointlike object and one enriched in resolved photon events where the photon acts as a source of partons. The cross section of the difference in the azimuthal angle, $\Delta\phi^{\rm jj}$, of the two highest $E_T$ jets has been measured. For the LO $2 \to 2$ process, the two jets are back-to-back. The data exhibit a significant cross section at low $\Delta\phi^{\rm jj}$ and for the direct photon events are reasonably well described by NLO QCD (not shown). However, the description for resolved photon events is poor as shown in Fig. 5a. This region is particularly sensitive to higher orders not present in the QCD calculation. Monte Carlo models are compared to the data in Fig. 5b; although the normalisation is poor, the shape of the distribution is very well described by the HERWIG simulation. This indicates that for the precise description of such processes, higher-order calculations or the implementation of additional parton showers in current NLO calculations are needed.

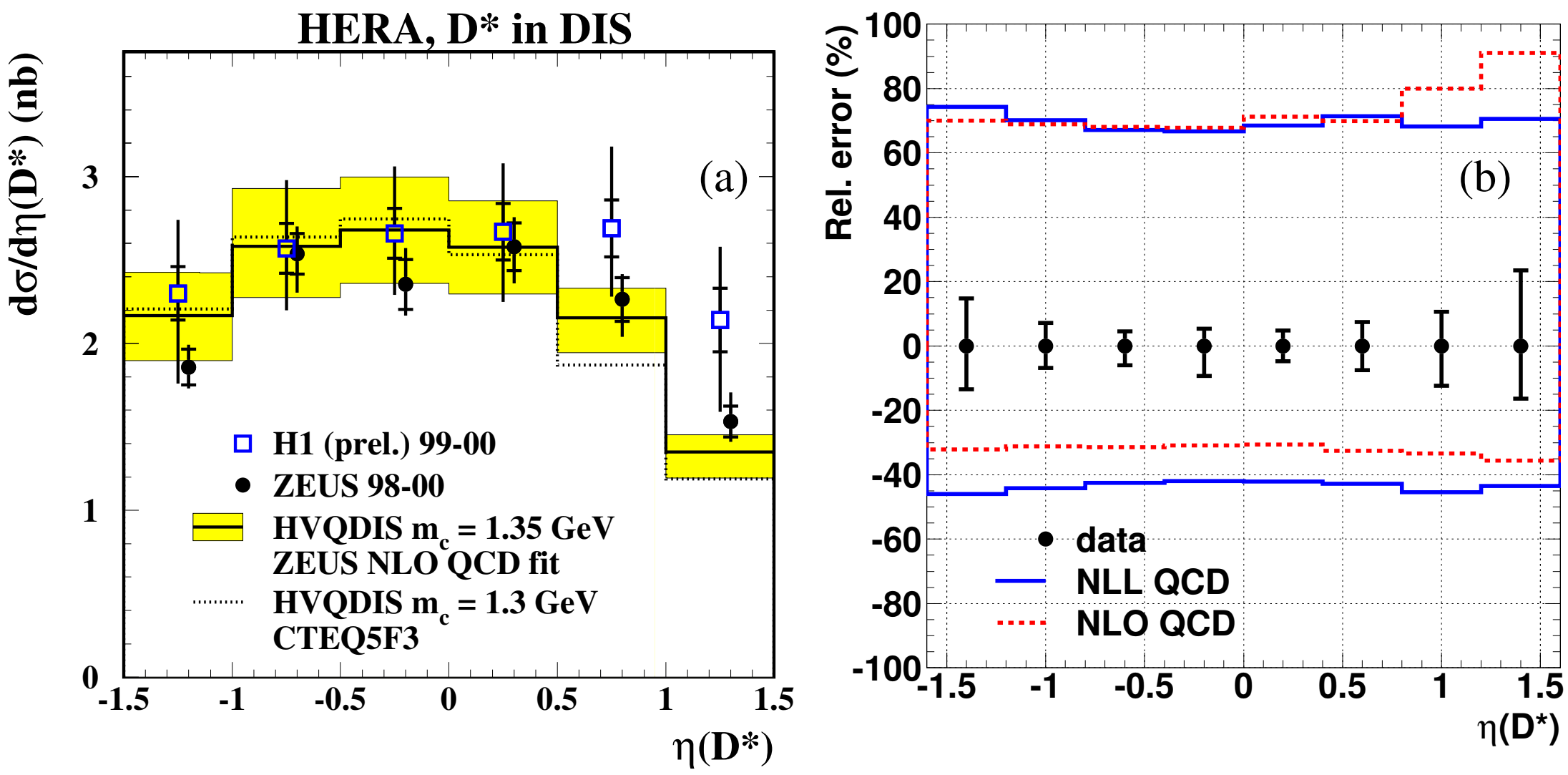

**Fig. 4:** Measurement of $D^*$ production compared with NLO QCD predictions: (a) the differential cross section in deep inelastic scattering and (b) the relative uncertainty in data and theory in photoproduction.

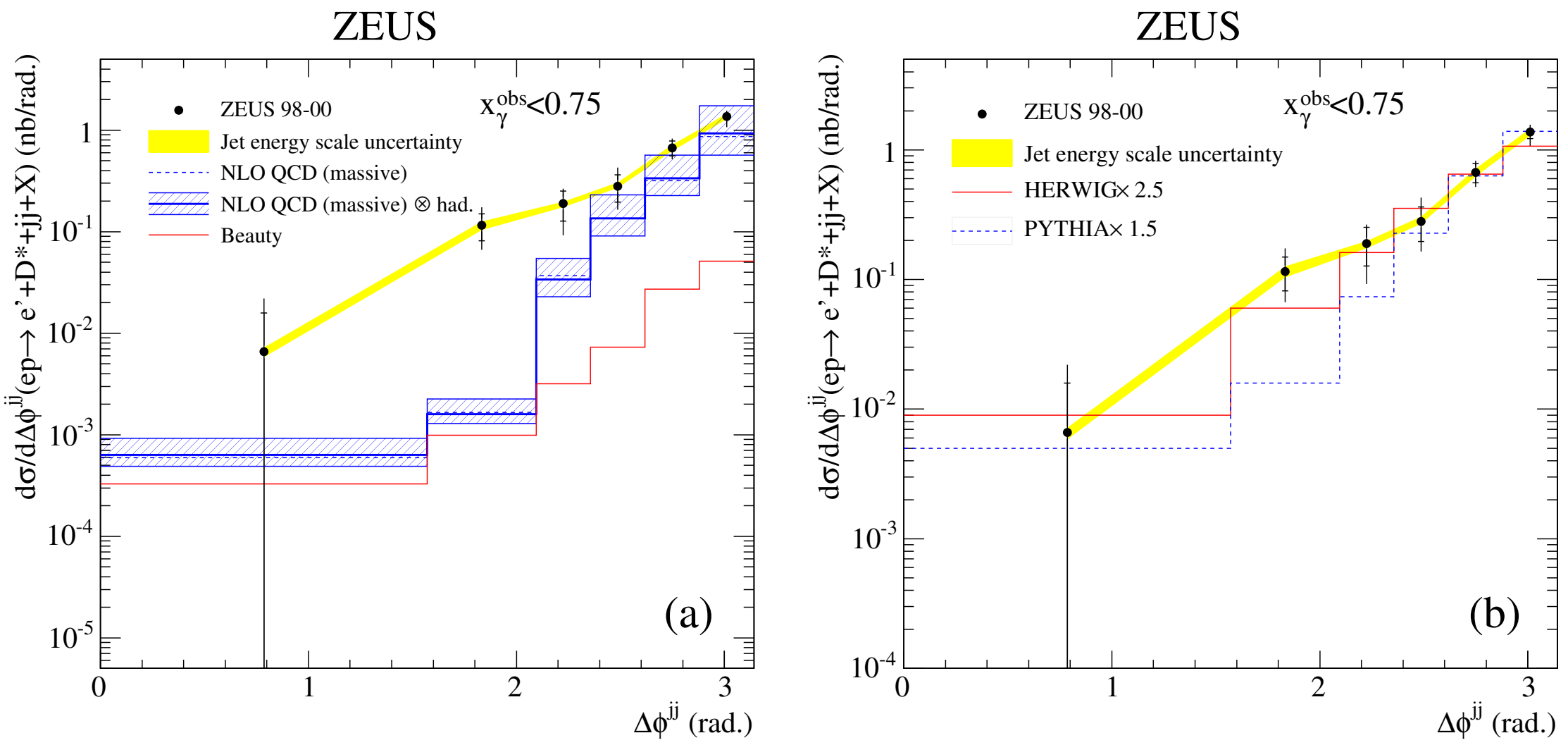

**Fig. 5:** Difference in the azimuthal angle of the two highest $E_T$ jets in charm photoproduction for a sample enriched in resolved photon events compared to (a) a NLO QCD calculation and (b) Monte Carlo models.

## 6 The structure of the proton

Open charm (and more recently beauty) production in deep inelastic scattering acts as a powerful probe of the structure of the proton, particularly the gluon and heavy quark densities. Such a direct measurement of the gluon density complements its extraction in QCD fits to inclusive data. The cross section for the production of a heavy quark pair can be written in terms of the heavy quark contribution to the proton structure functions:

$$\frac{d^2\sigma^{Q\bar{Q}}\left(x,Q^2\right)}{dxdQ^2} = \frac{2\pi\alpha^2}{xQ^4}\left\{\left[1+(1-y)^2\right]F_2^{Q\bar{Q}}\left(x,Q^2\right) - y^2 F_L^{Q\bar{Q}}\left(x,Q^2\right)\right\}$$

The value of the charm contribution, $F_2^{c\bar{c}}$, has traditionally been extracted by measuring $D^*$ mesons within the acceptance of the detector and extrapolating to the full phase space.

The values of $F_2^{c\bar{c}}$ extracted from the measured $D^*$ cross sections [20–22] are shown in Fig. 6a compared with NLO QCD. New measurements of $F_2^{c\bar{c}}$ have been recently performed using an inclusive sample of high $p_T$ tracks [12]. This data is more inclusive than the $D^*$ measurements probing much lower $p_T$ and thereby having much reduced extrapolation factors (a factor of 1.2 rather than 2–3 as for the $D^*$ measurements). These results confirm the previous data and add extra information on $F_2^{c\bar{c}}$. The results on $F_2^{c\bar{c}}$ demonstrate a large gluon density in the proton as exhibited by the scaling violations versus $Q^2$ and are well described by such a parton density function. At high $Q^2$, charm contributes up to about 30% of the inclusive cross section. It is hoped with higher statistics and a better control over the systematics that the charm cross section data can be used in QCD fits to constrain the gluon (or heavy quark) density in the proton.

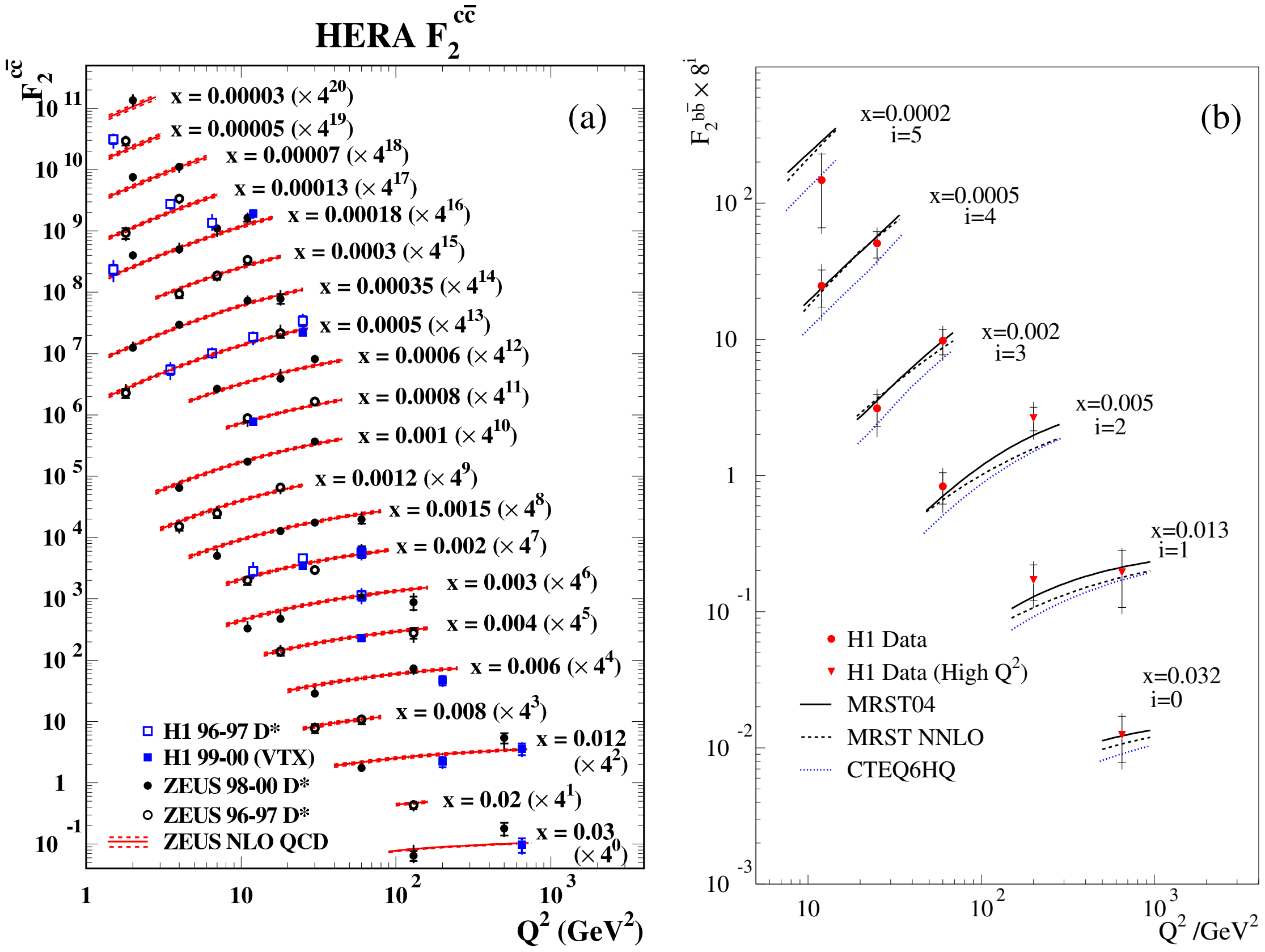

**Fig. 6:** (a) Charm contribution, $F_2^{c\bar{c}}$, and (b) beauty contribution, $F_2^{b\bar{b}}$, to the proton structure function, $F_2$, versus $Q^2$ for fixed $x$.

Applying the same technique of using high $p_T$ tracks, the H1 collaboration have made measurements of $F_2^{b\bar{b}}$ which are shown in Fig. 6. The results are consistent with scaling violations and are well described by new parton density functions. The differences between the different parametrisations are not insignificant and future measurements should be able to discriminate between them. For the $Q^2$ range measured, beauty production contributes up to 3% of the inclusive cross section.

## 7 Universality of charm fragmentation

Heavy quark fragmentation has been extensively studied in $e^+e^-$ collisions. The clean environment, control over the centre-of-mass energy and back-to-back dijet system provide an ideal laboratory for ac-curate measurement of fragmentation parameters. The measured parameters, e.g. fragmentation function and fraction of charm quarks hadronising to a particular meson, are used as inputs to models and NLO QCD calculations of $ep$ collisions. Therefore, the validity of using fragmentation parameters extracted from $e^+e^-$ data in $ep$ data needs to be verified. The strangeness suppression factor, $\gamma_s$, the ratio of neutral and charged $D$-meson production rates, $R_{u/d}$, the fraction of charged $D$ mesons produced in a vector state, $P_v^d$ and the fragmentation fractions, $f(c \to D, \Lambda)$, have been measured in deep inelastic scattering [23] and in photoproduction [24]. The results are shown in Fig. 7 compared with values obtained in $e^+e^-$ collisions. The data obtained in different processes are consistent with each other and thereby consistent with the concept of universal fragmentation. The measurements in photoproduction also have precision competitive with the combined $e^+e^-$ data. The data therefore provide extra constraints and demonstrate that the fragmentation at a hadron collider in the central part of the detector looks like that in an $e^+e^-$ collision. This will provide useful input for future models to be used at the LHC.

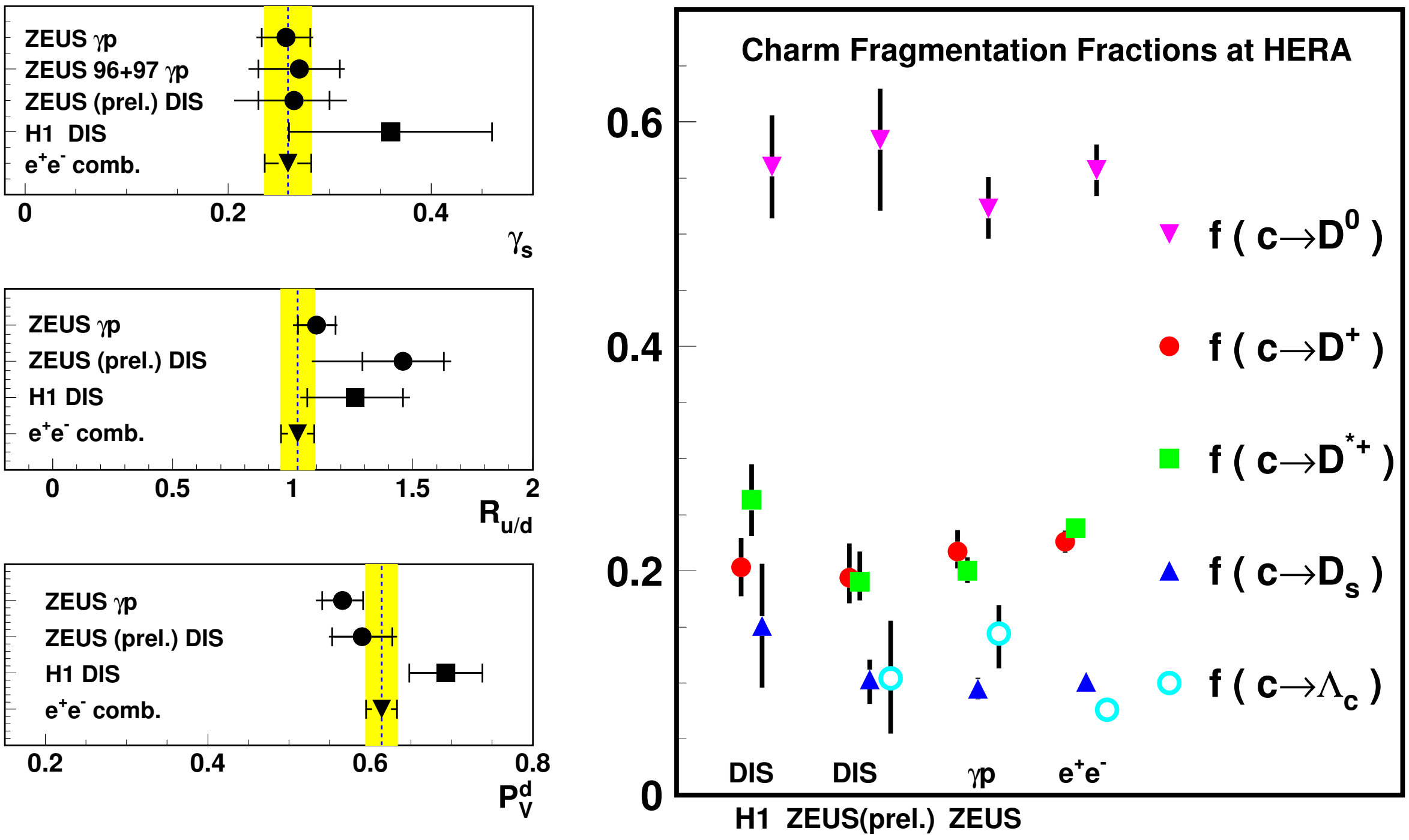

**Fig. 7:** Comparisons of fragmentation parameters, $\gamma_s$, $R_{u/d}$ and $P_v^d$ and $f(c \to D, \Lambda)$ in photoproduction, deep inelastic scattering and $e^+e^-$ collisions.

The charm fragmentation function to $D^*$ mesons has been measured by both the H1 and ZEUS collaborations [25] and compared to $e^+e^-$ data. Although the definitions of the fragmentation function and the energies are different, the general trends are the same. However, a consistent fit to all data within a given Monte Carlo or NLO calculation is needed to clarify this situation. Measurements at the Tevatron would also contribute significantly to this area.

## 8 Conclusions

An increasing number of high precision measurements of heavy quark production from HERA have recently become available. They are providing valuable information on the parton densities, the overall production rates and the concept of the universality of fragmentation. Precise and well-defined measurements have allowed phenomenological improvements to be made. Generally QCD describes the production of heavy quarks; in particular, due in part to the advances made in the HERA measurements, the prediction for the production of beauty quarks is no longer well below the data. There are some details still lacking which await to be confronted with higher order calculations or NLO calculations interfaced with parton showers and hadronisation. There is also ongoing work in tuning Monte Carlo predictions to all known data which demonstrates the need to have global calculations which can predict all processes under study. In the next few years in the run up to the LHC, HERA will produce a lot more data and more will be known about heavy quark production.

# From HERA to the LHC

*John Ellis*
CERN, Geneva

**Abstract**
Some personal comments are given on some of the exciting interfaces between the physics of HERA and the LHC. These include the quantitative understanding of perturbative QCD, the possible emergence of saturation phenomena and the Colour-Glass Condensate at small $x$ and large $Q^2$, the link between forward physics and ultra-high-energy cosmic rays, and new LHC opportunities opened up by the discovery of rapidity-gap events at HERA, including the search for new physics such as Higgs bosons in double-diffraction events.

## 1 Preview

There are many exciting interfaces between physics at HERA and the LHC, and I cannot do justice to all of them in this talk. Therefore, in this talk I focus on a few specific subjects that interest me personally, starting with the LHC's 'core business', namely the search for new physics at the TeV scale, notably the Higgs boson(s) and supersymmetry [1]. Identifying any signals for such new physics will require understanding of the Standard Model backgrounds, and QCD in particular. I then continue by discussing some other topics of specific interest to the DESY community.

- The understanding of QCD will be important for making accurate studies of any such new physics. Perturbative QCD at moderate $x$ and large $p_T$ is quite well understood, with dramatic further progress now being promised by novel calculational techniques based on string theory [2].
- Novel experimental phenomena are now emerging at RHIC at small $x$, following harbingers at HERA. The parton density saturates, and a powerful organizational framework is provided by the Colour-Glass Condensate (CGC) [3]. Forward measurements at the LHC will provide unique opportunities for following up on this HERA/RHIC physics.
- Forward physics at the LHC will also provide valuable insight into the interpretation of ultra-high-energy cosmic rays (UHECRs) [4]. One of the principal uncertainties in determining their energy scale is the modeling of the hadronic showers they induce, and the LHC will be the closest laboratory approximation to UHECR energies.
- Looking further forward, there is increasing interest in exploring at the LHC the new vistas in hard and soft diffraction opened up by the discovery of rapidity-gap events at HERA [5]. One particularly interesting possibility is quasi-exclusive diffractive production of Higgs bosons or other new particles at the LHC [6]. This is particularly interesting in supersymmetric extensions of the Standard Model, notably those in which CP is violated [7].

## 2 Prospects in Higgs Physics

Many studies have given confidence that the Standard Model Higgs boson will be found at the LHC, if it exists [8]. Moreover, there are some chances that it might be found quite quickly, in particular if its mass is between about 160 GeV and 600 GeV. However, discovering the Higgs boson will take rather longer if its mass is below about 130 GeV, as suggested in the minimal supersymmetric extension of the Standard Model (MSSM) [9]. In this case, the Higgs signal would be composed of contributions from several different production and decay channels, notably including $gg \to H \to \gamma\gamma$.

Understanding the gluon distribution at $x \sim 10^{-2}$ is therefore a high priority, and one to which HERA measurements of processes involving gluons have been playing key roles [10]. Perturbative corrections to the $gg \to H$ production process need to be understood theoretically, as do the corrections to $H \to \gamma\gamma$ decay. Resummation of the next-to-next-to-leading logarithms has by now reduced these uncertainties to the 10% level, and further improvements may be possible with the string-inspired calculational techniques now being introduced [11].

Fig. 1 shows estimates of the accuracy with which various Higgs couplings may be determined at the LHC, also if the luminosity may be increased by an order of magnitude (SLHC) [12] [see also [13]]. There are interesting prospects for measuring the couplings to $\tau\tau$, $\bar{b}b$, $WW$, $ZZ$ and $\bar{t}t$ as well as the total Higgs decay width, though not with great accuracy. Measurements at the ILC would clearly be much more powerful for this purpose [13].

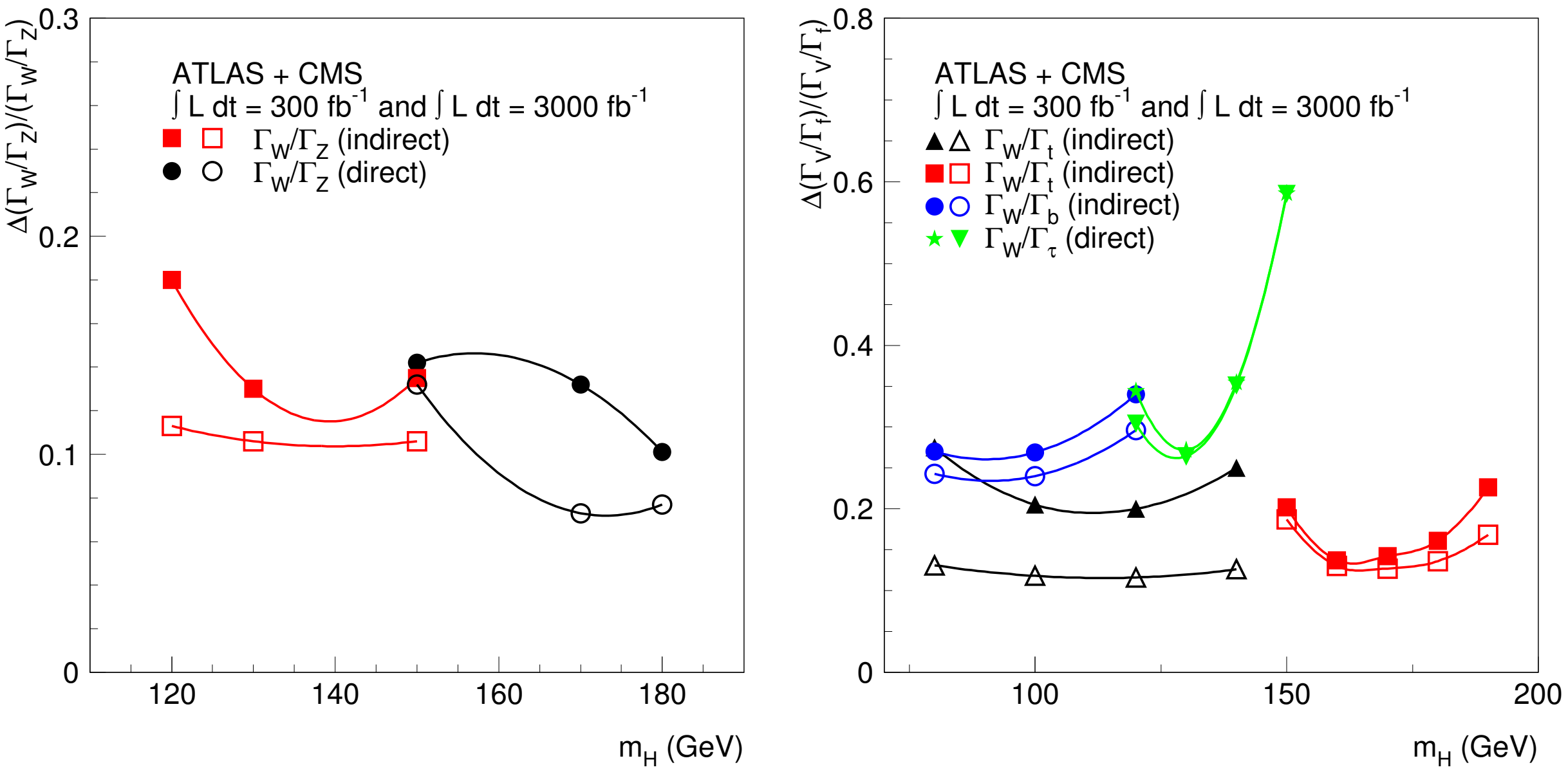

**Fig. 1:** Illustrations of the accuracy with which Higgs couplings could be measured at the LHC with the planned luminosity and with a possible upgrade by a factor of ten (SLHC) [12].

## 3 Theorists are Hedging their Bets

The prospect of imminent Higgs discovery is leading theorists to place their last bets on the LHC roulette wheel, and many are hedging their bets by proposing and discussing alternatives to the Standard Model or the MSSM. Composite Higgs models are not greatly favoured, since they have a strong tendency to conflict with the precision electroweak data [14]. This problem has led some theorists to question the interpretation of the electroweak data, which are normally taken to favour $m_H < 300$ GeV, debating their consistency and even arguing that some data should perhaps be discounted [15]. Personally, I see no strong reason to doubt the hints from the electroweak data. An alternative corridor leading towards higher Higgs masses is provided by including higher-dimensional operators in the electroweak data analysis [16]: this would require some fine-tuning, but cannot be excluded. An even more extreme alternative that has been re-explored recently is that of Higgsless models [17]. However, these lead to strong $WW$ scattering and conflict with the available electroweak data. These problems are alleviated, but not solved, by postulating extra dimensions at the TeV scale [18].

One of the least unappetizing alternatives to the supersymmetric Higgs paradigm is offered by little Higgs models [19]. Their key idea is to embed the Standard Model in a larger gauge group, from which the Higgs boson emerges as a relatively light pseudo-Goldstone boson. The one-loop quadratic divergence due to the top quark:

$$\delta m^2_{H,top}(SM) \sim (115 \text{ GeV})^2 \left(\frac{\Lambda}{400 \text{ GeV}}\right)^2$$

is cancelled by the contribution of a new heavy $T$ quark:

$$\delta m^2_{H,top}(LH) \sim \frac{6 G_F m_t^2}{\sqrt{2}\pi^2} m_T^2 \log \frac{\Lambda}{m_T}$$

Additionally, there are new gauge bosons and exotic Higgs representations. The Standard-Model-like Higgs boson is expected to be relatively light, possibly below $\sim 150$ GeV, whereas the other new particles are expected to be heavier:

$$\begin{aligned} M_T &< 2\,\text{TeV}(m_h/200\text{GeV})^2 \\ M'_W &< 6\,\text{TeV}(m_h/200\text{GeV})^2 \\ M_{H++} &< 10\,\text{TeV} \end{aligned}$$

Certainly the new $T$ quark, probably the $W'$ boson and possibly even the doubly-charged Higgs boson will be accessible to the LHC. Thus little Higgs models have quite rich phenomenology, as well being decently motivated. However, they are not as complete as supersymmetry, and would require more new physics at energies $> 10$ TeV.

Depending on the mass scale of this new physics, there may be some possibility for distinguishing a little Higgs model from the Standard Model by measurements of the $gg \to H \to \gamma\gamma$ process at the LHC. However, the ILC would clearly have better prospects in this regard [13].

## 4 Supersymmetry

No apologies for repeating the supersymmetric mantra: it resolves the naturalness aspect of the hierarchy problem by cancelling systematically the quadratic divergences in all loop corrections to the Higgs mass and hence stabilizes the electroweak scale [20], it enables the gauge couplings to unify [21], it predicts $m_H < 150$ GeV [9] as suggested by the precision electroweak data [14], it stabilizes the Higgs potential for low Higgs masses [22], and it provides a plausible candidate [23] for the dark matter that astrophysicists and cosmologists claim clutters up the Universe.

However, all we have from accelerators at the moment are lower limits on the possible supersymmetric particle masses, most notably from the absence of sparticles at LEP: $m_{\tilde{\ell}}, m_{\chi^\pm} > 100$ GeV and the Tevatron collider: $m_{\tilde{g}}, m_{\tilde{q}} > 300$ GeV, the LEP lower limit $m_H > 114.4$ GeV, and the consistency of $b \to s\gamma$ decay with the Standard Model. However, if we assume that the astrophysical cold dark matter is largely composed of the lightest supersymmetric particle (LSP), and require its density to lie within the range allowed by WMAP et al [24]:

$$0.094 < \Omega_\chi h^2 < 0.129,$$

we obtain upper as well as lower limits on the possible sparticle masses. The anomalous magnetic moment of the muon, $g_\mu - 2$, provides intermittent hints on the supersymmetric mass scale [25]: these are lower limits if you do not believe there is any significant discrepancy with the Standard Model prediction, but also an upper limit if you do not believe that the Standard Model can fit the data, as is indicated by the current interpretation of the $e^+e^-$ data used to calculate the Standard Model prediction.

If one compares the production of the lightest neutral Higgs boson in the constrained MSSM (CMSSM) in which all the soft supersymmetry-breaking scalar masses $m_0$ and gaugino masses $m_{1/2}$ are assumed to be universal, the *good news* is that the rate for $gg \to h \to \gamma\gamma$ is expected to be within 10% of the Standard Model value, as seen in Fig. 2(a) [26]. On the other hand, the bad news is the rates are so similar that it will be difficult to distinguish a CMSSM Higgs boson from its Standard Model counterpart. This would be much easier at the ILC, as seen in Fig. 2(b) [27].

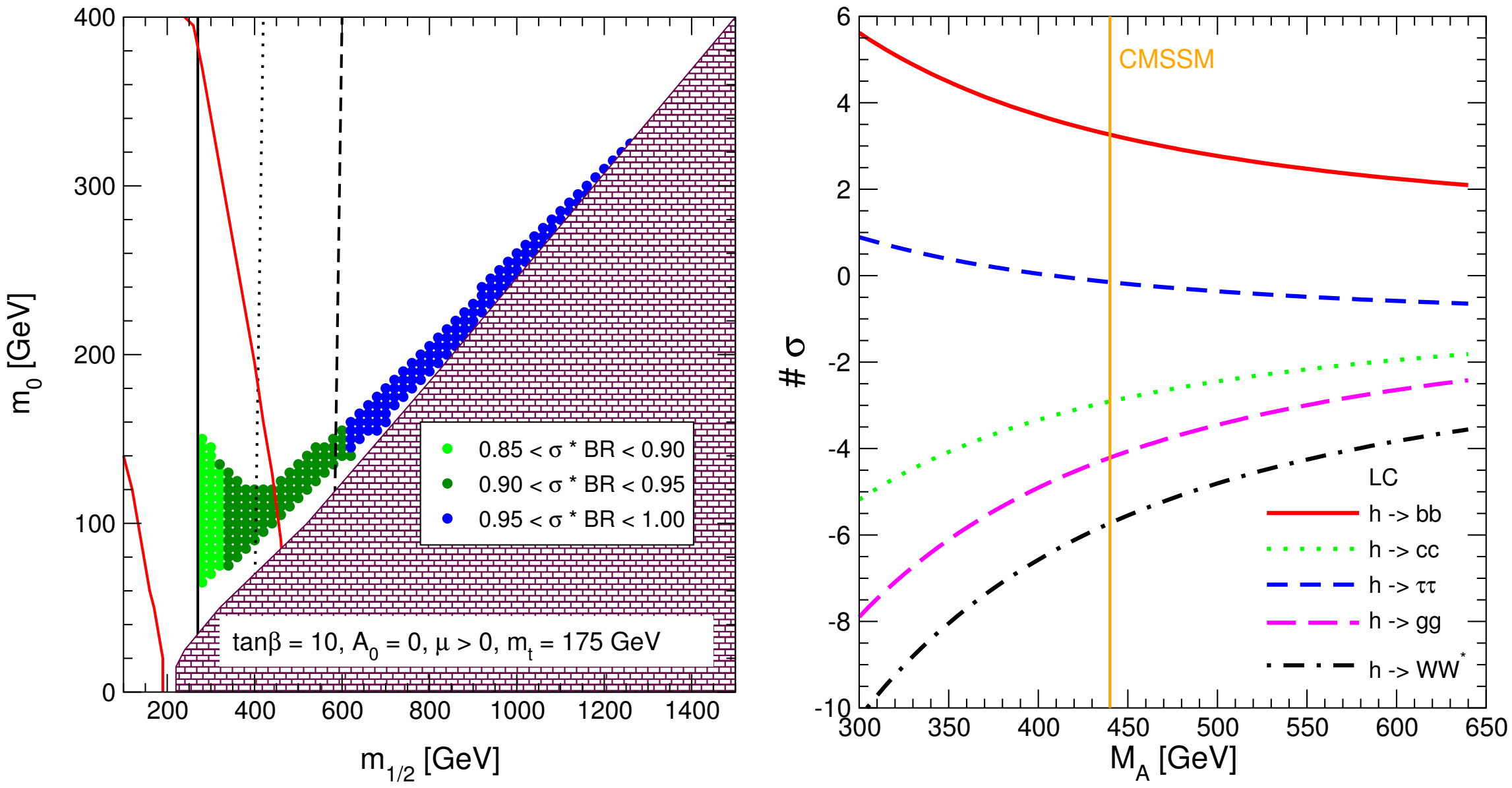

**Fig. 2:** Left panel: The cross section for production of the lightest CP-even MSSM Higgs boson in gluon fusion and its decay into a photon pair, $\sigma(gg \to h) \times \mathcal{B}(h \to \gamma\gamma)$, normalized to the Standard Model value with the same Higgs mass, is given in the $(m_{1/2}, m_0)$ plane for $\mu > 0$, $\tan\beta = 10$, assuming $A_0 = 0$ and $m_t = 175$ GeV [26]. The diagonal (red) solid lines are the $\pm 2 - \sigma$ contours for $g_\mu - 2$. The near-vertical solid, dotted and dashed (black) lines are the $m_h = 113, 115, 117$ GeV contours. The (brown) bricked regions are excluded since in these regions the LSP is the charged $\tilde{\tau}_1$. Right panel: The numbers of standard deviations by which the predictions of the MSSM with non-universal Higgs masses may be distinguished from those of the Standard Model in different channels by measurements at the ILC [27]. The predictions with the CMSSM values of $M_A$ and $\mu$ are indicated by light vertical (orange) lines. The other parameters have been chosen as $m_{1/2} = 300$ GeV, $m_0 = 100$ GeV, $\tan\beta = 10$ and $A_0 = 0$.

One of the distinctive possibilities opened up by the MSSM is the possibility of CP violation in the Higgs sector, induced radiatively by phases in the gaugino masses and the soft supersymmetry-breaking trilinear couplings. Fig. 3 displays CP-violating asymmetries that might be observable in the $gg, \bar{b}b \to \tau^+\tau^-$ and $W^+W^- \to \tau^+\tau^-$ processes at the LHC, in one particular CP-violating scenario with large three-way mixing between all three of the neutral MSSM Higgs bosons [28].

A typical supersymmetric event at the LHC is expected to contain high-$p_T$ jets and leptons, as well as considerable missing transverse energy. Studies show that the LHC should be able to observe squarks and gluinos weighing up to about 2.5 TeV [8], covering most of the possibilities for astrophysical dark matter. As seen in Fig. 4(a) [1], the dark matter constraint restricts $m_{1/2}$ and $m_0$ to narrow strips extending to an upper limit $m_{1/2} \sim 1$ TeV. As seen in Fig. 4(b), whatever the value of $m_{1/2}$ along one of these strips, the LHC should be able to observe several distinct species of sparticle [1]. In a favourable case, such as the benchmark point B in Fig. 4(a) (also known as SPS Point 1a), experiments at the LHC should be able to measure the CMSSM parameters with sufficient accuracy to calculate the supersymmetric relic density $\Omega_\chi h^2$ (blue histogram) with errors comparable to the present astrophysical error (yellow band) as seen in Fig. 4(c) [1]. Fig. 4(d) summarizes the scapabilities of the LHC and other accelerators to detect various numbers of sparticle species. We see that the LHC is almost guaranteed to discover supersymmetry if it is relevant to the naturalness of the mass hierarchy. However, there are some variants of the CMSSM, in particular at the tips of the WMAP strips for large $\tan\beta$, that might escape detection at the LHC.

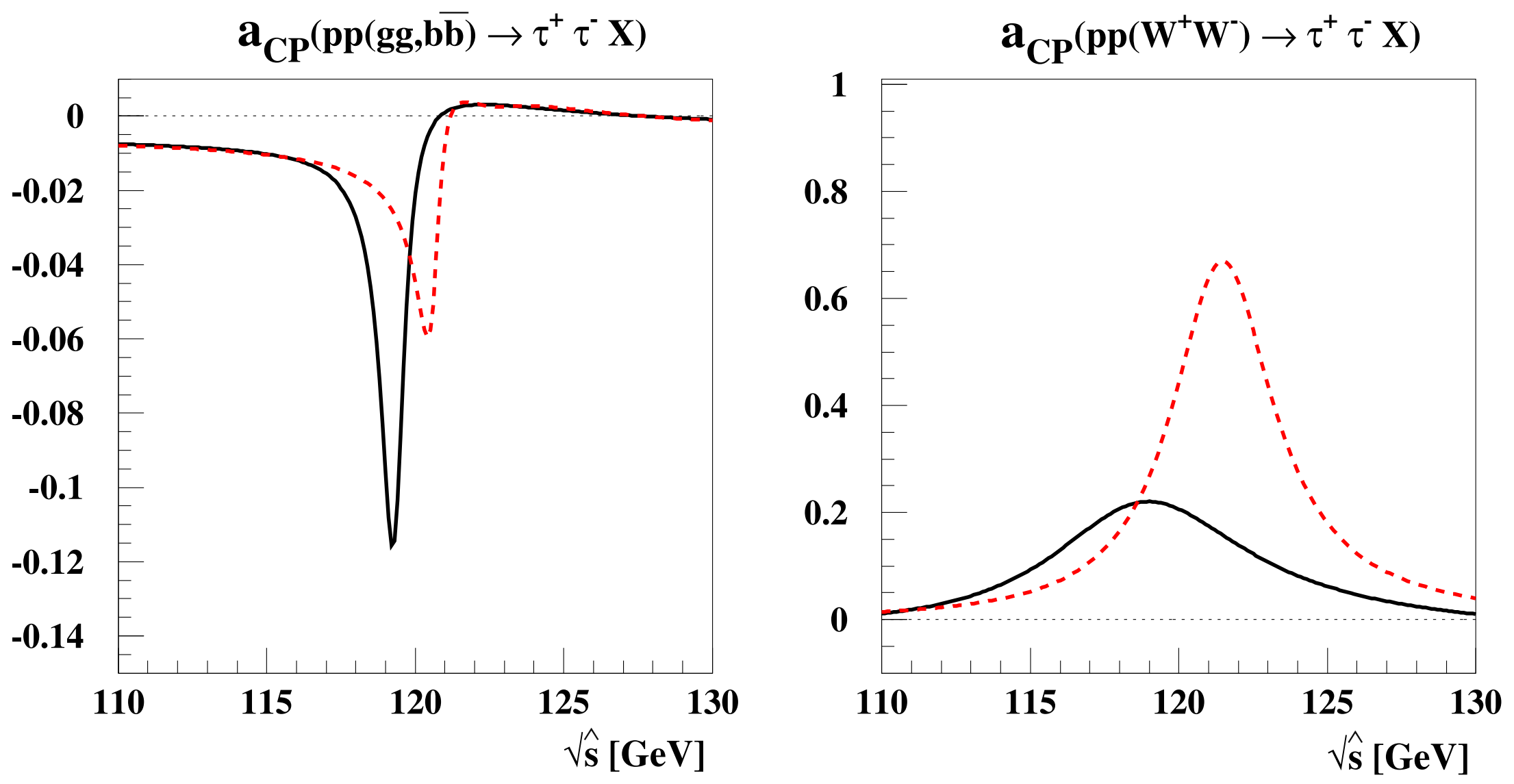

**Fig. 3:** Numerical estimates of differential CP asymmetries as functions of the effective reduced centre-of-mass energy $\sqrt{\hat{s}}$ in a CP-violating three-Higgs mixing scenario with gaugino phase $\Phi_3 = -90°$ (solid lines) and $\Phi_3 = -10°$ (dashed lines) [28].

As we also see in Fig. 4(d), linear colliders would be able to observe a complementary subset of sparticles, particularly sleptons, charginos and neutralinos [1]. A linear collider with a centre-of-mass energy of 1 TeV would have comparable physics reach to the LHC, but a higher centre-of-mass energy, such as the 3 TeV option offered by CLIC [29], would be needed to complete the detection and accurate measurement of all the sparticles in most variants of the CMSSM.

We have recently evaluated whether precision low-energy observables currently offer any hint about the mass scale of supersymmetric particles, by exploring their sensitivities to $m_{1/2}$ along WMAP lines for different values of the trilinear supersymmetry-breaking parameter $A_0$ and the ratio of Higgs v.e.v's, $\tan\beta$ [31]. The measurements of $m_W$ and $\sin^2\theta_W$ each currently favour $m_{1/2} \sim 300$ GeV for $\tan\beta = 10$ and $m_{1/2} \sim 600$ GeV for $\tan\beta = 50$. The agreement of $b \to s\gamma$ decay with the Standard Model is compatible with a low value of $m_{1/2}$ for $\tan\beta = 10$ but prefers a larger value for $\tan\beta = 50$, whereas $B_s \to \mu^+\mu^-$ decay currently offers no useful information on the scale of supersymmetry breaking [30]. The current disagreement of the measured value of the anomalous magnetic moment of the muon, $g_\mu - 2$, also favours independently $m_{1/2} \sim 300$ GeV for $\tan\beta = 10$ and $m_{1/2} \sim 600$ GeV for $\tan\beta = 50$. Putting all these indications together, as seen in Fig. 5, we see a preference for $m_{1/2} \sim 300$ GeV when $\tan\beta = 10$, and a weaker preference for $m_{1/2} \sim 600$ GeV when $\tan\beta = 50$ [31]. At the moment, this preference is far from definitive, and $m_{1/2} \to \infty$ is excluded at lass than 3 $\sigma$, but it nevertheless offers some hope that supersymmetry might lurk not far away.

As seen in Fig. 6, the likelihood function for $m_{1/2}$ can be converted into the corresponding likelihood functions for the masses of various species of sparticles. The preferred squark and gluino masses lie below 1000 GeV for $\tan\beta = 10$, with somewhat heavier values for $\tan\beta = 50$, though still well within the reach of the LHC [31].

## 5 Gravitino Dark Matter

The above analysis assumed that the lightest supersymmetric particle (LSP) is the lightest neutralino $\chi$, assuming implicitly that the gravitino is sufficiently heavy and/or rare to have been neglected. This implicit assumption may or may not be true in a minimal supergravity model, where the gravitino mass $m_{3/2} = m_0$, as seen in Fig. 7[1] [32]. In this model, the gravitino mass is fixed throughout the $(m_{1/2}, m_0)$ plane: there is a familiar WMAP strip where the $\chi$ is the LSP, but there is also a wedge of parameter space where the LSP is the gravitino. There is no way known to detect such astrophysical gravitino dark matter (GDM), since the gravitino has very weak interactions.

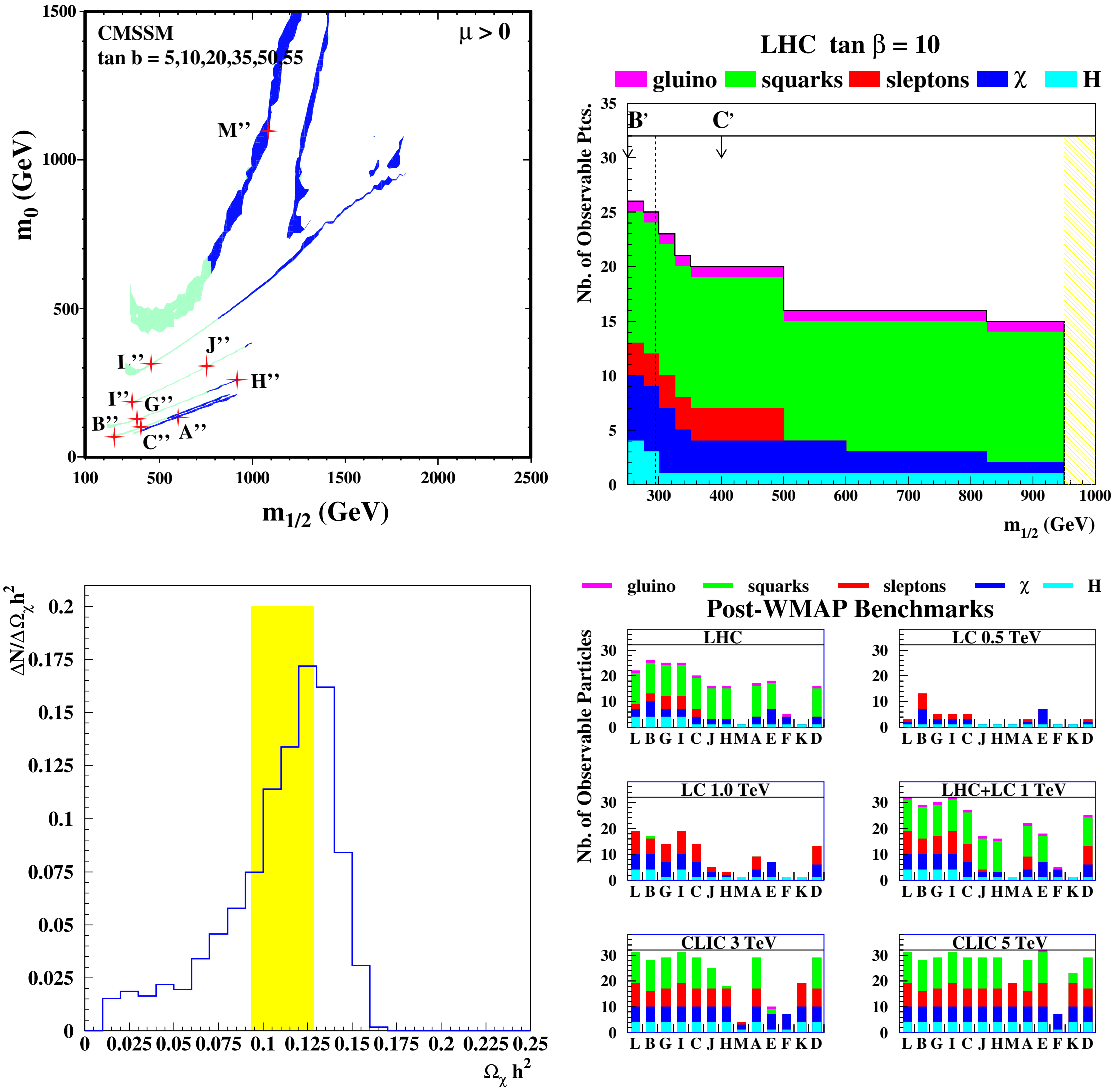

**Fig. 4:** Top left panel: The strips of CMSSM parameter space allowed by WMAP and other constraints, with specific benchmark scenarios indicated by (red) crosses. Top right panel: The numbers of MSSM particle species observable at the LHC as a function of $m_{1/2}$ along the WMAP strip for $\tan\beta = 10$ [26]. Bottom left panel: The accuracy with which the relic dark matter density could be calculated using LHC measurements at benchmark point B, compared with the uncertainty provided by WMAP and other astrophysical data. Bottom right panel: The numbers of MSSM particle species observable in the benchmark scenarios at the LHC and $e^+e^-$ colliders with different centre-of-mass energies [27].

However, the LHC may have prospects for detecting GDM indirectly [33–35]. In the GDM region, the lighter stau, $\tilde{\tau}_1$, is expected to be the next-to-lightest sparticle (NLSP), and may be metastable with a lifetime measurable in hours, days, weeks, months or even years. The $\tilde{\tau}_1$ would be detectable in CMS or ATLAS as a slow-moving charged particle. Staus that are sufficiently slow-moving might be stopped in the detector itself, in some external detection volume designed to observe and measure their late decays into GDM [33, 34], or in the walls of the caverns surrounding the detectors [35].

[1]Minimal supergravity also relates the trilinear and bilinear supersymmetry-breaking parameters: $A_0 = B_0 + 1$, thereby fixing $\tan\beta$ as a function of $m_{1/2}, m_0$ and $A_0$, see the contours in Fig. 7(b).

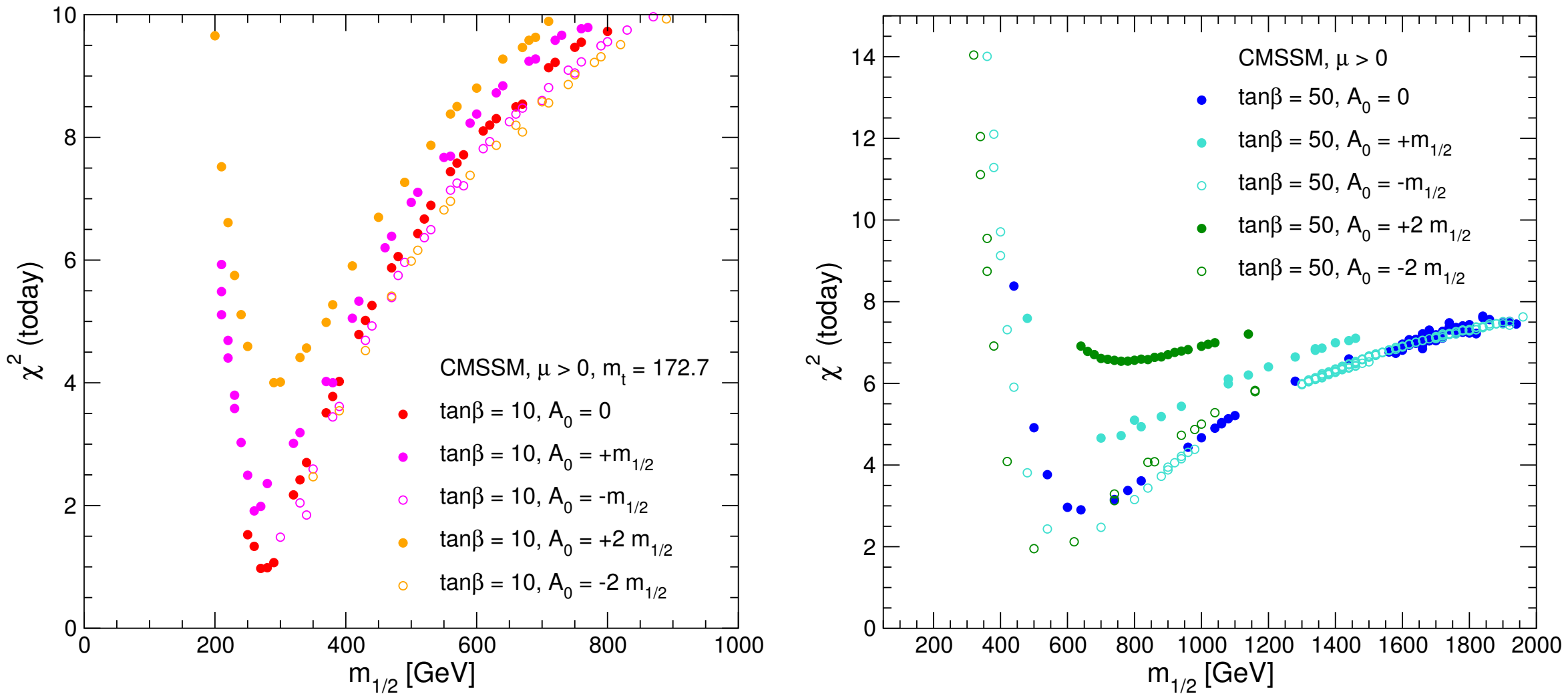

**Fig. 5:** The results of $\chi^2$ fits based on the current experimental results for the precision observables $M_W$, $\sin^2\theta_{weff}$, $(g-2)_\mu$ and $b \to s\gamma$ are shown as functions of $m_{1/2}$ in the CMSSM parameter space with WMAP constraints for different values of $A_0$ and (left) $m_t = 172.7 \pm 2.9$ GeV and $\tan\beta = 10$ and (right) $m_t = 178.0 \pm 4.3$ GeV and $\tan\beta = 50$ [31].

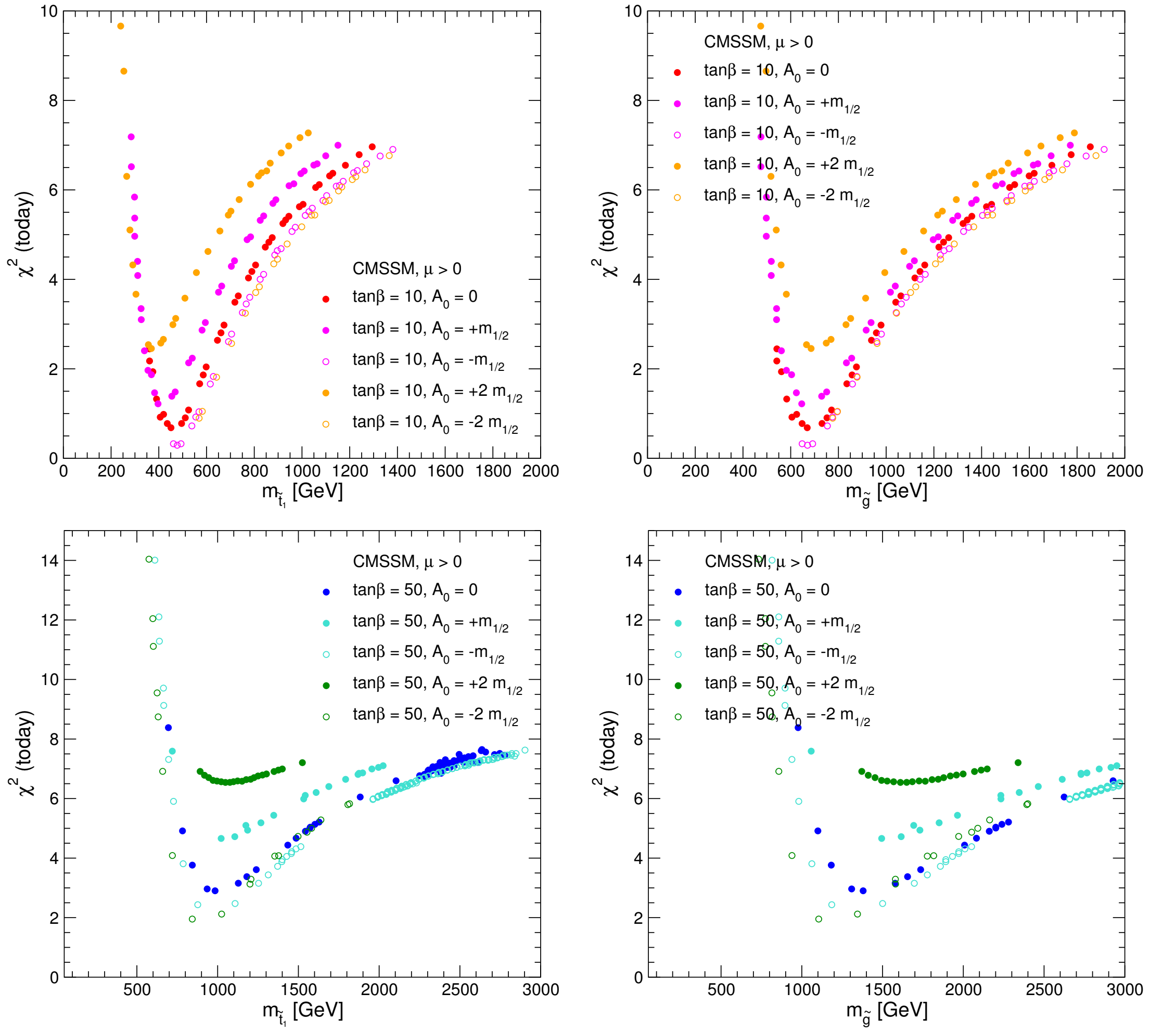

**Fig. 6:** The $\chi^2$ contours in the CMSSM with $\tan\beta = 10$ for the lighter stop (left) and gluino (right) masses, assuming $\tan\beta = 10$ (top) and $\tan\beta = 50$ (bottom) [31].

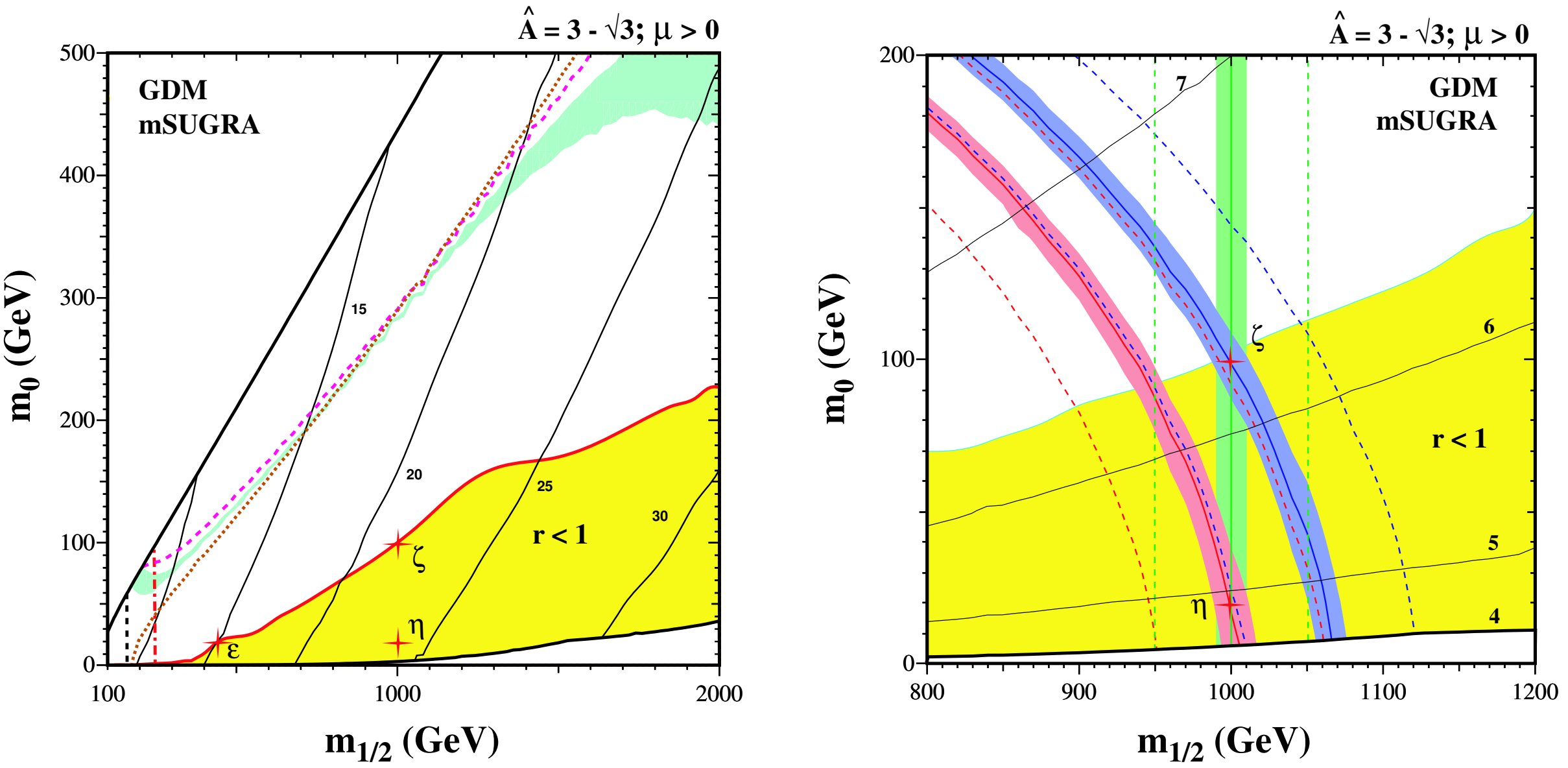

**Fig. 7:** Left panel: The allowed region in the $(m_{1/2}, m_0)$ planes for minimal supergravity (mSUGRA) with a gravitino LSP [light (yellow) shaded regions labelled $r < 1$], for $A \equiv \hat{A}m_0 : \hat{A} = 3 - \sqrt{3}$. The red crosses denote the benchmark GDM models $\epsilon, \zeta$ and $\eta$ [35]. Right panel: The potential impacts on the determination of GDM parameters in the mSUGRA framework of prospective measurement errors of 1 % and 5 % for $m_{\tilde{\tau}_1}$ (diagonal bands and lines) and $m_{1/2}$ (vertical bands and lines), shown as constraints in the $(m_{1/2}, m_0)$ plane [35]. The smaller errors would enable the benchmark scenarios $\zeta$ and $\eta$ to be distinguished, and the possible NLSP lifetime to be estimated. The near-horizontal thin solid lines are labelled by the logarithm of the NLSP lifetime in seconds.

## 6 The LHC and Ultra-High-Energy Cosmic Rays

Historically, the two experiments with (until recently) the largest statistics for ultra-high-energy cosmic rays (UHECRs), AGASA [36] and HiRes [37], have not agreed on their energy spectra above about $10^{19}$ eV and, specifically, whether there is a significant number of events beyond the GZK cutoff due to interactions of primary UHECRs with the cosmic microwave background radiation. The Auger experiment now has the second-largest statistics but does not yet have sufficient data to settle the issue [38], though these should soon be forthcoming. If there are super-GZK events, they might be due either to nearby astrophysical sources that have not yet been identified, or (more speculatively) to the decays of metastable superheavy particles [39]. Normalizing the energies of UHECRs requires understanding of the development of extensive air showers. At the moment, this is not very well known, and models of shower development are not even able to tell us the composition of cosmic rays with lower energies between $10^{15}$ and $10^{19}$ eV [4].

The LHC is the accelerator that comes closest to reproducing the UHECR energy range, with a centre-of-mass energy corresponding to $4 \times 10^{17}$ eV, in the range where the cosmic-ray composition is still uncertain. This uncertainty would be reduced by better modelling of hadronic showers, which would in turn benefit from measurements in the forward direction [4].

Unfortunately, the LHC is currently not equipped to make good measurements in this kinematic region, where most of the centre-of-mass energy is deposited. More instrumentation in the forward direction would be most welcome in both CMS and ATLAS. This region is also of fundamental importance for our understanding of QCD, as I now explain.

## 7 Back to Forward QCD

We discussed earlier the success of perturbative QCD, and the accuracy with which it could be used to calculate high-$p_T$ physics, thanks to the structure functions provided by HERA data [10], in particular. The simple parton description is expected, however, to break down at 'small' $x$ and 'large' $Q^2$, due to saturation effects. At small $x$, there is a large probability to emit an extra gluon $\sim \alpha_s \ln(1/x)$, and the number of gluons grows in a limited transverse area. When the transverse density becomes large, partons of size $1/Q$ may start to overlap, and non-linear effects may appear, such as the annihilation of low-$x$ partons. The Malthusian growth in the number of gluons seen at HERA is eventually curbed by these annihilation effects when $\ln(1/x)$ exceeds some critical $x$-dependent saturation value of $Q^2$. At larger values of $x$, the parton evolution with $Q^2$ is described by the usual DGLAP equations, and the evolution with $\ln(1/x)$ is described by the BFKL equation. However, at lower values of $x$ and large $Q^2$, a new description is need for the saturated configuration, for which the most convincing proposal is the Colour-Glass Condensate (CGC) [3].

According to the CGC proposal, the proton wave function participating in interactions at low $x$ and $Q^2$ is to be regarded as a classical colour field that fluctuates more slowly than the collision time-scale. This possibility may be probed in Gold-Gold collisions at RHIC and proton-proton collisions at the LHC: the higher beam energy of LHC compensates approximately for the higher initial parton density in Gold-Gold collisions at RHIC. At central rapidities $y \sim 0$, effects of the CGC are expected to appear only when the parton transverse momentum $< 1$ GeV. However, CGC effects are expected to appear at larger parton transverse momenta in the forward direction when $y \sim 3$. Lead-Lead collisions at the LHC should reveal even more important saturation effects [40].

What is the experimental evidence for parton saturation? First evidence came from HERA, and Fig. 8(a) displays an extraction of the saturation scale from HERA data [41]. At RHIC, in proton-nucleus collisions one expects the suppression of hard particles at large rapidity and small angle compared to proton-proton collisions, whereas one expects an enhancement at small rapidity, the nuclear 'Cronin effect'. The data [42] from the BRAHMS collaboration at RHIC shown in Fig. 8(b) are quite consistent with CGC expectations [43], but it remains to be seen whether this approach can be made more quantitative than older nuclear shadowing ideas.

## 8 New Physics in Diffraction?

HERA has revealed a menagerie of different diffractive phenomena, opening up a Pandora's box of possible new physics at the LHC. Classically one had soft diffraction dissociation in peripheral proton-proton collisions, in which one (or both) of the colliding protons would dissociate into a low-mass system (or systems). HERA discovered an additional class of diffractive events [5], which may be interpreted [44] as a small colour dipole produced by an incoming virtual photon penetrates the proton and produces a high-mass system. Additionally, one expects at the LHC soft double diffraction, in which a peripheral proton-proton collision produces a low-mass central system separated from each beam by a large rapidity gap. Events with mixed hard and soft diffraction are also possible at the LHC, as are events with multiple large rapidity gaps. The LHC will certainly provide good prospects for deepening our understanding of diffraction, building upon the insights being gained from HERA.

Double diffraction also offers the possibility of searching for new physics in a relatively clean experimental environment containing, in addition to Higgs boson or other new particle, only a couple of protons or their low-mass diffraction-dissociation products[2]. The leading-order cross-section formula (nominal values of the diffractive parameters are quoted in the brackets) is [6]:

$$M^2 \frac{\partial^2 \mathcal{L}}{\partial y \partial M^2} = 4.0 \times 10^{-4} \left[ \frac{\int_{lnQ_{min}}^{ln\mu} F_g(x_1, x_2, Q_T, \mu) dlnQ_T}{\mathrm{GeV}^{-2}} \right]^2 \left( \frac{\hat{S}^2}{0.02} \right) \left( \frac{4}{b\mathrm{GeV}^2} \right)^2 \left( \frac{R_g}{1.2} \right)^4 .$$

[2]New physics might also be produced in other classes of diffractive events, but with less distinctive signatures.

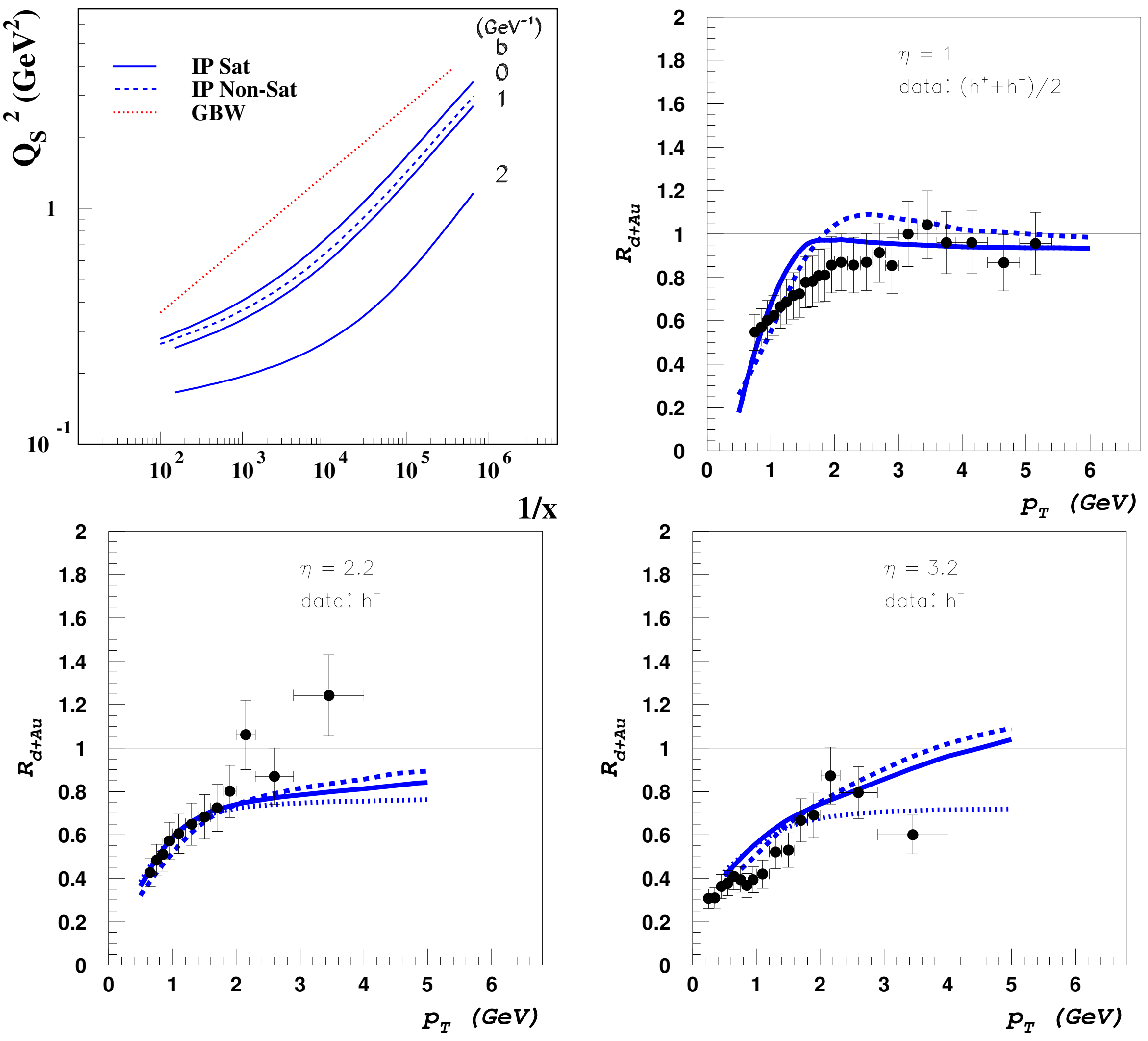

**Fig. 8:** Top left panel: The parton saturation scale as a function of Bjorken $x$, extracted from HERA data in [41]. Other three panels: Nuclear modification factor $R_{dAu}$ of charged particles for rapidities $\eta = 1, 2.2, 3.2$ [42], compared with calculations from [43].

The gluon collision factor is currently inferred from HERA data via different parameterizations of the integrated gluon distribution function, and has an uncertainty of a factor of about two [6]. Further analyses of HERA data, as well as future LHC data, would enable the determination to be refined.

The observation of diffractive Higgs production at the LHC would be a challenge in the Standard Model, but the cross section is expected to be considerably larger in the MSSM, particularly at large $\tan\beta$. One of the enticing possibilities offered by supersymmetry is a set of novel mechanisms for CP violation induced by phases in the soft supersymmetry-breaking parameters [7]. These would show up in the MSSM Higgs sector, generating three-way mixing among the neutral MSSM Higgs bosons. This might be observable in inclusive Higgs production at the LHC [7], but could be far more dramatic in double diffraction. Fig. 9(a) displays the mass spectrum expected in double diffraction in one particular three-way mixing scenario [45]: it may exhibit one or more peaks that do not coincide with the Higgs masses. Analogous structures may also be seen in CP-violating asymmetries in $H_i \to \tau^+\tau^-$ decay, as seen in Fig. 9(b). These structures could not be resolved in conventional inclusive Higgs production at the LHC, but may be distinguished in exclusive double diffraction by exploiting the excellent missing-mass resolution $\sim 2$ GeV that could be provided by suitable forward spectrometers [46].

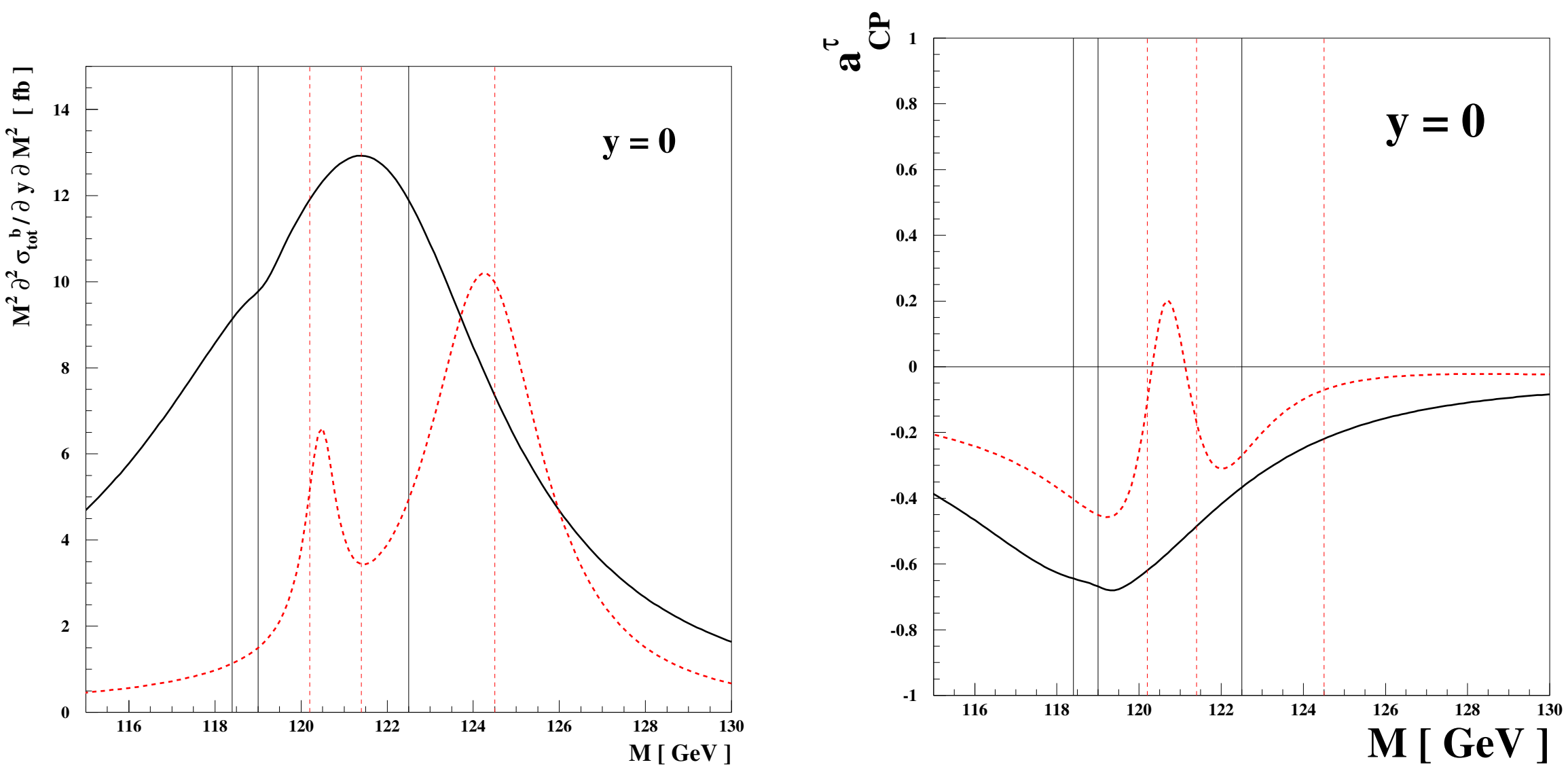

**Fig. 9:** Left panel: The hadron-level cross section for the double-diffractive producion of Higgs bosons decaying into $b$ quarks. CP-violating three-way mixing scenarios have been taken, with the gluino phase $\Phi_3 = -90°$ (solid lines) and $\Phi_3 = -10°$ (dotted line). The vertical lines indicate the three Higgs-boson pole-mass positions. Right panel: The CP-violating asymmetry $a^\tau_{\rm CP}$ observable in three-way mixing scenarios when Higgs bosons decay into $\tau$ leptons, using the same line styles [45].

## 9 Summary

We do not know what the LHC will find - maybe there will be no supersymmetry and we will observe mini-black-hole production instead! However, whatever the physics scenario, HERA physics will provide crucial inputs, for example via measuring the parton distributions that will be crucial for searches for new physics such as the Higgs boson, or via the observation of saturation effects that will be important for forward physics, or via measurements of diffraction.

Forward physics is a potentially exciting area of LHC physics that is not covered by the present detectors. HERA and RHIC suggest that parton saturation and the Colour Glass Condensate may be observable here, understanding of forward physics is essential for the modelling of cosmic-ray air showers and hence determining the spectrum of ultra-high-energy cosmic rays, and diffractive events related to those observed by HERA may be a valuable tool for discovering new physics such as Higgs production. There is still plenty of room at the LHC for novel experimental contributions [46].

# Part II

# Working Group 1: Parton Density Functions

## List of participants in working group

S. Forte, A. Glazov, S. Moch, S. Alekhin, G. Altarelli, J. Andersen, R. D. Ball, J. Blümlein, H. Böttcher, T. Carli, M. Ciafaloni, D. Colferai, A. Cooper-Sarkar, G. Corcella, L. Del Debbio, G. Dissertori, J. Feltesse, A. Guffanti, C. Gwenlan, J. Huston, G. Ingelman, M. Klein, T. Laštovička, G. Laštovička-Medin, J. I. Latorre, L. Magnea, A. Piccione, J. Pumplin, V. Ravindran, B. Reisert, J. Rojo, A. Sabio Vera, G. P. Salam, F. Siegert, A. Staśto, H. Stenzel, C. Targett-Adams, R.S. Thorne, A. Tricoli, J.A.M. Vermaseren, A. Vogt

C. Anastasiou, M. Cacciari, G. Chachamis, V. Chekelian, J. Cole, T. Falter, S. Ferrag, E. Gardi, M. Grazzini, V. Guzey, B. Heinemann, H. Jung, V. Kolhinen, K. Kutak, E. Lobodzinska, M. Lublinsky, F. Maltoni, S. Munier, K. Peters, C. Salgado, T. Schorner, M. Skrzypek, G. Steinbrueck, J. Stirling, A. Tonazzo, H. Weigert, M. Whalley

# Introduction to Parton Distribution Functions

CONVENERS:
*M. Dittmar* [1], *S. Forte* [2], *A. Glazov* [3], *S. Moch* [4]
CONTRIBUTING AUTHORS:
*S. Alekhin* [5], *G. Altarelli* [6,7], *J. Andersen* [8], *R. D. Ball* [9], *J. Blümlein* [4], *H. Böttcher* [4], *T. Carli* [10], *M. Ciafaloni* [11], *D. Colferai* [11], *A. Cooper-Sarkar* [12], *G. Corcella* [6], *L. Del Debbio* [6,9], *G. Dissertori* [1], *J. Feltesse*[13], *A. Guffanti* [4], *C. Gwenlan* [12], *J. Huston* [14], *G. Ingelman* [15], *M. Klein* [4], *T. Laštovička* [10], *G. Laštovička-Medin* [17], *J. I. Latorre* [16], *L. Magnea* [18], *A. Piccione* [18], *J. Pumplin* [14], *V. Ravindran* [19], *B. Reisert* [20], *J. Rojo* [16], *A. Sabio Vera* [21], *G. P. Salam* [22], *F. Siegert* [10], *A. Staśto* [23], *H. Stenzel* [24], *C. Targett-Adams* [25], *R.S. Thorne* [8], *A. Tricoli*[12], *J.A.M. Vermaseren* [26], *A. Vogt* [27]

[1] Institute for Particle Physics, ETH-Zürich Hönggerberg, CH 8093 Zürich, Switzerland
[2] Dipartimento di Fisica, Universitá di Milano, INFN Sezione di Milano, Via Celoria 16, I 20133 Milan, Italy
[3] DESY, Notkestrasse 85, D 22603 Hamburg, Germany
[4] DESY, Platanenallee 6, D 15738 Zeuthen, Germany
[5] Institute for High Energy Physics, 142284 Protvino, Russia
[6] CERN, Physics Department, Theory Unit, CH 1211 Geneva 23, Switzerland
[7] Dipartimento di Fisica "E.Amaldi", Università Roma Tre and INFN, Sezione di Roma Tre, via della Vasca Navale 84, I 00146 Roma, Italy
[8] Cavendish Laboratory, University of Cambridge, Madingley Road, Cambridge, CB3 0HE, UK
[9] School of Physics, University of Edinburgh, Edinburgh EH9 3JZ, UK
[10] CERN, Physics Department, CH 1211 Geneva 23, Switzerland
[11] Dipartimento di Fisica, Università di Firenze and INFN, Sezione di Firenze, I 50019 Sesto Fiorentino, Italy
[12] Department of Physics, Nuclear and Astrophysics Lab., Keble Road, Oxford, OX1 3RH, UK
[13] DSM/DAPNIA, CEA, Centre d'Etudes de Saclay, F 91191 Gif-sur-Yvette, France
[14] Department of Physics and Astronomy, Michigan State University, E. Lansing, MI 48824, USA
[15] High Energy Physics, Uppsala University, Box 535, SE 75121 Uppsala, Sweden
[16] Departament d'Estructura i Constituents de la Matèria, Universitat de Barcelona, Diagonal 647, E 08028 Barcelona, Spain
[17] University of Podgorica, Cetinjski put bb, CNG 81000 Podgorica, Serbia and Montenegro
[18] Dipartimento di Fisica Teorica, Università di Torino and INFN Sezione di Torino, Via P. Giuria 1, I 10125 Torino, Italy
[19] Harish-Chandra Research Institute, Chhatnag Road, Jhunsi, Allahabad, India
[20] FNAL, Fermi National Accelerator Laboratory, Batavia, IL 60126, USA
[21] II. Institut für Theoretische Physik, Universität Hamburg, Luruper Chaussee 149, D 22761 Hamburg, Germany
[22] LPTHE, Universities of Paris VI and VII and CNRS, F 75005, Paris, France
[23] H. Niewodniczański Institute of Nuclear Physics, PL 31-342 Kraków, Poland
[24] II. Physikalisches Institut, Universität Giessen, Heinrich-Buff-Ring 16, D 35392 Giessen, Germany
[25] Department of Physics and Astronomy, UC London, Gower Street, London, WC1E 6BT, UK
[26] NIKHEF Theory Group, Kruislaan 409, NL 1098 SJ Amsterdam, The Netherlands
[27] IPPP, Department of Physics, Durham University, Durham DH1 3LE, UK

M. DITTMAR, S. FORTE, A. GLAZOV, S. MOCH, S. ALEKHIN, G. ALTARELLI, ...

**Abstract**

We provide an assessment of the impact of parton distributions on the determination of LHC processes, and of the accuracy with which parton distribution functions (PDFs) can be extracted from data, in particular from current and forthcoming HERA experiments. We give an overview of reference LHC processes and their associated PDF uncertainties, and study in detail $W$ and $Z$ production at the LHC. We discuss the precision which may be obtained from the analysis of existing HERA data, tests of consistency of HERA data from different experiments, and the combination of these data. We determine further improvements on PDFs which may be obtained from future HERA data (including measurements of $F_L$), and from combining present and future HERA data with present and future hadron collider data. We review the current status of knowledge of higher (NNLO) QCD corrections to perturbative evolution and deep-inelastic scattering, and provide reference results for their impact on parton evolution, and we briefly examine non-perturbative models for parton distributions. We discuss the state-of-the art in global parton fits, we assess the impact on them of various kinds of data and of theoretical corrections, by providing benchmarks of Alekhin and MRST parton distributions and a CTEQ analysis of parton fit stability, and we briefly present proposals for alternative approaches to parton fitting. We summarize the status of large and small $x$ resummation, by providing estimates of the impact of large $x$ resummation on parton fits, and a comparison of different approaches to small $x$ resummation, for which we also discuss numerical techniques.

The physics of parton distributions, especially within the context of deep-inelastic scattering (DIS), has been an active subject of detailed theoretical and experimental investigations since the origins of perturbative quantum chromodynamics (QCD), which, thanks to asymptotic freedom, allows one to determine perturbatively their scale dependence [1–5].

Since the advent of HERA, much progress has been made in determining the Parton Distribution Functions (PDFs) of the proton. A good knowledge of the PDFs is vital in order to make predictions for both Standard Model and beyond the Standard Model processes at hadronic colliders, specifically the LHC. Furthermore, PDFs must be known as precisely as possible in order to maximize the discovery potential for new physics at the LHC. Conversely, LHC data will lead to an improvement in the knowledge of PDFs.

The main aim of this document is to provide a state-of-the art assessment of the impact of parton distributions on the determination of LHC processes, and of the accuracy with which parton distributions can be extracted from data, in particular current and forthcoming HERA data.

In Ref. [6] we set the stage by providing an overview of relevant LHC processes and a discussion of their experimental and theoretical accuracy. In Ref. [7] we turn to the experimental determination of PDFs, and in particular examine the improvements to be expected from forthcoming measurements at HERA, as well as from analysis methods which allow one to combine HERA data with each other, and also with data from existing (Tevatron) and forthcoming (LHC) hadron colliders. In Ref. [8] we discuss the state of the art in the extraction of parton distributions of the data by first reviewing recent progress in higher-order QCD corrections and their impact on the extraction of PDFs, and then discussing and comparing the determination of PDFs from global fits. Finally, in Ref. [9] we summarize the current status of resummed QCD computations which are not yet used in parton fits, but could lead to an improvement in the theoretical precision of PDF determinations.

In addition to summarizing the state of the art, we also provide several new results, benchmarks and predictions obtained within the framework of the HERA–LHC workshop.

## Acknowledgements

A. Glazov thanks E. Rizvi, M. Klein, M. Cooper-Sarkar and C. Pascaud for help and many fruitful discussions. C. Gwenlan, A. Cooper-Sarka and C. Targett-Adams thank M. Klein and R. Thorne for providing the $F_L$ predictions, as well as for useful discussions. T. Carli, G. Salam and F. Siegert thank Z. Nagy, M. H. Seymour, T. Schörner-Sadenius, P. Uwer and M. Wobisch for useful discussions on the grid technique, A. Vogt for discussion on moment-space techniques and Z. Nagy for help and support with NLOJET++. R. Thorne thanks S. Alekhin for collaboration on the project of obtaining the benchmark parton distributions, for providing his benchmark partons and for many useful exchanges. J. Huston and J. Pumplin thank W.K. Tung and D. Stump for collaboration on the research work presented.

S. Moch acknowledges partial support by the Helmholtz Gemeinschaft under contract VH-NG-105 and by DFG Sonderforschungsbereich Transregio 9, Computergestützte Theoretische Physik. J. Blümlein acknowledges partial support by DFG Sonderforschungsbereich Transregio 9, Computergestützte Theoretische Physik. C. Gwenlan acknowledges support by PPARC. F. Siegert acknowledges support by the CERN Summer Student Programme. R.S. Thorne acknowledges support by the Royal Society as University Research Fellow. J. Huston and J. Pumplin acknowledge support by the National Science Foundation. J.R. Andersen acknowledges support from PPARC under contract PPA/P/S/2003/00281. A. Sabio Vera acknowledges support from the Alexander von Humboldt Foundation.

# LHC final states and their potential experimental and theoretical accuracies

*Amanda Cooper-Sarkar, Michael Dittmar, Günther Dissertori Claire Gwenlan, Hasko Stenzel, Alessandro Tricoli*

## 1 LHC final states and their potential experimental and theoretical accuracies [1]

### 1.1 Introduction

Cross section calculations and experimental simulations for many LHC reactions, within the Standard Model and for many new physics scenarios have been performed during the last 20 years. These studies demonstrate how various final states might eventually be selected above Standard Model backgrounds and indicate the potential statistical significance of such measurements. In general, these studies assumed that the uncertainties from various sources, like the PDF uncertainties, the experimental uncertainties and the various signal and background Monte Carlo simulations will eventually be controlled with uncertainties small compared to the expected statistical significance. This is the obvious approach for many so called discovery channels with clean and easy signatures and relatively small cross sections.

However, during the last years many new and more complicated signatures, which require more sophisticated selection criteria, have been discussed. These studies indicate the possibility to perform more ambitious searches for new physics and for precise Standard Model tests, which would increase the physics potential of the LHC experiments. Most of these studies concentrate on the statistical significance only and potential systematic limitations are rarely discussed.

In order to close this gap from previous LHC studies, questions related to the systematic limits of cross section measurements from PDF uncertainties, from imperfect Standard Model Monte Carlo simulations, from various QCD uncertainties and from the efficiency and luminosity uncertainties were discussed within the PDF working group of this first HERA-LHC workshop. The goal of the studies presented during the subgroup meetings during the 2004/5 HERA LHC workshop provide some answers to questions related to these systematic limitations. In particular, we have discussed potential experimental and theoretical uncertainties for various Standard Model signal cross sections at the LHC. Some results on the experimental systematics, on experimental and theoretical uncertainties for the inclusive W, Z and for diboson production, especially related to uncertainties from PDF's and from higher order QCD calculations are described in the following sections.

While it was not possible to investigate the consequences for various aspects of the LHC physics potential in detail, it is important to keep in mind that many of these Standard Model reactions are also important backgrounds in the search for the Higgs and other exotic phenomena. Obviously, the consequences from these unavoidable systematic uncertainties need to be investigated in more detail during the coming years.

### 1.2 Measuring and interpreting cross sections at the LHC [2]

The LHC is often called a machine to make discoveries. However, after many years of detailed LHC simulations, it seems clear that relatively few signatures exist, which do not involve cross section measurements for signals and the various backgrounds. Thus, one expects that cross section measurements for a large variety of well defined reactions and their interpretation within or perhaps beyond the Standard Model will be one of the main task of the LHC physics program.

While it is relatively easy to estimate the statistical precision of a particular measurement as a function of the luminosity, estimates of potential systematic errors are much more complicated. Furthmore, as almost nobody wants to know about systematic limitations of future experiments, detailed studies are not rewarding. Nevertheless, realistic estimates of such systematic errors are relevant, as they might allow the LHC community to concentrate their efforts on the areas where current systematic errors, like the ones which are related to uncertainties from Parton Distribution Functions (PDF) or the ones from missing higher order QCD calculations, can still be improved during the next years.

[1]Subsection coordinator: Michael Dittmar
[2]Contributing author: Michael Dittmar

In order to address the question of systematics, it is useful to start with the basics of cross section measurements. Using some clever criteria a particular signature is separated from the data sample and the surviving $N_{\mathrm{observed}}$ events can be counted. Backgrounds, $N_{\mathrm{background}}$, from various sources have to be estimated either using the data or some Monte Carlo estimates. The number of signal events, $N_{\mathrm{signal}}$, is then obtained from the difference. In order to turn this experimental number of signal events into a measurement one has to apply a correction for the efficiency. This experimental number can now be compared with the product of the theoretical production cross section for the considered process and the corresponding Luminosity. For a measurement at a hadron collider, like the LHC, processes are calculated on the basis of quark and gluon luminosities which are obtained from the proton-proton luminosity “folded” with the PDF’s.

In order to estimate potential systematic errors one needs to examine carefully the various ingredients to the cross section measurement and their interpretation. First, a measurement can only be as good as the impact from of the background uncertainties, which depend on the optimized signal to background ratio. Next, the experimental efficiency uncertainty depends on many subdetectors and their actual real time performance. While this can only be known exactly from real data, one can use the systematic error estimates from previous experiments in order to guess the size of similar error sources for the future LHC experiments. We are furthermore confronted with uncertainties from the PDF’s and from the proton-proton luminosity. If one considers all these areas as essentially experimental, then one should assign uncertainties originating from imperfect knowledge of signal and background cross sections as theoretical.

Before we try to estimate the various systematic errors in the following subsections, we believe that it is important to keep in mind that particular studies need not to be much more detailed than the largest and limiting uncertainty, coming from either the experimental or the theoretical area. Thus, one should not waste too much time, in order to achieve the smalled possible uncertainty in one particular area. Instead, one should try first to reduce the most important error sources and if one accepts the “work division” between experimental and theoretical contributions, then one should simply try to be just a little more accurate than either the theoretical or the experimental colleagues.

### *1.2.1 Guessing experimental systematics for ATLAS and CMS*

In order to guess experimental uncertainties, without doing lengthy and detailed Monte Carlo studies, it seems useful to start with some simple and optimistic assumptions about ATLAS and CMS[3].

First of all, one should assume that both experiments can actually operate as planned in their proposals. As the expected performance goals are rather similar for both detectors the following list of measurement capabilities looks as a reasonable first guess.

– Isolated electrons, muons and photons with a transverse momentum above 20 GeV and a pseudorapidity $\eta$ with $|\eta| \leq 2.5$ are measured with excellent accuracy and high (perhaps as large as 95% for some reactions) “homogeneous” efficiency. Within the pseudo rapidity coverage one can assume that experimentalists will perhaps be able, using the large statistics from leptonic W and Z decays, to control the efficiency for electrons and muons with a 1% accuracy. For simplicity, one can also assume that these events will allow to control measurements with high energy photons to

[3]Up to date performance of the ATLAS and CMS detectors and further detailed references can be found on the corresponding homepages `http://atlas.web.cern.ch/Atlas/` and `http://cmsinfo.cern.ch/Welcome.html/`

a similar accuracy. For theoretical studies one might thus assume that high $p_t$ electrons, muons and photons and $|\eta| \leq 2.5$ are measured with a systematic uncertainty of $\pm$ 1% for each lepton (photon).

- Jets are much more difficult to measure. Optimistically one could assume that jets can be seen with good efficiency and angular accuracy if the jet transverse momentum is larger than 30 GeV and if their pseudo rapidity fulfills $|\eta| \leq 4.5$. The jet energy resolution is not easy to quantify, but numbers could be given using some "reasonable" assumptions like $\Delta E/E \approx 100 - 150\%/\sqrt{E}$. For various measurements one want to know the uncertainty of the absolute jet energy scale. Various tools, like the decays of $W \to q\bar{q}$ in $t\bar{t}$ events or the photon-jet final state, might be used to calibrate either the mean value or the maximum to reasonably good accuracy. We believe that only detailed studies of the particular signature will allow a quantitative estimate of the uncertainties related to the jet energy scale measurements.
- The tagging of b–flavoured jets can be done, but the efficiency depends strongly on the potential backgrounds. Systematic efficiency uncertainties for the b–tagging are difficult to quantify but it seems that, in the absence of a new method, relative b-tagging uncertainties below $\pm$ 5% are almost impossible to achieve.

With this baseline LHC detector capabilities, it seems useful to divide the various high $q^2$ LHC reactions into essentially five different non overlapping categories. Such a devision can be used to make some reasonable accurate estimates of the different systematics.

- Drell–Yan type lepton pair final states. This includes on– and off–shell W and Z decays.
- $\gamma$–jet and $\gamma\gamma X$ final states.
- Diboson events of the type $WW$, $WZ$, $ZZ$, $W\gamma$ with leptonic decays of the $W$ and $Z$ bosons. One might consider to include the Standard Model Higgs signatures into this group of signatures.
- Events with top quarks in the final state, identified with at least one isolated lepton.
- Hadronic final states with up to n(=2,3 ..) Jets and different $p_t$ and mass.

With this "grouping" of experimental final states, one can now start to analyze the different potential error sources. Where possible, one can try to define and use relative measurements of various reactions such that some systematic errors will simply cancel.

Starting with the resonant W and Z production with leptonic decays, several million of clean events will be collected quickly, resulting in relative statistical errors well below $\pm 1\%$. Theoretical calculations for these reactions are well advanced and these reactions are among the best understood LHC final states allowing to build the most accurate LHC Monte Carlo generators. Furthermore, some of the experimental uncertainties can be reduced considerably if ratio measurements of cross section, such as $W^+/W^-$ and $Z/W$, are performed. The similarities in the production mechanism should also allow to reduce theoretical uncertainties for such ratios. The experimental counting accuracy of W and Z events, which includes background and efficiency corrections, might achieve eventually uncertainties of 1% or slightly better for cross section ratios.

Furthermore, it seems that the shape of the $p_t$ distribution of the Z, using the decay into electron pairs ($pp \to ZX \to e{+}e^- X$), can be determined with relative accuracies of much less than 1%. This distribution, shown in figure 1, can be used to tune the Monte Carlo description of this particular process. This tuning of the Monte Carlo can than be used almost directly to predict theoretically also the W $p_t$ spectrum, and the $p_t$ spectrum for high mass Drell-Yan lepton pair events. Once an accurate model description of these Standard Model reactions is achieved one might use these insights also to predict the $p_t$ spectrum of other well defined final states.

From all the various high $q^2$ reactions, the inclusive production of W and Z events is known to be the theoretically best understood and best experimentally measurable LHC reaction. Consequently, the idea to use these simple well defined final states as the LHC cross section normalisation tool, or standard candle was described first in reference [1]. This study indicated that the W and Z production might result in a precise and simple parton luminosity monitor. In addition, these reactions can also be used to improve the relative knowledge of the PDF's. In fact, if one gives up on the idea to measure absolute cross sections, the relative parton luminosity can in principle be determined with relative uncertainties well below $\pm5\%$, the previously expected possible limit for any absolut proton-proton luminosity normalisation procdure.

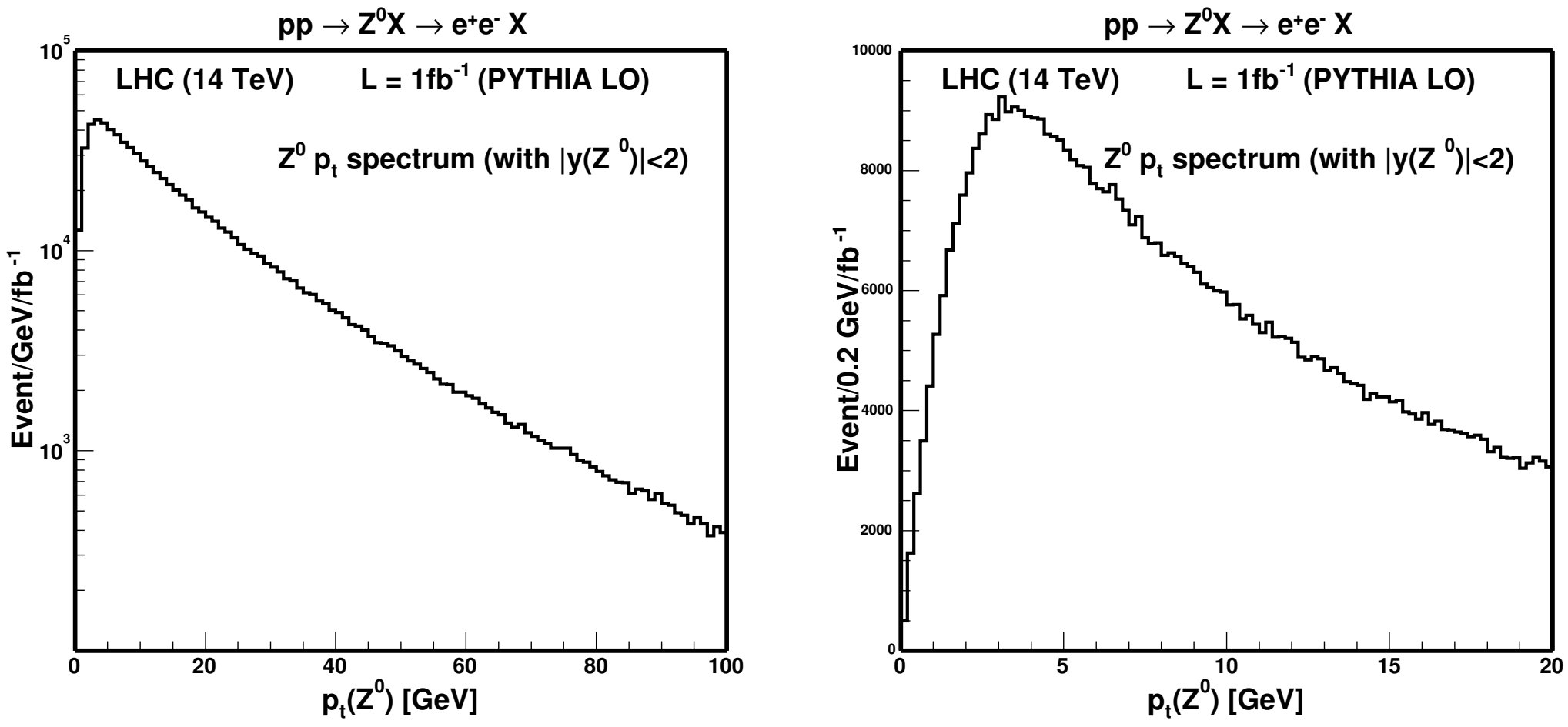

**Fig. 1:** Simple simulation of a potential measurement of the Z $p_t$ spectrum, possible with a luminosity of only 1 fb$^{-1}$. Who will be able to predict this $p_t$ spectrum in all its beauty and with similar accuracy?

In summary, one can estimate that it should be possible to reduce experimental uncertainties for Drell-Yan processes to systematic uncertainties below $\pm5\%$, optimistically one might envisage an event counting accuracy of perhaps $\pm1\%$, limited mainly from the lepton identification efficiency.

The next class of final states, which can be measured exclusively with leptons, are the diboson pair events with subsequent leptonic decays. Starting with the ZZ final state, we expect that the statistical accuracy will dominate the measurement for several years. Nevertheless, the systematic uncertainties of the measurement, based on four leptons, should in principle be possible with relative errors of a few % only.

The production of WZ and WW involves unmeasurable neutrinos. Thus, experimentally only an indirect and incomplete determination of the kinematics of the final states is possible and very detailed simulations with precise Monte Carlo generators are required for the interpretation of these final state. It seems that a measurement of the event counting with an accuracy below $\pm5\%$, due to efficiency uncertainties from the selection alone, to be highly non trivial. Nevertheless, if the measurements and the interpretations can be done relative to the W and Z resonance production, some uncertainties from the lepton identification efficiency, from the PDF and from the theoretical calculation can perhaps be reduced. Without going into detailed studies for each channel, one could try to assume that a systematic uncertainty of $\pm5\%$ might be defined as a goal. Similar characteristics and thus limitations can be expected for other diboson signatures.

The production cross section of top antitop quark pairs is large and several million of semileptonic tagged and relatively clean events ($pp \rightarrow t\bar{t} \rightarrow WbWb$ identified with one leptonic $W$ decay) can be expected. However, the signature involves several jets, some perhaps tagged as b–flavoured, and missing transverse momentum from the neutrino(s). The correct association of the various jets to the corresponding top quark is known to be extremely difficult, leading to large combinatorial backgrounds. Thus, it seems that, even if precise Monte Carlo generators will become eventually available, that systematic uncertainties smaller than 5-10% should not be expected. Consequently,we assume that top antitop backgrounds for a wide class of signals can not be determined with uncertainties smaller than 5-10%.

Measurements of so called “single” top quarks are even more difficult, as the cross section is smaller and larger backgrounds exist. Systematic errors will therefore always be larger than the one guessed for top-antitop pair production.

Finally, we can address the QCD jet production. Traditionally one measures and interprets the so called jet cross section as a function of $p_t$ jet and the mass of the multi jet system using various rapidity intervals. With the steeply falling $p_t$ jet spectrum and essentially no background, one will determine the differential spectrum such that only the slope has to be measured with good relative accuracy. If one is especially interested into the super high mass or high $p_t$ events, then we expect that migrations due to jet mis-measurements and non Gaussian tails in the jet energy measurements will limit any measurement. A good guess might be that the LHC experiments can expect absolut normalisation uncertainties similar to the ones achieved with CDF and D0, corresponding to uncertainties of about $\pm$ 10-20%.

Are the above estimated systematic limits for the various measurements pessimistic, optimistic or simply realistic? Of course, only real experiments will tell during the coming LHC years. However, while some of these estimates will need perhaps some small modification, they could be used as a limit waiting to be improved during the coming years. Thus, some people full of ideas might take these numbers as a challenge, and discover and develop new methods that will improve these estimates. This guess of systematic limitations for LHC experiments could thus be considered as a “provocation”, which will stimulate activities to prove them wrong. In fact, if the experimental and theoretical communities could demonstrate why some of these “pessimistic” numbers are wrong the future real LHC measurements will obviously benefit from the required efforts to develope better Monte Carlo programs and better experimental methods.

The following summary from a variety of experimental results from previous high energy collider experiments might help to quantify particular areas of concern for the LHC measurements. These previous measurements can thus be used as a starting point for an LHC experimenter, who can study and explain why the corresponding errors at LHC will be smaller or larger.

### *1.2.2 Learning from previous collider experiments*

It is broadly accepted, due to the huge hadronic interaction rate and the short bunch crossing time, that the experimental conditions at the LHC will be similar or worse than the ones at the Tevatron collider. One experimental answer was to improve the granularity, speed and accuracy of the different detector elements accordingly. Still, no matter how well an experiment can be realized, the LHC conditions to do experiments will be much more difficult than at LEP or any hypothetical future high energy $e^+e^-$ collider. One important reason is the large theoretical uncertainty, which prevents to make signal and background Monte Carlos with accuracies similar to the ones which were used at LEP.

Thus, we can safely expect that systematic errors at LHC experiments will be larger than the corresponding ones from LEP and that the Tevatron experience can be used as a first guess.

- Measurement of $\sigma\times$BR for W and Z production from CDF [2] and D0 [3]:
  The CDF collaboration has presented a high statistics measurement with electrons and muons. Similar systematic errors of about $\pm$ 2% were achieved for efficiency and thus the event counting with electrons and muons. The error was reduced to $\pm$ 1.4% for the ratio measurement where some lepton identification efficiencies cancel. Similar errors about $\times$ 1.5-2 larger have been obtained by the corresponding measurements from the D0 experiment.
- Measurement of the cross section for $p\bar{p} \rightarrow Z\gamma(\gamma)$ from D0 [4]:

A total of 138 $ee\gamma$ and 152 $\mu\mu\gamma$ candidate events were selected. The background was estimated to be about 10% with a systematic uncertainty of ± 10-15%, mainly from $\gamma$-jet misidentification. Using Monte Carlo and a large sample of inclusive Z events, the efficiency uncertainty has been estimated to be $\approx 5\%$ and when the data were used in comparison with the Standard Model prediction another uncertainty of 3.3% originating from PDF's was added.

- Measurement of the $p\bar{p} \to t\bar{t}$ production cross section from CDF [5]
  A recent CDF measurement, using 197 pb$^{-1}$, obtained a cross section (in pb) of 7.0 +2.4 (-2.1) from statistics. This should be comapred with +1.7 (-1.2) from systematics, which includes $\pm 0.4$ from the luminosity measurement. Thus, uncertainties from efficiency and background are roughly $\pm 20\%$. It is expected that some of the uncertainties can be reduced with the expected 10 fold luminosity increase such that the systematic error will eventually decrease to about ± 10%, sufficient to be better than the expected theoretical error of ± 15%.
- A search for Supersymmetry with b-tagged jets from CDF [6]:
  This study, using single and double b-tagged events was consistent with background only. However, it was claimed that the background uncertainty was dominated by the systematic error, which probably originated mostly from the b tagging efficiency and the misidentification of b-flavoured jets. The numbers given were 16.4± 3.7 events (3.15 from systematics) for the single b-tagged events and 2.6±0.7 events (0.66 from systematics) for the double b-tagged events. These errors originate mainly from the b-tagging efficiency uncertainties, which are found to roughly ± 20-25% for this study of rare events.
- Some "random" selection of recent $e^+e^-$ measurements:
  A recent measurement from ALEPH (LEP) of the W branching ratio to $q\bar{q}$ estimated a systematic uncertainty of about ± 0.2% [7]. This small uncertainty was possible because many additional constraints could be used.
  OPAL has reported a measurement of $R_b$ at LEP II energies, with a systematic uncertainty of ± 3.7%. Even though this uncertainty could in principle be reduced with higher statistics, one can use it as an indication on how large efficiency uncertainties from b-tagging are already with clean experimental conditions [8]
  Recently, ALEPH and DELPHI have presented cross section measurements for $e^+e^- \to \gamma\gamma$ with systematic errors between 2.2% (ALEPH) [9] and 1.1% (DELPHI) [10]. In both cases, the efficiency uncertainty, mainly from conversions, for this in principle easy signal was estimated to be roughly 1%. In the case of ALEPH an uncertainty of about ±0.8% was found for the background correction.

Obviously, these measurements can only be used, in absence of anything better, as a most optimistic guess for possible systematic limitations at a hadron collider. One might conclude that the systematics from LEP experiments give (1) an optimistic limit for comparable signatures at the LHC and (2) that the results from CDF and D0 should indicate systematics which might be obtained realistically during the early LHC years.

Thus, in summary the following list might be used as a first order guess on achievable LHC systematics[4].

- "Isolated" muons, electrons and photons can be measured with a small momentum (energy) uncertainty and with an almost perfect angular resolution. The efficiency for $p_t \geq 20$ GeV and $|\eta| \leq 2.5$ will be "high" and can be controlled optimistically to ± 1%. Some straight forward selection criteria should reduce jet background to small or negligible levels.
- "Isolated" jets with a $p_t \geq 30$ GeV and $|\eta| \leq 4.5$ can be seen with high (veto) efficiency and a small uncertainty from the jet direction measurement. However, it will be very difficulty to

[4] Reality will hopefully show new brilliant ideas, which combined with hard work will allow to obtain even smaller uncertainties.

measure the absolute jet energy scale and Non-Gaussian tails will limit the systematics if the jet energy scale is important.

– Measurements of the missing transverse momentum depend on the final state but will in general be a sum of the errors from the lepton and the jet accuracies.

Using these assumptions, the following "optimistic" experimental systematic errors can be used as a guideline:

1. Efficiency uncertainties for isolated leptons and photons with a $p_t$ above 20 GeV can be estimated with a $\pm 1\%$ accuracy.
2. Efficiencies for tagging jets will be accurate to a few percent and the efficiency to tag b-flavoured jets will be known at best within $\pm 5\%$.
3. Backgrounds will be known, combining theoretical uncertainties and some experimental determinations, at best with a $\pm$5-10% accuracy. Thus, discovery signatures without narrow peaks require signal to background ratios larger than 0.25-0.5, if 5 $\sigma$ discoveries are claimed. Obviously, for accurate cross section measurements, the signal to background ratio should be much larger.
4. In case of ratio measurements with isolated leptons, like $pp \to W^+/pp \to W^-$, relative errors between 0.5-1% should be possible. Furthermore, it seems that the measurement of the shape of Z $p_t$ spectrum, using Z$\to e^+e^-$, will be possible with a systematic error much smaller than 1%. As the Z cross section is huge and clean we expect that this signature will become the best measurable final state and should allow to test a variety of production models with errors below $\pm$ 1%, thus challenging future QCD calculations for a long time.

## 1.3 Uncertainties on $W$ and $Z$ production at the LHC[5]

### 1.3.1 *Introduction*

At leading order (LO), $W$ and $Z$ production occur by the process, $q\bar{q} \to W/Z$, and the momentum fractions of the partons participating in this subprocess are given by $x_{1,2} = \frac{M}{\sqrt{s}} \exp(\pm y)$, where $M$ is the centre of mass energy of the subprocess, $M = M_W$ or $M_Z$, $\sqrt{s}$ is the centre of mass energy of the reaction ($\sqrt{s} = 14$ TeV at the LHC) and $y = \frac{1}{2} \ln \frac{(E+p_l)}{(E-p_l)}$ gives the parton rapidity. The kinematic plane for LHC parton kinematics is shown in Fig. 2. Thus, at central rapidity, the participating partons have small momentum fractions, $x \sim 0.005$. Moving away from central rapidity sends one parton to lower $x$ and one to higher $x$, but over the measurable rapidity range, $|y| < 2.5$, $x$ values remain in the range, $10^{-4} < x < 0.1$. Thus, in contrast to the situation at the Tevatron, valence quarks are not involved, the scattering is happening between sea quarks. Furthermore, the high scale of the process $Q^2 = M^2 \sim 10,000$ GeV$^2$ ensures that the gluon is the dominant parton, see Fig. 2, so that these sea quarks have mostly been generated by the flavour blind $g \to q\bar{q}$ splitting process. Thus the precision of our knowledge of $W$ and $Z$ cross-sections at the LHC is crucially dependent on the uncertainty on the momentum distribution of the gluon.

HERA data have dramatically improved our knowledge of the gluon, as illustrated in Fig. 3, which shows $W$ and $Z$ rapidity spectra predicted from a global PDF fit which does not include the HERA data, compared to a fit including HERA data. The latter fit is the ZEUS-S global fit [11], whereas the former is a fit using the same fitting analysis but leaving out the ZEUS data. The full PDF uncertainties for both fits are calculated from the error PDF sets of the ZEUS-S analysis using LHAPDF [12] (see the contribution of M.Whalley to these proceedings). The predictions for the $W/Z$ cross-sections, decaying to the lepton decay mode, are summarised in Table 1. The uncertainties in the predictions for these cross-sections have decreased from $\sim 16\%$ pre-HERA to $\sim 3.5\%$ post-HERA. The reason for this can be seen clearly in Fig. 4, where the sea and gluon distributions for the pre- and post-HERA fits are shown for several different $Q^2$ bins, together with their uncertainty bands. It is the dramatically increased precision in the low-$x$ gluon PDF, feeding into increased precision in the low-$x$ sea quarks, which has led to the increased precision on the predictions for $W/Z$ production at the LHC.

[5]Contributing authors: Alessandro Tricoli, Amanda Cooper-Sarkar, Claire Gwenlan

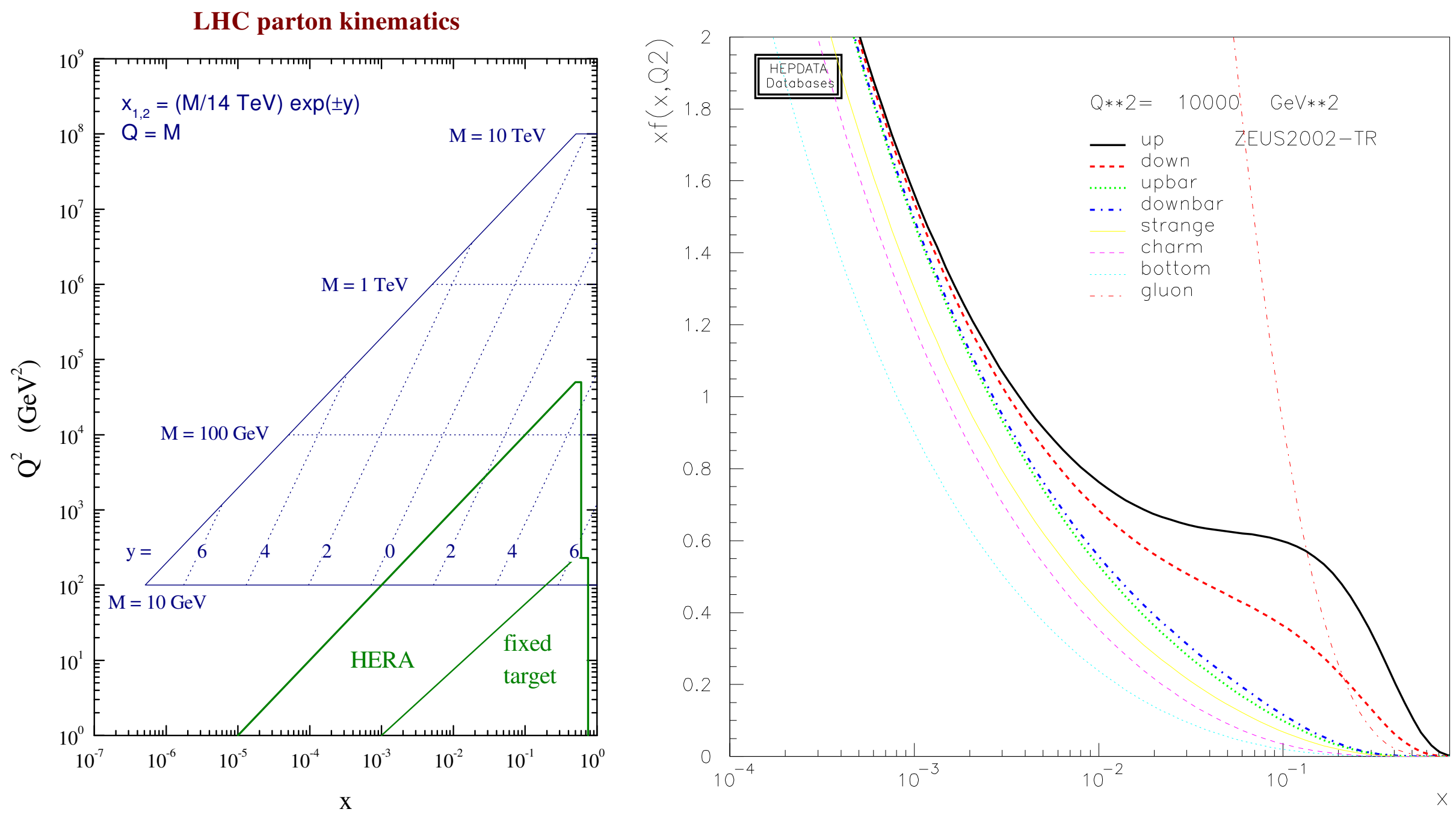

**Fig. 2:** Left plot: The LHC kinematic plane (thanks to James Stirling). Right plot: PDF distributions at $Q^2 = 10,000$ GeV$^2$.

**Table 1:** LHC $W/Z$ cross-sections for decay via the lepton mode, for various PDFs

| PDF Set | $\sigma(W^+).B(W^+ \to l^+\nu_l)$ | $\sigma(W^-).B(W^- \to l^-\bar{\nu}_l)$ | $\sigma(Z).B(Z \to l^+l^-)$ |
|---|---|---|---|
| ZEUS-S no HERA | $10.63 \pm 1.73$ nb | $7.80 \pm 1.18$ nb | $1.69 \pm 0.23$ nb |
| ZEUS-S | $12.07 \pm 0.41$ nb | $8.76 \pm 0.30$ nb | $1.89 \pm 0.06$ nb |
| CTEQ6.1 | $11.66 \pm 0.56$ nb | $8.58 \pm 0.43$ nb | $1.92 \pm 0.08$ nb |
| MRST01 | $11.72 \pm 0.23$ nb | $8.72 \pm 0.16$ nb | $1.96 \pm 0.03$ nb |

Further evidence for the conclusion that the uncertainties on the gluon PDF at the input scale ($Q_0^2 = 7$ GeV$^2$, for ZEUS-S) are the major contributors to the uncertainty on the $W/Z$ cross-sections at $Q^2 = M_W(M_Z)$, comes from decomposing the predictions down into their contributing eigenvectors. Fig 5 shows the dominant contributions to the total uncertainty from eigenvectors 3, 7, and 11 which are eigenvectors which are dominated by the parameters which control the low-$x$, mid-$x$ and high-$x$, gluon respectively.

The post-HERA level of precision illustrated in Fig. 3 is taken for granted in modern analyses, such that $W/Z$ production have been suggested as 'standard-candle' processes for luminosity measurement. However, when considering the PDF uncertainties on the Standard Model (SM) predictions it is necessary not only to consider the uncertainties of a particular PDF analysis, but also to compare PDF analyses. Fig. 6 compares the predictions for $W^+$ production for the ZEUS-S PDFs with those of the CTEQ6.1 [13] PDFs and the MRST01 [14] PDFs[6]. The corresponding $W^+$ cross-sections, for decay to leptonic mode

[6]MRST01 PDFs are used because the full error analysis is available only for this PDF set.

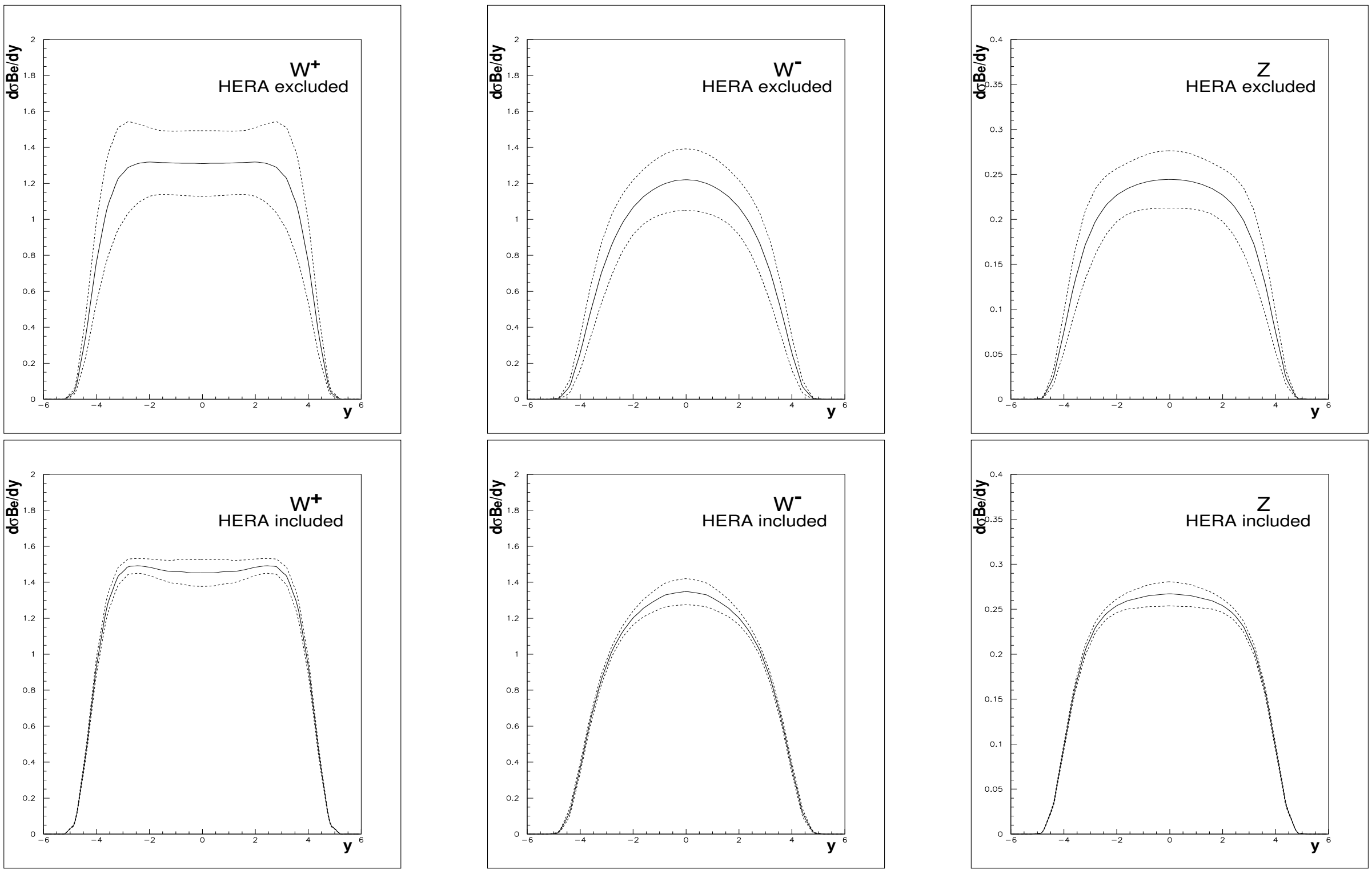

**Fig. 3:** LHC $W^+, W^-, Z$ rapidity distributions and their PDF uncertainties (the full line shows the central value and the dashed lines show the spread of the uncertainty): Top Row: from the ZEUS-S global PDF analysis not including HERA data; left plot $W^+$; middle plot $W^-$; right plot $Z$: Bottom Row: from the ZEUS-S global PDF analysis including HERA data; left plot $W^+$; middle plot $W^-$; right plot $Z$

are given in Table 1. Comparing the uncertainty at central rapidity, rather than the total cross-section, we see that the uncertainty estimates are rather larger: $5.2\%$ for ZEUS-S; $8.7\%$ for CTEQ6.1M and about $3.6\%$ for MRST01. The difference in the central value between ZEUS-S and CTEQ6.1 is $3.5\%$. Thus the spread in the predictions of the different PDF sets is comparable to the uncertainty estimated by the individual analyses. Taking all of these analyses together the uncertainty at central rapidity is about $8\%$.

Since the PDF uncertainty feeding into the $W^+, W^-$ and $Z$ production is mostly coming from the gluon PDF, for all three processes, there is a strong correlation in their uncertainties, which can be removed by taking ratios. Fig. 7 shows the $W$ asymmetry

$$A_W = (W^+ - W^-)/(W^+ + W^-).$$

for CTEQ6.1 PDFs, which have the largest uncertainties of published PDF sets. The PDF uncertainties on the asymmetry are very small in the measurable rapidity range. An eigenvector decomposition indicates that sensitivity to high-$x$ $u$ and $d$ quark flavour distributions is now evident at large $y$. Even this residual flavour sensitivity can be removed by taking the ratio

$$A_{ZW} = Z/(W^+ + W^-)$$

as also shown in Fig. 7. This quantity is almost independent of PDF uncertainties. These quantities have been suggested as benchmarks for our understanding of Standard Model Physics at the LHC. However, whereas the $Z$ rapidity distribution can be fully reconstructed from its decay leptons, this is not possible for the $W$ rapidity distribution, because the leptonic decay channels which we use to identify the $W$'s

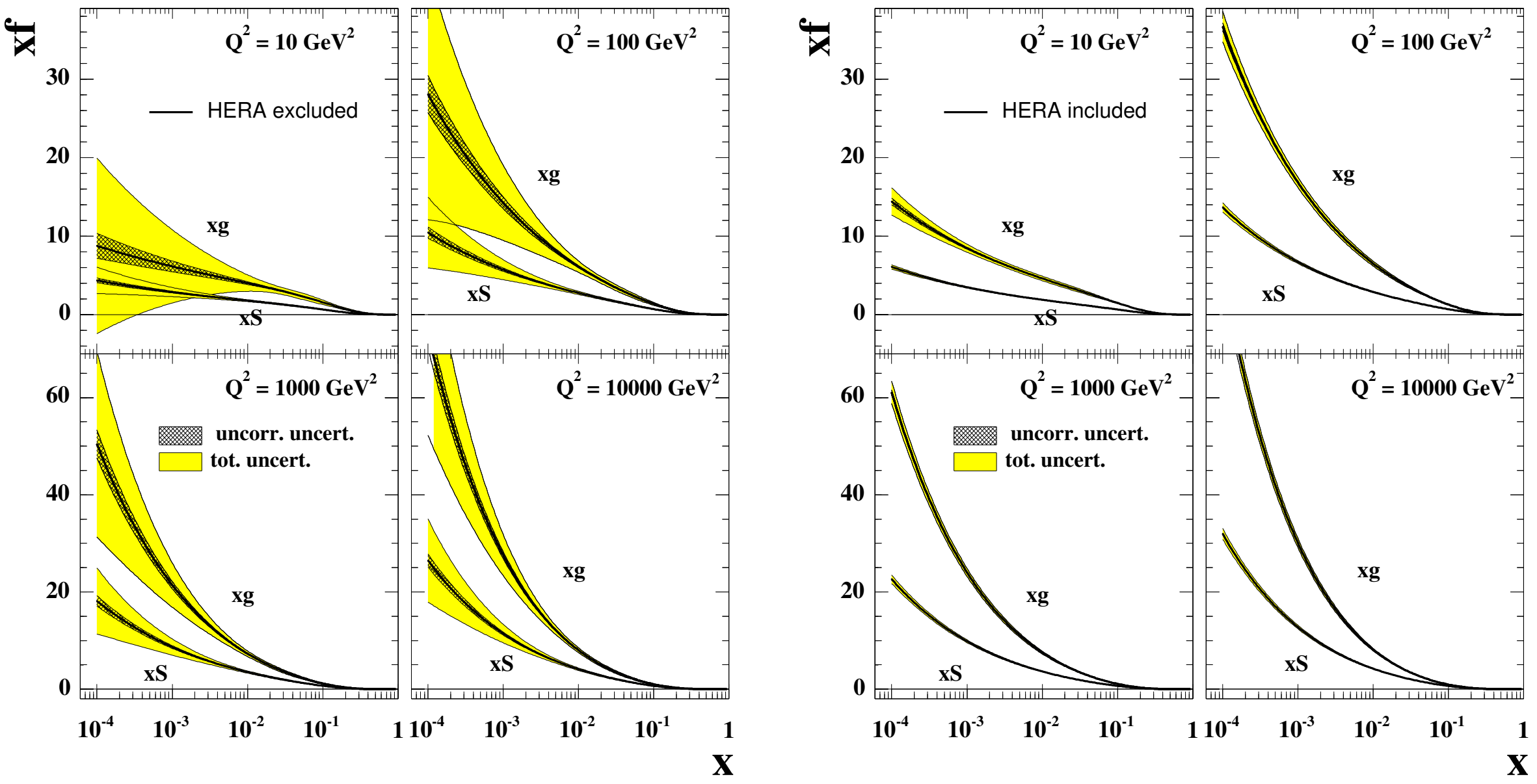

**Fig. 4:** Sea ($xS$) and gluon ($xg$) PDFs at various $Q^2$: left plot; from the ZEUS-S global PDF analysis not including HERA data; right plot: from the ZEUS-S global PDF analysis including HERA data. The inner cross-hatched error bands show the statistical and uncorrelated systematic uncertainty, the outer error bands show the total uncertainty including experimental correlated systematic uncertainties, normalisations and model uncertainty.

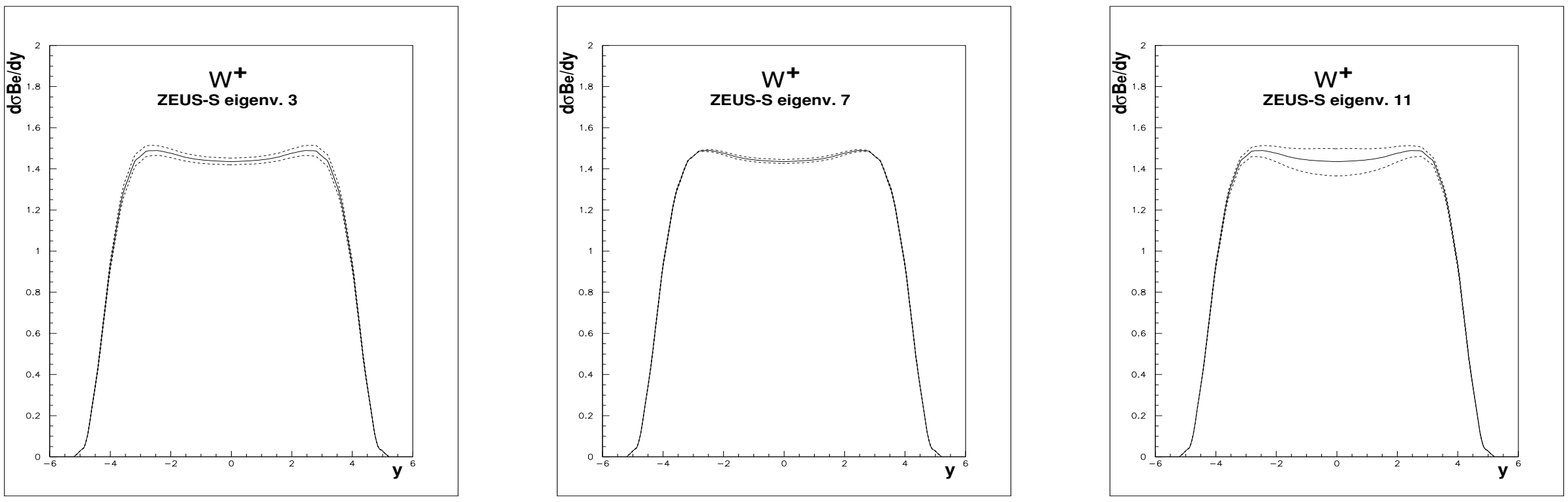

**Fig. 5:** LHC $W^+$ rapidity distributions and their PDF uncertainties due to the eigenvectors 3,7 and 11 of the ZEUS-S analysis.

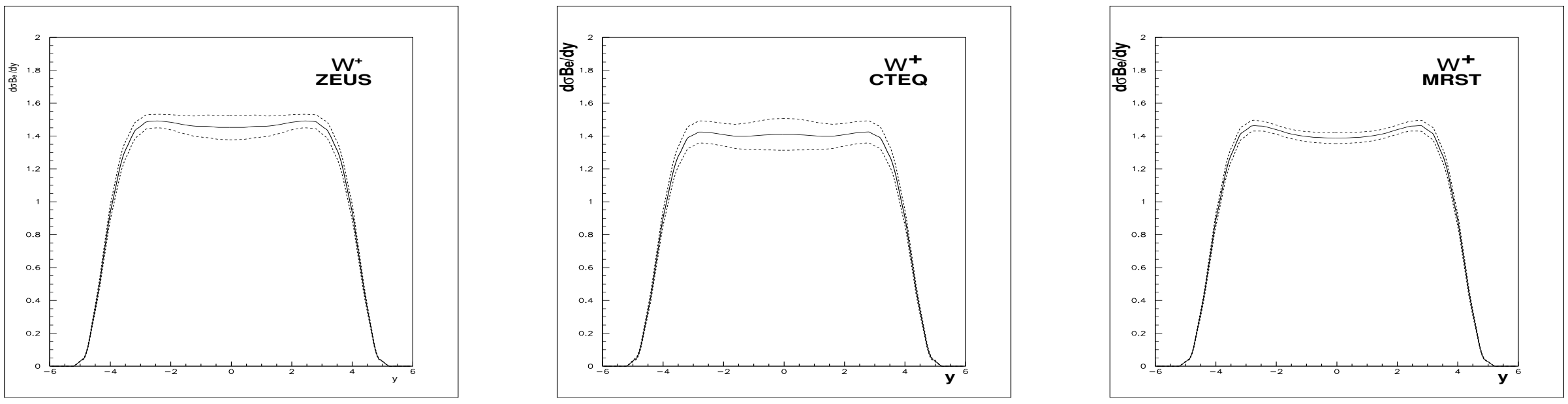

**Fig. 6:** LHC $W^+$ rapidity distributions and their PDF uncertainties: left plot, ZEUS-S PDFs; middle plot, CTEQ6.1 PDFs; right plot: MRST01 PDFs.

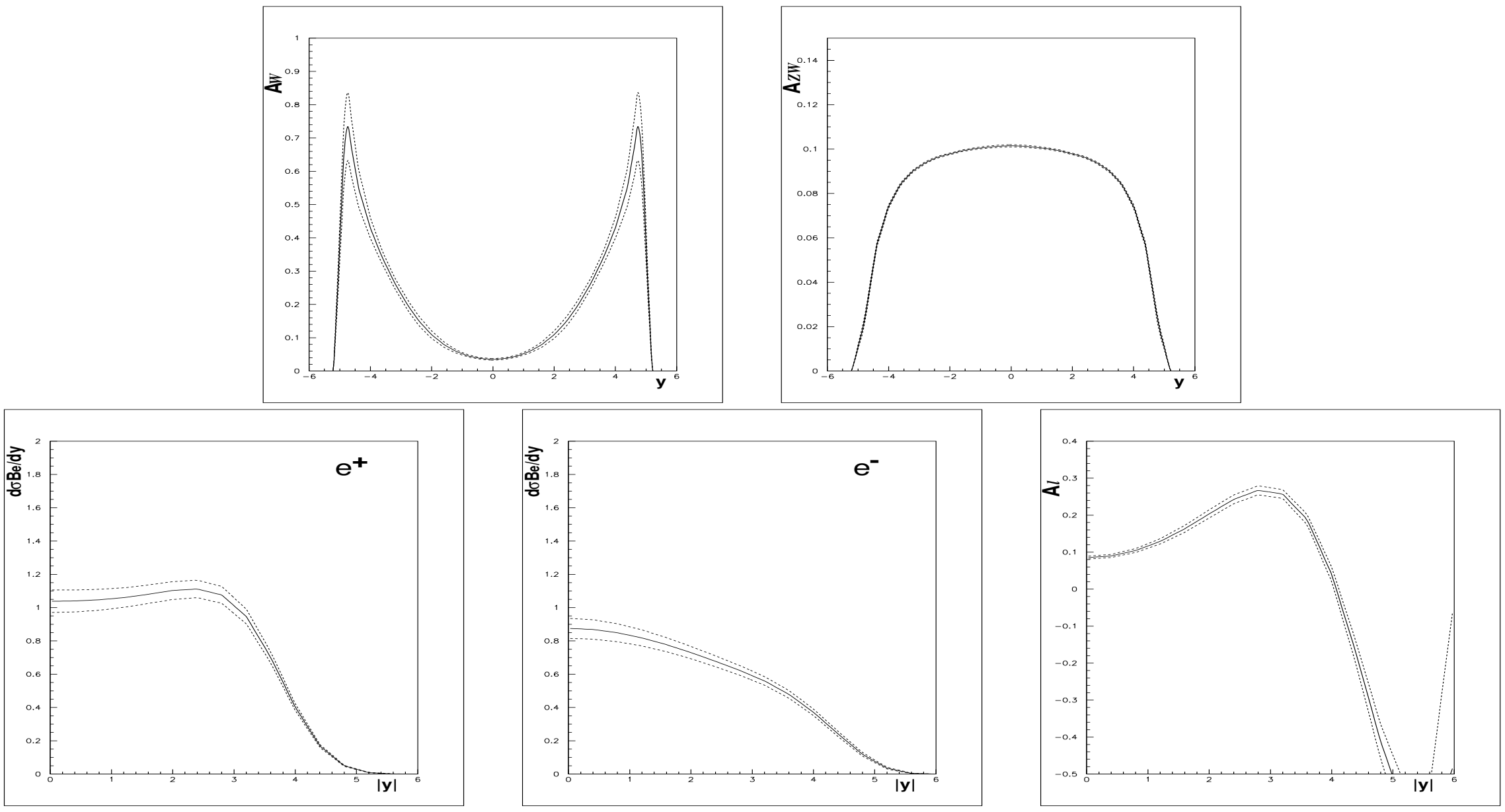

**Fig. 7:** Predictions for $W$, $Z$ production at the LHC from the CTEQ6.1 PDFs. Top row: left plot, the $W$ asymmetry, $A_W$; right plot, the ratio, $A_{ZW}$: Bottom row: left plot, decay $e^+$ rapidity spectrum; middle plot, decay $e^-$ rapidity spectrum; right plot, lepton asymmetry, $A_e$

have missing neutrinos. Thus we actually measure the $W$'s decay lepton rapidity spectra rather than the $W$ rapidity spectra. The lower half of Fig. 7 shows the rapidity spectra for positive and negative leptons from $W^+$ and $W^-$ decay and the lepton asymmetry,

$$A_l = (l^+ - l^-)/(l^+ + l^-).$$

A cut of, $p_{tl} > 25$ GeV, has been applied on the decay lepton, since it will not be possible to trigger on leptons with small $p_{tl}$. A particular lepton rapidity can be fed from a range of $W$ rapidities so that the contributions of partons at different $x$ values is smeared out in the lepton spectra, but the broad features of the $W$ spectra and the sensitivity to the gluon parameters remain. The lepton asymmetry shows the change of sign at large $y$ which is characteristic of the $V - A$ structure of the lepton decay. The cancellation of the uncertainties due to the gluon PDF is not so perfect in the lepton asymmetry as in the $W$ asymmetry. Nevertheless in the measurable rapidity range sensitivity to PDF parameters is small. Correspondingly, the PDF uncertainties are also small ( $4\%$) and this quantity provides a suitable Standard Model benchmark.

In summary, these preliminary investigations indicate that PDF uncertainties on predictions for the $W$, $Z$ rapidity spectra, using standard PDF sets which describe all modern data, have reached a precision of $\sim 8\%$. This may be good enough to consider using these processes as luminosity monitors. The predicted precision on ratios such as the lepton ratio, $A_l$, is better ($\sim 4\%$) and this measurement may be used as a SM benchmark. It is likely that this current level of uncertainty will have improved before the LHC turns on- see the contribution of C. Gwenlan ( [15]) to these proceedings. The remainder of this contribution will be concerned with the question: how accurately can we measure these quantities and can we use the early LHC data to improve on the current level of uncertainty?

### *1.3.2 k-factor and PDF re-weighting*

To investigate how well we can really measure $W$ production we need to generate samples of Monte-Carlo (MC) data and pass them through a simulation of a detector. Various technical problems arise.

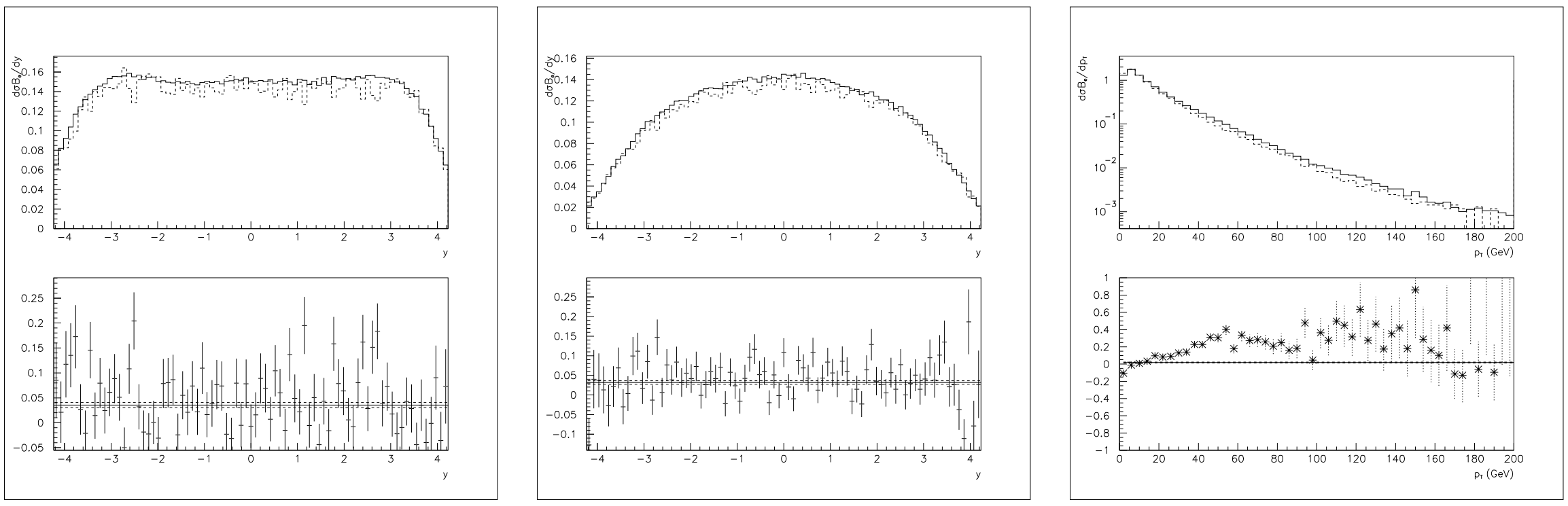

**Fig. 8:** Top Row: $W$ rapidity and $p_t$ spectra for events generated with HERWIG + k-Factors (full line), compared to those generated by MC@NLO (dashed line); left plot $W^+$ rapidity; middle plot $W^-$ rapidity; right plot $W^-$ $p_t$. Bottom row: the fractional differences of the spectra generated by HERWIG + k-factors and those generated by MC@NLO. The full line represents the weighted mean of these difference spectra and the dashed lines show its uncertainty

Firstly, many physics studies are done with HERWIG (6.505) [16], which generates events at LO with parton showers to account for higher order effects. Distributions can be corrected from LO to NLO by k-factors which are applied as a function of the variable of interest. The use of HERWIG is gradually being superceded by MC@NLO (2.3) [17] but this is not yet implemented for all physics processes. Thus it is necessary to investigate how much bias is introduced by using HERWIG with k-factors. Secondly, to simulate the spread of current PDF uncertainties, it is necessary to run the MC with all of the eigenvector error sets of the PDF of interest. This would be unreasonably time-consuming. Thus the technique of PDF reweighting has been investigated.

One million $W \to e\nu_e$ events were generated using HERWIG (6.505). This corresponds to 43 hours of LHC running at low luminosity, $10fb^{-1}$. The events are split into $W^+$ and $W^-$ events according to their Standard Model cross-section rates, $58\%$: $42\%$ (the exact split depends on the input PDFs). These events are then weighted with k-factors, which are analytically calculated as the ratio of the NLO to LO cross-section as a function of rapidity for the same input PDF [18]. The resultant rapidity spectra for $W^+, W^-$ are compared to rapidity spectra for $\sim$ 107,700 events generated using MC@NLO(2.3) in Fig 8[7]. The MRST02 PDFs were used for this investigation. The accuracy of this study is limited by the statistics of the MC@NLO generation. Nevertheless it is clear that HERWIG with k-factors does a good job of mimicking the NLO rapidity spectra. However, the normalisation is too high by $3.5\%$. This is not suprising since, unlike the analytic code, HERWIG is not a purely LO calculation, parton showering is also included. This normalisation difference is not too crucial since in an analysis on real data the MC will only be used to correct data from the detector level to the generator level. For this purpose, it is essential to model the shape of spectra to understand the effect of experimental cuts and smearing but not essential to model the overall normalisation perfectly. However, one should note that HERWIG with k-factors is not so successful in modelling the shape of the $p_t$ spectra, as shown in the right hand plot of Fig. 8. This is hardly surprising, since at LO the $W$ have no $p_t$ and non-zero $p_t$ for HERWIG is generated by parton showering, whereas for MC@NLO non-zero $p_t$ originates from additional higher order processes which cannot be scaled from LO, where they are not present.

Suppose we generate $W$ events with a particular PDF set: PDF set 1. Any one event has the hard scale, $Q^2 = M_W^2$, and two primary partons of flavours $flav_1$ and $flav_2$, with momentum fractions $x_1, x_2$ according to the distributions of PDF set 1. These momentum fractions are applicable to the hard process before the parton showers are implemented in backward evolution in the MC. One can then evaluate the probability of picking up the same flavoured partons with the same momentum fractions from an alternative PDF set, PDF set 2, at the same hard scale. Then the event weight is given by

$$\mathrm{PDF(re-weight)} = \frac{\mathrm{f_{PDF_2}(x_1, flav_1, Q^2).f_{PDF_2}(x_2, flav_2, Q^2)}}{\mathrm{f_{PDF_1}(x_1, flav_1, Q^2).f_{PDF_1}(x_2, flav_2, Q^2)}} \tag{1}$$

where $xf_{PDF}(x, flav, Q^2)$ is the parton momentum distribution for flavour, $flav$, at scale, $Q^2$, and momentum fraction, $x$. Fig. 9 compares the $W^+$ and $W^-$ spectra for a million events generated using MRST02 as PDF set 1 and re-weighting to CTEQ6.1 as PDF set 2, with a million events which are directly generated with CTEQ6.1. Beneath the spectra the fractional difference between these distributions is shown. These difference spectra show that the reweighting is good to better than $1\%$, and there is no evidence of a $y$ dependent bias. This has been checked for reweighting between MRST02, CTEQ6.1 and ZEUS-S PDFs. Since the uncertainties of any one analysis are similar in size to the differences between the analyses it is clear that the technique can be used to produce spectra for the eigenvector error PDF sets of each analysis and thus to simulate the full PDF uncertainties from a single set of MC generated events. Fig. 9 also shows a similar comparison for $p_t$ spectra.

[7] In MC@NLO the hard emissions are treated by NLO computations, whereas soft/collinear emissions are handled by the MC simulation. In the matching procedure a fraction of events with negative weights is generated to avoid double counting. The event weights must be applied to the generated number of events before the effective number of events can be converted to an equivalent luminosity. The figure given is the effective number of events.

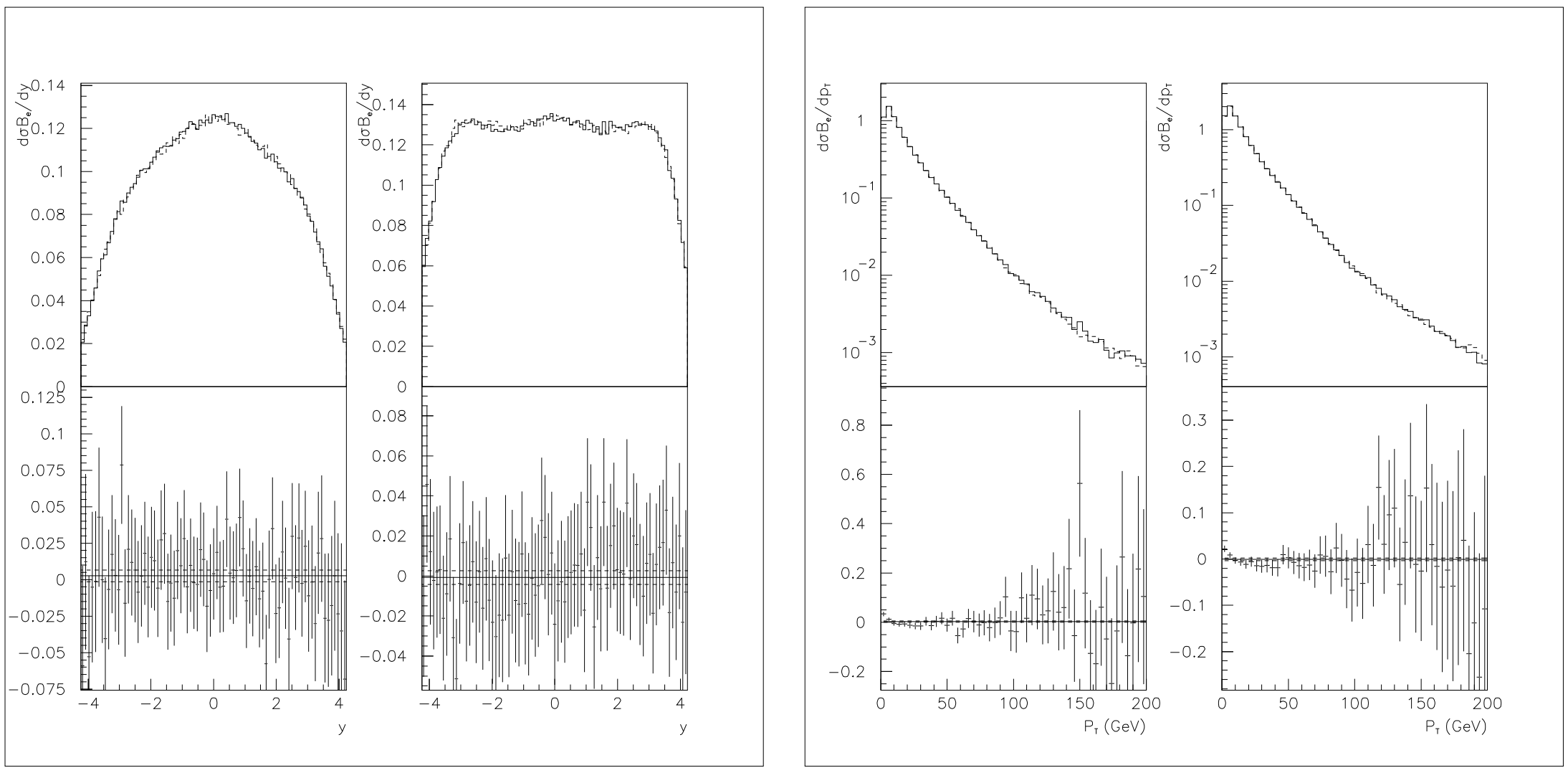

**Fig. 9:** Left side: $W^-$ (left) and $W^+$ (right) rapidity spectra, for events generated with MRST02 PDFs reweighted to CTEQ6.1 PDFs (full line), compared to events generated directly with CTEQ6.1 PDFs (dashed line). The fractional difference between these spectra are also shown beneath the plots. The full line represents the weighted mean of these difference spectra and the dashed lines show its uncertainty. Right side: the same for $p_t$ spectra.

### *1.3.3 Background Studies*

To investigate the accuracy with which $W$ events can be measured at the LHC it is necessary to make an estimate of the importance of background processes. We focus on $W$ events which are identified through their decay to the $W \to e\, \nu_e$ channel. There are several processes which can be misidentified as $W \to e\nu_e$. These are: $W \to \tau\nu_\tau$, with $\tau$ decaying to the electron channel; $Z \to \tau^+\tau^-$ with at least one $\tau$ decaying to the electron channel (including the case when both $\tau$'s decay to the electron channel, but one electron is not identified); $Z \to e^+e^-$ with one electron not identified. We have generated one million events for each of these background processes, using HERWIG and CTEQ5L, and compared them to one million signal events generated with CTEQ6.1. We apply event selection criteria designed to eliminate the background preferentially. These criteria are:

**Table 2:** Reduction of signal and background due to cuts

| Cut | $W \to e\nu_e$ | | $Z \to \tau^+\tau^-$ | | $Z \to e^+e^-$ | | $W \to \tau\nu_\tau$ | |
|---|---|---|---|---|---|---|---|---|
| | $e^+$ | $e^-$ | $e^+$ | $e^-$ | $e^+$ | $e^-$ | $e^+$ | $e^-$ |
| ATLFAST cuts | 382,902 | 264,415 | 5.5% | 7.9% | 34.7% | 50.3% | 14.8% | 14.9% |
| $\|\eta\| < 2.4$ | 367,815 | 255,514 | 5.5% | 7.8% | 34.3% | 49.4% | 14.7% | 14.8% |
| $p_{te} > 25$ GeV | 252,410 | 194,562 | 0.6% | 0.7% | 12.7% | 16.2% | 2.2% | 2.3% |
| $p_{tmiss} > 25$ GeV | 212,967 | 166,793 | 0.2% | 0.2% | 0.1% | 0.2% | 1.6% | 1.6% |
| No jets with $P_t > 30$ GeV | 187,634 | 147,415 | 0.1% | 0.1% | 0.1% | 0.1% | 1.2% | 1.2% |
| $p_t^{recoil} < 20$ GeV | 159,873 | 125,003 | 0.1% | 0.1% | 0.0% | 0.0% | 1.2% | 1.2% |

– ATLFAST cuts (see Sec. 1.3.5)
– pseudorapidity, $|\eta| < 2.4$, to avoid bias at the edge of the measurable rapidity range
– $p_{te} > 25$ GeV, high $p_t$ is necessary for electron triggering
– missing $E_t > 25$ GeV, the $\nu_e$ in a signal event will have a correspondingly large missing $E_t$
– no reconstructed jets in the event with $p_t > 30$ GeV, to discriminate against QCD background
– recoil on the transverse plane $p_t^{recoil} < 20$ GeV, to discriminate against QCD background

Table 2 gives the percentage of background with respect to signal, calculated using the known relative cross-sections of these processes, as each of these cuts is applied. After, the cuts have been applied the background from these processes is negligible. However, there are limitations on this study from the fact that in real data there will be further QCD backgrounds from $2 \to 2$ processes involving $q, \bar{q}, g$ in which a final state $\pi^0 \to \gamma\gamma$ decay mimics a single electron. A preliminary study applying the selection criteria to MC generated QCD events suggests that this background is negligible, but the exact level of QCD background cannot be accurately estimated without passing a very large number of events though a full detector simulation, which is beyond the scope of the current contribution.

### *1.3.4 Charge misidentification*

Clearly charge misidentification could distort the lepton rapidity spectra and dilute the asymmetry $A_l$.

$$A_{true} = \frac{A_{raw} - F^+ + F^-}{1 - F^- - F^+}$$

where $A_{raw}$ is the measured asymmetry, $A_{true}$ is the true asymmetry, $F^-$ is the rate of true $e^-$ misidentified as $e^+$ and $F^+$ is the rate of true $e^+$ misidentified as $e^-$. To make an estimate of the importance of charge misidentification we use a sample of $Z \to e^+e^-$ events generated by HERWIG with CTEQ5L and passed through a full simulation of the ATLAS detector. Events with two or more charged electromagnetic objects in the EM calorimeter are then selected and subject to the cuts; $|\eta| < 2.5$, $p_{te} > 25$ GeV, as usual and, $E/p < 2$, for bremsstrahlung rejection. We then look for the charged electromagnetic pair with invariant mass closest to $M_Z$ and impose the cut, $60 < M_Z < 120$ GeV. Then we tag the charge of the better reconstructed lepton of the pair and check to see if the charge of the second lepton is the same as the first. Assuming that the pair really came from the decay of the $Z$ this gives us a measure of charge misidentification. Fig 10 show the misidentification rates $F^+$, $F^-$ as functions of pseudorapidity[8]. These rates are very small. The quantity $A_l$, can be corrected for charge misidentification applying Barlow's method for combining asymmetric errors [19]. The level of correction is $0.3\%$ in the central region and $0.5\%$ in the more forward regions.

[8]These have been corrected for the small possibility that the better reconstructed lepton has had its charge misidentified as follows. In the central region, $|\eta| < 1$, assume the same probability of misidentification of the first and second leptons, in the more forward regions assume the same rate of first lepton misidentification as in the central region.

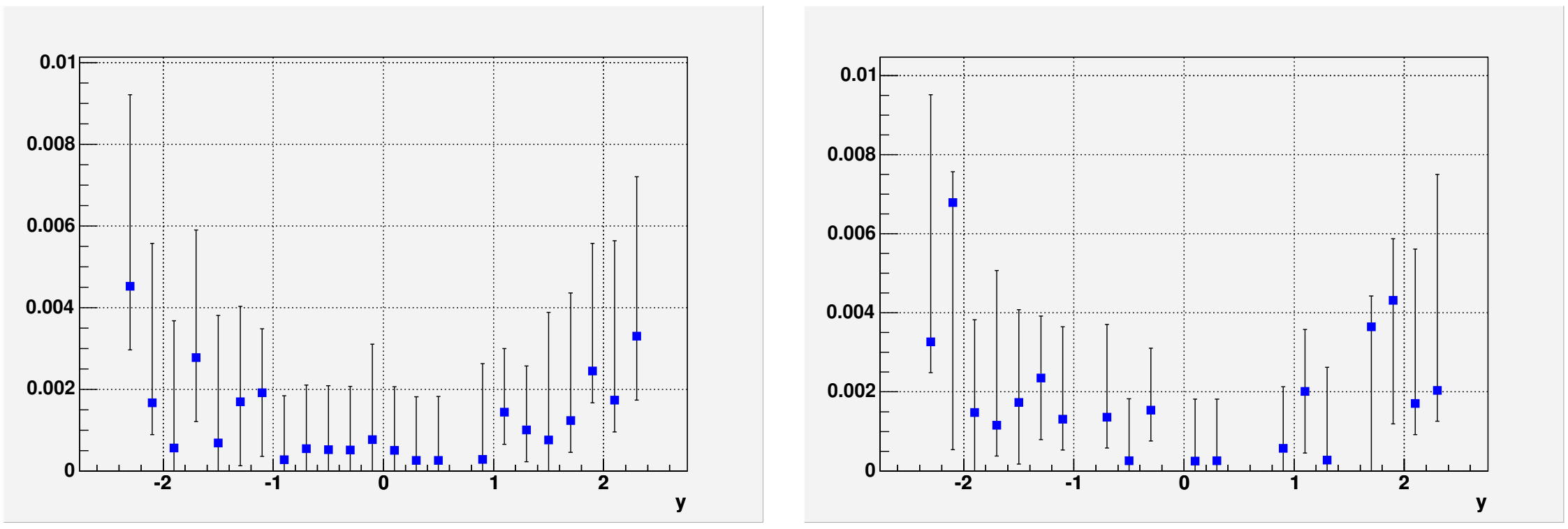

**Fig. 10:** The rates of charge misidentification as a function of rapidity for $e^-$ misidentified as $e^+$ (left), $e^+$ misidentifed as $e^-$ (right).

### *1.3.5 Compare events at the generator level to events at the detector level*

We have simulated one million signal, $W \to e\nu_e$, events for each of the PDF sets CTEQ6.1, MRST2001 and ZEUS-S using HERWIG (6.505). For each of these PDF sets the eigenvector error PDF sets have been simulated by PDF reweighting and k-factors have been applied to approximate an NLO generation. The top part of Fig. 11 shows the $e^\pm$ and $A_l$ spectra at this generator level, for all of the PDF sets superimposed. The events are then passed through the ATLFAST fast simulation of the ATLAS detector. This applies loose kinematic cuts: $|\eta| < 2.5$, $p_{te} > 5$ GeV, and electron isolation criteria. It also smears the 4-momenta of the leptons to mimic momentum dependent detector resolution. We then apply the selection cuts described in Sec. 1.3.3. The lower half of Fig. 11 shows the $e^\pm$ and $A_l$ spectra at the detector level after application of these cuts, for all of the PDF sets superimposed. The level of precision of each PDF set, seen in the analytic calculations of Fig. 6, is only slightly degraded at detector level, so that a net level of PDF uncertainty at central rapidity of $\sim 8\%$ is maintained. The anticipated cancellation of PDF uncertainties in the asymmetry spectrum is also observed, within each PDF set, and the spread between PDF sets suggests that measurements which are accurate to better than $\sim 5\%$ could discriminate between PDF sets.

### *1.3.6 Using LHC data to improve precision on PDFs*

The high cross-sections for $W$ production at the LHC ensure that it will be the experimental systematic errors, rather than the statistical errors, which are determining. We have imposed a random $4\%$ scatter on our samples of one million $W$ events, generated using different PDFs, in order to investigate if measurements at this level of precision will improve PDF uncertainties at central rapidity significantly if they are input to a global PDF fit. Fig. 12 shows the $e^+$ and $e^-$ rapidity spectra for events generated from the ZEUS-S PDFs ($|\eta| < 2.4$) compared to the analytic predictions for these same ZEUS-S PDFs. The lower half of this figure illustrates the result if these events are then included in the ZEUS-S PDF fit. The size of the PDF uncertainties, at $y = 0$, decreases from $5.8\%$ to $4.5\%$. The largest improvement is in the PDF parameter $\lambda_g$ controlling the low-$x$ gluon at the input scale, $Q_0^2$: $xg(x) \sim x^{\lambda_g}$ at low-$x$, $\lambda_g = -0.199 \pm 0.046$, before the input of the LHC pseudo-data, compared to, $\lambda_g = -0.196 \pm 0.029$, after input. Note that whereas the relative normalisations of the $e^+$ and $e^-$ spectra are set by the PDFs, the absolute normalisation of the data is free in the fit so that no assumptions are made on our ability to measure luminosity. Secondly, we repeat this procedure for events generated using the CTEQ6.1 PDFs. As shown in Fig. 13, the cross-section for these events is on the lower edge of the uncertainty band of the ZEUS-S predictions. If these events are input to the fit the central value shifts and the uncertainty decreases. The value of the parameter $\lambda_g$ becomes, $\lambda_g = -0.189 \pm 0.029$, after input of these pseudo-data. Finally to simulate the situation which really faces experimentalists we generate events with CTEQ6.1, and pass them through the ATLFAST detector simulation and cuts. We then correct back from detector level to generator level using a different PDF set- in this case the ZEUS-S PDFs- since in practice we will not know the true PDFs. Fig. 14 shows that the resulting corrected data look pleasingly like CTEQ6.1, but they are more smeared. When these data are input to the PDF fit the central values shift and errors decrease just as for the perfect CTEQ6.1 pseudo-data. The value of $\lambda_g$ becomes, $\lambda = -0.181 \pm 0.030$, after input of these pseudo-data. Thus we see that the bias introduced by the correction procedure from detector to generator level is small compared to the PDF uncertainty.

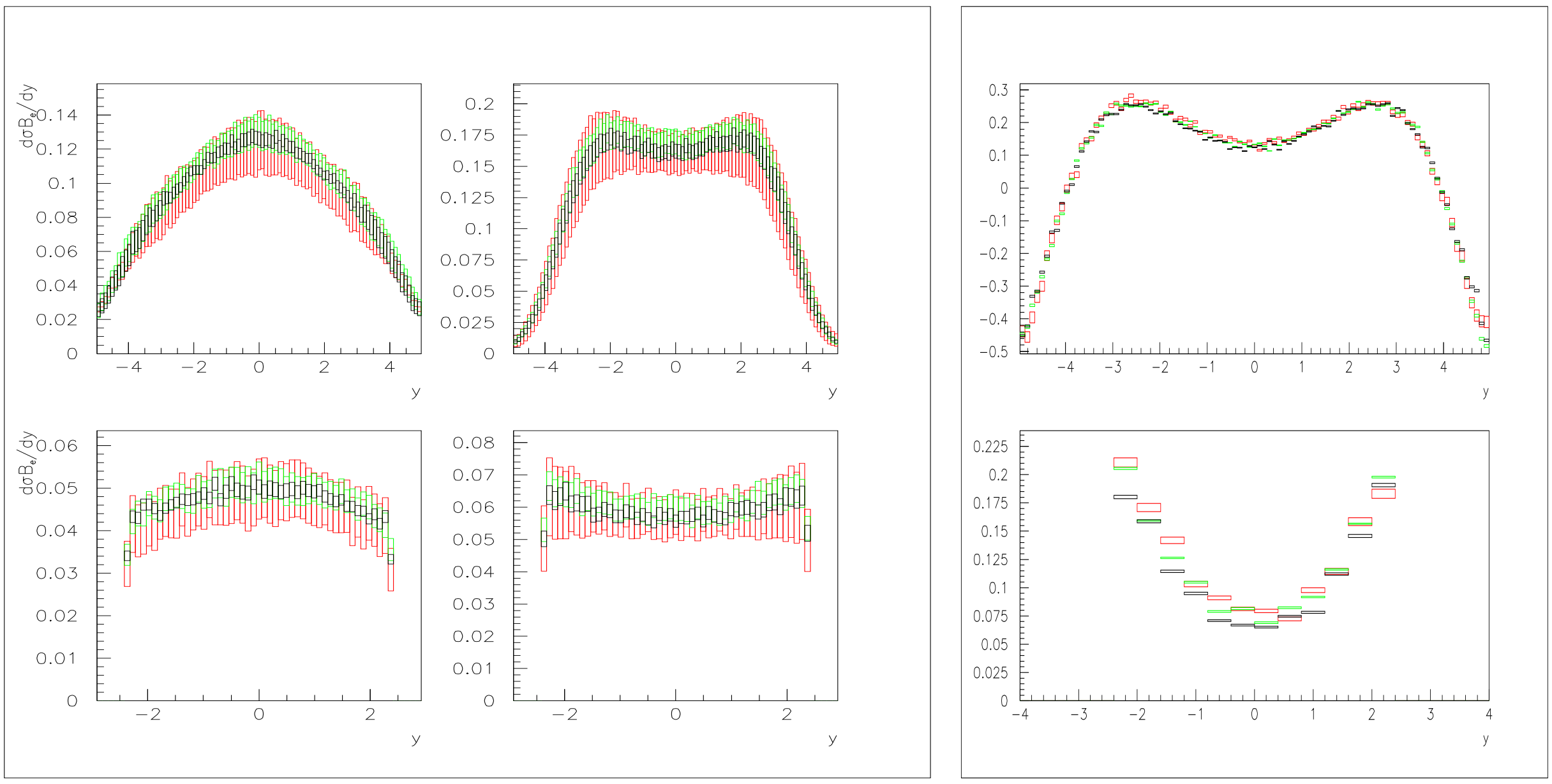

**Fig. 11:** Top row: $e^-$, $e^+$ and $A_e$ rapidity spectra for the lepton from the $W$ decay, generated using HERWIG + k factors and CTEQ6.1 (red), ZEUS-S (green) and MRST2001 (black) PDF sets with full uncertainties. Bottom row: the same spectra after passing through the ATLFAST detector simulation and selection cuts.

### *1.3.7 Conclusions and a warning: problems with the theoretical predictions at small-$x$?*

We have investigated the PDF uncertainty on the predictions for $W$ and $Z$ production at the LHC, taking into account realistic expectations for measurement accuracy and the cuts on data which will be needed to identify signal events from background processes. We conclude that at the present level of PDF uncertainty the decay lepton asymmetry, $A_l$, will be a useful standard model benchmark measurement, and that the decay lepton spectra can be used as a luminosity monitor which will be good to $\sim 8\%$. However, we have also investigated the measurement accuracy necessary for early measurements of these decay lepton spectra to be useful in further constraining the PDFs. A systematic measurement error of less than $\sim 4\%$ would provide useful extra constraints.

However, a caveat is that the current study has been performed using standard PDF sets which are extracted using NLO QCD in the DGLAP [20–23] formalism. The extension to NNLO is straightforward, giving small corrections $\sim 1\%$. PDF analyses at NNLO including full accounting of the PDF uncertainties are not extensively available yet, so this small correction is not pursued here. However, there may be much larger uncertainties in the theoretical calculations because the kinematic region involves low-$x$. There may be a need to account for $ln(1/x)$ resummation (first considered in the BFKL [24–26] formalism) or high gluon density effects. See reference [27] for a review.

The MRST group recently produced a PDF set, MRST03, which does not include any data for $x < 5 \times 10^{-3}$. The motivation behind this was as follows. In a global DGLAP fit to many data sets there is always a certain amount of tension between data sets. This may derive from the use of an inappropriate theoretical formalism for the kinematic range of some of the data. Investigating the effect of kinematic cuts on the data, MRST found that a cut, $x > 5 \times 10^{-3}$, considerably reduced tension between the remaining data sets. An explanation may be the inappropriate use of the DGLAP formalism at small-$x$. The MRST03 PDF set is thus free of this bias BUT it is also only valid to use it for $x > 5 \times 10^{-3}$. What is needed is an alternative theoretical formalism for smaller $x$. However, the MRST03 PDF set may be used as a toy PDF set, to illustrate the effect of using very different PDF sets on our predictions. A comparison of Fig. 15 with Fig. 3 or Fig. 6 shows how different the analytic predictions are from the conventional ones, and thus illustrates where we might expect to see differences due to the need for an alternative formalism at small-$x$.

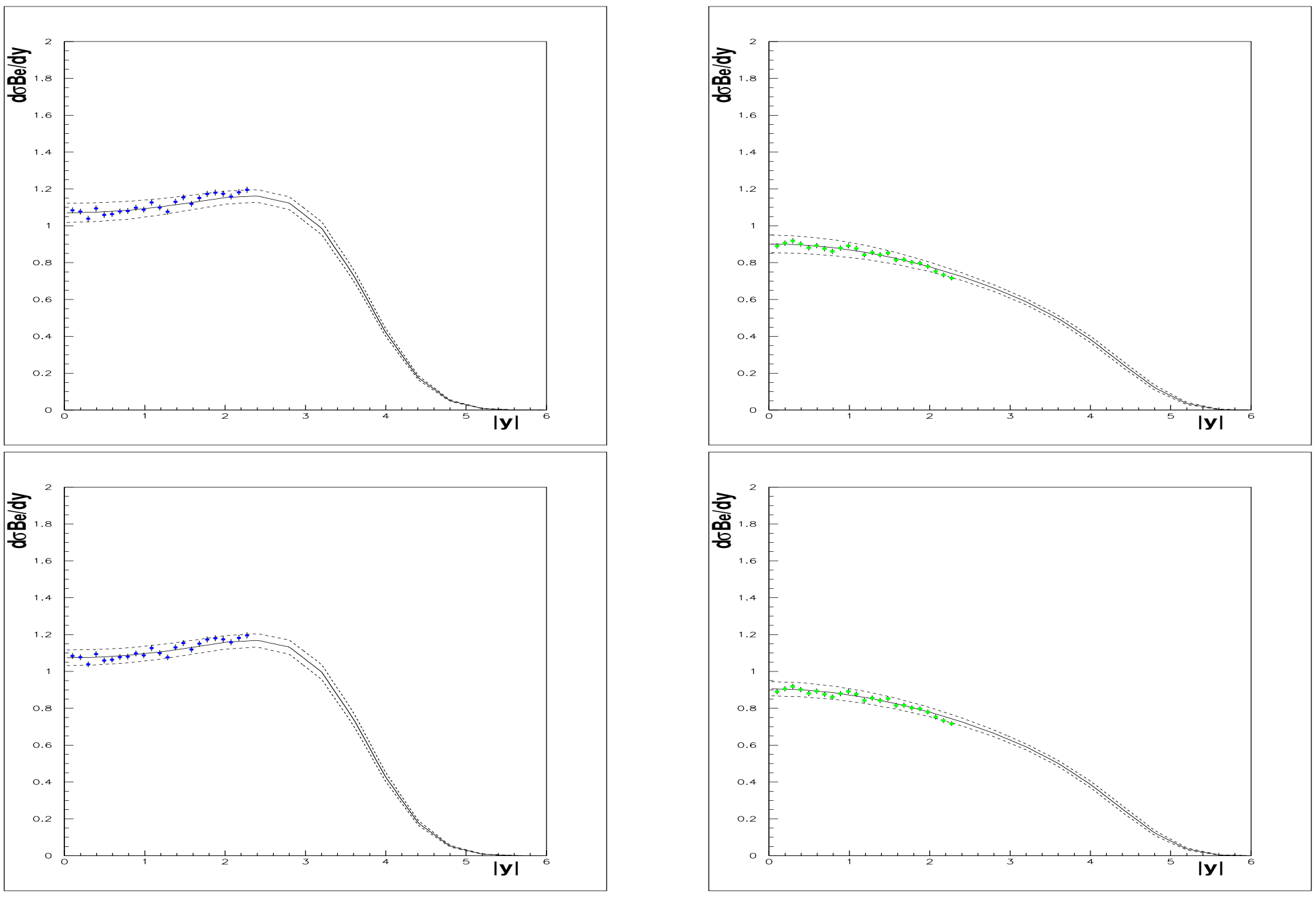

**Fig. 12:** Top row: $e^+$ and $e^-$ rapidity spectra generated from ZEUS-S PDFs compared to the analytic prediction using ZEUS-S PDFs. Bottom row: the same lepton rapidity spectra compared to the analytic prediction AFTER including these lepton pseudo-data in the ZEUS-S PDF fit.

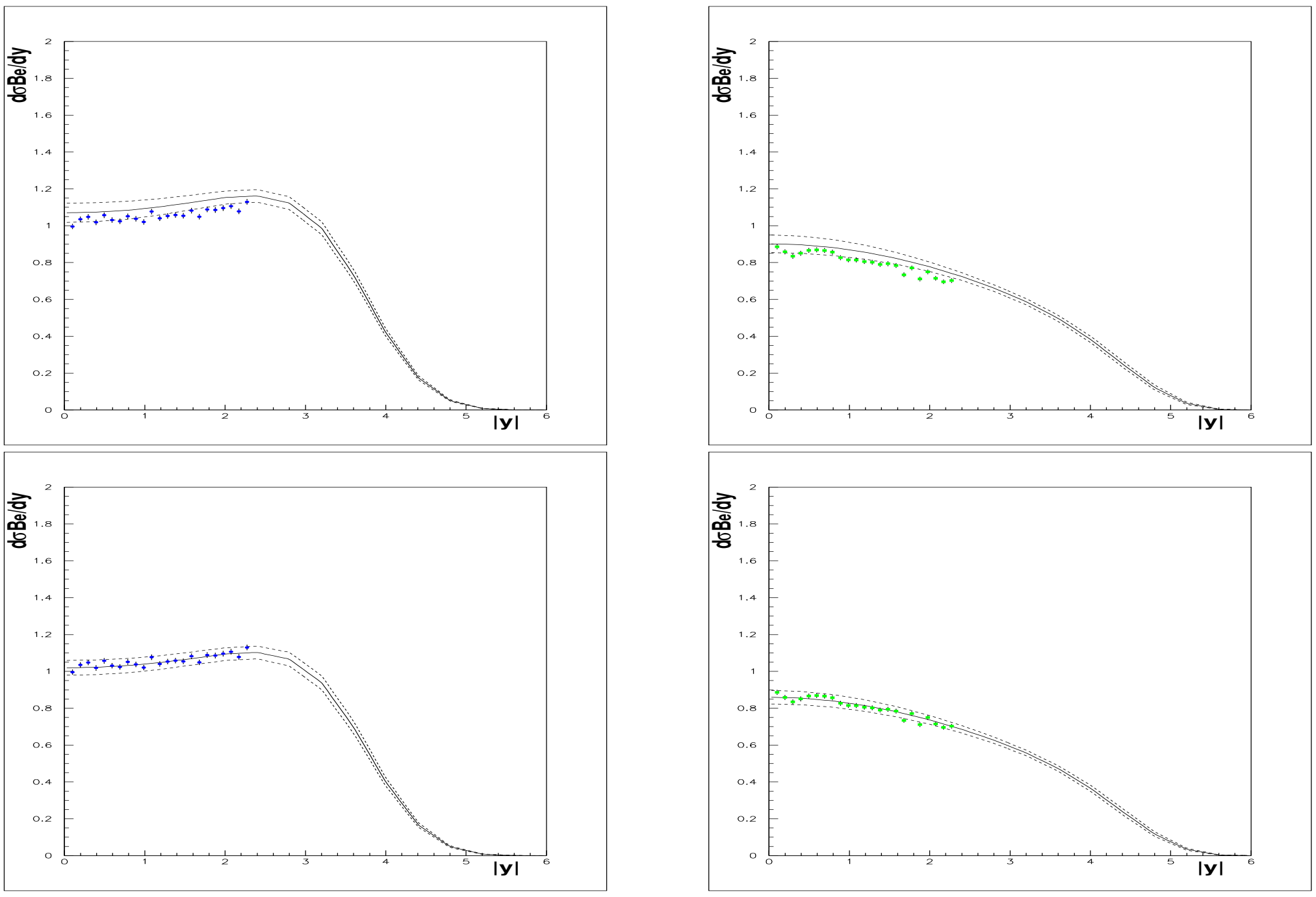

**Fig. 13:** Top row: $e^+$ and $e^-$ rapidity spectra generated from CTEQ6.1 PDFs compared to the analytic prediction using ZEUS-S PDFs. Bottom row: the same lepton rapidity spectra compared to the analytic prediction AFTER including these lepton pseudo-data in the ZEUS-S PDF fit.

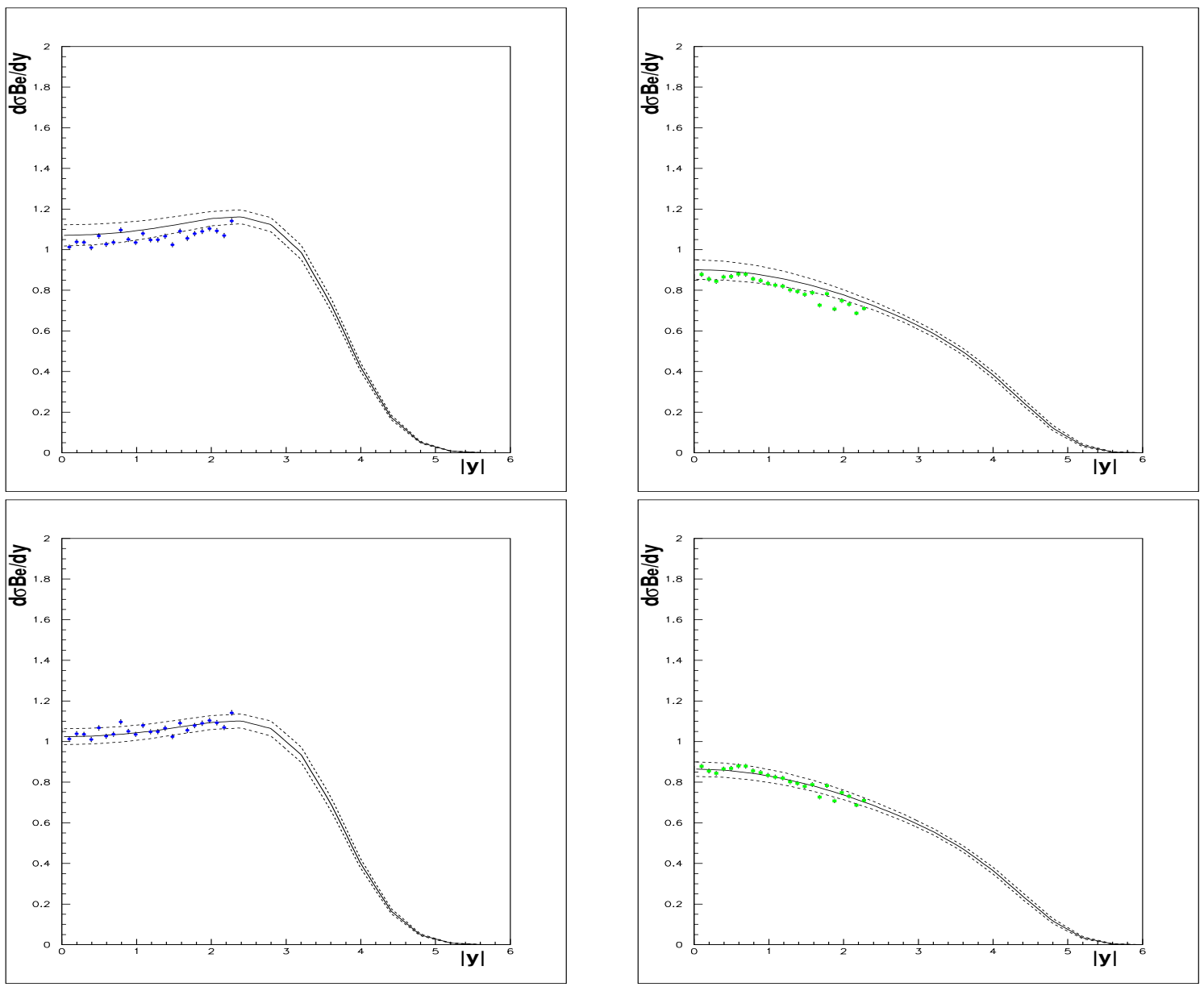

**Fig. 14:** Top row: $e^+$ and $e^-$ rapidity spectra generated from CTEQ6.1 PDFs, which have been passed through the ATLFAST detector simulation and corrected back to generator level using ZEUS-S PDFs, compared to the analytic prediction using ZEUS-S PDFs. Bottom row: the same lepton rapidity spectra compared to the analytic prediction AFTER including these lepton pseudo-data in the ZEUS-S PDF fit.

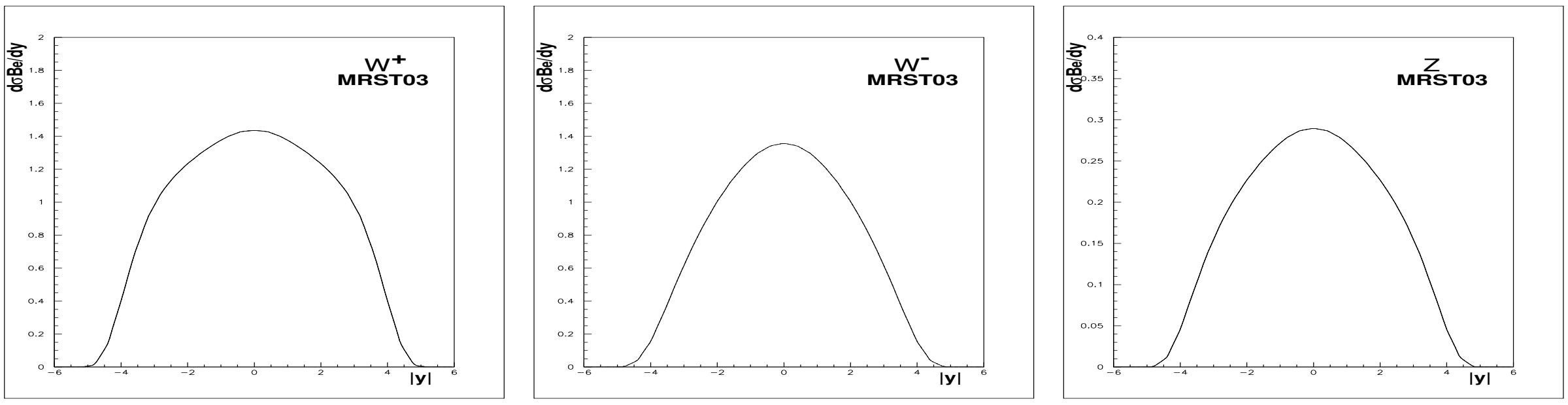

**Fig. 15:** LHC $W^+$, $W^-$, $Z$ rapidity distributions for MRST03 PDFs: left plot $W^+$; middle plot $W^-$; right plot $Z$

### 1.4 $W$ and $Z$ production at the LHC [9]

The study of the production at the LHC of the electroweak bosons $W$ and $Z$ with subsequent decays in leptonic final states will provide several precision measurements of Standard Model parameters such as the mass of the $W$ boson or the weak mixing angle from the $Z$ boson forward-backward asymmetry. Given their large cross section and clean experimental signatures, the bosons will furthermore serve as calibration tool and luminosity monitor. More challenging, differential cross sections in rapidity or transverse momentum may be used to further constrain parton distribution functions. Eventually these measurements for single inclusive boson production may be applied to boson pair production in order to derive precision predictions for background estimates to discovery channels like $H \to W^+W^-$.

[9]Contributing author: Hasko Stenzel

This contribution is devoted to the estimation of current uncertainties in the calculations for Standard Model cross sections involving $W$ and $Z$ bosons with particular emphasis on the PDF and perturbative uncertainties. All results are obtained at NLO with MCFM [28] version 4.0 interfaced to LHAPDF [12] for a convenient selection of various PDF families and evaluation of their intrinsic uncertainties. The cross sections are evaluated within a typical experimental acceptance and for momentum cuts summarised in Table 3. The electromagnetic decays of $W$ and $Z$ are considered (massless leptons) and the missing transverse energy is assigned to the neutrino momentum sum (in case of $W$ decays). Jets in the processes $W/Z + jets$ are produced in an inclusive mode with at least one jet in the event reconstructed with the $k_T$-algorithm. MCFM includes one- and two-jet processes at NLO and three-jet processes at LO. In the case of boson pair production the cuts of Table 3 can only be applied to the two leading leptons, hence a complete acceptance is assumed for additional leptons e.g. from $ZZ$ or $WZ$ decays.

**Table 3:** Experimental acceptance cuts used for the calculation of cross-sections.

| Observable | cut |
|---|---|
| $p_T^{\mathrm{lept}}$ | $> 25$ GeV |
| $p_T^{\mathrm{jet}}$ | $> 25$ GeV |
| $\lvert\eta_{\mathrm{lept}}\rvert$ | $< 3.0$ |
| $\lvert\eta_{\mathrm{jet}}\rvert$ | $< 4.0$ |
| $R(\mathrm{lepton} - \mathrm{jet})$ | $> 0.8$ |
| $R(\mathrm{lepton} - \mathrm{lepton})$ | $> 0.2$ |
| $E_T^{\mathrm{miss}}$ | $> 25$ GeV |

The calculations with MCFM are carried out for a given fixed set of electroweak input parameters using the effective field theory approach [28]. The PDF family CTEQ61 provided by the CTEQ collaboration [29] is taken as nominal PDF input while MRST2001E given by the MRST group [30] is considered for systematic purposes. The difference between CTEQ61 and MRST2001E alone can't be considered as systematic uncertainty but merely as cross-check. The systematic uncertainty is therefore estimated for each family separately with the family members, 40 for CTEQ61 and 30 for MRST2001E, which are variants of the nominal PDF obtained with different assumptions while maintaining a reasonable fit of the input data. The value of $\alpha_s$ is not a free input parameter for the cross section calculation but taken from the corresponding value in the PDF.

Important input parameters are renormalisation and factorisation scales. The central results are obtained with $\mu_R = \mu_F = M_V$, $V = W, Z$ for single boson production and $\mu_R = \mu_F = M_V + M_V'$ for pair production ($V'$ being the second boson in the event). Missing higher orders are estimated by a variation of the scales in the range $1/2 \leq x_{\mu R} \leq 2$ and independently $1/2 \leq x_{\mu F} \leq 2$ where $\mu = x_\mu \cdot M_V$, following prescriptions applied to other processes [31], keeping in mind that the range of variation of the scales is purely conventional.

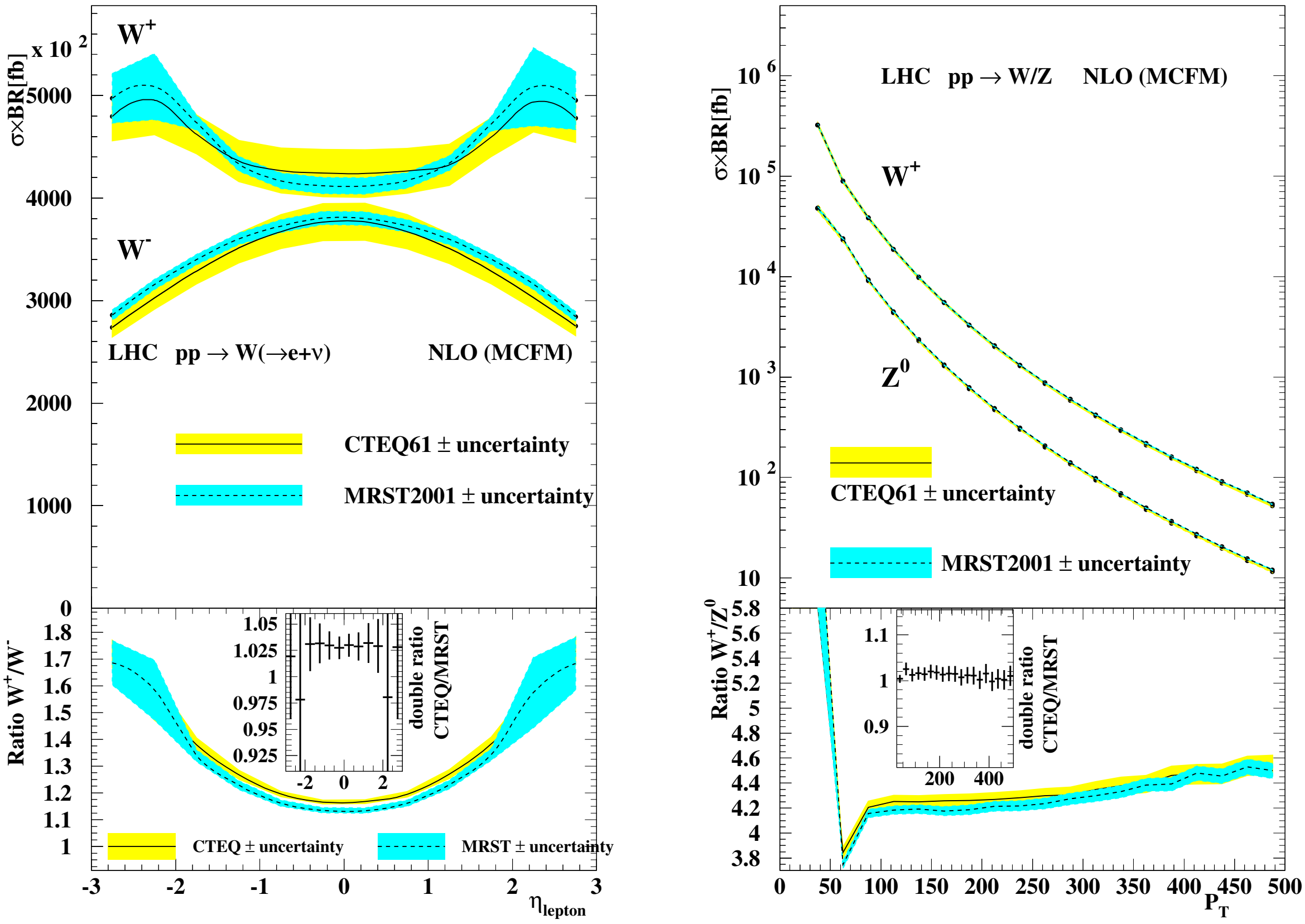

**Fig. 16:** Left: pseudo-rapidity distribution of the decay lepton from inclusive $W$ production and right: $p_T$ spectra of $W$ and $Z$. The bands represent the PDF-uncertainty. The lower inserts show on the left side the ratio $W^+/W^-$ resp. the double-ratio CTEQ/MRST and on the right side the ratios for $W^+/Z^0$.

### *1.4.1 Single W and Z cross sections*

Detailed studies of single $W$ and $Z$ production including detector simulation are presented elsewhere in these proceedings, here these channels are mainly studied for comparison with the associated production with explicitly reconstructed jets and with pair production. The selected process is inclusive in the sense that additional jets, present in the NLO calculation, are not explicitly reconstructed. The experimentally required lepton isolation entailing a jet veto in a restricted region of phase space is disregarded at this stage.

As an example the pseudo-rapidity distribution of the lepton from $W$ decays and the $p_T$ spectra for $Z$ and $W^+$ are shown in Fig. 16. The cross section for $W^+$ is larger than for $W^-$ as a direct consequence of the difference between up- and down-quark PDFs, and this difference survives in the pseudo-rapidity distribution of the decay lepton with a maximum around $|\eta|$=2.5. In the central part the PDF uncertainty, represented by the bands in Fig. 16, amounts to about 5$\%$ for CTEQ and 2$\%$ for MRST, and within the uncertainty CTEQ and MRST are fully consistent. Larger differences are visible in the peaks for the $W^+$, where at the same time the PDF uncertainty increases. In the ratio $W^+/W^-$ the PDF uncertainty is reduced to about 1-2$\%$ in the central region and a difference of about 3$\%$ is observed between CTEQ and MRST, as can be seen from the double-ratio CTEQ/MRST. The uncertainty of the double ratio is calculated from the CTEQ uncertainty band alone.

In the case of $Z$ production the rapidity and $p_T$ spectra can be fully reconstructed from the $e^+e^-$ pair. A measurement of the $Z$ $p_T$ spectrum may be used to tune the Monte Carlo description of $W$ $p_T$, which is relevant for measurements of the $W$ mass. The $p_T$ spectra are shown in the right part of Fig. 16. The total yield for $W^+$ is about six times larger than for $Z^0$ but for $p_T > 150$ GeV the ratio stabilises around 4.5. At small values of $p_T$ the fixed-order calculation becomes trustless and should be supplemented by resummed calculations. The PDF uncertainties for the $p_T$ spectra themselves are again about 5% and about 2% in the ratio, CTEQ and MRST being consistent over the entire $p_T$ range.

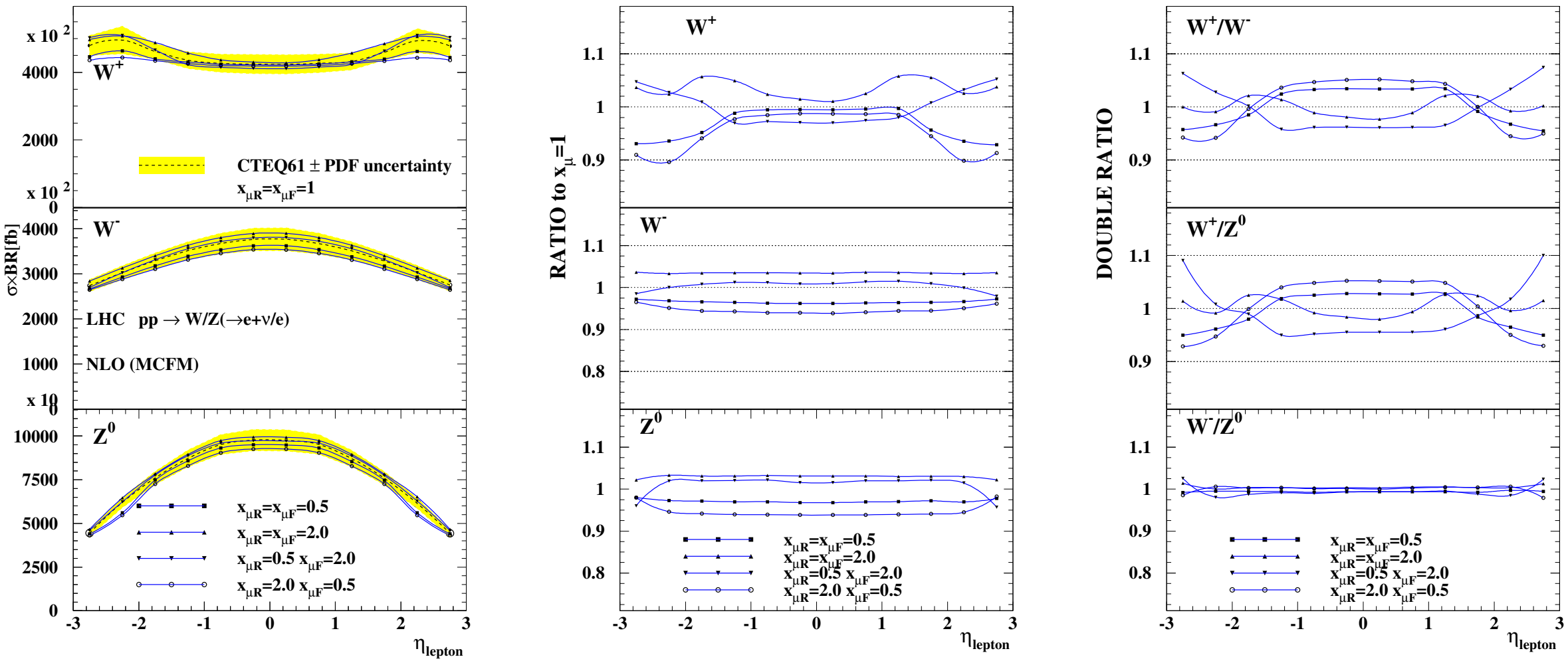

**Fig. 17:** Left: pseudo-rapidity distribution of the decay lepton from inclusive $W/Z$ production for different values of $x_{\mu R}$ and $x_{\mu F} = 1$, centre: the ratio of predictions with respect to $x_\mu = 1$ and right: double ratio $V/V'$ of cross sections for actual scale settings normalised to the nominal scale.

The perturbative uncertainties are estimated by variations of the renormalisation and factorisation scales in by a factor of two. The scale variation entails a global change in the total cross section of the order of 5%. The $\eta$ distribution of leptons from $W/Z$ decays are shown in Fig. 17, comparing the nominal cross section with $x_{\mu R} = x_{\mu F} = 1$, to alternative scale settings. The nominal cross section is drawn with its PDF uncertainty band, illustrating that the perturbative uncertainties are of the same size. For $W^-$ and $Z^0$ the shape of the distribution is essentially unaltered, but for $W^+$ the region around the maxima is changed more than the central part, leading to a shape deformation. The scale variation uncertainty is strongly correlated for $W^-$ and $Z^0$ and cancels in the ratio $W^-/Z^0$, but for $W^+$ it is almost anti-correlated with $W^-$ and $Z^0$ and partly enhanced in the ratio.

Globally the perturbative uncertainty is dominated by the asymmetric scale setting $x_{\mu R} = 2, x_{\mu R} = 1/2$ for which a change of $-5\%$ is observed, the largest upward shift of $3.5\%$ is obtained for $x_{\mu R} = 2, x_{\mu R} = 2$, locally the uncertainty for $W^+$ can be much different. It can be expected that the perturbative uncertainties are reduced for NNLO calculations to the level of $1\%$.

The integrated cross sections and systematic uncertainties within the experimental acceptance are summarised in Table 4.

### *1.4.2 $W/Z + jet$ production*

In the inclusive production of $W/Z + jet$ at least one jet is requested to be reconstructed, isolated from any lepton by $R > 0.8$. Additional jets are in case of overlap eventually merged at reconstruction level by the $k_T$-prescription. Given the presence of a relatively hard ($p_T > 25$ GeV) jet, it can be expected that PDF- and perturbative uncertainties are different than for single boson production. The study of this process at the LHC, other than being a stringent test of perturbative QCD, may in addition contribute to a better understanding of the gluon PDF.

The first difference with respect to single boson production appears in the lepton pseudo-rapidities, shown in Fig. 18. The peaks in the lepton spectrum from $W^+$ disappeared, the corresponding spectrum from $W^-$ is stronger peaked at central rapidity while the ratio $W^+/W^-$ with jets is essentially the same as without jets. The PDF uncertainties are slightly smaller (4.2-4.4%) compared to single bosons. The jet pseudo-rapidities are shown in the right part of Fig. 18, they are much stronger peaked in the central region but the ratio $W^+/W^-$ for jets is similar to the lepton ratio.

**Table 4:** Total cross-sections and systematic uncertainties within the experimental acceptance.

| | $W^+$ | $W^-$ | $Z^0$ |
|---|---|---|---|
| CTEQ61 [pb] | 5438 | 4002 | 923.9 |
| $\Delta_{\rm PDF}^{\rm CTEQ}$ [pb] | $\pm282$ | $\pm221$ | $\pm49.1$ |
| $\Delta_{\rm PDF}^{\rm CTEQ}$ [%] | $\pm5.2$ | $\pm5.5$ | $\pm5.3$ |
| MRST [pb] | 5480 | 4110 | 951.1 |
| $\Delta_{\rm PDF}^{\rm MRST}$ [pb] | $\pm103$ | $\pm83.4$ | $\pm17.4$ |
| $\Delta_{\rm PDF}^{\rm MRST}$ [%] | $\pm1.9$ | $\pm2.1$ | $\pm1.9$ |
| $\Delta_{\rm pert}$ [%] | $+3.5$ | $+3.5$ | $+3.1$ |
| | $-5.2$ | $-5.4$ | $-5.5$ |

**Table 5:** Total cross-sections and systematic uncertainties within the experimental acceptance for $W/Z + jet$ processes.

| | $W^+ + jet$ | $W^- + jet$ | $Z^0 + jet$ |
|---|---|---|---|
| CTEQ61 [pb] | 1041 | 784.5 | 208.1 |
| $\Delta_{\rm PDF}^{\rm CTEQ}$ [pb] | $\pm44.1$ | $\pm34.3$ | $\pm9.01$ |
| $\Delta_{\rm PDF}^{\rm CTEQ}$ [%] | $\pm4.2$ | $\pm4.4$ | $\pm4.3$ |
| MRST [pb] | 1046 | 797.7 | 211.3 |
| $\Delta_{\rm PDF}^{\rm MRST}$ [pb] | $\pm17.6$ | $\pm14.8$ | $\pm3.67$ |
| $\Delta_{\rm PDF}^{\rm MRST}$ [%] | $\pm1.7$ | $\pm1.9$ | $\pm1.8$ |
| $\Delta_{\rm pert}$ [%] | $+8.7$ | $+8.9$ | $+7.6$ |
| | $-9.8$ | $-10.0$ | $-9.1$ |

The transverse momenta of associated jets from $W/Z + jet$ production is shown in Fig. 19, the spectra are steeply falling and the ratio $W^+/W^-$ is increasing from 1.3 at low $p_T$ to almost 2 at 500 GeV $p_T$.

The perturbative uncertainties are investigated in the same way as for the single boson production and are shown in Fig. 20. The scale variation entails here a much larger uncertainty between 8 and 10%, almost twice as large as for single bosons. In contrast to the latter case, the scale variation is correlated for $W$ and $Z$ and cancels in the ratio $W^+/W^-$, with an exception for $W^-$ where a bump appears at $|\eta| = 1.8$ for $x_{\mu R} = 2$.

The total cross sections and their systematic uncertainties are summarised in Table 5.

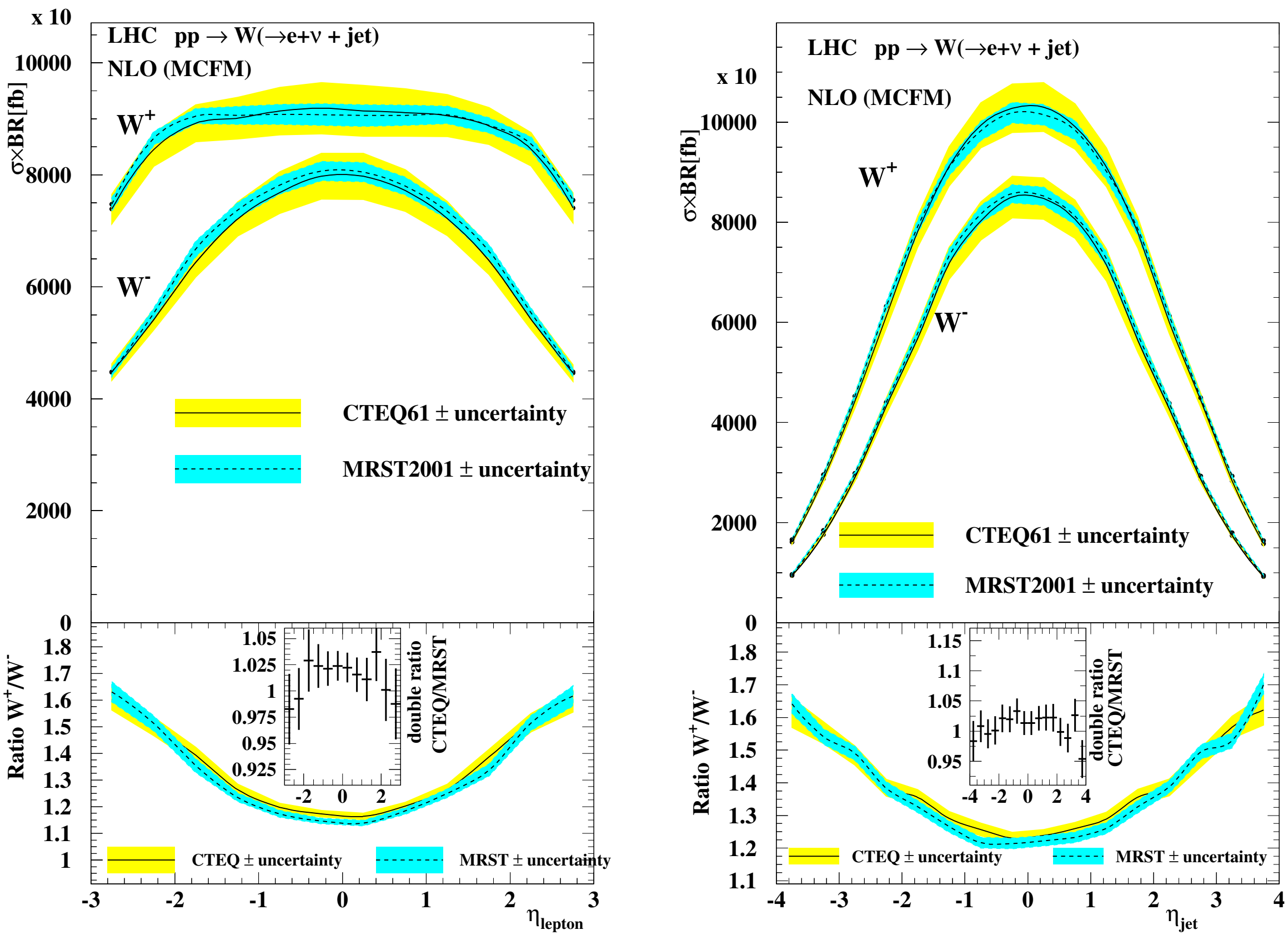

**Fig. 18:** Left: pseudo-rapidity distribution of the decay lepton from inclusive $W$+jet production and right: pseudo-rapidity of the associated leading jet. The bands represent the PDF-uncertainty.

### *1.4.3 Vector Boson pair production*

In the Standard Model the non-resonant production of vector bosons pairs in the continuum is suppressed by factors of $10^4$-$10^5$ with respect to single Boson production. The cross sections for $WW$, $WZ$ and $ZZ$ within the experimental acceptance range from 500 fb ($WW$) to 10 fb ($ZZ$). Given the expected limited statistics for these processes, the main goal of their experimental study is to obtain the best estimate of the background they represent for searches of the Higgs boson or new physics yielding boson pairs.

The selection of boson pairs follows in extension the single boson selection cuts applied to 2, 3 or 4 isolated leptons. Again real gluon radiation and virtual loops have been taken into account at NLO but without applying lepton-jet isolation cuts. Lepton-lepton separation is considered only for the two leading leptons.

The pseudo-rapidity and transverse momentum distributions taking the $e^+$ from $W^+W^-$ production as example are shown in Fig. 21. The pseudo-rapidity is strongly peaked and the cross section at $\eta = 0$ twice as large as at $|\eta| = 3$. The PDF uncertainties are smaller than for single bosons, between 3.5 and 4 $\%$.

The same shape of lepton distributions is also found for the other lepton and for the other pair production processes, as shown for the $W^-Z^0$ case in Fig. 22.

The rapidity distribution of the leading $Z^0$ from $ZZ$ production is shown in the left part of Fig. 23. With both $Z$'s being fully reconstructed, the invariant mass of the $ZZ$ system can be compared in the right part of Fig. 23 to the invariant mass spectrum of the Higgs decaying into the same final state for an intermediate mass of $m_H = 200$ GeV. In this case a clear peak appears at low invariant masses above the continuum, and the mass spectrum is also harder at high masses in presence of the Higgs.

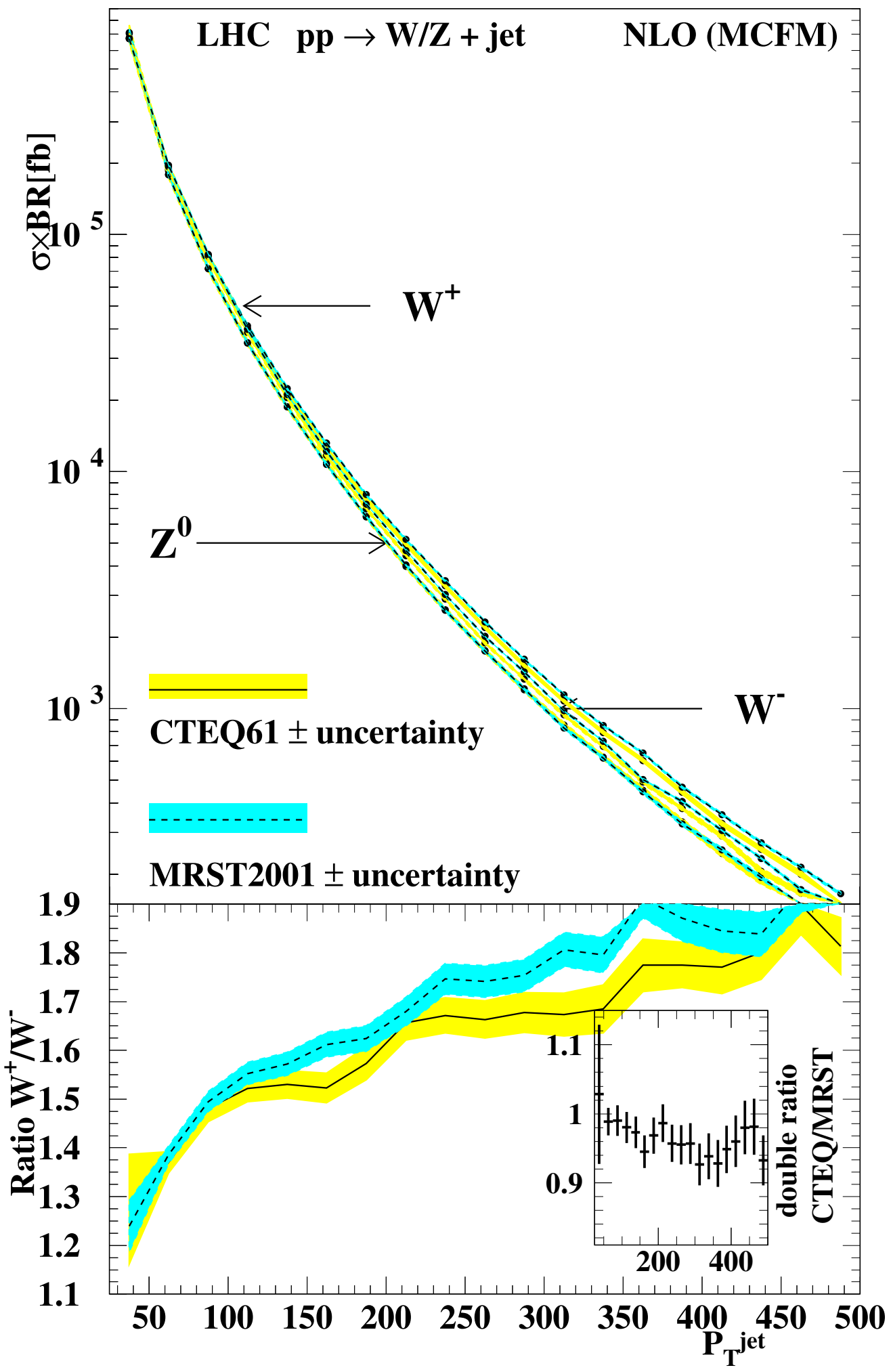

**Fig. 19:** Transverse momentum distribution of the jet from inclusive $W/Z + jet$ production

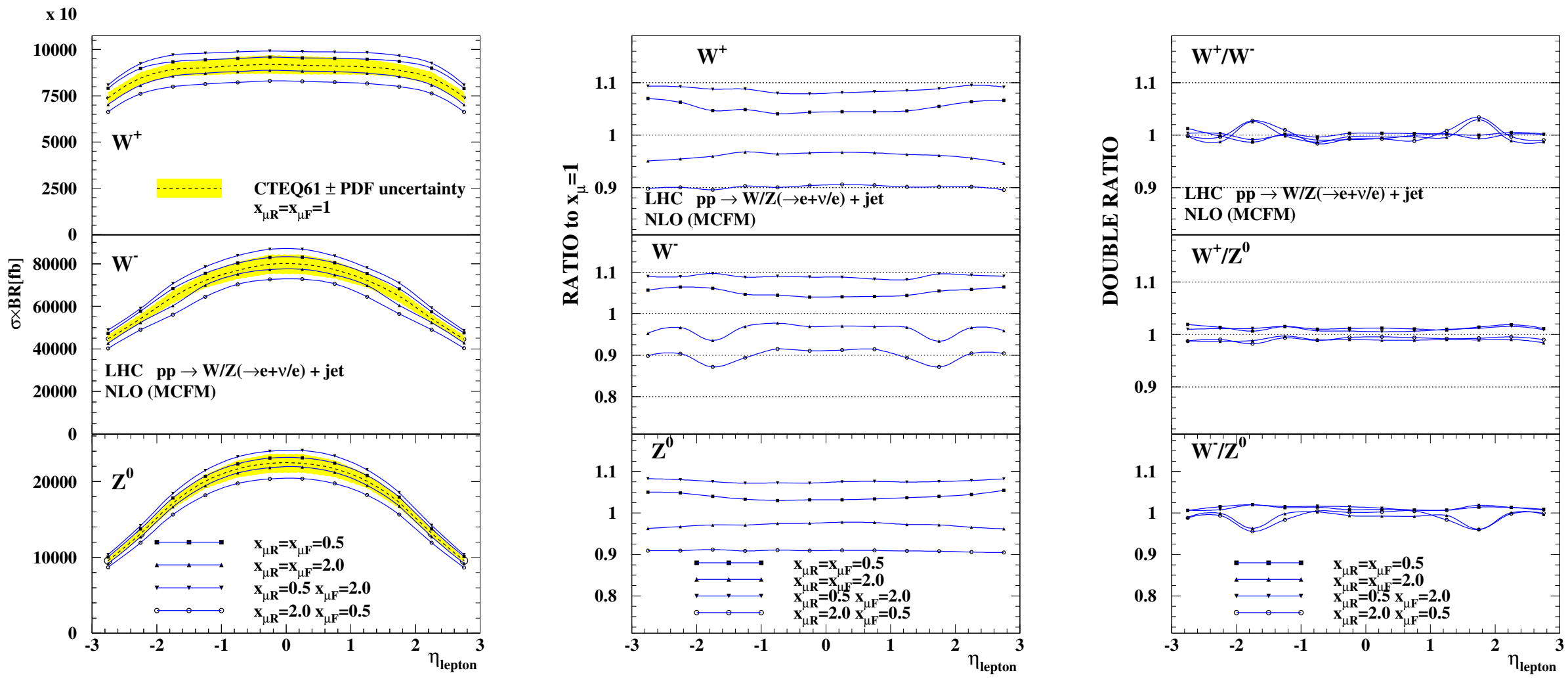

**Fig. 20:** Left: pseudo-rapidity distribution of the decay lepton from inclusive $W/Z + jet$ production for different values of $x_{\mu R}$ and $x_{\mu F} = 1$, centre: the ratio of predictions with respect to $x_{\mu} = 1$ and right: double ratio $V/V'$ of cross sections for actual scale settings normalised to the nominal scale.

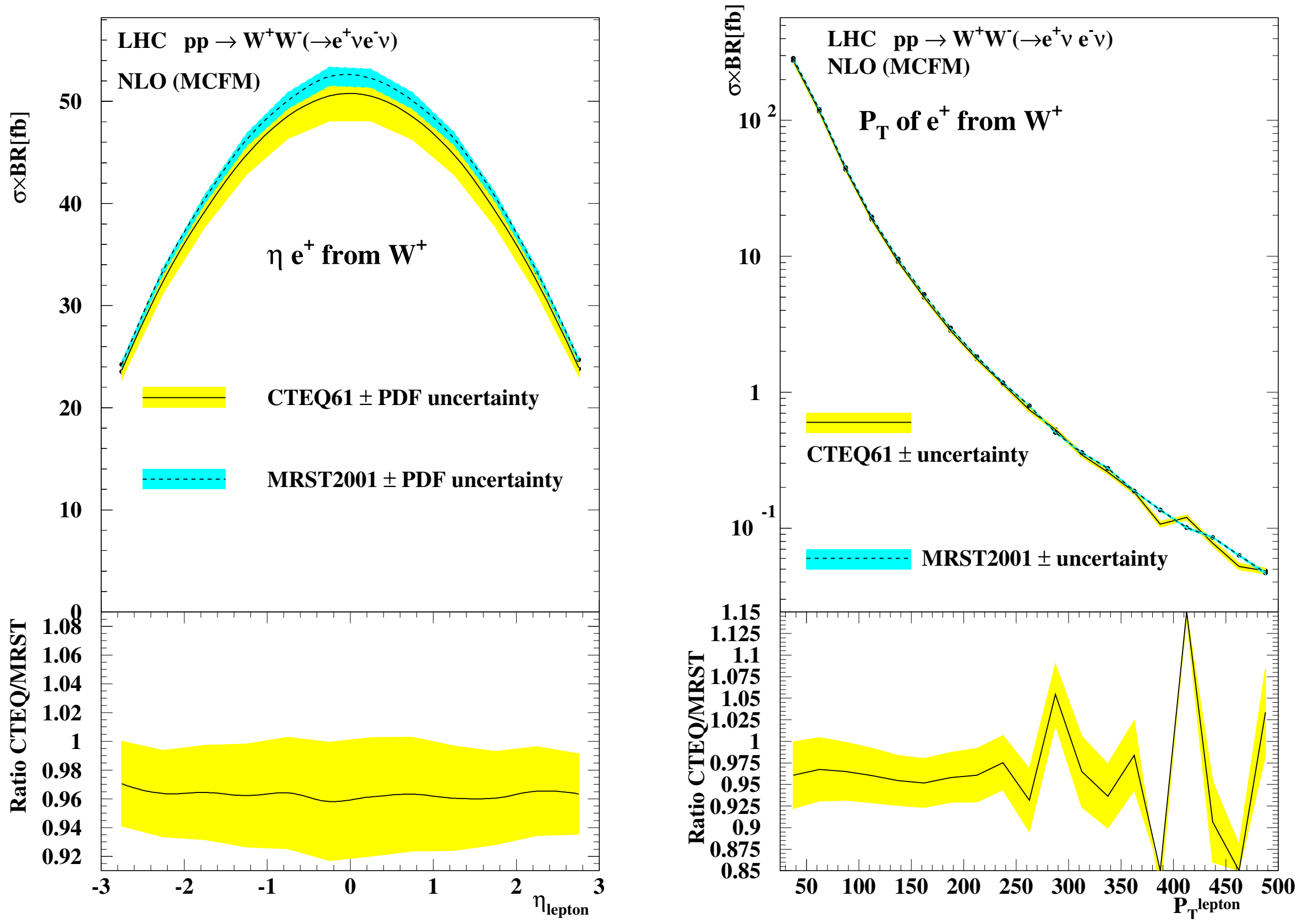

**Fig. 21:** Left: pseudo-rapidity distribution of the decay lepton from inclusive $WW$ production and right: transverse momentum of the decay lepton.

The perturbative uncertainties, obtained as for the other processes, are shown in Fig. 24 for the lepton distributions. The systematic uncertainties range from 3.3 to 4.9 % and are slightly smaller than for single bosons, given the larger scale $\mu = 2M_V$ and better applicability of perturbative QCD. The perturbative uncertainty is essentially constant across the pseudo-rapidity and largely correlated between different pair production processes.

The ratio of boson pair production to single $Z$ production is of particular interest, as similar quark configurations contribute to both process types, though evidently in a somewhat different $x, Q^2$ regime. This ratio is shown in Fig. 25 for the lepton distribution, given the different shapes of pseudo-rapidity is not flat but its PDF uncertainty is reduced to the level of 2 %. The perturbative uncertainties of the $VV/Z$ ratio, however, are only reduced for the $ZZ/Z$ case and even slightly larger for other ratios because the scale variations have partly an opposite effect on the cross sections for $Z$ and e.g. $WW$ production.

The total cross sections and their systematic uncertainties are summarised in Table 6.

### 1.5 Study of next-to-next-to-leading order QCD predictions for W and Z production at LHC[10]

It has been in 2004 that the first differential next-to-next-to-leading order (NNLO) QCD calculation for vector boson production in hadron collisions was completed by Anastasiou *et al.* [32]. This group has calculated the rapidity dependence for W and Z production at NNLO. They have shown that the perturbative expansion stabilizes at this order in perturbation theory and that the renormalization and factorization scale uncertainties are drastically reduced, down to the level of one per-cent. It is therefore interesting to perform a more detailed study of these NNLO predictions for various observables which can be measured at LHC, as well as to investigate their systematic uncertainties.

[10]Contributing author:Günther Dissertori

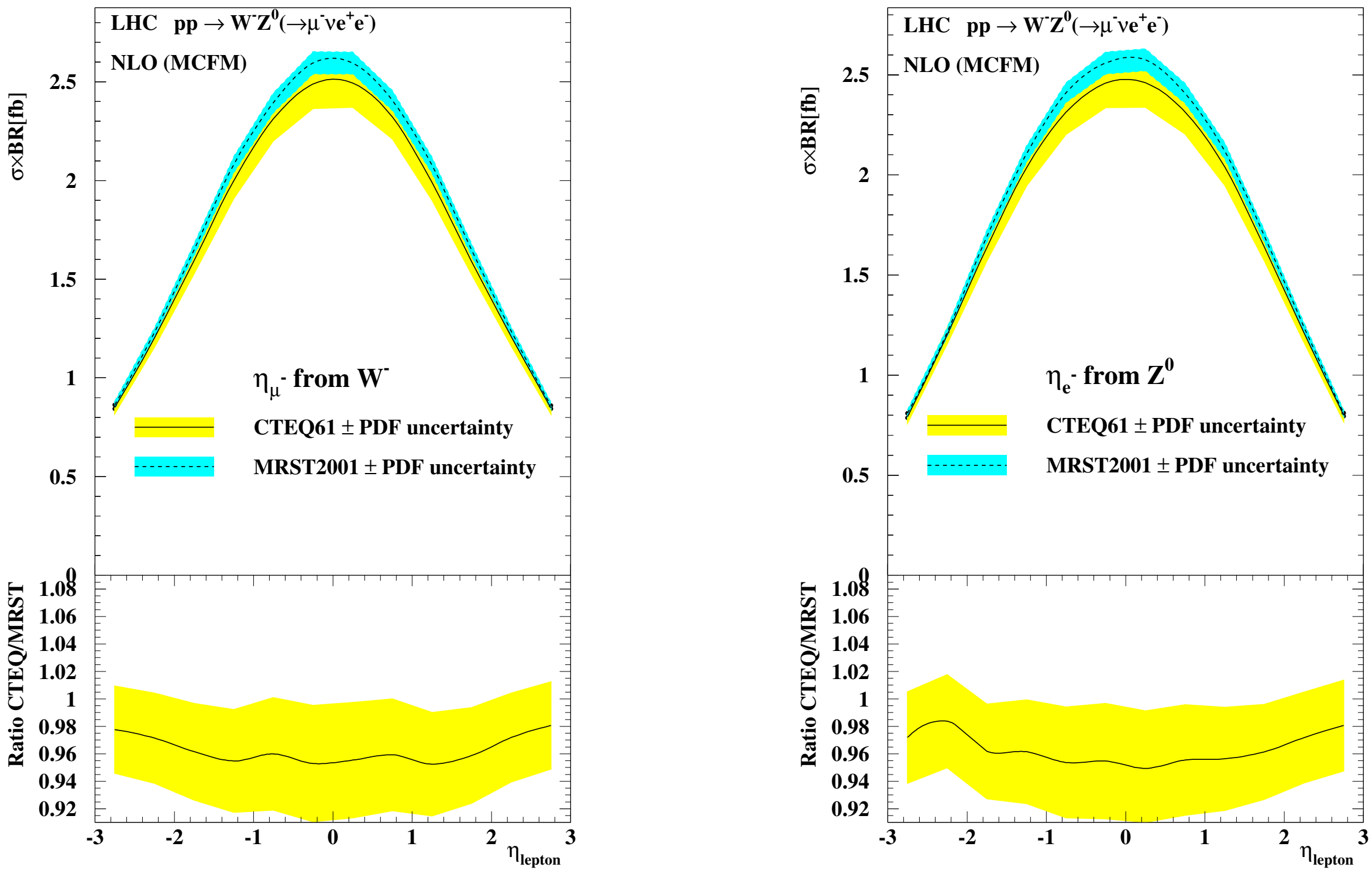

**Fig. 22:** Left: pseudo-rapidity distribution of the decay lepton of the $W^-$from inclusive $W^-Z^0$ production and right: pseudo-rapidity distribution of a decay lepton of the $Z^0$.

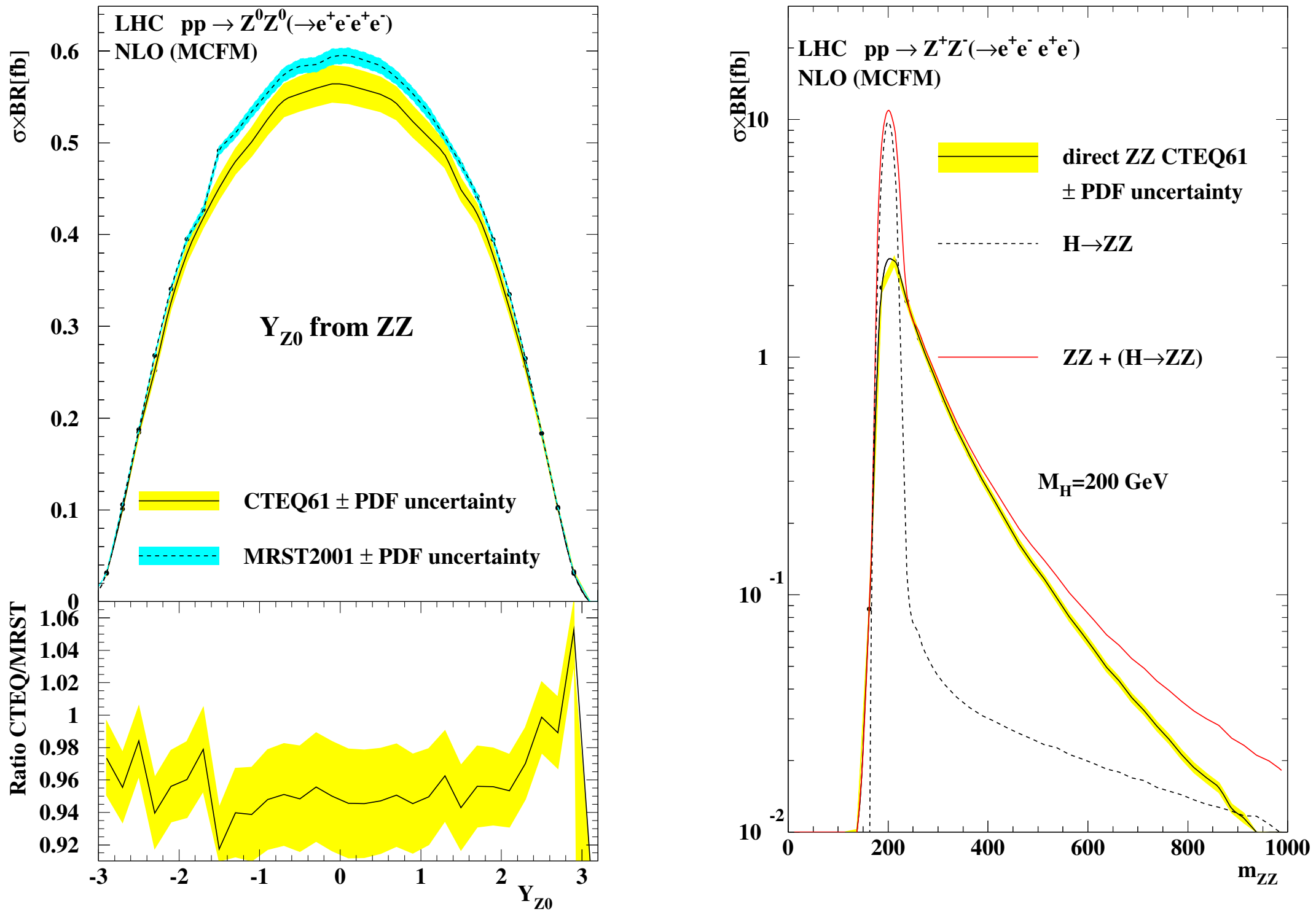

**Fig. 23:** Left: rapidity distribution of the leading $Z$ from inclusive $ZZ$ production and right: invariant mass of the $ZZ$ pair for non-resonant continuum production compared to resonant pair production via the SM Higgs decay.

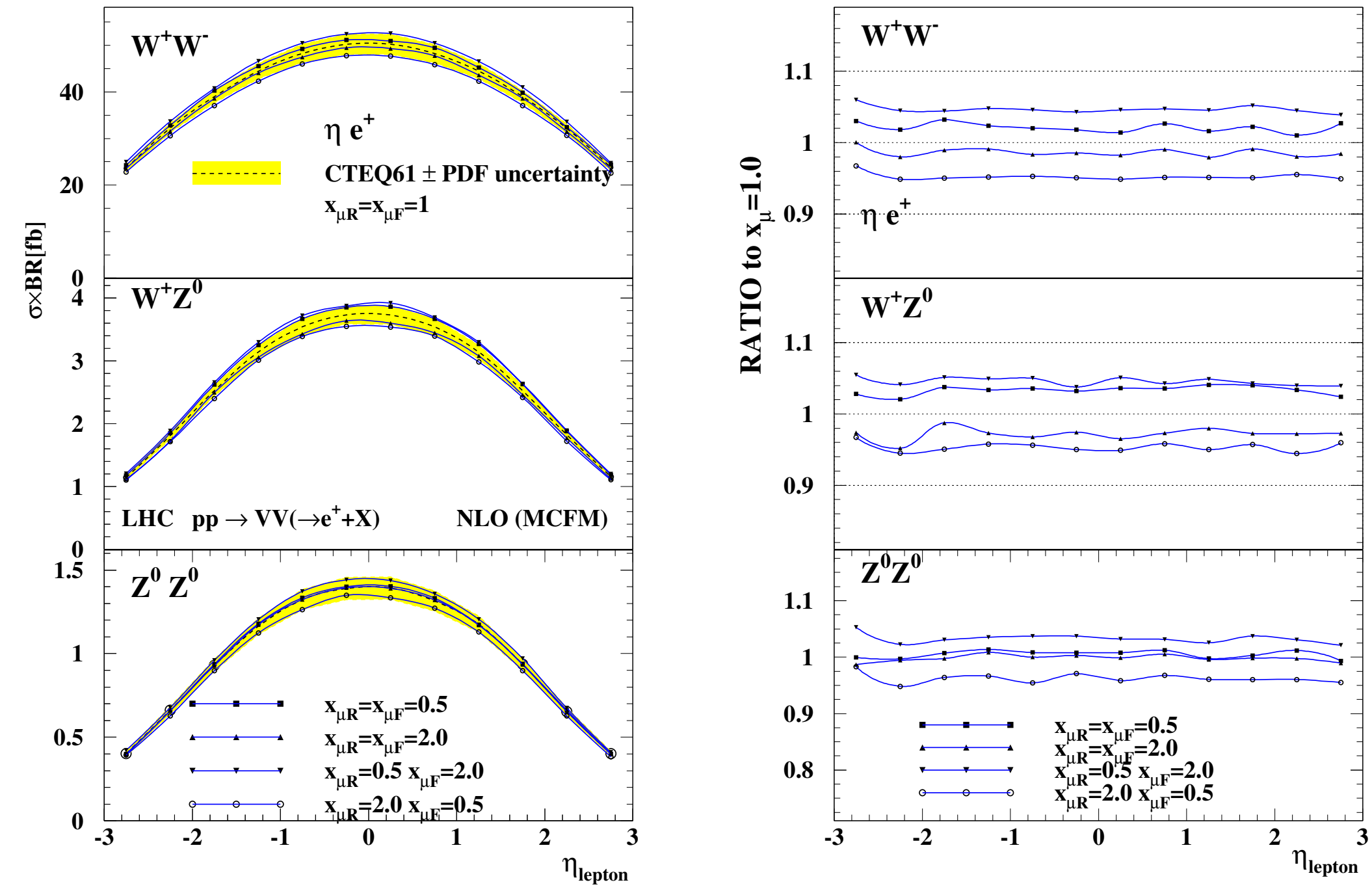

**Fig. 24:** Left: pseudo-rapidity distributions of leptons from various boson pair production processes and different scale settings and right: ratio of predictions relative to $x_\mu = 1$.

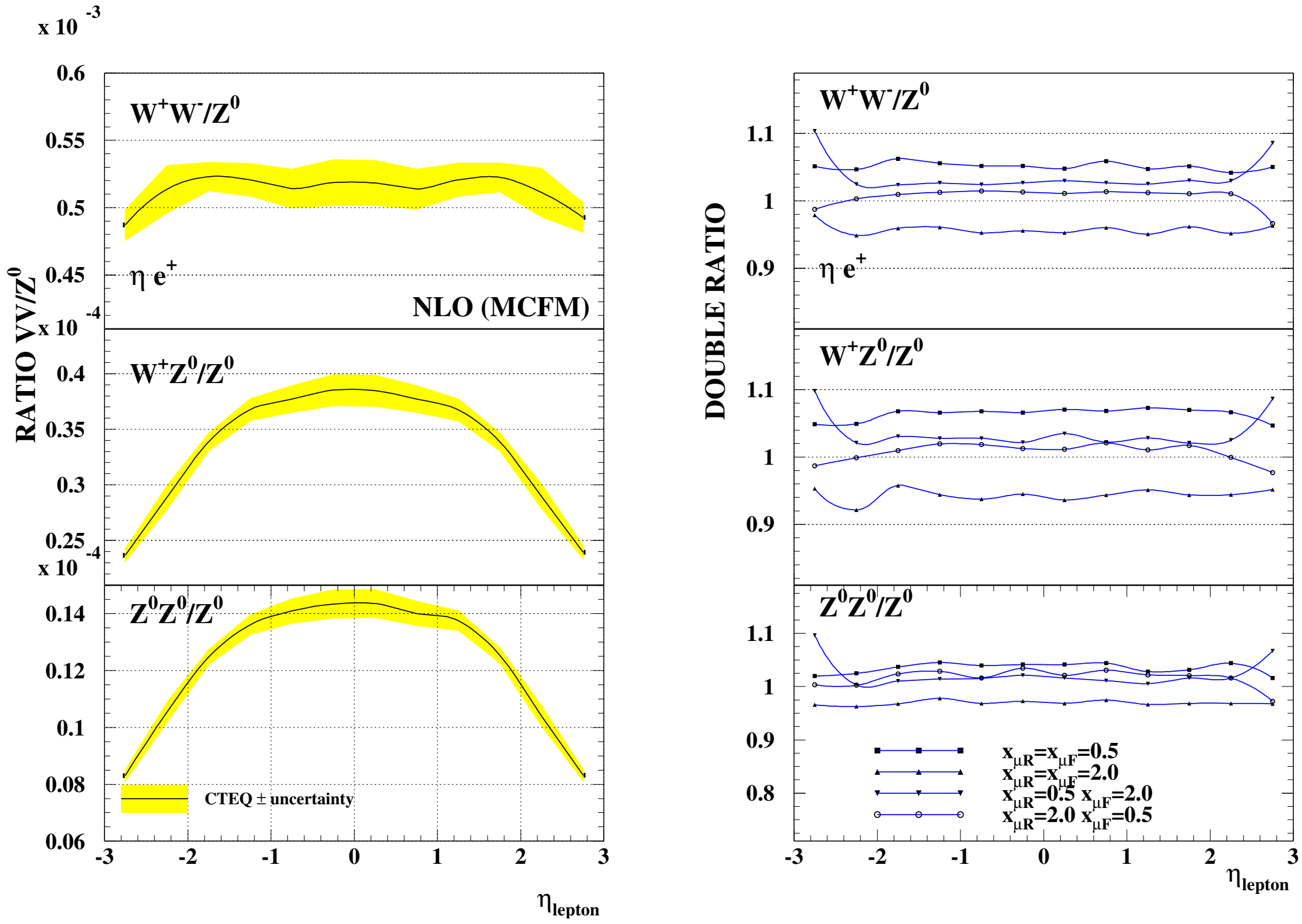

**Fig. 25:** Left: the ratio of pseudo-rapidity distributions of leptons from boson pair production processes normalised to single $Z$ production and right: the double ratio $VV/Z$ of predictions for different scales relative to $x_\mu = 1$.

**Table 6:** Total cross-sections and systematic uncertainties within the experimental acceptance for pair production processes.

| | $WW$ | $ZZ$ | $W^+Z^0$ | $W^-Z^0$ |
|---|---|---|---|---|
| CTEQ61 [fb] | 475.7 | 11.75 | 31.81 | 20.77 |
| $\Delta_{\rm PDF}^{\rm CTEQ}$ [fb] | $\pm 17.0$ | $\pm 0.48$ | $\pm 1.12$ | $\pm 0.80$ |
| $\Delta_{\rm PDF}^{\rm CTEQ}$ [%] | $\pm 3.6$ | $\pm 4.1$ | $\pm 3.5$ | $\pm 3.8$ |
| MRST [fb] | 494.2 | 12.34 | 32.55 | 21.62 |
| $\Delta_{\rm PDF}^{\rm MRST}$ [fb] | $\pm 6.3$ | $\pm 0.19$ | $\pm 0.49$ | $\pm 0.41$ |
| $\Delta_{\rm PDF}^{\rm MRST}$ [%] | $\pm 1.3$ | $\pm 1.6$ | $\pm 1.5$ | $\pm 1.9$ |
| $\Delta_{\rm pert}$ [%] | $+4.6$ | $+3.3$ | $+4.6$ | $+4.8$ |
| | $-4.9$ | $-3.8$ | $-4.7$ | $-4.7$ |

In the study presented here we have calculated both the differential (in rapidity) and inclusive cross sections for W, Z and high-mass Drell-Yan (Z/$\gamma^*$) production. Here "inclusive" refers to the results obtained by integrating the differential cross sections over a rapidity range similar to the experimentally accessible region, which might be more relevant than the complete cross section which also includes the large-rapidity tails.

Such a prediction would then be compared to the experimental measurements at LHC, which will allow for precise tests of the Standard Model as well as to put strong constraints on the parton distribution functions (PDFs) of the proton. It is clear that in the experiment only the rapidity and transverse momenta of the leptons from the vector boson decays will be accessible, over a finite range in phase space. In order to compute the rapidity of the vector boson by taking into account the finite experimental lepton acceptance, Monte Carlo simulations have to be employed which model vector boson production at the best possible precision in QCD, as for example the program MC@NLO [17]. The so computed acceptance corrections will include further systematic uncertainties, which are not discussed here.

### *1.5.1 Parameters and analysis method*

The NNLO predictions have been implemented in the computer code VRAP [33], which has been modified in order to include ROOT [34] support for producing ntuples, histograms and plots. The code allows to specify the collision energy (14 TeV in our case), the exchanged vector boson ($\gamma^*$, Z, $Z/\gamma^*$, $W^+$, $W^-$), the scale $Q$ of the exchanged boson ($M_{\rm Z}, M_{\rm W}$ or off-shell, e.g. $Q = 400\,{\rm GeV}$), the renormalization and factorization scales, the invariant mass of the di-lepton system (fixed or integrated over a specified range), the value of the electro-magnetic coupling ($\alpha_{\rm QED} = 1/128$ or $\alpha_{\rm QED}(Q)$) and the number of light fermions considered. Regarding the choice of pdfs, the user can select a pdf set from the MRST2001 fits [35] or from the ALEKHIN fits [36], consistent at NNLO with variable flavour scheme. We have chosen the MRST2001 NNLO fit, mode 1 with $\alpha_s(M_{\rm Z}) = 0.1155$ [35], as reference set.

The program is run to compute the differential cross section $d\sigma/dY$, $Y$ being the boson rapidity, at a fixed number of points in $Y$. This result is then parametrized using a spline interpolation, and the thus found function can be integrated over any desired rapidity range, such as $|Y| < 2, |Y| < 2.5$ or $|Y| < 3$, as well as over finite bins in rapidity. For the study of on-shell production the integration range over the di-lepton invariant mass $M_{ll}$ was set to $M_V - 3\Gamma_V < M_{ll} < M_V + 3\Gamma_V$, with $M_V$ and $\Gamma_V$ the vector boson mass and width. This simulates an experimental selection over a finite signal range.

The systematic uncertainties have been divided into several categories: The PDF uncertainty is estimated by taking the maximum deviation from the reference set when using different PDFs from within the MRST2001 set or the ALEKHIN set. The latter difference is found to give the maximal variation in all of the investigated cases. The renormalization and factorization scales $\mu = \mu_{\rm R}, \mu_{\rm F}$ have been varied between $0.5 < \mu/Q < 2$, both simultaneously as well as fixing one to $\mu = Q$ and varying the other. The maximum deviation from the reference setting $\mu = Q$ is taken as uncertainty. The observed difference when using either a fixed or a running electro-magnetic coupling constant is also studied as possible systematic uncertainty due to higher-order QED effects. Since it is below the one per-cent level, it is not discussed further. Finally, in the case of Z production it has been checked that neglecting photon exchange and interference contributions is justified in view of the much larger PDF and scale uncertainties.

### *1.5.2 Results for W and Z production*

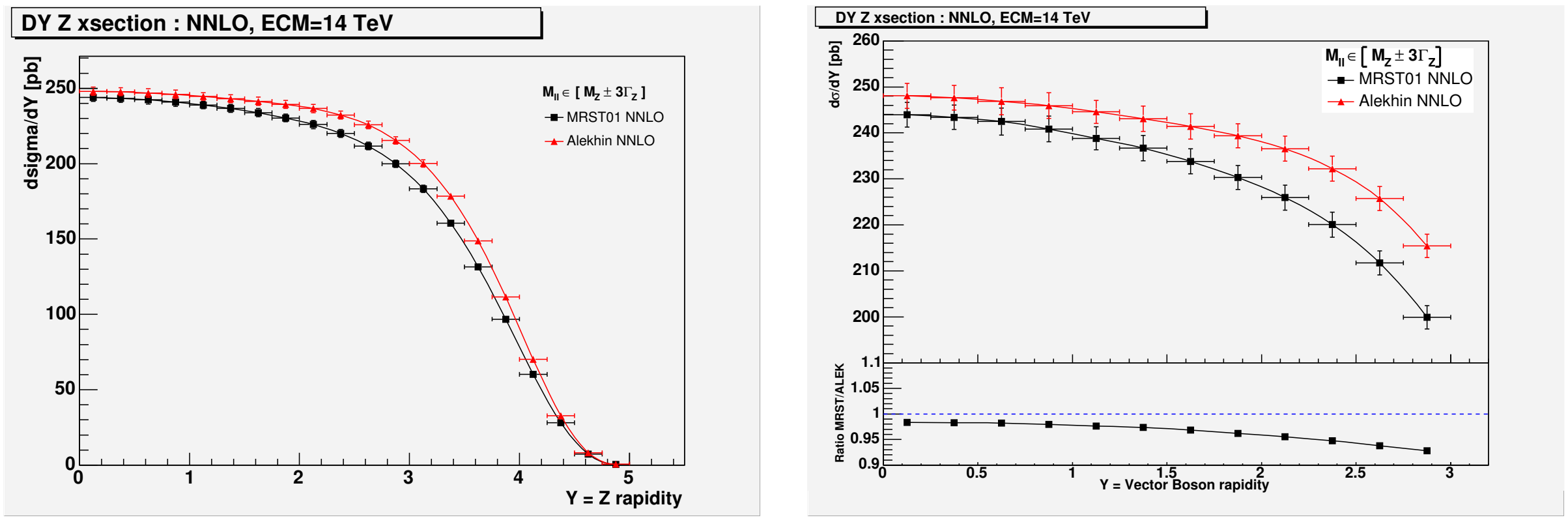

**Fig. 26:** Left : Drell-Yan Z production cross section ($\times$ BR) at LHC energies, as a function of the Z rapidity, for two different PDF choices. Right : Zoom into a restricted rapidity region, with the ratio of the predictions for the two different PDF sets as lower inset. The error bars indicate the scale uncertainties.

In Fig. 26 the results for Z production at LHC are shown for two different choices of PDF set, as a function of the boson rapidity. It can be seen that the predictions differ by about 2% at central rapidity, and the difference increases to about 5% at large rapidity. A similar picture is obtained when integrating the differential cross section up to rapidities of 2, 2.5 and 3 (Table 7). The more of the high-rapidity tail is included, the larger the uncertainty due to the PDF choice. From Table 1 it can also be seen that the scale uncertainties are slightly below the one per-cent level. It is worth noting that the choice of the integration range over the di-lepton invariant mass can have a sizeable impact on the cross section. For example, increasing the range from the standard value to $66\,{\rm GeV} < M_{\rm Z} < 116\,{\rm GeV}$ increases the cross section by 8%.

**Table 7:** NNLO QCD results for W and Z production at the LHC for the integration over different rapidity ranges. Also given are the relative uncertainties due to the choice of the PDFs and of the renormalization and factorization scale. The numbers include the branching ratio $Z(W) \to ee(e\nu)$.

| Channel | Z prod. | | | W prod. | | |
|---|---|---|---|---|---|---|
| range | $\lvert Y\rvert < 2$ | $\lvert Y\rvert < 2.5$ | $\lvert Y\rvert < 3$ | $\lvert Y\rvert < 2$ | $\lvert Y\rvert < 2.5$ | $\lvert Y\rvert < 3$ |
| cross section [nb] | 0.955 | 1.178 | 1.384 | 9.388 | 11.648 | 13.800 |
| $\Delta$ PDF [%] | 2.44 | 2.95 | 3.57 | 5.13 | 5.47 | 5.90 |
| $\Delta$ scale [%] | 0.85 | 0.87 | 0.90 | 0.99 | 1.02 | 1.05 |

The results for W production (Table 7) have been obtained by first calculating separately the cross sections for $W^+$ and $W^-$ production, and then adding these up. Again we observe an increase of the PDF uncertainty when going to larger rapidity ranges. Compared to the Z production, here the PDF uncertainties are larger, between 5 and 6%, whereas the scale uncertainties are of the same level, $\approx 1\%$. It is interesting to note that the PDF uncertainty for $W^-$ production is about 10 - 20% (relative) lower than that for $W^+$.

A considerable reduction in systematic uncertainty can be obtained by calculating cross section ratios. Two options have been investigated, namely the ratios $\sigma(W^+)/\sigma(W^-)$ and $\sigma(W)/\sigma(Z)$. As can be seen from Figure 27, the PDF uncertainties are reduced to the 0.7% level in the former ratio, and to about $2\%$ in the latter. The scale uncertainties are reduced to the $0.15\%$ level in both cases. Taking such ratios has also the potential advantage of reduced experimental systematic uncertainties, such as those related to the acceptance corrections.

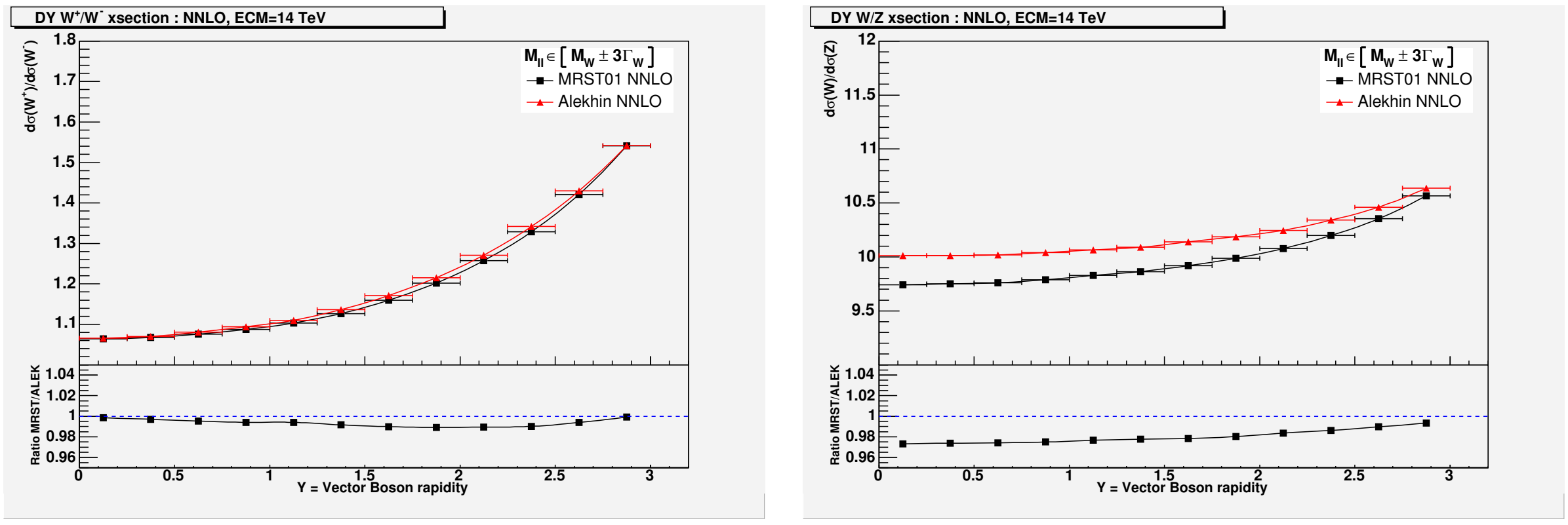

**Fig. 27:** Ratio of the production cross sections for $W^+$, $W^-$ (left), and W, Z (right), as a function of rapidity, for two different PDF sets. The inserts show the ratios of the results for the two PDF choices.

### *1.5.3 Results for high-mass Drell-Yan processes*

Similarly to on-shell W and Z production we have also analyzed the high-mass Drell-Yan process, namely $Z/\gamma^*$ production at a scale of $Q = 400$ GeV. In this case the di-lepton invariant mass has been integrated over the range $M_{ll} = 400 \pm 50$ GeV. Here the PDF uncertainties are found between $3.7\%$ and $5.1\%$ for the various integration ranges over rapidity, somewhat larger than for on-shell production. However, by normalizing the high-mass production cross section to the on-shell case, the PDF uncertainties are considerably reduced, being 1.2 - $1.5\%$.

The systematic uncertainties related to the renormalization and factorization scale are reduced ($\Delta$ scale $\approx 0.2\%$) when going to the high-mass exchange, as expected from perturbative QCD with a decreasing strong coupling constant. In this case a normalization of the cross section to the on-shell case does not give an improvement. However, since the scale uncertainties are well below the PDF uncertainties, this is less of an issue for the moment.

### *1.5.4 Summary*

We have studied NNLO QCD predictions for W and Z production at LHC energies. We have identified the choice of PDF set as the dominant systematic uncertainty, being between 3 and $6\%$. The choice of the renormalization and factorization scale leads to much smaller uncertainties, at or below the $1\%$ level. In particular we have shown that the systematic uncertainties can be sizeably reduced by taking ratios of cross sections, such as $\sigma(W^+)/\sigma(W^-)$, $\sigma(W)/\sigma(Z)$ or $\sigma(Z/\gamma^*, Q = 400\,\mathrm{GeV})/\sigma(Z/\gamma^*, Q = M_Z)$. For such ratios it can be expected that also part of the experimental uncertainties cancel. With theoretical uncertainties from QCD at the few per-cent level the production of W and Z bosons will most likely be the best-known cross section at LHC.

Concerning the next steps, it should be considered that at this level of precision it might become relevant to include also higher-order electro-weak corrections. In addition, since experimentally the boson rapidity will be reconstructed from the measured lepton momenta, a detailed study is needed to evaluate the precision at which the acceptance correction factors for the leptons from the boson decays can be obtained. For this Monte Carlo programs such as MC@NLO should be employed, which combine next-to-leading-order matrix elements with parton showers and correctly take account of spin correlations.

# Experimental determination of Parton Distributions

*T. Carli, A. Cooper-Sarkar, J. Feltesse, A. Glazov, C. Gwenlan, M. Klein, T. Laštovička G. Laštovička-Medin, S. Moch, B. Reisert G. Salam, F. Siegert*

## 1 Introduction [1]

With HERA currently in its second stage of operation, it is possible to assess the potential precision limits of HERA data and to estimate the potential impact of the measurements which are expected at HERA-II, in particular with respect to the PDF uncertainties.

Precision limits of the structure function analyses at HERA are examined in [1]. Since large amounts of luminosity are already collected, the systematic uncertainty becomes most important. A detailed study of error sources with particular emphasis on correlated errors for the upcoming precision analysis of the inclusive DIS cross section at low $Q^2$ using 2000 data taken by the H1 experiment is presented. A new tool, based on the ratio of cross sections measured by different reconstruction methods, is developed and its ability to qualify and unfold various correlated error sources is demonstrated.

An important issue is the consistency of the HERA data. In section 3, the H1 and ZEUS published PDF analyses are compared, including a discussion of the different treatments of correlated systematic uncertainties. Differences in the data sets and the analyses are investigated by putting the H1 data set through both PDF analyses and by putting the ZEUS and H1 data sets through the same (ZEUS) analysis, separately. Also, the HERA averaged data set (section 4) is put through the ZEUS PDF analysis and the result is compared to that obtained when putting the ZEUS and H1 data sets through this analysis together, using both the Offset and Hessian methods of treating correlated systematic uncertainties.

The HERA experimental data can not only be cross checked with respect to each other but also combined into one common dataset, as discussed in section 4. In this respect, a method to combine measurements of the structure functions performed by several experiments in a common kinematic domain is presented. This method generalises the standard averaging procedure by taking into account point-to-point correlations which are introduced by the systematic uncertainties of the measurements. The method is applied to the neutral and charged current DIS cross section data published by the H1 and ZEUS collaborations. The averaging improves in particular the accuracy due to the cross calibration of the H1 and ZEUS measurements.

The flavour decomposition of the light quark sea is discussed in [2]. For low $x$ and thus low $Q^2$ domain at HERA only measurement of the photon exchange induced structure functions $F_2$ and $F_L$ is possible, which is insufficient to disentangle individual quark flavours. A general strategy in this case is to assume flavour symmetry of the sea. [2] considers PDF uncertainties if this assumption is released. These uncertainties can be significantly reduced if HERA would run in deuteron-electron collision mode.

The impact of projected HERA-II data on PDFs is estimated in section 7. In particular, next-to-leading order (NLO) QCD predictions for inclusive jet cross sections at the LHC centre-of-mass energy are presented using the estimated PDFs. A further important measurement which could improve understanding of the gluon density at low $x$ and, at the same time, provide consistency checks of the low $Q^2$ QCD evolution is the measurement of the longitudinal structure function $F_L$. Perspectives of this measurement are examined in section 5, while the impact of this measurement is also estimated in section 7.

Further improvements for consistently including final-state observables in global QCD analyses are discussed in section 8. There, a method for "a posteriori" inclusion of PDFs, whereby the Monte Carlo run calculates a grid (in $x$ and $Q$) of cross section weights that can subsequently be combined with an arbitrary PDF. The procedure is numerically equivalent to using an interpolated form of the PDF. The main novelty relative to prior work is the use of higher-order interpolation, which substantially improves the tradeoff between accuracy and memory use. An accuracy of about $0.01\%$ has been reached for the single inclusive cross-section in the central rapidity region $|y| < 0.5$ for jet transverse momenta from $100$ to $5000\mathrm{GeV}$. This method will make it possible to consistently include measurements done at HERA, Tevatron and LHC in global QCD analyses.

[1] Subsection coordinators: A. Glazov, S. Moch

## 2 Precision Limits for HERA DIS Cross Section Measurement [2]

The published precision low $Q^2$ cross section data [3] of the H1 experiment became an important data set in various QCD fit analyses [3–6]. Following success of these data the H1 experiment plans to analyse a large data sample, taken during 2000 running period[3], in order to reach precision limits of low $Q^2$ inclusive cross sections measurements at HERA. The precision is expected to approach $1\%$ level.

The aim of this contribution is to calculate realistic error tables for 2000 H1 data and pursue paths how to reach such a high precision. Correlated error sources are studied in particular and a new tool, based on the ratio of cross sections measured by different reconstruction methods, is developed. All errors, including correlated errors, are treated in the same manner as in [3]. Error tables are provided and used in QCD fit analysis, see Sec 7, in order to study the impact of the new data on PDFs. The new data are expected to reach higher precision level than [3] due to the following reasons

- Larger data statistics - Statistical errors will decrease by factor of $1.5 - 2$, compared to [3], depending on the kinematic region.
- Very large Monte Carlo simulations (MC) - Due to a progress in computing a number of simulated events can be significantly increased in order to minimise statistical error of MC, to understand uncorrelated errors and to estimate correlated errors more precisely.
- During past years increasing knowledge, arriving from various H1 analyses, enabled better understanding of the detector and its components as well as improving quality of MC.
- Data taking in 2000 was particularly smooth. Both HERA and H1 were running at peak performance for HERA-I running period.

This contribution uses existing 2000 data and MC ntuples along with the full analysis chain. It applies all preliminary technical work done on these data, including calibration, alignment, trigger studies etc. Quoted errors are assumed to be achieved in the final version of analysis yet the analysis has not been finalised, all the numbers in the paper are preliminary and may change in the publication.

The uncertainties of the cross section measurement are divided into a number of different types. Namely, these are *statistical uncertainties* of the data, *uncorrelated systematics* and *correlated systematics*. The term 'correlated' refers to the fact that cross section measurements in kinematic bins are affected in a correlated way while different correlated systematic error sources are considered uncorrelated among each other. The classification of the systematic errors into types is sometimes straightforward (MC statistics is uncorrelated error source) but sometimes is rather arbitrary (radiative corrections are assumed to be uncorrelated error source). The main goal of this classification is to preserve correlation between data points while keeping the treatement as simple as possible.

The cross section uncertainties depend on the method used to reconstruct event kinematics. There are various methods existing, involving a measurement of the scattered electron as well as of the hadronic finale state. In the following two of them, so called *electron method* and *sigma method*, are employed [7]. The electron method uses only the measurement of the scattered electron, namely its energy and polar angle, while the sigma method uses both the scattered electron and the hadronic final state. An advantage of the sigma method is a proper treatment of QED radiation from the incoming beam electron (ISR).

[2]Contributing authors: G. Laštovička-Medin, A. Glazov, T. Laštovička
[3]Data statistics will be increased further by adding data taken in year 1999.

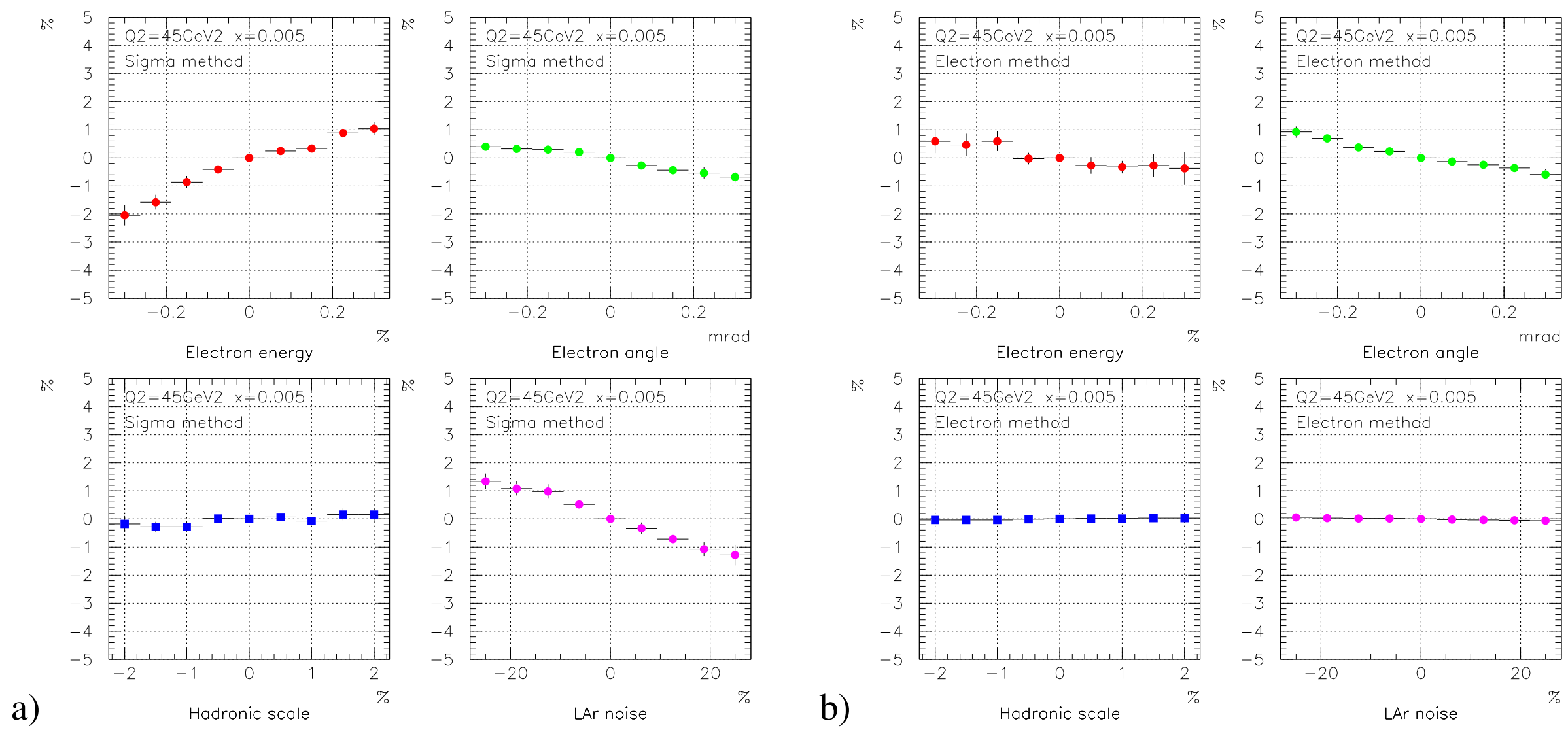

**Fig. 1:** A scan of the cross section measurement change in % depending on a variation of (from top-left) electron energy, electron polar angle, hadronic final state calibration scale and noise level in LAr calorimeter (bottom-right). The sigma method (a) and the electron method (b) were used to reconstruct kinematics of events.

The *statistical uncertainty* of the data is typically 0.5-1%, depending on the kinematic region analysed and the definition of the kinematic bins. In the following we adapt the bin definition used in [3], apart from merging bins at low $y$ which was done in the published data in order to increase statistics.

The *uncorrelated systematics* consists from various contributions. A cross section uncertainty due to the Monte Carlo statistics is the one with very good potential to be minimised. In the following we assume 100 million simulated events to be used in analysis of 2000 data. Estimates were calculated with available 12 million simulated events and corresponding statistical errors scaled by a factor of $\sqrt{100/12}$. As a result the uncertainty is very small and typically on the level of few permile.

Additional contributions to the uncorrelated systematics are efficiencies. We assume for trigger efficiency 0.3% and backward tracker tracker efficiency 0.3% uncertainty. Radiative corrections are expected to affect the final cross section by 0.4%.

Effect of *correlated uncertainties* on the cross section measurement is studied in the following manner. Particular source of correlated uncertainty, for instance the scattered electron energy measurement, is varied by assumed error and the change of the measured cross section is quoted as the corresponding cross section measurement error. An example of cross section change on various correlated error source is shown in Fig. 1 for bin of $Q^2 = 45\,\mathrm{GeV}^2$ and $x = 0.005$. The kinematics of events was reconstructed with the sigma method (a) and the electron method (b). Errors are calculated as so called standard errors of the mean in calculation of which the available Monte Carlo sample was split into nine sub-samples. It is clearly seen that the cross section measurement with the sigma method in this kinematic bin is particularly sensitive to the electron energy measurement (top-left) and to noise description in LAr calorimeter (bottom-right). On the contrary, the electron polar angle measurement and the calibration of the hadronic final state play a little role. The electron method is mainly sensitive to the electron energy measurement. The importance of the systematic sources vary from bin to bin.

There are five individual sources contributing to the correlated cross section uncertainties:

– Uncertainties of 0.15% at $E_e = 27\,\mathrm{GeV}$ and 1% at 7 GeV are assigned to the electron energy scale for the backward calorimeter. The uncertainty is treated as a linear function of $E_e$ interpolating between the results at 27 GeV and 7 GeV.
– The uncertainty on the scattered electron polar angle measurement is 0.3 mrad . The corresponding

**Table 1:** An example of the error table for $Q^2 = 25\,\mathrm{GeV}^2$ for 2000 data, large Monte Carlo sample and suppressed systematic errors compared to [1], see text for details. Absolute errors are shown. The table format is identical to the one published in [1].

| $Q^2$ | x | y | $\sigma_r$ | R | $F_2$ | Tot.(%) | Sta. | Uncorr. | Corr. | $E_e$ | θ | Ehad | Noise | γp |
|---|---|---|---|---|---|---|---|---|---|---|---|---|---|---|
| 25 | 0.0005 | 0.493 | 1.391 | 0.261 | 1.449 | 0.88 | 0.47 | 0.63 | 0.41 | 0.19 | 0.21 | 0.22 | 0.15 | 0.13 |
| 25 | 0.0008 | 0.308 | 1.251 | 0.261 | 1.268 | 0.91 | 0.43 | 0.62 | 0.51 | 0.34 | 0.37 | 0.02 | 0.04 | 0 |
| 25 | 0.0013 | 0.19 | 1.138 | 0.248 | 1.143 | 0.94 | 0.44 | 0.62 | 0.56 | 0.45 | 0.33 | 0.03 | 0.02 | 0 |
| 25 | 0.002 | 0.123 | 1.041 | 0.236 | 1.042 | 0.9 | 0.45 | 0.62 | 0.47 | 0.13 | 0.45 | 0.03 | 0.05 | 0 |
| 25 | 0.0032 | 0.077 | 0.842 | 0.254 | 0.843 | 1.42 | 0.5 | 0.63 | 1.17 | 0.74 | 0.36 | 0.17 | 0.8 | 0 |
| 25 | 0.005 | 0.049 | 0.745 | 0.243 | 0.745 | 1.17 | 0.52 | 0.63 | 0.83 | 0.59 | 0.42 | 0.25 | 0.33 | 0 |
| 25 | 0.008 | 0.031 | 0.667 | 0.225 | 0.667 | 1.22 | 0.56 | 0.64 | 0.87 | 0.43 | 0.35 | 0.66 | 0.09 | 0 |
| 25 | 0.013 | 0.019 | 0.586 | 0.214 | 0.586 | 2.02 | 0.65 | 0.66 | 1.8 | 0.67 | 0.57 | 1.43 | 0.65 | 0 |
| 25 | 0.02 | 0.012 | 0.569 | 0.159 | 0.569 | 5.77 | 0.86 | 0.71 | 5.66 | 0.83 | 0.52 | 3.51 | 4.33 | 0 |
| 25 | 0.032 | 0.008 | 0.553 | 0.065 | 0.553 | 10.64 | 1.34 | 0.88 | 10.52 | 0.93 | 0.64 | 3.86 | 9.72 | 0 |

**Table 2:** An example of the full error table for $Q^2 = 25\,\mathrm{GeV}^2$, published H1 data. The definition of kinematic bins is not identical to that in Table 1, some bins were merged to enlarge statistics.

| $Q^2$ | x | y | $\sigma_r$ | R | $F_2$ | Tot.(%) | Sta. | Uncorr. | Corr. | $E_e$ | θ | Ehad | Noise | γp |
|---|---|---|---|---|---|---|---|---|---|---|---|---|---|---|
| 25 | 0.0005 | 0.553 | 1.345 | 0.248 | 1.417 | 2.41 | 1.04 | 1.81 | 1.21 | -1.04 | -0.37 | 0.25 | 0.04 | -0.41 |
| 25 | 0.0008 | 0.346 | 1.242 | 0.243 | 1.263 | 1.94 | 0.67 | 1.62 | 0.85 | -0.6 | -0.6 | 0.04 | 0.02 | -0.07 |
| 25 | 0.0013 | 0.213 | 1.091 | 0.238 | 1.097 | 1.78 | 0.66 | 1.36 | 0.93 | -0.64 | -0.69 | 0 | 0 | 0 |
| 25 | 0.002 | 0.138 | 0.985 | 0.236 | 0.987 | 2.89 | 0.76 | 1.43 | 2.4 | 1.78 | -0.7 | 0.17 | 1.34 | 0 |
| 25 | 0.0032 | 0.086 | 0.879 | 0.234 | 0.88 | 2.78 | 0.79 | 1.46 | 2.23 | 1.8 | -0.77 | -0.23 | 0.92 | 0 |
| 25 | 0.005 | 0.055 | 0.754 | 0.234 | 0.754 | 2.38 | 0.85 | 1.49 | 1.64 | 1.01 | -0.58 | 0.16 | 1.03 | 0 |
| 25 | 0.008 | 0.034 | 0.663 | 0.234 | 0.663 | 2.52 | 0.92 | 1.54 | 1.78 | 1.11 | -0.68 | -0.72 | 0.84 | 0 |
| 25 | 0.0158 | 0.018 | 0.547 | 0.226 | 0.547 | 3.71 | 0.85 | 1.49 | 3.29 | 1.36 | -0.88 | -2.44 | -1.42 | 0 |
| 25 | 0.05 | 0.005 | 0.447 | 0.148 | 0.447 | 7.54 | 1.28 | 3.35 | 6.64 | 0.99 | -0.68 | -3.28 | -5.62 | 0 |

error on the cross section measurement is typically well below 1% but may be larger at lowest values of $Q^2$.

- The uncertainty on the hadronic energy scale comprises a number of systematic error sources corresponding to the $E - p_z$ decomposition: an uncertainty of the hadronic energy scale calibration of 2% for the central and forward calorimeter, an uncertainty of 3% for the fraction carried by tracks and a 5% uncertainty of the hadronic energy scale measured in backward calorimeter.
- The uncertainty on the hadronic energy scale is further affected by the subtracted noise in the calorimetery. The noise is described to the level of 10% and the corresponding error is propagated to the cross section uncertainty. The largest influence is in the low $y$ region, which is measured with the sigma method.
- The uncertainty due to the photoproduction background at large $y$ is estimated from the normalisation error of the PHOJET simulations to about 10%. At low and medium values of $y \lesssim 0.5$ it is negligible.

The total systematic error is calculated from the quadratic summation over all sources of the uncorrelated and correlated systematic uncertainties. The total error of the DIS cross section measurement is obtained from the statistical and systematical errors added in quadrature.

An example of the full error table for kinematic bin of $Q^2 = 25\,\mathrm{GeV}^2$ is shown in Table 1. For a comparison the corresponding part of the published data from [3] is presented in Table 2. One can see that precision about 1% can be reached especially in four lowest $x$ bins, where the electron method was used to reconstruct the event kinematics. The key contributions to the seen improvement in the cross section measurement precision are the electron energy measurement, very large Monte Carlo statistics, well understood noise in LAr calorimeter and precisely controlled efficiencies entering the analysis.

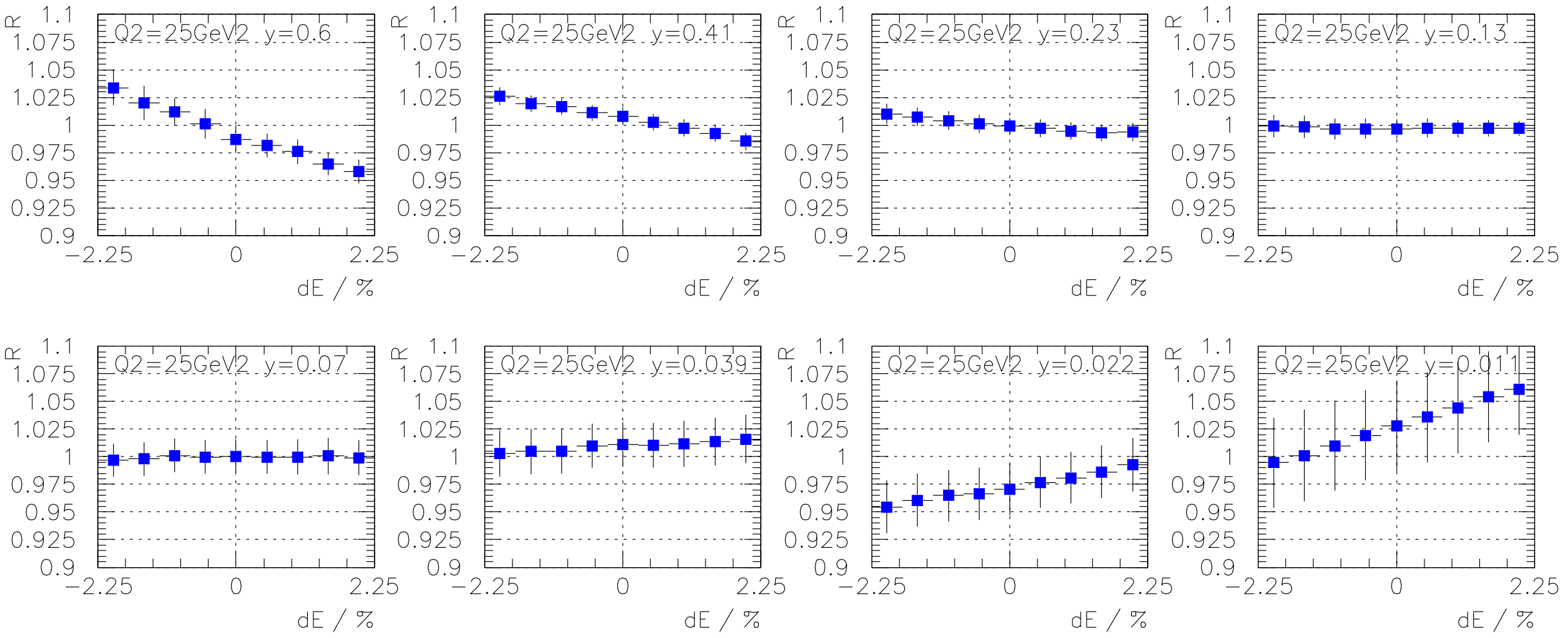

**Fig. 2:** A scan of the cross section ratio R in bins of $Q^2$ and $y$ as a function of the hadronic final state calibration variation.

Full error table, covering the kinematic region of $5 \leq Q^2 \leq 150\,\mathrm{GeV}^2$ and $0.01 \leq y \leq 0.6$ was produced. The electron method was applied for kinematic bins at $y > 0.1$ while the sigma method otherwise. The measurement of the proton structure function $F_2$ was simulated using fractal parametrisation [8] for central values, accounting for all sources of correlated and uncorrelated errors. This table was used to estimate effect of precise low $Q^2$ data on the determination of proton PDFs from QCD fits.

The fact that different kinematics reconstruction methods are affected differently by the correlated systematic uncertainties may be employed as a tool to estimate these uncertainties. We define

$$R_i = \frac{\sigma_r^{el,i}}{\sigma_r^{\Sigma,i}} \tag{1}$$

to be the cross section measurement ratio, where the reduced cross section $\sigma_r^{el,i}$ and $\sigma_r^{\Sigma,i}$ is measured using the electron method and the sigma method, respectively. Kinematic bins, indexed by $i$, cover a region of the analysis phase space where both reconstruction methods are applicable for the measurement. The statistical error of $R_i$ measurement is again evaluated by splitting the sample to a number of sub-samples and calculating the standard error of the mean. An example of a scan of the cross section ratio $R_i$ dependence on the hadronic final state calibration variation in a bin of $Q^2 = 25\,\mathrm{GeV}^2$ and various inelasticity $y$ is shown in Fig. 2.

An error of a particular correlated uncertainty source $j$ can be estimated by searching for lowest $\chi^2 = \sum_i (R_i(\alpha_j) - 1)^2/\sigma_i^2$, where summation runs over kinematic bins, $\sigma_i$ is the error of $R_i$ measurement and $\alpha_j$ is the variation of the source $j$. However, since there is a number of correlated error sources the correct way to find correlated uncertainties is account for all of them.

Unfolding of the correlated error sources can be linearised and directly solved by minimising the following function:

$$\mathcal{L} = \sum_i \frac{1}{\sigma_i^2}(R_i + \sum_j \alpha_j \frac{\partial R_i}{\partial \alpha_j} - 1)^2. \tag{2}$$

The partial derivatives $\frac{\partial R_i}{\partial \alpha_j}$ for systematic source $\alpha_j$ are obtained from linear fits to distributions as shown in Fig. 2. Parameters $\alpha_j$ and their respective errors are obtained by matrix inversion technique.

The procedure was tested on available Monte Carlo sample for 2000 H1 data. Half of the sample, six million events, was used to simulate data. Full analysis chain was applied to measure the cross section and thus $R_i$. Kinematic bins were selected according to $15 \leq Q^2 \leq 60\,\mathrm{GeV}^2$ and $0.011 \leq y \leq 0.6$, i.e.

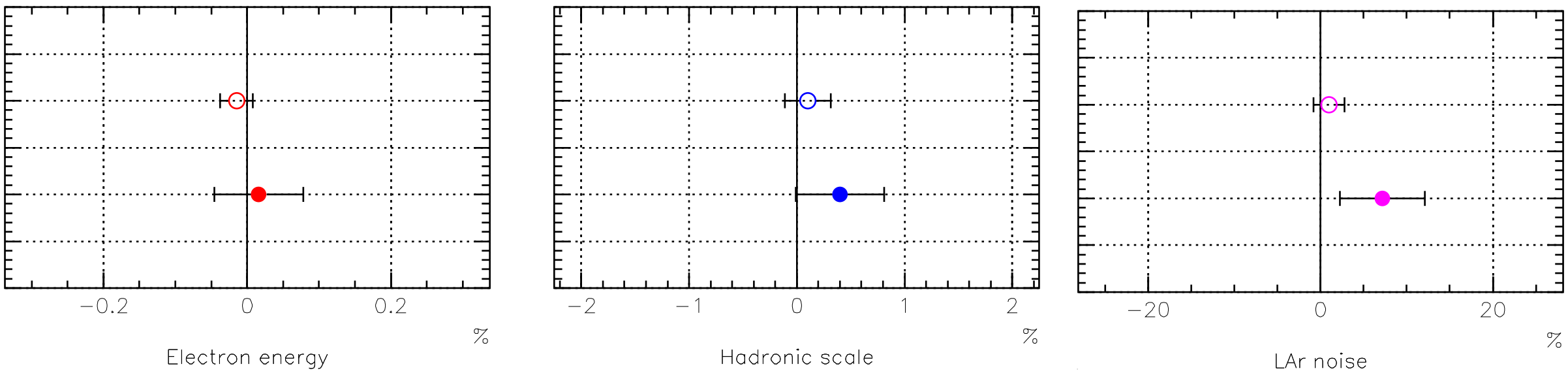

**Fig. 3:** Errors on the electron energy measurement (top-left), hadronic scale calibration (top-right) and noise in LAr calorimeter (bottom-left). Open points correspond to $\chi^2$ scan in one correlated error source. Closed points show the result of complete unfolding, taking into account correlations.

in the main region of the data. The results are shown in Fig. 3. Closed points correspond to unfolded errors of the electron energy measurement (top-left), hadronic final state calibration and noise in the LAr calorimeter (bottom-left). There is no sensitivity observed to the electron polar angle measurement. All values are within statistical errors compatible with zero, as expected. For the final analysis the statistical errors are expected to be approximately three times smaller due to the significantly larger statistics than used for the presented study. This will enable the method to gain sufficient control over systematic correlated errors. Apart from being able to evaluate calibration of the scattered electron and of the hadronic final state, it gives a very good handle on the LAr calorimeter noise.

For a comparison, open points in Fig. 3 correspond to a $\chi^2$ scan in one correlated error source. The statistical errors are smaller, as expected, and compatible with zero. However, the unfolding method is preferred since it takes into account all correlated error sources correctly.

In summary, a study of the DIS cross section uncertainties realistically achievable at HERA has been performed. For $x \in 0.001-0.01$ a precision of $1\%$ can be reached across for a wide range of $Q^2 \in 5-150$ GeV$^2$, allowing improved estimate of $W$, $Z$ production cross section in the central rapidity region of LHC. The accuracy of the DIS cross section measurement can be verified using different kinematic reconstruction methods available at the HERA collider.

## 3 Comparison and combination of ZEUS and H1 PDF analyses [4]

Parton Density Function (PDF) determinations are usually global fits [4,5,9], which use fixed target DIS data as well as HERA data. In such analyses the high statistics HERA NC $e^+p$ data, which span the range $6.3 \times 10^{-5} < x < 0.65, 2.7 < Q^2 < 30,000$GeV$^2$, have determined the low-$x$ sea and gluon distributions, whereas the fixed target data have determined the valence distributions and the higher-$x$ sea distributions. The $\nu$-Fe fixed target data have been the most important input for determining the valence distributions, but these data suffer from uncertainties due to heavy target corrections. Such uncertainties are also present for deuterium fixed target data, which have been used to determine the shape of the high-$x$ $d$-valence quark.

HERA data on neutral and charged current (NC and CC) $e^+p$ and $e^-p$ inclusive double differential cross-sections are now available, and have been used by both the H1 and ZEUS collaborations [10, 11] in order to determine the parton distributions functions (PDFs) using data from within a single experiment. The HERA high $Q^2$ cross-section data can be used to determine the valence distributions, thus eliminating uncertainties from heavy target corrections. The PDFs are presented with full accounting for uncertainties from correlated systematic errors (as well as from statistical and uncorrelated sources). Peforming an analysis within a single experiment has considerable advantages in this respect, since the global fits have found significant tensions between different data sets, which make a rigorous statistical

[4]Contributing authors: A. Cooper-Sarkar, C. Gwenlan

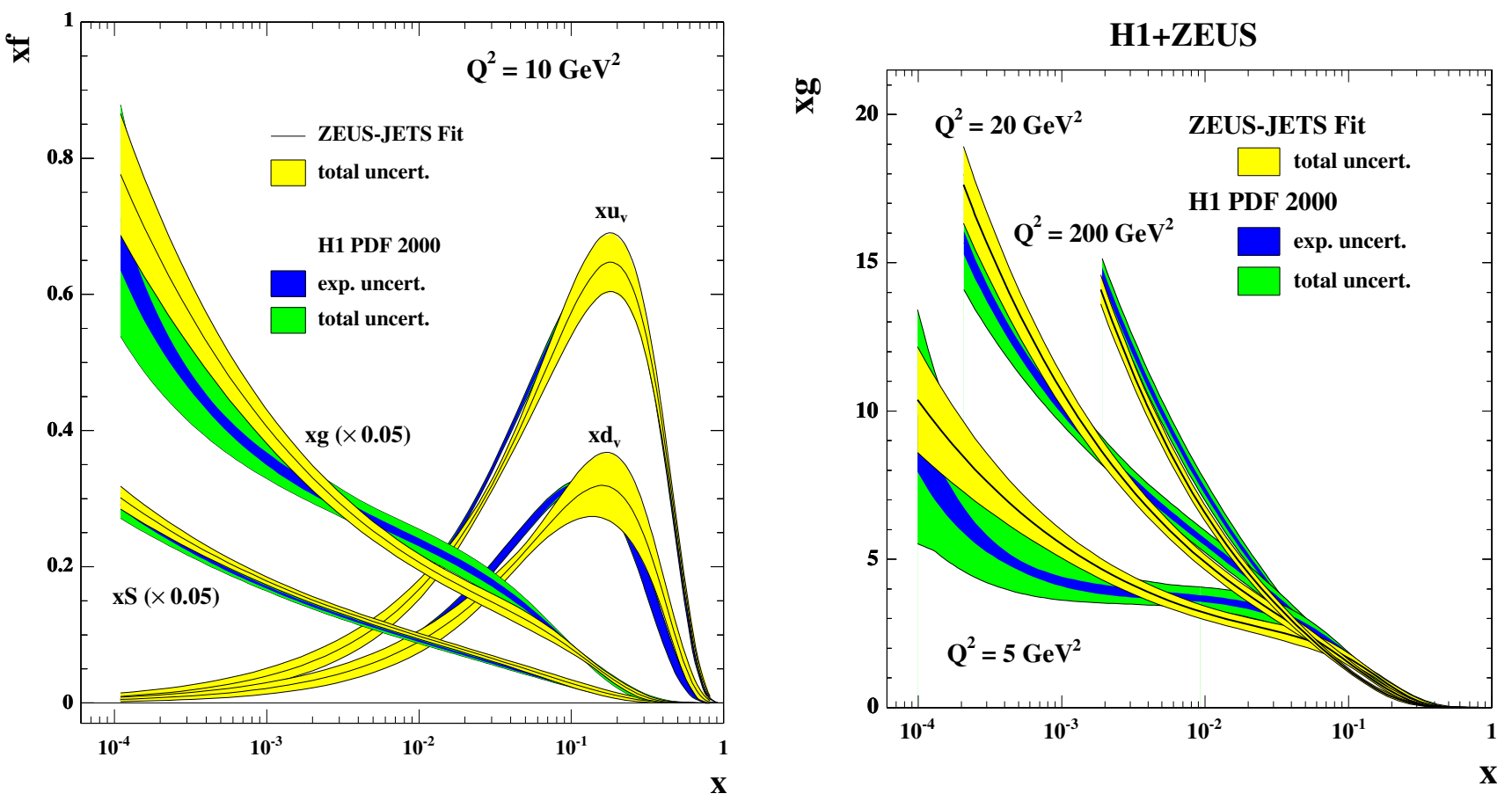

**Fig. 4:** Left plot: Comparison of PDFs from ZEUS and H1 analyses at $Q^2 = 10\text{GeV}^2$. Right plot: Comparison of gluon from ZEUS and H1 analyses, at various $Q^2$. Note that the ZEUS analysis total uncertainty includes both experimental and model uncertainties.

treatment of uncertainties difficult.

Fig. 4 compares the results of the H1 and ZEUS analyses. Whereas the extracted PDFs are broadly compatible within errors, there is a noticeable difference in the shape of the gluon PDFs. Full details of the analyses are given in the relevant publications, in this contribution we examine the differences in the two analyses, recapping only salient details.

### 3.1 Comparing ZEUS and H1 published PDF analyses

The kinematics of lepton hadron scattering is described in terms of the variables $Q^2$, the invariant mass of the exchanged vector boson, Bjorken $x$, the fraction of the momentum of the incoming nucleon taken by the struck quark (in the quark-parton model), and $y$ which measures the energy transfer between the lepton and hadron systems. The differential cross-section for the NC process is given in terms of the structure functions by

$$\frac{d^2\sigma(e^\pm p)}{dxdQ^2} = \frac{2\pi\alpha^2}{Q^4 x}\left[Y_+\, F_2(x,Q^2) - y^2\, F_L(x,Q^2) \mp Y_-\, xF_3(x,Q^2)\right], \tag{3}$$

where $Y_\pm = 1 \pm (1-y)^2$. The structure functions $F_2$ and $xF_3$ are directly related to quark distributions, and their $Q^2$ dependence, or scaling violation, is predicted by pQCD. At $Q^2 \leq 1000$ GeV$^2$ $F_2$ dominates the charged lepton-hadron cross-section and for $x \leq 10^{-2}$, $F_2$ itself is sea quark dominated but its $Q^2$ evolution is controlled by the gluon contribution, such that HERA data provide crucial information on low-$x$ sea-quark and gluon distributions. At high $Q^2$, the structure function $xF_3$ becomes increasingly important, and gives information on valence quark distributions. The CC interactions enable us to separate the flavour of the valence distributions at high-$x$, since their (LO) cross-sections are given by,

$$\frac{d^2\sigma(e^+p)}{dxdQ^2} = \frac{G_F^2 M_W^4}{(Q^2+M_W^2)^2 2\pi x} x\left[(\bar{u}+\bar{c}) + (1-y)^2(d+s)\right],$$

$$\frac{d^2\sigma(e^-p)}{dxdQ^2} = \frac{G_F^2 M_W^4}{(Q^2+M_W^2)^2 2\pi x} x\left[(u+c) + (1-y)^2(\bar{d}+\bar{s})\right].$$

For both HERA analyses the QCD predictions for the structure functions are obtained by solving the DGLAP evolution equations [12–15] at NLO in the $\overline{\text{MS}}$ scheme with the renormalisation and factorization scales chosen to be $Q^2$. These equations yield the PDFs at all values of $Q^2$ provided they are input as functions of $x$ at some input scale $Q_0^2$. The resulting PDFs are then convoluted with coefficient functions, to give the structure functions which enter into the expressions for the cross-sections. For a full explanation of the relationships between DIS cross-sections, structure functions, PDFs and the QCD improved parton model see ref. [16].

The HERA data are all in a kinematic region where there is no sensitivity to target mass and higher twist contributions but a minimum $Q^2$ cut must be imposed to remain in the kinematic region where perturbative QCD should be applicable. For ZEUS this is $Q^2 > 2.5$ GeV$^2$, and for H1 it is $Q^2 > 3.5$ GeV$^2$. Both collaborations have included the sensitivity to this cut as part of their model errors.

In the ZEUS analysis, the PDFs for $u$ valence, $xu_v(x)$, $d$ valence, $xd_v(x)$, total sea, $xS(x)$, the gluon, $xg(x)$, and the difference between the $d$ and $u$ contributions to the sea, $x(\bar{d} - \bar{u})$, are each parametrized by the form

$$p_1 x^{p_2} (1-x)^{p_3} P(x), \tag{4}$$

where $P(x) = 1 + p_4 x$, at $Q_0^2 = 7\text{GeV}^2$. The total sea $xS = 2x(\bar{u} + \bar{d} + \bar{s} + \bar{c} + \bar{b})$, where $\bar{q} = q_{sea}$ for each flavour, $u = u_v + u_{sea}, d = d_v + d_{sea}$ and $q = q_{sea}$ for all other flavours. The flavour structure of the light quark sea allows for the violation of the Gottfried sum rule. However, there is no information on the shape of the $\bar{d} - \bar{u}$ distribution in a fit to HERA data alone and so this distribution has its shape fixed consistent with the Drell-Yan data and its normalisation consistent with the size of the Gottfried sum-rule violation. A suppression of the strange sea with respect to the non-strange sea of a factor of 2 at $Q_0^2$, is also imposed consistent with neutrino induced dimuon data from CCFR. Parameters are further restricted as follows. The normalisation parameters, $p_1$, for the $d$ and $u$ valence and for the gluon are constrained to impose the number sum-rules and momentum sum-rule. The $p_2$ parameter which constrains the low-$x$ behaviour of the $u$ and $d$ valence distributions is set equal, since there is no information to constrain any difference. When fitting to HERA data alone it is also necessary to constrain the high-$x$ sea and gluon shapes, because HERA-I data do not have high statistics at large-$x$, in the region where these distributions are small. The sea shape has been restricted by setting $p_4 = 0$ for the sea, but the gluon shape is constrained by including data on jet production in the PDF fit. Finally the ZEUS analysis has 11 free PDF parameters. ZEUS have included reasonable variations of these assumptions about the input parametrization in their analysis of model uncertainties. The strong coupling constant was fixed to $\alpha_s(M_Z^2) = 0.118$ [17]. Full account has been taken of correlated experimental systematic errors by the Offset Method, as described in ref [9, 18].

For the H1 analysis, the value of $Q_0^2 = 4\text{GeV}^2$, and the choice of quark distributions which are parametrized is different. The quarks are considered as $u$-type and $d$-type with different parametrizations for, $xU = x(u_v + u_{sea} + c)$, $xD = x(d_v + d_{sea} + s)$, $x\bar{U} = x(\bar{u} + \bar{c})$ and $x\bar{D} = x(\bar{d} + \bar{s})$, with $q_{sea} = \bar{q}$, as usual, and the the form of the quark and gluon parametrizations given by Eq. 4. For $x\bar{D}$ and $x\bar{U}$ the polynomial, $P(x) = 1.0$, for the gluon and $xD$, $P(x) = (1 + p_4 x)$, and for $xU$, $P(x) = (1 + p_4 x + p_5 x^3)$. The parametrization is then further restricted as follows. Since the valence distributions must vanish as $x \to 0$, the low-$x$ parameters, $p_1$ and $p_2$ are set equal for $xU$ and $x\bar{U}$, and for $xD$ and $x\bar{D}$. Since there is no information on the flavour structure of the sea it is also necessary to set $p_2$ equal for $x\bar{U}$ and $x\bar{D}$. The normalisation, $p_1$, of the gluon is determined from the momentum sum-rule and the $p_4$ parameters for $xU$ and $xD$ are determined from the valence number sum-rules. Assuming that the strange and charm quark distributions can be expressed as $x$ independent fractions, $f_s$ and $f_c$, of the $d$ and $u$ type sea, gives the further constraint $p_1(\bar{U}) = p_1(\bar{D})(1 - f_s)/(1 - f_c)$. Finally there are 10 free parameters. H1 has also included reasonable variations of these assumptions in their analysis of model uncertainties. The strong coupling constant was fixed to $\alpha_s(M_Z^2) = 0.1185$ and this is sufficiently similar to the ZEUS choice that we can rule it out as a cause of any significant difference. Full account has been taken of correlated experimental systematic errors by the Hessian Method, see ref. [18].

For the ZEUS analysis, the heavy quark production scheme used is the general mass variable flavour number scheme of Roberts and Thorne [19]. For the H1 analysis, the zero mass variable flavour number scheme is used. It is well known that these choices have a small effect on the steepness of the gluon at very small-$x$, such that the zero-mass choice produces a slightly less steep gluon. However, there is no effect on the more striking differences in the gluon shapes at larger $x$.

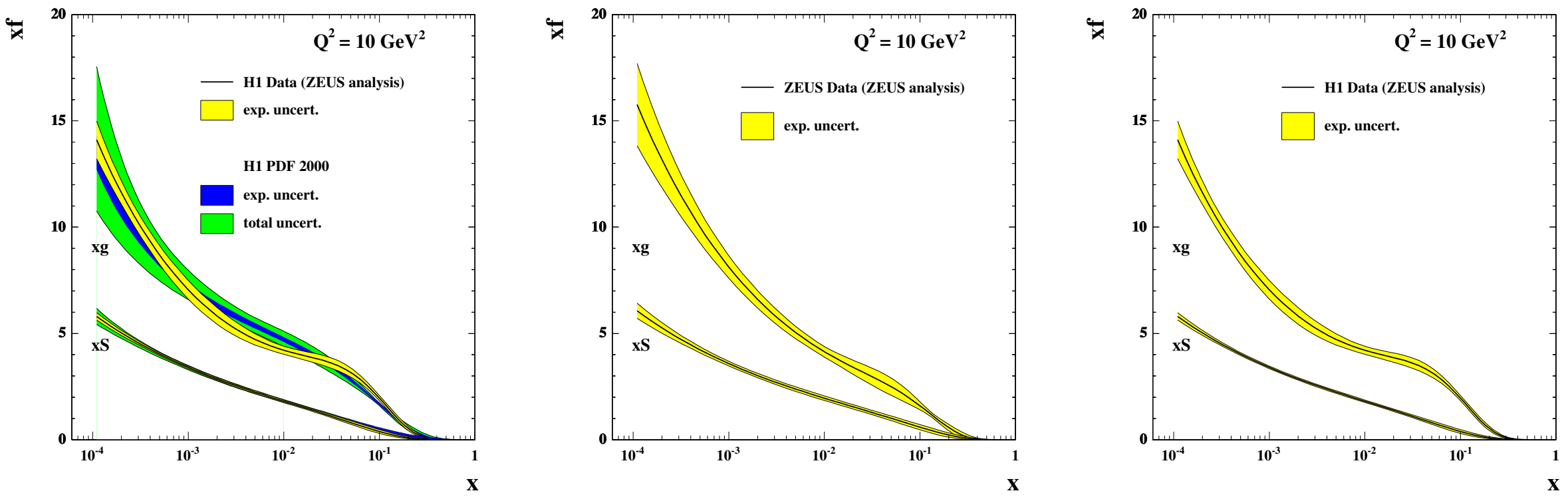

**Fig. 5:** Sea and gluon distributions at $Q^2 = 10\text{GeV}^2$ extracted from different data sets and different analyses. Left plot: H1 data put through both ZEUS and H1 analyses. Middle plot: ZEUS data put through ZEUS analysis. Right plot: H1 data put through ZEUS analysis.

There are two differences in the analyses which are worth further investigation. The different choices for the form of the PDF parametrization at $Q_0^2$ and the different treatment of the correlated experimental uncertainties.

### 3.2 Comparing different PDF analyses of the same data set and comparing different data sets using the same PDF analysis.

So far we have compared the results of putting two different data sets into two different analyses. Because there are many differences in the assumptions going into these analyses it is instructive to consider:(i) putting both data sets through the same analysis and (ii) putting one of the data sets through both analyses. For these comparisons, the ZEUS analysis does NOT include the jet data, so that the data sets are more directly comparable, involving just the inclusive double differential cross-section data. Fig. 5 compares the sea and gluon PDFs, at $Q^2 = 10\text{GeV}^2$, extracted from H1 data using the H1 PDF analysis with those extracted from H1 data using the ZEUS PDF analysis. These alternative analyses of the same data set give results which are compatible within the model dependence error bands. Fig. 5 also compares the sea and gluon PDFs extracted from ZEUS data using the ZEUS analysis with those extracted from H1 data using the ZEUS analysis. From this comparison we can see that the different data sets lead to somewhat different gluon shapes even when put through exactly the same analysis. Hence the most of the difference in shape of the ZEUS and H1 PDF analyses can be traced back to a difference at the level of the data sets.

### 3.3 Comparing the Offset and Hessian method of assessing correlated experimental uncertainties

Before going further it is useful to discuss the treatment of correlated systematic errors in the ZEUS and H1 analyses. A full discussion of the treatment of correlated systematic errors in PDF analyses is given in ref [16], only salient details are recapped here. Traditionally, experimental collaborations have evaluated an overall systematic uncertainty on each data point and these have been treated as uncorrelated, such that they are simply added to the statistical uncertainties in quadrature when evaluating $\chi^2$. However, modern deep inelastic scattering experiments have very small statistical uncertainties, so that the contribution of systematic uncertainties becomes dominant and consideration of point to point correlations between systematic uncertainties is essential.

For both ZEUS and H1 analyses the formulation of the $\chi^2$ including correlated systematic uncertainties is constructed as follows. The correlated uncertainties are included in the theoretical prediction, $F_i(p,s)$, such that

$$F_i(p,s) = F_i^{\rm NLOQCD}(p) + \sum_\lambda s_\lambda \Delta_{i\lambda}^{\rm sys}$$

where, $F_i^{\rm NLOQCD}(p)$, represents the prediction from NLO QCD in terms of the theoretical parameters $p$, and the parameters $s_\lambda$ represent independent variables for each source of systematic uncertainty. They have zero mean and unit variance by construction. The symbol $\Delta_{i\lambda}^{\rm sys}$ represents the one standard deviation correlated systematic error on data point $i$ due to correlated error source $\lambda$. The $\chi^2$ is then formulated as

$$\chi^2 = \sum_i \frac{[F_i(p,s) - F_i({\rm meas})]^2}{\sigma_i^2} + \sum_\lambda s_\lambda^2 \qquad (5)$$

where, $F_i({\rm meas})$, represents a measured data point and the symbol $\sigma_i$ represents the one standard deviation uncorrelated error on data point $i$, from both statistical and systematic sources. The experiments use this $\chi^2$ in different ways. ZEUS uses the Offset method and H1 uses the Hessian method.

Traditionally, experimentalists have used 'Offset' methods to account for correlated systematic errors. The $\chi^2$ is formluated without any terms due to correlated systematic errors ($s_\lambda = 0$ in Eq. 5) for evaluation of the central values of the fit parameters. However, the data points are then offset to account for each source of systematic error in turn (i.e. set $s_\lambda = +1$ and then $s_\lambda = -1$ for each source $\lambda$) and a new fit is performed for each of these variations. The resulting deviations of the theoretical parameters from their central values are added in quadrature. (Positive and negative deviations are added in quadrature separately.) This method does not assume that the systematic uncertainties are Gaussian distributed. An equivalent (and much more efficient) procedure to perform the Offset method has been given by Pascaud and Zomer [20], and this is what is actually used. The Offset method is a conservative method of error estimation as compared to the Hessian method. It gives fitted theoretical predictions which are as close as possible to the central values of the published data. It does not use the full statistical power of the fit to improve the estimates of $s_\lambda$, since it choses to mistrust the systematic error estimates, but it is correspondingly more robust.

The Hessian method is an alternative procedure in which the systematic uncertainty parameters $s_\lambda$ are allowed to vary in the main fit when determining the values of the theoretical parameters. Effectively, the theoretical prediction is not fitted to the central values of the published experimental data, but these data points are allowed to move collectively, according to their correlated systematic uncertainties. The theoretical prediction determines the optimal settings for correlated systematic shifts of experimental data points such that the most consistent fit to all data sets is obtained. Thus, in a global fit, systematic shifts in one experiment are correlated to those in another experiment by the fit. In essence one is allowing the theory to calibrate the detectors. This requires great confidence in the theory, but more significantly, it requires confidence in the many model choices which go into setting the boundary conditions for the theory (such as the parametrization at $Q_0^2$).

The ZEUS analysis can be performed using the Hessian method as well as the Offset method and Fig. 6 compares the PDFs, and their uncertainties, extracted from ZEUS data using these two methods. The central values of the different methods are in good agreement but the use of the Hessian method results in smaller uncertainties, for a the standard set of model assumptions, since the input data can be shifted within their correlated systematic uncertainties to suit the theory better. However, model uncertainties are more significant for the Hessian method than for the Offset method. The experimental uncertainty band for any one set of model choices is set by the usual $\chi^2$ tolerance, $\Delta\chi^2 = 1$, but the acceptability of a different set of choices is judged by the hypothesis testing criterion, such that the $\chi^2$ should be approximately in the range $N \pm \sqrt{(2N)}$, where $N$ is the number of degrees of freedom. The PDF parameters obtained for the different model choices can differ by much more than their experimental uncertainties, because each model choice can result in somewhat different values of the systematic uncertainty parameters, $s_\lambda$, and thus a different estimate of the shifted positions of the data points. This results in a larger spread of model uncertainty than in the Offset method, for which the data points cannot move. Fig 4 illustrates the comparability of the ZEUS (Offset) total uncertainty estimate to the H1 (Hessian) experimental plus model uncertainty estimate.

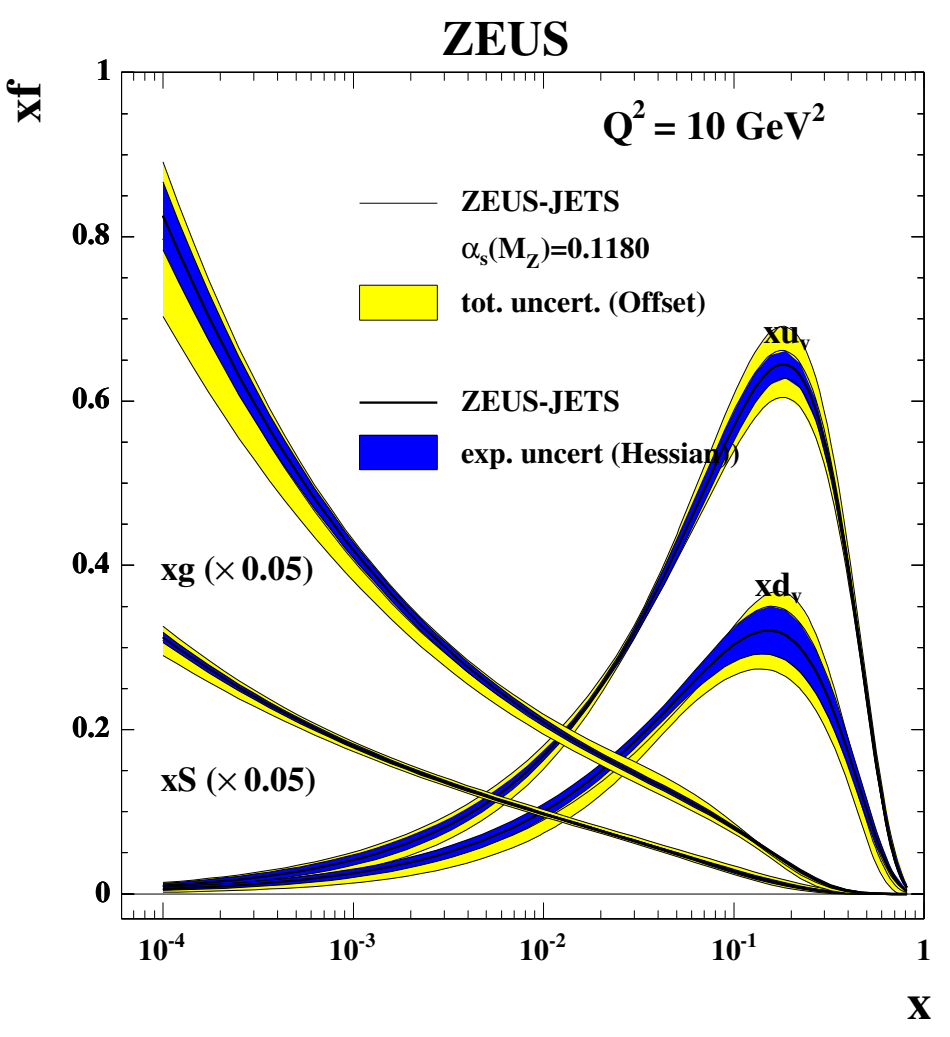

**Fig. 6:** PDFs at $Q^2 = 10\mathrm{GeV}^2$, for the ZEUS analysis of ZEUS data performed by the Offset and the Hessian methods.

Another issue which arises in relation to the Hessian method is that the data points should not be shifted far outside their one standard deviation systematic uncertainties. This can indicate inconsistencies between data sets, or parts of data sets, with respect to the rest of the data. The CTEQ collaboration have considered data inconsistencies in their most recent global fit [4]. They use the Hessian method but they increase the resulting uncertainty estimates, by increasing the $\chi^2$ tolerance to $\Delta\chi^2 = 100$, to allow for both model uncertainties and data inconsistencies. In setting this tolerance they have considered the distances from the $\chi^2$-minima of individual data sets to the global minimum for all data sets. These distances by far exceed the range allowed by the $\Delta\chi^2 = 1$ criterion. Strictly speaking such variations can indicate that data sets are inconsistent but the CTEQ collaboration take the view that all of the current world data sets must be considered acceptable and compatible at some level, even if strict statistical criteria are not met, since the conditions for the application of strict criteria, namely Gaussian error distributions, are also not met. It is not possible to simply drop "inconsistent" data sets, as then the partons in some regions would lose important constraints. On the other hand the level of "inconsistency" should be reflected in the uncertainties of the PDFs. This is achieved by raising the $\chi^2$ tolerance. This results in uncertainty estimates which are comparable to those achieved by using the Offset method [18].

### 3.4 Using both H1 and ZEUS data in the same PDF analysis

Using data from a single experiment avoids questions of data consistency, but to get the most information from HERA it is necessary to put ZEUS and H1 data sets into the same analysis together, and then questions of consistency arise. Fig 7 compares the sea and gluon PDFs and the $u$ and $d$ valence PDFs extracted from the ZEUS PDF analysis of ZEUS data alone, to those extracted from the ZEUS PDF analysis of both H1 and ZEUS data. It is noticeable that, for the low-$x$ sea and gluon PDFs, combining the data sets does not bring a reduction in uncertainty equivalent to doubling the statistics. This is because the data which determine these PDFs are systematics limited. In fact there is some degree of tension between the ZEUS and the H1 data sets, such that the $\chi^2$ per degree of freedom rises for both data sets when they are fitted together. The Offset method of treating the systematic errors reflects this tension such that the overall uncertainty is not much improved when H1 data are added to ZEUS data. However, the uncertainty on the high-$x$ valence distributions is reduced by the input of H1 data, since the data are still statistics limited at high $x$.

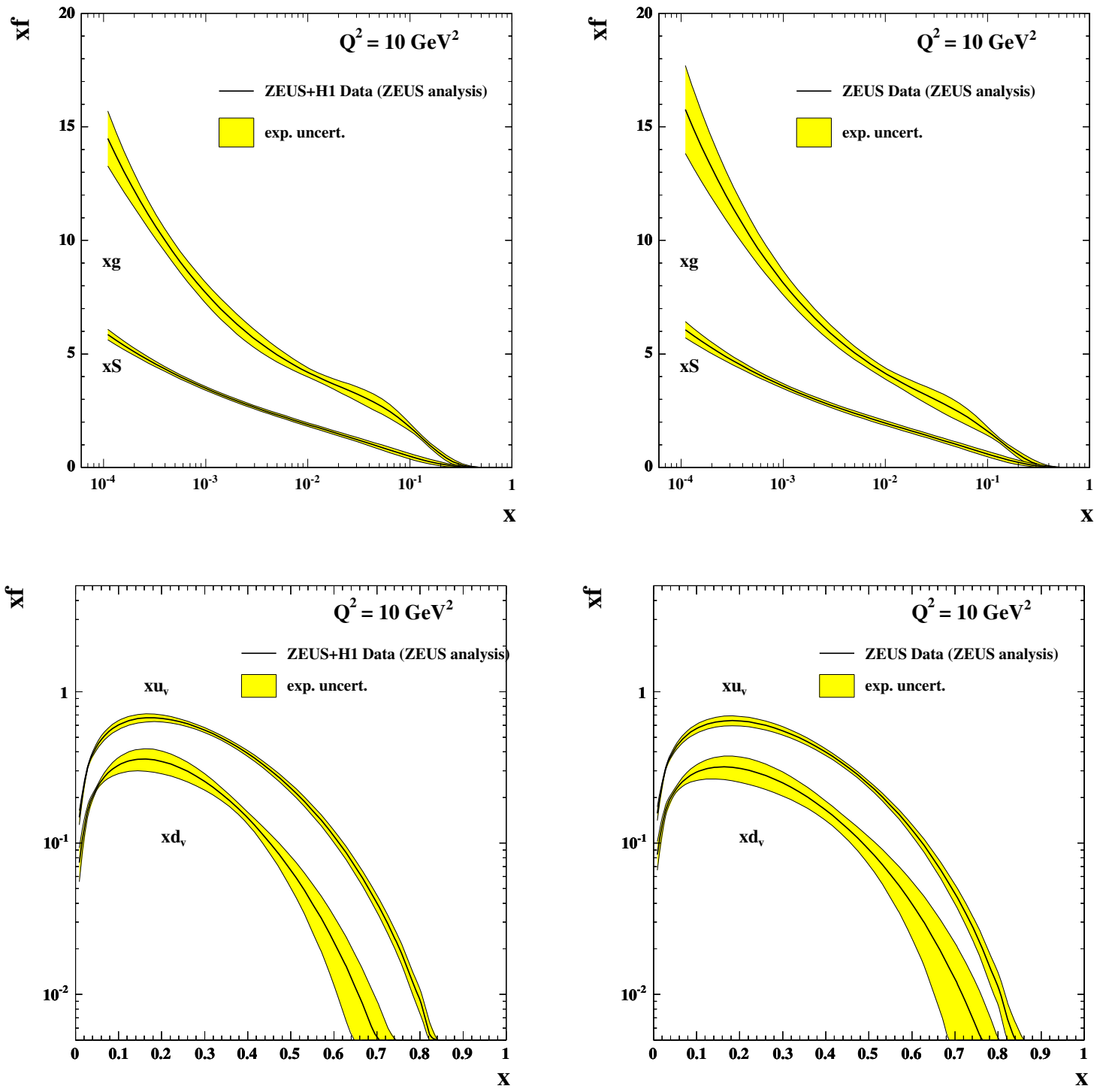

**Fig. 7:** Top plots: Sea and gluon distributions at $Q^2 = 10\text{GeV}^2$ extracted from H1 and ZEUS data using the ZEUS analysis (left) compared to those extracted from ZEUS data alone using the ZEUS analysis (right). Bottom Plots: Valence distributions at $Q^2 = 10\text{GeV}^2$, extracted from H1 and ZEUS data using the ZEUS analysis (left) compared to those extracted from ZEUS data alone using the ZEUS analysis (right).

### 3.5 Combining the H1 and ZEUS data sets before PDF analysis

Thus there could be an advantage in combining ZEUS and H1 data in a PDF fit if the tension between the data sets could be resolved. It is in this context the question of combining these data into a single data set arises. The procedure for combination is detailed in the contribution of S. Glazov to these proceedings (section 4). Essentially, since ZEUS and H1 are measuring the same physics in the same kinematic region, one can try to combine them using a 'theory-free' Hessian fit in which the only assumption is that there is a true value of the cross-section, for each process, at each $x, Q^2$ point. The systematic uncertainty parameters, $s_\lambda$, of each experiment are fitted to determine the best fit to this assumption. Thus each experiment is calibrated to the other. This works well because the sources of systematic uncertainty in each experiment are rather different. Once the procedure has been performed the resulting systematic uncertainties on each of the combined data points are significantly smaller than the statistical errors. Thus one can legitimately make a fit to the combined data set in which these statistical and systematic uncertainties are simply combined in quadrature. The result of making such a fit, using the ZEUS analysis, is shown in Fig. 8. The central values of the ZEUS and H1 published analyses are also shown for comparison. Looking back to Fig. 7 one can see that there has been a dramatic reduction in the level of uncertainty compared to the ZEUS Offset method fit to the separate ZEUS and H1 data sets. This result is very promising. A preliminary study of model dependence, varying the form of the polynomial, $P(x)$, used in the PDF paremtrizations at $Q_0^2$, also indicates that model dependence is relatively small.

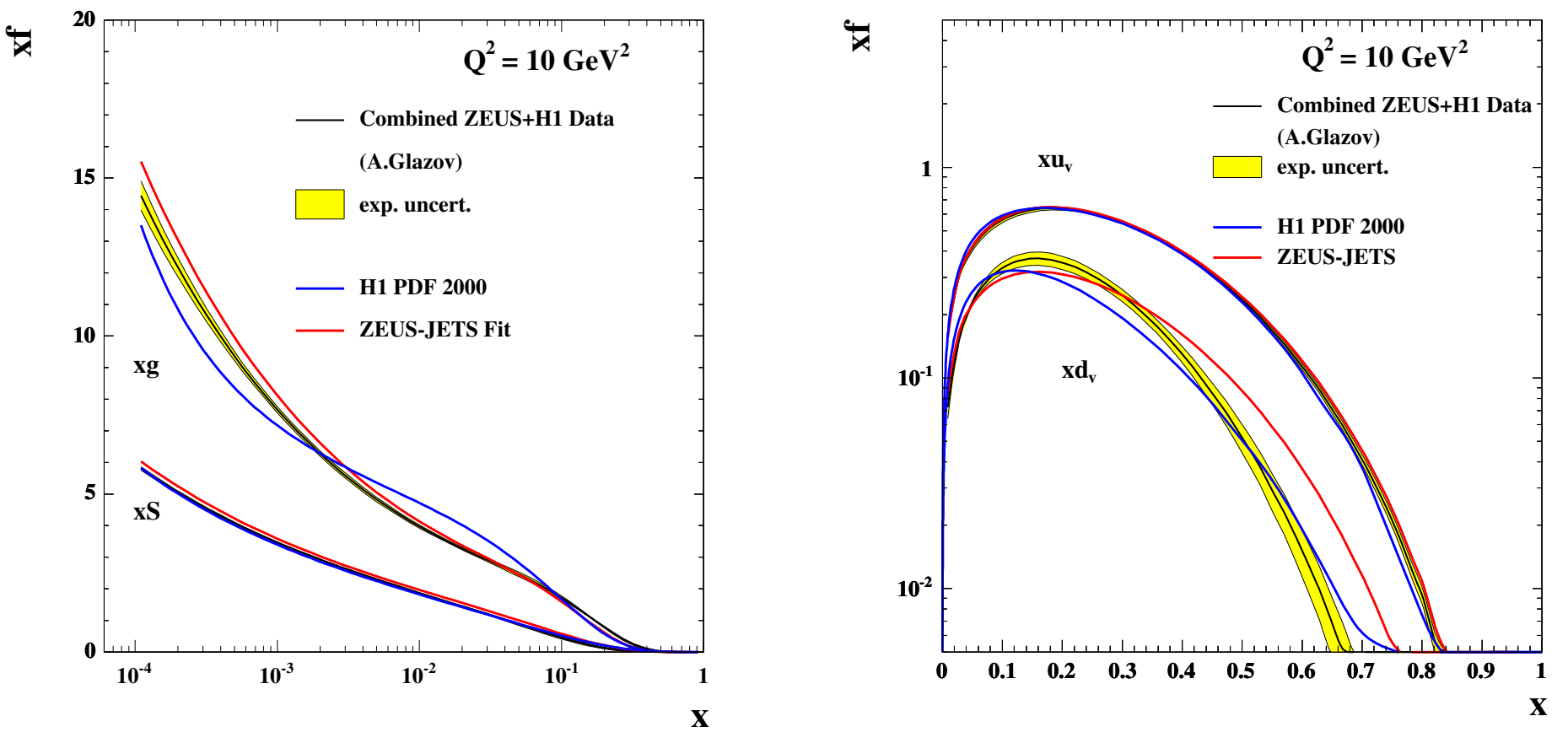

**Fig. 8:** Left plot: Sea and gluon distributions at $Q^2 = 10\text{GeV}^2$, extracted from the combined H1 and ZEUS data set using the ZEUS analysis. Right plot: Valence distributions at $Q^2 = 10\text{GeV}^2$, extracted from the combined H1 and ZEUS data set using the ZEUS analysis.

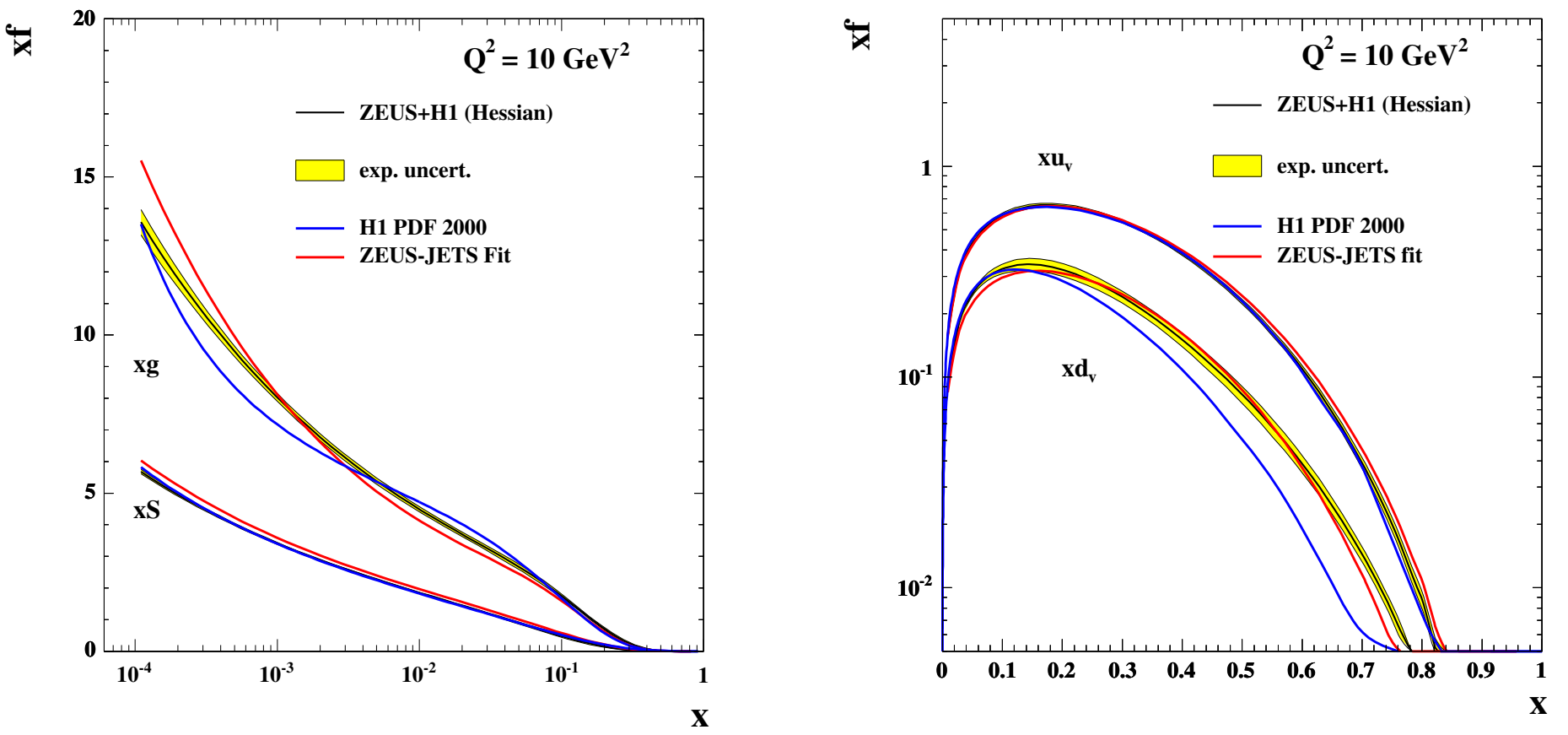

**Fig. 9:** Left plot: Sea and gluon distributions at $Q^2 = 10\text{GeV}^2$, extracted from the H1 and ZEUS data sets using the ZEUS analysis done by Hessian method. Right plot: Valence distributions at $Q^2 = 10\text{GeV}^2$, extracted from the H1 and ZEUS data sets using the ZEUS analysis done by Hessian method.

The tension between ZEUS and H1 data could have been resolved by putting them both into a PDF fit using the Hessian method to shift the data points. That is, rather than calibrating the two experiments to each other in the 'theory-free' fit, we could have used the theory of pQCD to calibrate each experiment. Fig. 9 shows the PDFs extracted when the ZEUS and H1 data sets are put through the ZEUS PDF analysis procedure using the Hessian method. The uncertainties on the resulting PDFs are comparable to those found for the fit to the combined data set, see Fig. 8. However, the central values of the resulting PDFs are rather different- particularly for the less well known gluon and $d$ valence PDFs. For both of the fits shown in Figs. 8 and 9 the values of the systematic error parameters, $s_\lambda$, for each experiment have been allowed to float so that the data points are shifted to give a better fit to our assumptions, but the values of the systematic error parameters chosen by the 'theory-free' fit and by the PDF fit are rather different. A representaive sample of these values is given in Table 3. These discrepancies might be somewhat alleviated by a full consideration of model errors in the PDF fit, or of appropriate $\chi^2$ tolerance when combining the ZEUS and H1 experiments in a PDF fit, but these differences should make us wary about the uncritical use of the Hessian method.

**Table 3:** Systematic shifts for ZEUS and H1 data as determine by a joint pQCD PDF fit, and as determined by the theory-free data combination fit

| Syatematic uncertainty $s_\lambda$ | in PDF fit | in Theory-free fit |
|---|---|---|
| ZEUS electron efficiency | 1.68 | 0.31 |
| ZEUS electron angle | -1.26 | -0.11 |
| ZEUS electron energy scale | -1.04 | 0.97 |
| ZEUS hadron calorimeter energy scale | 1.05 | -0.58 |
| H1 electron energy scale | -0.51 | 0.61 |
| H1 hadron energy scale | -0.26 | -0.98 |
| H1 calorimeter noise | 1.00 | -0.63 |
| H1 photoproduction background | -0.36 | 0.97 |

## 4 Averaging of DIS Cross Section Data [5]

The QCD fit procedures (Alekhin [6], CTEQ [4], MRST [5], H1 [11], ZEUS [9]) use data from a number of individual experiments directly to extract the parton distribution functions (PDF). All programs use both the central values of measured cross section data as well as information about the correlations among the experimental data points.

The direct extraction procedure has several shortcomings. The number of input datasets is large containing several individual publications. The data points are correlated because of common systematic uncertainties, within and also across the publications. Handling of the experimental data without additional expert knowledge becomes difficult. Additionally, as it is discussed in Sec. 3, the treatment of the correlations produced by the systematic errors is not unique. In the Lagrange Multiplier method [20] each systematic error is treated as a parameter and thus fitted to QCD. Error propogation is then used to estimate resulting uncertainties on PDFs. In the so-called "offset" method (see e.g. [9]) the datasets are shifted in turn by each systematic error before fitting. The resulting fits are used to form an envelope function to estimate the PDF uncertainty. Each method has its own advantages and shortcomings, and it is difficult to select the standard one. Finally, some global QCD analyses use non-statistical criteria to estimate the PDF uncertainties ($\Delta\chi^2 \gg 1$). This is driven by the apparent discrepancy between different experiments which is often difficult to quantify. Without a model independent consistency check of the data it might be the only safe procedure.

These drawbacks can be significantly reduced by averaging of the input structure function data in a model independent way before performing a QCD analysis of that data. One combined dataset of deep inelastic scattering (DIS) cross section measurements is much easier to handle compared to a scattered set of individual experimental measurements, while retaining the full correlations between data points. The averaging method proposed here is unique and removes the drawback of the offset method, which fixes the size of the systematic uncertainties. In the averaging procedure the correlated systematic uncertainties are floated coherently allowing in some cases reduction of the uncertainty. In addition, study of a global $\chi^2/dof$ of the average and distribution of the pulls allows a model independent consistency check between the experiments. In case of discrepancy between the input datasets, localised enlargement of the uncertainties for the average can be performed.

A standard way to represent a cross section measurement of a single experiment is given in the case of the $F_2$ structure function by:

$$\chi^2_{exp}(\left\{F_2^{i,true}\right\},\{\alpha_j\}) \;=\; \sum_i \frac{\left[F_2^{i,true} - \left(F_2^i + \sum_j \frac{\partial F_2^i}{\partial \alpha_j}\alpha_j\right)\right]^2}{\sigma_i^2} + \sum_j \frac{\alpha_j^2}{\sigma_{\alpha_j}^2}. \qquad (6)$$

Here $F_2^i$ ($\sigma_i^2$) are the measured central values (statistical and uncorrelated systematic uncertainties) of the

[5] Contributing author: A. Glazov

$F_2$ structure function[6], $\alpha_j$ are the correlated systematic uncertainty sources and $\partial F_2^i/\partial\alpha_j$ are the sensitivities of the measurements to these systematic sources. Eq. 6 corresponds to the correlated probability distribution functions for the structure function $F_2^{i,true}$ and for the systematic uncertainties $\alpha_j$. Eq. 6 resembles Eq. 5 where the theoretical predictions for $F_2$ are substituted by $F_2^{i,true}$.

The $\chi^2$ function Eq. 6 by construction has a minimum $\chi^2 = 0$ for $F_2^{i,true} = F_2^i$ and $\alpha_j = 0$. One can show that the total uncertainty for $F_2^{i,true}$ determined from the formal minimisation of Eq. 6 is equal to the sum in quadrature of the statistical and systematic uncertainties. The reduced covariance matrix $cov(F_2^{i,true}, F_2^{j,true})$ quantifies the correlation between experimental points.

In the analysis of data from more than one experiment, the $\chi^2_{tot}$ function is taken as a sum of the $\chi^2$ functions Eq. 6 for each experiment. The QCD fit is then performed in terms of parton density functions which are used to calculate predictions for $F_2^{i,true}$.

Before performing the QCD fit, the $\chi^2_{tot}$ function can be minimised with respect to $F_2^{i,true}$ and $\alpha_j$. If none of correlated sources is present, this minimisation is equivalent to taking an average of the structure function measurements. If the systematic sources are included, the minimisation corresponds to a generalisation of the averaging procedure which contains correlations among the measurements.

Being a sum of positive definite quadratic functions, $\chi^2_{tot}$ is also a positive definite quadratic and thus has a unique minimum which can be found as a solution of a system of linear equations. Although this system of the equations has a large dimension it has a simple structure allowing fast and precise solution.

A dedicated program has been developed to perform this averaging of the DIS cross section data (`http://www.desy.de/~glazov/f2av.tar.gz`). This program can calculate the simultaneous averages for neutral current (NC) and charged current (CC) electron- and positron-proton scattering cross section data including correlated systematic sources. The output of the program includes the central values and uncorrelated uncertainties of the average cross section data. The correlated systematic uncertainties can be represented in terms of (i) covariance matrix, (ii) dependence of the average cross section on the original systematic sources together with the correlation matrix for the systematic sources, (iii) and finally the correlation matrix of the systematic sources can be diagonalised, in this case the form of $\chi^2$ for the average data is identical to Eq. 6 but the original systematic sources are not preserved.

The first application of the averaging program has been a determination of the average of the published H1 and ZEUS data [3, 11, 21–28]. Nine individual NC and CC cross section measurements are included from H1 and seven are included from ZEUS. Several sources of systematic uncertainties are correlated between datasets, the correlations among H1 and ZEUS datasets are taken from [11] and [10], respectively. No correlations are assumed between H1 and ZEUS systematic uncertainties apart from a common $0.5\%$ luminosity measurement uncertainty. The total number of data points is 1153 (552 unique points) and the number of correlated systematic sources, including normalisation uncertainties, is 43.

The averaging can take place only if most of the data from the experiments are quoted at the same $Q^2$ and $x$ values. Therefore, before the averaging the data points are interpolated to a common $Q^2, x$ grid. This interpolation is based on the H1 PDF 2000 QCD fit [11]. The interpolation of data points in principle introduces a model dependency. For H1 and ZEUS structure function data both experiments employ rather similar $Q^2, x$ grids. About $20\%$ of the input points are interpolated, for most of the cases the correction factors are small (few percent) and stable if different QCD fit parametrizations [4, 5] are used.

The cross section data have also been corrected to a fixed center of mass energy squared $S = 101570$ GeV$^2$. This has introduced a small correction for the data taken at $S = 90530$ GeV$^2$. The correction is based on H1-2000 PDFs, it is only significant for high inelasticity $y > 0.6$ and does not exceed $6\%$.

[6]The structure function is measured for different $Q^2$ (four momentum transfer squared) and Bjorken-$x$ values which are omitted here for simplicity.

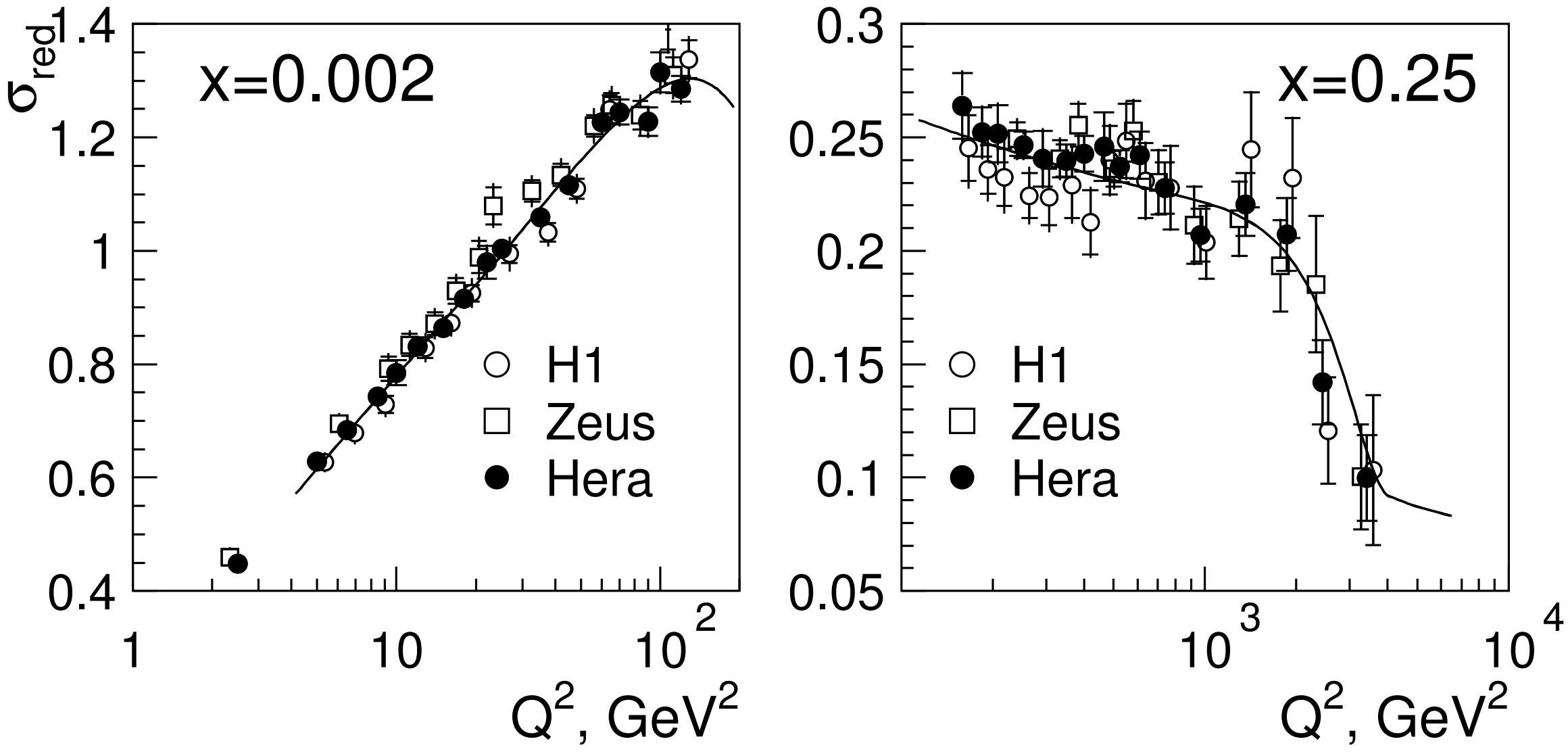

**Fig. 10:** $Q^2$ dependence of the NC reduced cross section for $x = 0.002$ and $x = 0.25$ bins. H1 data is shown as open circles, ZEUS data is shown as open squares and the average of H1 and ZEUS data is shown as filled circles. The line represents the expectation from the H1 PDF 2000 QCD fit.

The HERA data sets agree very well: $\chi^2/dof$ for the average is $521/601$. The distribution of pulls does not show any significant tensions across the kinematic plane. Some systematic trends can be observed at low $Q^2 < 50$ GeV$^2$, where ZEUS NC data lie systematically higher than the H1 data, although this difference is within the normalisation uncertainty. An example of the resulting average DIS cross section is shown in Fig. 10, where the data points are displaced in $Q^2$ for clarity.

A remarkable side feature of the averaging is a significant reduction of the correlated systematic uncertainties. For example the uncertainty on the scattered electron energy measurement in the H1 backward calorimeter is reduced by a factor of three. The reduction of the correlated systematic uncertainties thus leads to a significant reduction of the total errors, especially for low $Q^2 < 100$ GeV$^2$, where systematic uncertainties limit the measurement accuracy. For this domain the total errors are often reduced by a factor two compared to the total errors of the individual H1 and ZEUS measurements.

The reduction of the correlated systematic uncertainties is achieved since the dependence of the measured cross section on the systematic sources is significantly different between H1 and ZEUS experiments. This difference is due mostly to the difference in the kinematic reconstruction methods used by the two collaborations, and to a lesser extent to the individual features of the H1 and ZEUS detectors. For example, the cross section dependence on the scattered electron energy scale has a very particular behaviour for H1 data which relies on kinematic reconstruction using only the scattered electron in one region of phase space. ZEUS uses the double angle reconstruction method where the pattern of this dependence is completely different leading to a measurement constraint.

In summary, a generalised averaging procedure to include point-to-point correlations caused by the systematic uncertainties has been developed. This averaging procedure has been applied to H1 and ZEUS DIS cross section data. The data show good consistency. The averaging of H1 and ZEUS data leads to a significant reduction of the correlated systematic uncertainties and thus a large improvement in precision for low $Q^2$ measurements. The goal of the averaging procedure is to obtain HERA DIS cross section set which takes into account all correlations among the experiments.

# 5 The longitudinal structure function $F_L$ [7]

## 5.1 Introduction

At low $x$ the sea quarks are determined by the accurate data on $F_2(x, Q^2)$ . The charm contribution to $F_2$ is directly measured while there is no separation of up and down quarks at low $x$ which are assumed to have the same momentum distribution, see [2]. Within this assumption, and setting the strange sea to be a fraction of the up/down sea, the proton quark content at low $x$ is determined. The gluon distribution $xg(x, Q^2)$ , however, is determined only by the derivative $\partial F_2/\partial \ln Q^2$ which is not well measured [3]. It is thus not surprising that rather different gluon distributions are obtained in global NLO analyses, as is illustrated in Figure 11. The figure displays the result of recent fits by MRST and CTEQ on the gluon distribution at low and high $Q^2$. It can be seen that there are striking differences at the initial scale, $Q^2 = 5\,\mathrm{GeV}^2$, which at high $Q^2$ get much reduced due to the evolution mechanism. The ratio of these distributions, however, exhibits differences at lower $x$ at the level of 10% even in the LHC Higgs and $W$ production kinematic range, see Figure 12. One also observes a striking problem at large $x$ which is beyond the scope of this note, however. In a recent QCD analysis it was observed [3] that the dependence of the gluon distribution at low $x$, $xg \propto x^{b_G}$, is correlated to the value of $\alpha_s(M_Z^2)$ , see Figure 13.

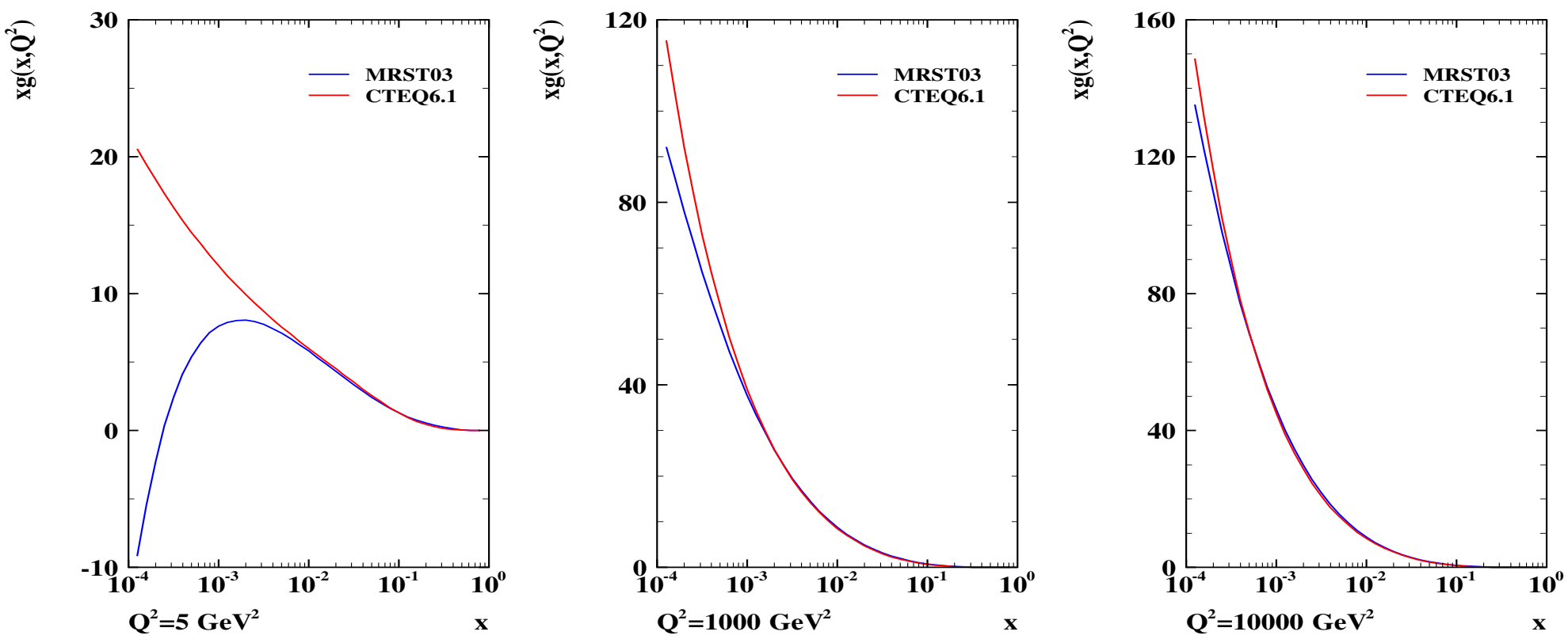

**Fig. 11:** Gluon momentum distributions determined by MRST and CTEQ in NLO QCD, as a function of $x$ for $Q^2 = 5\,\mathrm{GeV}^2$, close to the initial scale of the fits, and at higher $Q^2$ as the result of the DGLAP evolution.

In the Quark-Parton Model the longitudinal structure function $F_L(x, Q^2)$ is zero [29]. In DGLAP QCD, to lowest order, $F_L$ is given by [30]

$$F_L(x, Q^2) = \frac{\alpha_s}{4\pi} x^2 \int_x^1 \frac{dz}{z^3} \cdot \left[ \frac{16}{3} F_2(z, Q^2) + 8 \sum e_q^2 \left(1 - \frac{x}{z}\right) zg(z, Q^2) \right] \quad (7)$$

with contributions from quarks and from gluons. Approximately this equation can be solved [31] and the gluon distribution appears as a measurable quantity,

$$xg(x) = 1.8[\frac{3\pi}{2\alpha_s} F_L(0.4x) - F_2(0.8x)] \simeq \frac{8.3}{\alpha_s} F_L, \quad (8)$$

determined by measurements of $F_2$ and $F_L$ . Since $F_L$ , at low $x$, is not much smaller than $F_2$ , to a good approximation $F_L$ is a direct measure for the gluon distribution at low $x$.

Apart from providing a very useful constraint to the determination of the gluon distribution, see also Sect. 7, a measurement of $F_L(x, Q^2)$ is of principal theoretical interest. It provides a crucial test of QCD to high orders. A significant departure of an $F_L$ measurement from the prediction which is based on the measurement of $F_2(x, Q^2)$ and $\partial F_2/\partial \ln Q^2$ only, would require theory to be modified. There are known reasons as to why the theoretical description of gluon radiation at low $x$ may differ from conventional DGLAP evolution: the neglect of $\ln(1/x)$, in contrast to BFKL evolution, or the importance of NLL resummation effects on the gluon splitting function (see [32]). Furthermore recent calculations of deep inelastic scattering to NNLO predict very large effects from the highest order on $F_L$ contrary to $F_2$ [33].

[7] Contributing authors: J. Feltesse, M. Klein

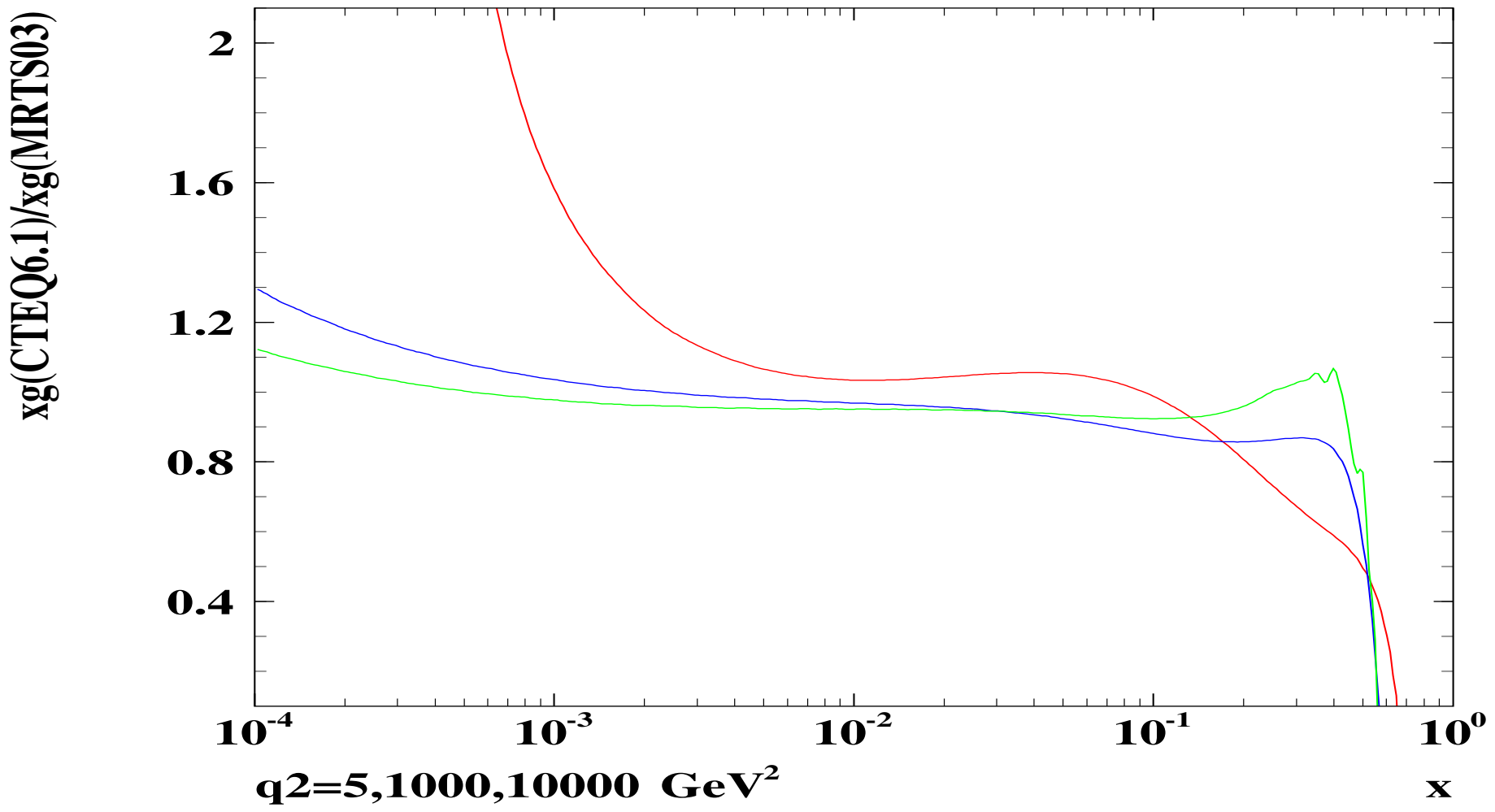

**Fig. 12:** Ratio of the gluon distributions of CTEQ to MRST as a function of $x$ for low and large $Q^2$.

Within the framework of the colour dipole model there exists a testable prediction for $F_L(x, Q^2)$, and the longitudinal structure function, unlike $F_2$, may be subject to large higher twist effects [34].

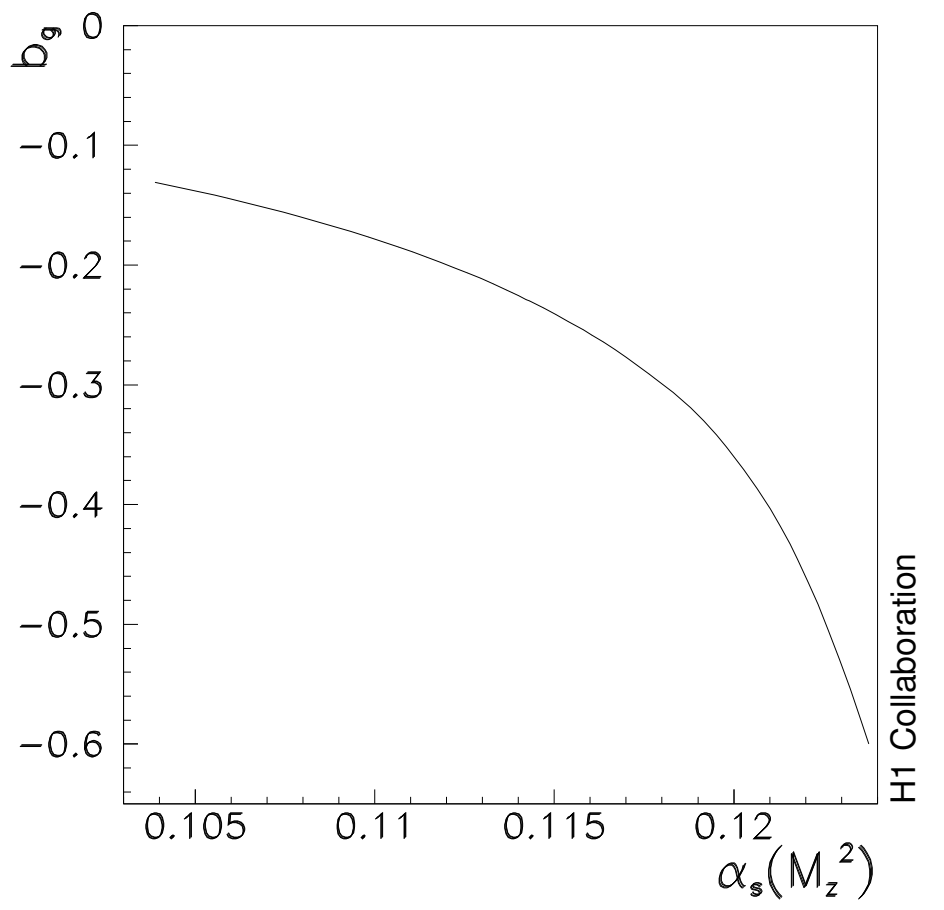

**Fig. 13:** Correlation of the low $x$ behaviour of the gluon distribution, characterised by the power $x^{-b_g}$, with the strong coupling constant $\alpha_s$ as obtained in the H1 NLO QCD fit to H1 and BCDMS data.

## 5.2 Indirect Determinations of $F_L$ at Low $x$

So far first estimates on $F_L(x, Q^2)$ at low $x$ have been obtained by the H1 Collaboration. These result from data on the inclusive $ep \to eX$ scattering cross section

$$\frac{Q^4 x}{2\pi\alpha^2 Y_+} \cdot \frac{d^2\sigma}{dxdQ^2} = [F_2(x, Q^2) - f(y) \cdot F_L(x, Q^2)] = \sigma_r \tag{9}$$

obtained at fixed, large energy, $s = 4E_eE_p$. The cross section is defined by the two proton structure functions, $F_2$ and $F_L$, with $Y_+ = 1 + (1-y)^2$ and $f(y) = y^2/Y_+$. At fixed $s$ the inelasticity $y$ is fixed by $x$ and $Q^2$ as $y = Q^2/sx$. Thus one can only measure a combination $F_2 - f(y)F_L$. Since HERA accesses a large range of $y$, and $f(y)$ is large only at large $y > 0.4$, assumptions have been made on $F_L$ to extract $F_2$ at larger $y$. Since the cross section measurement accuracy has reached the few per cent level [3], the effect of the $F_L$ assumption on $F_2$ at lowest $x$ has been non-negligible. The determination of $F_2(x, Q^2)$ has thus been restricted to a region in which $y < 0.6$. The proton structure function $F_2(x, Q^2)$ is known over a few orders of magnitude in $x$ rather well, from HERA and at largest $x$ from fixed target data. Thus H1 did interpret the cross section at higher $y$ as a determination of $F_L(x, Q^2)$ imposing assumptions about the behaviour of $F_2(x, Q^2)$ at lowest $x$. These were derived from QCD fits to the H1 data [35] or at lower $Q^2$, where QCD could not be trusted, from the derivative of $F_2$ [36]. Recently, with the established $x$ behaviour [37] of $F_2(x, Q^2) = c(Q^2)x^{-\lambda(Q^2)}$, a new method [36] has been used to determine $F_L$ . This "shape method" is based on the observation that the shape of $\sigma_r$, Eq. 9, at high $y$ is driven by $f \propto y^2$ and sensitivity to $F_L$ is restricted to a very narrow range of $x$ corresponding to $y = 0.3 - 0.9$. Assuming that $F_L(x, Q^2)$ in this range, for each bin in $Q^2$, does not depend on $x$, one obtains a simple relation, $\sigma_r = cx^{-\lambda} - fF_L$. which has been used to determine $F_L(x, Q^2)$ . Figure 14 shows the existing, preliminary data on $F_L(x, Q^2)$ at low $Q^2$ from the H1 Collaboration in comparison with predictions from NLO DGLAP QCD fits to HERA and further cross section data. One can see that the accuracy and the $x$ range of these $F_L(x, Q^2)$ determinations are rather limited although the data have some discriminative power already.

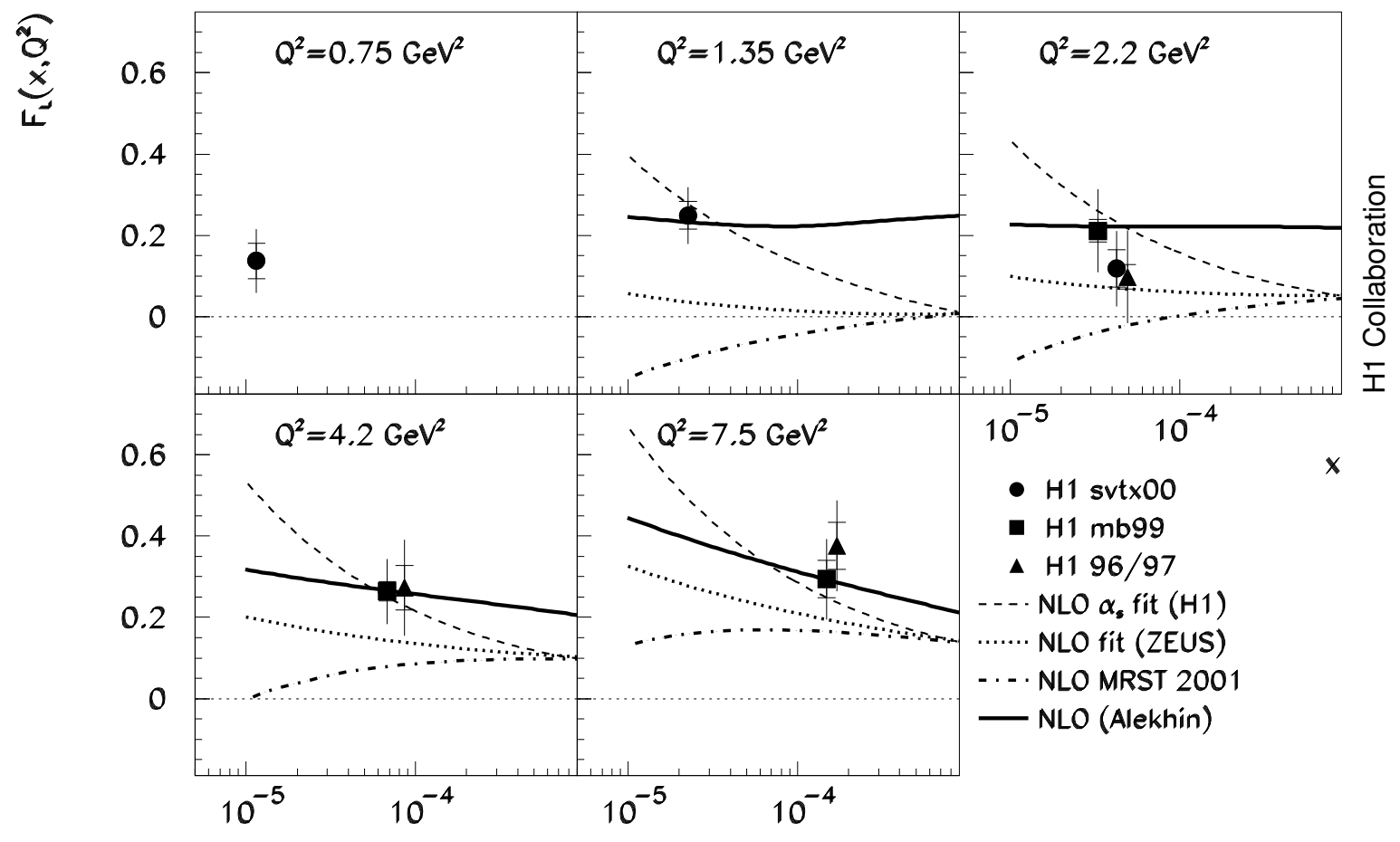

**Fig. 14:** Data on the longitudinal structure function obtained using assumptions on the behaviour of the other structure function $F_2$ in comparison with NLO QCD fit predictions. The data labeled svtx00 and mb99 data are preliminary.

### 5.3 Backgrounds and Accuracy

The longitudinal structure function contribution to $\sigma_r$ represents a small correction of the cross section in a small part of the kinematic range only. The demands for the $F_L$ measurement are extremely high: the cross section needs to be measured at the per cent level and the scattered electron be uniquely identified up to high $y$. The method of unfolding $F_2$ and $F_L$ consists in a measurement of $\sigma_r$ at fixed $x$ and $Q^2$ with varying $s$. This allows both structure functions to be determined from a straight line variation of $\sigma_r$ as a function of $f(y)$, see [38].

At large $y$, corrresponding to low $x$, and low $Q^2$ the scattering kinematics at HERA resembles that of a fixed target scattering experiment: the electron scattered off quarks at very low $x$ ("at rest") is going in the backward detector region, i.e. in the direction of the electron beam. The scattered electron is accompanied by part of the hadronic final state which is related to the struck quark. High inelasticities $y \simeq 1 - E'_e/E_e$ demand to identify scattered electrons down to a few GeV of energy $E'_e$. Thus a considerable background is to be isolated and removed which stems from hadrons or photons, from the $\pi_0 \to \gamma\gamma$ decay. These particles may originate both from a genuine DIS event but to a larger extent stem from photoproduction processes, in which the scattered electron escapes mostly non recognised in electron beam direction. Removal of this background in H1 is possible by requiring a track associated to the Spacal cluster, which rejects photons, and by measuring its charge which on a statistical basis removes the remaining part of the background as was demonstrated before [3, 36].

The scattered electron kinematics, $E'_e$ and $\theta_e$, can be accurately reconstructed using the high resolution Spacal calorimeter energy determination and the track measurements in the Backward Silicon Tracker (BST) and the Central Jet Drift Chamber (CJC). Reconstruction of the hadronic final state allows the energy momentum constraint to be imposed, using the "$E - p_z$" cut, which removes radiative corrections, and the Spacal energy scale to be calibrated at large $E'_e$ using the double angle method. At low energies $E'_e$ the Spacal energy scale can be calibrated to a few % using the $\pi_0$ mass constraint and be cross checked with the BST momentum measurement and with QED Compton events. The luminosity is measured to 1-2%. Any common normalisation uncertainty may be removed, or further constrained, by comparing cross section data at very low $y$ where the contribution of $F_L$ is negligible.

Subsequently two case studies are presented which illustrate the potential of measuring $F_L$ directly in unfolding it from the large $F_2$ contribution to the cross section, a study using a set of 3 low proton beam energies and a simulation for just one low $Ep$ data set combined with standard 920 GeV data. Both studies use essentially the same correlated systematic errors and differ slightly in the assumptions on the background and efficiency uncertainties which regard the errors on cross section ratios. The following assumptions on the correlated systematics are used: $\delta E'_e/E'_e = 0.003$ at large $E_e$ linearly rising to 0.03 at 3 GeV; $\delta\theta_e = 0.2$ mrad in the BST acceptance region and 1 mrad at larger angles; $\delta E_h/E_h = 0.02$. These and further assumed systematic uncertainties represent about the state of analysis reached so far in inclusive low $Q^2$ cross section measurements of H1.

### 5.4 Simulation Results

A simulation has been performed for $E_e = 27.6$ GeV and for four different proton beam energies, $E_p = 920,\ 575,\ 465$ and $400$ GeV assuming luminosities of 10, 5, 3 and 2 pb$^{-1}$ , respectively. The beam energies are chosen such that the cross section data are equidistant in $f(y)$. If the luminosity scales as expected as $E_p^2$, the low $E_p$ luminosities are equivalent to 35 pb$^{-1}$ at standard HERA settings. Further systematic errors regard the residual radiative corrections, assumed to be 0.5%, and the photoproduction background, 1-2% depending on $y$. This assumption on the background demands an improvement by a factor of about two at high $y$ which can be expected from a high statistics subtraction of background using the charge assignment of the electron scattering candidate. An extra uncorrelated efficiency correction is assumed of 0.5%. The resulting cross section measurements are accurate to 1-2%. For each $Q^2$ and $x$ point this choice provides up to four cross section measurements. The two structure functions are then obtained from a fit to $\sigma_r = F_2 + f(y)F_L$ taking into account the correlated systematics. This separation provides also accurate data of $F_2$, independently of $F_L$ . The simulated data on $F_L$ span nearly one order of magnitude in $x$ and are shown in Figure 15. For the chosen luminosity the statistical and systematic errors on $F_L$ are of similar size. The overall accuracy on $F_L(x, Q^2)$ , which may be obtained according to the assumed experimental uncertainties, is thus estimated to be of the order of 10-20%.

Based on recent information about aspects of the machine conditions in a low proton beam energy mode, a further case study was performed [39] for only one reduced proton beam energy. In this simulation, for the standard electron beam energy of $E_e = 27.6$ GeV, proton beam energies of $E_p = 920$ and 460 GeV were chosen with luminosities of 30 and 10 $pb^{-1}$, respectively. According to [40] it would take about three weeks to change the configuration of the machine and to tune the luminosity plus 10 weeks to record 10 $pb^{-1}$ of good data with High Voltage of trackers on. Uncertainties besides the correlated errors specified above are assumed for photo-production background subtraction varying from 0% at y=0.65 to 4% at y = 0.9, and of 0.5% for the residual radiative corrections. An overall uncertainty of 1% is assumed on the measurement of the cross section at low beam energy settings, which covers relative uncertainties on electron identification, trigger efficiency, vertex efficiency, and relative luminosity.

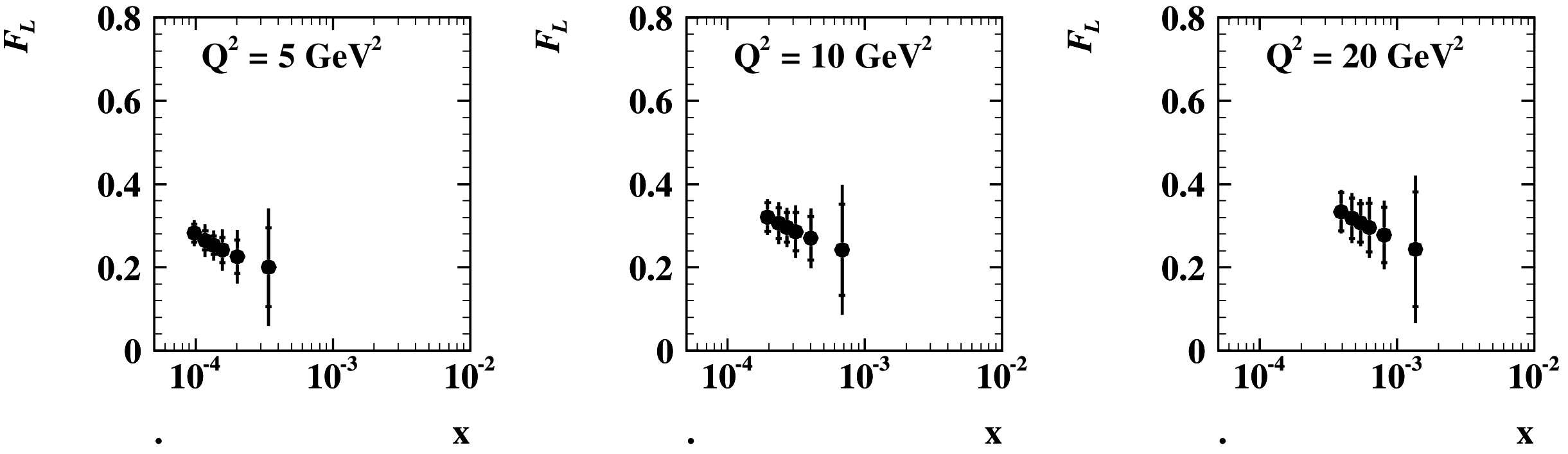

**Fig. 15:** Simulated measurement of the longitudinal structure function $F_L(x, Q^2)$ using the H1 backward apparatus to reconstruct the scattered electron up to maximum inelasticities of $y = 0.9$ corresponding to a mimimum electron energy of $E_e'$ of about 3 GeV. The inner error bar is the statistical error. The full error bar denotes the statistical and systematic uncertainty added in quadrature.

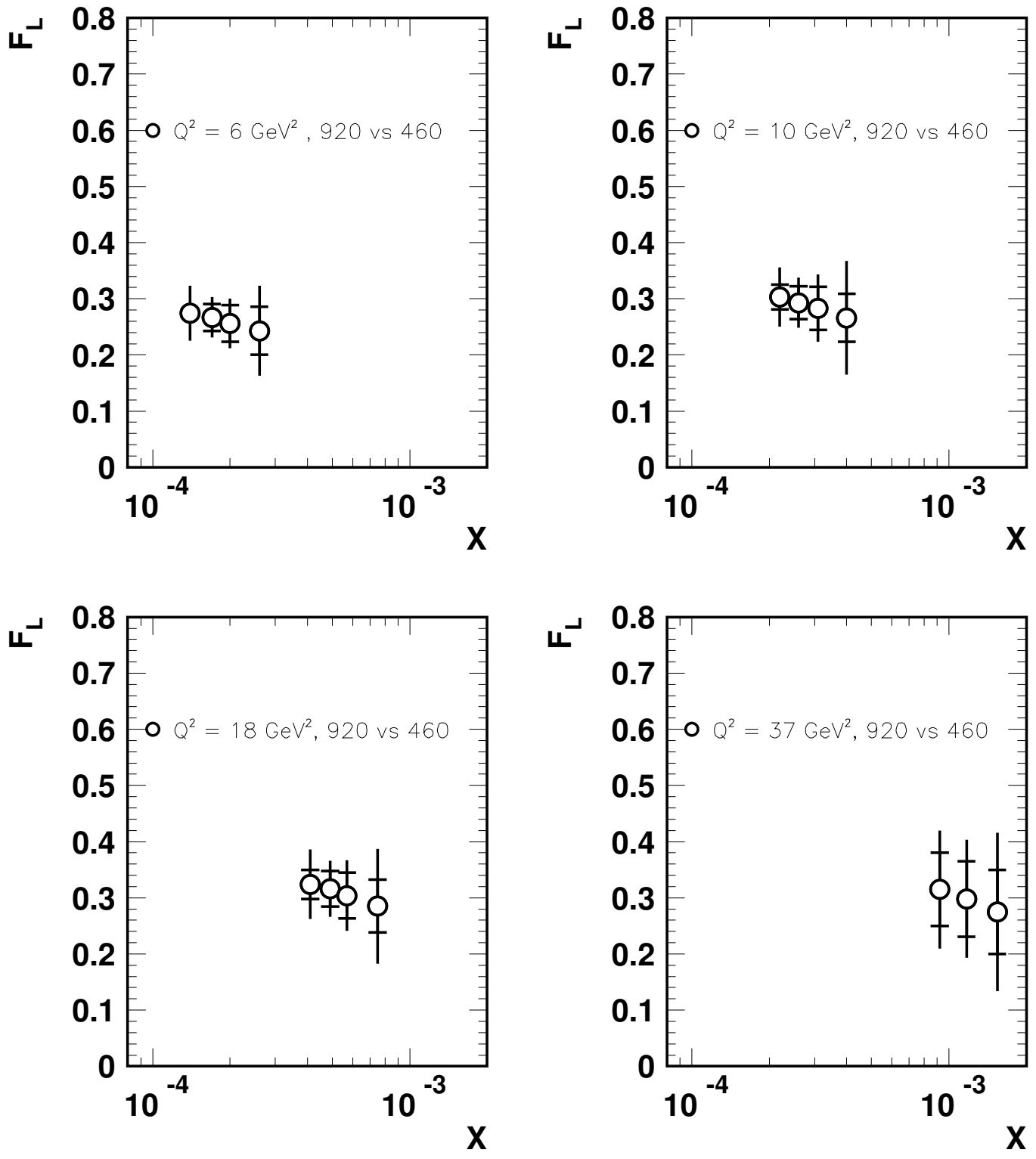

**Fig. 16:** Simulated measurement of the longitudinal structure function $F_L(x, Q^2)$ for data at 920 GeV (30 $pb^{-1}$) and 460 GeV (10 $pb^{-1}$). The inner error bar is the statistical error. The full error bar denotes the statistical and systematic uncertainty added in quadrature.

To evaluate the errors two independent methods have been considered an analytic calculation and a fast Monte-Carlo simulation technique. The two methods provide statistical and systematic errors which are in excellent agreement. The overall result of this simulation of $F_L$ is displayed in Figure 16. In many bins the overall precision on $F_L(x, Q^2)$ is around or below 20%. It is remarkable that the overall precision would stay below 25% even if the statistical error or the larger source of systematic uncertainty would turn out to be twice larger than assumed to be in this study.

### 5.5 Summary

It has been demonstrated with two detailed studies that a direct measurement of the longitudinal structure function $F_L(x, Q^2)$ may be performed at HERA at the five sigma level of accuracy, in the $x$ range from $10^{-4}$ to $10^{-3}$ in four bins of $Q^2$. This measurement requires about three months of running and tuning time at reduced proton beam energy. In addition it would provide the first measurement of the diffractive longitudinal structure function at the three sigma level (see the contribution of P. Newman in the summary of Working Group 4). The exact choice of the parameters of such a measurement are subject to further studies. In conclusion an accurate measurement of $F_L(x, Q^2)$ is feasible, it requires efficient detectors, dedicated beam time and analysis skills. It would be the right sign of completion to have measured $F_2$ first, in 1992 and onwards, and to conclude the HERA data taking with a dedicated measurement of the second important structure function $F_L(x, Q^2)$ , which is related to the gluon density in the low $x$ range of the LHC.

## 6 Determination of the Light Quark Momentum Distributions at Low $x$ at HERA [8]

Based on the data taken in the first phase of HERA's operation (1993-2000), the HERA collider experiments have measured a complete set of neutral (NC) and charged (CC) current double differential $e^{\pm}p$ inclusive scattering cross sections, based on about 120 $\text{pb}^{-1}$ of positron-proton and 15 $\text{pb}^{-1}$ of electron-proton data. The NC and CC deep inelastic scattering (DIS) cross sections for unpolarised $e^{\pm}p$ scattering are determined by structure functions and quark momentum distributions in the proton as follows:

$$\sigma^{\pm}_{NC} \sim Y_+ F_2 \mp Y_- xF_3, \tag{10}$$

$$F_2 \simeq e_u^2 x(U + \overline{U}) + e_d^2 x(D + \overline{D}), \tag{11}$$

$$xF_3 \simeq 2x[a_u e_u(U - \overline{U}) + a_d e_d(D - \overline{D})], \tag{12}$$

$$\sigma^{+}_{CC} \sim x\overline{U} + (1 - y)^2 xD, \tag{13}$$

$$\sigma^{-}_{CC} \sim xU + (1 - y)^2 x\overline{D}. \tag{14}$$

Here $y = Q^2/sx$ is the inelasticity, $s = 4E_e E_p$ and $Y_{\pm} = 1 \pm (1 - y)^2$. The parton distribution $U = u + c + b$ is the sum of the momentum distributions of the up-type quarks with charge $e_u = 2/3$ and axial vector coupling $a_u = 1/2$, while $D = d + s$ is the sum of the momentum distributions of the down type quarks with charge $e_d = -1/3$, $a_d = -1/2$. Similar relationships hold for the anti-quark distributions $\overline{U}$ and $\overline{D}$.

As is illustrated in Fig. 17 the H1 experiment [11] has determined all four quark distributions and the gluon distribution $xg$. The accuracy achieved so far by H1, for $x = 0.01, 0.4$ and $0.65$, is $1\%$, $3\%, 7\%$ for the sum of up quark distributions and $2\%$, $10\%, 30\%$ for the sum of down quark distributions, respectively. The extracted parton distributions are in reasonable agreement with the results obtained in global fits by the MRST [5] and CTEQ [4] collaborations. The H1 result is also consistent with the pdfs determined by the ZEUS Collaboration [10] which uses jet data to improve the accuracy for the gluon distribution and imports a $\overline{d} - \overline{u}$ asymmetry fit result from MRST. New data which are being taken (HERA II) will improve the accuracy of these determinations further. At the time this is written, the available data per experiment have been grown to roughly 150 $\text{pb}^{-1}$ for both $e^+p$ and $e^-p$ scattering, and more is still to come. These data will be particularly important to improve the accuracy at large $x$, which at HERA is related to high $Q^2$.

As is clear from the above equations, the NC and CC cross section data are sensitive directly to only these four quark distribution combinations. Disentangling the individual quark flavours (up, down, strange, charm and beauty) requires additional assumptions. While informations on the $c$ and $b$ densities are being obtained from measurements of $F_2^{c\overline{c}}$ and $F_2^{b\overline{b}}$ of improving accuracy, the determination of the strange quark density at HERA is less straightforward and may rest on $sW^+ \to c$ and strange ($\Phi$) particle production [41]. The relative contributions from the heavy quarks become increasingly important with $Q^2$, as is illustrated in Fig. 18.

[8] Contributing authors: M. Klein, B. Reisert

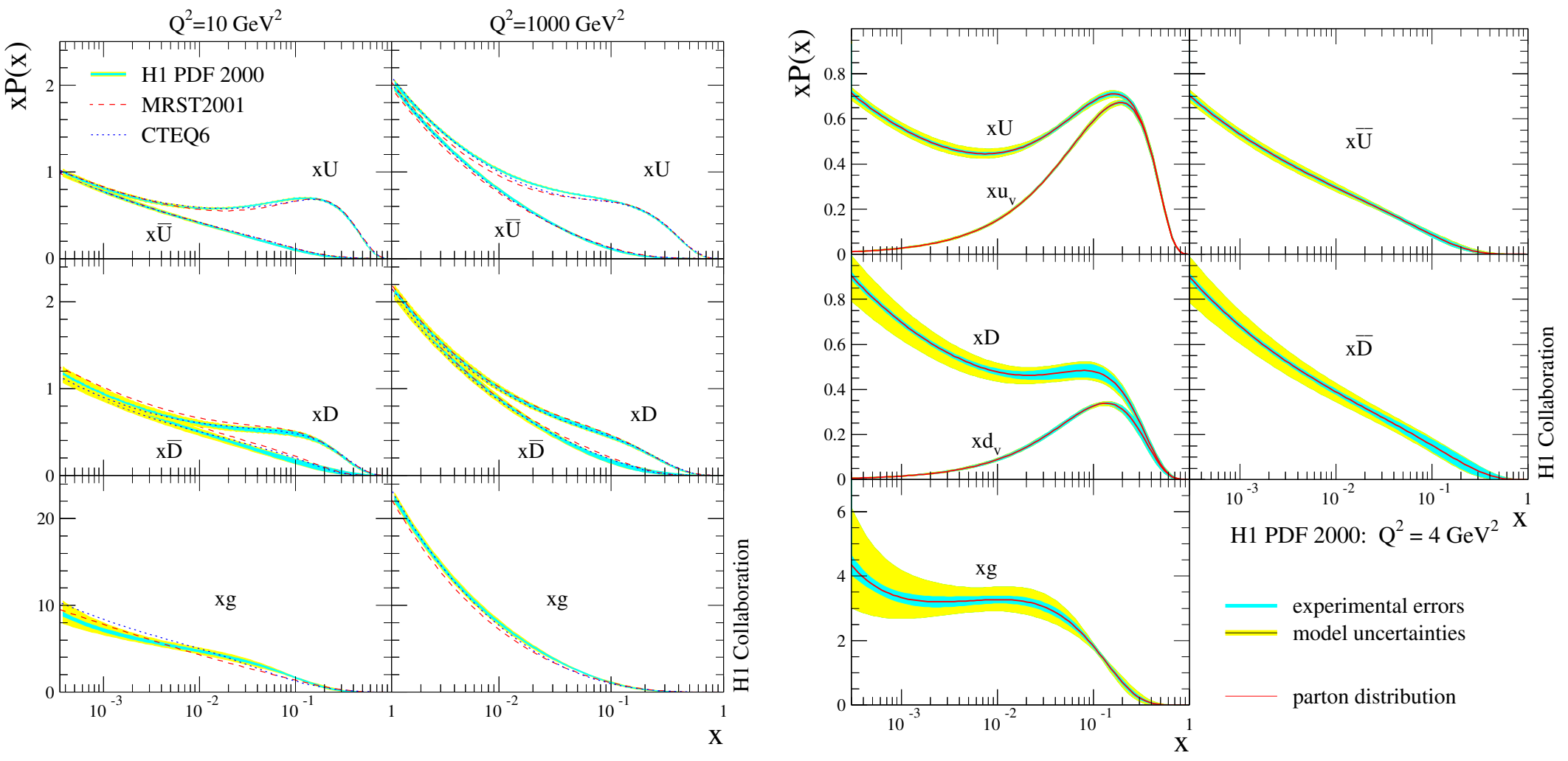

**Fig. 17:** Determination of the sum of up, anti-up, down and anti-downquark distributions and of the gluon distribution in the proton based on the H1 neutral and charged current cross section data. Left: for $Q^2$ of 10 and 1000 GeV$^2$ compared with results from MRST and CTEQ; Right: the parton distributions with their experimental and model uncertainties as determined by H1 at the starting scale $Q_0^2 = 4\,\mathrm{GeV}^2$.

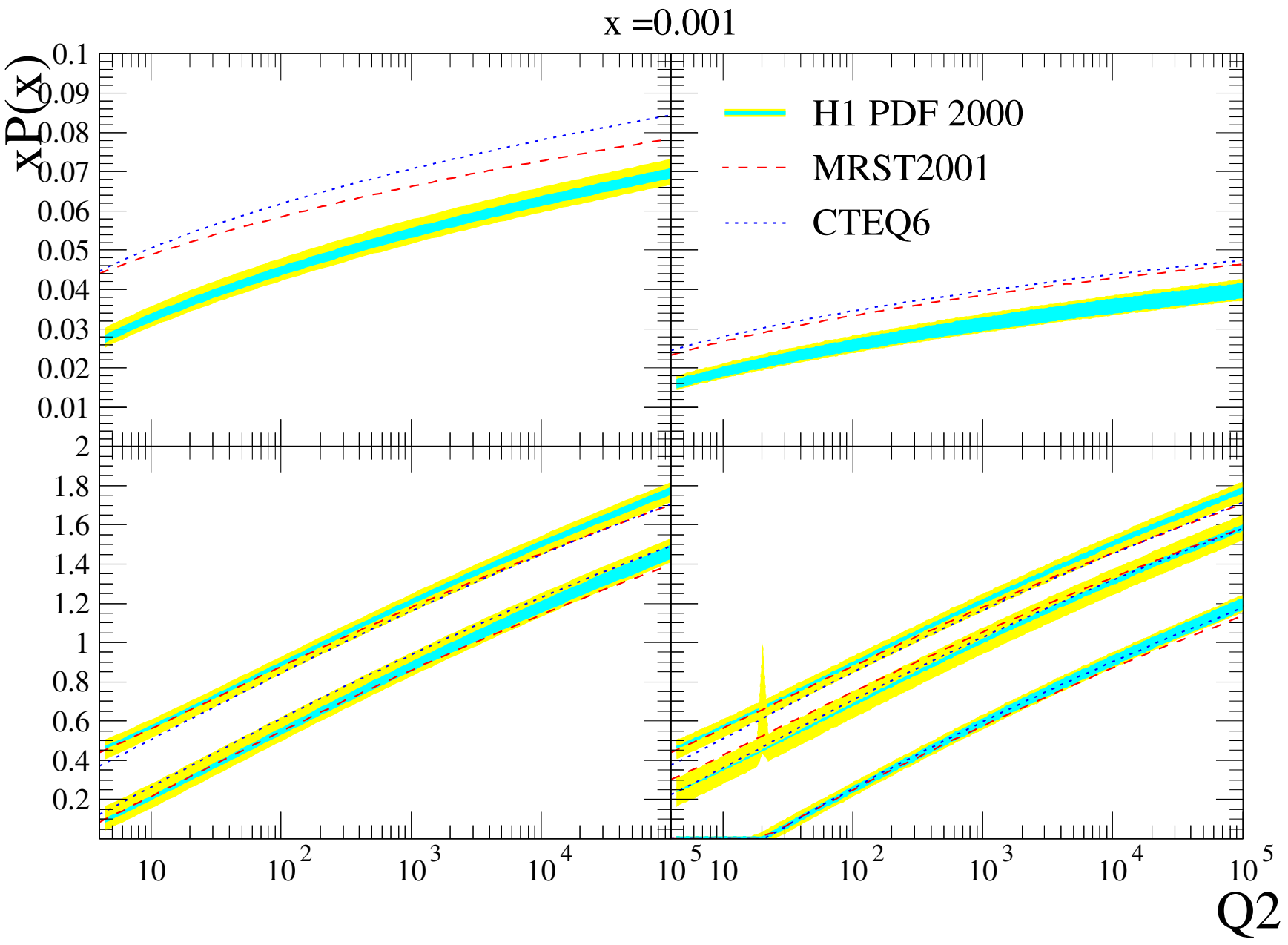

**Fig. 18:** Parton distributions and their uncertainties as determined by H1 extrapolated to the region of the LHC, for $x = 0.001$ near to the rapidity plateau. Top left: $u$ valence; top right: $d$ valence; bottom left: $\overline{u}$ and below $c$; bottom right, in decreasing order: $\overline{d}$, $s$, $b$. The results are compared with recent fits to global data by MRST and CTEQ. Note that at such small $x$ the valence quark distributions are very small. With increasing $Q^2$ the relative importance of the heavy quarks compared to the light quarks increases while the absolute difference of the quark distributions is observed to be rather independent of $Q^2$. The beauty contribution to the cross section thus becomes sizeable, amounting to about 5% for $pp \to HW$.

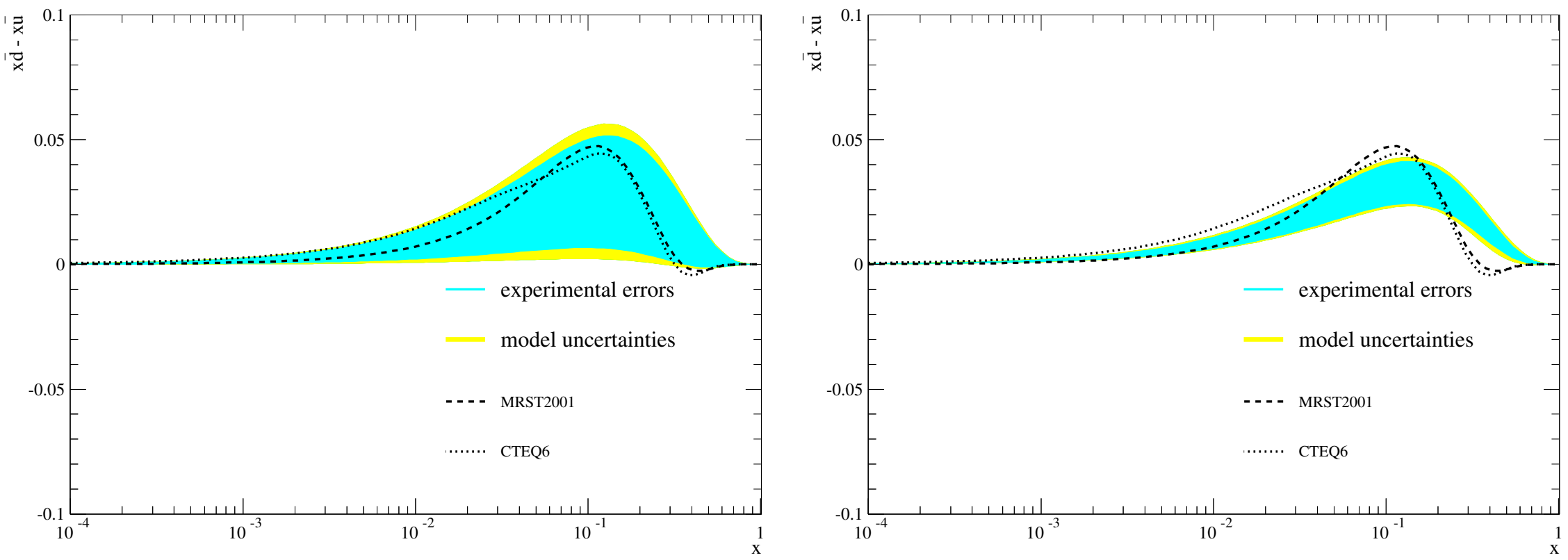

**Fig. 19:** Determination of the difference $x(\overline{d} - \overline{u})$ in the H1 PDF 2000 fit to the H1 data (left) and the H1 and the BCDMS $\mu p$ and $\mu D$ data (right). The sea quark difference is enforced to tend to zero at low $x$. The global fit results of MRST and CTEQ include Drell Yan data which suggest a sea quark asymmetry at $x \sim 0.1$.

The larger $x$ domain is dominated by the valence quarks. At HERA the valence quark distributions are not directly determined but extracted from the differences $u_v = U - \overline{U}$ and $d_v = D - \overline{D}$. Note that this implies the assumption that sea and anti-quarks are equal which in non-perturbative QCD models may not hold. A perhaps more striking assumption is inherent in these fits and regards the sea quark asymmetries at low $x$ which is the main subject of the subsequent discussion.

Fig. 19 shows the difference $x\overline{d} - x\overline{u}$ as determined in the H1 PDF 2000 fit based on the H1 data alone (left) and using in addition the BCDMS proton and deuteron data (right). One observes a trend of these fits to reproduce the asymmetry near $x \sim 0.1$ which in the MRST and CTEQ fits, shown in Fig. 19, is due to fitting the Drell Yan data from the E866/NuSea experiment [42]. While this enhancement is not very stable in the H1 fit [43] and not significant either, with the BCDMS data an asymmetry is observed which reflects the violation of the Gottfried sum rule.

In the H1 fit [11] the parton distributions at the initial scale $Q^2 = 4\,\mathrm{GeV}^2$ are parameterised as $xP = A_p x^{B_P}(1-x)^{C_P} \cdot f_P(x)$. The function $f_P$ is a polynomial in $x$ which is determined by requiring "$\chi^2$ saturation" of the fits, i.e. starting from $f_P = 1$ additional terms $D_P x$, $E_P x^2$ etc. are added and only considered if they cause a significant improvement in $\chi^2$, half integer powers were considered in [43]. The result for fitting the H1 data has been as follows: $f_g = (1 + D_g x)$, $f_U = (1 + D_U x + F_U x^3)$, $f_D = (1 + D_D x)$ and $f_{\overline{U}} = f_{\overline{D}} = 1$. The parton distributions at low $x$ are thus parameterised as $xP \rightarrow A_P x^{B_P}$. The strange (charm) anti-quark distribution is coupled to the total amount of down (up) anti-quarks as $\overline{s} = f_c\overline{D}$ ($\overline{c} = f_c\overline{U}$). Two assumptions have been made on the behaviour of the quark and anti-quark distributions at low $x$. It has been assumed that quark and anti-quark distributions are equal and, moreover, that the sea is flavour symmetric. This implies that the slopes $B$ of all four quark distributions are set equal $B_U = B_D = B_{\overline{U}} = B_{\overline{D}}$. Moreover, the nomalisations of up and down quarks are the same, i.e. $A_{\overline{U}}(1 - f_c) = A_{\overline{D}}(1 - f_s)$, which ensures that $\overline{d}/\overline{u} \rightarrow 1$ as $x$ tends to zero. The consequence of this assumption is illustrated in Fig. 19. While the DIS data suggest some asymmetry at larger $x$, the up-down quark asymmetry is enforced to vanish at lower $x$. This results in a rather fake high accuracy in the determination of the four quark distributions at low $x$, despite the fact that at low $x$ there is only one combination of them measured, which is $F_2 = x[4(U + \overline{U}) + (D + \overline{D})]/9$. If one relaxes both the conditions on the slopes and normalisations, the fit to the H1 data decides to completely remove the down quark contributions as is seen in Fig. 20 (left plot).

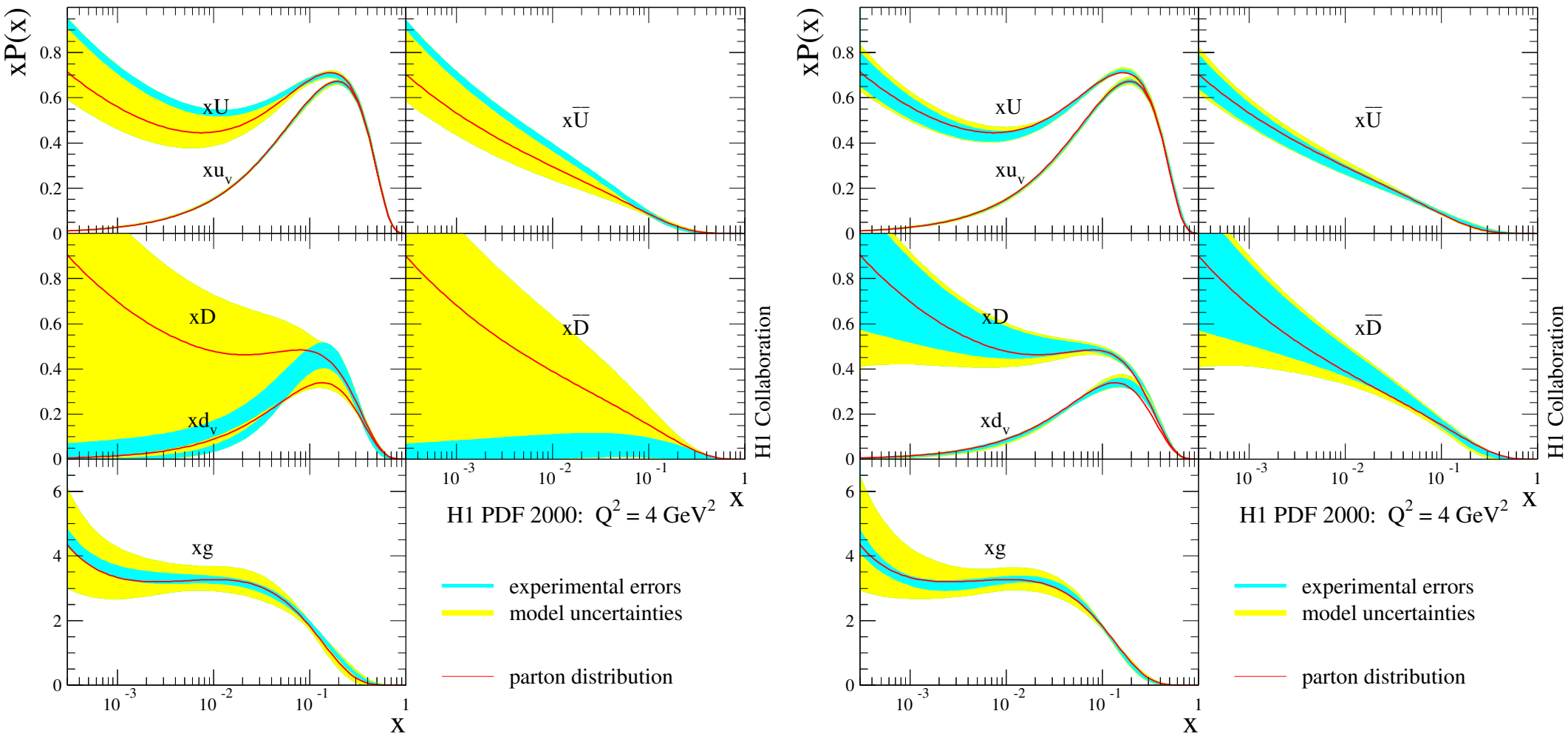

**Fig. 20:** Determinations of the quark and gluon momentum distributions releasing the constraint $x\overline{d} = x\overline{u}$ at low $x$, from the H1 NC and CC data alone (left) and from the H1 $ep$ and the BCDMS $\mu p$ and $\mu D$ data (right). Since at low $x < 0.01$ there is no further constraint than that given from $F_2$ the uncertainties of $\overline{U}$ and in particular of $\overline{D}$ become sizeable.

In DIS the up and down quark asymmetry can be constrained using deuteron data because the nucleon structure function determines a different linear combination according to $F_2^N = 5x(U + \overline{U} + D + \overline{D})/18 + x(c + \overline{c} - s - \overline{s})/6$ with $N = (p+n)/2$. Unfortunately, there are only data at rather large $x$ available. The effect of including the BCDMS data on the low $x$ behaviour of the parton distributions is illustrated in Fig. 20 (right plot). It restores some amount of down quarks at low $x$ , the errors, however, in particular of the down quarks, are still very large. The result is a large sea quark asymmetry uncertainty, which is shown in Fig. 21. At HERA a proposal had been made [44] to operate the machine in electron-deuteron mode. Measuring the behaviour at low $x$ would not require high luminosity. Such data would constrain [9] a possible sea quark asymmetry with very high accuracy, as is also shown in Fig. 21.

Deuterons at HERA would require a new source and modest modifications to the preaccelerators. The H1 apparatus could be used in its standard mode with a forward proton detector added to take data at half the beam energy. Tagging the spectator protons with high accuracy at HERA, for the first time in DIS, one could reconstruct the electron-neutron scattering kinematics essentially free of nuclear corrections [44]. Since the forward scattering amplitude is related to diffraction one would also be able to constrain shadowing to the per cent level [47]. The low $x$ measurements would require small luminosity amounts, of less than $50\,\mathrm{pb}^{-1}$. Long awaited constraints of the $d/u$ ratio at large $x$ and $Q^2$ would require extended running, as would CC data. Besides determining the parton distributions better, the measurement of the singlet $F_2^N$ structure function would give important constraints on the evolution and theory at low $x$ [48]. It would also result in an estimated factor of two improvement on the measurement of $\alpha_s$ at HERA [49]. For the development of QCD, of low $x$ physics in particular, but as well for understanding physics at the LHC and also for superhigh energy neutrino astrophysics, HERA $eD$ data remain to be important.

[9] Constraints on the sea quark distributions may also be obtained from $W^+/W^-$ production at the TeVatron. However, the sensitivity is limited to larger $x \geq 0.1$ [45] since $W's$ produced in collisions involving sea quarks of smaller $x$ will be boosted so strongly, that their decay products are not within the acceptance of the collider detectors. $W^+$ and $W^-$ production at the LHC has been discussed in [46].

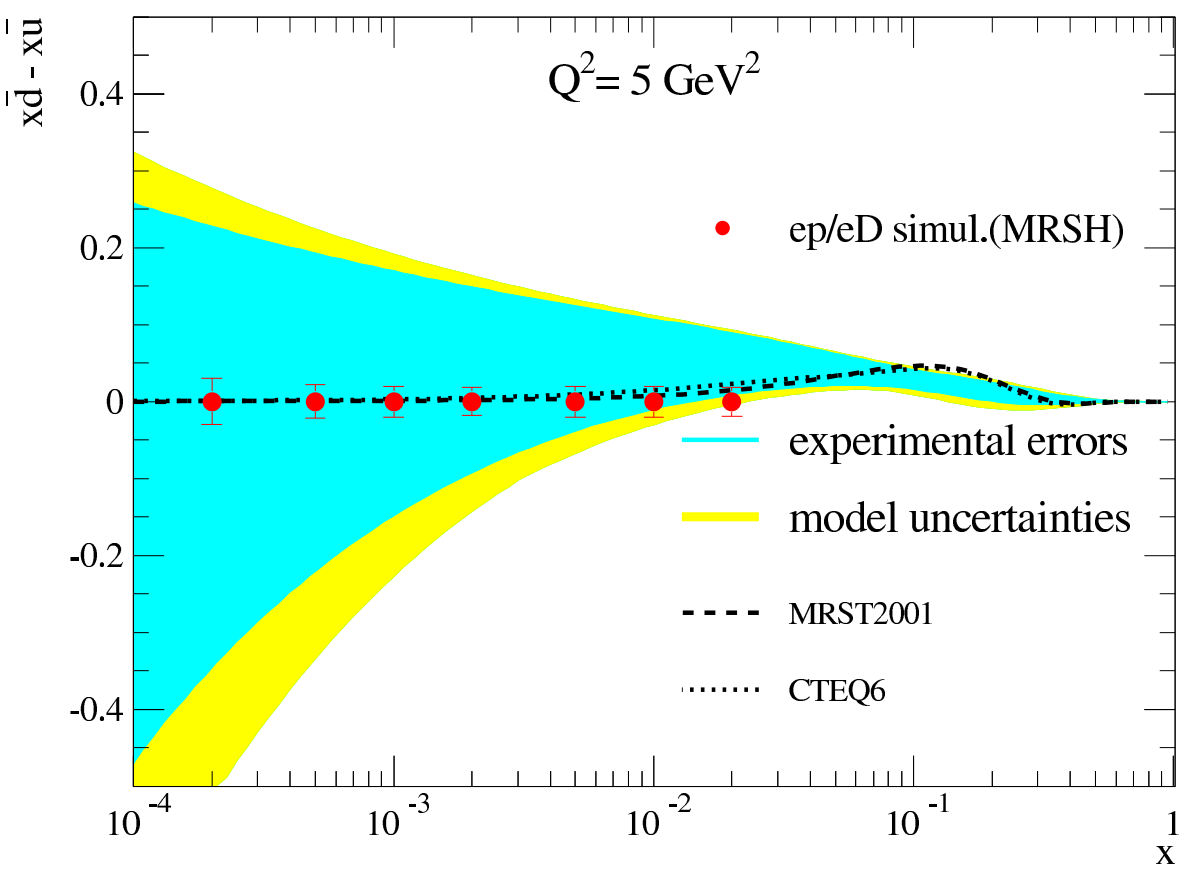

**Fig. 21:** Simulation of the difference of sea quark distributions, here assumed to be zero, at low $x$ based on additional 20 pb$^{-1}$ of electron-deuteron data at HERA. The error band represents the uncertainty of the H1 NLO QCD fit to the H1 $ep$ and the BCDMS $\mu p$ and $\mu d$ data without the constraint $\bar{d} = \bar{u}$ at low $x$. The dashed curves represent calculations using recent global fits by MRST and by CTEQ.

## 7 Impact of future HERA data on the determination of proton PDFs using the ZEUS NLO QCD fit [10]

### 7.1 PDF fits to HERA data

Recently, the ZEUS Collaboration have performed a combined NLO QCD fit to inclusive neutral and charged current DIS data [23–28] as well as high precision jet data in DIS [50] and $\gamma p$ scattering [51]. This is called the ZEUS-JETS PDF fit [10]. The use of only HERA data eliminates the uncertainties from heavy-target corrections and removes the need for isospin symmetry assumptions. It also avoids the difficulties that can sometimes arise from combining data-sets from several different experiments, thereby allowing a rigorous statistical treatment of the PDF uncertainties. Furthermore, PDF uncertainties from current global fits are, in general, limited by (irreducible) experimental systematics. In contrast, those from fits to HERA data alone, are largely limited by the statistical precision of existing measurements. Therefore, the impact of future data from HERA is likely to be most significant in fits to only HERA data.

### 7.2 The ZEUS NLO QCD fit

The ZEUS-JETS PDF fit has been used as the basis for all results shown in this contribution. The most important details of the fit are summarised here. A full description may be found elsewhere [10]. The fit includes the full set of ZEUS inclusive neutral and charged current $e^{\pm}p$ data from HERA-I (1994-2000), as well as two sets of high precision jet data in $e^{+}p$ DIS ($Q^2 >> 1$ GeV$^2$) and $\gamma p$ ($Q^2 \sim 0$) scattering. The inclusive data used in the fit, span the kinematic range $6.3 \times 10^{-5} < x < 0.65$ and $2.7 < Q^2 < 30000$ GeV$^2$.

The PDFs are obtained by solving the NLO DGLAP equations within the $\overline{\mathrm{MS}}$ scheme. These equations yield the PDFs at all values of $Q^2$ provided they are input as functions of $x$ at some starting scale $Q_0^2$. The resulting PDFs are convoluted with coefficient functions to give predictions for structure functions and, hence, cross sections. In the ZEUS fit, the $xu_v(x)$ ($u$-valence), $xd_v(x)$ ($d$-valence), $xS(x)$ (total sea-quark), $xg(x)$ (gluon) and $x(\bar{d}(x) - \bar{u}(x))$ PDFs are parameterised at a starting scale of $Q_0^2 = 7$ GeV$^2$ by the form,

$$xf(x) = p_1 x^{p_2} (1-x)^{p_3} P(x), \tag{15}$$

[10]Contributing authors: C. Gwenlan, A. Cooper-Sarkar, C. Targett-Adams.

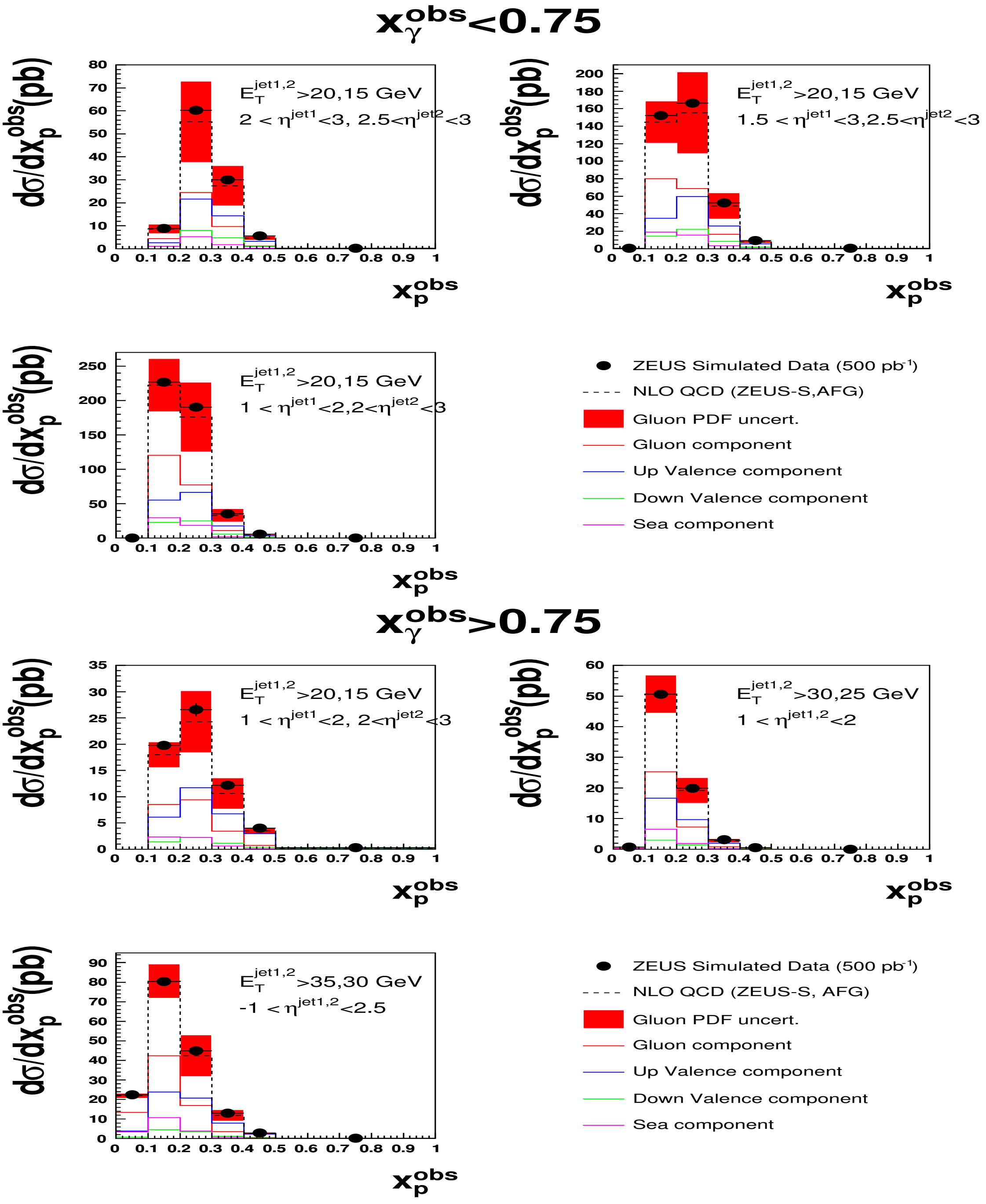

**Fig. 22:** The optimised jet cross sections included in the HERA-II projected fit. The solid points show the simulated data generated using the NLO QCD programme of Frixione-Ridolfi, using the CTEQ5M1 proton and the AFG photon PDFs. The error bars show the statistical uncertainties, which correspond to $500\ \mathrm{pb}^{-1}$ of HERA data. Systematic uncertainties have been neglected. The dashed line shows the NLO QCD prediction using the ZEUS-S proton and AFG photon PDFs. The shaded band shows the contribution to the cross section uncertainty arising from the uncertainty in the gluon distribution in the proton.

**Table 4:** The data-sets included in the ZEUS-JETS and HERA-II projected PDF fits. The first column lists the type of data and the second gives the kinematic coverage. The third column gives the integrated luminosities of the HERA-I measurements included in the ZEUS-JETS fit. The fourth column gives the luminosities assumed in the HERA-II projection. Note that the 96-97 NC and the 94-97 CC measurements have not had their luminosity scaled for the HERA-II projection.

| **data sample** | **kinematic coverage** | **HERA-I** $\mathcal{L}$ ($\mathrm{pb}^{-1}$) | **HERA-II** $\mathcal{L}$ ($\mathrm{pb}^{-1}$) **(assumed)** |
|---|---|---|---|
| 96-97 NC $e^+p$ [23] | $2.7 < Q^2 < 30000$ GeV$^2$; $6.3 \cdot 10^{-5} < x < 0.65$ | 30 | 30 |
| 94-97 CC $e^+p$ [24] | $280 < Q^2 < 17000$ GeV$^2$; $6.3 \cdot 10^{-5} < x < 0.65$ | 48 | 48 |
| 98-99 NC $e^-p$ [25] | $200 < Q^2 < 30000$ GeV$^2$; $0.005 < x < 0.65$ | 16 | 350 |
| 98-99 CC $e^-p$ [26] | $280 < Q^2 < 17000$ GeV$^2$; $0.015 < x < 0.42$ | 16 | 350 |
| 99-00 NC $e^+p$ [27] | $200 < Q^2 < 30000$ GeV$^2$; $0.005 < x < 0.65$ | 63 | 350 |
| 99-00 CC $e^+p$ [28] | $280 < Q^2 < 17000$ GeV$^2$; $0.008 < x < 0.42$ | 61 | 350 |
| 96-97 inc. DIS jets [50] | $125 < Q^2 < 30000$ GeV$^2$; $E_{\mathrm{T}}^{Breit} > 8$ GeV | 37 | 500 |
| 96-97 dijets in $\gamma p$ [51] | $Q^2 \lesssim 1$ GeV$^2$; $E_{\mathrm{T}}^{jet1,2} > 14, 11$ GeV | 37 | 500 |
| optimised jets [52] | $Q^2 \lesssim 1$ GeV$^2$; $E_{\mathrm{T}}^{jet1,2} > 20, 15$ GeV | - | 500 |

where $P(x) = (1 + p_4 x)$. No advantage in the $\chi^2$ results from using more complex polynomial forms. The normalisation parameters, $p_1(u_v)$ and $p_1(d_v)$, are constrained by quark number sum rules while $p_1(g)$ is constrained by the momentum sum rule. Since there is no information to constrain any difference in the low-$x$ behaviour of the $u$- and $d$-valence quarks, $p_2(u_v)$ has been set equal to $p_2(d_v)$. The data from HERA are currently less precise than the fixed target data in the high-$x$ regime. Therefore, the high-$x$ sea and gluon distributions are not well constrained in current fits to HERA data alone. To account for this, the sea shape has been restricted by setting $p_4(S) = 0$. The high-$x$ gluon shape is constrained by the inclusion of HERA jet data. In fits to only HERA data, there is no information on the shape of $\bar{d} - \bar{u}$. Therefore, this distribution has its shape fixed consistent with Drell-Yan data and its normalisation set consistent with the size of the Gottfried sum rule violation. A suppression of the strange sea with respect to the non-strange sea of a factor of 2 at $Q_0^2$ is also imposed, consistent with neutrino induced dimuon data from CCFR. The value of the strong coupling has been fixed to $\alpha_s(M_{\mathrm{Z}}) = 0.1180$. After all constraints, the ZEUS-JETS fit has 11 free parameters. Heavy quarks were treated in the variable flavour number scheme of Thorne & Roberts [19]. Full account was taken of correlated experimental systematic uncertainties, using the Offset Method [9, 18].

The results of two separate studies are presented. The first study provides an estimate of how well the PDF uncertainties may be known by the end of HERA-II, within the currently planned running scenario, while the second study investigates the impact of a future HERA measurement of $F_L$ on the gluon distribution. All results presented, are based on the recent ZEUS-JETS PDF analysis [10].

### 7.3 PDF uncertainty estimates for the end of HERA running

The data from HERA-I are already very precise and cover a wide kinematic region. However, HERA-II is now running efficiently and is expected to provide a substantial increase in luminosity. Current estimates suggest that, by the end of HERA running (in mid-2007), an integrated luminosity of 700 $\mathrm{pb}^{-1}$ should be achievable. This will allow more precise measurements of cross sections that are curently statistically limited: in particular, the high-$Q^2$ NC and CC data, as well as high-$Q^2$ and/or high-$E_{\mathrm{T}}$ jet data. In addition to the simple increase in luminosity, recent studies [52] have shown that future jet cross section measurements, in kinematic regions optimised for sensitivity to PDFs, should have a significant impact on the gluon uncertainties. In this contribution, the effect on the PDF uncertainties, of both the higher precision expected from HERA-II and the possibility of optimised jet cross section measurements, has been estimated in a new QCD fit. This fit will be referred to as the “HERA-II projection”.

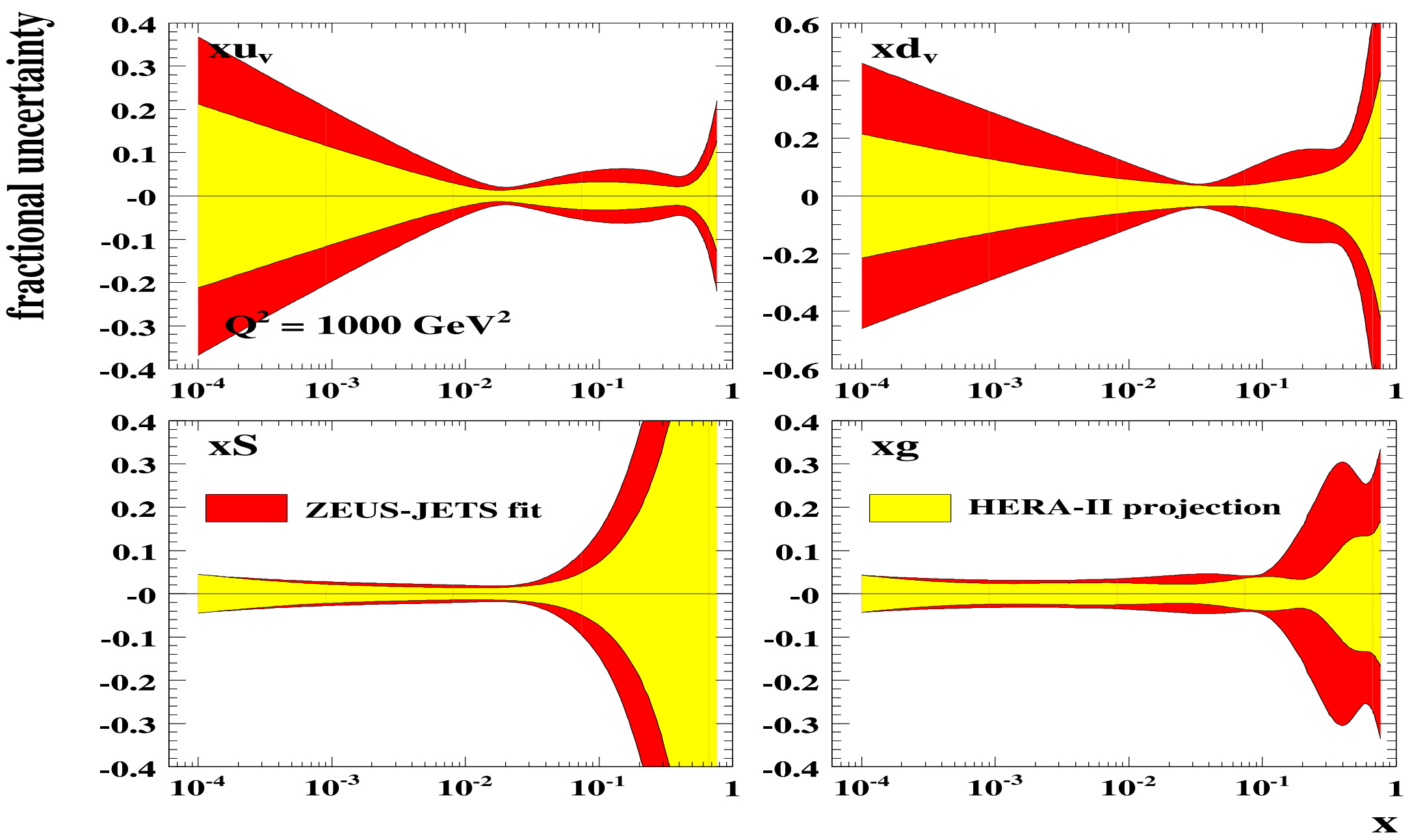

**Fig. 23:** The fractional PDF uncertainties, as a function of $x$, for the $u$-valence, $d$-valence, sea-quark and gluon distributions at $Q^2 = 1000$ GeV$^2$. The red shaded bands show the results of the ZEUS-JETS fit and the yellow shaded bands show the results of the HERA-II projected fit.

In the HERA-II projected fit, the statistical uncertainties on the currently available HERA-I data have been reduced. For the high-$Q^2$ inclusive data, a total integrated luminosity of 700 $\mathrm{pb}^{-1}$ was assumed, equally divided between $e^+$ and $e^-$. For the jet data, an integrated luminosity of 500 $\mathrm{pb}^{-1}$ was assumed. The central values and systematic uncertainties were taken from the published data in each case. In addition to the assumed increase in precision of the measurements, a set of optimised jet cross sections were also included, for forward dijets in $\gamma p$ collisions, as defined in a recent study [52]. Since no real data are yet available, simulated points were generated using the NLO QCD program of Frixione-Ridolfi [53], using the CTEQ5M1 [4] proton and AFG [54] photon PDFs. The statistical uncertainties were taken to correspond to 500 $\mathrm{pb}^{-1}$. For this study, systematic uncertainties on the optimised jet cross sections were ignored. The simulated optimised jet cross section points, compared to the predictions of NLO QCD using the ZEUS-S proton PDF [9], are shown in Fig. 22.

Table 4 lists the data-sets included in the ZEUS-JETS and HERA-II projected fits. The luminosities of the (real) HERA-I measurements and those assumed for the HERA-II projection are also given.

The results are summarised in Fig. 23, which shows the fractional PDF uncertainties, for the $u$- and $d$-valence, sea-quark and gluon distributions, at $Q^2 = 1000$ GeV$^2$. The yellow bands show the results of the ZEUS-JETS fit while the red bands show those for the HERA-II projection. Note that the same general features are observed for all values of $Q^2$. In fits to only HERA data, the information on the valence quarks comes from the high-$Q^2$ NC and CC cross sections. The increased statistical precision of the high-$Q^2$ data, as assumed in the HERA-II projected fit, gives a significant improvement in the valence uncertainties over the whole range of $x$. For the sea quarks, a significant improvement in the uncertainties at high-$x$ is also observed. In contrast, the low-$x$ uncertainties are not visibly reduced. This is due to the fact that the data constraining the low-$x$ region tends to be at lower-$Q^2$, which are already systematically limited. This is also the reason why the low-$x$ gluon uncertainties are not significantly reduced. However, the mid-to-high-$x$ gluon, which is constrained by the jet data, is much improved in the HERA-II projected fit. Note that about half of the observed reduction in the gluon uncertainties is due to the inclusion of the simulated optimised jet cross sections.

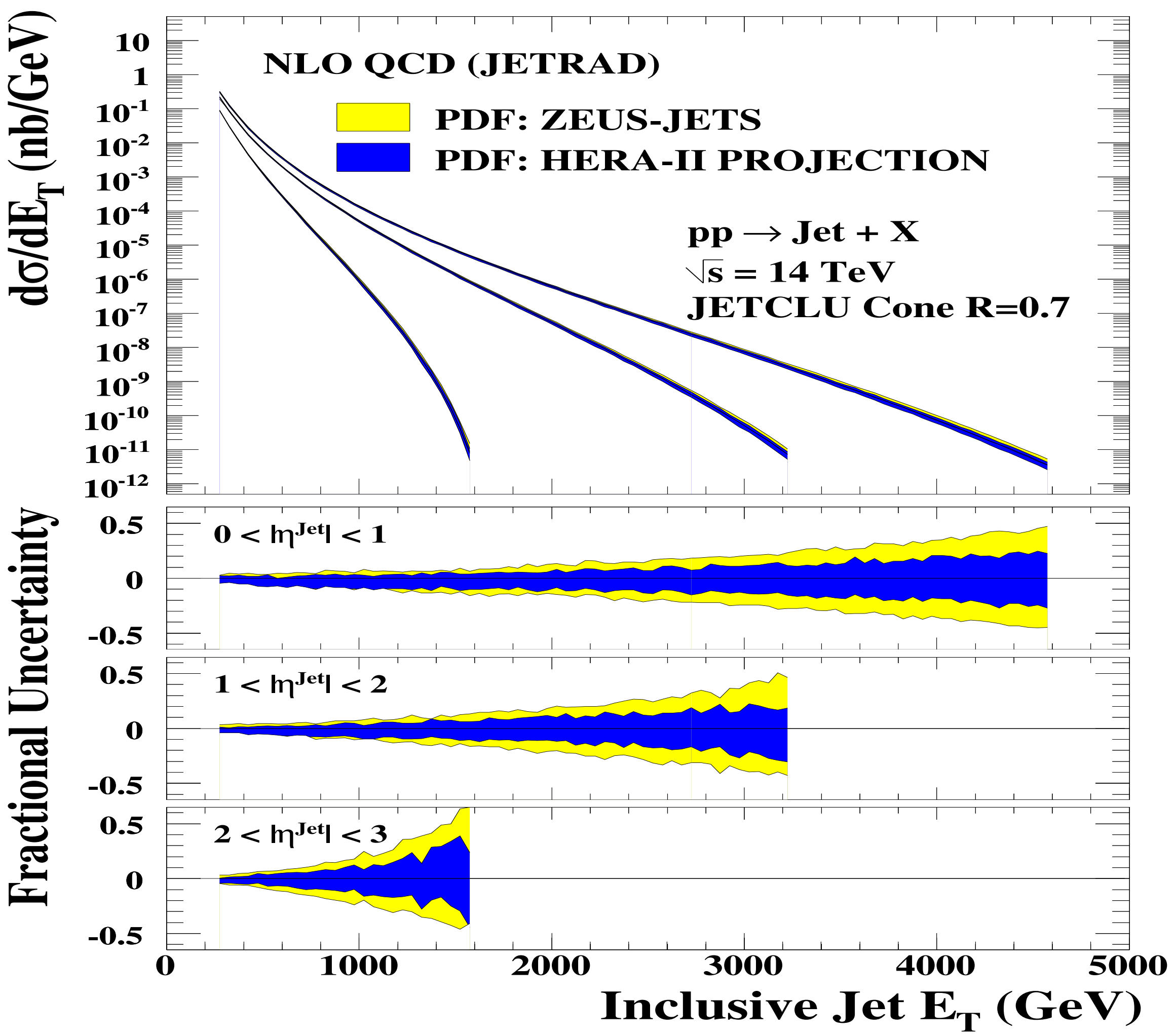

**Fig. 24:** NLO QCD inclusive jet cross section predictions at $\sqrt{s}$=14 TeV in three regions of pseudo-rapidity. The yellow and blue bands show the PDF uncertainties from the ZEUS-JETS and HERA-II projected fits, respectively.

*Inclusive jet cross sections at the LHC*

The improvement to the high-$x$ partons, observed in the HERA-II projection compared to the ZEUS-JETS fit, will be particularly relevant for high-scale physics at the LHC. This is illustrated in Fig. 24, which shows NLO QCD predictions from the JETRAD [55] programme for inclusive jet production at $\sqrt{s} = 14$ TeV. The results are shown for both the ZEUS-JETS and the HERA-II projected PDFs. The uncertainties on the cross sections, resulting from the PDFs, have been calculated using the LHAPDF interface [56]. For the ZEUS-JETS PDF, the uncertainty reaches $\sim 50\%$ at central pseudo-rapidities, for the highest jet transverse energies shown. The prediction using the HERA-II projected PDF shows a marked improvement at high jet tranverse energy.

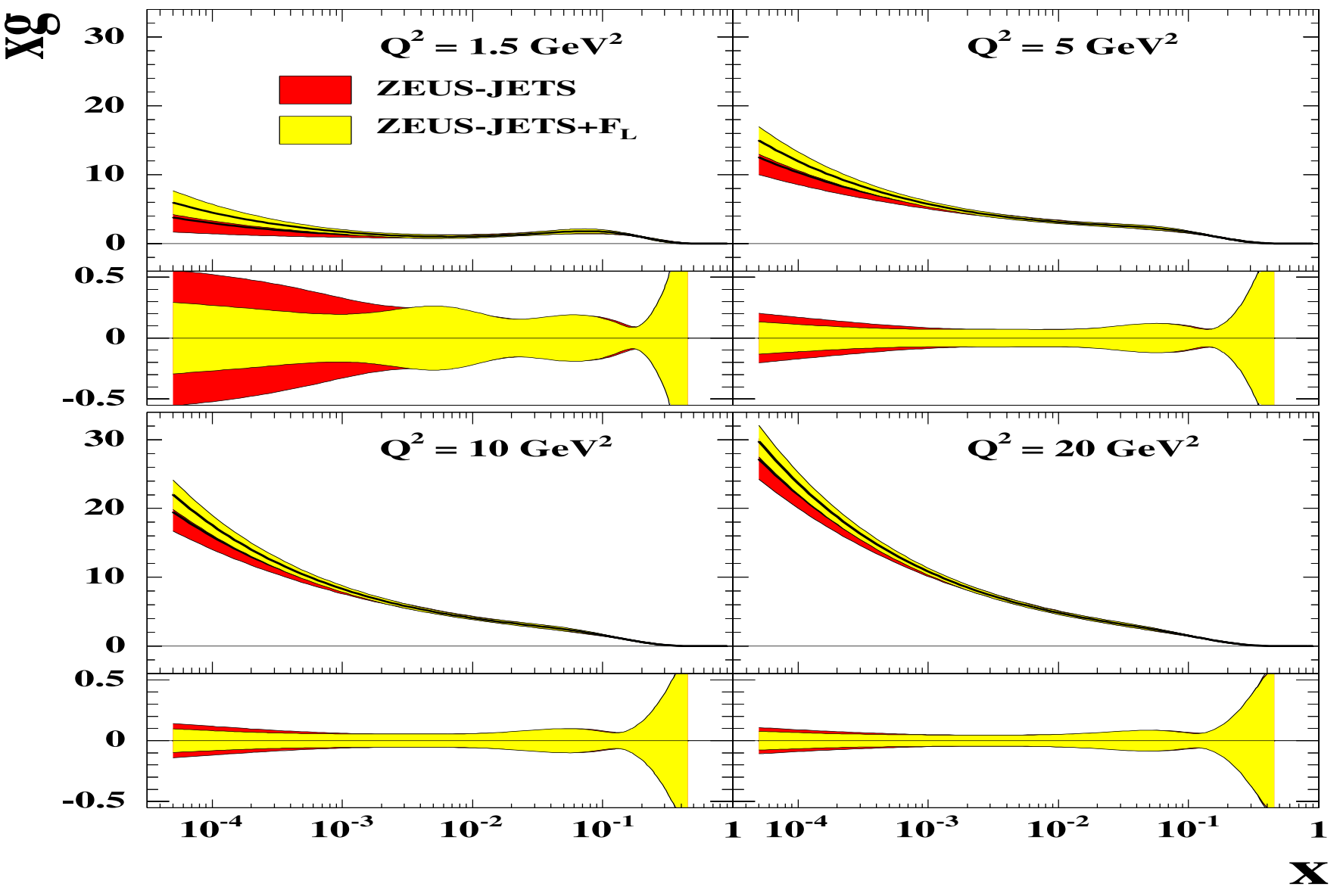

**Fig. 25:** The gluon PDFs, showing also the fractional uncertainty, for fits with and without inclusion of the simulated $F_L$ data, for $Q^2 = 1.5$, 5, 10 and 20 GeV$^2$. The red shaded bands show the results of the ZEUS-JETS fit and the yellow shaded band show the results of the ZEUS-JETS+$F_L$ fit.

### 7.4 Impact of a future HERA measurement of $F_L$ on the gluon PDF

The longitudinal structure function, $F_L$, is directly related to the gluon density in the proton. In principle, $F_L$ can be extracted by measuring the NC DIS cross section at fixed $x$ and $Q^2$, for different values of $y$ (see Eqn. 3). A precision measurement could be achieved by varying the centre-of-mass energy, since $s = Q^2/xy \approx 4E_eE_p$, where $E_e$ and $E_p$ are the electron and proton beam energies, respectively. Studies [38] (Sec. 5) have shown that this would be most efficiently achieved by changing the proton beam energy. However, such a measurement has not yet been performed at HERA.

There are several reasons why a measurement of $F_L$ at low-$x$ could be important. The gluon density is not well known at low-$x$ and so different PDF parameterisations can give quite different predictions for $F_L$ at low-$x$. Therefore, a precise measurement of the longitudinal sturcture function could both pin down the gluon PDF and reduce its uncertainties. Furthermore, predictions of $F_L$ also depend upon the nature of the underlying theory (e.g. order in QCD, resummed calculation etc). Therefore, a measurement of $F_L$ could also help to discriminate between different theoretical models.

*Impact on the gluon PDF uncertainties*

The impact of a possible future HERA measurement of $F_L$ on the gluon PDF uncertainties has been investigated, using a set of simulated $F_L$ data-points [38]. (see Sec. 5). The simulation was performed using the GRV94 [57] proton PDF for the central values, and assuming $E_e = 27.6$ GeV and $E_p = 920, 575, 465$ and 400 GeV, with luminosities of 10, 5, 3 and 2 pb$^{-1}$, respectively. Assuming that the luminosity scales simply as $E_p^2$, this scenario would nominally cost 35 pb$^{-1}$ of luminosity under standard HERA conditions. However, this estimate takes no account of time taken for optimisation of the machine with each change in $E_p$, which could be considerable. The systematic uncertainties on the simulated data-points were calculated assuming a $\sim 2\%$ precision on the inclusive NC cross section measurement. A more comprehensive description of the simulated data is given in contribution for this proceedings, see Sec. 5.

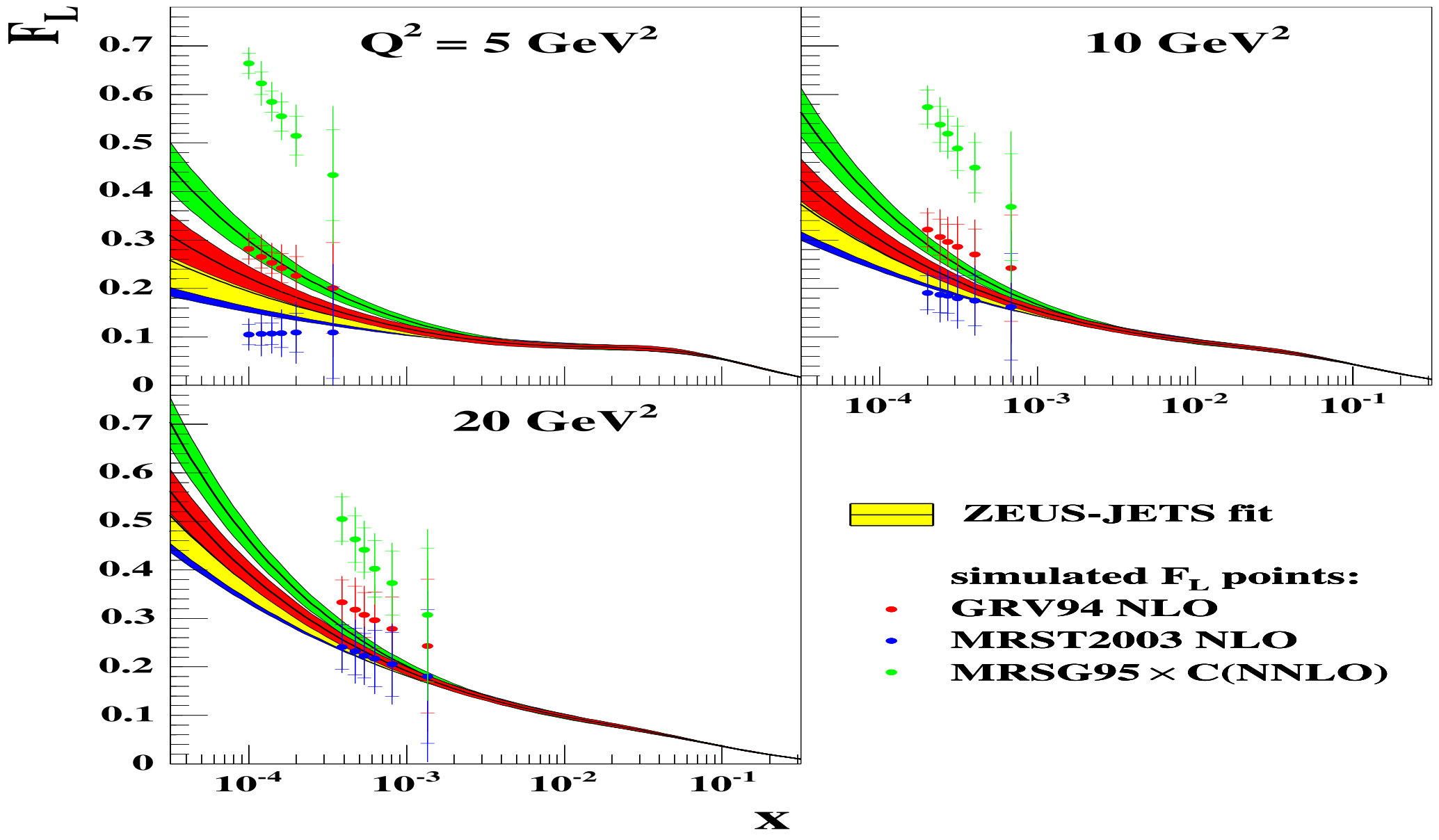

**Fig. 26:** The distribution of the longitudinal structure function $F_L$ at $Q^2$ =5, 10 and 20 GeV$^2$. The blue, red and green points show the simulated $F_L$ data-points, respectively labelled maximum, middle and minimum in Table 5. The blue, red and green shaded bands show the NLO QCD predictions, in the case where the data-points of the corresponding colour have been included in the fit. For comparison, the yellow shaded band shows the prediction of the ZEUS-JETS fit.

The simulated data were included in the ZEUS-JETS fit. Figure 25 shows the gluon distribution and fractional uncertainties for fits with and without inclusion of the simulated $F_L$ data. The results indicate that the gluon uncertainties are reduced at low-$x$, but the improvement is only significant at relatively low $Q^2 \lesssim 20$ GeV$^2$.

*Discrimination between theoretical models*

In order to assess whether a HERA measurement of $F_L$ could discriminate between theoretical models, two more sets of $F_L$ data-points have been simulated [58], using different theoretical assumptions. The first of the two sets was generated using the MRSG95 [59] proton PDF, which has a large gluon density. The PDFs were then convoluted with the NNLO order coefficient functions, which are large and positive. This gives the “maximum” set of $F_L$ data-points. In contrast, the second set has been generated using the MRST2003 [60] proton PDF, which has a negative gluon at low-$x$ and low-$Q^2$, thus providing a “minimum” set of $F_L$ data. The original set of $F_L$ points described in the previous subsection lies between these two extremes. The details of all three sets are summarised in Table 5.

Figure 26 shows the results of including, individually, each set of simulated $F_L$ data into the ZEUS NLO QCD fit. The results show that the NLO fit is relatively stable to the inclusion of the extreme sets of data. This indicates that a measurement of $F_L$ could discriminate between certain theoretical models. However, it should be noted that the maximum and minimum models studied here were chosen specifically to give the widest possible variation in $F_L$. There are many other alternatives that would lie between these extremes and the ability of an $F_L$ measurement to discriminate between them would depend both on the experimental precision of the measurement itself, as well as the theoretical uncertainties on the models being tested.

**Table 5:** Summary of the PDFs used to generate the simulated $F_L$ data-points. The extreme maximum $F_L$ points were generated using the MRSG95 PDF, and convoluted with NNLO coefficient functions. The middle points were generated using the GRV94 PDF, and the extreme minimum points were generated using the MRST2003 PDF, which has a negative gluon at low-$x$.

| | PDF | QCD order of coefficient functions |
|---|---|---|
| Maximum $F_L$ | MRSG95 | NNLO |
| Middle $F_L$ | GRV94 | NLO |
| Minimum $F_L$ | MRST2003 | NLO |

# 8 A Method to Include Final State Cross-sections Measured in Proton-Proton Collisions to Global NLO QCD Analysis [11]

The Large Hadron Collider (LHC), currently under construction at CERN, will collide protons on protons with an energy of 7 TeV. Together with its high collision rate the high available centre-of-mass energy will make it possible to test new interactions at very short distances that might be revealed in the production cross-sections of Standard Model (SM) particles at very high transverse momentum ($P_T$) as deviation from the SM theory.

The sensitivity to new physics crucially depends on experimental uncertainties in the measurements and on theoretical uncertainties in the SM predictions. It is therefore important to work out a strategy to minimize both the experimental and theoretical uncertainties from LHC data. For instance, one could use single inclusive jet or Drell-Yan cross-sections at low $P_T$ to constrain the PDF uncertainties at high $P_T$. Typical residual renormalisation and factorisation scale uncertainties in next-to-leading order (NLO) calculations for single inclusive jet-cross-section are about $5 - 10\%$ and should hopefully be reduced as NNLO calculations become available. The impact of PDF uncertainties on the other hand can be substantially larger in some regions, especially at large $P_T$, and for example at $P_T = 2000$ GeV dominate the overall uncertainty of $20\%$. If a suitable combination of data measured at the Tevatron and LHC can be included in global NLO QCD analyses, the PDF uncertainties can be constrained.

The aim of this contribution is to propose a method for consistently including final-state observables in global QCD analyses.

For inclusive data like the proton structure function $F_2$ in deep-inelastic scattering (DIS) the perturbative coefficients are known analytically. During the fit the cross-section can therefore be quickly calculated from the strong coupling ($\alpha_s$) and the PDFs and can be compared to the measurements. However, final state observables, where detector acceptances or jet algorithms are involved in the definition of the perturbative coefficients (called "weights" in the following), have to be calculated using NLO Monte Carlo programs. Typically such programs need about one day of CPU time to calculate accurately the cross-section. It is therefore necessary to find a way to calculate the perturbative coefficients with high precision in a long run and to include $\alpha_s$ and the PDFs "a posteriori".

To solve this problem many methods have been proposed in the past [3,10,61–64]. In principle the highest efficiencies can be obtained by taking moments with respect to Bjorken-$x$ [61,62], because this converts convolutions into multiplications. This can have notable advantages with respect to memory consumption, especially in cases with two incoming hadrons. On the other hand, there are complications such as the need for PDFs in moment space and the associated inverse Mellin transforms.

Methods in $x$-space have traditionally been somewhat less efficient, both in terms of speed (in the 'a posteriori' steps — not a major issue here) and in terms of memory consumption. They are, however, somewhat more transparent since they provide direct information on the $x$ values of relevance. Furthermore they can be used with any PDF. The use of $x$-space methods can be further improved by using methods developed originally for PDF evolution [65,66].

[11]Contributing authors: T. Carli, G. Salam, F. Siegert.

### 8.1 PDF-independent representation of cross-sections

*Representing the PDF on a grid*

We make the assumption that PDFs can be accurately represented by storing their values on a two-dimensional grid of points and using $n^{\mathrm{th}}$-order interpolations between those points. Instead of using the parton momentum fraction $x$ and the factorisation scale $Q^2$, we use a variable transformation that provides good coverage of the full $x$ and $Q^2$ range with uniformly spaced grid points:[12]

$$y(x) = \ln\frac{1}{x} \quad \text{and} \quad \tau(Q^2) = \ln\ln\frac{Q^2}{\Lambda^2}. \tag{16}$$

The parameter $\Lambda$ is to be chosen of the order of $\Lambda_{\mathrm{QCD}}$, but not necessarily identical. The PDF $q(x, Q^2)$ is then represented by its values $q_{i_y,i_\tau}$ at the 2-dimensional grid point $(i_y\,\delta y, i_\tau\,\delta\tau)$, where $\delta y$ and $\delta\tau$ denote the grid spacings, and obtained elsewhere by interpolation:

$$q(x, Q^2) = \sum_{i=0}^{n}\sum_{\iota=0}^{n'} q_{k+i,\kappa+\iota}\; I_i^{(n)}\left(\frac{y(x)}{\delta y} - k\right) I_\iota^{(n')}\left(\frac{\tau(Q^2)}{\delta\tau} - \kappa\right), \tag{17}$$

where $n$, $n'$ are the interpolation orders. The interpolation function $I_i^{(n)}(u)$ is 1 for $u = i$ and otherwise is given by:

$$I_i^{(n)}(u) = \frac{(-1)^{n-i}}{i!(n-i)!}\frac{u(u-1)\ldots(u-n)}{u-i}. \tag{18}$$

Defining $\mathrm{int}(u)$ to be the largest integer such that $\mathrm{int}(u) \le u$, $k$ and $\kappa$ are defined as:

$$k(x) = \ \mathrm{int}\left(\tfrac{y(x)}{\delta y} - \tfrac{n-1}{2}\right), \quad \kappa(x) = \mathrm{int}\left(\frac{\tau(Q^2)}{\delta\tau} - \frac{n'-1}{2}\right). \tag{19}$$

Given finite grids whose vertex indices range from $0\ldots N_y - 1$ for the $y$ grid and $0\ldots N_\tau - 1$ for the $\tau$ grid, one should additionally require that eq. (17) only uses available grid points. This can be achieved by remapping $k \to \max(0, \min(N_y - 1 - n, k))$ and $\kappa \to \max(0, \min(N_\tau - 1 - n', \kappa))$.

*Representing the final state cross-section weights on a grid (DIS case)*

Suppose that we have an NLO Monte Carlo program that produces events $m = 1\ldots N$. Each event $m$ has an $x$ value, $x_m$, a $Q^2$ value, $Q^2_m$, as well as a weight, $w_m$, and a corresponding order in $\alpha_{\mathrm{s}}$, $p_m$. Normally one would obtain the final result $W$ of the Monte Carlo integration from:[13]

$$W = \sum_{m=1}^{N} w_m\left(\frac{\alpha_s(Q^2_m)}{2\pi}\right)^{p_m} q(x_m, Q^2_m). \tag{20}$$

Instead one introduces a weight grid $W^{(p)}_{i_y,i_\tau}$ and then for each event updates a portion of the grid with:

$i = 0\ldots n,\ \iota = 0\ldots n'$ :

$$W^{(p_m)}_{k+i,\kappa+\iota} \to W^{(p_m)}_{k+i,\kappa+\iota} + w_m\, I_i^{(n)}\left(\frac{y(x_m)}{\delta y} - k\right) I_\iota^{(n')}\left(\frac{\tau(Q^2_m)}{\delta\tau} - \kappa\right), \tag{21}$$
$$\text{where} \quad k \equiv k(x_m),\ \kappa \equiv \kappa(Q^2_m).$$

[12] An alternative for the $x$ grid is to use $y = \ln 1/x + a(1-x)$ with $a$ a parameter that serves to increase the density of points in the large $x$ region.

[13] Here, and in the following, renormalisation and factorisation scales have been set equal for simplicity.

The final result for $W$, for an arbitrary PDF, can then be obtained *subsequent* to the Monte Carlo run:

$$W = \sum_p \sum_{i_y} \sum_{i_\tau} W^{(p)}_{i_y,i_\tau} \left( \frac{\alpha_s\left(Q^{2(i_\tau)}\right)}{2\pi} \right)^p q\left(x^{(i_y)}, Q^{2(i_\tau)}\right) , \tag{22}$$

where the sums index with $i_y$ and $i_\tau$ run over the number of grid points and we have have explicitly introduced $x^{(i_y)}$ and $Q^{2(i_\tau)}$ such that:

$$y(x^{(i_y)}) = i_y\, \delta y \quad \text{and} \quad \tau\left(Q^{2(i_\tau)}\right) = i_\tau\, \delta\tau. \tag{23}$$

*Including renormalisation and factorisation scale dependence*

If one has the weight matrix $W^{(p)}_{i_y,i_\tau}$ determined separately order by order in $\alpha_\mathrm{s}$, it is straightforward to vary the renormalisation $\mu_R$ and factorisation $\mu_F$ scales a posteriori (we assume that they were kept equal in the original calculation).

It is helpful to introduce some notation relating to the DGLAP evolution equation:

$$\frac{dq(x,Q^2)}{d\ln Q^2} = \frac{\alpha_s(Q^2)}{2\pi}(P_0 \otimes q)(x,Q^2) + \left(\frac{\alpha_s(Q^2)}{2\pi}\right)^2 (P_1 \otimes q)(x,Q^2) + \ldots, \tag{24}$$

where the $P_0$ and $P_1$ are the LO and NLO matrices of DGLAP splitting functions that operate on vectors (in flavour space) $q$ of PDFs. Let us now restrict our attention to the NLO case where we have just two values of $p$, $p_\mathrm{LO}$ and $p_\mathrm{NLO}$. Introducing $\xi_R$ and $\xi_F$ corresponding to the factors by which one varies $\mu_R$ and $\mu_F$ respectively, for arbitrary $\xi_R$ and $\xi_F$ we may then write:

$$\begin{aligned} W(\xi_R,\xi_F) = \sum_{i_y}\sum_{i_\tau} & \left( \frac{\alpha_s\left(\xi_R^2 Q^{2(i_\tau)}\right)}{2\pi} \right)^{p_\mathrm{LO}} W^{(p_\mathrm{LO})}_{i_y,i_\tau} q\left(x^{(i_y)}, \xi_F^2 Q^{2(i_\tau)}\right) + \\ \left( \frac{\alpha_s\left(\xi_R^2 Q^{2(i_\tau)}\right)}{2\pi} \right)^{p_\mathrm{NLO}} & \left[ \left( W^{(p_\mathrm{NLO})}_{i_y,i_\tau} + 2\pi\beta_0 p_\mathrm{LO} \ln \xi_R^2\, W^{(p_\mathrm{LO})}_{i_y,i_\tau} \right) q\left(x^{(i_y)}, \xi_F^2 Q^{2(i_\tau)}\right) \right. \\ & \left. - \ln \xi_F^2\, W^{(p_\mathrm{LO})}_{i_y,i_\tau} (P_0 \otimes q)\left(x^{(i_y)}, \xi_F^2 Q^{2(i_\tau)}\right) \right] , \end{aligned} \tag{25}$$

where $\beta_0 = (11N_c - 2n_f)/(12\pi)$ and $N_c$ ($n_f$) is the number of colours (flavours). Though this formula is given for $x$-space based approach, a similar formula applies for moment-space approaches. Furthermore it is straightforward to extend it to higher perturbative orders.

*Representing the weights in the case of two incoming hadrons*

In hadron-hadron scattering one can use analogous procedures with one more dimension. Besides $Q^2$, the weight grid depends on the momentum fraction of the first ($x_1$) and second ($x_2$) hadron.

In the case of jet production in proton-proton collisions the weights generated by the Monte Carlo program as well as the PDFs can be organised in seven possible initial state combinations of partons:

$$\begin{aligned} \mathrm{gg}: \quad F^{(0)}(x_1,x_2;Q^2) &= G_1(x_1)G_2(x_2) & (26)\\ \mathrm{qg}: \quad F^{(1)}(x_1,x_2;Q^2) &= \left(Q_1(x_1) + \overline{Q}_1(x_1)\right) G_2(x_2) & (27)\\ \mathrm{gq}: \quad F^{(2)}(x_1,x_2;Q^2) &= G_1(x_1)\left(Q_2(x_2) + \overline{Q}_2(x_2)\right) & (28)\\ \mathrm{qr}: \quad F^{(3)}(x_1,x_2;Q^2) &= Q_1(x_1)Q_2(x_2) + \overline{Q}_1(x_1)\overline{Q}_2(x_2) - D(x_1,x_2) & (29)\\ \mathrm{qq}: \quad F^{(4)}(x_1,x_2;Q^2) &= D(x_1,x_2) & (30) \end{aligned}$$

$$\begin{aligned} \mathrm{q\bar{q}}: \quad F^{(5)}(x_1, x_2; Q^2) &= \overline{D}(x_1, x_2) & (31) \\ \mathrm{q\bar{r}}: \quad F^{(6)}(x_1, x_2; Q^2) &= Q_1(x_1)\overline{Q}_2(x_2) + \overline{Q}_1(x_1)Q_2(x_2) - \overline{D}(x_1, x_2), & (32) \end{aligned}$$

where $g$ denotes gluons, $q$ quarks and $r$ quarks of different flavour $q \neq r$ and we have used the generalized PDFs defined as:

$$G_H(x) = f_{0/H}(x, Q^2), \qquad Q_H(x) = \sum_{i=1}^{6} f_{i/H}(x, Q^2), \;\; \overline{Q}_H(x) = \sum_{i=-6}^{-1} f_{i/H}(x, Q^2),$$

$$D(x_1, x_2) = \sum_{\substack{i=-6 \\ i\neq 0}}^{6} f_{i/H_1}(x_1, Q^2) f_{i/H_2}(x_2, Q^2), \tag{33}$$

$$\overline{D}(x_1, x_2, \mu_F^2) = \sum_{\substack{i=-6 \\ i\neq 0}}^{6} f_{i/H_1}(x_1, Q^2) f_{-i/H_2}(x_2, Q^2),$$

where $f_{i/H}$ is the PDF of flavour $i = -6 \ldots 6$ for hadron $H$ and $H_1$ ($H_2$) denotes the first or second hadron[14].

The analogue of eq. 22 is then given by:

$$W = \sum_p \sum_{l=0}^{6} \sum_{i_{y_1}} \sum_{i_{y_2}} \sum_{i_\tau} W^{(p)(l)}_{i_{y_1}, i_{y_2}, i_\tau} \left( \frac{\alpha_s \left( Q^{2(i_\tau)} \right)}{2\pi} \right)^p F^{(l)} \left( x_1^{(i_{y_1})}, x_2^{(i_{y_1})}, Q^{2(i_\tau)} \right). \tag{34}$$

*Including scale depedence in the case of two incoming hadrons*

It is again possible to choose arbitrary renormalisation and factorisation scales, specifically for NLO accuracy:

$$\begin{aligned} W(\xi_R, \xi_F) = & \sum_{l=0}^{6} \sum_{i_{y_1}} \sum_{i_{y_2}} \sum_{i_\tau} \left( \frac{\alpha_s \left( \xi_R^2 Q^{2(i_\tau)} \right)}{2\pi} \right)^{p_{\mathrm{LO}}} W^{(p_{\mathrm{LO}})(l)}_{i_{y_1}, i_{y_2}, i_\tau} F^{(l)} \left( x_1^{(i_{y_1})}, x_2^{(i_{y_1})}, \xi_F^2 Q^{2(i_\tau)} \right) + \\ & \left( \frac{\alpha_s \left( \xi_R^2 Q^{2(i_\tau)} \right)}{2\pi} \right)^{p_{\mathrm{NLO}}} \left[ \left( W^{(p_{\mathrm{NLO}})(l)}_{i_{y_1}, i_{y_2}, i_\tau} + 2\pi\beta_0 p_{\mathrm{LO}} \ln \xi_R^2 \, W^{(p_{\mathrm{LO}})(l)}_{i_{y_1}, i_{y_2}, i_\tau} \right) F^{(l)} \left( x_1^{(i_{y_1})}, x_2^{(i_{y_1})}, \xi_F^2 Q^{2(i_\tau)} \right) \right. \\ & \left. - \ln \xi_F^2 \, W^{(p_{\mathrm{LO}})(l)}_{i_{y_1}, i_{y_2}, i_\tau} \left( F^{(l)}_{q_1 \to P_0 \otimes q_1} \left( x_1^{(i_{y_1})}, x_2^{(i_{y_1})}, \xi_F^2 Q^{2(i_\tau)} \right) + F^{(l)}_{q_2 \to P_0 \otimes q_2} \left( x_1^{(i_{y_1})}, x_2^{(i_{y_1})}, \xi_F^2 Q^{2(i_\tau)} \right) \right) \right], \end{aligned} \tag{35}$$

where $F^{(l)}_{q_1 \to P_0 \otimes q_1}$ is calculated as $F^{(l)}$, but with $q_1$ replaced wtih $P_0 \otimes q_1$, and analogously for $F^{(l)}_{q_2 \to P_0 \otimes q_2}$.

## 8.2 Technical implementation

To test the scheme discussed above we use the NLO Monte Carlo program NLOJET++ [67] and the CTEQ6 PDFs [4]. The grid $W^{(p)(l)}_{i_{y_1}, i_{y_2}, i_\tau}$ of eq. 34 is filled in a NLOJET++ user module. This module has access to the event weight and parton momenta and it is here that one specifies and calculates the physical observables that are being studied (e.g. jet algorithm).

Having filled the grid we construct the cross-section in a small standalone program which reads the weights from the grid and multiplies them with an arbitrary $\alpha_\mathrm{s}$ and PDF according to eq. 34. This program runs very fast (in the order of seconds) and can be called in a PDF fit.

[14] In the above equation we follow the standard PDG Monte Carlo numbering scheme [17] where gluons are denoted as 0, quarks have values from 1-6 and anti-quarks have the corresponding negative values.

The connection between these two programs is accomplished via a C++ class, which provides methods e.g. for creating and optimising the grid, filling weight events and saving it to disk. The classes are general enough to be extendable for the use with other NLO calculations.

The complete code for the NLOJET++ module, the C++ class and the standalone job is available from the authors. It is still in a development, testing and tuning stage, but help and more ideas are welcome.

*The C++ class*

The main data members of this class are the grids implemented as arrays of three-dimensional ROOT histograms, with each grid point at the bin centers[15]:

$$\mathrm{TH3D[p][l][iobs](x_1, x_2, Q^2)}, \tag{36}$$

where the $l$ and $p$ are explained in eq. 34 and $iobs$ denotes the observable bin, e.g. a given $P_T$ range[16].

The C++ class initialises, stores and fills the grid using the following main methods:

- *Default constructor:* Given the pre-defined kinematic regions of interest, it initializes the grid.
- *Optimizing method:* Since in some bins the weights will be zero over a large kinematic region in $x_1, x_2, Q^2$, the optimising method implements an automated procedure to adapt the grid boundaries for each observable bin. These boundaries are calculated in a first (short) run. In the present implementation, the optimised grid has a fixed number of grid points. Other choices, like a fixed grid spacing, might be implemented in the future.
- *Loading method:* Reads the saved weight grid from a ROOT file
- *Saving method:* Saves the complete grid to a ROOT file, which will be automatically compressed.

*The user module for NLOJET++*

The user module has to be adapted specifically to the exact definition of the cross-section calculation. If a grid file already exists in the directory where NLOJET++ is started, the grid is not started with the default constructor, but with the optimizing method (see 8.2). In this way the grid boundaries are optimised for each observable bin. This is necessary to get very fine grid spacings without exceeding the computer memory. The grid is filled at the same place where the standard NLOJET++ histograms are filled. After a certain number of events, the grid is saved in a root-file and the calculation is continued.

*The standalone program for constructing the cross-section*

The standalone program calculates the cross-section in the following way:

1. Load the weight grid from the ROOT file
2. Initialize the PDF interface[17], load $q(x, Q^2)$ on a helper PDF-grid (to increase the performance)
3. For each observable bin, loop over $i_{y_1}, i_{y_2}, i_\tau, l, p$ and calculate $F^l(x_1, x_2, Q^2)$ from the appropriate PDFs $q(x, Q^2)$, multiply $\alpha_s$ and the weights from the grid and sum over the initial state parton configuration $l$, according to eq. 34.

[15]ROOT histograms are easy to implement, to represent and to manipulate. They are therefore ideal in an early development phase. An additional advantage is the automatic file compression to save space. The overhead of storing some empty bins is largely reduced by optimizing the $x_1$, $x_2$ and $Q^2$ grid boundaries using the NLOJET++ program before final filling. To avoid this residual overhead and to exploit certain symmetries in the grid, a special data class (e.g. a sparse matrix) might be constructed in the future.

[16]For the moment we construct a grid for each initial state parton configuration. It will be easy to merge the $qg$ and the $gq$ initial state parton configurations in one grid. In addition, the weights for some of the initial state parton configurations are symmetric in $x_1$ and $x_2$. This could be exploited in future applications to further reduce the grid size.

[17]We use the C++ wrapper of the LHAPDF interface [56].

### 8.3 Results

We calculate the single inclusive jet cross-section as a function of the jet transverse momentum ($P_T$) for jets within a rapidity of $|y| < 0.5$. To define the jets we use the seedless cone jet algorithm as implemented in NLOJET++ using the four-vector recombination scheme and the midpoint algorithm. The cone radius has been put to $R = 0.7$, the overlap fraction was set to $f = 0.5$. We set the renormalisation and factorization scale to $Q^2 = P_{T,max}^2$, where $P_{T,max}$ is the $P_T$ of the highest $P_T$ jet in the required rapidity region[18].

In our test runs, to be independent from statistical fluctuations (which can be large in particular in the NLO case), we fill in addition to the grid a reference histogram in the standard way according to eq. 20.

The choice of the grid architecture depends on the required accuracy, on the exact cross-section definition and on the available computer resources. Here, we will just sketch the influence of the grid architecture and the interpolation method on the final result. We will investigate an example where we calculate the inclusive jet cross-section in $N_{\rm obs} = 100$ bins in the kinematic range $100 \leq P_T \leq 5000\,\mathrm{GeV}$. In future applications this can serve as guideline for a user to adapt the grid method to his/her specific problem. We believe that the code is transparent and flexible enough to adapt to many applications.

As reference for comparisons of different grid architectures and interpolation methods we use the following:

– *Grid spacing in $y(x)$:* $10^{-5} \leq x_1, x_2 \leq 1.0$ with $N_y = 30$
– *Grid spacing in $\tau(Q^2)$:* $100\,\mathrm{GeV} \leq Q \leq 5000\,\mathrm{GeV}$ with $N_\tau = 30$
– *Order of interpolation:* $n_y = 3$, $n_\tau = 3$

The grid boundaries correspond to the user setting for the first run which determines the grid boundaries for each observable bin. In the following we call this grid architecture $30^2$x30x100$(3,3)$. Such a grid takes about 300 Mbyte of computer memory. The root-file where the grid is stored has about 50 Mbyte.

The result is shown in Fig. 27a). The reference cross-section is reproduced everywhere to within $0.05\%$. The typical precision is about $0.01\%$. At low and high $P_T$ there is a positive bias of about $0.04\%$. Also shown in Fig. 27a) are the results obtained with different grid architectures. For a finer $x$ grid ($50^2$x30x100$(3,3)$) the accuracy is further improved (within $0.005\%$) and there is no bias. A finer ($30^2$x60x100$(3,3)$) as well as a coarser ($30^2$x10x100$(3,3)$) binning in $Q^2$ does not improve the precision.

Fig. 27b) and Fig. 27c) show for the grid ($30^2$x30x100) different interpolation methods. With an interpolation of order $n = 5$ the precision is $0.01\%$ and the bias at low and high $P_T$ observed for the $n = 3$ interpolation disappears. The result is similar to the one obtained with finer $x$-points. Thus by increasing the interpolation order the grid can be kept smaller. An order $n = 1$ interpolation gives a systematic negative bias of about $1\%$ becoming even larger towards high $P_T$.

Depending on the available computer resources and the specific problem, the user will have to choose a proper grid architecture. In this context, it is interesting that a very small grid $10^2$x10x100$(5,5)$ that takes only about 10 Mbyte computer memory reaches still a precision of $0.5\%$, if an interpolation of order $n = 5$ is used (see Fig. 27d)).

We have developed a technique to store the perturbative coefficients calculated by an NLO Monte Carlo program on a grid allowing for a-posteriori inclusion of an arbitrary parton density function (PDF) set. We extended a technique that was already successfully used to analyse HERA data to the more demanding case of proton-proton collisions at LHC energies.

[18]Note that beyond LO the $P_{T,max}$ will in general differ from the $P_T$ of the other jets, so when binning an inclusive jet cross section, the $P_T$ of a given jet may not correspond to the renormalisation scale chosen for the event as a whole. For this reason we shall need separate grid dimensions for the jet $P_T$ and for the renormalisation scale. Only in certain moment-space approaches [62] has this requirement so far been efficiently circumvented.

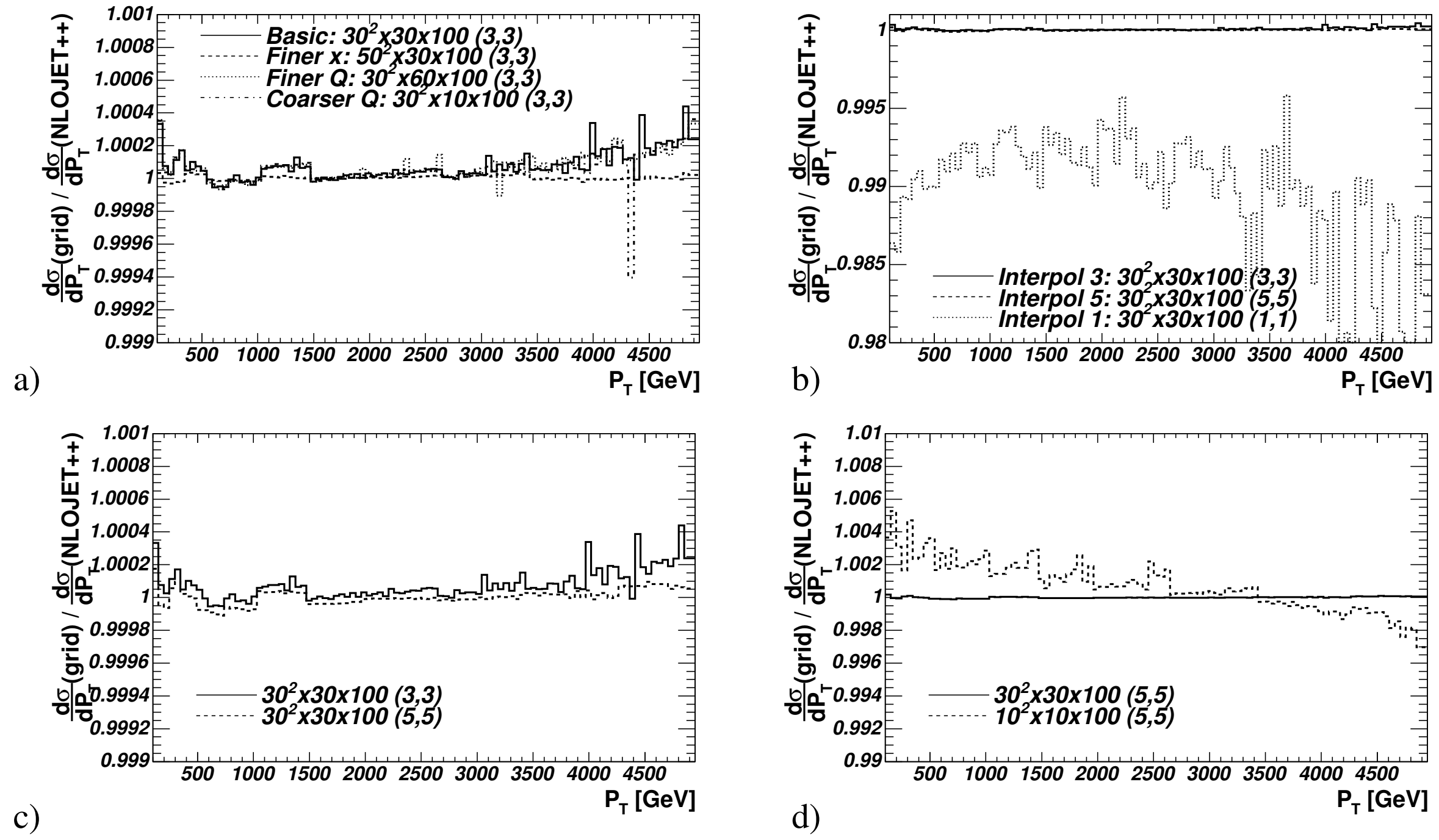

**Fig. 27:** Ratio between the single inclusive jet cross-section with 100 $P_T$ bins calculated with the grid technique and the reference cross-section calculated in the standard way. Shown are the standard grid, grids with finer $x$ and $Q^2$ sampling (a) with interpolation of order 1, 3 and 5 (b) (and on a finer scale in c)) and a small grid (d).

The technique can be used to constrain PDF uncertainties, e.g. at high momentum transfers, from data that will be measured at LHC and allows the consistent inclusion of final state observables in global QCD analyses. This will help increase the sensitivity of LHC to find new physics as deviations from the Standard Model predictions.

Even for the large kinematic range for the parton momentum fractions $x_1$ and $x_2$ and of the squared momentum transfer $Q^2$ accessible at LHC, grids of moderate size seem to be sufficient. The single inclusive jet cross-section in the central region $|y| < 0.5$ can be calculated with a precision of $0.01\%$ in a realistic example with 100 bins in the transverse jet energy range $100 \leq P_T \leq 5000\,\mathrm{GeV}$. In this example, the grid occupies about 300 Mbyte computer memory. With smaller grids of order 10 Mbyte the reachable accuracy is still $0.5\%$. This is probably sufficient for all practical applications.

# DGLAP evolution and parton fits

*S. I. Alekhin, J. Blümlein, H. Böttcher, L. Del Debbio, S. Forte, A. Glazov, A. Guffanti, J. Huston, G. Ingelman J. I. Latorre, S. Moch, A. Piccione, J. Pumplin, V. Ravindran, J. Rojo G.P. Salam, R.S. Thorne, J.A.M. Vermaseren, A. Vogt*

## 1 DGLAP evolution and parton fits [1]

### 1.1 Introduction

The high-precision data from HERA and the anticipated data from LHC open the possibility for a precise determination of parton distributions. This, however, requires an improvement in the theoretical description of DIS and hard hadronic scattering processes, as well as an improvement of the techniques used to extract parton distributions from the data.

The determination of perturbative QCD corrections has undergone substantial progress recently. The key ingredient of a complete next-to-next-to-leading order (NNLO) prediction in perturbative QCD are the recently calculated three-loop splitting functions which govern the scale dependence of PDFs. Extensions in the accuracy of the perturbative predictions yet beyond NNLO are given by the three-loop coefficient functions for $F_2$, while the coefficient functions for $F_L$ at this order are actually required to complete the NNLO predictions. Section 2 briefly discusses the recent results and their phenomenological implications. Certain mathematical aspects, which are important in the calculation of higher order corrections in massless QCD are presented in section 3. In particular, algebraic relations in Mellin-$N$ space are pointed out, which are of importance for harmonic sums, harmonic polylogarithms and multiple $\zeta$-values.

These calculation of the PDF evolution to NNLO in perturbative QCD are used in section 4 to provide an update and extension of a set of benchmark tables for the evolution of parton distributions of hadrons. These benchmark tables were first presented in the report of the QCD/SM working group at the 2001 Les Houches workshop, but based on approximate NNLO splitting functions, which are superseded by the exact results which are now available. In addition, section 4 now includes also reference tables for the case of polarized PDF evolution.

Whereas in principle the $x$-shapes of PDFs at low scales can be determined from first principles using non-perturbative methods, in practice at present this is only possible using models (briefly touched in in section 5). Therefore, an accurate determination of PDFs requires a global QCD fit to the data, which is the subject of sections 6–8.

Section 6 discusses in particular the impact on parton fits of NNLO corrections on the one hand, and of the inclusion of Drell-Yan data and future LHC data on the other hand. It then presents values for a benchmark fit together with a table of correlation coefficients for the parameter obtained in the fit. This benchmark fit is then re-examined in sec. 7, along with a comparison between PDFs and the associated uncertainty obtained using the approaches of Alekhin and the MRST group. The differences between these benchmark partons and the actual global fit partons are also discussed, and used to explore complications inherent in extracting PDFs with uncertainties. Finally, in section 8 the stability of PDF determinations in NLO global analyses is re-investigated and the results of the CTEQ PDF group on this issue are summarized.

An alternative approach to a completely bias-free parameterization of PDFs is presented in section 9. There, a neural network approach to global fits of parton distribution functions is introduced and work on unbiased parameterizations of deep-inelastic structure functions with faithful estimation of their uncertainties is reviewed together with a summary of the current status of neural network parton distribution fits.

[1]Subsection coordinators: A. Glazov, S. Moch

## 2 Precision Predictions for Deep-Inelastic Scattering [2]

With high-precision data from HERA and in view of the outstanding importance of hard scattering processes at the LHC, a quantitative understanding of deep-inelastic processes is indispensable, necessitating calculations beyond the standard next-to-leading order of perturbative QCD.

In this contribution we review recent results for the complete next-to-next-to-leading order (NNLO, N$^2$LO) approximation of massless perturbative QCD for the structure functions $F_1$, $F_2$, $F_3$ and $F_L$ in DIS. These are based on the second-order coefficient functions [1–5], the three-loop splitting functions which govern the evolution of unpolarized parton distributions of hadrons [6, 7] and the three-loop coefficient functions for $F_L = F_2 - 2xF_1$ in electromagnetic (photon-exchange) DIS [8, 9]. Moreover we discuss partial N$^3$LO results for $F_2$, based on the corresponding three-loop coefficient functions also presented in Ref. [9]. For the splitting functions $P$ and coefficient functions $C$ we employ the convention

$$P(\alpha_s) = \sum_{n=0} \left(\frac{\alpha_s}{4\pi}\right)^{n+1} P^{(n)}, \qquad C(\alpha_s) = \sum_{n=0} \left(\frac{\alpha_s}{4\pi}\right)^{n} C^{(n)} \tag{1}$$

for the expansion in the running coupling constant $\alpha_s$. For the longitudinal structure function $F_L$ the third-order corrections are required to complete the NNLO predictions, since the leading contribution to the coefficient function $C_L$ is of first order in the strong coupling constant $\alpha_s$.

In the following we briefly display selected results to demonstrate the quality of precision predictions for DIS and their effect on the evolution. The exact (analytical) results to third order for the quantities in Eq. (1) are too lengthy, about $\mathcal{O}(100)$ pages in normalsize fonts and will not be reproduced here. Also the method of calculation is well documented in the literature [5–7, 9–11]. In particular, it proceeds via the Mellin transforms of the functions of the Bjorken variable $x$,

$$A(N) = \int_0^1 dx\, x^{N-1} A(x)\,. \tag{2}$$

Selected mathematical aspects of Mellin transforms are discussed in section 3.

### 2.1 Parton evolution

The well-known $2n_f - 1$ scalar non-singlet and $2 \times 2$ singlet evolution equations for $n_f$ flavors read

$$\frac{d}{d\ln\mu_f^2}\, q_{\rm ns}^i = P_{\rm ns}^i \otimes q_{\rm ns}^i\,, \qquad i = \pm, {\rm v}\,, \tag{3}$$

for the quark flavor asymmetries $q_{\rm ns}^{\pm}$ and the valence distribution $q_{\rm ns}^{\rm v}$, and

$$\frac{d}{d\ln\mu_f^2}\begin{pmatrix} q_{\rm s} \\ g \end{pmatrix} = \begin{pmatrix} P_{\rm qq} & P_{\rm qg} \\ P_{\rm gq} & P_{\rm gg} \end{pmatrix} \otimes \begin{pmatrix} q_{\rm s} \\ g \end{pmatrix} \tag{4}$$

for the singlet quark distribution $q_{\rm s}$ and the gluon distribution $g$, respectively. Eqs. (3) and (4) are governed by three independent types of non-singlet splitting functions, and by the $2 \times 2$ matrix of singlet splitting functions. Here $\otimes$ stands for the Mellin convolution. We note that benchmark numerical solutions to NNLO accuracy of Eqs. (3) and (4) for a specific set of input distributions are given in section 4. Phenomenological QCD fits of parton distributions in data analyses are extensively discussed in sections 6–8. An approach based on neural networks is described in section 9.

Let us start the illustration of the precision predictions by looking at the parton evolution and at large Mellin-$N$ (large Bjorken-$x$) behavior. Fig. 1 shows the stability of the perturbative expansion which is very benign and indicates, for $\alpha_{\mathrm{s}} \lesssim 0.2$, corrections of less than 1% beyond NNLO. On the left we show the results for the perturbative expansion of $P_{\mathrm{ns}}$ in Mellin space, cf. Eqs. (1), (2). We employ four active flavors, $n_f = 4$, and an order-independent value for the strong coupling constant,

$$\alpha_{\mathrm{s}}(\mu_0^2) \; = \; 0.2 \; , \tag{5}$$

which corresponds to $\mu_0^2 \simeq 25 \ldots 50\,\mathrm{GeV}^2$ for $\alpha_{\mathrm{s}}(M_Z^2) = 0.114 \ldots 0.120$ beyond the leading order. On the right of Fig. 1 the perturbative expansion of the logarithmic derivative, cf. Eqs. (1), (3), is illustrated at the standard choice $\mu_r = \mu_f$ of the renormalization scale. We use the schematic, but characteristic model distribution,

$$xq_{\mathrm{ns}}(x, \mu_0^2) \;\; = \;\; x^{0.5}(1-x)^3 \, . \tag{6}$$

[2] Contributing authors: S. Moch, J.A.M. Vermaseren, A. Vogt

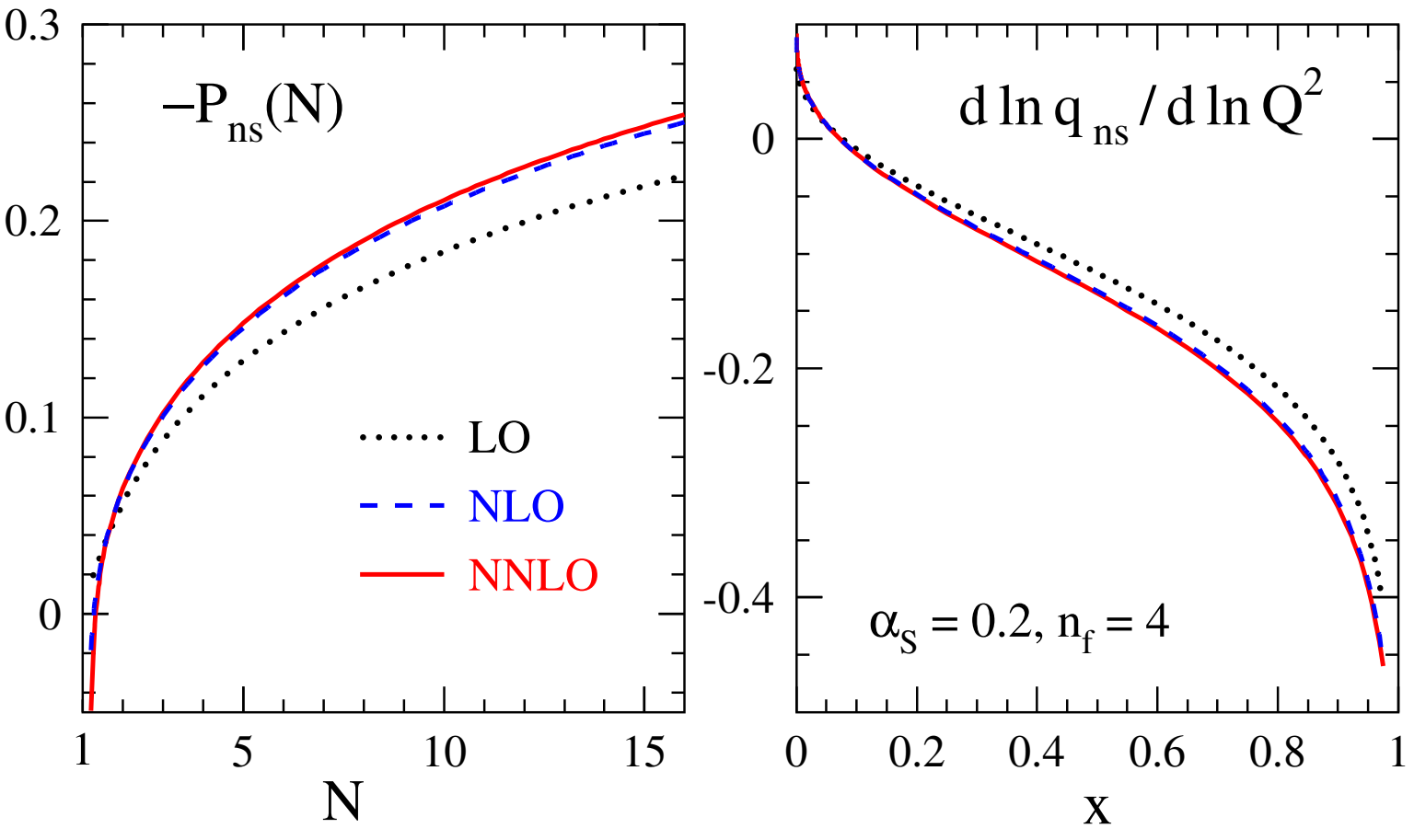

**Fig. 1:** On the left we show the perturbative expansion of $P^{\mathrm{v}}_{\mathrm{ns}}(N)$, and on the right the resulting perturbative expansion of the logarithmic scale derivative $d \ln q_{\mathrm{ns}}/d \ln \mu_f^2$ is displayed for a model input. See the text for details.

The normalization of $q_{\mathrm{ns}}$ is irrelevant at this point, as we consider the logarithmic scale derivative only.

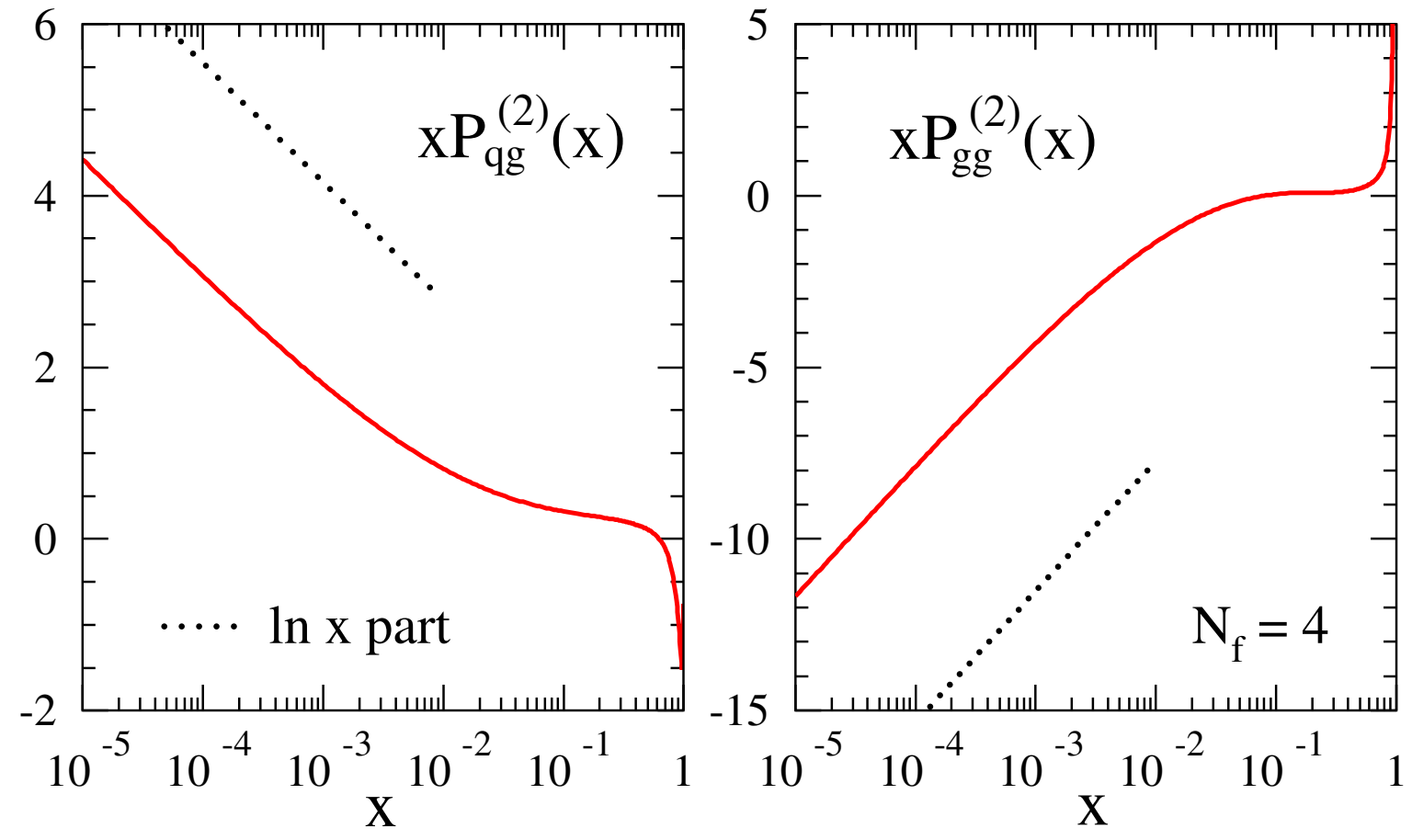

**Fig. 2:** The three-loop gluon-quark (left) and gluon-gluon (right) splitting functions together with the leading small-$x$ contribution (dotted line).

Next, let us focus on the three-loop splitting functions at small momentum fractions $x$, where the splitting functions $P_{\mathrm{ig}}$ in the lower row of the $2 \times 2$ matrix in Eq. (4), representing $\mathrm{g} \to \mathrm{i}$ splittings, are most important. In Fig. 2 we show, again for $n_f = 4$, the three-loop splitting functions $P^{(2)}_{\mathrm{qg}}$ and $P^{(2)}_{\mathrm{gg}}$ together with the leading small-$x$ term indicated separately for $x < 0.01$. In the present singlet case the leading logarithmic small-$x$ limits $\sim x^{-1} \ln x$ of Refs. [12, 13] are confirmed together with the general structure of the BFKL limit [14–16]. The same holds for the leading small-$x$ terms $\ln^4 x$ in the non-singlet sector [17, 18], with the qualification that a new, unpredicted leading logarithmic contribution is found for the color factor $d^{abc}d_{abc}$ entering at three loops for the first time.

It is obvious from Fig. 2 (see also Refs. [5–7, 11]) that the leading $x \to 0$-terms alone are insufficient for collider phenomenology at HERA or the LHC as they do not provide good approximations of the full results at experimentally relevant small values of $x$. Resummation of the small-$x$ terms and various phenomenological improvements are discussed in detail in [19].

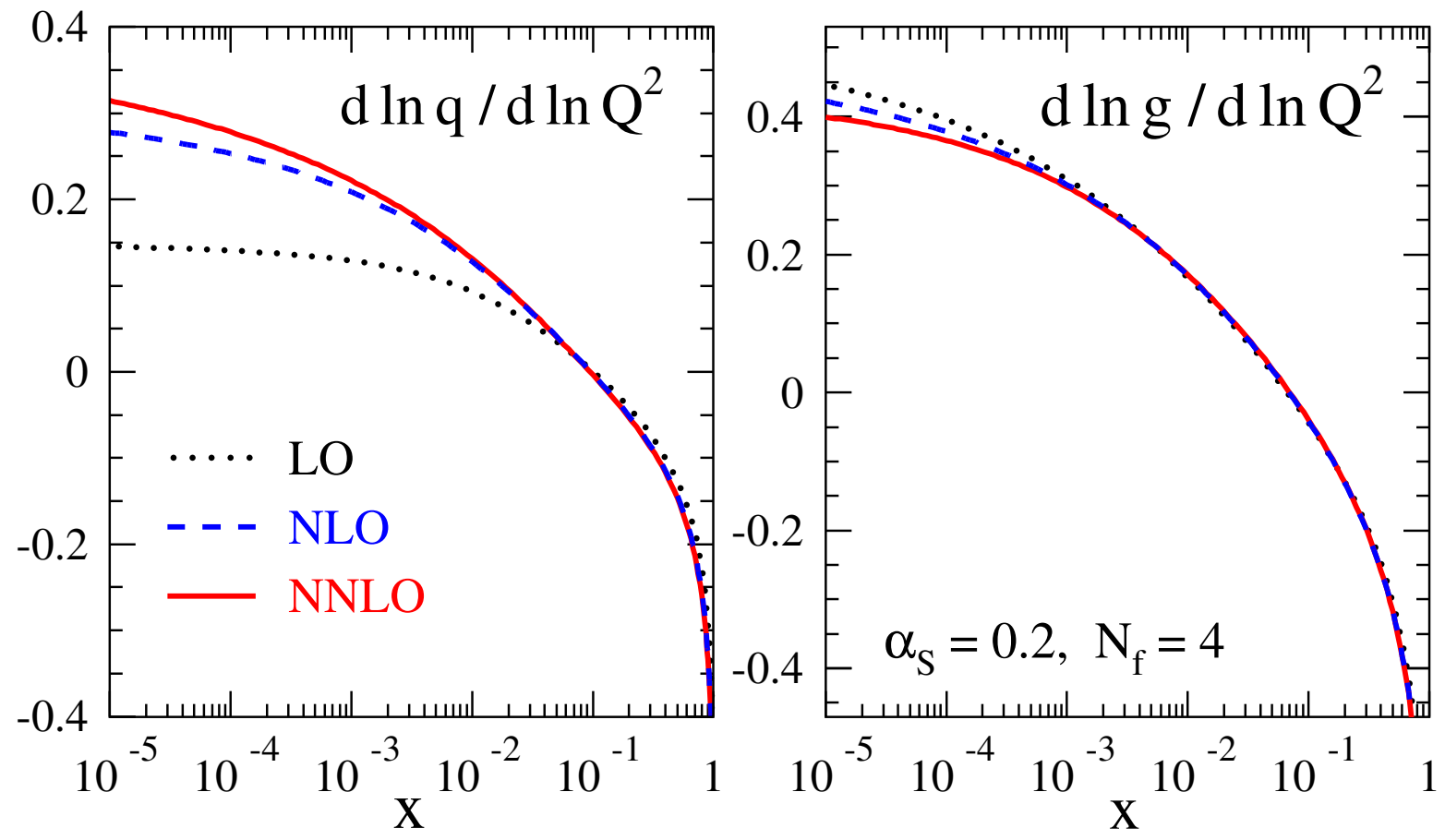

**Fig. 3:** The perturbative expansion of the scale derivatives (4) of the singlet distributions (7).

In the same limit of small $x$, it is instructive to look at the evolution of parton distributions. Again, we choose the reference scale of Eq. (5), $n_f = 4$ and the sufficiently realistic model distributions

$$\begin{aligned} xq_{\rm s}(x,\mu_0^2) &= 0.6\, x^{-0.3}(1-x)^{3.5}\,(1+5.0\, x^{0.8}) \\ xg(x,\mu_0^2) &= 1.6\, x^{-0.3}(1-x)^{4.5}\,(1-0.6\, x^{0.3}) \end{aligned} \qquad (7)$$

irrespective of the order of the expansion to facilitate direct comparisons of the various contributions. Of course, this order-independence does not hold for actual data-fitted parton distributions like those in sections 6–8. In Fig. 3 we display the perturbative expansion of the scale derivative for the singlet quark and gluon densities at $\mu_f^2 = \mu_0^2$ for the initial conditions specified in Eqs. (5) and (7). For the singlet quark distribution the total NNLO corrections, while reaching 10% at $x = 10^{-4}$, remain smaller than the NLO results by a factor of eight or more over the full $x$-range. For the gluon distribution already the NLO corrections are small and the NNLO contribution amounts to only 3% for $x$ as low as $10^{-4}$. Thus, we see in Fig. 3 that the perturbative expansion is very stable. It appears to converge rapidly at $x > 10^{-3}$, while relatively large third-order corrections are found for very small momenta $x \lesssim 10^{-4}$.

### 2.2 Coefficient functions

While the previous considerations were addressing the evolution of parton distributions, we now turn to the further improvements of precision predictions due to the full third-order coefficient functions for the structure functions $F_2$ and $F_L$ in electromagnetic DIS [8, 9]. The results for $F_L$ complete the NNLO description of unpolarized electromagnetic DIS, and the third-order coefficient functions for $F_2$ form, at not too small values of the Bjorken variable $x$, the dominant part of the next-to-next-to-next-to-leading order (N$^3$LO) corrections. Thus, they facilitate improved determinations of the strong coupling $\alpha_{\rm s}$ from scaling violations.

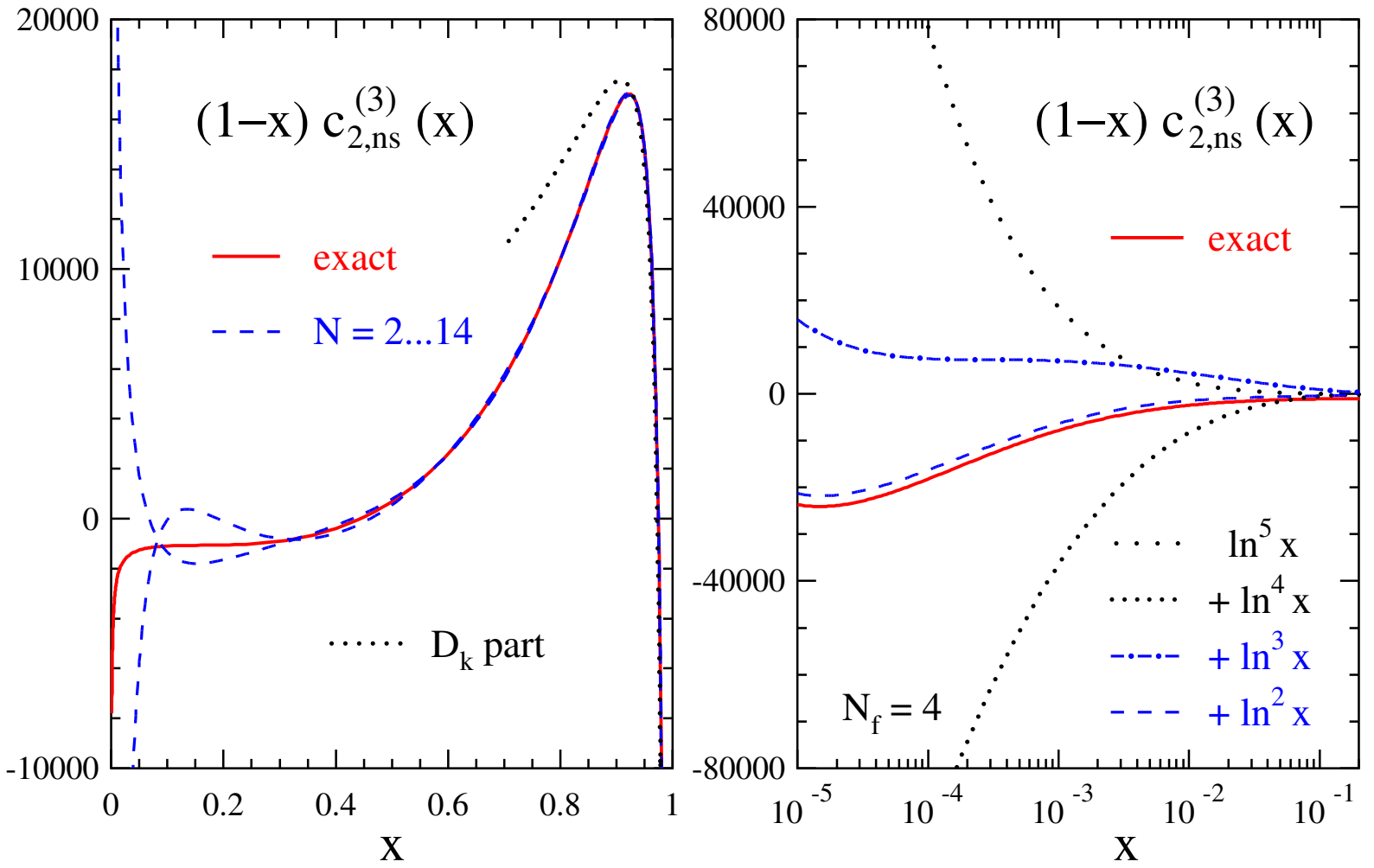

**Fig. 4:** The three-loop non-singlet coefficient function $c^{(3)}_{2,\text{ns}}(x)$ in the large-$x$ (left) and the small-$x$ (right) region, multiplied by $(1-x)$ for display purposes.

Let us start with the three-loop coefficient functions for $F_2$ in the non-singlet case. In Fig. 4 we display the three-loop non-singlet coefficient function $c^{(3)}_{2,\text{ns}}(x)$ for $n_f = 4$ flavors. We also show the soft-gluon enhanced terms $\mathcal{D}_k$ dominating the large-$x$ limit,

$$\mathcal{D}_k \;=\; \frac{\ln^{2k-1}(1-x)}{(1-x)_+}\,, \tag{8}$$

and the small-$x$ approximations obtained by successively including enhanced logarithms $\ln^k x$. However the latter are insufficient for an accurate description of the exact result. The dashed band in Fig. 4 shows the uncertainty of previous estimates [20] mainly based on the calculation of fixed Mellin moments [21–23]. For a detailed discussion of the soft-gluon resummation of the the $\mathcal{D}_k$ terms, we refer to [19].

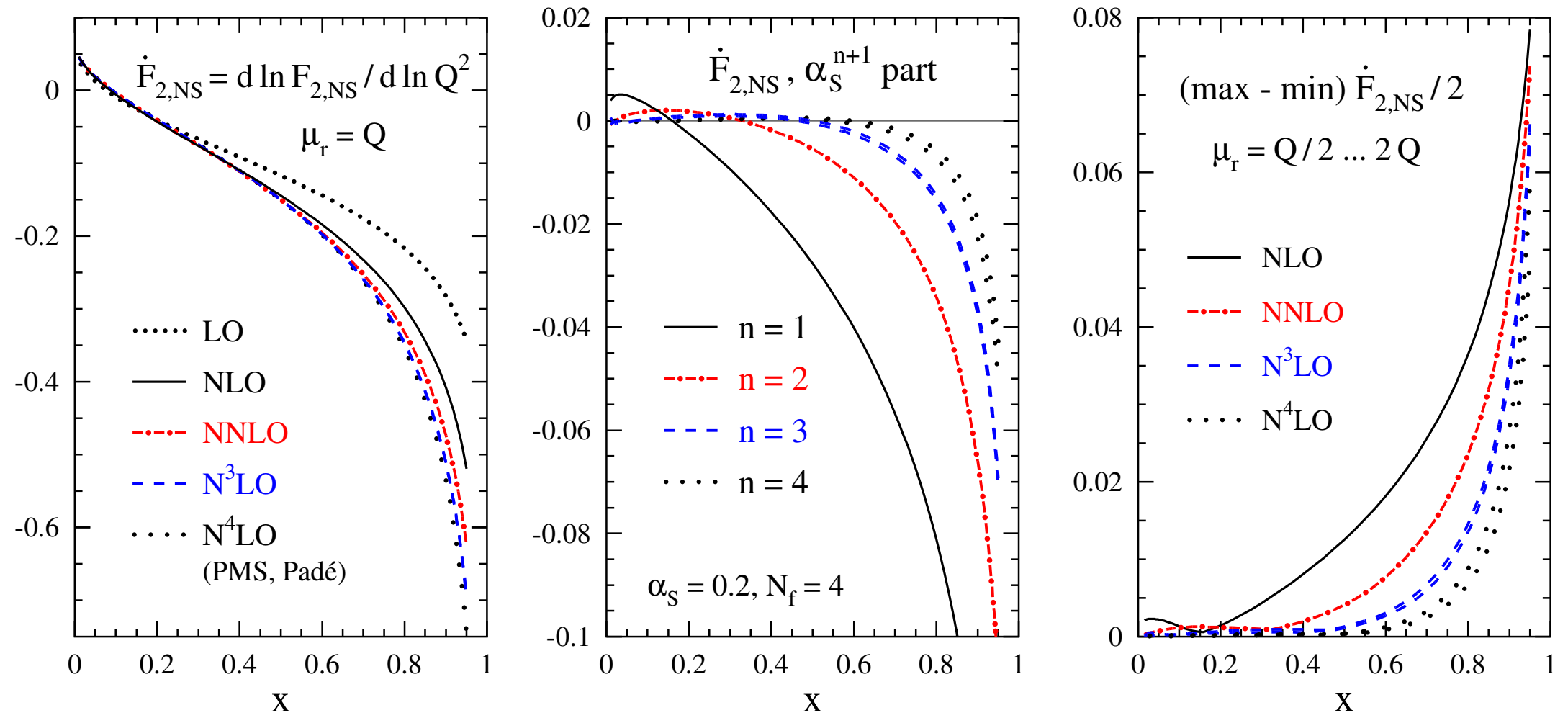

**Fig. 5:** The perturbative expansion of the logarithmic scale derivative of the non-singlet structure function $F_{2,\text{ns}}$. The results up to NNLO are exact, while those at N$^3$LO are very good approximations. The N$^4$LO corrections have been estimated by various methods.

Building on the coefficient functions, it is interesting to study the perturbative expansion of the logarithmic scale derivative for the non-singlet structure function $F_{2,\text{ns}}$. To that end we use in Fig. 5 again the input shape Eq. (6) (this time for $F_{2,\mathrm{ns}}$ itself) irrespective of the order of the expansion, $n_f = 4$ flavors and the reference scale of Eq. (5). The N$^4$LO approximation based on Padé summations of the perturbation series can be expected to correctly indicate at least the rough size of the four-loop corrections, see Ref. [9] for details. From Fig. 5 we see that the three-loop results for $F_2$ can be employed to effectively extend the main part of DIS analyses to the N$^3$LO at $x > 10^{-2}$ where the effect of the unknown fourth-order splitting functions is expected to be very small. This has, for example, the potential for a 'gold-plated' determination of $\alpha_s(M_Z)$ with an error of less than 1% from the truncation of the perturbation series. On the right hand side of Fig. 5 the scale uncertainty which is conventionally estimated by

$$\Delta\dot{f} \quad \equiv \quad \frac{1}{2}\left(\max\left[\dot{f}(x,\mu_r^2)\right] - \min\left[\dot{f}(x,\mu_r^2)\right]\right) , \tag{9}$$

is plotted, where the scale varies $\mu_r \in [Q/2, 2Q]$.

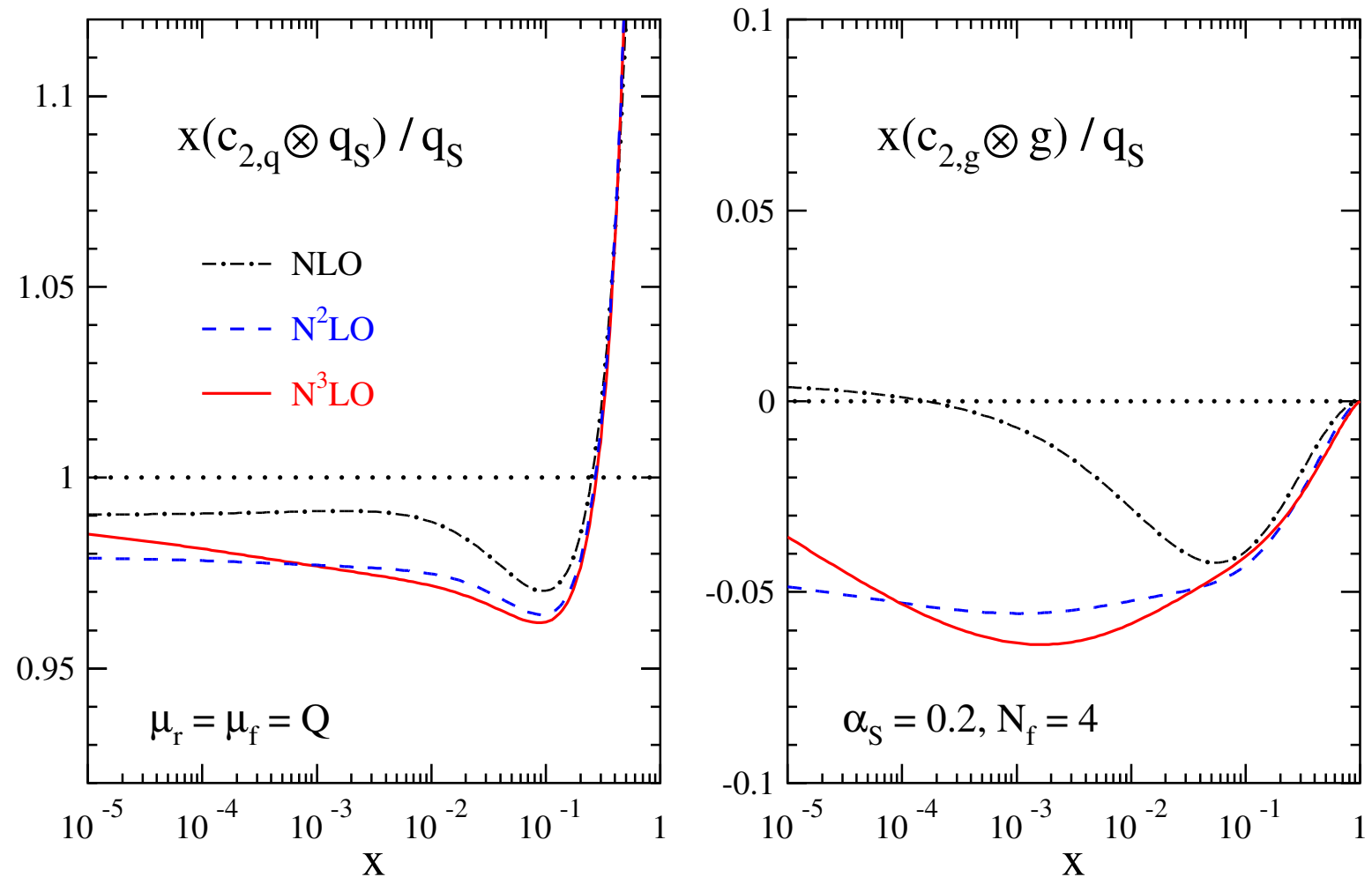

**Fig. 6:** The perturbative expansion up to three loops (N$^3$LO) of the quark (left) and gluon (right) contributions to singlet structure function $F_2$.

In the singlet case, we can study the quark and gluon contributions to the structure function $F_2$. In Fig. 6 we plot the perturbative expansion up to N$^3$LO of the quark and gluon contributions to structure function $F_{2,\mathrm{s}}$ at the scale (5) using the distributions (7). All curves have been normalized to the leading-order result $F_{2,\mathrm{s}}^{\mathrm{LO}} = \langle e^2 \rangle \, q_{\mathrm{s}}$ . Fig. 6 nicely illustrates the perturbative stability of the structure function $F_2$.

Finally, we address the longitudinal structure function $F_L$ at three loops. In the left part of Fig. 7 we plot the singlet-quark and gluon coefficient functions $c_{L,\mathrm{q}}$ and $c_{L,\mathrm{g}}$ for $F_L$ up to the third order for four flavors and the $\alpha_{\mathrm{s}}$-value of Eq. (5). The curves have been divided by $a_{\mathrm{s}} = \alpha_s/(4\pi)$ to account for the leading contribution being actually of first order in the strong coupling constant $\alpha_{\mathrm{s}}$. Both the second-order and the third-order contributions are rather large over almost the whole $x$-range. Most striking, however, is the behavior at very small values of $x$, where the anomalously small one-loop parts are negligible against the (negative) constant two-loop terms, which in turn are completely overwhelmed by the (positive) new three-loop corrections $xc^{(3)}_{L,a} \sim \ln x + \mathit{const}$ , which we have indicated in Fig. 7.

To assess the effect for longitudinal structure function $F_L$, we convolute in Fig. 7 on the right the coefficient functions with the input shapes Eq. (7) for $n_f = 4$ flavors and the reference scale of Eq. (5). A comparison of the left and right plots in Fig. 7 clearly reveals the smoothening effect of the Mellin convolutions. For the chosen input conditions, the (mostly positive) NNLO corrections to the flavor-singlet $F_L$ amount to less than 20% for $5 \cdot 10^{-5} < x < 0.3$. In data fits we expect that the parton distributions, in particular the gluon distribution, will further stabilize the overall NNLO/NLO ratio. Thus, at not too small scales, $F_L$ is a quantity of good perturbative stability, for the $x$-values accessible at HERA, see Ref. [8] for more details.

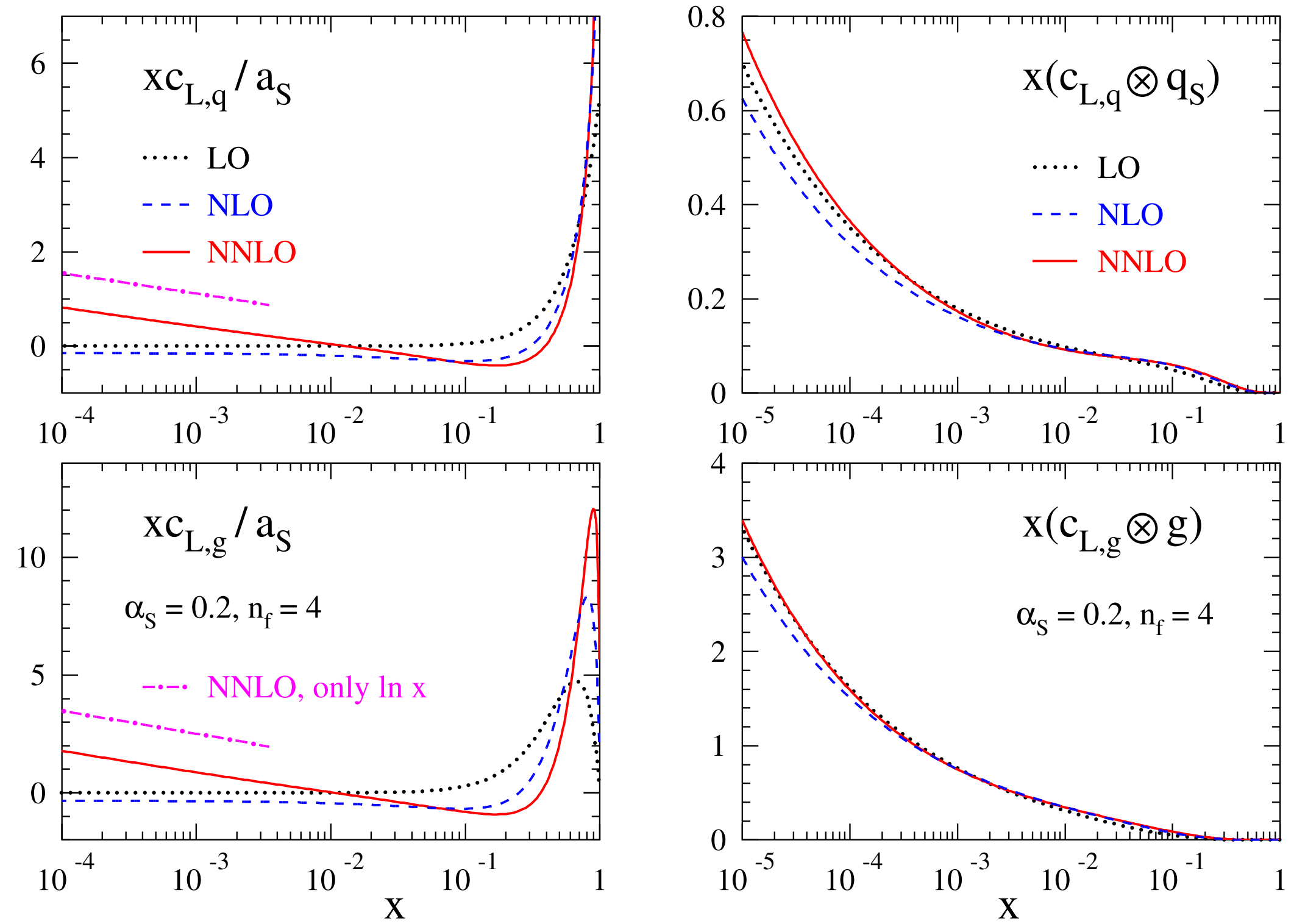

**Fig. 7:** The perturbative expansion to N$^2$LO of the longitudinal singlet-quark and gluon coefficient functions to third order multiplied by $x$ for display purposes (left) and of the quark and gluon contributions to singlet structure function $F_L$ (right).

**Table 1:** Number of alternating and non-alternating harmonic sums in dependence of their weight, [28].

| Weight | Number of | | | | | |
|---|---|---|---|---|---|---|
| | Sums | a-basic sums | Sums $\neg\{-1\}$ | a-basic sums | Sums $i > 0$ | a-basic sums |
| 1 | 2 | 2 | 1 | 1 | 1 | 1 |
| 2 | 6 | 3 | 3 | 2 | 2 | 1 |
| 3 | 18 | 8 | 7 | 4 | 4 | 2 |
| 4 | 54 | 18 | 17 | 7 | 8 | 3 |
| 5 | 162 | 48 | 41 | 16 | 16 | 6 |
| 6 | 486 | 116 | 99 | 30 | 32 | 9 |
| 7 | 1458 | 312 | 239 | 68 | 64 | 18 |

## 3 Mathematical Structure of Higher Order Corrections [3]

The QCD anomalous dimensions and Wilson coefficients for structure functions are single scale quantities and may be expressed in simple form in Mellin space in terms of polynomials of harmonic sums and ration functions of the Mellin variable. Unlike the case in various calculations using representations in momentum-fraction ($z$-) space the use of multiple nested harmonic sums leads to a synchronization in language. Furthermore, significant simplifications w.r.t. the number of functions needed can be achieved. This is due to algebraic [24,25] relations between these quantities, which in a similar way are also present between harmonic polylogarithms [26] and multiple $\zeta$-values [27]. These relations result from the the specific index pattern of the objects considered and their multiplication relation and do not refer to further more specific properties. In Table 1 we illustrate the level of complexity which one meets in case of harmonic sums. To three-loop order weight w=6 harmonic sums occur. The algebraic relations for the whole class of harmonic sums lead to a reduction by a factor of $\sim 4$ (column 3). As it turns out, physical pseudo-observables, as anomalous dimensions and Wilson-coefficients in the $\overline{\rm MS}$ scheme, to 2-, resp. 3-loop order depend on harmonic sums only, in which the index $\{-1\}$ never occurs. The algebraic reduction for this class is illustrated in column 5. We also compare the complexity of only non-alternating harmonic sums and their algebraic reduction, which is much lower. This class of sums is, however, not wide enough to describe the above physical quantities. In addition to the algebraic relations of harmonic sums structural relations exist, which reduces the basis further [28]. Using all these relations one finds that 5 basic functions are sufficient to describe all 2-loop Wilson coefficients for deep-inelastic scattering [29] and further 8 [30] for the 3-loop anomalous dimensions. Their analytic continuations to complex values of the Mellin variable are given in [31,32]. These functions are the (regularized) Mellin transforms of :

$$\begin{array}{llll}\frac{\ln(1+x)}{1+x}, & \frac{\mathrm{Li}_2(x)}{1\pm x}, & \frac{S_{1,2}(x)}{1\pm x}, & \frac{\mathrm{Li}_4(x)}{x\pm 1}, \\ \frac{S_{1,3}(x)}{1+x}, & \frac{S_{2,2}(x)}{x\pm 1}, & \frac{\mathrm{Li}_2^2(x)}{1+x}, & \frac{S_{2,2}(-x)-\mathrm{Li}_2^2(-x)/2}{x\pm 1}\,. \end{array} \tag{10}$$

[3]Contributing authors: J. Blümlein, H. Böttcher, A. Guffanti, V. Ravindran

It is remarkable, that the numerator-functions in (10) are Nielsen integrals [33] and polynomials thereof, although one might expect harmonic polylogarithms [26] outside this class in general. The representation of the Wilson coefficients and anomalous dimensions in the way described allows for compact expressions and very fast and precise numerical evaluation well suited for fitting procedures to experimental data.

### 3.1 Two-loop Processes at LHC in Mellin Space

Similar to the case of the Wilson coefficients in section 3 one may consider the Wilson coefficients for inclusive hard processes at hadron colliders, as the Drell–Yan process to $O(\alpha_s^2)$ [34–36], scalar or pseudoscalar Higgs-boson production to $O(\alpha_s^3)$ in the heavy-mass limit [37–42], and the 2-loop time-like Wilson coefficients for fragmentation [43–45]. These quantities have been analyzed in [46,47] w.r.t. their general structure in Mellin space. The cross section for the Drell–Yan process and Higgs production is given by

$$\sigma\left(\frac{\hat{s}}{s},Q^2\right)=\int_x^1\frac{dx_1}{x_1}\int_{x/x_1}^1\frac{dx_2}{x_2}f_a(x_1,\mu^2)f_b(x_2,\mu^2)\hat{\sigma}\left(\frac{x}{x_1x_2},\frac{Q^2}{\mu^2}\right)\,, \tag{3.11}$$

with $x=\hat{s}/s$. Here, $f_c(x,\mu^2)$ are the initial state parton densities and $\mu^2$ denotes the factorization scale. The Wilson coefficient of the process is $\hat{\sigma}$ and $Q^2$ is the time-like virtuality of the $s$-channel boson. Likewise, for the fragmentation process of final state partons into hadrons in $pp$–scattering one considers the double differential final state distribution

$$\frac{d^2\sigma^H}{dx d\cos\theta}=\frac{3}{8}(1+\cos^2\theta)\frac{d\sigma_T^H}{dx}+\frac{3}{4}\sin^2\theta\frac{d\sigma_L^H}{dx}\,. \tag{3.12}$$

Here,

$$\frac{d\sigma_k^H}{dx} = \int_x^1\frac{dz}{z}\Bigg[\sigma_{\rm tot}^{(0)}\left\{D_S^H\left(\frac{x}{z},M^2\right)C_{k,q}^{\rm S}(z,Q^2/M^2)+D_g^H\left(\frac{x}{z},M^2\right)C_{k,q}^{\rm S}(z,Q^2/M^2)\right\}$$

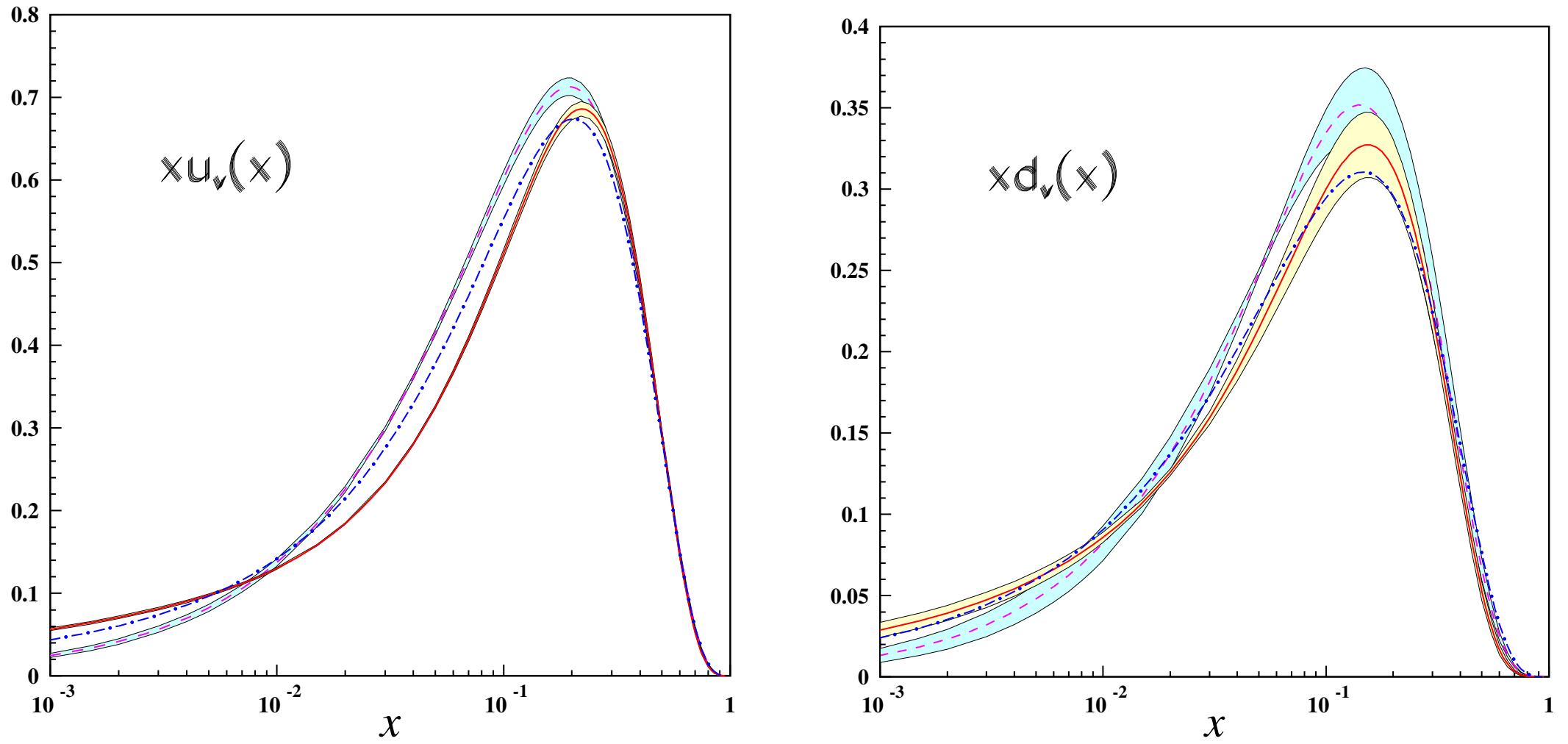

**Fig. 8:** $xu_v$ and $xd_v$ at $Q_0^2 = 4\text{GeV}^2$ (full lines) [48]; dashed lines [50]; dash-dotted lines [51].

$$+ \sum_{p=1}^{N_f} \sigma_p^{(0)} D_{\text{NS},p}^H \left(\frac{x}{z}, M^2\right) C_{k,q}^{\text{NS}}(z, Q^2/M^2)\Bigg] . \quad (3.13)$$

In the subsystem cross-sections $\sigma$ the initial state parton distributions are included. $D_k^H$ denote the non-perturbative fragmentation functions and $C_{k,i}^{\text{S,NS}}(z, Q^2/M^2)$ the respective time-like Wilson coefficients describing the fragmentaion process for a parton $i$ into the hadron $H$.

Although these Wilson coefficients are not directly related to the 2-loop Wilson coefficients for deeply inelastic scattering, one finds for these functions at most the same set of basic functions as given above. Again one obtains very fast and concise numerical programs also for these processes working in Mellin space, which will be well suited for inclusive analyses of experimental collider data at LHC in the future.

### 3.2 Non-Singlet Parton Densities at $O(\alpha_s^3)$

The precision determination of the QCD-scale $\Lambda_{\text{QCD}}$ and of the idividual parton densities is an important issue for the whole physics programme at LHC since all measurements rely on the detailed knwoledge of this parameter and distribution functions. In Ref. [48] first results were reported of a world data analysis for charged lepton-$p(d)$ scattering w.r.t. the flavor non-singlet sector at $O(\alpha_s^3)$ accuracy. The flavor non-singlet distributions $xu_v(x, Q^2)$ and $xd_v(x, Q^2)$ were determined along with fully correlated error bands giving parameterizations both for the values and errors of these distributions for a wide range in $x$ and $Q^2$. In Figure 8 these distributions including their error are shown. The value of the strong coupling constant $\alpha_s(M_Z^2)$ was determined as $0.1135 + 0.0023 - 0.0026$ (exp.) The full analysis is given in [49], including the determination of higher twist contributions in the large $x$ region both for $F_2^p(x, Q^2)$ and $F_2^d(x, Q^2)$.

### 3.3 Scheme-invariant evolution for unpolarzed DIS structure functions

The final HERA-II data on unpolarized DIS structure functions, combined with the present world data from other experiments, will allow to reduce the experimental error on the strong coupling constant, $\alpha_s(M_Z^2)$, to the level of $1\%$ [52]. On the theoretical side the NLO analyzes have intrinsic limitations which allow no better than $5\%$ accuracy in the determination of $\alpha_s$ [53]. In order to match the expected experimental accuracy, analyzes of DIS structure functions need then to be carried out at the NNLO-level. To perform a full NNLO analysis the knowledge of the 3-loop $\beta$-function coefficient, $\beta_2$, the 2-

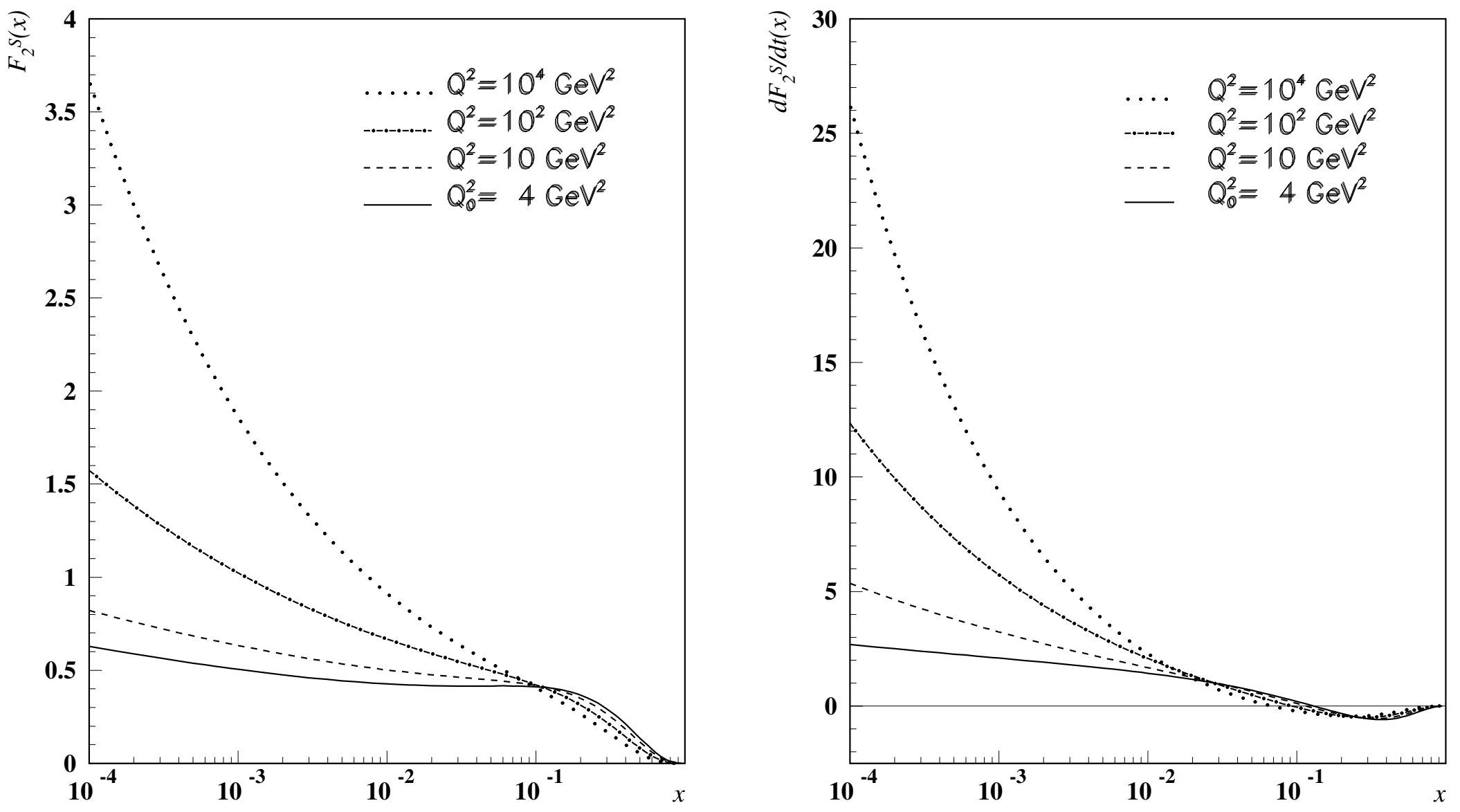

**Fig. 9:** NNLO scheme invariant evolution for the singlet part of the structure function $F_2$ and its slope $\partial F_2/\partial t$ for four massless flavours, [54].

resp. 3-loop Wilson coefficients and the 3-loop anomalous dimensions is required. With the calculation of the latter [6, 7], the whole scheme-independent set of quantities is known, thus allowing a complete NNLO study of DIS structure functions.

Besides the standard approach solving the QCD evolution equations for parton densities in the $\overline{\text{MS}}$ scheme it appears appealing to study scheme-invariant evolution equations [54]. Within this approach the input distributions at a scale $Q_0^2$ are measured experimentally. The only parameter to be determined by a fit to data is the QCD-scale $\Lambda_{\text{QCD}}$. To perform an analysis in the whole kinematic region the non-singlet [48] contribution has to be separated from the singlet terms of two measured observables. In practice these can be chosen to be $F_2(x, Q^2)$ and $\partial F_2(x, Q^2)/\partial \ln(Q^2)$ or $F_2(x, Q^2)$ and $F_L(x, Q^2)$ if the latter structure function is measured well enough. Either $\partial F_2(x, Q^2)/\partial \ln(Q^2)$ or $F_L(x, Q^2)$ play a role synonymous to the gluon distribution while $F_2(x, Q^2)$ takes the role of the singlet-quark distribution compared to the standard analysis. These equations do no longer describe the evolution of universal quantities depending on the choice of a scheme but of process-dependent quantities which are observables and thus factorization scheme-indedependent. Since the respective evolution kernels are calculated in perturbation theory the dependence on the renormalization scale remains and becomes smaller with the order in the coupling constant included.

Physical evolution kernels have been studied before in [55–57]. The 3-loop scheme-invariant evolution equations were solved in the massless case in [54]. This analysis is extended including the heavy flavor contributions at present [49]. The large complexity of the evolution kernels can only be handeled in Mellin space since in $z$-space various inverse and direct Mellin convolutions would be required numerically, causing significant accuracy and run-time problems. The inclusion of the heavy flavor contributions is possible using the parameterizations [58].

In Fig. 8 we present the scheme invariant evolution for the structure functions $F_2$ and $\partial F_2/\partial t$ to NNLO with $t = -2/\beta_0 \ln(\alpha_s(Q^2)/\alpha_s(Q_0^2))$. The input distribution at the reference scale are not extracted from data, but rather built up as a convolution of Wilson coefficients and PDFs, the latter being parametrised according to [59].

Scheme-invariant evolution equations allow a widely un-biased approach to determine the initial conditions for QCD evolution, which in general is a source of systematic effects which are difficult to control. On the other hand, their use requires to consider all correlations of the input measurements in a detailed manner experimentally. At any scale $Q^2$ mappings are available to project the observables evolved onto the quark-singlet and the gluon density in whatever scheme. In this way the question whether sign changes in the unpolarized gluon distribution in the $\overline{\mathrm{MS}}$ scheme do occur or do not occur in the small $x$ region can be answered uniquely. As in foregoing analyses [48, 60] correlated error propagation throughout the evolution is being performed.

## 4 Updated reference results for the evolution of parton distributions [4]

In this contribution we update and extend our benchmark tables, first presented in the report of the QCD/SM working group at the 2001 Les Houches workshop [59], for the evolution of parton distributions of hadrons in perturbative QCD. Since then the complete next-to-next-to-leading order (NNLO) splitting functions have been computed [6,7], see also section 2. Thus we can now replace the NNLO results of 2001 which were based on the approximate splitting functions of Ref. [61]. Furthermore we now include reference tables for the polarized case treated in neither Ref. [59] nor the earlier study during the 1995/6 HERA workshop [62]. Since the spin-dependent NNLO splitting functions are still unknown, we have to restrict ourselves to the polarized leading-order (LO) and next-to-leading-order (NLO) evolution.

As in Ref. [59], we employ two entirely independent and conceptually different FORTRAN programs. At this point, the $x$-space code of G.S. is available from the author upon request, while the Mellin-space program of A.V. has been published in Ref. [63]. The results presented below correspond to a direct iterative solution of the N$^{\mathrm{m}}$LO evolution equations for the parton distributions $f_p(x,\mu_{\mathrm{f}}^2) \equiv p(x,\mu_{\mathrm{f}}^2)$, where $p = q_i,\, \bar{q}_i\,, g$ with $i = 1, \ldots, N_{\mathrm{f}}$,

$$\frac{d f_p(x,\mu_{\mathrm{f}}^2)}{d \ln \mu_{\mathrm{f}}^2} \;=\; \sum_{l=0}^{m} a_{\mathrm{s}}^{l+1}(\mu_{\mathrm{r}}^2) \int_x^1 \frac{dy}{y} \sum_{p'} P_{pp'}^{(l)}\!\left(\frac{x}{y}, \frac{\mu_{\mathrm{f}}^2}{\mu_{\mathrm{r}}^2}\right) f_{p'}(y,\mu_{\mathrm{f}}^2) \tag{4.14}$$

with the strong coupling, normalized as $a_{\mathrm{s}} \equiv \alpha_{\mathrm{s}}/(4\pi)$, given in terms of

$$\frac{d\, a_{\mathrm{s}}}{d \ln \mu_r^2} \;=\; \beta_{\mathrm{N^mLO}}(a_{\mathrm{s}}) \;=\; -\sum_{l=0}^{m} a_{\mathrm{s}}^{l+2}\, \beta_l \tag{4.15}$$

with $\beta_0 = 11 - 2/3\, N_{\mathrm{f}}$ etc. $\mu_{\mathrm{r}}$ and $\mu_{\mathrm{f}}$ represent the renormalization and mass-factorization scales in the $\overline{\mathrm{MS}}$ scheme. The reader is referred to Refs. [59, 63] for the scale dependence of the splitting functions $P^{(l)}$ and a further discussion of our solutions of Eqs. (4.14) and (4.15).

For the unpolarized case we retain the initial conditions as set up at the Les Houches meeting: The evolution is started at

$$\mu_{\mathrm{f},0}^2 \;=\; 2\ \mathrm{GeV}^2\ . \tag{4.16}$$

Roughly along the lines of the CTEQ5M parametrization [64], the input distributions are chosen as

$$\begin{aligned}
xu_{\mathrm{v}}(x,\mu_{\mathrm{f},0}^2) &= 5.107200\, x^{0.8}\, (1-x)^3 \\
xd_{\mathrm{v}}(x,\mu_{\mathrm{f},0}^2) &= 3.064320\, x^{0.8}\, (1-x)^4 \\
xg\,(x,\mu_{\mathrm{f},0}^2) &= 1.700000\, x^{-0.1} (1-x)^5 \\
x\bar{d}\,(x,\mu_{\mathrm{f},0}^2) &= .1939875\, x^{-0.1} (1-x)^6 \\
x\bar{u}\,(x,\mu_{\mathrm{f},0}^2) &= (1-x)\, x\bar{d}\,(x,\mu_{\mathrm{f},0}^2) \\
xs\,(x,\mu_{\mathrm{f},0}^2) &= x\bar{s}\,(x,\mu_{\mathrm{f},0}^2) \;=\; 0.2\, x(\bar{u}+\bar{d})(x,\mu_{\mathrm{f},0}^2)
\end{aligned} \tag{4.17}$$

where, as usual, $q_{i,\mathrm{v}} \equiv q_i - \bar{q}_i$. The running couplings are specified by Eq. (4.15) and

$$\alpha_{\mathrm{s}}(\mu_{\mathrm{r}}^2 = 2\ \mathrm{GeV}^2) \;=\; 0.35\ . \tag{4.18}$$

[4]Contributing authors: G.P. Salam, A. Vogt

For simplicity initial conditions (4.17) and (4.18) are employed regardless of the order of the evolution and the (fixed) ratio of the renormalization and factorization scales.

For the evolution with a fixed number $N_{\rm f} > 3$ of quark flavours the quark distributions not specified in Eq. (4.17) are assumed to vanish at $\mu_{\rm f,0}^2$, and Eq. (4.18) is understood to refer to the chosen value of $N_{\rm f}$. For the evolution with a variable $N_{\rm f} = 3 \ldots 6$, Eqs. (4.16) and (4.17) always refer to three flavours. $N_{\rm f}$ is then increased by one unit at the heavy-quark pole masses taken as

$$m_{\rm c} \;=\; \mu_{\rm f,0}\,, \quad m_{\rm b} \;=\; 4.5\;{\rm GeV}^2\,, \quad m_{\rm t} \;=\; 175\;{\rm GeV}^2\,, \tag{4.19}$$

i.e., Eqs. (4.14) and (4.15) are solved for a fixed number of flavours between these thresholds, and the respective matching conditions are invoked at $\mu_{\rm f}^2 = m_{\rm h}^2,\ h = c,\, b,\, t$. The matching conditions for the unpolarized parton distributions have been derived at NNLO in Ref. [65], and were first implemented in an evolution program in Ref. [66]. Note that, while the parton distributions are continuous up to NLO due to our choice of the matching scales, $\alpha_{\rm s}$ is discontinuous at these flavour thresholds already at this order for $\mu_{\rm r} \neq \mu_{\rm f}$, see Refs. [67,68]. Again the reader is referred to Refs. [59,63] for more details.

Since the exact NNLO splitting functions $P^{(2)}$ are rather lengthy and not directly suitable for use in a Mellin-space program (see, however, Ref. [32]), the reference tables shown below have been computed using the parametrizations (4.22)–(4.24) of Ref. [6] and (4.32)–(4.35) of Ref. [7]. Likewise, the operator matrix element $\widetilde{A}_{\rm hg}^{\rm S,2}$ entering the NNLO flavour matching is taken from Eq. (3.5) of Ref. [63]. The relative error made by using the parametrized splitting functions is illustrated in Fig. 10. It is generally well below $10^{-4}$, except for the very small sea quark distributions at very large $x$.

Eqs. (4.16), (4.18) and (4.19) are used for the (longitudinally) polarized case as well, where Eq. (4.17) replaced by the sufficiently realistic toy input [63]

$$\begin{aligned} xu_{\rm v} &= +1.3\, x^{0.7}\, (1-x)^3\, (1+3x) \\ xd_{\rm v} &= -0.5\, x^{0.7}\, (1-x)^4\, (1+4x) \\ xg &= +1.5\, x^{0.5}\, (1-x)^5 \\ x\bar{d} &= x\bar{u} \;=\; -0.05\, x^{0.3}\, (1-x)^7 \\ xs &= x\bar{s} \;=\; +0.5\, x\bar{d}\ . \end{aligned} \tag{4.20}$$

As Eq. (4.17) in the unpolarized case, this input is employed regardless of the order of the evolution.

As in Ref. [59], we have compared the results of our two evolution programs, under the conditions specified above, at 500 $x$-$\mu_{\rm f}^2$ points covering the range $10^{-8} \leq x \leq 0.9$ and $2\;{\rm GeV}^2 \leq \mu_{\rm f}^2 \leq 10^6\;{\rm GeV}^2$. A representative subset of our results at $\mu_{\rm f}^2 = 10^4\;{\rm GeV}^4$, a scale relevant to high-$E_T$ jets and close to $m_{\rm W}^2$, $m_{\rm Z}^2$ and, possibly, $m_{\rm Higgs}^2$, is presented in Tables 2 − 6. These results are given in terms of the valence distributions, defined below Eq. (4.17), $L_\pm \equiv \bar{d} \pm \bar{u}$, and the quark–antiquark sums $q_+ \equiv q - \bar{q}$ for $q = s,\, c$ and, for the variable-$N_{\rm f}$ case, $b$.

For compactness an abbreviated notation is employed throughout the tables, i.e., all numbers $a \cdot 10^b$ are written as $a^b$. In the vast majority of the $x$-$\mu_{\rm f}^2$ points our results are found to agree to all five figures displayed, except for the tiny NLO and NNLO sea-quark distributions at $x = 0.9$, in the tables. Entries where the residual offsets between our programs lead to a different fifth digit after rounding are indicated by the subscript '$*$'. In these cases the number with the smaller modulus is given in the tables.

The approximate splitting functions [61], as mentioned above employed in the previous version [59] of our reference tables, have been used in (global) NNLO fits of the unpolarized parton distributions [51,69], which in turn have been widely employed for obtaining NNLO cross sections, in particular for $W$ and Higgs production. The effect of replacing the approximate results by the full splitting functions [6, 7] is illustrated in Figure 11. Especially at scales relevant to the above-mentioned processes, the previous approximations introduce an error of less than 0.2% for $x \gtrsim 10^{-3}$, and less than 1% even down to $x \simeq 10^{-5}$. Consequently the splitting-function approximations used for the evolution the parton distributions of Refs. [51,69] are confirmed to a sufficient accuracy for high-scale processes at the LHC.

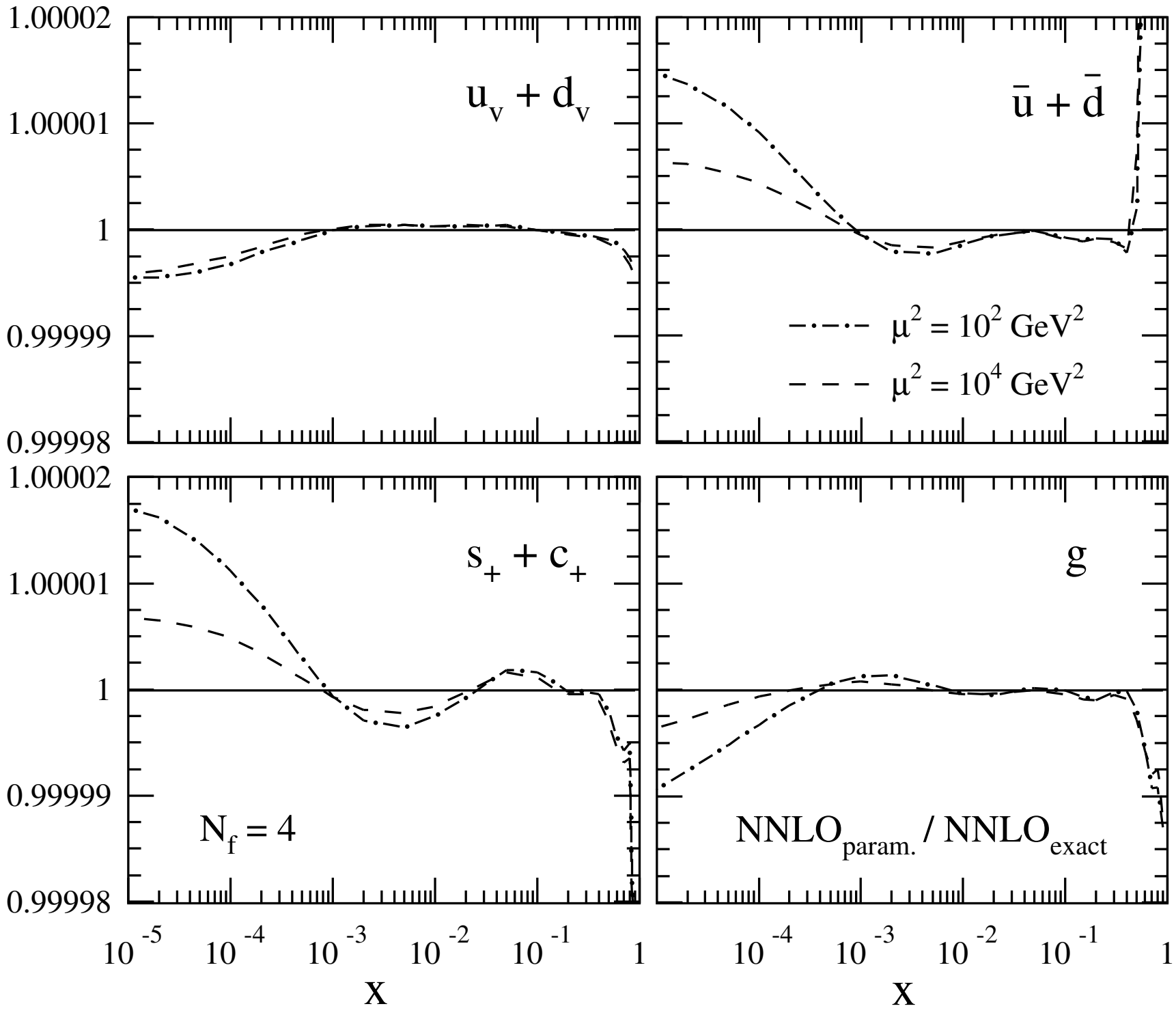

**Fig. 10:** Relative effects of using the parametrized three-loop splitting functions of Refs. [6,7], instead of the exact expressions from the same source, on the NNLO evolution for the input (4.16)–(4.18) at two representative values of $\mu = \mu_{\rm r} = \mu_{\rm f}$.

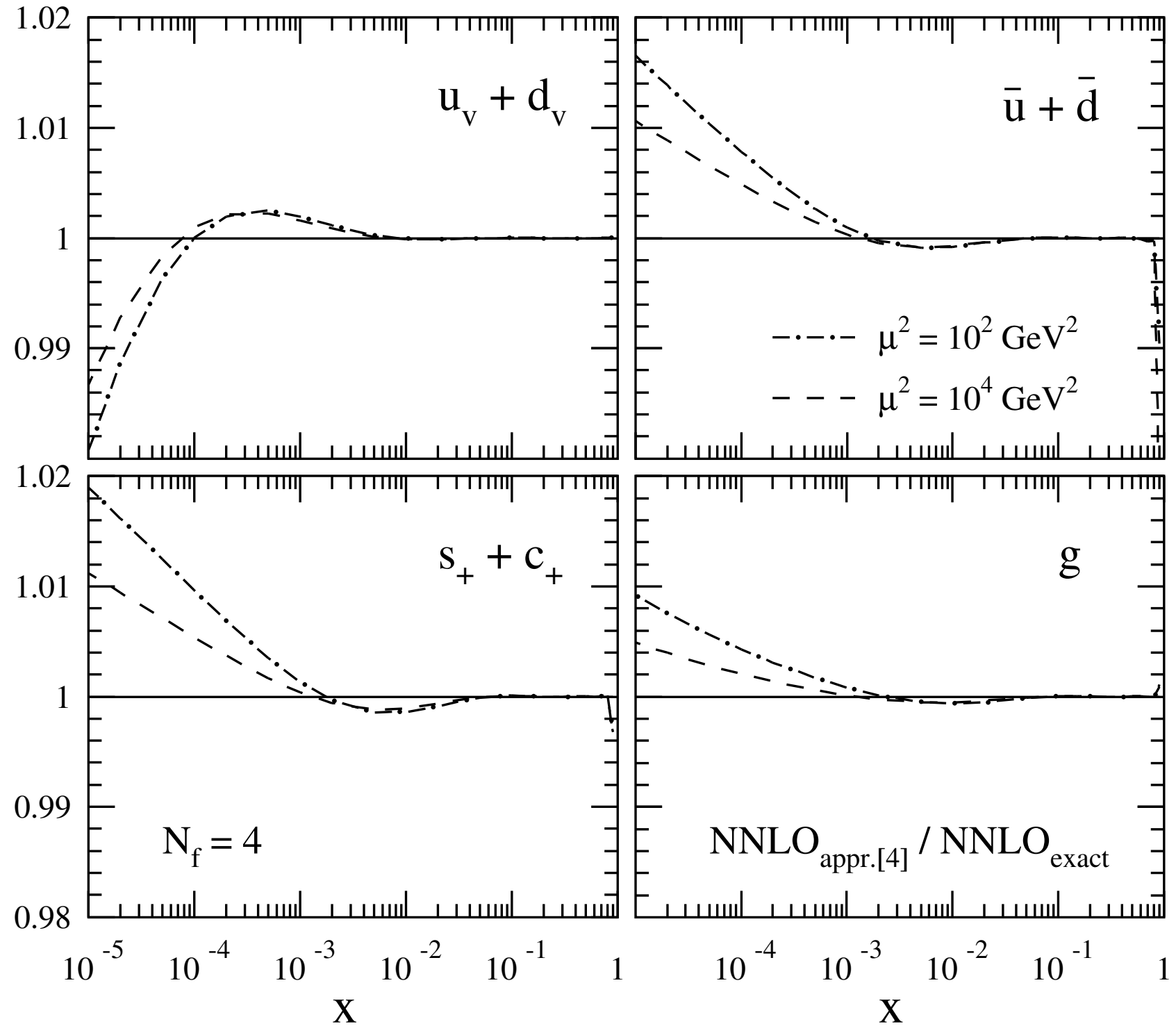

**Fig. 11:** Relative errors made by using the previous average approximations [61] for the three-loop splitting functions (used, e.g., in Refs. [51,69]) instead of the full results [6,7], on the NNLO evolution of the input (4.16)–(4.18) at $\mu_{\rm r} = \mu_{\rm f}$.

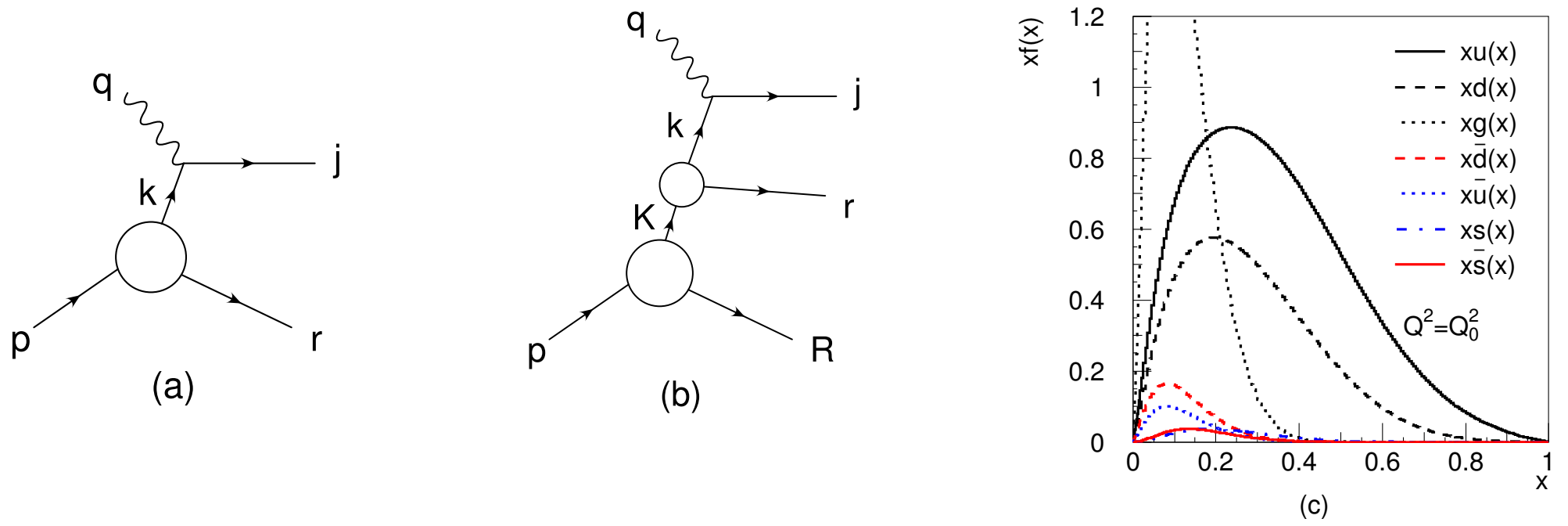

**Fig. 12:** Probing (a) a valence parton in the proton and (b) a sea parton in a hadronic fluctuation (letters are four-momenta) resulting in (c) parton distributions at the starting scale $Q_0^2$.

The unchanged unpolarized LO and NLO reference tables of Ref. [59] are not repeated here. Note that the one digit of the first (FFN) $\alpha_s$ value was mistyped in the header of Table 1 in that report [5], the correct value can be found in Table 3 below.

## 5 Non-perturbative $x$-shape of PDFs [6]

The $x$-shape of parton density functions at a low scale $Q_0^2$ is due to the dynamics of the bound state proton and is hence an unsolved problem of non-perturbative QCD. Usually this is described by parameterizations of data using more or less arbitrary functional forms. More understanding can be obtained by a recently developed physical model [70], which is phenomenologically successful in describing data.

The model gives the four-momentum $k$ of a single probed valence parton (Fig. 12a) by assuming that, in the nucleon rest frame, the shape of the momentum distribution for a parton of type $i$ and mass $m_i$ can be taken as a Gaussian $f_i(k) = N(\sigma_i, m_i) \exp\left\{-\left[(k_0 - m_i)^2 + k_x^2 + k_y^2 + k_z^2\right]/2\sigma_i^2\right\}$, which may be motivated as a result of the many interactions binding the parton in the nucleon. The width of the distribution should be of order hundred MeV from the Heisenberg uncertainty relation applied to the nucleon size, *i.e.* $\sigma_i = 1/d_N$. The momentum fraction $x$ of the parton is then defined as the light-cone fraction $x = k_+/p_+$ and is therefore invariant under longitudinal boosts (e.g. to the infinite momentum frame). Constraints are imposed on the final-state momenta to obtain a kinematically allowed final state, which also ensures that $0 < x < 1$ and $f_i(x) \to 0$ for $x \to 1$.

The sea partons are obtained using a hadronic basis for the non-perturbative dynamics of the bound state proton and considering hadronic fluctuations

$$|p\rangle = \alpha_0 |p_0\rangle + \alpha_{p\pi^0}|p\pi^0\rangle + \alpha_{n\pi^+}|n\pi^+\rangle + \ldots + \alpha_{\Lambda K}|\Lambda K^+\rangle + \ldots \quad (5.21)$$

Probing a parton $i$ in a hadron $H$ of a baryon-meson fluctuation $|BM\rangle$ (Fig. 12b) gives a sea parton with light-cone fraction $x = x_H\, x_i$ of the target proton. The momentum of the probed hadron is given by a similar Gaussian, but with a separate width parameter $\sigma_H$. Also here, kinematic constraints ensure physically allowed final states.

Using a Monte Carlo method the resulting valence and sea parton $x$-distributions are obtained without approximations. These apply at a low scale $Q_0^2$ and the distributions at higher $Q^2$ are obtained using perturbative QCD evolution at next-to-leading order. To describe all parton distributions (Fig. 12c),

[5] We thank H. Böttcher and J. Blümlein for pointing this out to us.
[6] Contributing author: G. Ingelman

**Table 2:** Reference results for the $N_\mathrm{f} = 4$ next-next-to-leading-order evolution for the initial conditions (4.16)–(4.18). The corresponding value of the strong coupling is $\alpha_\mathrm{s}(\mu_\mathrm{r}^2 = 10^4\ \mathrm{GeV}^2) = 0.110141$. The valence distributions $s_v$ and $c_v$ are equal for the input (4.17). The notation is explained below Eq. (4.17) and in the paragraph below Eq. (4.20).

| NNLO, $N_\mathrm{f} = 4\,,\ \mu_\mathrm{f}^2 = 10^4\ \mathrm{GeV}^2$ | | | | | | | | |
|---|---|---|---|---|---|---|---|---|
| $x$ | $xu_v$ | $xd_v$ | $xL_-$ | $2xL_+$ | $xs_v$ | $xs_+$ | $xc_+$ | $xg$ |
| $\mu_\mathrm{r}^2 = \mu_\mathrm{f}^2$ | | | | | | | | |
| $10^{-7}$ | $1.5287^{-4}$ | $1.0244^{-4}$ | $5.7018^{-6}$ | $1.3190^{+2}$ | $3.1437^{-5}$ | $6.4877^{+1}$ | $6.4161^{+1}$ | $9.9763^{+2}$ |
| $10^{-6}$ | $6.9176^{-4}$ | $4.4284^{-4}$ | $2.5410^{-5}$ | $6.8499^{+1}$ | $9.4279^{-5}$ | $3.3397^{+1}$ | $3.2828^{+1}$ | $4.9124^{+2}$ |
| $10^{-5}$ | $3.0981^{-3}$ | $1.8974^{-3}$ | $1.0719^{-4}$ | $3.3471^{+1}$ | $2.2790^{-4}$ | $1.6059^{+1}$ | $1.5607^{+1}$ | $2.2297^{+2}$ |
| $10^{-4}$ | $1.3722^{-2}$ | $8.1019^{-3}$ | $4.2558^{-4}$ | $1.5204^{+1}$ | $3.6644^{-4}$ | $7.0670^{+0}$ | $6.7097^{+0}$ | $9.0668^{+1}$ |
| $10^{-3}$ | $5.9160^{-2}$ | $3.4050^{-2}$ | $1.6008^{-3}$ | $6.3230^{+0}$ | $1.4479^{-4}$ | $2.7474^{+0}$ | $2.4704^{+0}$ | $3.1349^{+1}$ |
| $10^{-2}$ | $2.3078^{-1}$ | $1.2919^{-1}$ | $5.5688^{-3}$ | $2.2752^{+0}$ | $-5.7311^{-4}$ | $8.5502^{-1}$ | $6.6623^{-1}$ | $8.1381^{+0}$ |
| 0.1 | $5.5177^{-1}$ | $2.7165^{-1}$ | $1.0023^{-2}$ | $3.9019^{-1}$ | $-3.0627^{-4}$ | $1.1386^{-1}$ | $5.9773^{-2}$ | $9.0563^{-1}$ |
| 0.3 | $3.5071^{-1}$ | $1.3025^{-1}$ | $3.0098^{-3}$ | $3.5358^{-2}$ | $-3.1891^{-5}$ | $9.0480^{-3}$ | $3.3061^{-3}$ | $8.4186^{-2}$ |
| 0.5 | $1.2117^{-1}$ | $3.1528^{-2}$ | $3.7742^{-4}$ | $2.3867^{-3}$ | $-2.7215^{-6}$ | $5.7965^{-4}$ | $1.7170^{-4}$ | $8.1126^{-3}$ |
| 0.7 | $2.0077^{-2}$ | $3.0886^{-3}$ | $1.3434^{-5}$ | $5.4244^{-5}$ | $-1.0106^{-7}$ | $1.2936^{-5}$ | $3.5304^{-6}$ | $3.8948^{-4}$ |
| 0.9 | $3.5111^{-4}$ | $1.7783^{-5}$ | $8.651^{-9}$ | $2.695^{-8}$ | $-1.476^{-10}$ | $7.132^{-9}$ | $2.990^{-9}$ | $1.2136^{-6}$ |
| $\mu_\mathrm{r}^2 = 2\,\mu_\mathrm{f}^2$ | | | | | | | | |
| $10^{-7}$ | $1.3416^{-4}$ | $8.7497^{-5}$ | $4.9751^{-6}$ | $1.3020^{+2}$ | $2.1524^{-5}$ | $6.4025^{+1}$ | $6.3308^{+1}$ | $1.0210^{+3}$ |
| $10^{-6}$ | $6.2804^{-4}$ | $3.9406^{-4}$ | $2.2443^{-5}$ | $6.6914^{+1}$ | $6.5149^{-5}$ | $3.2602^{+1}$ | $3.2032^{+1}$ | $4.9626^{+2}$ |
| $10^{-5}$ | $2.9032^{-3}$ | $1.7575^{-3}$ | $9.6205^{-5}$ | $3.2497^{+1}$ | $1.5858^{-4}$ | $1.5570^{+1}$ | $1.5118^{+1}$ | $2.2307^{+2}$ |
| $10^{-4}$ | $1.3206^{-2}$ | $7.7673^{-3}$ | $3.9093^{-4}$ | $1.4751^{+1}$ | $2.5665^{-4}$ | $6.8388^{+0}$ | $6.4807^{+0}$ | $9.0162^{+1}$ |
| $10^{-3}$ | $5.8047^{-2}$ | $3.3434^{-2}$ | $1.5180^{-3}$ | $6.1703^{+0}$ | $1.0388^{-4}$ | $2.6695^{+0}$ | $2.3917^{+0}$ | $3.1114^{+1}$ |
| $10^{-2}$ | $2.2930^{-1}$ | $1.2857^{-1}$ | $5.4626^{-3}$ | $2.2492^{+0}$ | $-3.9979^{-4}$ | $8.4058^{-1}$ | $6.5087^{-1}$ | $8.0993^{+0}$ |
| 0.1 | $5.5428^{-1}$ | $2.7326^{-1}$ | $1.0072^{-2}$ | $3.9297^{-1}$ | $-2.1594^{-4}$ | $1.1439^{-1}$ | $5.9713^{-2}$ | $9.0851^{-1}$ |
| 0.3 | $3.5501^{-1}$ | $1.3205^{-1}$ | $3.0557^{-3}$ | $3.6008^{-2}$ | $-2.2632^{-5}$ | $9.2227^{-3}$ | $3.3771^{-3}$ | $8.5022^{-2}$ |
| 0.5 | $1.2340^{-1}$ | $3.2166^{-2}$ | $3.8590^{-4}$ | $2.4459^{-3}$ | $-1.9420^{-6}$ | $5.9487^{-4}$ | $1.7699^{-4}$ | $8.2293^{-3}$ |
| 0.7 | $2.0597^{-2}$ | $3.1751^{-3}$ | $1.3849^{-5}$ | $5.5722^{-5}$ | $-7.2616^{-8}$ | $1.3244^{-5}$ | $3.5361^{-6}$ | $3.9687^{-4}$ |
| 0.9 | $3.6527^{-4}$ | $1.8544^{-5}$ | $9.050^{-9}$ | $2.663^{-8}$ | $-1.075^{-10}$ | $6.713^{-9}$ | $2.377^{-9}$ | $1.2489^{-6}$ |
| $\mu_\mathrm{r}^2 = 1/2\,\mu_\mathrm{f}^2$ | | | | | | | | |
| $10^{-7}$ | $1.7912^{-4}$ | $1.2521^{-4}$ | $6.4933_*^{-6}$ | $1.2714^{+2}$ | $4.9649^{-5}$ | $6.2498^{+1}$ | $6.1784^{+1}$ | $9.2473^{+2}$ |
| $10^{-6}$ | $7.7377^{-4}$ | $5.1222^{-4}$ | $2.8719^{-5}$ | $6.7701^{+1}$ | $1.4743^{-4}$ | $3.2999^{+1}$ | $3.2432^{+1}$ | $4.6863^{+2}$ |
| $10^{-5}$ | $3.3184^{-3}$ | $2.0760^{-3}$ | $1.1977^{-4}$ | $3.3644^{+1}$ | $3.5445^{-4}$ | $1.6147^{+1}$ | $1.5696^{+1}$ | $2.1747^{+2}$ |
| $10^{-4}$ | $1.4184^{-2}$ | $8.4455^{-3}$ | $4.6630^{-4}$ | $1.5408^{+1}$ | $5.6829^{-4}$ | $7.1705^{+0}$ | $6.8139^{+0}$ | $8.9820_*^{+1}$ |
| $10^{-3}$ | $5.9793^{-2}$ | $3.4418^{-2}$ | $1.6996^{-3}$ | $6.4042^{+0}$ | $2.2278^{-4}$ | $2.7892^{+0}$ | $2.5128^{+0}$ | $3.1336^{+1}$ |
| $10^{-2}$ | $2.3106^{-1}$ | $1.2914^{-1}$ | $5.7016^{-3}$ | $2.2876^{+0}$ | $-8.9125^{-4}$ | $8.6205^{-1}$ | $6.7377^{-1}$ | $8.1589^{+0}$ |
| 0.1 | $5.5039^{-1}$ | $2.7075^{-1}$ | $1.0031^{-2}$ | $3.8850^{-1}$ | $-4.7466^{-4}$ | $1.1332^{-1}$ | $5.9489^{-2}$ | $9.0795^{-1}$ |
| 0.3 | $3.4890^{-1}$ | $1.2949^{-1}$ | $2.9943^{-3}$ | $3.5090^{-2}$ | $-4.9304^{-5}$ | $8.9667^{-3}$ | $3.2670^{-3}$ | $8.4309^{-2}$ |
| 0.5 | $1.2026^{-1}$ | $3.1269^{-2}$ | $3.7428^{-4}$ | $2.3729^{-3}$ | $-4.1981^{-6}$ | $5.7783^{-4}$ | $1.7390^{-4}$ | $8.1099_*^{-3}$ |
| 0.7 | $1.9867^{-2}$ | $3.0534^{-3}$ | $1.3273^{-5}$ | $5.4635^{-5}$ | $-1.5541^{-7}$ | $1.3275^{-5}$ | $3.9930^{-6}$ | $3.8824^{-4}$ |
| 0.9 | $3.4524^{-4}$ | $1.7466^{-5}$ | $8.489^{-9}$ | $3.030^{-8}$ | $-2.255^{-10}$ | $8.863^{-9}$ | $4.803^{-9}$ | $1.2026^{-6}$ |

**Table 3:** As Table 2, but for the variable-$N_{\rm f}$ evolution using the flavour matching conditions of Ref. [65, 67, 68]. The corresponding values for the strong coupling $\alpha_{\rm s}(\mu_{\rm r}^2 = 10^4\ {\rm GeV}^2)$ are given by 0.115818, 0.115605 and 0.115410 for $\mu_{\rm r}^2/\mu_{\rm f}^2 = 0.5$, 1 and 2, respectively. For brevity the small, but non-vanishing valence distributions $s_v$, $c_v$ and $b_v$ are not displayed.

| NNLO, $N_{\rm f} = 3\ldots5$, $\mu_{\rm f}^2 = 10^4\ {\rm GeV}^2$ | | | | | | | | |
|---|---|---|---|---|---|---|---|---|
| $x$ | $xu_v$ | $xd_v$ | $xL_-$ | $2xL_+$ | $xs_+$ | $xc_+$ | $xb_+$ | $xg$ |
| $\mu_{\rm r}^2 = \mu_{\rm f}^2$ | | | | | | | | |
| $10^{-7}$ | $1.5978^{-4}$ | $1.0699^{-5}$ | $6.0090^{-6}$ | $1.3916^{+2}$ | $6.8509^{+1}$ | $6.6929^{+1}$ | $5.7438^{+1}$ | $9.9694^{+3}$ |
| $10^{-6}$ | $7.1787^{-4}$ | $4.5929^{-4}$ | $2.6569^{-5}$ | $7.1710^{+1}$ | $3.5003^{+1}$ | $3.3849^{+1}$ | $2.8332^{+1}$ | $4.8817^{+2}$ |
| $10^{-5}$ | $3.1907^{-3}$ | $1.9532^{-3}$ | $1.1116^{-4}$ | $3.4732^{+1}$ | $1.6690^{+1}$ | $1.5875^{+1}$ | $1.2896^{+1}$ | $2.2012^{+2}$ |
| $10^{-4}$ | $1.4023^{-2}$ | $8.2749^{-3}$ | $4.3744^{-4}$ | $1.5617^{+1}$ | $7.2747^{+0}$ | $6.7244^{+0}$ | $5.2597^{+0}$ | $8.8804^{+1}$ |
| $10^{-3}$ | $6.0019^{-2}$ | $3.4519^{-2}$ | $1.6296^{-3}$ | $6.4173^{+0}$ | $2.7954^{+0}$ | $2.4494^{+0}$ | $1.8139^{+0}$ | $3.0404^{+1}$ |
| $10^{-2}$ | $2.3244^{-1}$ | $1.3000^{-1}$ | $5.6100^{-3}$ | $2.2778^{+0}$ | $8.5749^{-1}$ | $6.6746^{-1}$ | $4.5073^{-1}$ | $7.7912^{+0}$ |
| 0.1 | $5.4993^{-1}$ | $2.7035^{-1}$ | $9.9596^{-3}$ | $3.8526^{-1}$ | $1.1230^{-1}$ | $6.4466^{-2}$ | $3.7280^{-2}$ | $8.5266^{-1}$ |
| 0.3 | $3.4622^{-1}$ | $1.2833^{-1}$ | $2.9572^{-3}$ | $3.4600^{-2}$ | $8.8410^{-3}$ | $4.0134^{-3}$ | $2.1047^{-3}$ | $7.8898^{-2}$ |
| 0.5 | $1.1868^{-1}$ | $3.0811^{-2}$ | $3.6760^{-4}$ | $2.3198^{-3}$ | $5.6309^{-4}$ | $2.3752^{-4}$ | $1.2004^{-4}$ | $7.6398^{-3}$ |
| 0.7 | $1.9486^{-2}$ | $2.9901^{-3}$ | $1.2957^{-5}$ | $5.2352^{-5}$ | $1.2504^{-5}$ | $5.6038^{-6}$ | $2.8888^{-6}$ | $3.7080^{-4}$ |
| 0.9 | $3.3522^{-4}$ | $1.6933^{-5}$ | $8.209^{-9}$ | $2.574^{-8}$ | $6.856^{-9}$ | $4.337^{-9}$ | $2.679^{-9}$ | $1.1721^{-6}$ |
| $\mu_{\rm r}^2 = 2\,\mu_{\rm f}^2$ | | | | | | | | |
| $10^{-7}$ | $1.3950^{-4}$ | $9.0954^{-5}$ | $5.2113^{-6}$ | $1.3549^{+2}$ | $6.6672^{+1}$ | $6.5348^{+1}$ | $5.6851^{+1}$ | $1.0084^{+3}$ |
| $10^{-6}$ | $6.4865^{-4}$ | $4.0691^{-4}$ | $2.3344^{-5}$ | $6.9214^{+1}$ | $3.3753^{+1}$ | $3.2772^{+1}$ | $2.7818^{+1}$ | $4.8816^{+2}$ |
| $10^{-5}$ | $2.9777^{-3}$ | $1.8020^{-3}$ | $9.9329^{-5}$ | $3.3385^{+1}$ | $1.6015^{+1}$ | $1.5306^{+1}$ | $1.2601^{+1}$ | $2.1838^{+2}$ |
| $10^{-4}$ | $1.3452^{-2}$ | $7.9078^{-3}$ | $4.0036^{-4}$ | $1.5035^{+1}$ | $6.9818^{+0}$ | $6.4880^{+0}$ | $5.1327^{+0}$ | $8.7550^{+1}$ |
| $10^{-3}$ | $5.8746^{-2}$ | $3.3815^{-2}$ | $1.5411^{-3}$ | $6.2321^{+0}$ | $2.7012^{+0}$ | $2.3747^{+0}$ | $1.7742^{+0}$ | $3.0060^{+1}$ |
| $10^{-2}$ | $2.3063^{-1}$ | $1.2923^{-1}$ | $5.4954^{-3}$ | $2.2490^{+0}$ | $8.4141^{-1}$ | $6.5083^{-1}$ | $4.4354^{-1}$ | $7.7495^{+0}$ |
| 0.1 | $5.5279^{-1}$ | $2.7222^{-1}$ | $1.0021^{-2}$ | $3.8897^{-1}$ | $1.1312^{-1}$ | $6.2917^{-2}$ | $3.7048^{-2}$ | $8.5897^{-1}$ |
| 0.3 | $3.5141^{-1}$ | $1.3051^{-1}$ | $3.0134^{-3}$ | $3.5398^{-2}$ | $9.0559^{-3}$ | $3.8727^{-3}$ | $2.0993^{-3}$ | $8.0226^{-2}$ |
| 0.5 | $1.2140^{-1}$ | $3.1590^{-2}$ | $3.7799^{-4}$ | $2.3919^{-3}$ | $5.8148^{-4}$ | $2.2376^{-4}$ | $1.1918^{-4}$ | $7.8098^{-3}$ |
| 0.7 | $2.0120^{-2}$ | $3.0955^{-3}$ | $1.3462^{-5}$ | $5.4194^{-5}$ | $1.2896^{-5}$ | $5.0329^{-6}$ | $2.8153^{-6}$ | $3.8099^{-4}$ |
| 0.9 | $3.5230^{-4}$ | $1.7849^{-5}$ | $8.687^{-9}$ | $2.568^{-8}$ | $6.513^{-9}$ | $3.390^{-9}$ | $2.407^{-9}$ | $1.2188^{-6}$ |
| $\mu_{\rm r}^2 = 1/2\,\mu_{\rm f}^2$ | | | | | | | | |
| $10^{-7}$ | $1.8906^{-4}$ | $1.3200^{-4}$ | $6.9268^{-6}$ | $1.3739^{+2}$ | $6.7627^{+1}$ | $6.5548^{+1}$ | $5.5295^{+1}$ | $9.4403^{+2}$ |
| $10^{-6}$ | $8.1001^{-4}$ | $5.3574^{-4}$ | $3.0345^{-5}$ | $7.2374^{+1}$ | $3.5337^{+1}$ | $3.3846^{+1}$ | $2.7870^{+1}$ | $4.7444^{+2}$ |
| $10^{-5}$ | $3.4428^{-3}$ | $2.1524^{-3}$ | $1.2531^{-4}$ | $3.5529^{+1}$ | $1.7091^{+1}$ | $1.6065^{+1}$ | $1.2883^{+1}$ | $2.1802^{+2}$ |
| $10^{-4}$ | $1.4580^{-2}$ | $8.6744^{-3}$ | $4.8276^{-4}$ | $1.6042^{+1}$ | $7.4886^{+0}$ | $6.8276^{+0}$ | $5.3044^{+0}$ | $8.9013^{+1}$ |
| $10^{-3}$ | $6.0912^{-2}$ | $3.5030^{-2}$ | $1.7393^{-3}$ | $6.5544^{+0}$ | $2.8656^{+0}$ | $2.4802^{+0}$ | $1.8362^{+0}$ | $3.0617^{+1}$ |
| $10^{-2}$ | $2.3327^{-1}$ | $1.3022^{-1}$ | $5.7588^{-3}$ | $2.2949^{+0}$ | $8.6723^{-1}$ | $6.7688^{-1}$ | $4.5597^{-1}$ | $7.8243^{+0}_{*}$ |
| 0.1 | $5.4798^{-1}$ | $2.6905^{-1}$ | $9.9470^{-3}$ | $3.8192^{-1}$ | $1.1124^{-1}$ | $6.7091^{-2}$ | $3.7698^{-2}$ | $8.4908^{-1}$ |
| 0.3 | $3.4291^{-1}$ | $1.2693^{-1}$ | $2.9239^{-3}$ | $3.4069^{-2}$ | $8.6867^{-3}$ | $4.3924^{-3}$ | $2.1435^{-3}$ | $7.8109^{-2}$ |
| 0.5 | $1.1694^{-1}$ | $3.0310^{-2}$ | $3.6112^{-4}$ | $2.2828^{-3}$ | $5.5537^{-4}$ | $2.7744^{-4}$ | $1.2416^{-4}$ | $7.5371^{-3}$ |
| 0.7 | $1.9076^{-2}$ | $2.9217^{-3}$ | $1.2635^{-5}$ | $5.2061^{-5}$ | $1.2677^{-5}$ | $7.2083^{-6}$ | $3.0908^{-6}$ | $3.6441^{-4}$ |
| 0.9 | $3.2404^{-4}$ | $1.6333^{-5}$ | $7.900^{-9}$ | $2.850^{-8}$ | $8.407^{-9}$ | $6.795^{-9}$ | $3.205^{-9}$ | $1.1411^{-6}$ |

**Table 4:** Reference results for the $N_{\mathrm{f}} = 4$ (FFN) and the variable-$N_{\mathrm{f}}$ (VFN) polarized leading-order evolution of the initial distributions (4.20), shown together with these boundary conditions. The respective values for $\alpha_{\mathrm{s}}(\mu_{\mathrm{r}}^2 = \mu_{\mathrm{f}}^2 = 10^4\ \mathrm{GeV}^2)$ read 0.117574 (FFN) and 0.122306 (VFN). The notation is the same as for the unpolarized case.

| $x$ | $xu_v$ | $-xd_v$ | $-xL_-$ | $-2xL_+$ | $xs_+$ | $xc_+$ | $xb_+$ | $xg$ |
|---|---|---|---|---|---|---|---|---|
| **Pol. input, $\mu_{\mathrm{f}}^2 = 2\ \mathrm{GeV}^2$** | | | | | | | | |
| $10^{-7}$ | $1.6366^{-5}$ | $6.2946^{-6}$ | $7.9433^{-5}$ | $1.5887^{-3}$ | $-3.9716^{-4}$ | $0.0^{+0}$ | $0.0^{+0}$ | $4.7434^{-4}$ |
| $10^{-6}$ | $8.2024^{-5}$ | $3.1548^{-5}$ | $1.5849^{-4}$ | $3.1698^{-3}$ | $-7.9244^{-4}$ | $0.0^{+0}$ | $0.0^{+0}$ | $1.5000^{-3}$ |
| $10^{-5}$ | $4.1110^{-4}$ | $1.5811^{-4}$ | $3.1621^{-4}$ | $6.3241^{-3}$ | $-1.5810^{-3}$ | $0.0^{+0}$ | $0.0^{+0}$ | $4.7432^{-3}$ |
| $10^{-4}$ | $2.0604^{-3}$ | $7.9245^{-4}$ | $6.3052^{-4}$ | $1.2610^{-2}$ | $-3.1526^{-3}$ | $0.0^{+0}$ | $0.0^{+0}$ | $1.4993^{-2}$ |
| $10^{-3}$ | $1.0326^{-2}$ | $3.9716^{-3}$ | $1.2501^{-3}$ | $2.5003^{-2}$ | $-6.2507^{-3}$ | $0.0^{+0}$ | $0.0^{+0}$ | $4.7197^{-2}$ |
| $10^{-2}$ | $5.1723^{-2}$ | $1.9886^{-2}$ | $2.3412^{-3}$ | $4.6825^{-2}$ | $-1.1706^{-2}$ | $0.0^{+0}$ | $0.0^{+0}$ | $1.4265^{-1}$ |
| 0.1 | $2.4582^{-1}$ | $9.1636^{-2}$ | $2.3972^{-3}$ | $4.7943^{-2}$ | $-1.1986^{-2}$ | $0.0^{+0}$ | $0.0^{+0}$ | $2.8009^{-1}$ |
| 0.3 | $3.6473^{-1}$ | $1.1370^{-1}$ | $5.7388^{-4}$ | $1.1478^{-2}$ | $-2.8694^{-3}$ | $0.0^{+0}$ | $0.0^{+0}$ | $1.3808^{-1}$ |
| 0.5 | $2.5008^{-1}$ | $5.7710^{-2}$ | $6.3457^{-5}$ | $1.2691^{-3}$ | $-3.1729^{-4}$ | $0.0^{+0}$ | $0.0^{+0}$ | $3.3146^{-2}$ |
| 0.7 | $8.4769^{-2}$ | $1.1990^{-2}$ | $1.9651^{-6}$ | $3.9301^{-5}$ | $-9.8254^{-6}$ | $0.0^{+0}$ | $0.0^{+0}$ | $3.0496^{-3}$ |
| 0.9 | $4.4680^{-3}$ | $2.1365^{-4}$ | $9.689^{-10}$ | $1.9378^{-8}$ | $-4.8444^{-9}$ | $0.0^{+0}$ | $0.0^{+0}$ | $1.4230^{-5}$ |
| **LO, $N_{\mathrm{f}} = 4\,,\ \mu_{\mathrm{f}}^2 = 10^4\ \mathrm{GeV}^2$** | | | | | | | | |
| $10^{-7}$ | $4.8350^{-5}_{*}$ | $1.8556^{-5}$ | $1.0385^{-4}$ | $3.5124^{-3}$ | $-1.2370^{-3}$ | $-7.1774^{-4}$ | $0.0^{+0}$ | $1.4116^{-2}$ |
| $10^{-6}$ | $2.3504^{-4}$ | $9.0090^{-5}$ | $2.0700^{-4}$ | $7.7716^{-3}$ | $-2.8508^{-3}$ | $-1.8158^{-3}$ | $0.0^{+0}$ | $4.2163^{-2}$ |
| $10^{-5}$ | $1.1220^{-3}$ | $4.2916^{-4}$ | $4.1147^{-4}$ | $1.6007^{-2}$ | $-5.9463^{-3}$ | $-3.8889^{-3}$ | $0.0^{+0}$ | $1.0922^{-1}$ |
| $10^{-4}$ | $5.1990^{-3}$ | $1.9818^{-3}$ | $8.0948^{-4}$ | $2.8757^{-2}$ | $-1.0331^{-2}$ | $-6.2836^{-3}$ | $0.0^{+0}$ | $2.4069^{-1}$ |
| $10^{-3}$ | $2.2900^{-2}$ | $8.6763^{-3}$ | $1.5309^{-3}$ | $4.0166^{-2}$ | $-1.2428^{-2}$ | $-4.7739^{-3}$ | $0.0^{+0}$ | $4.2181^{-1}$ |
| $10^{-2}$ | $9.1489^{-2}$ | $3.4200^{-2}$ | $2.4502^{-3}$ | $3.3928^{-2}$ | $-4.7126^{-3}$ | $7.5385^{-3}$ | $0.0^{+0}$ | $4.9485^{-1}$ |
| 0.1 | $2.6494^{-1}$ | $9.1898^{-2}$ | $1.5309^{-3}$ | $8.5427^{-3}$ | $3.3830^{-3}$ | $1.1037^{-2}$ | $0.0^{+0}$ | $2.0503^{-1}$ |
| 0.3 | $2.2668^{-1}$ | $6.2946^{-2}$ | $2.1104^{-4}$ | $6.6698^{-4}$ | $7.2173^{-4}$ | $1.7769^{-3}$ | $0.0^{+0}$ | $3.3980^{-2}$ |
| 0.5 | $9.7647^{-2}$ | $1.9652^{-2}$ | $1.4789^{-5}$ | $-1.8850^{-5}$ | $8.3371^{-5}$ | $1.5732^{-4}$ | $0.0^{+0}$ | $4.3802^{-3}$ |
| 0.7 | $1.9545^{-2}$ | $2.3809^{-3}$ | $2.7279^{-7}$ | $-4.1807^{-6}$ | $3.4543^{-6}$ | $4.8183^{-6}$ | $0.0^{+0}$ | $2.6355^{-4}$ |
| 0.9 | $4.1768^{-4}$ | $1.7059^{-5}$ | $5.494^{-11}$ | $-7.6712^{-9}$ | $4.1103^{-9}$ | $4.3850^{-9}$ | $0.0^{+0}$ | $9.8421^{-7}$ |
| **LO, $N_{\mathrm{f}} = 3\ldots5\,,\ \mu_{\mathrm{f}}^2 = 10^4\ \mathrm{GeV}^2$** | | | | | | | | |
| $10^{-7}$ | $4.9026^{-5}$ | $1.8815^{-5}$ | $1.0422^{-4}$ | $3.5315^{-3}$ | $-1.2447^{-3}$ | $-7.2356^{-4}$ | $-6.2276^{-4}$ | $1.3726^{-2}$ |
| $10^{-6}$ | $2.3818^{-4}$ | $9.1286^{-5}$ | $2.0774^{-4}$ | $7.8108^{-3}$ | $-2.8667^{-3}$ | $-1.8280^{-3}$ | $-1.5301^{-3}$ | $4.1011^{-2}$ |
| $10^{-5}$ | $1.1359^{-3}$ | $4.3445^{-4}$ | $4.1289^{-4}$ | $1.6070^{-2}$ | $-5.9705^{-3}$ | $-3.9060^{-3}$ | $-3.1196^{-3}$ | $1.0615^{-1}$ |
| $10^{-4}$ | $5.2567^{-3}$ | $2.0035^{-3}$ | $8.1206^{-4}$ | $2.8811^{-2}$ | $-1.0345^{-2}$ | $-6.2849^{-3}$ | $-4.5871^{-3}$ | $2.3343^{-1}$ |
| $10^{-3}$ | $2.3109^{-2}$ | $8.7537^{-3}$ | $1.5345^{-3}$ | $4.0125^{-2}$ | $-1.2390^{-2}$ | $-4.7174^{-3}$ | $-2.4822^{-3}$ | $4.0743^{-1}$ |
| $10^{-2}$ | $9.2035^{-2}$ | $3.4391^{-2}$ | $2.4501^{-3}$ | $3.3804^{-2}$ | $-4.6512^{-3}$ | $7.5994^{-3}$ | $6.4665^{-3}$ | $4.7445^{-1}$ |
| 0.1 | $2.6478^{-1}$ | $9.1762^{-2}$ | $1.5206^{-3}$ | $8.5181^{-3}$ | $3.3438^{-3}$ | $1.0947^{-2}$ | $6.5223^{-3}$ | $1.9402^{-1}$ |
| 0.3 | $2.2495^{-1}$ | $6.2376^{-2}$ | $2.0811^{-4}$ | $6.6195^{-4}$ | $7.0957^{-4}$ | $1.7501^{-3}$ | $9.2045^{-4}$ | $3.1960^{-2}$ |
| 0.5 | $9.6318^{-2}$ | $1.9353^{-2}$ | $1.4496^{-5}$ | $-1.8549^{-5}$ | $8.1756^{-5}$ | $1.5424^{-4}$ | $7.8577^{-5}$ | $4.1226^{-3}$ |
| 0.7 | $1.9147^{-2}$ | $2.3281^{-3}$ | $2.6556^{-7}$ | $-4.0936^{-6}$ | $3.3746^{-6}$ | $4.7024^{-6}$ | $2.4901^{-6}$ | $2.4888^{-4}$ |
| 0.9 | $4.0430^{-4}$ | $1.6480^{-5}$ | $5.285^{-11}$ | $-7.4351^{-9}$ | $3.9818^{-9}$ | $4.2460^{-9}$ | $2.6319^{-9}$ | $9.2939^{-7}$ |

**Table 5:** Reference results for the polarized next-to-leading-order polarized evolution of the initial distributions (4.20) with $N_\mathrm{f} = 4$ quark flavours. The corresponding value of the strong coupling is $\alpha_\mathrm{s}(\mu_\mathrm{r}^2 = 10^4\ \mathrm{GeV}^2) = 0.110902$. As in the leading-order case, the valence distributions $s_v$ and $c_v$ vanish for the input (4.20).

| Pol. NLO, $N_\mathrm{f} = 4$, $\mu_\mathrm{f}^2 = 10^4\ \mathrm{GeV}^2$ | | | | | | | |
|---|---|---|---|---|---|---|---|
| $x$ | $xu_v$ | $xd_v$ | $xL_-$ | $2xL_+$ | $xs_+$ | $xc_+$ | $xg$ |
| $\mu_\mathrm{r}^2 = \mu_\mathrm{f}^2$ | | | | | | | |
| $10^{-7}$ | $6.7336^{-5}$ | $-2.5747^{-5}$ | $-1.1434^{-4}$ | $-5.2002^{-3}$ | $-2.0528^{-3}$ | $-1.5034^{-3}$ | $2.6955^{-2}$ |
| $10^{-6}$ | $3.1280^{-4}$ | $-1.1938^{-4}$ | $-2.3497^{-4}$ | $-1.0725^{-2}$ | $-4.2774^{-3}$ | $-3.1845^{-3}$ | $6.5928^{-2}$ |
| $10^{-5}$ | $1.4180^{-3}$ | $-5.3982^{-4}$ | $-4.8579^{-4}$ | $-1.9994^{-2}$ | $-7.8594^{-3}$ | $-5.6970^{-3}$ | $1.4414^{-1}$ |
| $10^{-4}$ | $6.2085^{-3}$ | $-2.3546^{-3}$ | $-9.8473^{-4}$ | $-3.1788^{-2}$ | $-1.1749^{-2}$ | $-7.5376^{-3}$ | $2.7537^{-1}$ |
| $10^{-3}$ | $2.5741^{-2}$ | $-9.7004^{-3}$ | $-1.8276^{-3}$ | $-3.8222^{-2}$ | $-1.1427^{-2}$ | $-3.6138^{-3}$ | $4.3388^{-1}$ |
| $10^{-2}$ | $9.6288^{-2}$ | $-3.5778^{-2}$ | $-2.6427^{-3}$ | $-2.6437^{-2}$ | $-1.2328^{-3}$ | $1.0869^{-2}$ | $4.8281^{-1}$ |
| 0.1 | $2.5843^{-1}$ | $-8.9093^{-2}$ | $-1.4593^{-3}$ | $-7.5546^{-3}$ | $3.4258^{-3}$ | $1.0639^{-2}$ | $2.0096^{-1}$ |
| 0.3 | $2.1248^{-1}$ | $-5.8641^{-2}$ | $-1.9269^{-4}$ | $-1.2210^{-3}$ | $3.5155^{-4}$ | $1.3138^{-3}$ | $3.4126^{-2}$ |
| 0.5 | $8.9180^{-2}$ | $-1.7817^{-2}$ | $-1.3125^{-5}$ | $-9.1573^{-5}$ | $1.9823^{-5}$ | $8.5435^{-5}$ | $4.5803^{-3}$ |
| 0.7 | $1.7300^{-2}$ | $-2.0885^{-3}$ | $-2.3388^{-7}$ | $-1.9691^{-6}$ | $1.8480^{-7}$ | $1.3541^{-6}$ | $2.9526^{-4}$ |
| 0.9 | $3.4726^{-4}$ | $-1.4028^{-5}$ | $-4.407^{-11}$ | $-4.247^{-9}$ | $-1.903^{-9}$ | $-1.683^{-9}$ | $1.2520^{-6}$ |
| $\mu_\mathrm{r}^2 = 2\,\mu_\mathrm{f}^2$ | | | | | | | |
| $10^{-7}$ | $6.1781^{-5}$ | $-2.3641^{-5}$ | $-1.1137^{-4}$ | $-4.6947^{-3}$ | $-1.8092^{-3}$ | $-1.2695^{-3}$ | $2.2530^{-2}$ |
| $10^{-6}$ | $2.8974^{-4}$ | $-1.1068^{-4}$ | $-2.2755^{-4}$ | $-9.8528^{-3}$ | $-3.8580^{-3}$ | $-2.7838^{-3}$ | $5.7272_*^{-2}$ |
| $10^{-5}$ | $1.3281^{-3}$ | $-5.0612^{-4}$ | $-4.6740^{-4}$ | $-1.8799^{-2}$ | $-7.2908^{-3}$ | $-5.1629^{-3}$ | $1.2975^{-1}$ |
| $10^{-4}$ | $5.8891^{-3}$ | $-2.2361^{-3}$ | $-9.4412^{-4}$ | $-3.0787^{-2}$ | $-1.1292^{-2}$ | $-7.1363^{-3}$ | $2.5644^{-1}$ |
| $10^{-3}$ | $2.4777^{-2}$ | $-9.3502^{-3}$ | $-1.7632^{-3}$ | $-3.8610^{-2}$ | $-1.1658^{-2}$ | $-3.9083^{-3}$ | $4.1725^{-1}$ |
| $10^{-2}$ | $9.4371^{-2}$ | $-3.5129^{-2}$ | $-2.6087^{-3}$ | $-2.8767^{-2}$ | $-2.3430_*^{-3}$ | $9.7922_*^{-3}$ | $4.7804^{-1}$ |
| 0.1 | $2.6008^{-1}$ | $-8.9915^{-2}$ | $-1.4923^{-3}$ | $-8.3806^{-3}$ | $3.1932^{-3}$ | $1.0585^{-2}$ | $2.0495^{-1}$ |
| 0.3 | $2.1837^{-1}$ | $-6.0497^{-2}$ | $-2.0143^{-4}$ | $-1.2157^{-3}$ | $3.9810^{-4}$ | $1.4042^{-3}$ | $3.5366^{-2}$ |
| 0.5 | $9.3169^{-2}$ | $-1.8699^{-2}$ | $-1.3954^{-5}$ | $-7.9331^{-5}$ | $3.0091^{-5}$ | $9.9849^{-5}$ | $4.7690^{-3}$ |
| 0.7 | $1.8423^{-2}$ | $-2.2357^{-3}$ | $-2.5360^{-7}$ | $-1.0062^{-6}$ | $7.6483^{-7}$ | $2.0328^{-6}$ | $3.0796^{-4}$ |
| 0.9 | $3.8293^{-4}$ | $-1.5559^{-5}$ | $-4.952^{-11}$ | $-1.955^{-9}$ | $-7.298^{-10}$ | $-4.822^{-10}$ | $1.3247^{-6}$ |
| $\mu_\mathrm{r}^2 = 1/2\,\mu_\mathrm{f}^2$ | | | | | | | |
| $10^{-7}$ | $7.4443^{-5}$ | $-2.8435^{-5}$ | $-1.1815^{-4}$ | $-5.7829^{-3}$ | $-2.3341^{-3}$ | $-1.7739^{-3}$ | $3.2071^{-2}$ |
| $10^{-6}$ | $3.4143^{-4}$ | $-1.3016^{-4}$ | $-2.4482^{-4}$ | $-1.1668^{-2}$ | $-4.7305^{-3}$ | $-3.6168^{-3}$ | $7.5123^{-2}$ |
| $10^{-5}$ | $1.5256^{-3}$ | $-5.8002^{-4}$ | $-5.1085^{-4}$ | $-2.1193^{-2}$ | $-8.4295^{-3}$ | $-6.2295_*^{-3}$ | $1.5788^{-1}$ |
| $10^{-4}$ | $6.5726^{-3}$ | $-2.4891^{-3}$ | $-1.0409^{-3}$ | $-3.2697^{-2}$ | $-1.2166^{-2}$ | $-7.8952^{-3}$ | $2.9079^{-1}$ |
| $10^{-3}$ | $2.6766^{-2}$ | $-1.0070^{-2}$ | $-1.9171^{-3}$ | $-3.7730^{-2}$ | $-1.1160^{-2}$ | $-3.2890_*^{-3}$ | $4.4380^{-1}$ |
| $10^{-2}$ | $9.8073^{-2}$ | $-3.6370^{-2}$ | $-2.6942^{-3}$ | $-2.4056^{-2}$ | $-1.2354_*^{-4}$ | $1.1929^{-2}$ | $4.8272^{-1}$ |
| 0.1 | $2.5628^{-1}$ | $-8.8133^{-2}$ | $-1.4304^{-3}$ | $-6.9572^{-3}$ | $3.5561^{-3}$ | $1.0604^{-2}$ | $1.9831^{-1}$ |
| 0.3 | $2.0709^{-1}$ | $-5.6988^{-2}$ | $-1.8541^{-4}$ | $-1.3308^{-3}$ | $2.5993^{-4}$ | $1.1855^{-3}$ | $3.3524^{-2}$ |
| 0.5 | $8.5835^{-2}$ | $-1.7089^{-2}$ | $-1.2463^{-5}$ | $-1.1920^{-4}$ | $2.6972_*^{-6}$ | $6.4995^{-5}$ | $4.5044^{-3}$ |
| 0.7 | $1.6405^{-2}$ | $-1.9723^{-3}$ | $-2.1859_*^{-7}$ | $-3.6817^{-6}$ | $-7.4795_*^{-7}$ | $3.4496^{-7}$ | $2.9100^{-4}$ |
| 0.9 | $3.2011^{-4}$ | $-1.2870^{-5}$ | $-4.000^{-11}$ | $-8.173^{-9}$ | $-3.886^{-9}$ | $-3.686^{-9}$ | $1.2230^{-6}$ |

**Table 6:** As Table 5, but for the variable-$N_{\rm f}$ evolution using Eqs. (4.16), (4.17) and (4.20). The corresponding values for the strong coupling $\alpha_{\rm s}(\mu_{\rm r}^2 = 10^4\ {\rm GeV}^2)$ are given by 0.116461, 0.116032 and 0.115663 for $\mu_{\rm r}^2/\mu_{\rm f}^2 = 0.5$, 1 and 2, respectively.

| Pol. NLO, $N_{\rm f} = 3\ \ldots 5\,,\ \mu_{\rm f}^2 = 10^4\ {\rm GeV}^2$ | | | | | | | | |
|---|---|---|---|---|---|---|---|---|
| $x$ | $xu_v$ | $-xd_v$ | $-xL_-$ | $-2xL_+$ | $xs_+$ | $xc_+$ | $xb_+$ | $xg$ |
| $\mu_{\rm r}^2 = \mu_{\rm f}^2$ | | | | | | | | |
| $10^{-7}$ | $6.8787_*^{-5}$ | $2.6297^{-5}$ | $1.1496^{-4}$ | $5.2176^{-3}$ | $-2.0592^{-3}$ | $-1.5076^{-3}$ | $-1.2411^{-3}$ | $2.5681^{-2}$ |
| $10^{-6}$ | $3.1881^{-4}$ | $1.2165^{-4}$ | $2.3638^{-4}$ | $1.0770^{-2}$ | $-4.2953^{-3}$ | $-3.1979^{-3}$ | $-2.4951^{-3}$ | $6.3021^{-2}$ |
| $10^{-5}$ | $1.4413^{-3}$ | $5.4856^{-4}$ | $4.8893^{-4}$ | $2.0077^{-2}$ | $-7.8934^{-3}$ | $-5.7228^{-3}$ | $-4.1488^{-3}$ | $1.3809^{-1}$ |
| $10^{-4}$ | $6.2902^{-3}$ | $2.3849^{-3}$ | $9.9100^{-4}$ | $3.1883^{-2}$ | $-1.1785^{-2}$ | $-7.5596^{-3}$ | $-4.8420^{-3}$ | $2.6411^{-1}$ |
| $10^{-3}$ | $2.5980^{-2}$ | $9.7872^{-3}$ | $1.8364^{-3}$ | $3.8224^{-2}$ | $-1.1416^{-2}$ | $-3.5879^{-3}$ | $-1.1723^{-3}$ | $4.1601^{-1}$ |
| $10^{-2}$ | $9.6750^{-2}$ | $3.5935^{-2}$ | $2.6452^{-3}$ | $2.6306^{-2}$ | $-1.1774^{-3}$ | $1.0917^{-2}$ | $8.1196^{-3}$ | $4.6178^{-1}$ |
| 0.1 | $2.5807^{-1}$ | $8.8905^{-2}$ | $1.4509^{-3}$ | $7.4778^{-3}$ | $3.4207^{-3}$ | $1.0591^{-2}$ | $6.1480^{-3}$ | $1.9143^{-1}$ |
| 0.3 | $2.1104^{-1}$ | $5.8186^{-2}$ | $1.9054^{-4}$ | $1.2026^{-3}$ | $3.4999^{-4}$ | $1.3015^{-3}$ | $7.2795^{-4}$ | $3.2621^{-2}$ |
| 0.5 | $8.8199^{-2}$ | $1.7601^{-2}$ | $1.2924^{-5}$ | $8.9668^{-5}$ | $1.9771^{-5}$ | $8.4378^{-5}$ | $5.2125^{-5}$ | $4.4207^{-3}$ |
| 0.7 | $1.7027^{-2}$ | $2.0531^{-3}$ | $2.2921^{-7}$ | $1.9243^{-6}$ | $1.8384^{-7}$ | $1.3298^{-6}$ | $1.2157^{-6}$ | $2.8887^{-4}$ |
| 0.9 | $3.3898^{-4}$ | $1.3676^{-5}$ | $4.284^{-11}$ | $4.260^{-9}$ | $-1.916^{-9}$ | $-1.701^{-9}$ | $-7.492^{-11}$ | $1.2435^{-6}$ |
| $\mu_{\rm r}^2 = 2\,\mu_{\rm f}^2$ | | | | | | | | |
| $10^{-7}$ | $6.2819_*^{-5}$ | $2.4035^{-5}$ | $1.1180^{-4}$ | $4.6896^{-3}$ | $-1.8050^{-3}$ | $-1.2637^{-3}$ | $-1.0544^{-3}$ | $2.1305^{-2}$ |
| $10^{-6}$ | $2.9408^{-4}$ | $1.1232^{-4}$ | $2.2855^{-4}$ | $9.8538^{-3}$ | $-3.8554^{-3}$ | $-2.7780^{-3}$ | $-2.2077^{-3}$ | $5.4411^{-2}$ |
| $10^{-5}$ | $1.3450^{-3}$ | $5.1245^{-4}$ | $4.6965^{-4}$ | $1.8815^{-2}$ | $-7.2936^{-3}$ | $-5.1597^{-3}$ | $-3.8359^{-3}$ | $1.2368^{-1}$ |
| $10^{-4}$ | $5.9485^{-3}$ | $2.2582^{-3}$ | $9.4866^{-4}$ | $3.0816^{-2}$ | $-1.1297^{-2}$ | $-7.1323^{-3}$ | $-4.7404^{-3}$ | $2.4503^{-1}$ |
| $10^{-3}$ | $2.4951^{-2}$ | $9.4134^{-3}$ | $1.7698^{-3}$ | $3.8618^{-2}$ | $-1.1654^{-2}$ | $-3.8925^{-3}$ | $-1.5608^{-3}$ | $3.9912^{-1}$ |
| $10^{-2}$ | $9.4706^{-2}$ | $3.5243^{-2}$ | $2.6108^{-3}$ | $2.8761^{-2}$ | $-2.3471^{-3}$ | $9.7827^{-3}$ | $7.5188^{-3}$ | $4.5698^{-1}$ |
| 0.1 | $2.5982^{-1}$ | $8.9780^{-2}$ | $1.4862^{-3}$ | $8.3807^{-3}$ | $3.1615^{-3}$ | $1.0522^{-2}$ | $6.1973^{-3}$ | $1.9561^{-1}$ |
| 0.3 | $2.1732^{-1}$ | $6.0165^{-2}$ | $1.9984^{-4}$ | $1.2086^{-3}$ | $3.9371^{-4}$ | $1.3919^{-3}$ | $7.6929^{-4}$ | $3.3906^{-2}$ |
| 0.5 | $9.2445^{-2}$ | $1.8539^{-2}$ | $1.3804^{-5}$ | $7.8411^{-5}$ | $2.9799^{-5}$ | $9.8805^{-5}$ | $5.7333^{-5}$ | $4.6166^{-3}$ |
| 0.7 | $1.8219^{-2}$ | $2.2090^{-3}$ | $2.5004_*^{-7}$ | $9.8927_*^{-7}$ | $7.5552^{-7}$ | $2.0057^{-6}$ | $1.4438^{-6}$ | $3.0231^{-4}$ |
| 0.9 | $3.7653^{-4}$ | $1.5285^{-5}$ | $4.855^{-11}$ | $2.005^{-9}$ | $-7.599^{-10}$ | $-5.171^{-10}$ | $3.809^{-10}$ | $1.3232^{-6}$ |
| $\mu_{\rm r}^2 = 1/2\,\mu_{\rm f}^2$ | | | | | | | | |
| $10^{-7}$ | $7.6699^{-5}$ | $2.9289^{-5}$ | $1.1912^{-4}$ | $5.8548^{-3}$ | $-2.3667^{-3}$ | $-1.8030^{-3}$ | $-1.4521^{-3}$ | $3.1009^{-2}$ |
| $10^{-6}$ | $3.5067^{-4}$ | $1.3364^{-4}$ | $2.4707^{-4}$ | $1.1806^{-2}$ | $-4.7934^{-3}$ | $-3.6731^{-3}$ | $-2.7846^{-3}$ | $7.2690^{-2}$ |
| $10^{-5}$ | $1.5611^{-3}$ | $5.9329^{-4}$ | $5.1593^{-4}$ | $2.1406^{-2}$ | $-8.5248^{-3}$ | $-6.3125^{-3}$ | $-4.4072^{-3}$ | $1.5274^{-1}$ |
| $10^{-4}$ | $6.6957^{-3}$ | $2.5346^{-3}$ | $1.0509^{-3}$ | $3.2903^{-2}$ | $-1.2252^{-2}$ | $-7.9608^{-3}$ | $-4.8402^{-3}$ | $2.8097^{-1}$ |
| $10^{-3}$ | $2.7125^{-2}$ | $1.0200^{-2}$ | $1.9310^{-3}$ | $3.7698^{-2}$ | $-1.1127^{-2}$ | $-3.2334^{-3}$ | $-7.5827^{-4}$ | $4.2756^{-1}$ |
| $10^{-2}$ | $9.8758^{-2}$ | $3.6602^{-2}$ | $2.6980^{-3}$ | $2.3675^{-2}$ | $5.1386^{-5}$ | $1.2092^{-2}$ | $8.6053^{-3}$ | $4.6241^{-1}$ |
| 0.1 | $2.5572^{-1}$ | $8.7847^{-2}$ | $1.4179^{-3}$ | $6.7523^{-3}$ | $3.5944^{-3}$ | $1.0578^{-2}$ | $6.0904^{-3}$ | $1.8838^{-1}$ |
| 0.3 | $2.0497^{-1}$ | $5.6318^{-2}$ | $1.8228^{-4}$ | $1.2965^{-3}$ | $2.6142^{-4}$ | $1.1713^{-3}$ | $6.8941^{-4}$ | $3.1884^{-2}$ |
| 0.5 | $8.4404^{-2}$ | $1.6775^{-2}$ | $1.2174^{-5}$ | $1.1604^{-4}$ | $2.8309^{-6}$ | $6.3682^{-5}$ | $4.7009^{-5}$ | $4.3221^{-3}$ |
| 0.7 | $1.6013^{-2}$ | $1.9215^{-3}$ | $2.1196_*^{-7}$ | $3.6047^{-6}$ | $-7.4260^{-7}$ | $3.1714^{-7}$ | $9.6419^{-7}$ | $2.8268^{-4}$ |
| 0.9 | $3.0848^{-4}$ | $1.2377^{-5}$ | $3.829^{-11}$ | $8.129^{-9}$ | $-3.873^{-9}$ | $-3.681^{-9}$ | $-6.816^{-10}$ | $1.2009^{-6}$ |

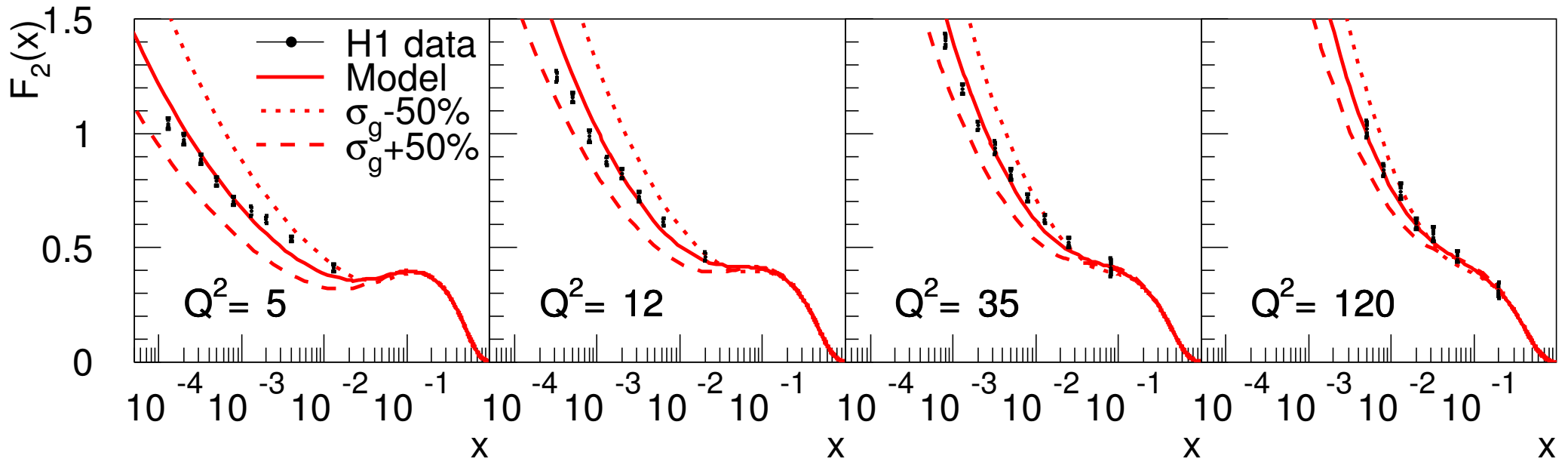

**Fig. 13:** $F_2(x, Q^2)$ from H1 compared to the model with $\pm 50\%$ variation of the width parameter $\sigma_g$ of the gluon distribution.

the model has only four shape parameters and three normalization parameters, plus the starting scale:

$$\begin{array}{cccc} \sigma_u = 230 \text{ MeV} & \sigma_d = 170 \text{ MeV} & \sigma_g = 77 \text{ MeV} & \sigma_H = 100 \text{ MeV} \\ \alpha^2_{\mathrm{p}\pi^0} = 0.45 & \alpha^2_{n\pi^+} = 0.14 & \alpha^2_{\Lambda K} = 0.05 & Q_0 = 0.75 \text{ GeV} \end{array} \tag{5.22}$$

These are determined from fits to data as detailed in [70] and illustrated in Fig. 13. The model reproduces the inclusive proton structure function and gives a natural explanation of observed quark asymmetries, such as the difference between the up and down valence distributions and between the anti-up and anti-down sea quark distributions. Moreover, its asymmetry in the momentum distribution of strange and anti-strange quarks in the nucleon is large enough to reduce the NuTeV anomaly to a level which does not give a significant indication of physics beyong the Standard Model.

Recent fits of PDF's at very low $x$ and $Q^2$ have revealed problems with the gluon density, which in some cases even becomes negative. The reason for this is that the DGLAP evolution, driven primarily by the gluon at small $x$, otherwise gives too large parton densities and thereby a poor fit to $F_2$ in the genuine DIS region at larger $Q^2$. It has been argued [71] that the root of the problem is the application of the formalism for DIS also in the low-$Q^2$ region, where the momentum transfer is not large enough that the parton structure of the proton is clearly resolved. The smallest distance that can be resolved is basically given by the momentum transfer of the exchanged photon through $d = 0.2/\sqrt{Q^2}$, where $d$ is in Fermi if $Q^2$ is in GeV$^2$. This indicates that partons are resolved only for $Q^2 \gtrsim 1\,\text{GeV}^2$. For $Q^2 \lesssim 1\,\text{GeV}^2$, there is no hard scale involved and a parton basis for the description is not justified. Instead, the interaction is here of a soft kind between the nearly on-shell photon and the proton. The cross section is then dominated by the process where the photon fluctuates into a virtual vector meson state which then interacts with the proton in a strong interaction. The quantum state of the photon can be expressed as $|\gamma\rangle = C_0|\gamma_0\rangle + \sum_V \frac{e}{f_V}|V\rangle + \int_{m_0} dm(\cdots)$. The sum is over $V = \rho^0, \omega, \phi \ldots$ as in the original vector meson dominance model (VDM), whereas the generalised vector meson dominance model (GVDM) also includes the integral over a continuous mass spectrum (not written out explicitly here).

Applied to $ep$ at low $Q^2$ this leads to the expression [71]

$$\begin{aligned} F_2(x, Q^2) \;=\; & \frac{(1-x)Q^2}{4\pi^2\alpha} \left\{ \sum_{V=\rho,\omega,\phi} r_V \left( \frac{m_V^2}{Q^2 + m_V^2} \right)^2 \left( 1 + \xi_V \frac{Q^2}{m_V^2} \right) \right. \\ & \left. + \, r_C \left[ (1-\xi_C) \frac{m_0^2}{Q^2 + m_0^2} + \xi_C \frac{m_0^2}{Q^2} \ln\left(1 + \frac{Q^2}{m_0^2}\right) \right] \right\} A_\gamma \frac{Q^{2\epsilon}}{x^\epsilon} \end{aligned} \tag{5.23}$$

where the hadronic cross-section $\sigma(ip \to X) = A_i s^\epsilon + B_i s^{-\eta} \approx A_i s^\epsilon \approx A_i(Q^2/x)^\epsilon$ has been used for the small-$x$ region of interest. The parameters involved are all essentially known from GVDM phe-nomenology. With $\epsilon = 0.091$, $\xi = 0.34$, $m_0 = 1.5$ GeV and $A_\gamma = 71\,\mu$b, this GVDM model gives a good fit ($\chi^2/\mathrm{d.o.f.} = 87/66 = 1.3$) as illustrated in Fig. 14. Using this model at very low $Q^2$ in combination with the normal parton density approach at larger $Q^2$ it is possible to obtain a good description of data over the full $Q^2$ range [71]. This involves, however, a phenomenological matching of these two approaches, since a theoretically well justified combination is an unsolved problem.

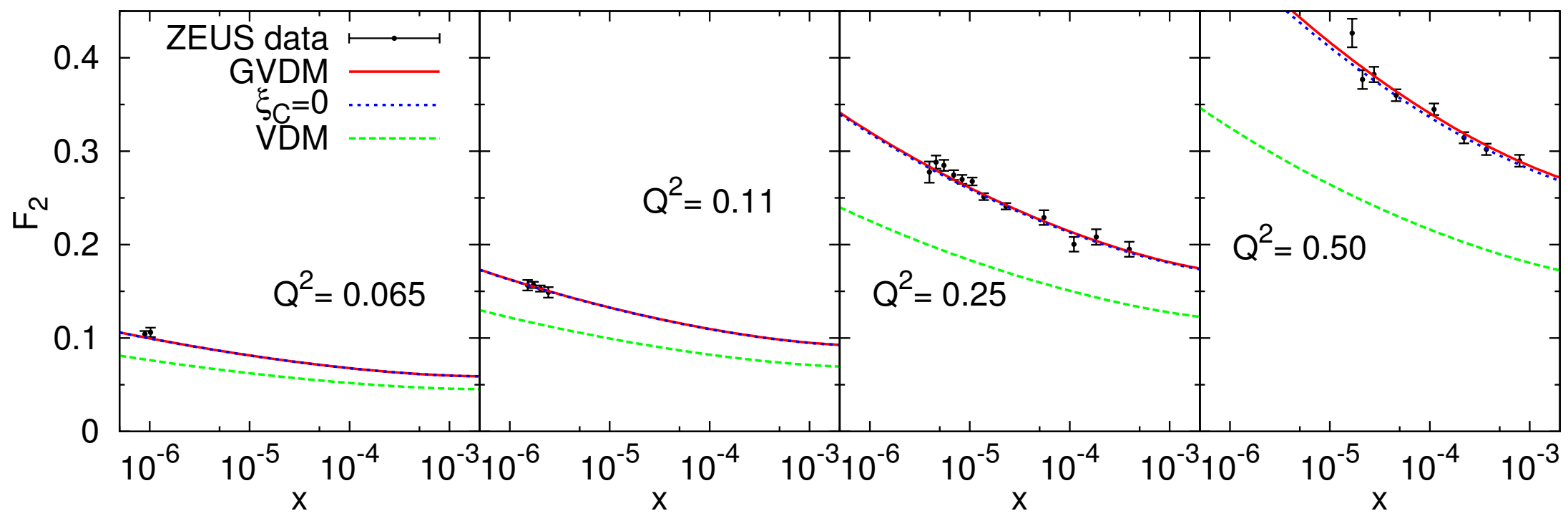

**Fig. 14:** $F_2$ data at low $Q^2$ from ZEUS compared to the full GVDM in eq. (5.23) (full curves), when excluding the longitudinal contribution of the continuum ($\xi_C = 0$) and excluding the continuous contribution altogether (setting $r_C = 0$) giving VDM.

Neglecting the GVDM component when fitting PDF's to data at small $Q^2$ may thus lead to an improper gluon distribution, which is not fully universal and therefore may give incorrect results when used for cross section calculations at LHC.

## 6 Towards precise determination of the nucleon PDFs [7]

The nucleon parton distribution functions (PDFs) available to the moment are extracted from the rather limited set of experimental distributions (the deep-inelastic scattering (DIS) structure functions, the Drell-Yan (DY) and jet production cross sections). Other high-energy processes potentially could provide additional constraints on PDFs, however insufficient theoretical understanding does not allow to use those data without risk of having uncontrolled theoretical inaccuracies. Even for the case of the existing global fits of the PDFs performed by the MRST and CTEQ groups missing next-to-next-to-leading (NNLO) order QCD corrections to the Drell-Yan and jet production cross sections are not small as compared to the accuracy of the corresponding data used and therefore might give non-negligible effect. In this section we outline progress in the QCD fits with consistent account of the NNLO corrections.

### 6.1 Impact of the NNLO evolution kernel fixation on PDFs

In order to allow account of the NNLO corrections in the fit of PDFs one needs analytical expressions for the 3-loop corrections to the QCD evolution kernel. Until recent times these expressions were known only in the approximate form of Ref. [61] derived from the partial information about the kernel, including the set of its Mellin moments and the low-$x$ asymptotics [12,22,23] However with the refined calculations of Ref. [6, 7] the exact expression for the NNLO kernel has been available. These improvement is of particular importance for analysis of the low-$x$ data including the HERA ones due to general rise of the high-order QCD correction at low $x$. We illustrate impact of the NNLO evolution kernel validation on PDFs using the case of fit to the global DIS data [72–77]. The exact NNLO corrections to the DIS coefficient functions are know [4, 78] that allowed to perform approximate NNLO fit of PDFs to these data [69] using the approximate NNLO corrections to the evolution kernel of Ref. [61]. Taking into account exact NNLO evolution kernel the analysis of Ref. [69] was updated recently to the exact NNLO case [79].

[7] Contributing author: S. I. Alekhin

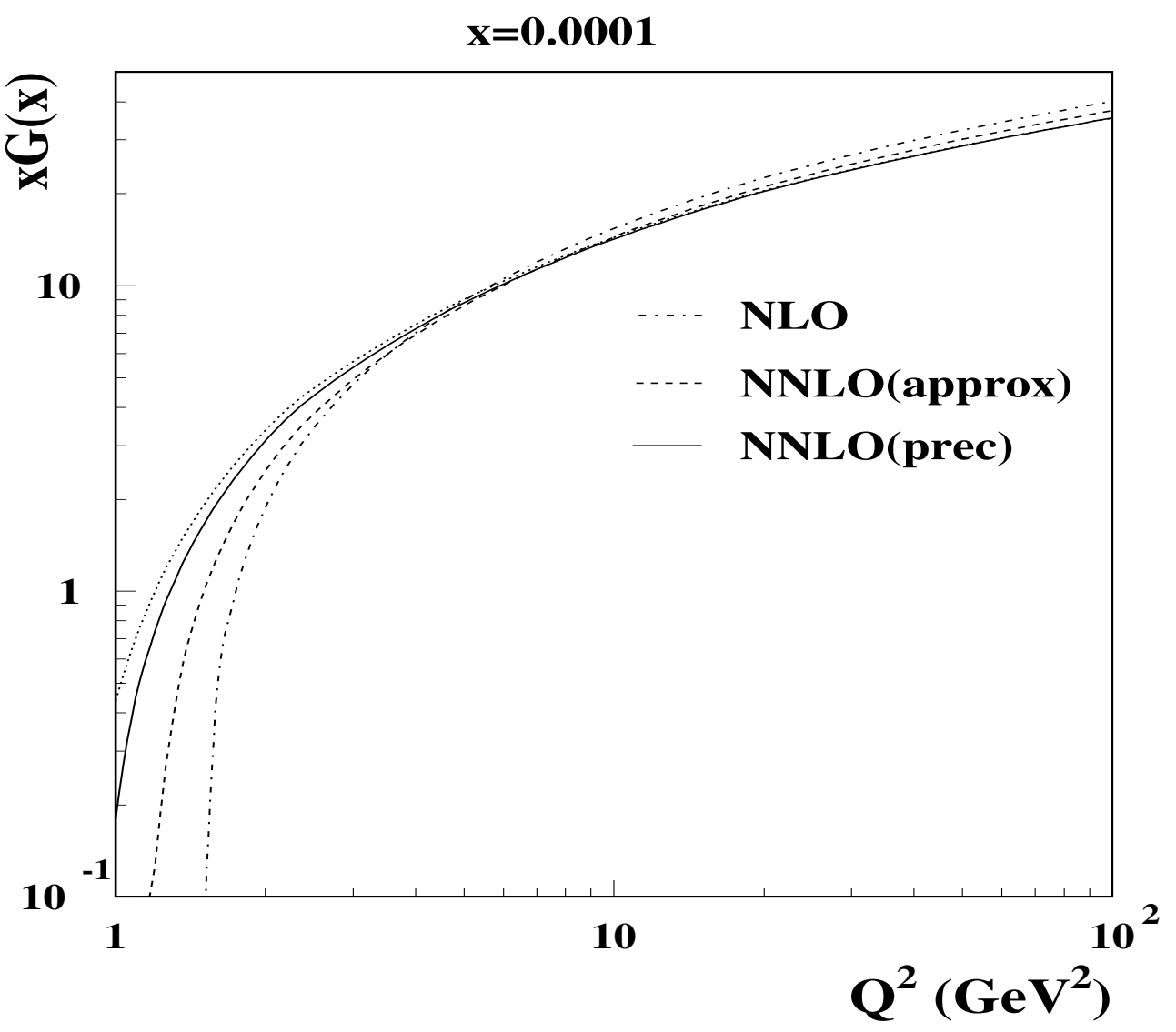

**Fig. 15:** The gluon distributions obtained in the different variants of PDFs fit to the DIS data (solid: the fit with exact NNLO evolution; dashes: the fit with approximate NNLO evolution; dots: the approximate NNLO gluons evolved with the exact NNLO kernel; dashed-dots: the NLO fit).

The gluon distributions at small $x$ obtained in these two variants of the fit are compared in Fig.15. With the exact NNLO corrections the QCD evolution of gluon distribution at small $x$ gets weaker and as a result at small $x/Q$ the gluon distribution obtained using the precise NNLO kernel is quite different from the approximate one. In particular, the approximate NNLO gluon distribution is negative at $Q^2 \lesssim 1.3$ GeV$^2$, while the precise one remains positive even below $Q^2 = 1$ GeV$^2$. For the NLO case the positivity of gluons at small $x/Q$ is even worse than for the approximate NNLO case due to the approximate NNLO corrections dampen the gluon evolution at small $x$ too, therefore account of the NNLO corrections is crucial in this respect. (cf. discussion of Ref. [80]). Positivity of the PDFs is not mandatory beyond the QCD leading order, however it allows probabilistic interpretation of the parton model and facilitates modeling of the soft processes, such as underlying events in the hadron-hadron collisions at LHC. The change of gluon distribution at small $x/Q$ as compared to the fit with approximate NNLO evolution is rather due the change in evolution kernel than due to shift in the fitted parameters of PDFs. This is clear from comparison of the exact NNLO gluon distribution to one obtained from the approximate NNLO fit and evolved to low $Q$ using the exact NNLO kernel (see Fig.15). In the vicinity of crossover in the gluon distribution to the negative values its relative change due to variation of the evolution kernel is quite big and therefore further fixation of the kernel at small $x$ discussed in Ref. [81] might be substantial for validation of the PDFs at low $x/Q$. For the higher-mass kinematics at LHC numerical impact of the NNLO kernel update is not dramatic. Change in the Higgs and $W/Z$ bosons production cross sections due to more precise definition of the NNLO PDFs is comparable to the errors coming from the PDFs uncertainties, i.e. at the level of several percent.

### 6.2 NNLO fit of PDFs to the combined DIS and Drell-Yan data

The DIS process provide very clean source of information about PDFs both from experimental and theoretical side, however very poorly constrains the gluon and sea distributions at $x \gtrsim 0.3$. The well known way to improve precision of the sea distributions is to combine DIS data with the Drell-Yan ones.

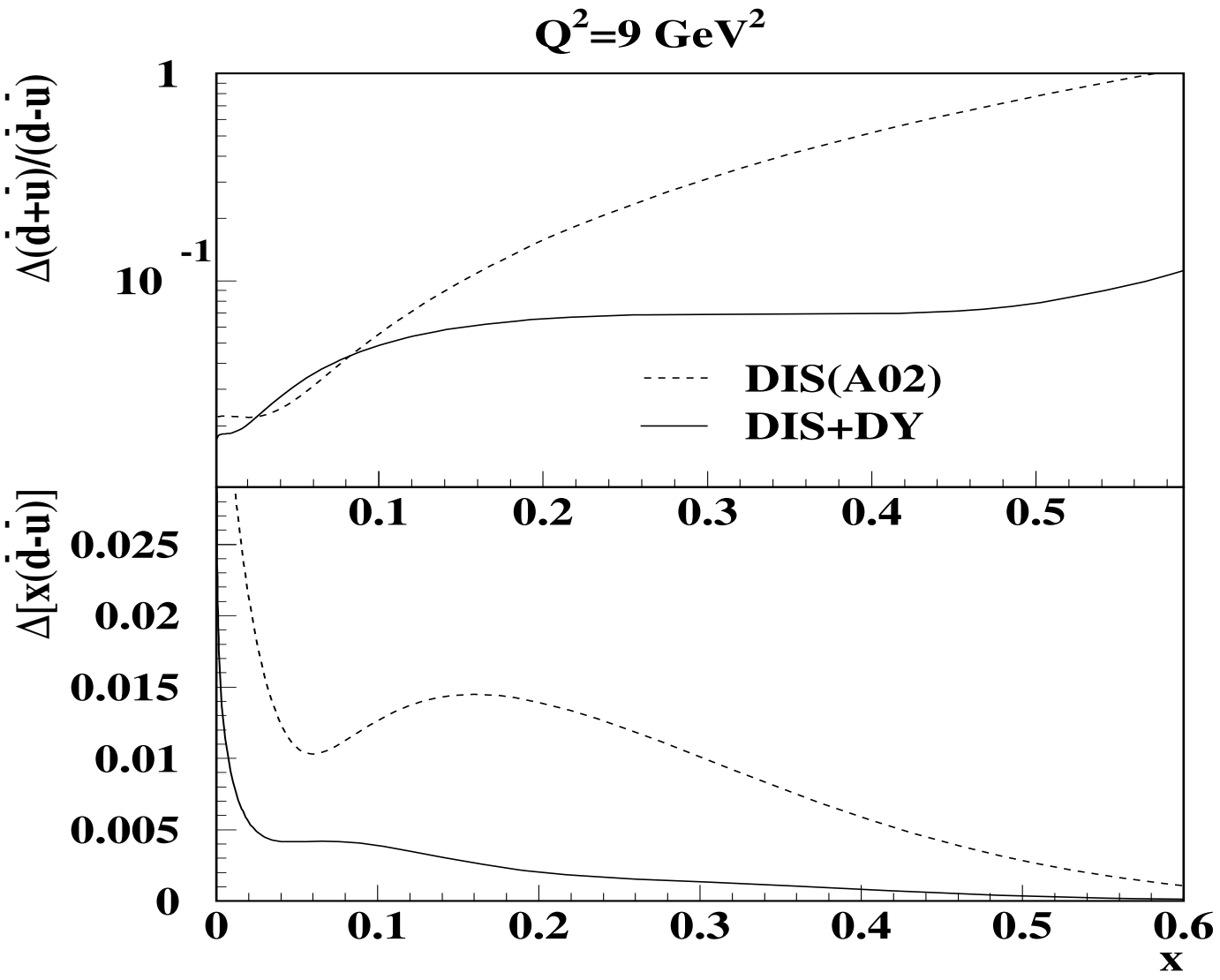

**Fig. 16:** Uncertainties in the non-strange sea distributions obtained from NNLO QCD fit to the DIS data combined with the fixed target Drell-Yan data (solid curves). The same uncertainties obtained in fit to the DIS data only [8] are given for comparison by dashes.

The cross section of process $NN \to l^+l^-$ reads

$$\sigma_{DY} \propto \sum_i \left[q_i(x_1)\bar{q}_i(x_2) + q_i(x_2)\bar{q}_i(x_1)\right] + \text{higher-order terms},$$

where $q(\bar{q})_i$ are the quarks(antiquarks) distribution and $x_{1,2}$ give the momentum fractions carried by each of the colliding partons. The quark distributions are determined by the DIS data with the precision of several percent in the wide region of $x$ and therefore precision of the sea distribution extracted from the combined fit to the DIS and DY data is basically determined by the latter. The Fermilab fixed-target experiments provide measurements of the DY cross sections for the isoscalar target [82] and the ratio of cross sections for the deuteron and proton targets [83] with the accuracy better than 20% at $x \lesssim 0.6$. Fitting PDFs to these data combined with the global DIS data of Ref. [72–77] we can achieve comparable precision in the sea distributions. Recent calculations of Ref. [84] allow to perform this fit with full account of the NNLO correction. Using these calculations the DY data of Refs. [82, 83] were included into the NNLO fit of Ref. [79] that leads to significant improvement in the precision of sea distributions (see Fig. 16). Due to the DY data on the deuteron/proton ratio the isospin asymmetry of sea is also improved. It is worth to note that the precision achieved for the total sea distribution is in good agreement to the rough estimates given above. The value of $\chi^2/\text{NDP}$ obtained in the fit is 1.1 and the spread of $\chi^2/\text{NDP}$ over separate experiments used in the fit is not dramatic, its biggest value is 1.4. We rescaled the errors in data for experiments with $\chi^2/\text{NDP} > 1$ in order to bring $\chi^2/\text{NDP}$ for this experiments to 1 and found that overall impact of this rescaling on the PDFs errors is marginal. This proofs sufficient statistical consistency of the data sets used in the fit and disfavors huge increase in the value of $\Delta\chi^2$ criterion suggested by the CTEQ collaboration for estimation of errors in the global fit of PDFs. A particular feature of the PDFs obtained is good stability with respect to the choice of factorization/renormalization scale in the DY cross section: Variation of this scale from $M_{\mu^+\mu^-}/2$ to $2M_{\mu^+\mu^-}$ leads to variation of PDFs comparable to their uncertainties due to errors in data.

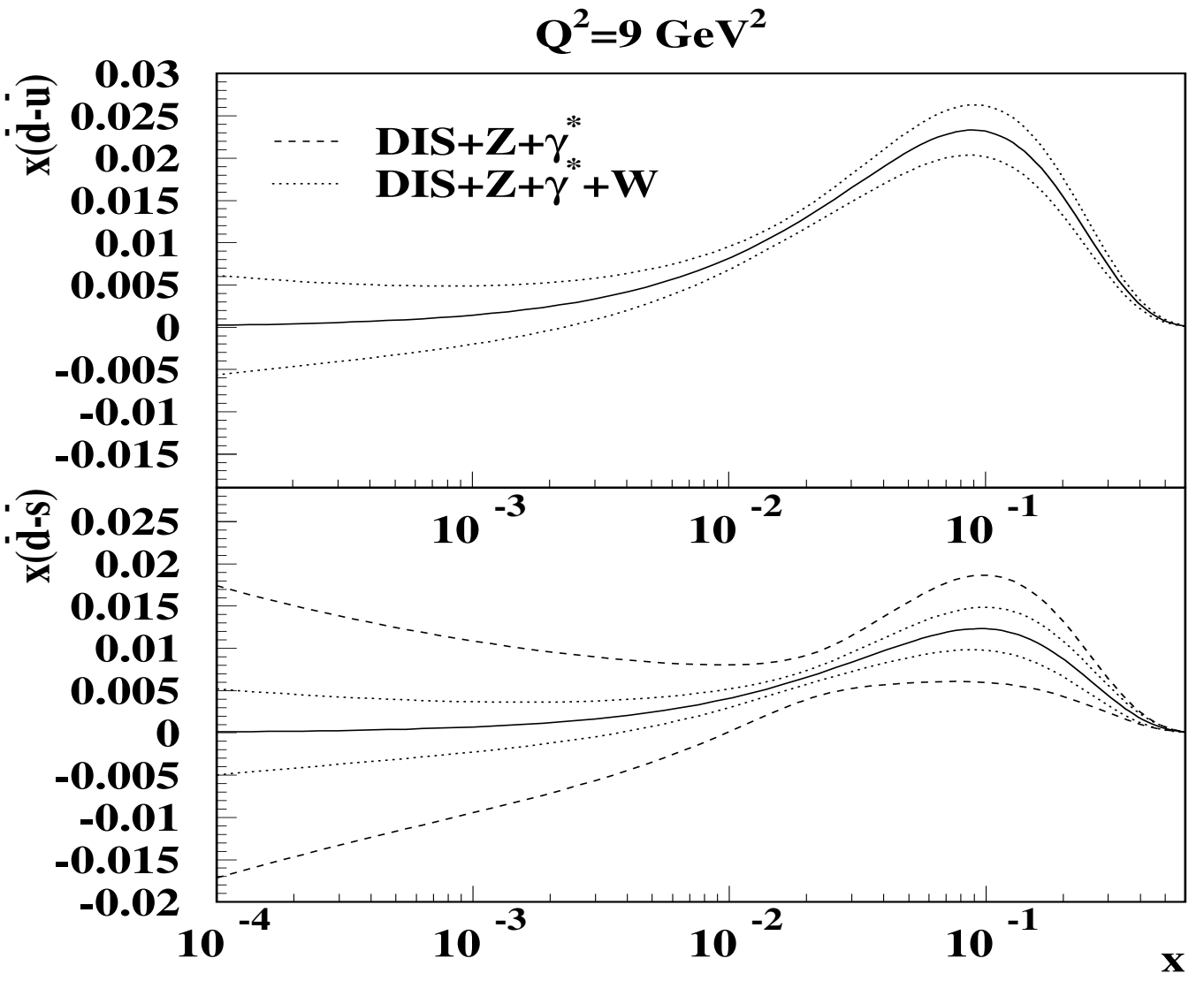

**Fig. 17:** The $1\sigma$ error band for $x(\bar{d}-\bar{u})$ (upper panel) and $x(\bar{d}-\bar{s})$ (lower panel) expected for the fit of PDFs to the LHC data combined with the global DIS ones. Dashed curves correspond to the case of $Z/\gamma*$-production, dots are for the combination $Z/\gamma*$- with the $W^+/W^-$-production. Solid curves are for the central values obtained from the reference fit to the global DIS data

.

### 6.3 LHC data and flavor separation of the sea at small $x$

Combination of the existing DIS and fixed-target DY data provide good constraint on the total sea quarks distribution and allows separation of the $\bar{u}$- and $\bar{d}$-quark distributions up to the values of $x$ sufficient for most practical applications at the LHC. At small $x$ the total sea is also well constrained by the precise HERA data on the inclusive structure functions, however $\bar{u}/\bar{d}$ separation is poor in this region due to lack of the deuteron target data at HERA. The problem of the sea flavor separation is regularly masked due to additional constraints imposed on PDFs. In particular, most often the Regge-like behavior of the sea isospin asymmetry $x(\bar{d}-\bar{u}) \propto x^{a_{ud}}$ is assumed with $a_{ud}$ selected around value of 0.5 motivated by the intercept of the meson trajectories. This assumption automatically provides constraint $\bar{d}=\bar{u}$ at $x \to 0$ and therefore leads to suppression of the uncertainties both in $\bar{u}$ and $\bar{d}$ at small $x$. If we do not assume the Regge-like behavior of $x(\bar{d}-\bar{u})$ its precision determined from the NNLO fit to the combined DIS and DY data of Section 1.2 is about 0.04 at $x = 10^{-4}$ furthermore this constraint is defined rather by assumption about the shape of PDFs at small $x$ than by data used in the fit. The strange sea distribution is known much worse than the non-strange ones. It is essentially defined only by the CCFR experiment from the cross section of dimuon production in the neutrino nucleus collisions [85]. In this experiment the strange sea distribution was probed at $x = 0.01 \div 0.2$ and the shape obtained is similar to one of the non-strange sea with the strangeness suppression factor about 0.5. This is in clear disagreement with the Regge-like constraint on $x(\bar{d}-\bar{s})$ or $x(\bar{u}-\bar{s})$ and therefore we cannot use even this assumption to predict the strange sea at small $x$.

The LHC data on $\mu^+\mu^-$ production cross section can be used for further validation of the sea distributions at small $x$. Study of this process at the lepton pair masses down to 15 GeV will allow to probe PDFs at $x$ down to $10^{-4}$, while with both leptons detected full kinematics can be reliably reconstructed. In order to check impact of the foreseen LHC data on the sea flavor separation we generated sample of pseudo-data for the process $pp \to \mu^+\mu^-X$ at $\sqrt{s} = 14$ TeV with integral luminosity of 10 1/fb corresponding to the first stage of the LHC operation. In order to meet typical limitations of the LHC detectors only events with the lepton pair absolute rapidity less than 2.5 were accepted; other detector effects were not taken into account. For generation of these pseudo-data we used PDFs obtained in the dedicated version of fit [79] with the sea distributions parameterized as $xS_{u,d,s} = \eta_{u,d,s}x^a(1-x)^{b_{u,d,s}}$ with the constraints $\eta_u = \eta_d = \eta_s$ and $b_s = (b_u + b_d)/2$ imposed. These constraints are necessary for stability of the fit in view of limited impact of the DIS data on the flavor separation and, besides, the former one guarantees SU(3) symmetry in the sea distributions at small $x$. The generated pseudo-data were added to the basic DIS data sample and the errors in PDFs parameters were re-estimated with no constraints on the sea distributions imposed at this stage. Since dimuon data give extra information about the PDFs products they allow to disentangle the strange distribution, if an additional constraint on the non-strange sea distributions is set. The dashed curves in the lower panel of Fig.17 give the $1\sigma$ bands for $x(\bar{d}-\bar{s})$ as they are defined by the LHC simulated data combined with the global DIS ones given $(\bar{d}-\bar{u})$ is fixed. One can see that $\bar{d}/\bar{s}$ (and $\bar{u}/\bar{s}$) separation at the level of several percents would be feasible down to x=$10^{-4}$ in this case. The supplementary constraint on $(\bar{d}-\bar{u})$ can be obtained from study of the $W$-boson charge asymmetry. To estimate impact of this process we simulated the single $W^+$- and $W^-$-production data similarly to the case of the $\mu^+\mu^-$-production and took into account this sample too. In this case one can achieve separation of all three flavors with the precision better than 0.01 (see Fig.17). Note that strange sea separation is also improved due to certain sensitivity of the $W$-production cross section to the strange sea contribution. The estimates obtained refer to the ideal case of full kinematical reconstruction of the $W$-bosons events. For the case of using the charge asymmetry of muons produced from the $W$-decays the precision of the PDFs would be worse. Account of the backgrounds and the detector effects would also deteriorate it, however these losses can be at least partially compensated by rise of the LHC luminosity at the second stage of operation.

**Table 7:** Values of the parameters obtained in the benchmark fit.

| | | |
|---|---|---|
| Valence | $a_u$ | 0.718±0.085 |
| | $b_u$ | 3.81±0.16 |
| | $\epsilon_u$ | −1.56±0.46 |
| | $\gamma_u$ | 3.30±0.49 |
| | $a_d$ | 1.71±0.20 |
| | $b_d$ | 10.00±0.97 |
| | $\epsilon_d$ | −3.83±0.23 |
| | $\gamma_d$ | 4.64±0.41 |
| Sea | $A_S$ | 0.211±0.016 |
| | $a_s$ | −0.048±0.039 |
| | $b_s$ | 2.20±0.20 |
| Glue | $a_G$ | 0.356±0.095 |
| | $b_G$ | 10.9±1.4 |
| | $\alpha_s(M_Z)$ | 0.1132±0.0015 |

### 6.4 Benchmarking of the PDFs fit

For the available nucleon PDFs the accuracy at percent level is reached in some kinematical regions. For this reason benchmarking of the codes used in these PDFs fits is becoming important issue. A tool for calibration of the QCD evolution codes was provided by Les Houches workshop [59]. To allow benchmarking of the PDFs errors calculation we performed a test fit suggested in Les Houches workshop too. This fit reproduces basic features of the existing global fits of PDFs, but is simplified a lot to facilitate its reproduction. We use for the analysis data on the proton DIS structure functions $F_2$ obtained by the BCDMS, NM, H1, and ZEUS collaborations and ratio of the deuteron and proton structure functions $F_2$ obtained by the NMC. The data tables with full description of experimental errors taken into account are available online[8]. Cuts for the momentum transferred $Q^2 > 9$ GeV$^2$ and for invariant mass of the hadronic system $W^2 > 15$ GeV$^2$ are imposed in order to avoid influence of the power corrections and simplify calculations. The contribution of the $Z$-boson exchange at large $Q$ is not taken into account for the same purpose. The PDFs are parameterized in the form

$$xp_i(x, 1\,GeV) = N_i x^{a_i}(1-x)^{b_i}(1 + \epsilon_i\sqrt{x} + \gamma_i x),$$

to meet choice common for many popular global fits of PDFs. Some of the parameters $\epsilon_i$ and $\gamma_i$ are set to zero since they were found to be consistent to zero within the errors. We assume isotopic symmetry for sea distribution and the strange sea is the same as the non-strange ones suppressed by factor of 0.5. Evolution of the PDFs is performed in the NLO QCD approximation within the $\overline{\text{MS}}$ scheme. The heavy quarks contribution is accounted in the massless scheme with the variable number of flavors (the thresholds for $c$- and $b$-quarks are 1.5 GeV and 4.5 GeV correspondingly). All experimental errors including correlated ones are taken into account for calculation of the errors in PDFs using the covariance matrix approach [86] and assuming linear propagation of errors. The results of the benchmark fit obtained with the code used in analysis of Refs. [69, 79] are given in Tables 7 and 8. The total number of the fitted PDF parameters left is 14. The normalization parameters $N_i$ for the gluon and valence quark distributions are calculated from the momentum and fermion number conservation. The remaining normalization parameter $A_S$ gives the total momentum carried by the sea distributions. Important note is that in view of many model assumptions made in the fit these results can be used mainly for the purposes of benchmarking rather for the phenomenological studies.

## 7 Benchmark Partons from DIS data and a Comparison with Global Fit Partons [9]

In this article I consider the uncertainties on partons arising from the errors on the experimental data that are used in a parton analysis. Various groups [87], [88], [69], [89], [76], [90], [91] have concentrated on the experimental errors and have obtained estimates of the uncertainties on parton distributions within a NLO QCD framework, using a variety of competing procedures. Here the two analyses, performed by myself and S. Alekhin (see Sec. 6) minimise the differences one obtains for the central values of the partons and the size of the uncertainties by fitting to exactly the same data sets with the same cuts, and using the same theoretical prescription. In order to be conservative we use only DIS data — BCDMS proton [73] and deuterium [74] fixed target data, NMC data on proton DIS and on the ratio $F_2^n(x,Q^2)/F_2^p(x,Q^2)$ [75], and H1 [76] and ZEUS [77] DIS data. We also apply cuts of $Q^2 = 9\text{GeV}^2$ and $W^2 = 15\text{GeV}^2$ in order to avoid the influence of higher twist. We each use NLO perturbative QCD in the $\overline{\text{MS}}$ renormalization and factorization scheme, with the zero-mass variable flavour number scheme and quark masses of $m_c = 1.5\text{GeV}$ and $m_b = 4.5\text{GeV}$. There is a very minor difference between $\alpha_S(\mu^2)$ used in the two fitting programs due to the different methods of implementing heavy quark thresholds (the differences being formally of higher order), as observed in the study by M. Whalley for this workshop [92]. If the couplings in the two approaches have the same value at $\mu^2 = M_Z^2$, then the MRST value is $\sim 1\%$ higher for $Q^2 \sim 20\text{GeV}^2$.

We each input our parton distributions at $Q_0^2 = 1\text{GeV}^2$ with a parameterization of the form

$$xf_i(x, Q_0^2) = A_i(1-x)^{b_i}(1 + \epsilon_i x^{0.5} + \gamma_i x)x^{a_i}. \tag{7.24}$$

The input sea is constrained to be $40\%$ up and anti-up quarks, $40\%$ down and anti-down quarks, and $20\%$ strange and antistrange. No difference between $\bar{u}$ and $\bar{d}$ is input. There is no negative term for the gluon, as introduced in [90], since this restricted form of data shows no strong requirement for it in order to obtain the best fit. Similarly we are able to set $\epsilon_g$, $\gamma_g$, $\epsilon_S$ and $\gamma_S$ all equal to zero. $A_g$ is set by the momentum sum rule and $A_{u_V}$ and $A_{d_V}$ are set by valence quark number. Hence, there are nominally 13 free parton parameters. However, the MRST fitting program exhibited instability in the error matrix due to a very high correlation between $u_V$ parameters, so $\epsilon_u$ was set at its best fit value of $\epsilon_u = -1.56$, while 12 parameters were free to vary. The coupling was also allowed to vary in order to obtain the best fit. The treatment of the errors on the data was exactly as for the published partons with uncertainties for each group, i.e. as in [69] and [93]. This means that all detail on correlations between errors is included for the Alekhin fit (see Sec. 6), assuming that these errors are distributed in the Gaussian manner. The errors in the MRST fit are treated as explained in the appendix of [93], and the correlated errors are not allowed to move the central values of the data to as great an extent for the HERA data, and cannot do so at all for the fixed target data, where the data used are averaged over the different beam energies. The Alekhin approach is more statistically rigorous. The MRST approach is more pragmatic, reducing the ability of the data to move relative to the theory comparison by use of correlated errors (other than normalization), and is in some ways similar to the offset method [91]. The danger of this movement of data relative to theory has been suggested by the joint analysis of H1 and ZEUS data at this workshop (see [94]), where letting the joint data sets determine the movement due to correlated errors gives different results from when the data sets are compared to theoretical results.

[8] https://mail.ihep.ru/~alekhin/benchmark/TABLE

[9] Contributing author: R.S. Thorne.

**Table 8:** Correlation coefficients for the parameters obtained in the benchmark fit.

| | $a_u$ | $b_u$ | $\epsilon_u$ | $\gamma_u$ | $a_d$ | $b_d$ | $\epsilon_d$ | $\gamma_d$ | $A_S$ | $a_s$ | $b_s$ | $a_G$ | $b_G$ | $\alpha_s(M_Z)$ |
|---|---|---|---|---|---|---|---|---|---|---|---|---|---|---|
| $a_u$ | 1.000 | 0.728 | -0.754 | -0.708 | 0.763 | 0.696 | -0.444 | 0.215 | -0.216 | -0.473 | -0.686 | 0.593 | 0.777 | -0.006 |
| $b_u$ | 0.728 | 1.000 | -0.956 | -0.088 | 0.377 | 0.620 | -0.420 | 0.387 | 0.175 | -0.182 | -0.713 | 0.067 | 0.505 | -0.337 |
| $\epsilon_u$ | -0.754 | -0.956 | 1.000 | 0.105 | -0.388 | -0.662 | 0.503 | -0.485 | -0.229 | 0.059 | 0.600 | -0.047 | -0.503 | 0.276 |
| $\gamma_u$ | -0.708 | -0.088 | 0.105 | 1.000 | -0.741 | -0.390 | 0.219 | 0.107 | 0.597 | 0.591 | 0.310 | -0.716 | -0.675 | -0.088 |
| $a_d$ | 0.763 | 0.377 | -0.388 | -0.741 | 1.000 | 0.805 | -0.622 | 0.248 | -0.367 | -0.509 | -0.528 | 0.652 | 0.664 | 0.101 |
| $b_d$ | 0.696 | 0.620 | -0.662 | -0.390 | 0.805 | 1.000 | -0.904 | 0.728 | 0.017 | -0.193 | -0.512 | 0.272 | 0.576 | -0.136 |
| $\epsilon_d$ | -0.444 | -0.420 | 0.503 | 0.219 | -0.622 | -0.904 | 1.000 | -0.896 | -0.132 | -0.019 | 0.245 | -0.038 | -0.362 | 0.173 |
| $\gamma_d$ | 0.215 | 0.387 | -0.485 | 0.107 | 0.248 | 0.728 | -0.896 | 1.000 | 0.346 | 0.240 | -0.107 | -0.241 | 0.120 | -0.228 |
| $A_S$ | -0.216 | 0.175 | -0.229 | 0.597 | -0.367 | 0.017 | -0.132 | 0.346 | 1.000 | 0.708 | 0.127 | -0.375 | -0.026 | 0.047 |
| $a_s$ | -0.473 | -0.182 | 0.059 | 0.591 | -0.509 | -0.193 | -0.019 | 0.240 | 0.708 | 1.000 | 0.589 | -0.595 | -0.241 | -0.011 |
| $b_s$ | -0.686 | -0.713 | 0.600 | 0.310 | -0.528 | -0.512 | 0.245 | -0.107 | 0.127 | 0.589 | 1.000 | -0.508 | -0.402 | -0.109 |
| $a_G$ | 0.593 | 0.067 | -0.047 | -0.716 | 0.652 | 0.272 | -0.038 | -0.241 | -0.375 | -0.595 | -0.508 | 1.000 | 0.565 | 0.587 |
| $b_G$ | 0.777 | 0.505 | -0.503 | -0.675 | 0.664 | 0.576 | -0.362 | 0.120 | -0.026 | -0.241 | -0.402 | 0.565 | 1.000 | -0.138 |
| $\alpha_s(M_Z)$ | -0.006 | -0.337 | 0.276 | -0.088 | 0.101 | -0.136 | 0.173 | -0.228 | 0.047 | -0.011 | -0.109 | 0.587 | -0.138 | 1.000 |

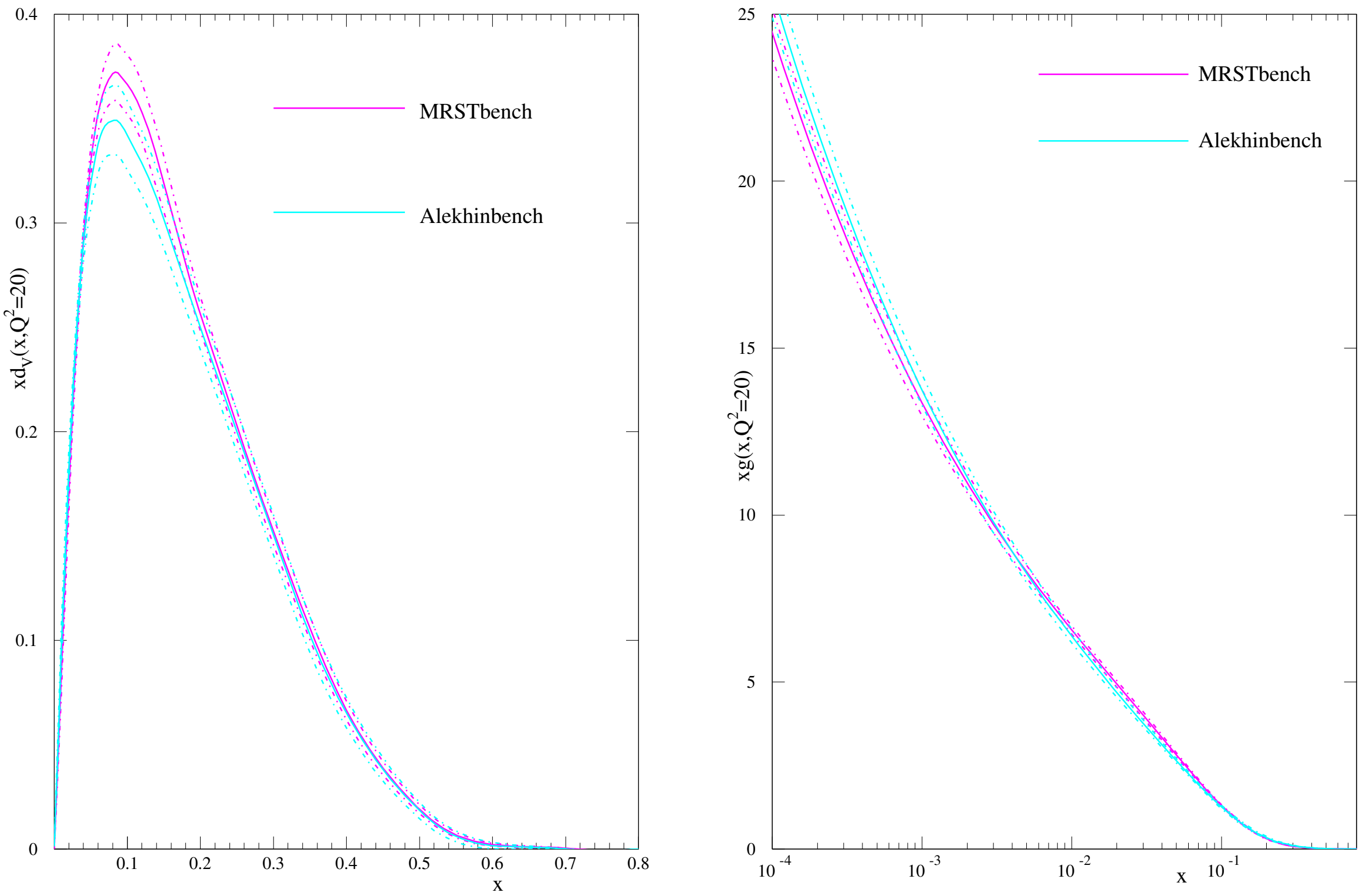

**Fig. 18:** Left plot: $xd_V(x, 20)$ from the MRST benchmark partons compared to that from the Alekhin benchmark partons. Right plot: $xg(x, 20)$ from the MRST benchmark partons compared to that from the Alekhin benchmark partons.

### 7.1 Comparison Between the Benchmark Parton Distributions.

I compare the results of the two approaches to fitting the restricted data chosen for the benchmarking. The input parameters for the Alekhin fit are presented in Sec. 6. Those for the MRST type fit are similar, but there are some differences which are best illustrated by comparing the partons at a typical $Q^2$ for the data, e.g. $Q^2 = 20\text{GeV}^2$. A comparison is shown for the $d_V$ quarks and the gluon in Fig. 18.

From the plots it is clear that there is generally good agreement between the parton distributions. The central values are usually very close, and nearly always within the uncertainties. The difference in the central values is mainly due to the different treatment of correlated errors, and partially due to the difference in the coupling definition. The uncertainties are similar in the two sets, but are generally about $1.2 - 1.5$ times larger for the Alekhin partons, due to the increased freedom in the use of the correlated experimental errors. The values of $\alpha_S(M_Z^2)$ are quite different, $\alpha_S(M_Z^2) = 0.1132 \pm 0.0015$ compared to $0.1110 \pm 0.0012$. However, as mentioned earlier, one expects a $1\%$ difference due to the different threshold prescriptions — the MRST $\alpha_S$ would be larger at $Q^2 \sim 20\mathrm{GeV}^2$, where the data are concentrated, so correspondingly to fit the data it receives a $1\%$ shift downwards for $Q^2 = M_Z^2$. Once this systematic effect is taken into account, the values of $\alpha_S(M_Z^2)$ are very compatible. Hence, there is no surprising inconsistency between the two sets of parton distributions.

### 7.2 Comparison of the Benchmark Parton Distributions and Global Fit Partons.

It is also illuminating to show the comparison between the benchmark partons and the published partons from a global fit. This is done below for the MRST01 partons. For example, $u_V(x, Q^2)$ and $\bar{u}(x, Q^2)$ are shown in Fig. 19. It is striking that the uncertainties in the two sets are rather similar. This is despite the fact that the uncertainty on the benchmark partons is obtained from allowing $\Delta\chi^2 = 1$ in the fit while that for the MRST01 partons is obtained from $\Delta\chi^2 = 50$.[10] This illustrates the great improvement in precision which is obtained due to the increase in data from the relaxation of the cuts and the inclusion of types of data other than DIS. For the $u_V$ partons, which are those most directly constrained by the DIS data in the benchmark fit, the comparison between the two sets of partons is reasonable, but hardly perfect — the central values differing by a few standard deviations. This is particularly important given that in this comparison the treatment of the data in the fit has been exactly the same in both cases. There is a minor difference in theoretical approach because of the simplistic treatment of heavy flavours in the benchmark fit. However, this would influence the gluon and sea quarks rather than valence quarks. Moreover, the region sensitive to this simplification would be $Q^2 \sim m_c^2$ (the lower charge weighting for bottom quarks greatly reducing the effect near $Q^2 = m_b^2$) which is removed by the $Q^2$ cut of $9\mathrm{GeV}^2$. Indeed, introducing the variable flavour number scheme usually used for the MRST partons modifies the benchmark partons only very minimally. Hence, if the statistical analysis is correct, the benchmark partons should agree with the global partons within their uncertainties (or at most 1.5 times their uncertainties, allowing for the effect of the correlated errors), which they do not. For the $\bar{u}$ partons the comparison is far worse, the benchmark partons being far larger at high $x$.

This disagreement in the high-$x$ $\bar{u}$ partons can be understood better if one also looks at the high-$x$ $d_V$ distribution shown in Fig. 20. Here the benchmark distribution is very much smaller than for MRST01. However, the increase in the sea distribution, which is common to protons and neutrons, at high-$x$ has allowed a good fit to the high-$x$ BCDMS deuterium data even with the very small high-$x$ $d_V$ distribution. In fact it is a better fit than in [93]. However, the fit can be shown to break down with the additional inclusion of high-$x$ SLAC data [72] on the deuterium structure function. More dramatically, the shape of the $\bar{u}$ is also completely incompatible with the Drell-Yan data usually included in the global fit, e.g. [82, 95]. Also in Fig. 20 we see that the $d_V$ distributions are very different at smaller $x$. The benchmark set is markedly inconsistent with NMC data on $F_2^n(x, Q^2)/F_2^p(x, Q^2)$ which is at small $x$, but below the cut of $Q^2 = 9\mathrm{GeV}^2$.

The gluon from the benchmark set is also compared to the MRST01 gluon in Fig. 21. Again there is an enormous difference at high $x$. Nominally the benchmark gluon has little to constrain it at high $x$. However, the momentum sum rule determines it to be very small in this region in order to get the best fit to HERA data, similar to the gluon from [76]. As such, the gluon has a small uncertainty and is many standard deviations from the MRST01 gluon. Indeed, the input gluon at high $x$ is so small that its value at higher $Q^2$ is dominated by the evolution of $u_V$ quarks to gluons, rather than by the input gluon. Hence, the uncertainty is dominated by the quark parton input uncertainty rather than its own, and since the up quark is well determined the uncertainty on the high-$x$ gluon is small for the benchmark partons. The smallness of the high-$x$ gluon results in the benchmark partons producing a very poor prediction indeed for the Tevatron jet data [96, 97], which are the usual data that constrain the high-$x$ gluon in global fits.

[10] Though it is meant to be interpreted as a one sigma error in the former case and a $90\%$ confidence limit in the latter.

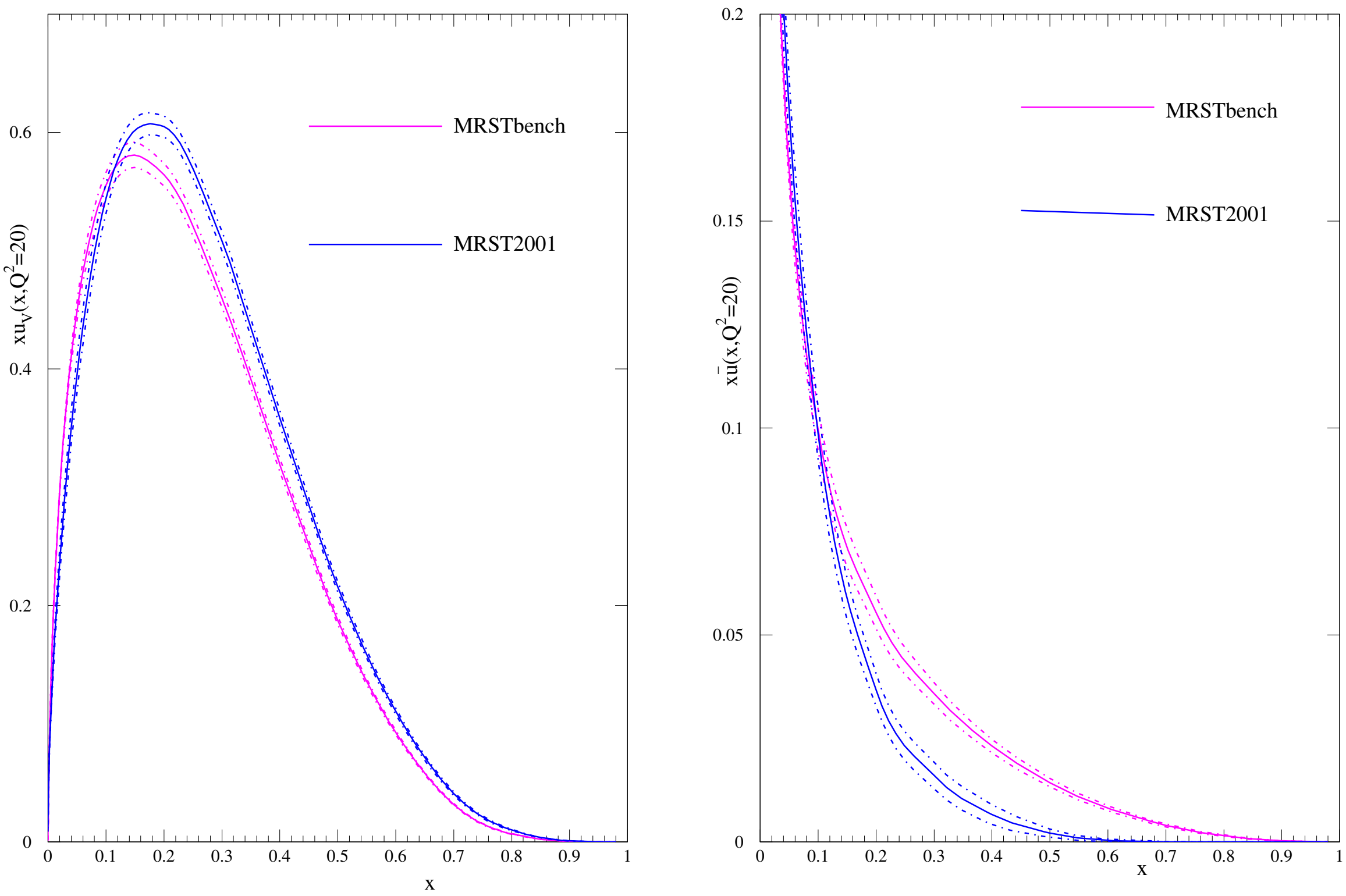

**Fig. 19:** Left plot: $xu_V(x, 20)$ from the MRST benchmark partons compared to that from the MRST01 partons. Right plot: $x\bar{u}(x, 20)$ from the MRST benchmark partons compared to that from the MRST01 partons with emphasis on large $x$.

It is also illustrative to look at small $x$. Here the benchmark gluon is only a couple of standard deviations from the MRST01 gluon, suggesting that its size is not completely incompatible with a good fit to the HERA small-$x$ data at $Q^2$ below the benchmark cut. However, the uncertainty in the benchmark gluon is much smaller than in the MRST01 gluon, despite the much smaller amount of low-$x$ data in the fit for the benchmark partons. This comes about as a result of the artificial choice made in the gluon input at $Q_0^2$. Since it does not have the term introduced in [93], allowing the freedom for the input gluon to be negative at very small $x$, the gluon is required by the fit to be valence-like. Hence, at input it is simply very small at small $x$. At higher $Q^2$ it becomes much larger, but in a manner driven entirely by evolution, i.e. it is determined by the input gluon at moderate $x$, which is well constrained. In this framework the small-$x$ gluon does not have any intrinsic uncertainty — its uncertainty is a reflection of moderate $x$. This is a feature of e.g. the CTEQ6 gluon uncertainty [89], where the input gluon is valence-like. In this case the percentage gluon uncertainty does not get any larger once $x$ reaches about 0.001. The alternative treatment in [93] gives the expected increase in the gluon uncertainty as $x \to 0$, since in this case the uncertainty is determined largely by that in the input gluon at small $x$. The valence-like input form for a gluon is an example of fine-tuning, the form being unstable to evolution in either direction. The artificial limit on the small-$x$ uncertainty is a consequence of this.

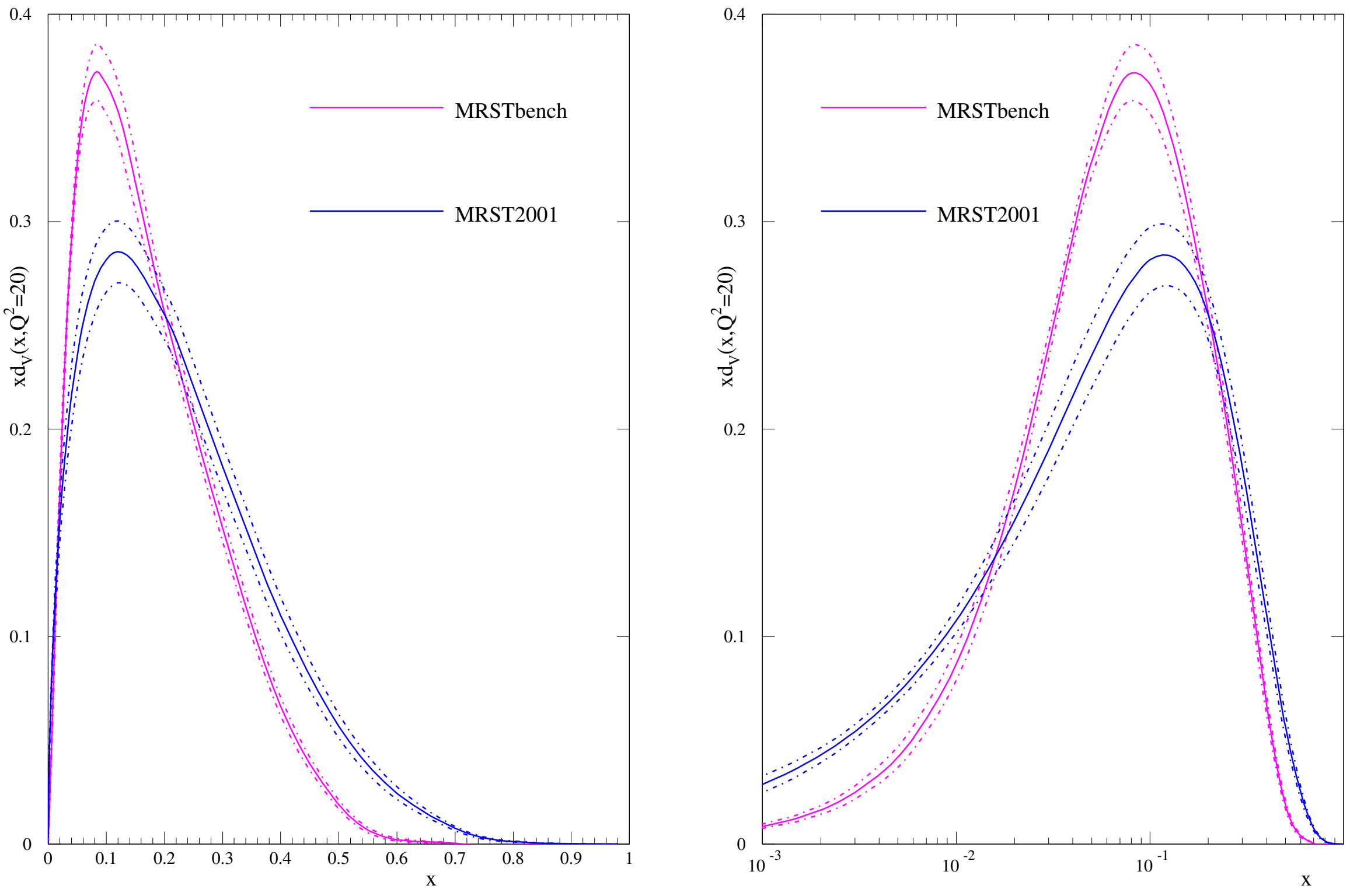

**Fig. 20:** Left plot: $xd_V(x, 20)$ from the MRST benchmark partons compared to that from the MRST01 partons. Right plot: $xd_V(x, 20)$ from the MRST benchmark partons compared to that from the MRST01 partons with emphasis on small $x$.

### 7.3 Conclusions.

I have demonstrated that different approaches to fitting parton distributions that use exactly the same data and theoretical framework produce partons that are very similar and have comparable uncertainties. There are certainly some differences due to the alternative approaches to dealing with experimental errors, but these are relatively small. However, the partons extracted using a very limited data set are completely incompatible, even allowing for the uncertainties, with those obtained from a global fit with an identical treatment of errors and a minor difference in theoretical procedure. This implies that the inclusion of more data from a variety of different experiments moves the central values of the partons in a manner indicating either that the different experimental data are inconsistent with each other, or that the theoretical framework is inadequate for correctly describing the full range of data. To a certain extent both explanations are probably true. Some data sets are not entirely consistent with each other (even if they are seemingly equally reliable). Also, there are a wide variety of reasons why NLO perturbative QCD might require modification for some data sets, or in some kinematic regions [98]. Whatever the reason for the inconsistency between the MRST benchmark partons and the MRST01 partons, the comparison exhibits the dangers in extracting partons from a very limited set of data and taking them seriously. It also clearly illustrates the problems in determining the true uncertainty on parton distributions.

## 8 Stability of PDF fits [11]

One of the issues raised at the workshop is the reliability of determinations of parton distribution functions (PDFs), which might be compromised for example by the neglect of NNLO effects or non-DGLAP evolution in the standard analysis, or hidden assumptions made in parameterizing the PDFs at nonper-turbative scales. We summarize the results of the CTEQ PDF group on this issue. For the full story see [80].

[11]Contributing authors: J. Huston, J. Pumplin.

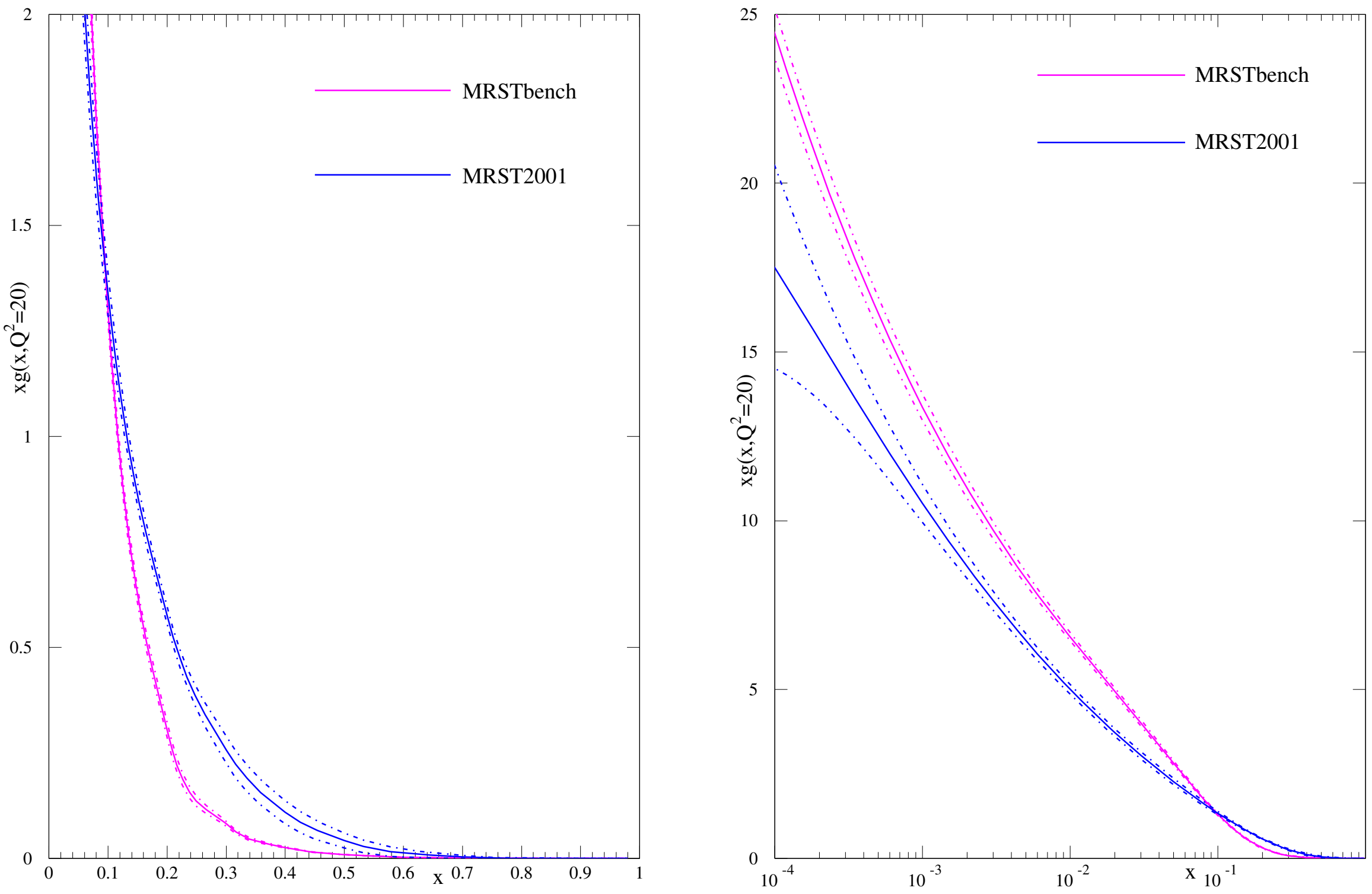

**Fig. 21:** Left plot: $xg(x, 20)$ from the MRST benchmark partons compared to that from the MRST2001 partons. Right plot: $xg(x, 20)$ from the MRST benchmark partons compared to that from the MRST2001 partons with emphasis on small $x$.

### 8.1 Stability of PDF determinations

The stability of NLO global analysis was seriously challenged by an analysis [98] which found a 20% variation in the cross section predicted for $W$ production at the LHC — a critical "standard candle" process for hadron colliders — when certain cuts on input data are varied. If this instability were confirmed, it would significantly impact the phenomenology of a wide range of physical processes for the Tevatron Run II and the LHC. The CTEQ PDF group therefore performed an independent study of this issue within their global analysis framework. In addition, to explore the dependence of the results on assumptions about the parameterization of PDFs at the starting scale $Q_0 = 1.3\,\mathrm{GeV}$, we also studied the effect of allowing a negative gluon distribution at small $x$ — a possibility that is favored by the MRST NLO analysis, and that is closely tied to the W cross section controversy.

The stability of the global analysis was investigated by varying the inherent choices that must be made to perform the analysis. These choices include the selection of experimental data points based on kinematic cuts, the functional forms used to parameterize the initial nonperturbative parton distribution functions, and the treatment of $\alpha_s$.

The stability of the results is most conveniently measured by differences in the global $\chi^2$ for the relevant fits. To quantitatively define a change of $\chi^2$ that characterizes a significant change in the quality of the PDF fit is a difficult issue in global QCD analysis. In the context of the current analysis, we have argued that an increase by $\Delta\chi^2 \sim 100$ (for $\sim 2000$ data points) represents roughly a 90% confidence level uncertainty on PDFs due to the uncertainties of the current input experimental data [89, 99–101].

**Table 9:** Comparisons of three fits with different choices of the cuts on input data at the $Q$ and $x$ values indicated. In these fits, a conventional positive-definite gluon parameterization was used.

| Cuts | $Q_{\rm min}$ | $x_{\rm min}$ | $N_{\rm pts}$ | $\chi^2_{1926}$ | $\chi^2_{1770}$ | $\chi^2_{1588}$ | $\sigma_W^{\rm LHC} \times B_{\ell\nu}$ [nb] |
|---|---|---|---|---|---|---|---|
| standard | 2 GeV | 0 | 1926 | 2023 | 1850 | 1583 | 20.02 |
| intermediate | 2.5 GeV | 0.001 | 1770 | – | 1849 | 1579 | 20.10 |
| strong | 3.162 GeV | 0.005 | 1588 | – | – | 1573 | 20.34 |

**Table 10:** Same as Table 9 except that the gluon parameterization is extended to allow negative values.

| Cuts | $Q_{\rm min}$ | $x_{\rm min}$ | $N_{\rm pts}$ | $\chi^2_{1926}$ | $\chi^2_{1770}$ | $\chi^2_{1588}$ | $\sigma_W^{\rm LHC} \times B_{\ell\nu}$ [nb] |
|---|---|---|---|---|---|---|---|
| standard | 2 GeV | 0 | 1926 | 2011 | 1845 | 1579 | 19.94 |
| intermediate | 2.5 GeV | 0.001 | 1770 | – | 1838 | 1574 | 19.80 |
| strong | 3.162 GeV | 0.005 | 1588 | – | – | 1570 | 19.15 |

In other words, PDFs with $\chi^2 - \chi^2_{\rm BestFit} > 100$ are regarded as not tolerated by current data.

The CTEQ6 and previous CTEQ global fits imposed "standard" cuts $Q > 2\,\rm GeV$ and $W > 3.5\,\rm GeV$ on the input data set, in order to suppress higher-order terms in the perturbative expansion and the effects of resummation and power-law ("higher twist") corrections. We examined the effect of stronger cuts on $Q$ to see if the fits are stable. We also examined the effect of imposing cuts on $x$, which should serve to suppress any errors due to deviations from DGLAP evolution, such as those predicted by BFKL. The idea is that any inconsistency in the global fit due to data points near the boundary of the accepted region will be revealed by an improvement in the fit to the data that remain after those near-boundary points have been removed. In other words, the decrease in $\chi^2$ for the subset of data that is retained, when the PDF shape parameters are refitted to that subset alone, measures the degree to which the fit to that subset was distorted in the original fit by compromises imposed by the data at low $x$ and/or low $Q$.

The main results of this study are presented in Table 9. Three fits are shown, from three choices of the cuts on input data as specified in the table. They are labeled 'standard', 'intermediate' and 'strong'. $N_{\rm pts}$ is the number of data points that pass the cuts in each case, and $\chi^2_{N_{\rm pts}}$ is the $\chi^2$ value for that subset of data. The fact that the changes in $\chi^2$ in each column are insignificant compared to the uncertainty tolerance is strong evidence that our NLO global fit results are very stable with respect to choices of kinematic cuts.

We extended the analysis to a series of fits in which the gluon distribution $g(x)$ is allowed to be negative at small $x$, at the scale $Q_0 = 1.3\,\rm GeV$ where we begin the DGLAP evolution. The purpose of this additional study is to determine whether the feature of a negative gluon PDF is a key element in the stability puzzle, as suggested by the findings of [98]. The results are presented in Table 10. Even in this extended case, we find no evidence of instability. For example, $\chi^2$ for the subset of 1588 points that pass the *strong* cuts increases only from 1570 to 1579 when the fit is extended to include the full standard data set.

Comparing the elements of Table 9 and Table 10 shows that our fits with $g(x) < 0$ have slightly smaller values of$\chi^2$: e.g., 2011 versus 2023 for the standard cuts. However, the difference $\Delta\chi^2 = 12$ between these values is again not significant according to our tolerance criterion.

### 8.2 W cross sections at the LHC

The last columns of Tables 9 and 10 show the predicted cross section for $W^+ + W^-$ production at the LHC. This prediction is also very stable: it changes by only $1.6\%$ for the positive-definite gluon parameterization, which is substantially less than the overall PDF uncertainty of $\sigma_W$ estimated previously with the standard cuts. For the negative gluon parameterization, the change is $4\%$–larger, but still less than the overall PDF uncertainty. These results are explicitly displayed, and compared to the MRST results in Fig. 22. We see that this physical prediction is indeed insensitive to the kinematic cuts used for the fits, and to the assumption on the positive definiteness of the gluon distribution.

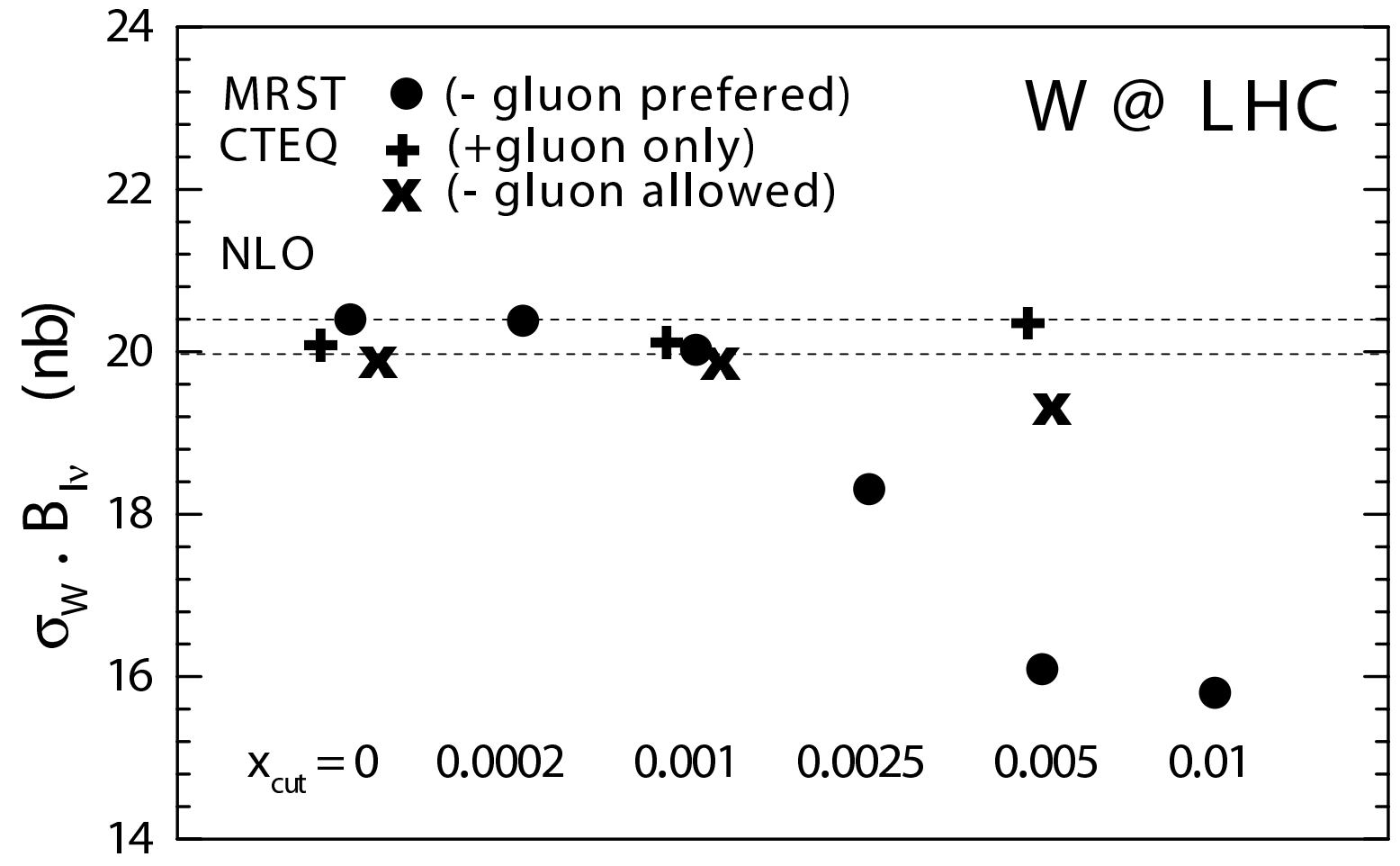

**Fig. 22:** Predicted total cross section of $W^+ + W^-$ production at the LHC for the fits obtained in our stability study, compared to the NLO results of Ref. [98]. The $Q$-cut values associated with the CTEQ points are given in the two tables. The overall PDF uncertainty of the prediction is $\sim 5\%$.

We also studied the stability of the prediction for $\sigma_W$ using the Lagrange Multiplier (LM) method of Refs. [99–101]. Specifically, we performed a series of fits to the global data set that are constrained to specific values of $\sigma_W$ close to the best-fit prediction. The resulting variation of $\chi^2$ versus $\sigma_W$ measures the uncertainty of the prediction. We repeated the constrained fits for each case of fitting choices (parameterization and kinematic cuts). In this way we gain an understanding of the stability of the uncertainty, in addition to the stability of the central prediction.

Figure 23 shows the results of the LM study for the three sets of kinematic cuts described in Table 9, all of which have a positive-definite gluon distribution. The $\chi^2$ shown along the vertical axis is normalized to its value for the best fit in each series. In all three series, $\chi^2$ depends almost quadratically on $\sigma_W$. We observe several features:

- The location of the minimum of each curve represents the best-fit prediction for $\sigma_W^{\rm LHC}$ for the corresponding choice of cuts. The fact that the three minima are close together displays the stability of the predicted cross section already seen in Table 9.
- Although more restrictive cuts make the global fit less sensitive to possible contributions from resummation, power-law and other nonperturbative effects, the loss of constraints caused by the removal of precision HERA data points at small $x$ and low $Q$ results directly in increased uncertainties on the PDF parameters and their physical predictions. This is shown in Fig. 23 by the increase of the width of the curves with stronger cuts. The uncertainty of the predicted $\sigma_W$ increases by more than a factor of 2 in going from the standard cuts to the strong cuts.

Figure 24 shows the results of the LM study for the three sets of kinematic cuts described in Table 10, all of which have a gluon distribution which is allowed to go negative.

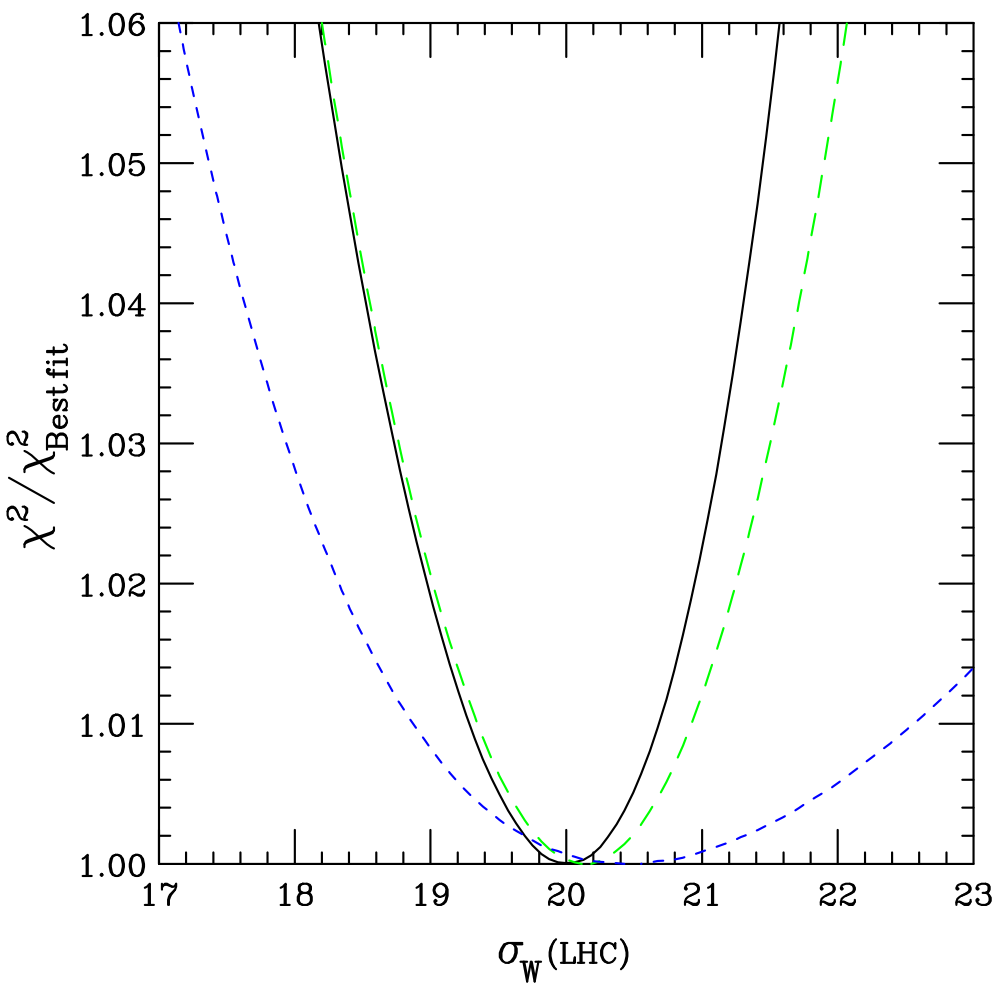

**Fig. 23:** Lagrange multiplier results for the $W$ cross section (in nb) at the LHC using a positive-definite gluon. The three curves, in order of decreasing steepness, correspond to the three sets of kinematic cuts labeled standard/intermediate/strong in Table 9.

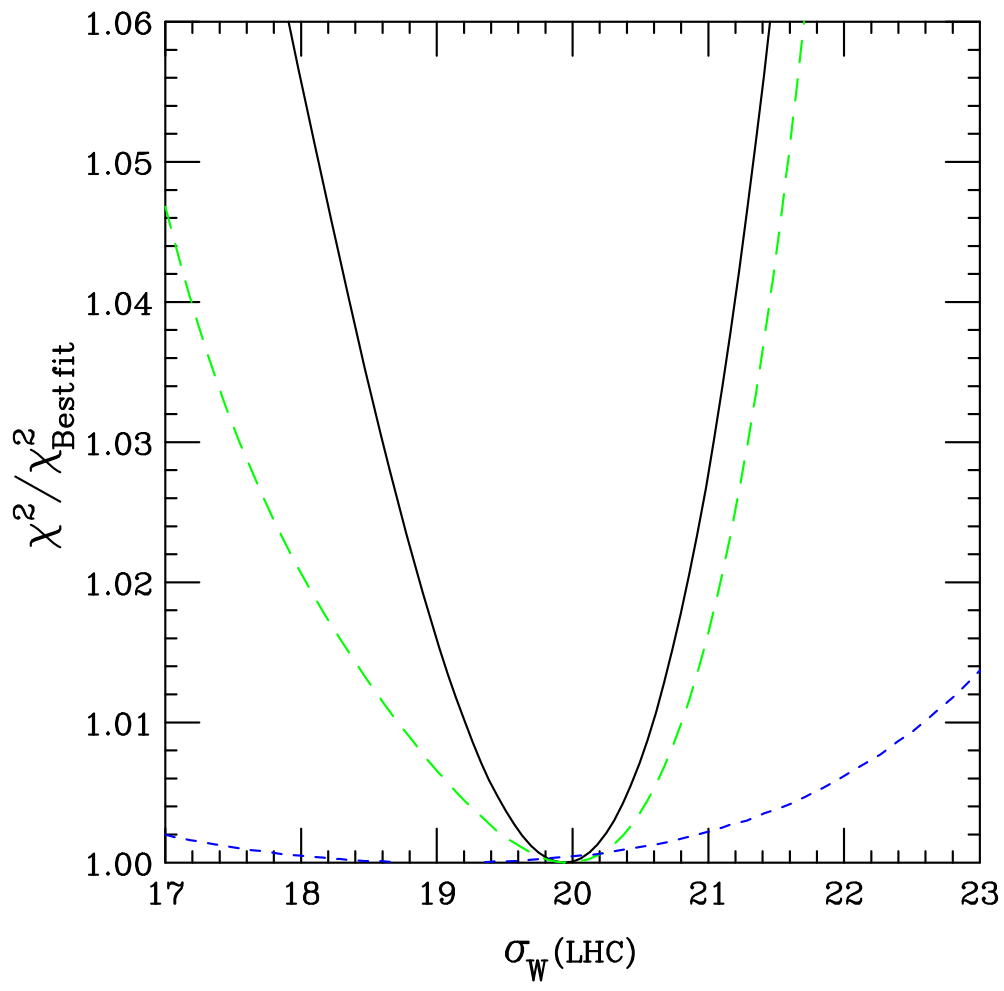

**Fig. 24:** Lagrange multiplier results for the $W$ cross section (in nb) at the LHC using a functional form where the gluon is not required to be positive-definite. The three curves, in order of decreasing steepness, correspond to the three sets of kinematic cuts labeled standard/intermediate/strong in Table 10.

We observe:

- Removing the positive definiteness condition necessarily lowers the value of $\chi^2$, because more possibilities are opened up in the $\chi^2$ minimization procedure. But the decrease is insignificant compared to other sources of uncertainty. Thus, a negative gluon PDF is allowed, but not required.
- The minima of the two curves occur at approximately the same $\sigma_W$. Allowing a negative gluon makes no significant change in the central prediction — merely a decrease of about 1 %, which is small compared to the overall PDF uncertainty.
- For the standard set of cuts, allowing a negative gluon PDF would expand the uncertainty range only slightly. For the intermediate and strong cuts, allowing a negative gluon PDF would significantly expand the uncertainty range.

We examined a number of aspects of our analysis that might account for the difference in conclusions between our stability study and that of [98]. A likely candidate seems to be that in order to obtain stability, it is necessary to allow a rather free parametrization of the input gluon distribution. This suspicion is seconded by recent work by MRST [102], in which a different gluon parametrization appears to lead to a best-fit gluon distribution that is close to that of CTEQ6. In summary, we found that the NLO PDFs and their physical predictions at the Tevatron and LHC are quite stable with respect to variations of the kinematic cuts and the PDF parametrization after all.

### 8.3 NLO and NNLO

In recent years, some preliminary next-to-next-leading-order (NNLO) analyses for PDFs have been carried out either for DIS alone [103], or in a global analysis context [51] — even if all the necessary hard cross sections, such as inclusive jet production, are not yet available at this order. Determining the parton distributions at NNLO is obviously desirable on theoretical grounds, and it is reasonable to plan for having a full set of tools for a true NNLO global analysis in place by the time LHC data taking begins. At the moment, however, NNLO fitting is not a matter of pressing necessity, since the difference between NLO and NNLO appears to be very small compared to the other uncertainties in the PDF analysis. This is demonstrated in Fig. 25, which shows the NLO and NNLO gluon distributions extracted by the MRST group. The difference between the two curves is much smaller than the other uncertainties measured by the 40 eigenvector uncertainty sets of CTEQ6.1, which is shown by the shaded region. The difference is also much smaller than the difference between CTEQ and MRST best fits. Similar conclusions [104] can be found using the NLO and NNLO fits by Alekhin.

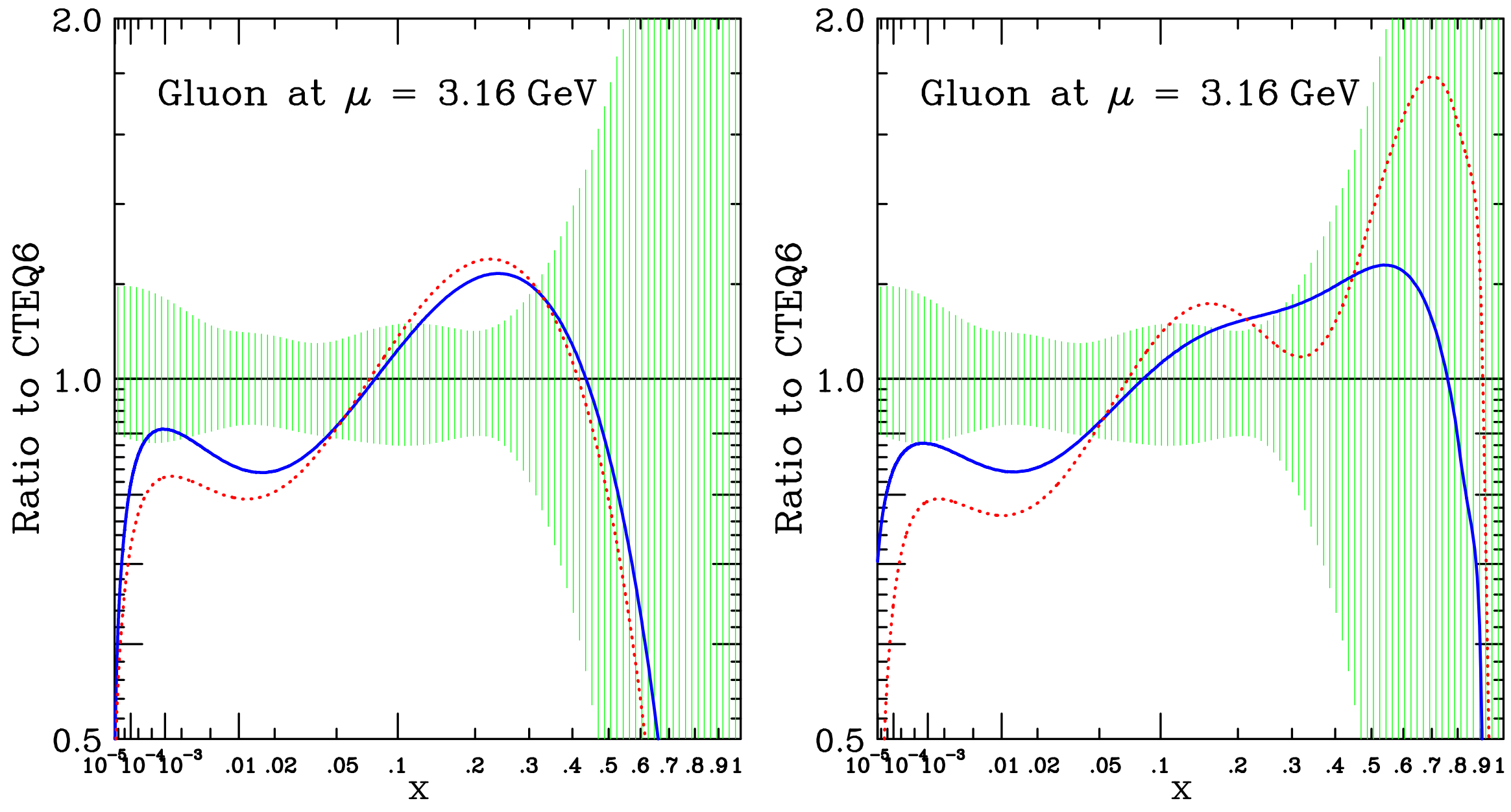

**Fig. 25:** Left: mrst2002 NLO (solid) and NNLO (dotted); Right: mrst2004 NLO (solid) and NNLO (dotted); Shaded region is uncertainty according to the 40 eigenvector sets of CTEQ6.1.

## 9 The neural network approach to parton distributions [12]

The requirements of precision physics at hadron colliders, as has been emphasized through this workshop, have recently led to a rapid improvement in the techniques for the determination of parton distribution functions (pdfs) of the nucleon. Specifically it is now mandatory to determine accurately the uncertainty on these quantities, and the different collaborations performing global pdf analysis [51, 69, 105] have performed estimations of these uncertainties using a variety of techniques. The main difficulty is that one is trying to determine the uncertainty on a function, that is, a probability measure in a space of functions, and to extract it from a finite set of experimental data, a problem which is mathematically ill-posed. It is also known that the standard approach to global parton fits have several shortcomings: the bias introduced by choosing fixed functional forms to parametrize the parton distributions (also known as *model dependence*), the problems to assess faithfully the pdf uncertainties, the combination of inconsistent experiments, and the lack of general, process-independent error propagation techniques. Although the problem of quantifying the uncertainties in pdfs has seen a huge progress since its paramount importance was raised some years ago, until now no unambiguous conclusions have been obtained.

In this contribution we present a novel strategy to address the problem of constructing unbiased parametrizations of parton distributions with a faithful estimation of their uncertainties, based on a combination of two techniques: Monte Carlo methods and neural networks. This strategy, introduced in [106, 107], has been first implemented to address the marginally simpler problem of parametrizing deep-inelastic structure functions $F(x, Q^2)$, which we briefly summarize now. In a first step we construct a Monte Carlo sampling of the experimental data (generating artificial data replicas), and then we train neural networks to each data replica, to construct a probability measure in the space of structure functions $\mathcal{P}\left[F(x,Q^2)\right]$. The probability measure constructed in this way contains all information from experimental data, including correlations, with the only assumption of smoothness. Expectation values and moments over this probability measure are then evaluated as averages over the trained network sample,

$$\left\langle \mathcal{F}\left[F(x,Q^2)\right]\right\rangle = \int \mathcal{D}F\mathcal{P}\left[F(x,Q^2)\right]\mathcal{F}\left[F(x,Q^2)\right] = \frac{1}{N_{\rm rep}}\sum_{k=1}^{N_{\rm rep}}\mathcal{F}\left(F^{({\rm net})(k)}(x,Q^2)\right) \ . \quad (9.25)$$

where $\mathcal{F}[F]$ is an arbitrary function of $F(x,Q^2)$.

[12]Contributing authors: L. Del Debbio, S. Forte, J. I. Latorre, A. Piccione, J. Rojo

The first step is the Monte Carlo sampling of experimental data, generating $N_{\rm rep}$ replicas of the original $N_{\rm dat}$ experimental data,

$$F_i^{({\rm art})(k)} = \left(1+r_N^{(k)}\sigma_N\right)\left[F_i^{({\rm exp})} + r_i^{s,(k)}\sigma_i^{stat} + \sum_{l=1}^{N_{sys}} r^{l,(k)}\sigma_i^{sys,l}\right] \ , \qquad i=1,\ldots,N_{\rm dat} \ , \quad (9.26)$$

where $r$ are gaussian random numbers with the same correlation as the respective uncertainties, and $\sigma^{stat}, \sigma^{sys}, \sigma_N$ are the statistical, systematic and normalization errors. The number of replicas $N_{\rm rep}$ has to be large enough so that the replica sample reproduces central values, errors and correlations of the experimental data.

The second step consists on training a neural network[13] on each of the data replicas. Neural networks are specially suitable to parametrize parton distributions since they are unbiased, robust approximants and interpolate between data points with the only assumption of smoothness. The neural network training consist on the minimization for each replica of the $\chi^2$ defined with the inverse of the experimental covariance matrix,

$$\chi^{2(k)} = \frac{1}{N_{\rm dat}}\sum_{i,j=1}^{N_{\rm dat}}\left(F_i^{({\rm art})(k)} - F_i^{({\rm net})(k)}\right){\rm cov}_{ij}^{-1}\left(F_j^{({\rm art})(k)} - F_j^{({\rm net})(k)}\right) \ . \quad (9.27)$$

Our minimization strategy is based on Genetic Algorithms (introduced in [108]), which are specially suited for finding global minima in highly nonlinear minimization problems.

The set of trained nets, once is validated through suitable statistical estimators, becomes the sought-for probability measure $\mathcal{P}\left[F(x,Q^2)\right]$ in the space of structure functions. Now observables with errors and correlations can be computed from averages over this probability measure, using eq. (9.25). For example, the average and error of a structure function $F(x,Q^2)$ at arbitrary $(x,Q^2)$ can be computed as

$$\left\langle F(x,Q^2)\right\rangle = \frac{1}{N_{\rm rep}}\sum_{k=1}^{N_{\rm rep}} F^{({\rm net})(k)}(x,Q^2), \quad \sigma(x,Q^2) = \sqrt{\left\langle F(x,Q^2)^2\right\rangle - \left\langle F(x,Q^2)\right\rangle^2} \ . \quad (9.28)$$

A more detailed account of the application of the neural network approach to structure functions can be found in [107], which describes the most recent NNPDF parametrization of the proton structure function[14].

Hence this strategy can be used also to parametrize parton distributions, provided one now takes into account perturbative QCD evolution. Therefore we need to define a suitable evolution formalism.

[13] For a more throughly description of neural network, see [106] and references therein

[14] The source code, driver program and graphical web interface for our structure function fits is available at `http://sophia.ecm.ub.es/f2neural`.

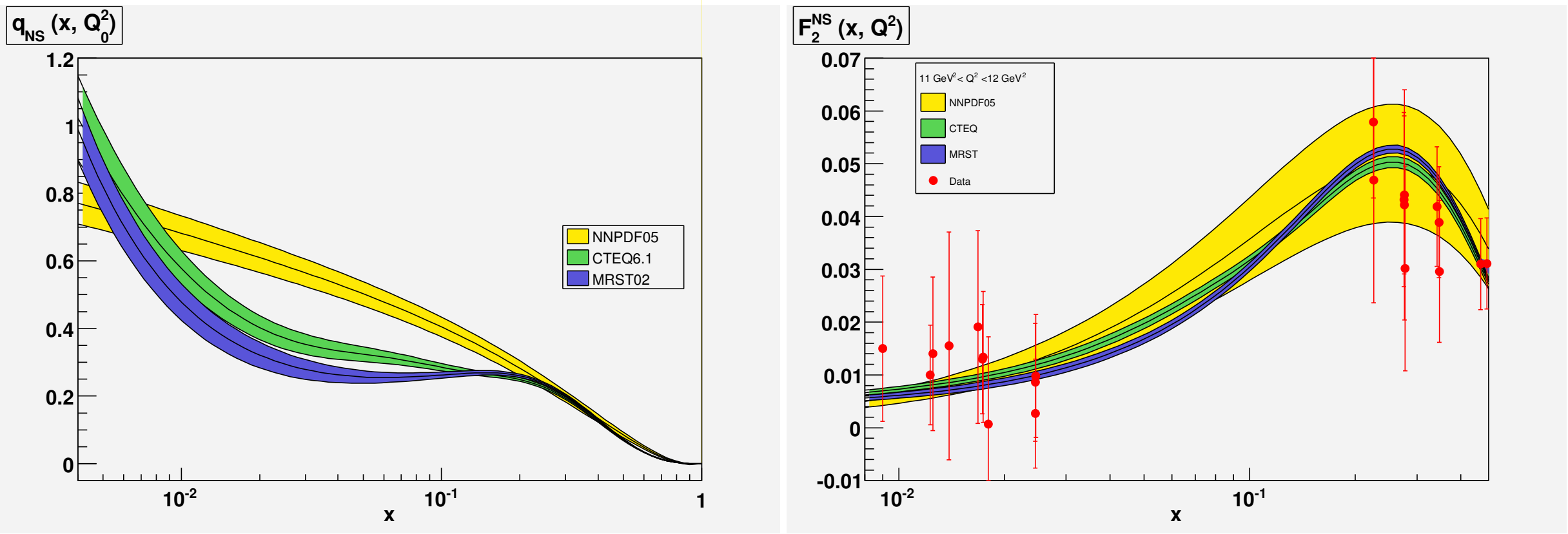

**Fig. 26:** Preliminary results for the NNPDF $q_{NS}$ fit at $Q_0^2 = 2\,\mathrm{GeV}^2$, and the prediction for $F_2^{NS}(x, Q^2)$ compared with the CTEQ and MRST results.

Since complex neural networks are not allowed, we must use the convolution theorem to evolve parton distributions in $x-$space using the inverse $\Gamma(x)$ of the Mellin space evolution factor $\Gamma(N)$, defined as

$$q(N, Q^2) = q(N, Q_0^2)\Gamma\left(N, \alpha_s\left(Q^2\right), \alpha_s\left(Q_0^2\right)\right) \ , \tag{9.29}$$

The only subtlety is that the x-space evolution factor $\Gamma(x)$ is a distribution, which must therefore be regulated at $x = 1$, yielding the final evolution equation,

$$q(x, Q^2) = q(x, Q_0^2)\int_x^1 dy\,\Gamma(y) + \int_x^1 \frac{dy}{y}\Gamma(y)\left(q\left(\frac{x}{y}, Q_0^2\right) - yq(x, Q_0^2)\right) \ , \tag{9.30}$$

where in the above equation $q(x, Q_0^2)$ is parametrized using a neural network. At higher orders in perturbation theory coefficient functions $C(N)$ are introduced through a modified evolution factor, $\tilde{\Gamma}(N) \equiv \Gamma(N)C(N)$. We have benchmarked our evolution code with the Les Houches benchmark tables [59] at NNLO up to an accuracy of $10^{-5}$. The evolution factor $\Gamma(x)$ and its integral are computed and interpolated before the neural network training in order to have a faster fitting procedure.

As a first application of our method, we extract the nonsinglet parton distribution $q_{NS}(x, Q_0^2) = \frac{1}{6}\left(u + \bar{u} - d - \bar{d}\right)(x, Q_0^2)$ from the nonsinglet structure function $F_2^{NS}(x, Q^2)$ as measured by the NMC [75] and BCDMS [73, 74] collaborations. The preliminary results of a NLO fit with fully correlated uncertainties [109] can be seen in fig. 26 compared to other pdfs sets. Our preliminary results appear to point in the direction that the uncertainties at small $x$ do not allow, provided the current experimental data, to determine if $q_{NS}(x, Q^2)$ grows at small $x$, as supported by different theoretical arguments as well as by other global parton fits. However, more work is still needed to confirm these results. Only additional nonsinglet structure function data at small $x$ could settle in a definitive way this issue[15].

Summarizing, we have described a general technique to parametrize experimental data in an bias-free way with a faithful estimation of their uncertainties, which has been successfully applied to structure functions and that now is being implemented in the context of parton distribution. The next step will be to construct a full set of parton distributions from all available hard-scattering data using the strategy described in this contribution.

[15]Like the experimental low $x$ deuteron structure function which would be measured in an hypothetical electron-deuteron run at HERA II, as it was pointed out during the workshop by M. Klein ( [110]) and C. Gwenlan

# Resummation

*G. Altarelli, J. Andersen, R. D. Ball, M. Ciafaloni, D. Colferai, G. Corcella, S. Forte, L. Magnea, A. Sabio Vera, G. P. Salam, A. Staśto*

## 1 Introduction [1]

An accurate perturbative determination of the hard partonic cross-sections (coefficient functions) and of the anomalous dimensions which govern parton evolution is necessary for the precise extraction of parton densities. Recent progress in the determination of higher order contributions to these quantities has been reviewed in [1]. As is well known, such high-order perturbative calculations display classes of terms containing large logarithms, which ultimately signal the breakdown of perturbation theory. Because these terms are scale–dependent and in general non universal, lack of their inclusion can lead to significant distortion of the parton densities in some kinematical regions, thereby leading to loss of accuracy if parton distributions extracted from deep-inelastic scattering (DIS) or the Drell-Yan (DY) processes are used at the LHC.

Logarithimic enhancement of higher order perturbative contribution may take place when more than one large scale ratio is present. In DIS and DY this happen in the two opposite limits when the center-of-mass energy of the partonic collision is much higher than the characteristic scale of the process, or close to the threshold for the production of the final state. These correspond respectively to the small $x$ and large $x$ kinematical regions, where $0 \leq x \leq 1$ is defined in terms of the invariant mass $M^2$ of the non-leptonic final state as $M^2 = \frac{(1-x)Q^2}{x}$. The corresponding perturbative contributions are respectively enhanced by powers of $\ln \frac{1}{x}$ and $\ln(1-x)$, or, equivalently, in the space of Mellin moments, by powers of $\frac{1}{N}$ and $\ln N$, where $N \to 0$ moments dominate as $x \to 0$ while $N \to \infty$ moments dominate as $x \to 1$.

The theoretical status of small $x$ and large $x$ resummation is somewhat different. Large $x$ logs are well understood and the corresponding perturbative corrections have been determined to all orders with very high accuracy. Indeed, the coefficients that determine their resummation can be extracted from fixed-order perturbative computations. Their resummation for DY and DIS was originally derived in [2, 3] and extended on very general grounds in [4]. The coefficients of the resulting exponentiation have now been determined so that resummation can now be performed exactly at $\mathrm{N^2LL}$ [5, 6], and to a very good approximation at $\mathrm{N^3LL}$ [7–9], including even some non-logarithmic terms [10]. On the other hand, small $x$ logs are due to the fact that at high energies, due to the opening of phase space, both collinear [11–13] and high-energy [14–17] logarithms contribute, and thus the coefficients required for their resummation can only be extracted from a simultaneous resolution of the DGLAP equation, which resums collinear logarithms, and the BFKL equation, which resum the high-energy logarithms. Although the determination of the kernels of these two equations has dramatically progressed in the last several years, thanks to the computation of the $\mathrm{N^2LO}$ DGLAP kernel [6, 18] and of the NLO BFKL kernel [14–17, 19, 20], the formalism which is needed to combine these two equations, as required for sucessful phenomenology, has only recently progressed to the point of being usable for realistic applications [21–30].

In practice, however, neither small $x$ nor large $x$ resummation is systematically incorporated in current parton fits, so data points for which such effects may be important must be discarded. This is especially unsatisfactory in the case of large $x$ resummation, where resummed results (albeit with a varying degree of logarithmic accuracy) are available for essentially all processes of interest for a global parton fit, in particular, besides DIS and DY, prompt photon production [31, 32], jet production [33, 34] and heavy quark electroproduction [35,36]. Even if one were to conclude that resummation is not needed, either because (at small $x$) it is affected by theoretical uncertainties or because (at large $x$) its effects are small, this conclusion could only be arrived at after a careful study of the impact of resummation on the determination of parton distributions, which is not available so far.

[1] Subsection coordinator: S. Forte

The purpose of this section is to provide a first assessment of the potential impact of the inclusion of small $x$ and large $x$ resummation on the determination of parton distributions. In the case of large $x$, this will be done by determining resummation effects on parton distributions extracted from structure functions within a simplified parton fit. In the case of small $x$, this will be done through a study of the impact of small $x$ resummation on splitting functions, as well as the theoretical uncertainty involved in the resummation process, in particular by comparing the results obtained within the approach of ref. [21–23] and that of ref. [24–30]. We will also discuss numerical approaches to the solution of the small-$x$ (BFKL) evolution equation.

## 2 Soft gluons

With the current level of theoretical control of soft gluon resummations, available calculations for DIS or DY should be fully reliable over most of the available phase space. Specifically, one expects current (resummed) predictions for DIS structure functions to apply so long as the leading power correction can be neglected, *i.e.* so long as $W^2 \sim (1-x)Q^2 >> \Lambda^2$, with $x = x_{Bj}$. Similarly, for the inclusive DY cross section, one would expect the same to be true so long as $(1-z)^2 Q^2 >> \Lambda^2$, where now $z = Q^2/\hat{s}$, with $\hat{s} = x_1 x_2 S$ the partonic center of mass energy squared. Indeed, as already mentioned, a consistent inclusion of resummation effects in parton fits is feasible with present knowledge: on the one hand, recent fits show that consistent parton sets can be obtained by making use of data from a single process (DIS) (see [37,38] and Ref. [39]), on the other hand, even if one adopts the philosophy of global fits, resummed calculations are available for all processes of interest.

In practice, however, currently available global parton fits are based on NLO, or $\mathrm{N^2LO}$ fixed-order perturbative calculations, so data points which would lie within the expected reach of resummed calculations cannot be fit consistently and must be discarded. The effect is that large-$x$ quark distributions become less constrained, which has consequences on the gluon distribution, as well as on medium-$x$ quark distributions, through sum rules and evolution. The pool of untapped information is growing, as more data at large values of $x$ have become available from, say, the NuTeV collaboration at Fermilab [40, 41]. A related issue is the fact that a growing number of QCD predictions for various processes of interest at the LHC are now computed including resummation effects in the hard partonic cross sections, which must be convoluted with parton densities in order to make predictions at hadron level. Such predictions are not fully consistent, since higher order effects are taken into account at parton level, but disregarded in defining the parton content of the colliding hadrons.

It is therefore worthwile to provide an assessment of the potential impact of resummation on parton distributions. Here, we will do this by computing resummation effects on quark distributions in the context of a simplified parton fit.

### 2.1 General Formalism in DIS

Deep Inelastic Scattering structure functions $F_i(x, Q^2)$ are given by the convolution of perturbative coefficient functions, typically given in the $\overline{\mathrm{MS}}$ factorization scheme, and parton densities. The coefficient functions $C_i^q$ for quark-initiated DIS present terms that become large when the Bjorken variable $x$ for the partonic process is close to $x = 1$, which forces gluon radiation from the incoming quark to be soft or collinear. At $\mathcal{O}(\alpha_s)$, for example, the coefficient functions can be written in the form

$$C_i^q \left( x, \frac{Q^2}{\mu_F^2}, \alpha_s(\mu^2) \right) = \delta(1-x) + \frac{\alpha_s(\mu^2)}{2\pi} H_i^q \left( x, \frac{Q^2}{\mu_F^2} \right) + \mathcal{O}\left(\alpha_s^2\right) . \qquad (1)$$

Treating all quarks as massless, the part of $H_i^q$ which contains terms that are logarithmically enhanced as $x \to 1$ reads

$$H_{i,\mathrm{soft}}^q\left(x, \frac{Q^2}{\mu_F^2}\right) = 2C_F\left\{\left[\frac{\ln(1-x)}{1-x}\right]_+ + \frac{1}{(1-x)_+}\left(\ln\frac{Q^2}{\mu_F^2} - \frac{3}{4}\right)\right\}. \tag{2}$$

In moment space, where soft resummation is naturally performed, the contributions proportional to $\alpha_s[\ln(1-x)/(1-x)]_+$ and to $\alpha_s[1/(1-x)]_+$ correspond to double $(\alpha_s \ln^2 N)$ and single $(\alpha_s \ln N)$ logarithms of the Mellin variable $N$. The Mellin transform of Eq. (2) in fact reads, at large $N$,

$$\hat{H}_{i,\mathrm{soft}}^q\left(N, \frac{Q^2}{\mu_F^2}\right) = 2C_F\left\{\frac{1}{2}\ln^2 N + \left[\gamma_E + \frac{3}{4} - \ln\frac{Q^2}{\mu_F^2}\right]\ln N\right\}. \tag{3}$$

All terms growing logarithmically with $N$, as well as all $N$-independent terms corresponding to contributions proportional to $\delta(1-x)$ in $x$-space, have been shown to exponentiate. In particular, the pattern of exponentiation of logarithmic singularities is nontrivial: one finds that the coefficient functions can be written as

$$\hat{C}_i^q\left(N, \frac{Q^2}{\mu_F^2}, \alpha_s(\mu^2)\right) = \mathcal{R}\left(N, \frac{Q^2}{\mu_F^2}, \alpha_s(\mu^2)\right)\Delta\left(N, \frac{Q^2}{\mu_F^2}, \alpha_s(\mu^2)\right), \tag{4}$$

where $\mathcal{R}(N, Q^2/\mu_F^2, \alpha_s(\mu^2))$ is a finite remainder, nonsingular as $N \to \infty$, while [4]

$$\ln\Delta\left(N, \frac{Q^2}{\mu_F^2}, \alpha_s(\mu^2)\right) = \int_0^1 dx \frac{x^{N-1}-1}{1-x}\left\{\int_{\mu_F^2}^{(1-x)Q^2} \frac{dk^2}{k^2} A\left[\alpha_s(k^2)\right] + B\left[\alpha_s\left(Q^2(1-x)\right)\right]\right\}. \tag{5}$$

In Eq. (5) the leading logarithms (LL), of the form $\alpha_s^n \ln^{n+1} N$, are generated at each order by the function $A$. Next-to-leading logarithms (NLL), on the other hand, of the form $\alpha_s^n \ln^n N$, require the knowledge of the function $B$. In general, resumming N$^k$LL to all orders requires the knowledge of the function $A$ to $k+1$ loops, and of the function $B$ to $k$ loops. In the following, we will adopt the common standard of NLL resummation, therefore we need the expansions

$$A(\alpha_s) = \sum_{n=1}^{\infty}\left(\frac{\alpha_s}{\pi}\right)^n A^{(n)} \; ; \; B(\alpha_s) = \sum_{n=1}^{\infty}\left(\frac{\alpha_s}{\pi}\right)^n B^{(n)} \tag{6}$$

to second order for $A$ and to first order for $B$. The relevant coefficients are

$$\begin{aligned} A^{(1)} &= C_F\,, \\ A^{(2)} &= \frac{1}{2}C_F\left[C_A\left(\frac{67}{18} - \frac{\pi^2}{6}\right) - \frac{5}{9}n_f\right], \\ B^{(1)} &= -\frac{3}{4}C_F\,. \end{aligned} \tag{7}$$

Notice that in Eq. (5) the term $\sim A(\alpha_s(k^2))/k^2$ resums the contributions of gluons that are both soft and collinear, and in fact the anomalous dimension $A$ can be extracted order by order from the residue of the singularity of the nonsinglet splitting function as $x \to 1$. The function $B$, on the other hand, is related to collinear emission from the final state current jet.

In [35, 36] soft resummation was extended to the case of heavy quark production in DIS. In the case of heavy quarks, the function $B(\alpha_s)$ needs to be replaced by a different function, called $S(\alpha_s)$ in [36], which is characteristic of processes with massive quarks, and includes effects of large-angle soft radiation. In the following, we shall consider values of $Q^2$ much larger than the quark masses and employ the resummation results in the massless approximation, as given in Eq. (5).

### 2.2 Simplified parton fit

We would like to use large-$x$ resummation in the DIS coefficient functions to extract resummed parton densities from DIS structure function data. Large-$x$ data typically come from fixed-target experiments: in the following, we shall consider recent charged-current (CC) data from neutrino-iron scattering, collected by the NuTeV collaboration [40, 41], and neutral-current (NC) data from the NMC [42] and BCDMS [43, 44] collaborations.

Our strategy will be to make use of data at different, fixed values of $Q^2$. We will extract from these data moments of the corresponding structure functions, with errors; since such moments factor into a product of moments of parton densities times moments of coefficient functions, computing parton moments with errors is straightforward. We then compare NLO to resummed partons in Mellin space, and subsequently provide a translation back to $x$-space by means of simple parametrization. Clearly, given the limited data set we are working with, our results will be affected by comparatively large errors, and we will have to make simplifying assumptions in order to isolate specific quark densities. Resummation effects are, however, clearly visible, and we believe that our fit provides a rough quantitative estimate of their size. A more precise quantitative analysis would have to be performed in the context of a global fit.

The first step is to construct a parametrization of the chosen data. An efficient and faithful parametrization of the NMC and BCDMS neutral-current structure functions was provided in [45, 46], where a large sample of Monte Carlo copies of the original data was generated, taking properly into account errors and correlations, and a neural network was trained on each copy of the data. One can then use the ensemble of networks as a faithful and unbiased representation of the probability distribution in the space of structure functions. We shall make use of the nonsinglet structure function $F_2^{\rm ns}(x, Q^2)$ extracted from these data, as it is unaffected by gluon contributions, and provides a combination of up and down quark densities which is independent of the ones we extract from charged current data (specifically, $F_2^{\rm ns}(x, Q^2)$ gives $u - d$).

As far as the NuTeV data are concerned, we shall consider the data on the CC structure functions $F_2$ and $F_3$. The structure function $F_3$ can be written as a convolution of the coefficient function $C_3^q$ with quark and antiquark distributions, with no gluon contribution, as

$$xF_3 = \frac{1}{2}\left(xF_3^{\nu} + xF_3^{\bar{\nu}}\right) = x\left[\sum_{q,q'} |V_{qq'}|^2 \,(q - \bar{q}) \otimes C_3^q\right]. \quad (8)$$

We consider data for $F_3$ at $Q^2 = 12.59$ and $31.62$ GeV$^2$, and, in order to compute moments, we fit them using the functional form

$$xF_3(x) = Cx^{-\rho}(1 - x)^{\sigma}(1 + kx)\,. \quad (9)$$

The best-fit values of $C$, $\rho$ and $\delta$, along with the $\chi^2$ per degree of freedom, are given in [47]. Here we show the relevant NuTeV data on $xF_3$, along with our best-fit curves, in Fig. 1.

The analysis of NuTeV data on $F_2$ is slightly complicated by the fact that gluon-initiated DIS gives a contribution, which, however, is not enhanced but suppressed at large $x$. We proceed therefore by taking the gluon density from a global fit, such as the NLO set CTEQ6M [48], and subtract from $F_2$ the gluon contribution point by point. We then write $F_2$ as

$$F_2 \equiv \frac{1}{2}\left(F_2^{\nu} + F_2^{\bar{\nu}}\right) = x\sum_{q,q'} |V_{qq'}|^2 \left[(q + \bar{q}) \otimes C_2^q + g \otimes C_2^g\right] \equiv F_2^q + F_2^g\,, \quad (10)$$

and fit only the quark-initiated part $F_2^q$, using the same parametrization as in Eq. (9). Fig. 2 shows the data on $F_2^q$ and the best fit curves, as determined in Ref. [47]. After the subtraction of the gluon contribution from $F_2$, the structure functions we are considering ($F_2^q$, $xF_3$ and $F_2^{\rm ns}$) are all given in factorized form as

$$F_i(x, Q^2) = x\int_x^1 \frac{d\xi}{\xi}\, q_i\left(\xi, \mu_F^2\right) C_i^q\left(\frac{x}{\xi}, \frac{Q^2}{\mu_F^2}, \alpha_s(\mu^2)\right)\,, \quad (11)$$

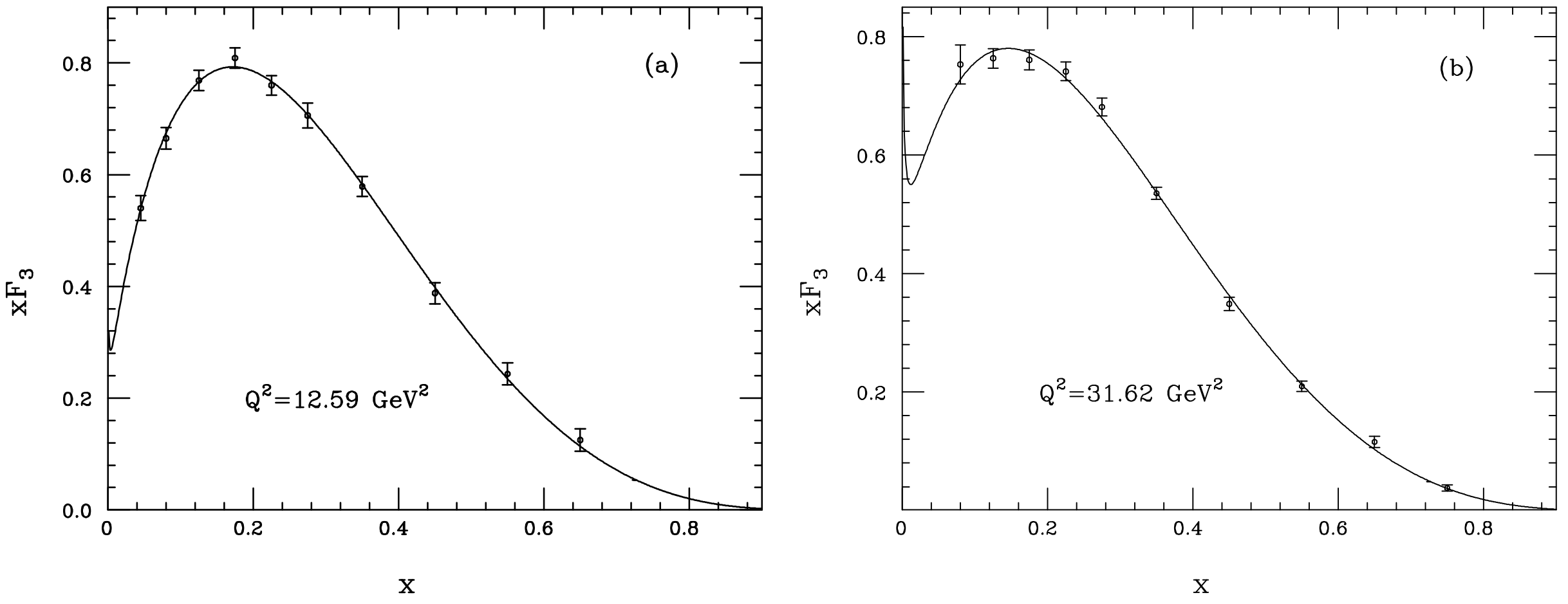

**Fig. 1:** NuTeV data on the structure function $xF_3$, at $Q^2 = 12.59$ GeV$^2$ (a) and at $Q^2 = 31.62$ GeV$^2$ (b), along with the best fit curve parametrized by Eq. (9).

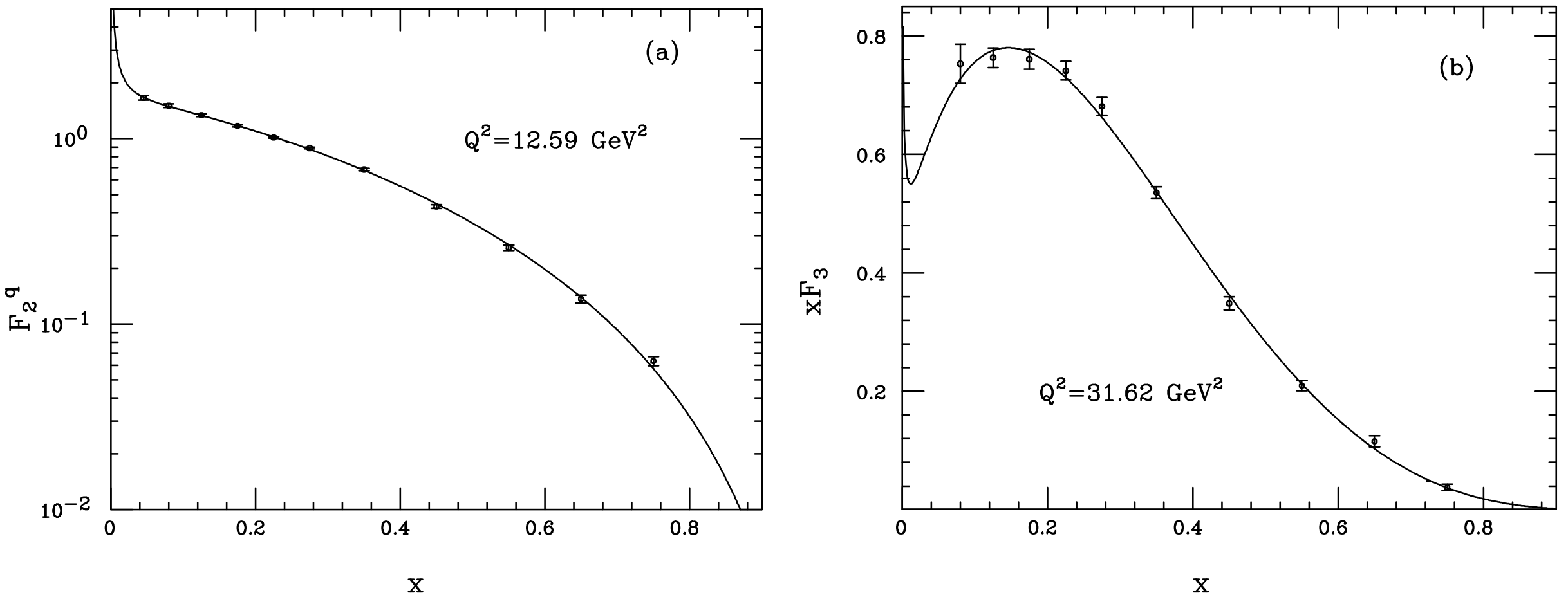

**Fig. 2:** NuTeV data on the quark-initiated contribution $F_2^q$ to the structure function $F_2$, for $Q^2 = 12.59$ GeV$^2$ (a), and $Q^2 = 31.62$ GeV$^2$ (b). The solid lines are the best-fit predictions.

where $C_i^q$ is the relevant coefficient function and $q_i$ is a combination of quark and antiquark distributions only. Hereafter, we shall take $\mu = \mu_F = Q$ for the factorization and renormalization scales. At this point, to identify individual quark distributions from this limited set of data, we need to make some simplifying assumptions. Following [47], we assume isospin symmetry of the sea, $\bar{u} = \bar{d}$, $s = \bar{s}$ and we further impose a simple proportionality relation expressing the antistrange density in terms of the other antiquarks, $\bar{s} = \kappa\bar{u}$. As in [47], we shall present results for $\kappa = \frac{1}{2}$. With these assumptions, we can explicit solve for the remaining three independent quark densities (up, down, and, say, strange), using the three data sets we are considering.

Taking the Mellin moments of Eq. (11), the convolution becomes an ordinary product and we can extract NLO or NLL-resummed parton densities, according to whether we use NLO or NLL coefficient functions. More precisely,

$$\hat{q}_i^{\rm NLO}(N,Q^2) = \frac{\hat{F}_i(N-1,Q^2)}{\hat{C}_i^{\rm NLO}\left(N,1,\alpha_s(Q^2)\right)} \ ; \ \ \hat{q}_i^{\rm res}(N,Q^2) = \frac{\hat{F}_i(N-1,Q^2)}{\hat{C}_i^{\rm res}\left(N,1,\alpha_s(Q^2)\right)} \ . \tag{12}$$

After extracting the combinations $q_i$, one can derive the individual quark densities, at NLO and including NLL large-$x$ resummation. We concentrate our analysis on the up quark distribution, since experimental errors on the structure functions are too large to see an effect of the resummation on the other quark densities, such as $d$ or $s$, with the limited data set we are using.

### 2.3 Impact of the resummation

We present results for moments of the up quark distribution in Figs. 3 and 4. Resummation effects become statistically significant around $N \sim 6-7$ at both values of $Q^2$. Notice that high moments of the resummed up density are *suppressed* with respect to the NLO density, as a consequence of the fact that resummation in the $\overline{\text{MS}}$ scheme enhances high moments of the coefficient functions.

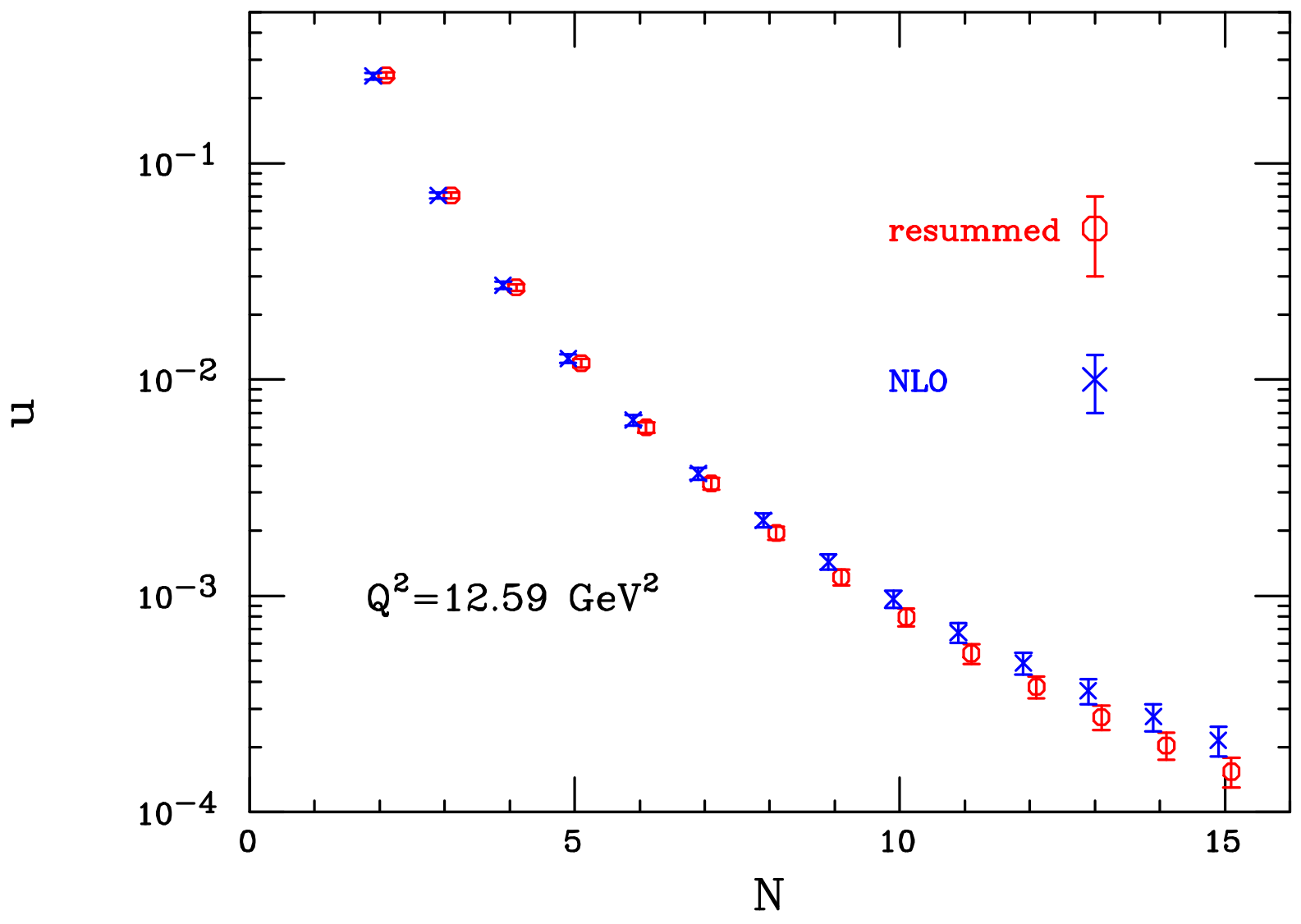

**Fig. 3:** NLO and resummed moments of the up quark distribution at $Q^2 = 12.59$ GeV$^2$

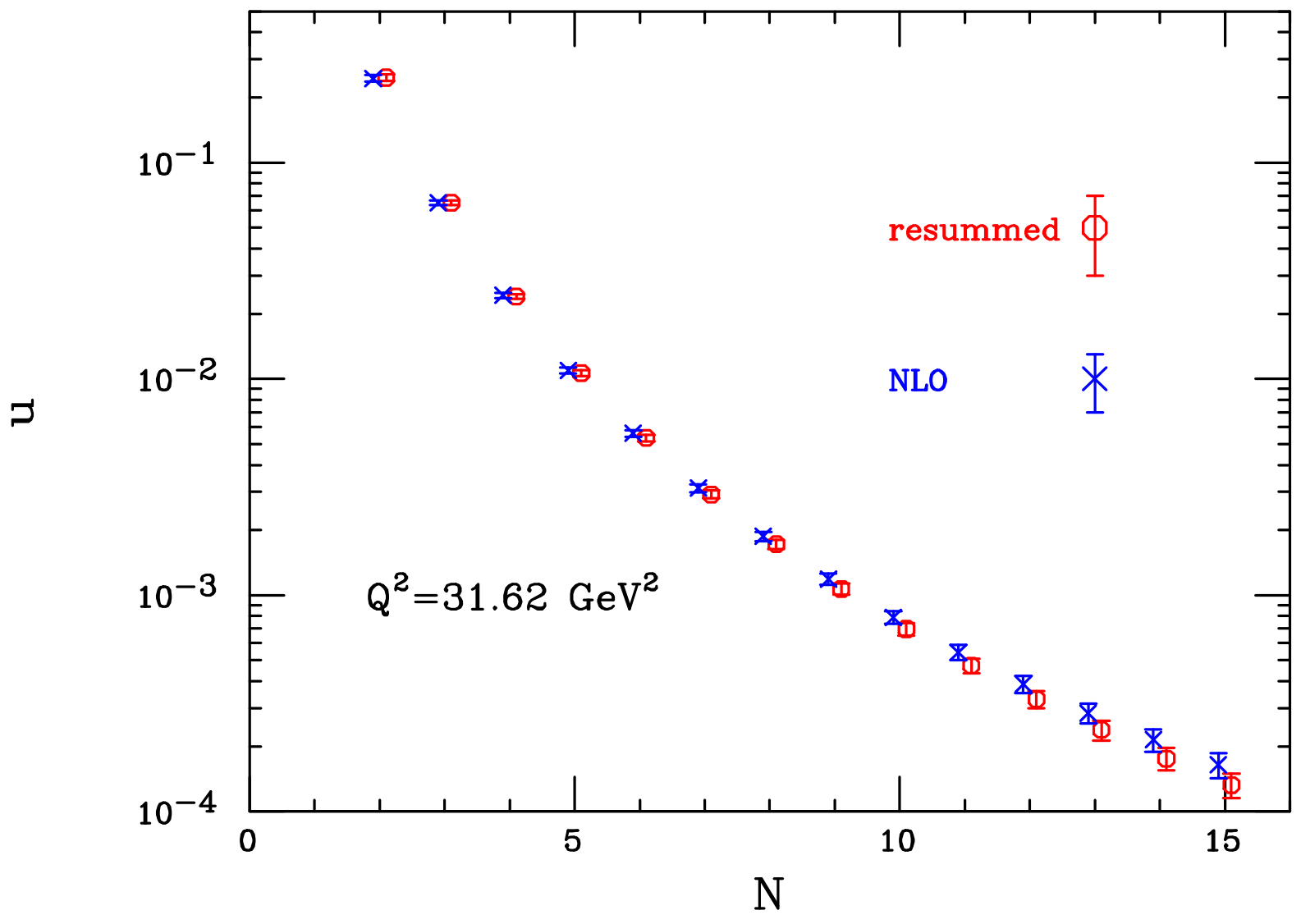

**Fig. 4:** As in Fig. 3, but at $Q^2 = 31.62$ GeV$^2$.

In order to illustrate the effect in the more conventional setting of $x$-space distributions, we fit our results for the moments to a simple parametrization of the form $u(x) = Dx^{-\gamma}(1-x)^{\delta}$. Our best fit values for the parameters, with statistical errors, are given in Table (1), and the resulting distributions are displayed in Fig. 5, with one standard deviation uncertainty bands. Once again, the effect of soft resummation is clearly visible at large $x$: it suppresses the quark densities extracted from the given structure function data with respect to the NLO prediction.

**Table 1:** Best fit values and errors for the up-quark $x$-space parametrization, at the chosen values of $Q^2$.

| $Q^2$ | PDF | $D$ | $\gamma$ | $\delta$ |
|---|---|---|---|---|
| 12.59 | NLO | $3.025 \pm 0.534$ | $0.418 \pm 0.101$ | $3.162 \pm 0.116$ |
| | RES | $4.647 \pm 0.881$ | $0.247 \pm 0.109$ | $3.614 \pm 0.128$ |
| 31.62 | NLO | $2.865 \pm 0.420$ | $0.463 \pm 0.086$ | $3.301 \pm 0.098$ |
| | RES | $3.794 \pm 0.583$ | $0.351 \pm 0.090$ | $3.598 \pm 0.104$ |

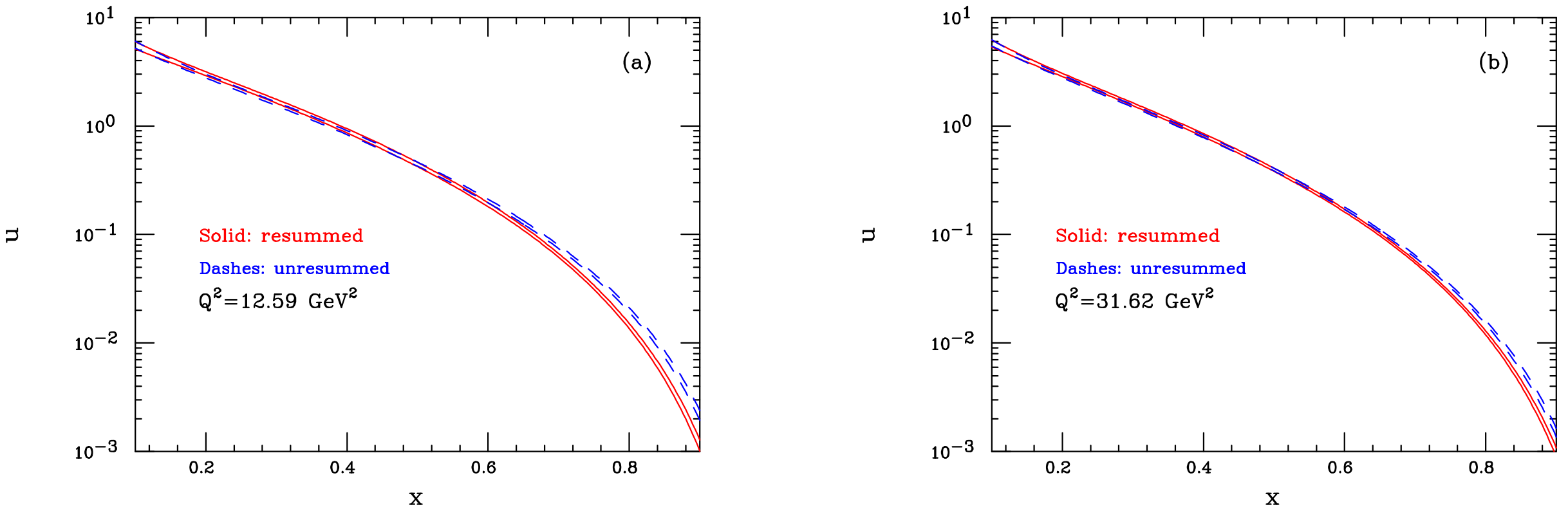

**Fig. 5:** NLO and resummed up quark distribution at $Q^2 = 12.59$ GeV$^2$ (a) and at $Q^2 = 31.62$ GeV$^2$, using the parametrization given in the text. The band corresponds to one standard deviation in parameter space.

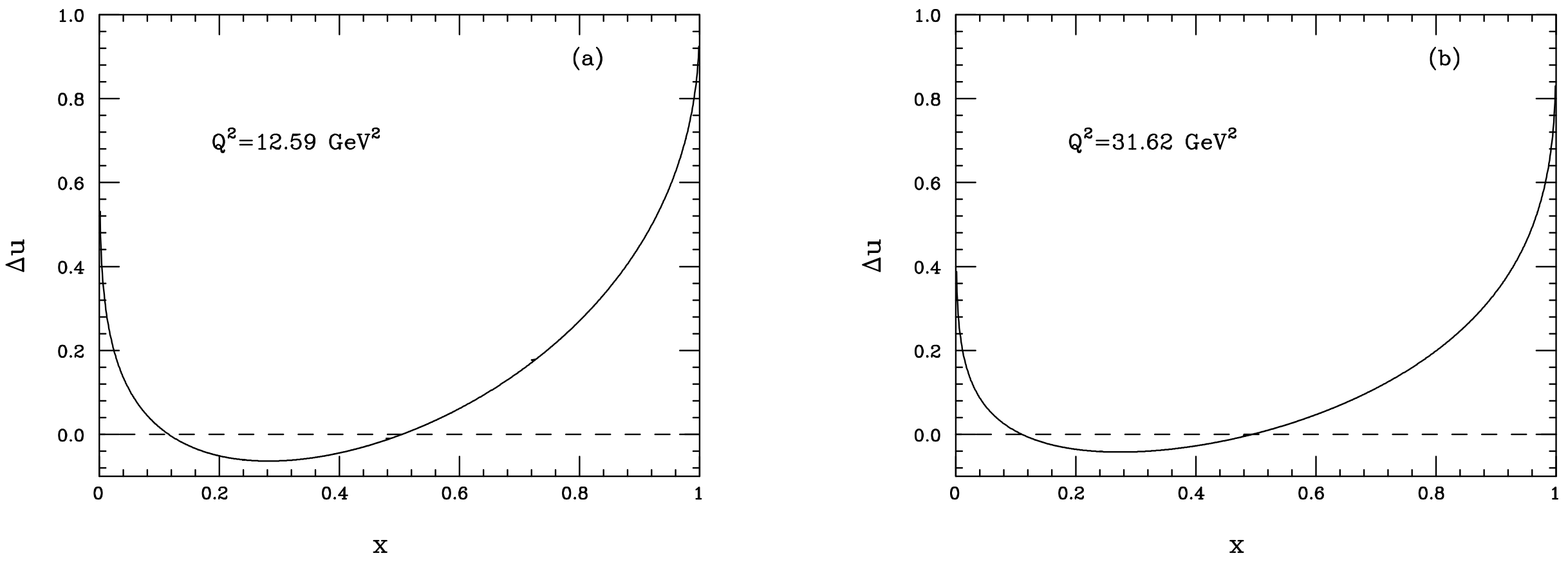

**Fig. 6:** Central value of relative change in the up quark distribution, $\Delta u(x) \equiv (u_{\rm NLO}(x) - u_{\rm res}(x))\,/u_{\rm NLO}(x)$, at $Q^2 = 12.59$ (a) and 31.62 GeV$^2$ (b).

In order to present the effect more clearly, we show in Fig. 6 the normalized deviation of the NLL-resummed prediction from the NLO one, i.e. $\Delta u(x) = (u_{\rm NLO}(x) - u_{\rm res}(x))\,/u_{\rm NLO}(x)$, at the two chosen values of $Q^2$ and for the central values of the best-fit parameters. We note a change in the sign of $\Delta u$ in the neighborhhod of the point $x = 1/2$: although our errors are too large for the effect to be statistically significant, it is natural that the suppression of the quark distribution at large $x$ be compensated by an enhancement at smaller $x$. In fact, the first moment of the coefficient function is unaffected by the resummation: thus $C_i^q$, being larger at large $x$, must become smaller at small $x$. The further sign change at $x \sim 0.1$, on the other hand, should not be taken too seriously, since our sample includes essentially no data at smaller $x$, and of course we are using an $x$-space parametrization of limited flexibility.

Finally, we wish to verify that the up-quark distributions extracted by our fits at $Q^2 = 12.59$ and $31.62$ GeV$^2$ are consistent with perturbative evolution. To achieve this goal, we evolve our $N$-space results at $Q^2 = 31.62$ GeV$^2$ down to 12.59 GeV$^2$, using NLO Altarelli–Parisi anomalous dimensions, and compare the evolved moments with the direct fit at 12.59 GeV$^2$. Figures 7 and 8 show that the results of our fits at 12.59 GeV$^2$ are compatible with the NLO evolution within the confidence level of one standard deviation. Note however that the evolution of resummed moments appears to give less consistent results, albeit within error bands: this can probably be ascribed to a contamination between pertubative resummation and power corrections, which we have not disentangled in our analysis.

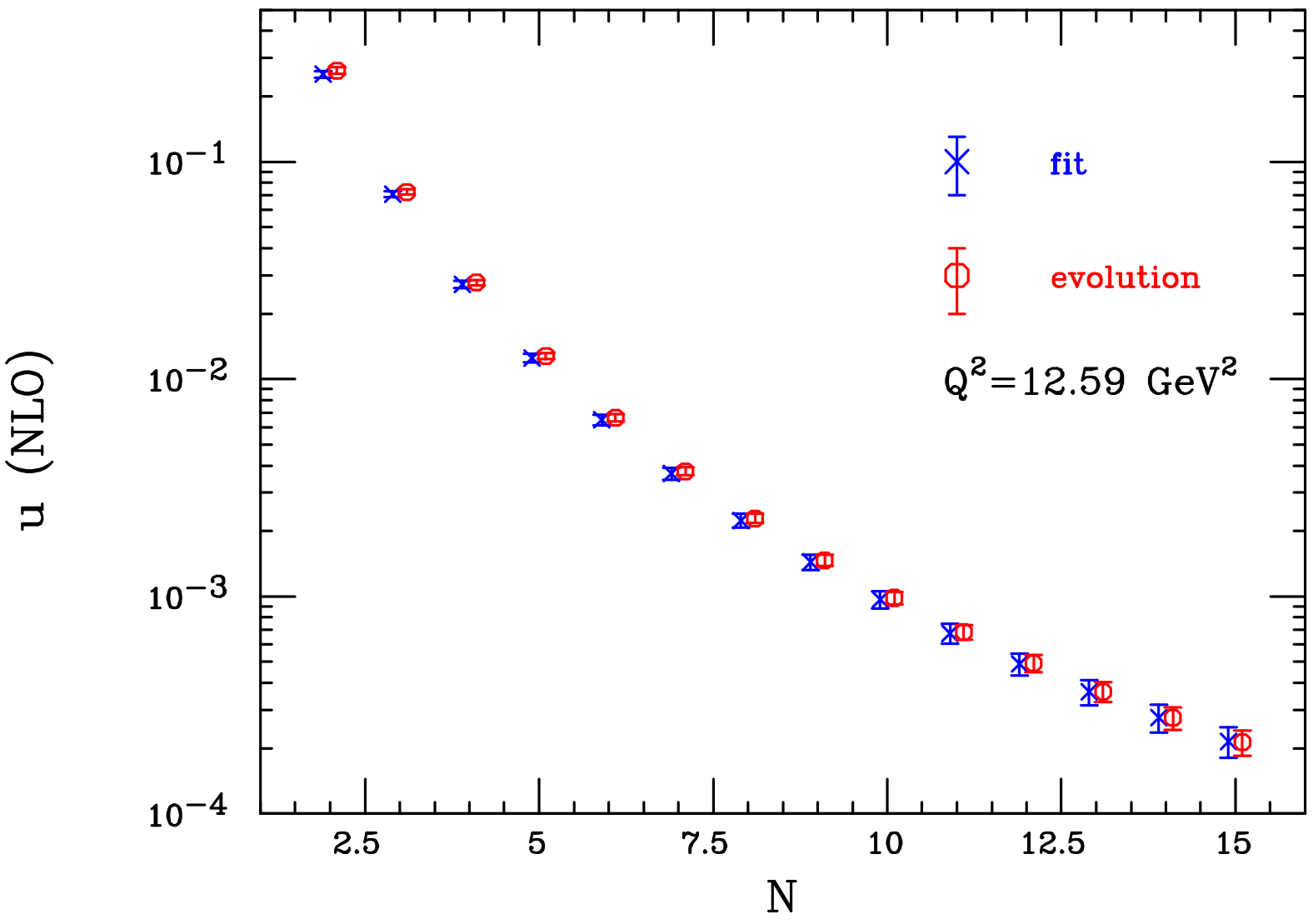

**Fig. 7:** Comparison of fitted moments of the NLO up quark distribution, at $Q^2 = 12.59$ GeV$^2$, with moments obtained via NLO evolution from $Q^2 = 31.62$ GeV$^2$.

Qualitatively, the observed effect on the up quark distribution is easily described, at least within the limits of a simple parametrization like the one we are employing: resummation increases the exponent $\delta$, responsible for the power-law decay of the distribution at large $x$, by about $10\%$ to $15\%$ at moderate $Q^2$. The exponent $\gamma$, governing the small-$x$ behavior, and the normalization $D$, are then tuned so that the first finite moment (the momentum sum rule) may remain essentially unaffected.

In conclusion, our results indicate that quark distributions are suppressed at large $x$ by soft gluon effects. Quantitatively, we observe an effect ranging between $10\%$ and $20\%$ when $0.6 < x < 0.8$ at moderate $Q^2$, where we expect power corrections not to play a significant role. Clearly, a more detailed quantitative understanding of the effect can be achieved only in the context of a broader and fully consistent fit. We would like however to notice two things: first, the effect of resummations propagates to smaller values of $x$, through the fact that the momentum sum rule is essentially unaffected by the resummation; similarly, evolution to larger values of $Q^2$ will shift the Sudakov suppression to smaller $x$. A second point is that, in a fully consistent treatment of hadronic cross section, there might be a partial compensation between the typical Sudakov enhancement of the partonic process and the Sudakov suppression of the quark distribution: the compensation would, however, be channel-dependent, since gluon-initiated partonic processes would be unaffected. We believe it would be interesting, and phenomenologically relevant, to investigate these issues in the context of a more comprehensive parton fit.

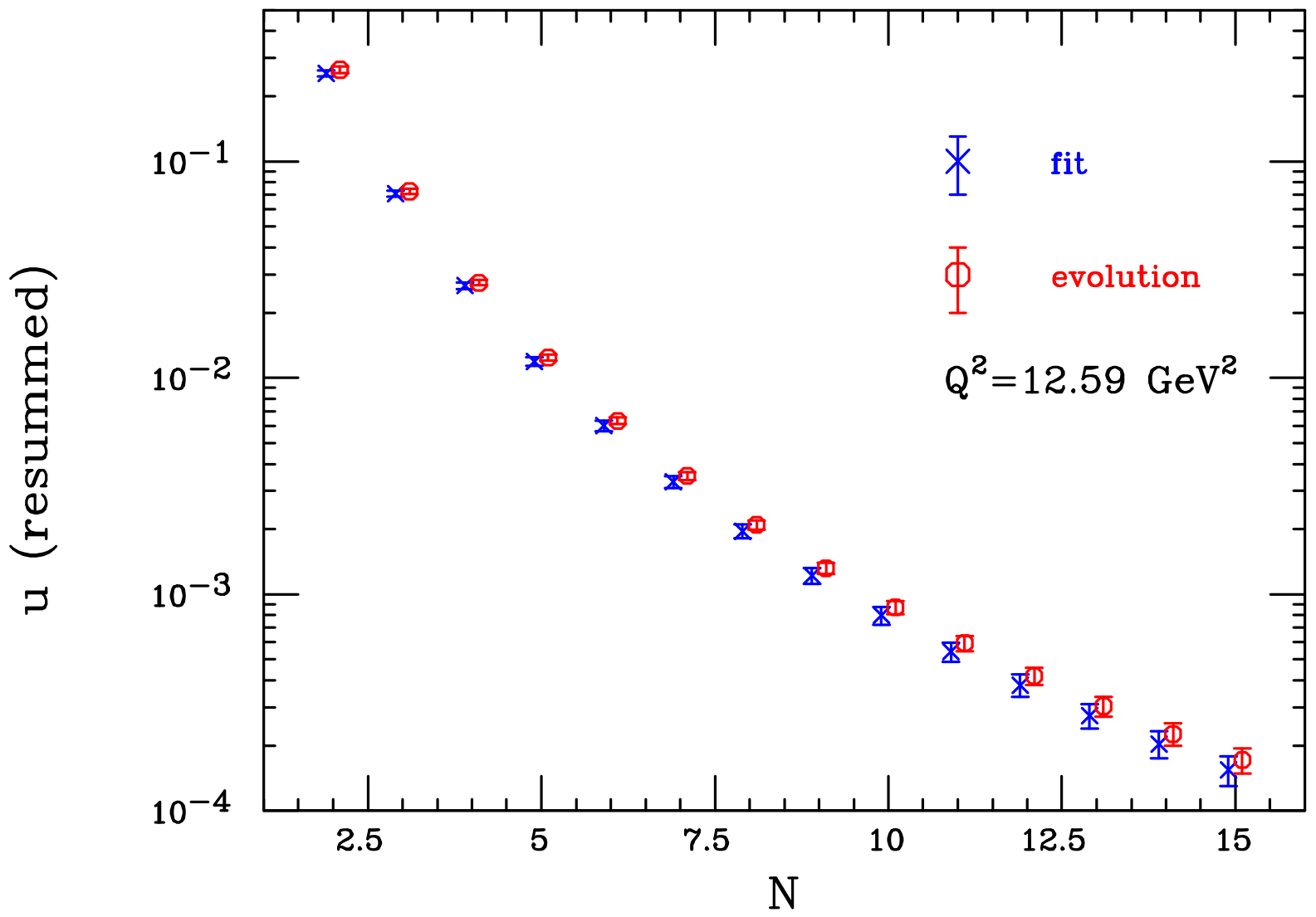

**Fig. 8:** As in Fig. 7, but comparing NLL-resummed moments of the up quark density.

## 3 Small $x$

Small $x$ structure functions are dominated by the flavour singlet contribution, whose coefficient functions and anomalous dimensions receive logarithmic enhancements, which make perturbation theory converge more slowly. In the small $x$, i.e. high energy limit, the cross section is quasi-constant and characterised by the effective expansion parameter $\langle\alpha_s(\boldsymbol{k}^2)\rangle \log\frac{1}{x}\log\frac{\boldsymbol{k}^2_{\rm max}}{\boldsymbol{k}^2_{\rm min}}$, where $x = Q^2/s$, $\boldsymbol{k}^2 \lesssim Q^2$ is the transverse momentum of the exchanged gluon, $s$ is the photon-proton centre of mass energy squared and $Q^2$ is the hard scale. Such expansion parameter can be large, due to both the double-logs and to the fact that $\langle\boldsymbol{k}^2\rangle$ may drift towards the non-perturbative region. Even assuming that truly non-perturbative effects are factored out — as is the case for structure functions — the problem remains of resumming the perturbative series with both kinds of logarithms [11–17]

In the BFKL approach one tries to resum the high-energy logarithms first, by an evolution equation in $\log 1/x$, whose $\boldsymbol{k}$-dependent evolution kernel is calculated perturbatively in $\alpha_s$. However, the leading kernel [14–17] overestimates the hard cross-section, and subleading ones [19, 20, 49] turn out to be large and of alternating sign, pointing towards an instability of the leading-log $x$ (L$x$) hierarchy. The problem is that, for any given value of the hard scales $Q, Q_0 \ll \sqrt{s}$ — think, for definiteness, of $\gamma^*(Q)$-$\gamma^*(Q_0)$ collisions —, the contributing kernels contain collinear enhancements in all $\boldsymbol{k}$-orderings of the exchanged gluons of type $\sqrt{s} \gg \cdots \boldsymbol{k}_1 \gg \boldsymbol{k}_2 \cdots$, or $\sqrt{s} \gg \cdots \boldsymbol{k}_2 \gg \boldsymbol{k}_1 \cdots$ and so on, to all orders in $\alpha_s$. Such enhancements are only partly taken into account by any given truncation of the L$x$ hierarchy, and they make it unstable. In the DGLAP evolution equation one resums collinear logarithms first, but fixed order splitting functions do contain [6, 18] high-energy logarithms also, and a further resummation is needed.

Two approaches to the simultaneous resummation of these two classes of logs have recently reached the stage where their phenomenological application can be envisaged. The renormalisation group improved (CCSS) approach [21–23, 50] is built up within the BFKL framework, by improving the whole hierarchy of subleading kernels in the collinear region, so as to take into account all the $\boldsymbol{k}$-orderings mentioned before, consistently with the RG. In the duality (ABF) approach [24–30, 51] one concentrates on the problem of obtaining an improved anomalous dimension (splitting function) for DIS which reduces to the ordinary perturbative result at large $N$ (large $x$), thereby automatically satisfying renormalization group constraints, while including resummed BFKL corrections at small $N$ (small $x$), determined through the renormalization-group improved (i.e. running coupling) version of the BFKL kernel.

We will briefly review the theoretical underpinnings of these two approaches in turn, and then compare phenomenological results obtained in both approaches. Note that we shall use the notation of the CCSS or ABF papers in the corresponding sections, in order to enable a simpler connection with the original literature, at the price of some notational discontinuity. In particular, $\ln\frac{1}{x}$ is called $Y$ by CCSS and $\xi$ by ABF; the Mellin variable conjugate to $\ln\frac{1}{x}$ is called $\omega$ by CCSS and $N$ by ABF; and the Mellin variable conjugated to $\ln\frac{Q^2}{\boldsymbol{k}^2}$ is called $\gamma$ by CCSS and $M$ by ABF.

### 3.1 The renormalisation group improved approach

The basic problem which is tackled in the CCSS approach [21–23, 50] is the calculation of the (azimuthally averaged) gluon Green function $\mathcal{G}(Y;k,k_0)$ as a function of the magnitudes of the external gluon transverse momenta $k\equiv|\boldsymbol{k}|,\ k_0\equiv|\boldsymbol{k}_0|$ and of the rapidity $Y\equiv\log\frac{s}{kk_0}$. This is not yet a hard cross section, because one needs to incorporate the impact factors of the probes [52–59]. Nevertheless, the Green function exhibits most of the physical features of the hard process, if we think of $k^2,\ k_0^2$ as external (hard) scales. The limits $k^2\gg k_0^2$ ($k_0^2\gg k^2$) correspond conventionally to the ordered (anti-ordered) collinear limit. By definition, in the $\omega$-space conjugate to $Y$ (so that $\hat{\omega}=\partial_Y$) one sets

$$\mathcal{G}_\omega(\boldsymbol{k},\boldsymbol{k}_0)\equiv[\omega-\mathcal{K}_\omega]^{-1}(\boldsymbol{k},\boldsymbol{k}_0)\,, \tag{13}$$

$$\omega\mathcal{G}_\omega(\boldsymbol{k},\boldsymbol{k}_0)=\delta^2(\boldsymbol{k}-\boldsymbol{k}_0)+\int d^2\boldsymbol{k}'\,\mathcal{K}_\omega(\boldsymbol{k},\boldsymbol{k}')\mathcal{G}_\omega(\boldsymbol{k}',\boldsymbol{k}_0)\,, \tag{14}$$

where $\mathcal{K}_\omega(\boldsymbol{k},\boldsymbol{k}')$ is a kernel to be defined, whose $\omega=0$ limit is related to the BFKL $Y$-evolution kernel discussed before.

In order to understand the RG constraints, it is useful to switch from $\boldsymbol{k}$-space to $\gamma$-space, where the variable $\gamma$ is conjugated to $t\equiv\log\boldsymbol{k}^2/\boldsymbol{k}_0^2$ at fixed $Y$, and to make the following kinematical remark: the ordered (anti-ordered) region builds up scaling violations in the Bjorken variable $x=\boldsymbol{k}^2/s$ ($x_0=\boldsymbol{k}_0^2/s$) and, if $x$ ($x_0$) is fixed instead of $kk_0/s=e^{-Y}$, the variable conjugated to $t$ is shifted [60] by an $\omega$-dependent amount, and becomes $\gamma+\frac{\omega}{2}\sim\partial_{\log\boldsymbol{k}^2}$ $\ (1-\gamma+\frac{\omega}{2}\sim\partial_{\log\boldsymbol{k}_0^2})$. Therefore, the characteristic function $\chi_\omega(\gamma)$ of $\mathcal{K}_\omega$ (with a factor $\alpha_s$ factored out) must be singular when either one of the variables is small, as shown (in the frozen $\alpha_s$ limit) by

$$\frac{1}{\omega}\chi_\omega(\gamma)\to\left[\frac{1}{\gamma+\frac{\omega}{2}}+\frac{1}{1-\gamma+\frac{\omega}{2}}+\cdots\right]\left[\gamma_{gg}^{(1)}(\alpha_s,\omega)+\cdots\right]\,, \tag{15}$$

where $\gamma_{gg}^{(1)}$ is the one-loop gluon anomalous dimension, and further orders may be added. Eq. (15) ensures the correct DGLAP evolution in either one of the collinear limits (because, e.g., $\gamma+\frac{\omega}{2}\sim\partial_{\log\boldsymbol{k}^2}$) and is $\omega$-dependent, because of the shifts. Since higher powers of $\omega$ are related to higher subleading powers of $\alpha_s$ [61], this $\omega$-dependence of the constraint (15) means that the whole hierarchy of subleading kernels is affected.

To sum up, the kernel $\mathcal{K}_\omega$ is constructed so as to satisfy the RG constraint (15) and to reduce to the exact L$x$ + NL$x$ BFKL kernels in the $\omega\to0$ limit; it is otherwise interpolated on the basis of various criteria (e.g., momentum conservation), which involve a "scheme" choice.

The resulting integral equation has been solved in [21–23] by numerical matrix evolution methods in $\boldsymbol{k}$- and $x$-space. Furthermore, introducing the integrated gluon density $g$, the resummed splitting function $P_{\rm eff}(x,Q^2)$ is defined by the evolution equation

$$\frac{\partial g(x,Q^2)}{\partial\log Q^2}=\int\frac{\mathrm{d}z}{z}\,P_{\rm eff}\big(z,\alpha_s(Q^2)\big)g\Big(\frac{x}{z},Q^2\Big)\,, \tag{16}$$

and has been extracted [21–23] by a numerical deconvolution method [62]. Note that in the RGI approach the running of the coupling is treated by adopting in (14) the off-shell dependence of $\alpha_s$ suggested by the BFKL and DGLAP limits, and then solving the ensuing integral equation numerically.

It should be noted that the RGI approach has the somewhat wider goal of calculating the off-shell gluon density (13), not only its splitting function. Therefore, a comparison with the ABF approach, to be discussed below, is possible in the "on-shell" limit, in which the homogeneous (eigenvalue) equation of RGI holds. In the frozen coupling limit we have simply

$$\chi_\omega(\alpha_s, \gamma - \tfrac{\omega}{2}) = \omega \ , \qquad (\chi_\omega \text{ is at scale } kk_0) \ . \tag{17}$$

In the same spirit as the ABF approach [24–30, 51], when solving Eq. (17) for either $\omega$ or $\gamma$, we are able to identify the effective characteristic function and its dual anomalous dimension

$$\omega = \chi_{\text{eff}}(\alpha_s, \gamma) \ ; \quad \gamma = \gamma_{\text{eff}}(\alpha_s, \omega) \ . \tag{18}$$

### 3.2 The duality approach

As already mentioned, in the ABF approach one constructs an improved anomalous dimension (splitting function) for DIS which reduces to the ordinary perturbative result at large $N$ (large $x$) given by:

$$\gamma(N, \alpha_s) = \alpha_s \gamma_0(N) \ + \ \alpha_s^2 \gamma_1(N) \ + \ \alpha_s^3 \gamma_2(N) \ \ldots . \tag{19}$$

while including resummed BFKL corrections at small $N$ (small $x$) which are determined by the aforementioned BFKL kernel $\chi(M, \alpha_s)$:

$$\chi(M, \alpha_s) = \alpha_s \chi_0(M) \ + \ \alpha_s^2 \chi_1(M) \ + \ \ldots, \tag{20}$$

which is the Mellin transform of the $\omega \to 0$, angular averaged kernel $\mathcal{K}$ eq. 14 with respect to $t = \ln \frac{k^2}{k_0^2}$. The main theoretical tool which enables this construction is the duality relation between the kernels $\chi$ and $\gamma$ [compare Eq. (18)]

$$\chi(\gamma(N, \alpha_s), \alpha_s) = N, \tag{21}$$

which is a consequence of the fact that the solutions of the BFKL and DGLAP equations coincide at leading twist [24, 51, 63]. Further improvements are obtained exploiting the symmetry under gluon interchange of the BFKL gluon-gluon kernel and through the inclusion of running coupling effects.

By using duality, one can construct a more balanced expansion for both $\gamma$ and $\chi$, the "double leading" (DL) expansion, where the information from $\chi$ is used to include in $\gamma$ all powers of $\alpha_s/N$ and, conversely $\gamma$ is used to improve $\chi$ by all powers of $\alpha_s/M$. A great advantage of the DL expansion is that it resums the collinear poles of $\chi$ at $M = 0$, enabling the imposition of the physical requirement of momentum conservation $\gamma(1, \alpha_s) = 0$, whence, by duality:

$$\chi(0, \alpha_s) = 1. \tag{22}$$

This procedure eliminates in a model independent way the alternating sign poles $+1/M, -1/M^2, .....$ that appear in $\chi_0$, $\chi_1$,.... These poles make the perturbative expansion of $\chi$ unreliable even in the central region of $M$: e.g., $\alpha_s \chi_0$ has a minimum at $M = 1/2$, while, at realistic values of $\alpha_s$, $\alpha_s \chi_0 + \alpha_s^2 \chi_1$ has a maximum.

At this stage, while the poles at $M = 0$ are eliminated, those at $M = 1$ remain, so that the DL expansion is still not finite near $M = 1$. The resummation of the $M = 1$ poles can be accomplished by exploiting the collinear-anticollinear symmetry, as suggested in the CCSS approach discussed above. In Mellin space, this symmetry implies that at the fixed-coupling level the kernel $\chi$ for evolution in $\ln \frac{s}{kk_0}$ must satisfy $\chi(M) = \chi(1 - M)$. This symmetry is however broken by the DIS choice of variables $\ln \frac{1}{x} = \ln \frac{s}{Q^2}$ and by the running of the coupling. In the fixed coupling limit the kernel $\chi_{\text{DIS}}$, dual to the DIS anomalous dimension, is related to the symmetric one $\chi_\sigma$ through the implicit equation [49]

$$\chi_{\text{DIS}}(M + 1/2\chi_\sigma(M)) = \chi_\sigma(M), \tag{23}$$

to be compared to eq. (17) of the CCSS approach.

Hence, the $M = 1$ poles can be resummed by performing the double-leading resummation of $M = 0$ poles of $\chi_{\rm DIS}$, determining the associated $\chi_\sigma$ through eq. (23), then symmetrizing it, and finally going back to DIS variables by using eq. (23) again in reverse. Using the momentum conservation eq. (22) and eq. (23), it is easy to show that $\chi_\sigma(M)$ is an entire function of M, with $\chi_\sigma(-1/2) = \chi_\sigma(3/2) = 1$ and has a minimum at $M = 1/2$. Through this procedure one obtains order by order from the DL expansion a symmetrized DL kernel $\chi_{\rm DIS}$, and its corresponding dual anomalous dimension $\gamma$. The kernel $\chi_{\rm DIS}$ has to all orders a minimum and satisfies a momentum conservation constraint $\chi_{\rm DIS}(0) = \chi_{\rm DIS}(2) = 1$.

The final ingredient of the ABF approach is a treatment of the running coupling corrections to the resummed terms. Indeed, their inclusion in the resummed anomalous dimension greatly softens the asymptotic behavior near $x = 0$. Hence, the dramatic rise of structure functions at small $x$, which characterized resummations based on leading–order BFKL evolution, and is ruled out phenomenologically, is replaced by a much milder rise. This requires a running coupling generalization of the duality Eq. (21), which is possible noting that in $M$ space the running coupling $\alpha_s(t)$ becomes a differential operator, since $t \to d/dM$. Hence, the BFKL evolution equation for double moments $G(N, M)$, which is an algebraic equation at fixed coupling, becomes a differential equation in $M$ for running coupling. In the ABF approach, one solves this differential equation analytically when the kernel is replaced by its quadratic approximation near the minimum. The solution is expressed in terms of an Airy function if the kernel is linear in $\alpha_s$, for example in the case of $\alpha_s\chi_0$, or of a Bateman function in the more general case of a non linear dependence on $\alpha_s$ as is the case for the DL kernels. The final result for the improved anomalous dimension is given in terms of the DL expansion plus the "Airy" or "Bateman" anomalous dimension, with the terms already included in the DL expansion subtracted away.

For example, at leading DL order, i.e. only using $\gamma_0(N)$ and $\chi_0(M)$, the improved anomalous dimension is

$$\gamma_I^{NL}(\alpha_s, N) = \Big[\alpha_s\gamma_0(N) + \alpha_s^2\gamma_1(N) + \gamma_s(\frac{\alpha_s}{N}) - \frac{n_c\alpha_s}{\pi N}\Big] + \gamma_A(c_0, \alpha_s, N) - \frac{1}{2} + \sqrt{\frac{2}{\kappa_0\alpha_s}[N - \alpha_s c_0]}. \tag{24}$$

The terms within square brackets give the LO DL approximation, i.e. they contain the fixed–coupling information from $\gamma_0$ and (through $\gamma_s$) from $\chi_0$. The "Airy" anomalous dimension $\gamma_A(c_0, \alpha_s, N)$ contains the running coupling resummation, i.e. it is the exact solution of the running coupling BFKL equation which corresponds to a quadratic approximation to $\chi_0$ near $M = 1/2$. The last two terms subtract the contributions to $\gamma_A(c_0, \alpha_s, N)$ which are already included in $\gamma_s$ and $\gamma_0$. In the limit $\alpha_s \to 0$ with $N$ fixed, $\gamma_I(\alpha_s, N)$ reduces to $\alpha_s\gamma_0(N) + O(\alpha_s^2)$. For $\alpha_s \to 0$ with $\alpha_s/N$ fixed, $\gamma_I(\alpha_s, N)$ reduces to $\gamma_s(\frac{\alpha_s}{N}) + O(\alpha_s^2/N)$, i.e. the leading term of the small $x$ expansion. Thus the Airy term is subleading in both limits. However, if $N \to 0$ at fixed $\alpha_s$, the Airy term replaces the leading singularity of the DL anomalous dimension, which is a square root branch cut, with a simple pole, located on the real axis at rather smaller $N$, thereby softening the small $x$ behaviour. The quadratic approximation is sufficient to give the correct asymptotic behaviour up to terms which are of subleading order in comparison to those included in the DL expression in eq. (24).

The running coupling resummation procedure can be applied to a symmetrized kernel, which possesses a minimum to all orders, and then extended to next-to-leading order [29, 30]. This entails various technical complications, specifically related to the nonlinear dependence of the symmetrized kernel on $\alpha_s$, to the need to include interference between running coupling effects and the small $x$ resummation, and to the consistent treatment of next-to-leading log $Q^2$ terms, in particular those related to the running of the coupling. It should be noted that even though the ABF appraoch is limited to the description of leading-twist evolution at zero-momentum transfer, it leads to a pair of systematic dual perturbative expansions for the $\chi$ and $\gamma$ kernels. Hence, comparison with the CCSS approach is possible for instance by comparing the NLO ABF kernel to the RG improved L$x$+NL$x$ CCSS kernel.

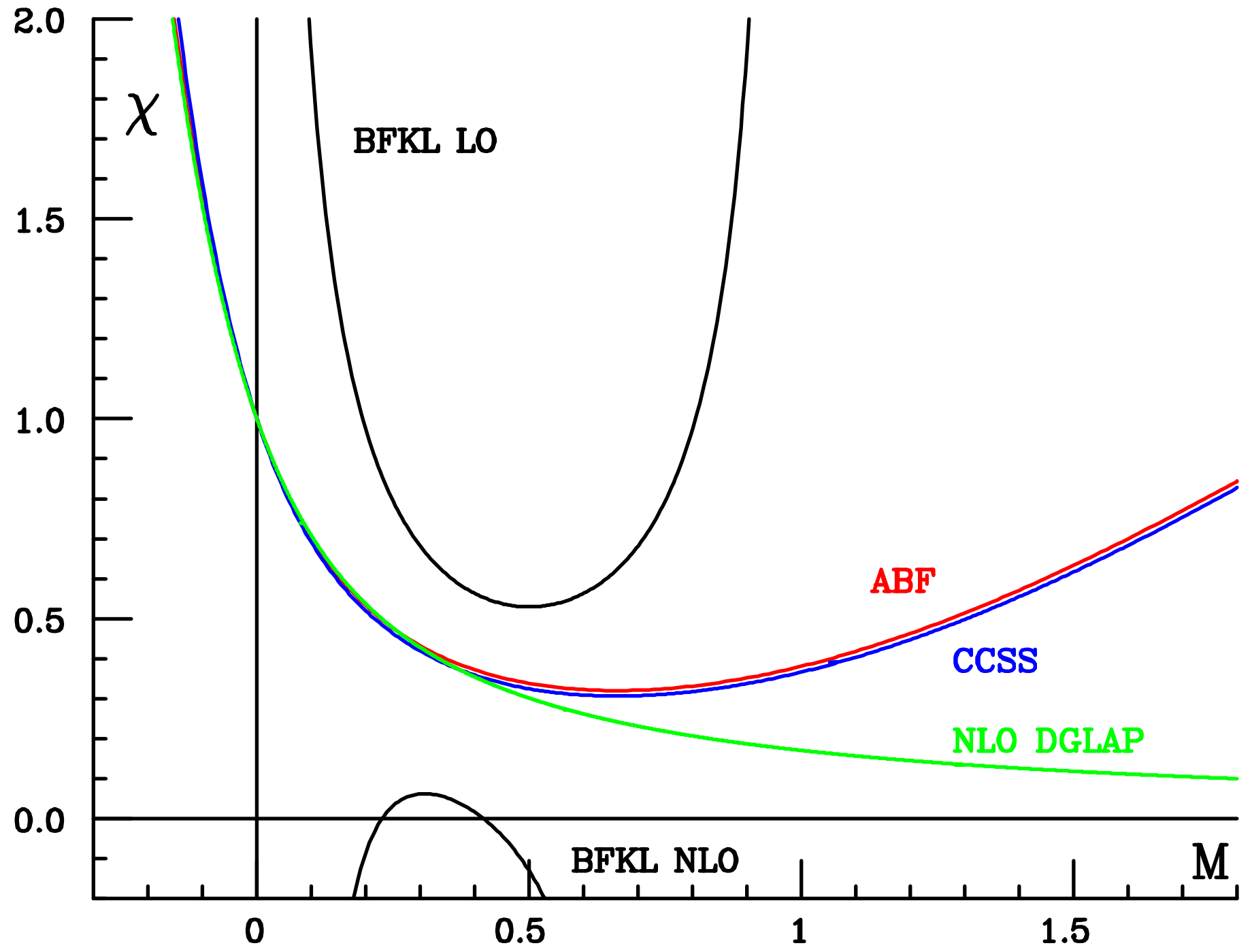

**Fig. 9:** The kernel $\chi$ (BFKL characteristic function) for fixed coupling ($\beta_0 = 0$) $\alpha_s = 0.2$ and $n_f = 0$. The BFKL curves are the LO and NLO truncations of eq. (20), the DGLAP curve is the dual eq. (21) of the NLO anomalous dimension eq. (19), while the CCSS and ABF curves are respectively the solution $\omega$ of eq. (17) and the solution $\chi_{\rm DIS}$ of eq. (23).

### 3.3 Comparison of results

Even though the basic underlying physical principles of the CCSS and ABF approaches are close, there are technical differences in the construction of the resummed RG-improved (CCSS) or symmetrized DL (ABF) kernel, in the derivation from it of an anomalous dimension and associated splitting function, and in the inclusion of running coupling effects. Therefore, we will compare results for the resummed fixed-coupling $\chi$ kernel (BFKL characteristic function), then the corresponding fixed-coupling splitting functions, and finally the running coupling splitting functions which provide the final result in both approaches. In order to assess the phenomenological impact on parton evolution we will finally compare the convolution of the splitting function with a "typical" gluon distribution.

In Fig. 9 we compare the solution, $\omega$, to the on-shell constraint, eq. (17) for the RGI CCSS result, and the solution $\chi_{\rm DIS}$ of eq. (23) for the symmetrized NLO DL ABF result. The pure L$x$ and NL$x$ (BFKL) and next-to-leading $\ln Q^2$ (DGLAP) are also shown. All curves are determined with frozen coupling ($\beta_0 = 0$), and with $n_f = 0$, in order to avoid complications related to the diagonalization of the DGLAP anomalous dimension matrix and to the choice of scheme for the quark parton distribution. The resummed CCSS and ABF results are very close, in that they coincide by construction at the momentum conservation points $M = \frac{1}{2}$ and $M = 2$, and differ only in the treatment of NLO DGLAP terms. In comparison to DGLAP, the resummed kernels have a minimum, related to the fact that both collinear and anticollinear logs are resummed. In comparison to BFKL, which has a minimum at LO but not NLO, the resummed kernels always have a perturbatively stable minimum, characterized by a lower intercept than leading–order BFKL: specifically, when $\alpha_s = 0.2$, $\lambda \sim 0.3$ instead of $\lambda \sim 0.5$. This corresponds to a softer small $x$ rise of the associated splitting function.

The fixed–coupling resummed splitting functions up to NLO are shown in figure 10, along with the unresummed DGLAP splitting functions up to NNLO.[2] In the CCSS approach the splitting function is determined by explicitly solving eq. (14) with the kernel corresponding to figure 9, and then applying the numerical deconvolution procedure of [62]. For $n_f = 0$ the NLO DGLAP splitting function has the property that it vanishes at small $x$ — this makes it relatively straightforward to combine not just LO DGLAP but also NLO DGLAP with the NLLx resummation. Both the CCSS NL$x$+LO and NL$x$+NLO curves are shown in Fig. 10. On the other hand, in the ABF approach the splitting function is the inverse Mellin transform of the anomalous dimension obtained using duality Eq. (21) from the symmetrized DL $\chi$ kernel. Hence, the LO and NLO resummed result respectively reproduce all information contained in the LO and NLO $\chi$ and $\gamma$ kernel with the additional constraint of collinear-anticollinear symmetry. Both the ABF LO and NLO results are shown in figure 10.

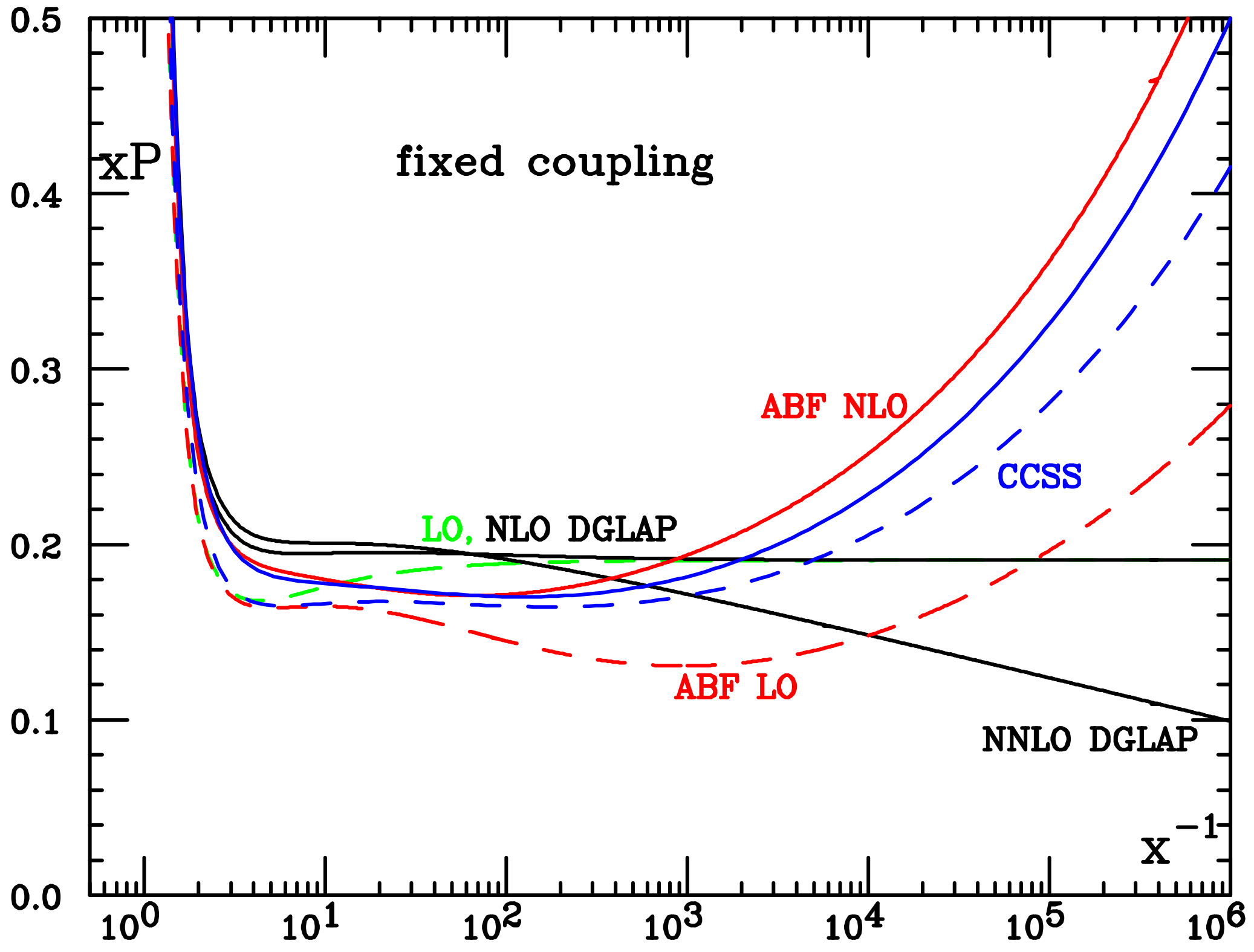

**Fig. 10:** The fixed coupling ($\beta_0 = 0$) $xP_{gg}(x)$ splitting function, evaluated with $\alpha_s = 0.2$ and $n_f = 0$. The dashed curves are LO for DGLAP, NL$x$+LO for CCSS and symmetrized LO DL for ABF, while the solid curves are NLO and NNLO for DGLAP, NL$x$+NLO for CCSS and symmetrized NLO DL for ABF.

In comparison to unresummed results, the resummed splitting functions display the characteristic rise at small $x$ of fixed-coupling leading-order BFKL resummation, though the small $x$ rise is rather milder ($\sim x^{-0.3}$ instead of $\sim x^{-0.5}$ for $\alpha_s = 0.2$). At large $x$ there is good agreement between the resummed results and the corresponding LO (dashed) or NLO (solid) DGLAP curves. At small $x$ the difference between the ABF LO and CCSS NL$x$+LO (dashed) curves is mostly due to the inclusion in CCSS of BFKL NL$x$ terms, as well as to differences in the symmetrization procedure. When comparing CCSS NL$x$+NLO with ABF NLO this difference is reduced, and , being only due the way the symmetrization is implemented, it might be taken as an estimate of the intrinsic ambiguity of the fixed–coupling resummation procedure. At intermediate $x$ the NLO resummed splitting functions is of a similar order of magnitude as the NLO DGLAP result even down to quite small $x$, but with a somewhat different shape, characterized by a shallow dip at $x \sim 10^{-2}$, until the small $x$ rise sets in for $x \sim 10^{-3}$. It has been suggested [66] that in the small $\alpha_s$ limit this dip can be explained as a consequence of the interplay between the $-{\alpha_s}^3 \ln x$ NNLO term of $xP_{gg}$ (also present in the resummation) and the first positive resummation effects which start with an ${\alpha_s}^4 \ln^3 1/x$ term. The unstable small $x$ drop of the NNLO DGLAP result appears to be a consequence of the unresummed $\frac{\alpha_s^3}{N^2}$ double pole in the NNLO anomalous dimension.

[2] Starting from NLO one needs also to specify a factorisation scheme. Small-$x$ results are most straightforwardly obtained in the $Q_0$ scheme, while fixed-order splitting functions are quoted in the $\overline{\text{MS}}$ scheme (for discussions of the relations between different schemes see [25, 50, 64, 65]).

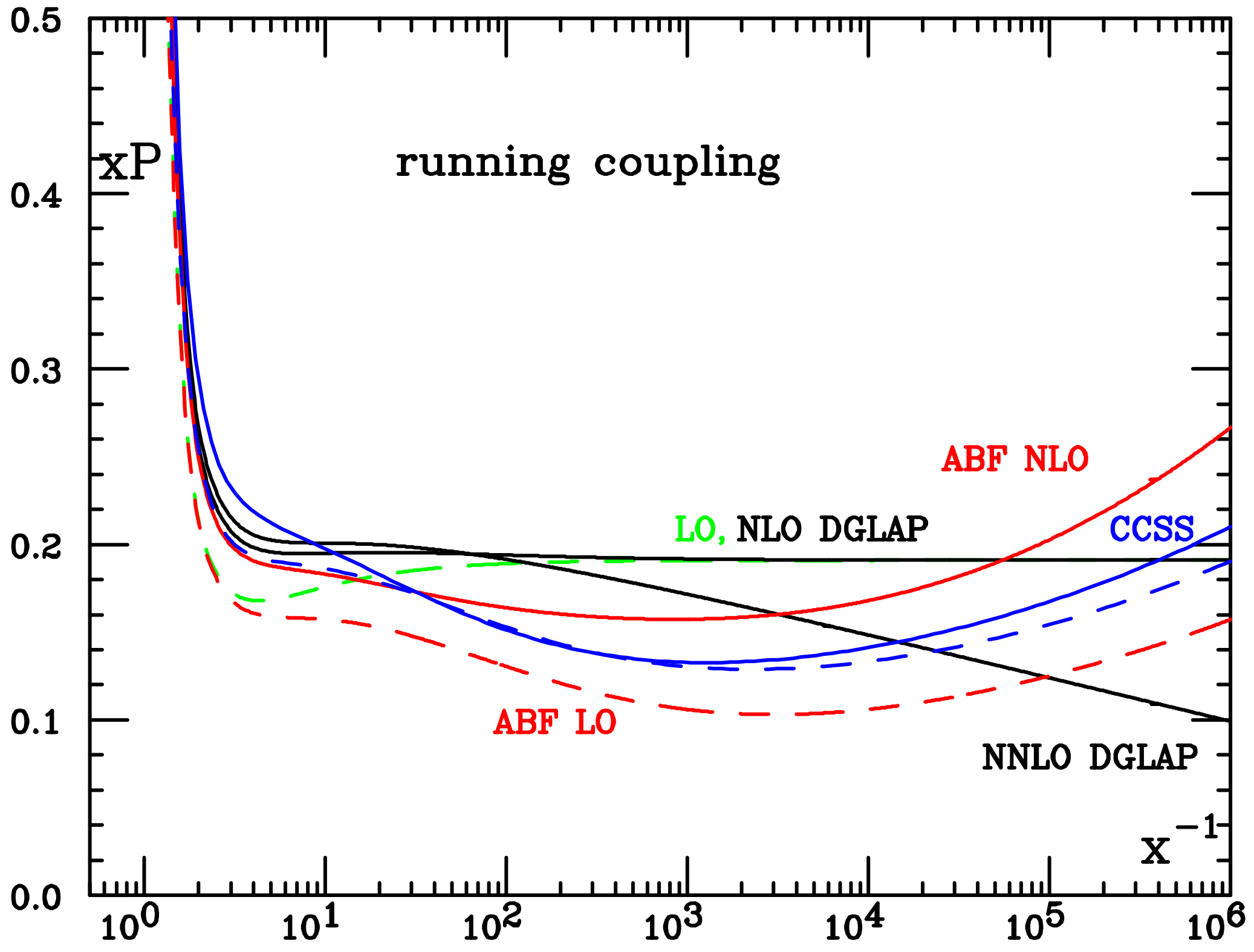

**Fig. 11:** The running coupling $xP_{gg}(x)$ splitting function, evaluated with $\alpha_s = 0.2$ and $n_f = 0$. The various curves correspond to the same cases as in figure 10.

The running-coupling resummed splitting functions are displayed in figure 11. Note that the unresummed curves are the same as in the fixed coupling case since their dependence on $\alpha_s$ is just through a prefactor of $\alpha_s^k$, whereas in the resummed case there is an interplay between the running of the coupling and the structure of the small-$x$ logs. All the resummed curves display a considerable softening of the small $x$ behaviour in comparison to their fixed-coupling counterparts, due to the softening of the leading small $x$ singularity in the running-coupling case [21, 26]. As a consequence, the various resummed results are closer to each other than in the fixed-coupling case, and also closer to the unresummed LO and NLO DGLAP results. The resummed perturbative expansion appears to be stable, subject to moderate theoretical ambiguity, and qualitatively close to NLO DGLAP.

Finally, to appreciate the impact of resummation it is useful to investigate not only the properties of the splitting function, but also its convolution with a physically reasonable gluon distribution. We take the following toy gluon

$$xg(x) = x^{-0.18}(1-x)^5\,, \tag{25}$$

and show in Fig. 12 the result of its convolution with various splitting functions of Fig. 11. The differences between resummed and unresummed results, and between the CCSS and ABF resummations are partly washed out by the convolution, even though the difference between the unresummed LO and NLO DGLAP results is clearly visible. In particular, differences between the fixed-order and resummed convolution start to become significant only for $x \lesssim 10^{-2} - 10^{-3}$, even though resummation effects started to be visible in the splitting functions at somewhat larger $x$.

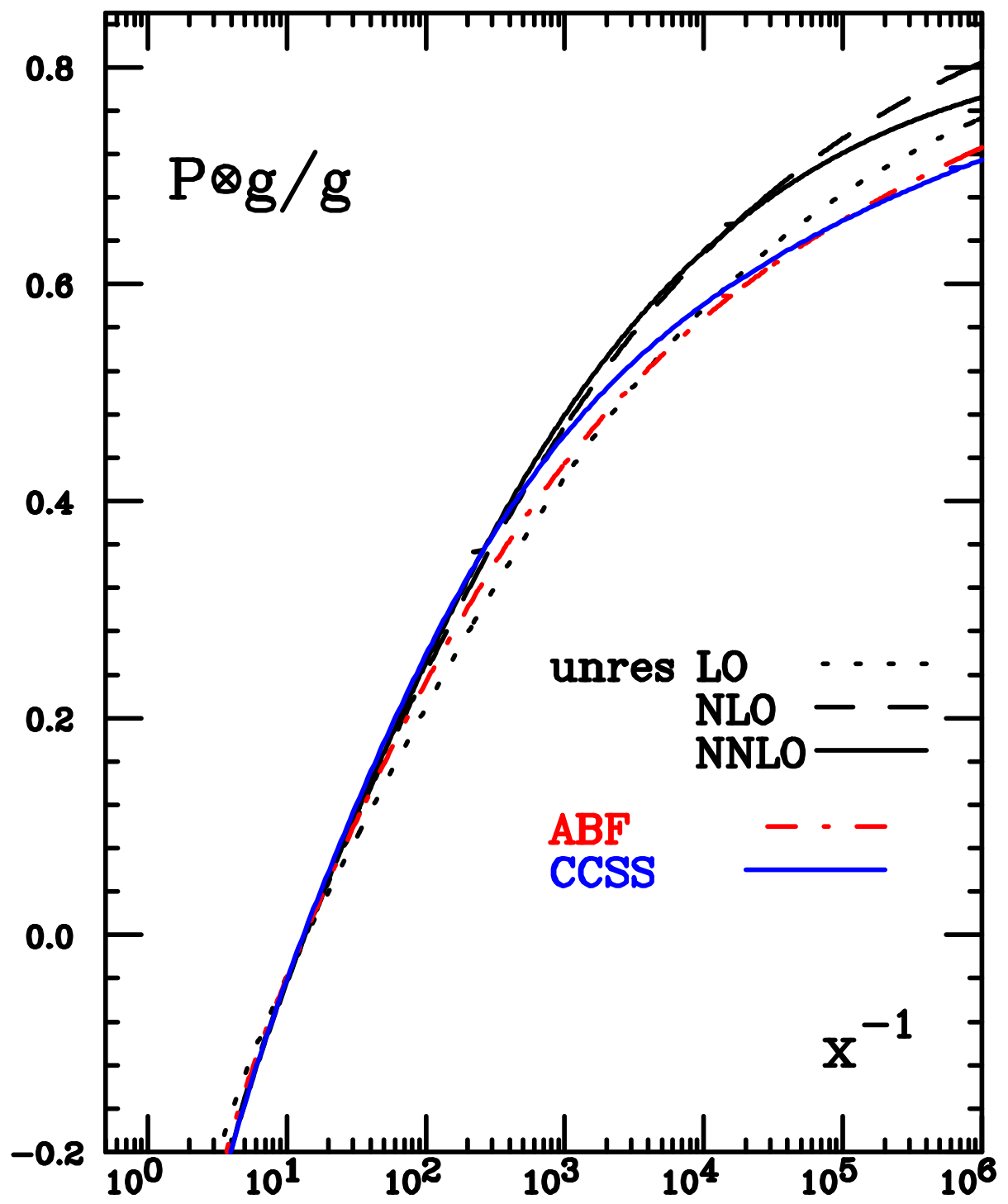

**Fig. 12:** Convolution of resummed and fixed-order $P_{gg}$ splitting functions with a toy gluon distribution, Eq. (25), normalised to the gluon distribution itself, with $\alpha_s = 0.2$ and $n_f = 0$. The resummed CCSS and ABF curves are obtained using respectively the CCSS NL$x$+NLO and the ABF NLO splitting function shown in Fig. 11.

It should be kept in mind that it is only the $gg$ entry of the singlet splitting function matrix that has so far been investigated at this level of detail and that the other entries may yet reserve surprises.

### 3.4 Explicit solution of the BFKL equation by Regge exponentiation

The CCSS approach of section 3.1 exploits a numerical solution of the BFKL equation in which the gluon Green's function is represented on a grid in $x$ and $k$. This method provides an efficient determination of the azimuthally averaged Green's function and splitting functions — for percent accuracy, up to $Y = 30$, it runs in a few seconds — for a wide range of physics choices, e.g. pure NL$x$, various NL$x$+NLO schemes. Here we will discuss an alternative framework suitable to solve numerically the NLL BFKL integral equation [67], based on Monte Carlo generation of events, which can also be applied to the study of different resummation schemes and DIS, but so far has been investigated for simpler NLL BFKL kernels and Regge–like configurations. This method has the advantage that it automatically provides information about azimuthal decorrelations as well as the pattern of final-state emissions.

This appproach relies on the fact that, as shown in Ref. [67], it is possible to trade the simple and double poles in $\epsilon$, present in $D = 4 + 2\epsilon$ dimensional regularisation, by a logarithmic dependence on an effective gluon mass $\lambda$. This $\lambda$ dependence numerically cancels out when the full NLL BFKL evolution is taken into account for a given center–of–mass energy, a consequence of the infrared finiteness of the full kernel. The introduction of this mass scale, differently to the original work of Ref. [49] was performed without angular averaging the NLL kernel.

With such reguralisation of the infrared divergencies it is then convenient to iterate the NLL BFKL equation for the $t$–channel partial wave, generating, in this way, multiple poles in the complex $\omega$–plane.

The positions of these singularities are set at different values of the gluon Regge trajectory depending on the transverse momenta of the Reggeized gluons entering the emission vertices. At this point it is possible to Mellin transform back to energy space and obtain an iterated form for the solution of the NLL BFKL equation:

$$f(\mathbf{k}_a,\mathbf{k}_b,\mathrm{Y}) = e^{\omega_0^\lambda(\mathbf{k}_a)\mathrm{Y}}\,\delta^{(2)}(\mathbf{k}_a-\mathbf{k}_b) + \sum_{n=1}^{\infty}\prod_{i=1}^{n}\int d^2\mathbf{k}_i\int_0^{y_{i-1}} dy_i\left[\frac{\theta\left(\mathbf{k}_i^2-\lambda^2\right)}{\pi\mathbf{k}_i^2}\xi\left(\mathbf{k}_i\right)+\widetilde{\mathcal{K}}_r\left(\mathbf{k}_a+\sum_{l=0}^{i-1}\mathbf{k}_l,\mathbf{k}_a+\sum_{l=1}^{i}\mathbf{k}_l\right)\right] \times e^{\omega_0^\lambda\left(\mathbf{k}_a+\sum_{l=1}^{i-1}\mathbf{k}_l\right)(y_{i-1}-y_i)}\,e^{\omega_0^\lambda\left(\mathbf{k}_a+\sum_{l=1}^{i}\mathbf{k}_l\right)y_n}\,\delta^{(2)}\left(\sum_{l=1}^{n}\mathbf{k}_l+\mathbf{k}_a-\mathbf{k}_b\right), \tag{26}$$

where the strong ordering in longitudinal components of the parton emission is encoded in the nested integrals in rapidity with an upper limit set by the logarithm of the total energy in the process, $y_0=\mathrm{Y}$. The first term in the expansion corresponds to two Reggeized gluons propagating in the $t$–channel with no additional emissions. The exponentials carry the dependence on the Regge gluon trajectory, *i.e.*

$$\omega_0^\lambda\left(\mathbf{q}\right) = -\bar{\alpha}_s\ln\frac{\mathbf{q}^2}{\lambda^2}+\frac{\bar{\alpha}_s^2}{4}\left[\frac{\beta_0}{2N_c}\ln\frac{\mathbf{q}^2}{\lambda^2}\ln\frac{\mathbf{q}^2\lambda^2}{\mu^4}+\left(\frac{\pi^2}{3}-\frac{4}{3}-\frac{5}{3}\frac{\beta_0}{N_c}\right)\ln\frac{\mathbf{q}^2}{\lambda^2}+6\zeta(3)\right], \tag{27}$$

corresponding to no–emission probabilities between two consecutive effective vertices. Meanwhile, the real emission is built out of two parts, the first one:

$$\xi\left(\mathrm{X}\right) \equiv \bar{\alpha}_s+\frac{\bar{\alpha}_s^2}{4}\left(\frac{4}{3}-\frac{\pi^2}{3}+\frac{5}{3}\frac{\beta_0}{N_c}-\frac{\beta_0}{N_c}\ln\frac{\mathrm{X}}{\mu^2}\right), \tag{28}$$

which cancels the singularities present in the trajectory order by order in perturbation theory, and the second one: $\widetilde{\mathcal{K}}_r$, which, although more complicated in structure, does not generate $\epsilon$ singularities when integrated over the full phase space of the emissions, for details see Ref. [67].

The numerical implementation and analysis of the solution as in Eq. (26) was performed in Ref. [68]. As in previous studies the intercept at NLL was proved to be lower than at leading–logarithmic (LL) accuracy. In this approach the kernel is not expanded on a set of functions derived from the LL eigenfunctions, and there are no instabilities in energy associated with a choice of functions breaking the $\gamma\leftrightarrow 1-\gamma$ symmetry, with $\gamma$ being the variable Mellin–conjugate of the transverse momenta. This is explicitly shown at the left hand side of Fig. 13 where the coloured bands correspond to uncertainties from the choice of renormalisation scale. Since the exponential growth at NLL is slower than at LL, there is little overlap between the two predictions, and furthermore these move apart for increasing rapidities. The NLL corrections to the intercept amount to roughly 50% and are stable with increasing rapidities.

In transverse momentum space the NLL corrections are stable when the two transverse scales entering the forward gluon Green's function are of similar magnitude. However, when the ratio between these scales departs largely from unity, the perturbative convergence is poor, driving, as it is well–known, the gluon Green's function into an oscillatory behaviour with regions of negative values along the period of oscillation. This behaviour is demonstrated in the second plot of Fig 13.

The way the perturbative expansion of the BFKL kernel is improved by simultaneous resummation of energy and collinear logs has been discussed in sections 3.1,3.2. In particular, the original approach based on the introduction in the NLL BFKL kernel of an all order resummation of terms compatible with renormalisation group evolution described in ref. [60] (and incorporated in the CCSS approach of section 3.1) can be implemented in the iterative method here explained [69] (the method of ref. [60] was combined with the imposition of a veto in rapidities in refs. [70–72]). The main idea is that the solution to the $\omega$–shift proposed in ref. [60]

$$\omega = \bar{\alpha}_s\left(1+\left(\mathrm{a}+\frac{\pi^2}{6}\right)\bar{\alpha}_s\right)\left(2\psi(1)-\psi\left(\gamma+\frac{\omega}{2}-\mathrm{b}\,\bar{\alpha}_s\right)-\psi\left(1-\gamma+\frac{\omega}{2}-\mathrm{b}\,\bar{\alpha}_s\right)\right)$$

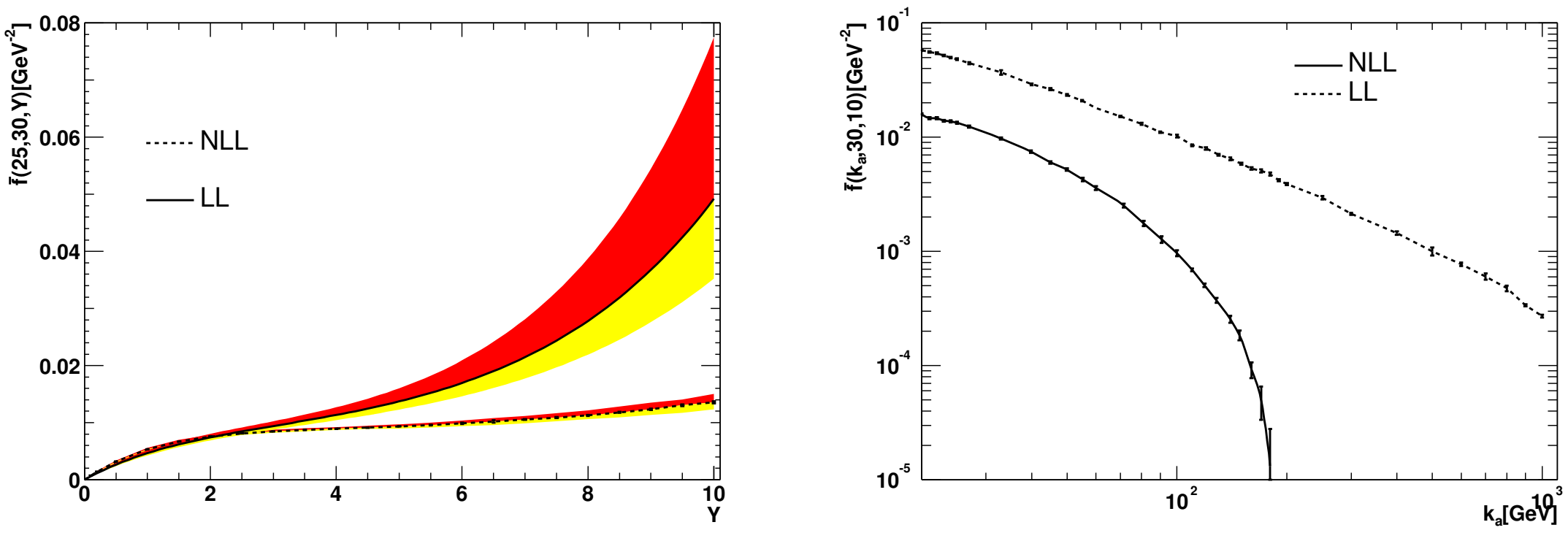

**Fig. 13:** Analysis of the gluon Green's function as obtained from the NLL BFKL equation. The plot to the left shows the evolution in rapidity of the gluon Green's function at LL and NLL for fixed $k_a = 25$ GeV and $k_b = 30$ GeV. The plot on the right hand side shows the dependence on $k_a$ for fixed $k_b = 30$ GeV and $Y = 10$.

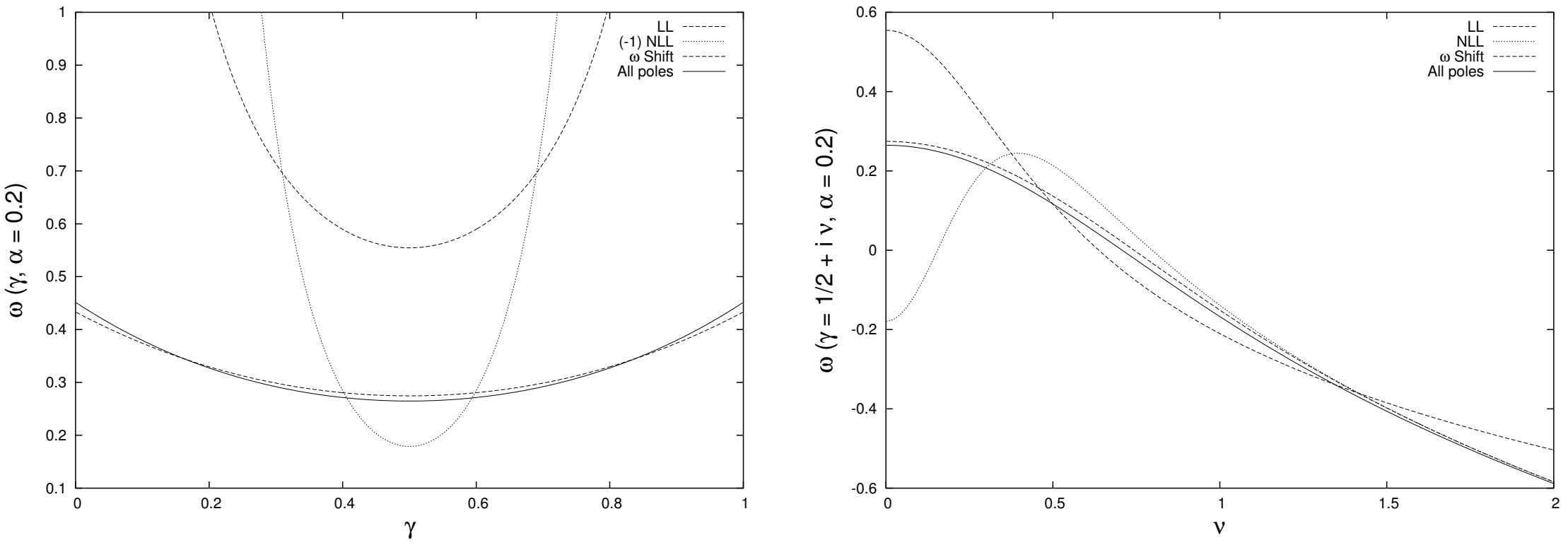

**Fig. 14:** The $\gamma$–representation of the LL and NLL scale invariant kernels together with the collinearly–improved kernel by an $\omega$–shift and the "all–poles" resummation.

$$+ \quad \bar{\alpha}_s^2 \left( \chi_1(\gamma) + \left( \frac{1}{2}\chi_0(\gamma) - \mathrm{b} \right) \left( \psi'(\gamma) + \psi'(1-\gamma) \right) - \left( \mathrm{a} + \frac{\pi^2}{6} \right) \chi_0(\gamma) \right), \tag{29}$$

can be very accurately approximated by the sum of the approximated solutions to the shift at each of the poles in $\gamma$ of the LL eigenvalue of the BFKL kernel. This provides an effective "solution" of Eq. (29) of the form [69]

$$\begin{aligned} \omega \quad &= \quad \bar{\alpha}_s\chi_0(\gamma) + \bar{\alpha}_s^2\chi_1(\gamma) + \left\{ \sum_{m=0}^{\infty} \left[ \left( \sum_{n=0}^{\infty} \frac{(-1)^n (2n)!}{2^n n!(n+1)!} \frac{\left(\bar{\alpha}_s + \mathrm{a}\,\bar{\alpha}_s^2\right)^{n+1}}{(\gamma + m - \mathrm{b}\,\bar{\alpha}_s)^{2n+1}} \right) \right. \right. \\ &- \left. \left. \frac{\bar{\alpha}_s}{\gamma + m} - \bar{\alpha}_s^2 \left( \frac{\mathrm{a}}{\gamma + m} + \frac{\mathrm{b}}{(\gamma+m)^2} - \frac{1}{2(\gamma+m)^3} \right) \right] + \{\gamma \to 1 - \gamma\} \right\}, \end{aligned} \tag{30}$$

where $\chi_0$ and $\chi_1$ are, respectively, the LL and NLL scale invariant components of the kernel in $\gamma$ representation with the collinear limit

$$\chi_1(\gamma) \simeq \frac{\mathrm{a}}{\gamma} + \frac{\mathrm{b}}{\gamma^2} - \frac{1}{2\gamma^3}, \quad \mathrm{a} = \frac{5}{12}\frac{\beta_0}{N_c} - \frac{13}{36}\frac{n_f}{N_c^3} - \frac{55}{36}, \quad \mathrm{b} = -\frac{1}{8}\frac{\beta_0}{N_c} - \frac{n_f}{6N_c^3} - \frac{11}{12}. \tag{31}$$

The numerical solution to Eq. (29) and the value of expression (30) are compared in Fig. 14. The stability of the perturbative expansion is recovered in all regions of transverse momenta with a prediction for the intercept of 0.3 at NLL for $\bar{\alpha}_s = 0.2$, a result valid up to the introduction of scale invariance breaking terms. The implementation of expression (30) in transverse momentum space is simple given that the transverse components decouple from the longitudinal in this form of the collinear resummation [69]. The prescription is to remove the term $-\frac{\bar{\alpha}_s^2}{4}\ln^2\frac{q^2}{k^2}$ from the real emission kernel, $\mathcal{K}_r\left(\vec{q},\vec{k}\right)$, and replace it with

$$\left(\frac{q^2}{k^2}\right)^{-\mathrm{b}\bar{\alpha}_s\frac{|k-q|}{k-q}}\sqrt{\frac{2\left(\bar{\alpha}_s+\mathrm{a}\,\bar{\alpha}_s^2\right)}{\ln^2\frac{q^2}{k^2}}}J_1\left(\sqrt{2\left(\bar{\alpha}_s+\mathrm{a}\,\bar{\alpha}_s^2\right)\ln^2\frac{q^2}{k^2}}\right)-\bar{\alpha}_s-\mathrm{a}\,\bar{\alpha}_s^2+\mathrm{b}\,\bar{\alpha}_s^2\frac{|k-q|}{k-q}\ln\frac{q^2}{k^2}, \quad (32)$$

with $J_1$ the Bessel function of the first kind. This prescription does not affect angular dependences and generates a well–behaved gluon Green's function as can be seen in Fig. 15 where the oscillations in the collinear and anticollinear regions of phase space are consistently removed. At present, work is in progress to study the effect of the running of the coupling in this analysis when the Bessel resummation is introduced in the iterative procedure of Ref. [67].

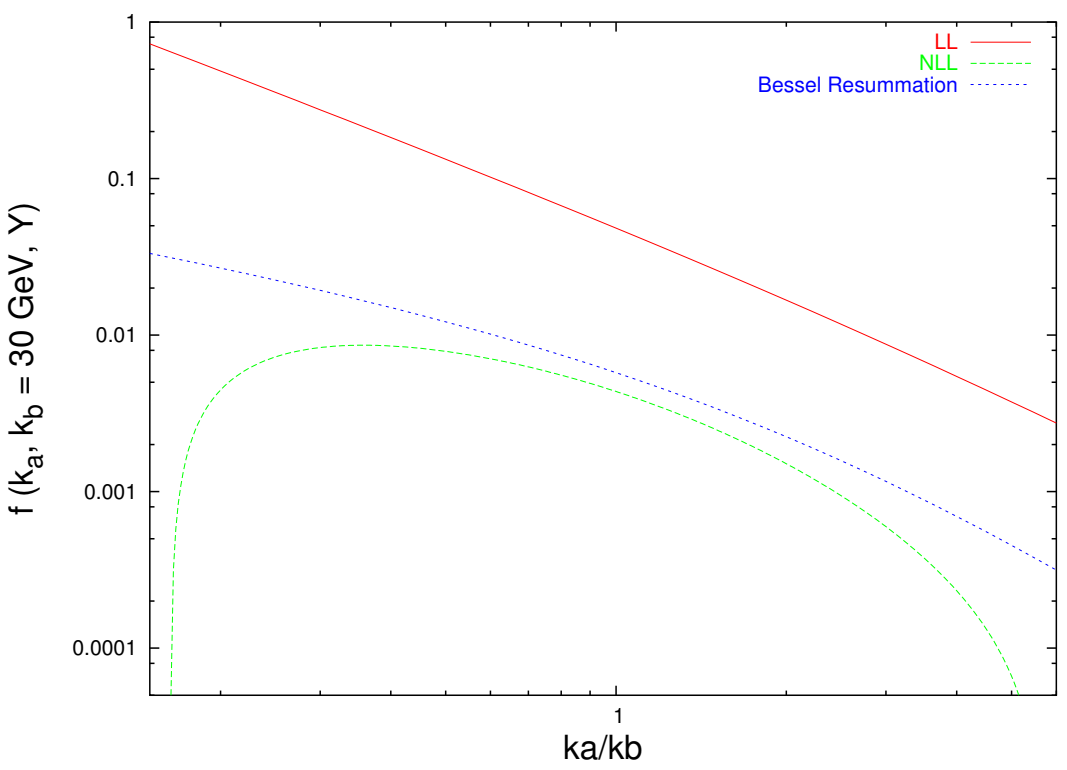

**Fig. 15:** The behaviour of the NLL gluon Green's function using the Bessel resummation.

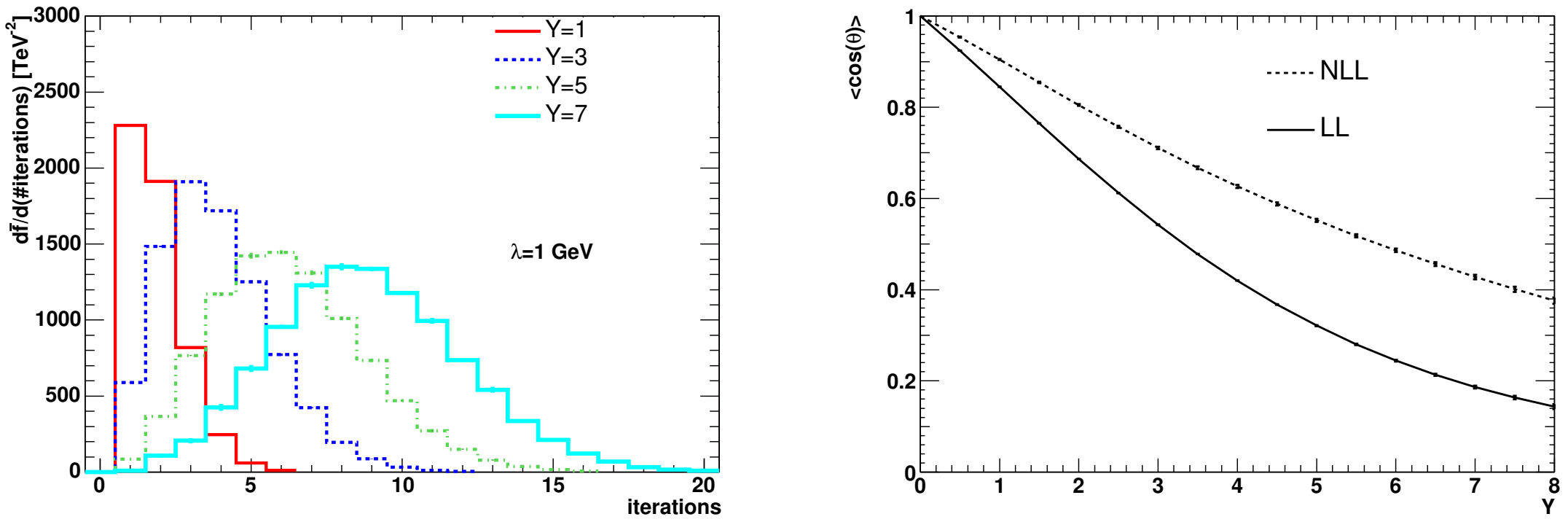

**Fig. 16:** Distribution in the number of iterations and angular dependence of the NLL gluon Green's function.

A great advantage of the iterative method here described is that the solution to the NLL BFKL equation is generated integrating the phase space using a Monte Carlo sampling of the different parton configurations. This allows for an investigation of the diffusion properties of the BFKL kernel as shown in ref. [73], and provides a good handle on the average multiplicities and angular dependences of the evolution. Multiplicities can be extracted from the Poisson–like distribution in the number of iterations of the kernel needed to reach a convergent solution, which is obtained numerically at the left hand side of Fig. 16 for a fixed value of the $\lambda$ parameter. On the right hand side of the figure a study of the azimuthal angular correlation of the gluon Green's function is presented at $Y=5$. This decorrelation will directly impact the prediction for the azimuthal angular decorrelation of two jets with a large rapidity separation, in a fully inclusive jet sample (i.e. no rapidity gaps). The increase of the angular correlation when the NLL terms are included is a characteristic feature of these corrections. This study is possible using this approach because the NLL kernel is treated in full, without angular averaging, so there is no need to use a Fourier expansion in angular variables.

# Part III

# Working Group 2: Multi-Jet Final States and Energy Flows

# List of participants

J. R. Andersen, A. Banfi, J. Bartels, R. Bellan, M. Boonekamp, K. Boreskov, N. Brook, A. Bruni, C. Buttar, J. Butterworth, T. Carli, S. Caron, G. Cerminara, S. Chekanov, J. Collins, G. Corcella, B. Cox, Z. Czyczula, M. Dasgupta, G. Davatz, Y. Delenda, A. De Roeck, M. Diehl, R. Field, J. Forshaw, P. Golonka, E. Gotsman, G. Grindhammer, C. Group, G. Gustafson, S. Höche, S. Jadach, H. Jung, A. Kaidalov, V. Khoze, M. Klasen, H. Kowalski, F. Krauss, N. Lavesson, V. Lendermann, E. Levin, L. Lönnblad, M. Lublinsky, S. Magill, M. Mangano, U. Maor, A. Martin, A. Mastroberardino, S. Maxfield, B. Mellado, A. Moraes, E. Naftali, A. Nikitenko, R. Orava, A. Prygarin, P. Richardson, E. Richter-Was, E. Rodrigues, C. Royon, M. Ryskin, A. Sabio Vera, G. Salam, A. Savin, F.-P. Schilling, T. Schoerner-Sadenius, S. Schumann, M. Seymour, A. Shälicke, T. Sjöstrand, M. Skrzypek, W.J. Stirling, P. Szczypka, M. Tasevsky, T. Teubner, N. Tuning, J. Turnau, R. Venugopalan, B. Ward, Z. Was, B. Waugh, S.A. Yost, G. Zanderighi

# Introduction to multi-jet final states and energy flows

*Craig Buttar*[1], *Jon Butterworth*[2], *Valery Khoze*[3], *Leif Lönnblad*[4] *and Niels Tuning*[5]
[1]Department of Physics and Astronomy, University of Glasgow, UK; [2]Department of Physics and Astronomy, University College London, UK; [3]IPPP, University of Durham, UK; [4] Department of Theoretical Physics, Lund University, Sweden; [5]NIKHEF, Amsterdam, The Netherlands.

**Abstract**
We summarize the activities of Working Group 2 of the HERA/LHC Workshop dealing with multi-jet final states and energy flows. Among the more specific topics considered were underlying event and minimum bias, rapidity gaps and survival probabilities, multi-jet topologies and multi-scale QCD, and parton shower–matrix element matching.

## 1 Introduction

In many ways, the LHC will become the best QCD machine ever built. It will allow us to study the production of hadrons and jets at unprecedented collision energies and will surely increase our understanding of QCD tremendously. Of course, some may argue that QCD already is a well understood and an integral part of the Standard Model, and the reason for building the LHC is to discover new phenomena, hopefully beyond the Standard Model.

However, the fact is that QCD is still not a completely understood theory. The qualitative aspects of asymptotic freedom and confinement may be under control, but the quantitative predictive power of the theory is still not at a satisfactory level. This is particularly true for the non-perturbative region, but also for the high-energy limit, where the hard scale of a process is much smaller than the total collision energy. The latter situation will be dominant in the bulk of events produced at the LHC. The triggers at the main LHC detectors will discard the majority of such events, but what is left will be processes with hard scales of around 100 GeV, which is still more than a hundred times smaller than the collision energy. And there will be significant amounts of minimum-bias data taken as well.

Except for a handful of gold-plated signals for new physics, any such search will be plagued by huge backgrounds stemming from pure QCD or other Standard Model processes involving jets. Hence, even if the study of QCD may seem to be a mundane preoccupation, it is of the utmost importance if we are to find and understand the few needles of new physics hopefully present in the immense LHC haystack.

Although the Tevatron may seem to be the obvious place to learn about QCD processes relevant for the LHC, the triggers there are typically tuned to high-scale processes, not far from the total collision energy. This means that HERA can give important additional insight, since there the situation is in some senses closer to that of the LHC, with the ratio of the typical hard scale and the total energy in DIS being $\sqrt{\langle Q^2\rangle/S} \sim 0.01$. In addition, HERA allows us to study such processes in a more controlled environment, where one side of the collision is well constrained by our relatively precise understanding of electroweak physics.

In our Working Group we have studied in some detail which lessons about multi-jet final states and general hadronic energy flows can be learned from HERA when preparing for the analysis of LHC data. And in this brief summary we will in a few pages try to distill the progress made by almost a hundred physicists as reported in more than fifty talks in this workshop and also in almost twenty separate contributions to these proceedings. The work was broadly divided into four categories: underlying events and minimum bias; rapidity gaps and survival probabilities; multi-jet topologies and multi-scale QCD; and matrix element–parton shower matching.

The first category may not represent the most striking feature of HERA physics, but it will surely be of great importance for the LHC. And it turns out that there are many possibilities to gain further understanding of underlying events in both photoproduction and DIS at HERA.

The study of rapidity gaps and, in particular, hard diffractive scattering gained momenta when it was observed at HERA, and the suggestion to use such processes to obtain clean signals of new physics at the LHC presents exciting prospects where the experience from HERA will be very important.

Multi-scale processes have already been presented as an important connection between HERA and the LHC. This is not least true for the LHCb experiment, where the understanding of the forward region is vital, a region which has been intensely studied at HERA. Also the recent theoretical development in QCD resummation techniques, which so far have mainly been applied to $e^+e^-$ annihilation, may provide important tools for understanding event shapes at the LHC, and the corresponding application to HERA data will be essential for this understanding.

Finally, the more technical issue of matching fixed-order tree-level matrix elements with parton shower generators as well as other theoretical improvements of such simulation programs will surely be vital for the successful understanding of data from the LHC and also here the comparison to HERA data will be essential for the tuning and validation.

It should be noted that all of these categories, presented in more detail below, have a fairly large overlap with other working groups in this workshop. The most obvious overlaps are the working groups for Diffraction and Monte Carlo simulations, but there is also overlap with the heavy flavour and parton distributions working groups.

## 2 Underlying events and minimum bias

An understanding of the underlying event is an interesting physics topic in its own right but is also crucial in developing robust analyses for LHC physics. The underlying event can enhance central jet production, reducing the effectiveness of the central jet veto in analyses such as the vector boson fusion Higgs channel, or reduce the isolation of leptons resulting in reduced efficiency for identifying isolated leptons. In particular for LHCb and ALICE, where the triggers typically do not mandate high-scale processes, a good understanding of underlying events and minimum-bias events is crucial.

In this workshop there were several contributions dealing with underlying events and multiple interactions. They are all described in a joint contribution to these proceedings [1]. There the event generator models in PYTHIA [2–5], HERWIG/JIMMY [6–8] and SHERPA [9] are presented together with results from tuning these and other models to available data. The contribution also includes a summary of the plenary talk by Gösta Gustafson on the theory and phenomenology underlying events and multiple scattering.

Of the models presented and studied in Ref. [1], the one implemented in PYTHIA is probably the most advanced. This model has recently been developed further, introducing a scheme for *interleaving* the multiple interaction with a transverse-momentum ordered parton shower [3]. In contrast, the default underlying event model in HERWIG is a simple parametrization of UA5 data [10]. However, HERWIG is easily interfaced to the multiple-interaction model in the JIMMY program, which is similar to the PYTHIA model in spirit, although many of the details differ. The JIMMY program has recently been improved, making the generation of events more efficient where the signal process is different from the additional multiple scattering processes. Also the SHERPA event generator is now equipped with multiple interactions. Again, this model is similar in spirit to that in PYTHIA. One interesting aspect which differs is the attempt to incorporate the multiple scatterings in the general CKKW (see Section 5 below) framework of SHERPA.

The CDF Collaboration has carried out studies of the underlying event in jet processes [11–13] and this was used to provide a tuning for PYTHIA. In Ref. [1] a new analysis is presented which has extended these studies by increasing the energy range of the leading jet from around 50 GeV to 450 GeV using $E_T$ from the calorimeter as well as particle $p_\perp$ measured in the tracker, and defining two-jet topologies as a subset of the leading jet to investigate the beam–beam and radiation components of the underlying event. Both PYTHIA tune-A and HERWIG/JIMMY were found to be in good agreement with the data, although both underestimate the transverse energy. The extension to higher energy scale shows that the underlying event activity increases with leading jet $p_\perp$ i.e., the hardness of the primary scatter, but by studying the maximum and minimum activity it is seen that this rise is largely due to bremsstrahlung from the primary scattering rather than secondary interactions between the beam remnants.

The CDF analysis was carried out primarily at 1.8 TeV although some of the early 546 GeV data has also been analysed. This has meant that there is only limited information on the energy dependence of the underlying event. To cover a wider range of energy, ATLAS have used minimum-bias data from the SppS and Tevatron covering 200 GeV to 1.8 TeV in addition to the CDF underlying event data to tune PYTHIA and HERWIG/JIMMY. Comparing the predictions of minimum-bias and underlying event distributions at the LHC using the tuned PYTHIA, the tuned HERWIG/JIMMY and PHOJET [14] shows large variations, emphasizing the need to understand the energy dependence of these processes better. The energy dependence was investigated further by LHCb, again using minimum-bias data to fit the parameters required for the model of energy dependence in PYTHIA.

Both the ATLAS and LHCb analyses have the implicit assumption that minimum bias and the underlying event have the same physics origin. While CDF data supports this, it would be helpful to probe the underlying event directly over a larger range of energy scales. HERA is in a prime position to make such a contribution by studying jets from photoproduction in an energy range corresponding to centre-of-mass energies in the region of 200 GeV, fitting well with the low-energy minimum-bias data. In photoproduction, resolved photons behave like hadrons so that HERA is effectively a hadron–hadron collider. Photoproduction data shows that particle flow and multi-jet measurements require models with multiple interactions to best describe the data but detailed studies of multiple interactions have not been made. However, studies of particle and energy flow in the transverse region similar to that carried out by CDF could be made at HERA.

An interesting question is whether there is also an underlying event present in DIS at HERA. As explained in Refs. [15, 16] it is possible to relate diffraction and saturation to multiple-interaction processes also for DIS using a QCD reformulation of the so-called AGK cutting rules [17]. And since diffractive processes have been clearly seen at high $Q^2$ at HERA, it is reasonable to expect that multiple interactions may also be present. A good place to search for such effects is in forward-jet production at HERA. In [18] preliminary results are presented indicating that multiple-interaction effects may indeed give a noticeable increase in the measured forward-jet cross-section in resolved virtual photon processes at small $x$ and moderate $Q^2$.

The connection between multiple interactions, saturation and diffraction was also discussed in the plenary talk by Gösta Gustafson. He pointed out a possible problem with the qualitative AGK predictions for the hadronic multiplicity in multiple-interaction events. Taking the tuning of PYTHIA to CDF data at face value, there is an indication that the colour flows of secondary interactions are not independent from the primary scattering. Rather, the different colour flows seem to combine in a way where the total string length is minimized, resulting in a multiplicity which does not grow proportionally to the number of scatterings. Currently there is no theoretical understanding of this phenomenon. Gustafson also pointed out the problem that all multiple-interaction models discussed here rely on collinear factorization of the individual scatterings in a region where we expect $k_\perp$ factorization to be the relevant formalism. In fact, using $k_\perp$ factorization, the soft divergencies in the partonic cross section present in the conventional models may be removed, which could make the extrapolation of the model predictions to high energy more constrained.

## 3 Rapidity gaps and survival probabilities

A characteristic signature of diffractive processes is the existence of a large rapidity gap (LRG) in the final state, defined as a region of (pseudo-) rapidity devoid of hadronic activity. A rapidity gap may be adjacent to a leading proton or may arise between the decay products of final hadronic systems. The appearance of the rapidity gaps is intimately related to the exchange in the $t$-channel of objects with vacuum quantum numbers (Pomeron in the Regge theory, di-gluon Pomeron in pQCD, photon or $W$ -mediator). The diffractive rapidity gap events have been studied in great detail at the ISR, SPS, HERA and the Tevatron. The LHC is the first collider which will have enough energy to allow the events with several (n = 2–4) LRGs.

The activity of our Working Group was focused mainly on the LRGs in the hard diffractive processes. For specifics of the photon and $W$-mediated reactions see, for example, Refs. [19–22].

An intensive discussion concerned the breakdown of factorization in hard hadronic diffractive processes. It is the consequence of unitarization effects, that both hard and Regge factorization are broken. This breakdown of factorization is experimentally seen [23] as the suppression of the single diffractive dijet cross section at the Tevatron as compared to the prediction based on HERA results. The observed suppression is in a quantitative agreement with the calculations [24] where the unitarization effects are described by multi-Pomeron exchange diagrams. The analysis of the current CDF diffractive dijet data with one or two rapidity gaps shows a good agreement with this approach. The situation with the factorization breaking in dijet photoproduction is not completely clear and further experimental and theoretical efforts are needed. A possible way to study this effect is to measure the ratio of diffractive and inclusive dijet photoproduction, see Ref. [25].

It is important to emphasize that the rapidity gap signal is very powerful but, at the same time, quite a fragile tool. We have to pay a price for ensuring such a clean environment. The gaps may easily fade away (filled by hadronic secondaries) on account of various sources of QCD 'radiation damage':

(i) soft or hard rescattering between the interacting hadrons (classic screening/unitarization effects or underlying event);
(ii) bremsstrahlung induced by the 'active' partons in the hard subprocesses;
(iii) radiation originating from the small transverse distances in two-gluon Pomeron dipoles.

An essential issue in the calculation of the rate of events with LRG concerns the size of the factor $W$ which determines the probability for the gaps to survive in the (hostile) QCD environment. As discussed in the contributions of Brian Cox [26] and Jeff Forshaw [27], this factor is a crucial ingredient for evaluation of the discovery potential of the LHC in the exclusive processes with double proton tagging.

Symbolically, the survival probability $W$ can be written as

$$W \ = \ S^2 T^2. \tag{1}$$

$S^2$ is the probability that the gaps are not filled by secondary particles generated by soft rescattering, i.e., that no other interactions occur except the hard production process. Following Bjorken [28,29], who first introduced such a factor in the context of rescattering, such a factor is often called the survival probability of LRG. The second factor, $T^2$, is the price to pay for not having gluon radiation in the hard production subprocess. It is related to Sudakov-suppression phenomena and is incorporated in the pQCD calculation via the skewed unintegrated parton densities. The physics of Sudakov suppression is discussed in more detail in the contribution of Jeff Forshaw to these Proceedings [27].

In some sense the soft survival factor $S^2$ is the 'Achilles heel' of the calculations of the rates of diffractive processes, since, in principle, $S^2$ could strongly depend on the phenomenological models for soft diffraction. This factor is not universal, but depends on the particular hard subprocess, as well as on the distribution of partons inside the proton in impact parameter space. It has a specific dependence on the characteristic momentum fractions carried by the active partons in the colliding hadrons [24].

However, the good news is that, as discussed in these Proceedings by Uri Maor et al. [30], the existing estimates of $S^2$ calculated by different groups for the same processes appear to be in a reasonably good agreement with each other. This is related to the fact that these approaches reproduce the existing data on high-energy soft interactions, and, thus, result in the similar profile of the optical density in the impact parameter space. Another reason results from the comparatively small role of the high-mass diffractive dissociation.

Note that it is possible to check the value of $S^2$ by observing double-diffractive dijet production [31]. The gap survival in the Higgs production via the $WW$-fusion process can be probed in $Z$ production which is driven by the same dynamics, and has a higher cross-section, see Refs. [32,33]. Let us emphasize that it is the presence of this factor which makes the calculation infrared stable, and pQCD applicable. Neglecting the Sudakov suppression would lead to a considerable overshooting of the cross section of the hard central exclusive processes at large momentum transfer.

## 4 Multi-jet topologies and multi-scale QCD

In this workshop work on a wide range of topics regarding jet production and multi-scale processes has been presented [34]. It is of great interest to know what the LHC will teach us in the area of QCD, but at the same time uncertainties on the theoretical predictions for processes at the LHC should be limited as far as possible beforehand. By using the knowledge attained at HERA, our models can be sharpened and our theories can be tested.

Predictions of the event topology of $gg \to H$ at the LHC have been investigated for various parton shower models — such as PYTHIA, HERWIG and ARIADNE, that have proven their validity at HERA — and uncertainties in the event selection have been estimated [35,36]. In the parton cascade as implemented in some of these programs, the parton emissions are calculated using the DGLAP approach, with the partons ordered in virtuality. DGLAP accurately describes high-energy collisions of particles at moderate values of the Bjorken-$x$ by resummation of the leading log terms of transverse momenta ($\alpha_s \ln Q^2$). However, to fixed order, the QCD scale used in the ladder is not uniquely defined. There are many examples were more than one hard scale plays a role in the hard scatter, such as the virtuality $Q$, the transverse momentum $E_T$ of the jet, or the mass of a produced object. Also, at low values of Bjorken-$x$ large logarithms appear ($\alpha_s \ln 1/x$), leading to large corrections.

The CCFM formalism takes this into account, describing the evolution in an angular ordered region of phase space, while reproducing DGLAP and BFKL in the appropriate asymptotic limits. The CASCADE program has implemented the CCFM formalism, describing the low-$x$ $F_2$ data and forward jet data at HERA. The predictions for the jet production at the LHC have been studied, both in the context of a $gg \to H$, as well as in the context of the forward event topology at LHCb [37].

In order to get reliable predictions for exclusive final-state processes, unintegrated parton density functions $f(x, Q^2, k_\perp)$ (uPDFs) become indispensable. For example, in the small-$x$ regime, when the transverse momenta of the partons are of the same order as their longitudinal momenta, the collinear approximation is no longer appropriate and $k_\perp$ factorization has to be applied, with the appropriate CCFM evolution equations. In this workshop various parametrizations for unintegrated gluon densities matched to HERA $F_2$ data were compared to each other [38]. It is, however, still questionable if these densities are constrained enough for reliable predictions for Higgs production cross-section. Final-state measurements like photoproduction of $D^*$+jet events could however constrain these uPDFs further. It is argued that it is important to reformulate perturbative QCD in terms of fully unintegrated parton densities, since neglecting parton transverse momentum leads to wrong results. The HERA $F_2$ data has also been fitted using non-linear BFKL evolution, expressed with a universal dipole cross section, which in turn can be related to the unintegrated gluon distribution.

Finally, a theoretical description of hard diffractive processes at HERA can provide information on the so-called generalized, or skewed, gluon distribution (depending on the $x$ of the emitted and absorbed gluon), providing for a theoretical description for diffractive Higgs production at the LHC.

The role of HERA is also emphasized in the area of resummed calculations, obtaining accurate QCD parameters such as the strong coupling, quark masses and parton distribution functions, which are vital inputs for predictions at the LHC. For example, event-shape distributions at HERA led to the finding of non-global logarithms, influencing observables at the LHC such as energy flows away from jets. Additionally, HERA data seem to confirm $1/Q$ power corrections (arising from gluon emission with transverse momentum $\sim \Lambda_{QCD}$), demonstrating that these corrections are not affected by the presence of the initial-state proton. HERA data is also used to study dijet $E_T$ and angular spectra, in order to test NLL perturbative predictions. Finally, we have discussed whether additional small-$x$ terms are needed to accommodate HERA DIS data, which at LHC energies would result in a broadening of the vector boson $p_T$ spectrum.

## 5 Parton shower/matrix element matching

The LHC is, of course, mainly a machine for discovering new physics. But irrespective of what new phenomena may exist, we know for sure that LHC events will contain huge numbers of hadrons, and that a large fraction of these events will have many hard jets produced by standard QCD processes. Such events are interesting in their own right, but they are also important backgrounds for almost any signal of new physics. Unfortunately the standard Parton Shower (PS)-based event generators of today are not well suited to describe events with more than a couple of hard jets. The alternative is to use matrix element (ME) generator programs; this typically can generate up to six hard partons according to the exact fixed-order tree-level matrix elements. But these generators are not well suited for describing the conversion of these hard partons into jets of hadrons.

To get properly generated events it is therefore important to interface the ME generators to realistic hadronization models; this requires that also soft and collinear partons are generated according to PS models to get reliable predictions for the intra- and inter-jet structure. When adding a PS to an event from a ME generator, it is important to avoid double-counting. Hence the PS must be *vetoed* to avoid generating parton emissions above the cutoff needed to avoid divergences in the ME generator. In addition the PS assumes that the emissions are ordered in some evolution variable (scale) and uses Sudakov form factors to ensure that there was no additional emission with a scale between two generated emissions. This also generates the virtual corrections to the splittings. The ME generators, of course, have no such ordering since all diagrams are added coherently. However, there is still a need for a cutoff in some scale to regulate soft and collinear divergencies, and to naively add a PS to events from a ME generator will therefore give a strong dependence on this cutoff.

A solution to this problem was presented by Catani et al. [39]. This so-called CKKW procedure is based on using a jet reconstruction algorithm on the ME-generated event to define an ordering of the emissions and then reweight the event according to Sudakov form factors obtained from the reconstructed scales. In this way it was shown that the dependence on the ME cutoff cancels to NLL accuracy. The procedure was originally developed for $e^+e^-$ annihilation where it was further developed in Ref. [40], but lately it has also been applied to hadron–hadron collisions [41–45] using several different parton shower models. In addition, an alternative procedure, called MLM, was developed by Mangano [46, 47] which is similar in spirit to CKKW, but which has a simpler interface between the ME and PS program.

There was some hope that during this workshop an implementation of CKKW for DIS would also be developed. This would be interesting, not least because the procedure would then be tested in a small-$x$ environment, and comparing with such HERA data as well as with high-scale Tevatron data should then give a more reliable understanding about the uncertainties when extrapolating to the LHC. Although some progress has been made on the application to DIS [48] there was not enough time to make a proper implementation. Instead the activities were focused on comparing the predictions of some of the programs (SHERPA [9] and MADGRAPH/MADEVENT [49]+ARIADNE [50] using CKKW, and ALPGEN [51]+PYTHIA [4] using MLM) for the case of W+jets production at the Tevatron and the LHC. This process is very interesting in its own right, but is also an important background for almost any signal of new physics at the LHC. The results are presented in these proceedings [52] and it was found that the models give fairly similar predictions for jet rates, but some differences were found, for example, for the rapidity correlation between jets and the W. The latter may be related to the fact that W production, especially at the LHC, can be considered to be a small-$x$ process ($m_W/\sqrt{S} \sim x \sim 0.005$) and we know that there are large differences between parton shower models in this region. This emphasizes again the importance of confronting the ME+PS matching procedures with HERA DIS data also.

Possible improvements to the QCD PS approach were discussed in three other contributions to these proceedings. All of these are based on experience of Monte Carlo programs for QED resummation. One of these contributions [53] describes a new algorithm for forward evolution of the initial-state parton cascade in which the type and energy of the final parton is predefined/constrained. Contrary to the widely used backward-evolution algorithms [54], this algorithm is similar to the one used in the LDCMC generator [55] and does not need a fully evolved PDF parametrization as input.

Using an operator formalism, another contribution [56] describes what we can learn about QCD parton showers from the popular PHOTOS generator, which combines in a clever way soft photon resummation and hard collinear photon resummation in QED. Finally there is a contribution [57] which describes a more ambitious attempt to combine ME+PS calculations for both QCD and QED, preserving the proper soft gluon limit and the standard factorization of collinear singularities. All of these contributions represents work which is still in a rather early stage. Nevertheless, they signal important efforts which may lead to interesting new Monte Carlo tools for the LHC era.

## 6 Conclusions and outlook

In this summary we hope to have made it clear that there is a rich flora of interesting topics relating to jets and hadronic energy flows where the understanding of results from HERA will be important for the upcoming analysis of LHC data. It should also be clear that although substantial progress has been made during this workshop, we have only started to botanize among these topics. Hence, as we now thank the participants of our Working Group for all the work they have contributed to the workshop, we would also like to remind them, and also other readers of these proceedings, that there is much work still to be done.

# The Underlying Event

*C.M.Buttar*[1], *J.M.Butterworth*[2], *R.D.Field*[3], *C.Group*[3], *G.Gustafson*[4], *S.Hoeche*[5], *F.Krauss*[5], *A.Moraes*[1], *M.H.Seymour*[6,7], *A.Schalicke*[8], *P.Szczypka*[9], *T.Sjöstrand*[4,7]
[1]Dept. of Physics and Astronomy, University of Glasgow, UK
[2]Dept. of Physics and Astronomy, University College London, UK
[3]Dept. of Physics, University of Florida, USA
[4]Dept. of Theoretical Physics, Lund University, Sweden
[5]Institute for Theoretical Physics, Dresden University of Technology, FRG
[6]School of Physics and Astronomy, University of Manchester, UK
[7]CERN, Switzerland
[8]DESY Zeuthen, FRG
[9]Dept. of Physics, University of Bristol, UK

**Abstract**
The contributions to working group II: "Multi-jet final states and energy flows" on the underlying event are summarized. The study of the underlying event in hadronic collisions is presented and Monte Carlo tunings based on this are described. New theoretical and Monte Carlo methods for describing the underlying event are also discussed.

## 1 Introduction

The underlying event is an important element of the hadronic environment within which all physics at the LHC, from Higgs searches to physics beyond the standard model, will take place. Many aspects of the underlying event will be constrained by LHC data when they arrive. However, the physics is so complex, spanning non-perturbative and perturbative QCD and including sensitivities to multi-scale and very low-x physics, that even after LHC switch-on many uncertainties will remain. For this reason, and also for planning purposes, it is critical to have to hand sensible models containing our best physical knowledge and intuition, tuned to all relevant available data.

In this summary of several contributions to the workshop, we first outline the available models in Section 2, most of which are in use at HERA and/or the Tevatron. Recent improvements, some of which were made during the workshop, are also discussed.

Next, current work on tuning these to data is discussed. The underlying event has been extensively studied by CDF and the latest results are presented in Section 3 and compared to predictions from the PYTHIA and HERWIG+JIMMY Monte Carlo generators. The CDF tunings are compared to other tunings based on CDF data and minimum bias data and used to predict the level of underlying events at the LHC in Sections 4 and 5. These reports are very much a snapshot of ongoing work, which will be continued in the follow-up meetings of this workshop and the TeV4LHC workshop.

One major issue in extrapolating the underlying event (UE) to LHC energies is the possible energy dependence of the transverse momentum cut-off between hard and soft scatters, $\hat{p}_T^{\min}$. The need for such a cut-off may be avoided by using the $k_\perp$ factorization scheme as discussed in Section 6, where soft emissions do not contribute to the total cross-section or to the parton density functions (PDFs), but do contribute to the properties of the event. The cross-section for a chain of partonic emission can be extracted from HERA data and can be used to predict the minijet rate or multiple interaction rate in pp or $\mathrm{p\bar{p}}$ collisions. The running of $\alpha_s$ still introduces a cut-off scale between soft and hard chains; however it has been shown that the total cross-section is insensitive to this cut-off and predictions for the mini-jet rate at the LHC are stable. The hadron multiplicity observed in the CDF underlying event data indicates that the string connections in the underlying event are made to minimise the string length. This is the opposite to what is observed in $e^+e^-$ collisions. The implications for this on the AGK cutting rules is discussed further in Section 6.

This summary ends with a section on conclusions and suggestions for future work.

## 2 Underlying event models

Several underlying event models are available, at varying stages of development and use. In this section we review the status of thosed discussed during the workshop.

### 2.1 Multiple Interactions in PYTHIA

The basic implementation of multiple interactions in PYTHIA is almost 20 years old, and many of the key aspects have been confirmed by comparisons with data. In recent years the model has been gradually improved, with junction-string topologies, with flavour-correlated multiparton densities, and with transverse-momentum-ordered showers interleaved with the multiple interactions. However, the "correct" description of colour flow still remains to be found.

The traditional PYTHIA [1,2] model for multiple interactions (MI) [3] is based on a few principles:

1. The naive perturbative QCD $2 \to 2$ cross section is divergent like $\mathrm{d}p_\perp^2/p_\perp^4$ for transverse momenta $p_\perp \to 0$. Colour screening, from the fact that the incoming coloured partons are confined in colour singlet states, should introduce a dampening of this divergence, e.g. by a factor $p_\perp^4/(p_{\perp 0}^2 + p_\perp^2)^2$, where $p_{\perp 0}$ is a free parameter, which comes out to be of the order of 2 GeV.
2. From the thus regularized integrated interaction rate $\sigma_{\mathrm{int}}(E_{\mathrm{cm}}, p_{\perp 0})$ and the nondiffractive cross section $\sigma_{\mathrm{nd}}(E_{\mathrm{cm}})$, the average number of interactions per event can be derived as $\langle n_{\mathrm{int}} \rangle = \sigma_{\mathrm{int}}/\sigma_{\mathrm{nd}}$. With no impact-parameter dependence, the actual number of interactions is given by a Poissonian with mean as above (modulo some corrections coming from $n_{\mathrm{int}} = 0$).
3. More realistically, since hadrons are extended objects, there should be more (average) activity in central collisions than in peripheral ones. By introducing a matter distribution inside a hadron, the overlap between the two incoming hadrons can be calculated as a function of impact parameter $b$. The number of interactions is now a Poissonian for each $b$ separately, with a mean proportional to the overlap. All events are required to contain at least one interaction; thereby the cross section is automatically dampened for large $b$. Empirically, the required hadronic impact parameter profile is more peaked at small $b$ than in a Gaussian distribution.
4. It is natural to consider the interactions in an event in order of decreasing $p_\perp$ values. Such a $p_\perp$ ordering has a natural interpretation in terms of formation-time arguments. The generation procedure can conveniently be written in a language similar to that used for parton showers, with the equivalent of a Sudakov form factor being used to pick the next smaller $p_\perp$, given the previous ones. It allows the hardest interaction to be described in terms of conventional PDFs, whereas subsequent ones have to be based on modified PDFs, at the very least reduced by energy–momentum conservation effects. This also reduces the tail of events with very many interactions.
5. Technical limitations lead to several simplifications, such that only the hardest interaction was allowed to develop initial- and final state interactions, and have flavours selected completely freely.
6. Colour correlations between different scatterings cannot be predicted by perturbation theory, but have a direct consequence on the structure of events. One of the most senstive quantities is $\langle p_\perp \rangle (n_{\mathrm{charged}})$. Data here suggest a very strong colour correlation, where the total string length is essentially minimized in the final state.

For a long period of time, only one significant change was made to this scenario:

7. Originally the $p_{\perp 0}$ parameter had been assumed energy-independent. In the wake of the HERA data [4], which led to newer PDF parametrizations having a steeper small-$x$ behaviour than previously assumed, it became necessary to let $p_{\perp 0}$ increase with energy to avoid too steep a rise of the multiplicity. Such an energy dependence can be motivated by colour screening effects [5]. A functional form $p_{\perp 0} \propto s^{\epsilon}$ with $\epsilon \sim 0.08$ is suggested by Pomeron arguments.

Several studies have been presented based on this framework. Some of the recent tuning activities are described elsewhere in this report. The PYTHIA Tune A [6] is a standard reference for much of the current Tevatron underlying-event and minimum-bias physics studies.

In recent years, an effort has been made to go beyond the framework outlined above. Several new or improved components have been introduced.

1. The fragmentation of junction-string topologies has been implemented [7] . Such topologies must be considered when at least two valence quarks are kicked out of an incoming proton beam particle. Here a proton is modelled as a Y-shaped topology, where each valence quarks sits at the end of one of the three legs going out from the middle, the junction. When some ends of this Y are kicked out, also the junction is set in motion. The junction carries no energy or momentum of its own, but it is around the junction that the baryon inheriting the original baryon number will be formed. The junction rest frame is defined by having $120^\circ$ between the three jets. A number of technical problems have to be overcome in realistic situations, where also gluons may be colour-connected on the three legs, thus giving more complicated space–time evolution patterns.
2. PDFs are more carefully modelled, to take into account the flavour structure of previous interactions [8], not only the overall energy–momentum constraints. Whenever a valence quark is kicked out, the remaining valence PDF of this flavour is rescaled to the new remaining number. When a sea quark is kicked out, an extra "companion" antiquark distribution contribution is inserted, thereby increasing the likelihood that also the antiquark is kicked out.
3. Also remnant flavours are more carefully considered, along with issues such as primordial $k_\perp$ values and remnant longitudinal momentum sharing.
4. A few further impact-parameter possibilities are introduced.
5. New transverse-momentum-ordered showers are introduced, both for initial- and final-state radiation (ISR and FSR) [9]. On the one hand, this appears to give an improved description of (hard) multijet production. On the other hand, it allows all evolution to be viewed in terms of a common "time" ordering given by decreasing $p_\perp$ values. This is especially critical for the description of MI and ISR, which are in direct competition, in the sense that both mechanisms take momentum out of the incoming beams and thereby require a rescaling of PDF's at later "times". This approach, with interleaved MI and ISR, is illustrated in Fig. 1.

Currently we still make use of two simplifications to the new $p_\perp$-ordered framework: (a) the inclusion of FSR is deferred until the MI and ISR have been considered in full, and (b) there is no intertwining, in which two seemingly separate higher-virtuality parton chains turns out to have a common origin when studied at lower $p_\perp$ scales. Fortunately there are good reasons why neither of those omitted aspects should be so important.

There is one big remaining unsolved issue in this model, however, namely that of colour flow. If colours are only connected via the fact that the incoming beam remnants are singlets, the correct $\langle p_\perp \rangle (n_{\text{charged}})$ behaviour cannot be reproduced whatever variation is tried. It appears necessary to assume that some final-state colour reconnection mechanism tends to reduce the total string length almost to the minimal possible, as was required for Tune A. The most physically reasonable approach, that is yet not too time-consuming to implement, remains to be found. It is possible that also diffractive topologies will need to become a part of this game.

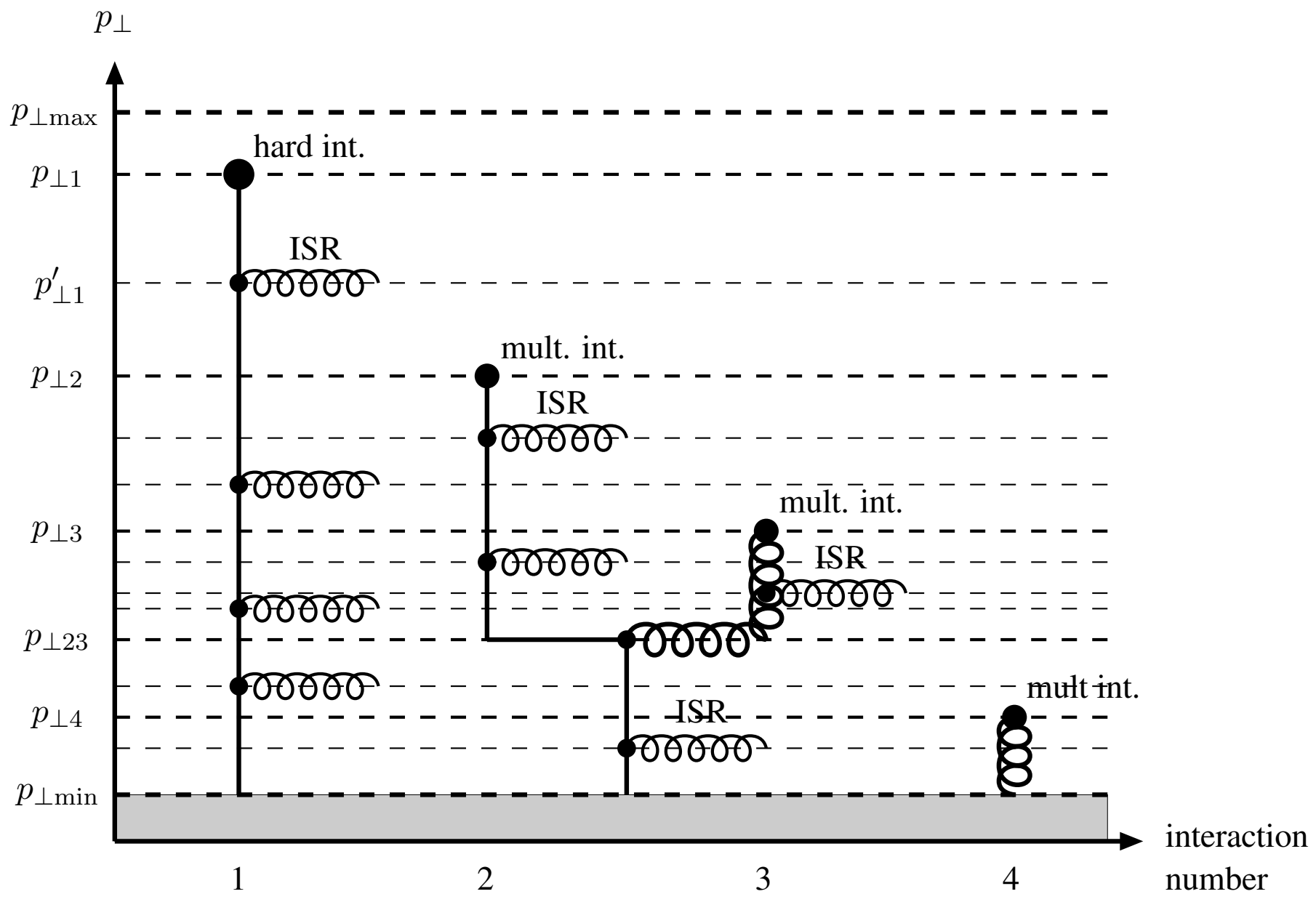

**Fig. 1:** Schematic figure illustrating one incoming hadron in an event with a hard interaction occurring at $p_{\perp 1}$ and three further interactions at successively lower $p_\perp$ scales, each associated with (the potentiality of) initial-state radiation, and further with the possibility of two interacting partons (2 and 3 here) having a common ancestor in the parton showers. Full lines represent quarks and spirals gluons. The vertical $p_\perp$ scale is chosen for clarity rather than realism; most of the activity is concentrated to small $p_\perp$ values.

Apart from this big colour issue, and the smaller ones of a complete interleaving/intertwining, PYTHIA now contains a very consistent and complete picture of both minimum-bias and underlying-event physics. It will be interesting to see how this framework fares in comparisons with data. However, if the models appears complex, this complexity is driven by necessity: all of the issues already brought up must be included in the "definitive" description, in one form or other, plus possibly some more not yet brought to light.

## 2.2 JIMMY

The basic ideas of the eikonal model implemented in JIMMY are discussed elsewhere [10]. The model derives from the observation that for partonic scatters above some minimum transverse momentum, $\hat{p}_T^{\min}$, the values of the hadronic momentum fraction $x$ which are probed decrease as the centre-of-mass energy, $s$, increases, and since the proton structure function rises rapidly at small $x$ [4], high parton densities are probed. Thus the perturbatively-calculated cross section grows rapidly with $s$. However, at such high densities, the probability of more than one partonic scattering in a single hadron-hadron event may become significant. Allowing such multiple scatters reduces the total cross section, and increases the activity in the final state of the collisions.

### *2.2.1 Model Assumptions*

The JIMMY model assumes some distribution of the matter inside the hadron in impact parameter ($b$) space, which is independent of the momentum fraction, $x$. The multiparton interaction rate is then calculated using the cross section for the hard subprocess, the conventional parton densities, and the area overlap function, $A(b)$. No assumption about the behaviour of the *total* cross section is used. For cross sections other than QCD $2 \to 2$ scatters, JIMMY makes use of approximate formulae, valid when all cross sections except QCD $2 \to 2$ are small, which is true in most cases of interest. This approximation is described in detail elsewhere [11].

### 2.2.2 *Standard* JIMMY

The starting point for the multiple scattering model is the assertion that, at fixed impact parameter, $b$, different scatters are independent and so obey Poisson statistics. It is then straightforward to show that the cross section for events in which there are $n$ scatters of type $\mathrm{a}$ is given by

$$\sigma_n = \int \mathrm{d}^2 b \, \frac{(A(b)\sigma_{\mathrm{a}})^n}{n!} \, \mathrm{e}^{-A(b)\sigma_{\mathrm{a}}}, \tag{1}$$

where $\sigma_{\mathrm{a}}$ is the parton–parton cross section and $A(b)$ is the matter density distribution, obeying

$$\int \mathrm{d}^2 b \, A(b) = 1. \tag{2}$$

It is straightforward to show that the inclusive cross section for scatters of type $\mathrm{a}$ is $\sigma_{\mathrm{a}}$ and the total cross section for events with at least one scatter of type $\mathrm{a}$ is

$$\sigma_{\mathrm{tota}} = \int \mathrm{d}^2 b \left(1 - \mathrm{e}^{-A(b)\sigma_{\mathrm{a}}}\right). \tag{3}$$

These can then be combined to give the probability that an event has exactly $n$ scatters of type $\mathrm{a}$, given that it has at least 1 scatter of type $\mathrm{a}$,

$$P_n = \frac{\int \mathrm{d}^2 b \, \frac{(A(b)\sigma_{\mathrm{a}})^n}{n!} \, \mathrm{e}^{-A(b)\sigma_{\mathrm{a}}}}{\int \mathrm{d}^2 b \left(1 - \mathrm{e}^{-A(b)\sigma_{\mathrm{a}}}\right)}, \qquad n \geq 1. \tag{4}$$

This is the probability distribution pretabulated (as a function of $\sqrt{s}$) by Jimmy.

Jimmy's procedure can then be summarized as:

1. Give all events cross section $\sigma_{\mathrm{tota}}$.
2. In a given event, choose $n$ according to Eq. (4).

It is interesting to note that Jimmy's procedure, despite integrating over $b$ once-and-for-all at initialization time, correctly reproduces the correlation between different scatters, whose physical origin is a $b$-space correlation: small cross section scatters are more likely to come from events with a large overlap and hence be accompanied by a larger-than-average number of large cross section scatters.

### 2.2.3 *Two Different Scattering Types*

We consider the possibility that there are two different scattering types, but that the cross section for the second type, $\sigma_{\mathrm{b}}$, is small enough that events with more than one scatter of type $\mathrm{b}$ are negligible. The probability distribution for number of scatters of type $\mathrm{a}$, $n$, in events with at least one of type $\mathrm{b}$ is given by [11]

$$P(n|m \geq 1) = \frac{\int \mathrm{d}^2 b \, \frac{(A(b)\sigma_{\mathrm{a}})^n}{n!} \, \mathrm{e}^{-A(b)\sigma_{\mathrm{a}}} \left(1 - \mathrm{e}^{-A(b)\sigma_{\mathrm{b}}}\right)}{\int \mathrm{d}^2 b \left(1 - \mathrm{e}^{-A(b)\sigma_{\mathrm{b}}}\right)}, \qquad n \geq 0. \tag{5}$$

Since $\sigma_{\mathrm{b}}$ is small, we can expand the exponentials and obtain

$$P(n|m \geq 1) \approx \int \mathrm{d}^2 b \, A(b) \, \frac{(A(b)\sigma_{\mathrm{a}})^n}{n!} \, \mathrm{e}^{-A(b)\sigma_{\mathrm{a}}}, \qquad n \geq 0. \tag{6}$$

Note that this expression is independent of $\sigma_{\rm b}$. It is therefore ideal for implementing into JIMMY. It is useful to rewrite this equation, as follows. We redefine $n$ to be the total number of scatters, including the one of type b (i.e. "new $n$"="old $n$"+1) and rewrite, to obtain

$$P_n \approx \frac{\int {\rm d}^2 b\, n \,\frac{(A(b)\sigma_{\rm a})^n}{n!}\, {\rm e}^{-A(b)\sigma_{\rm a}}}{\sigma_{\rm a}}, \qquad n \geq 1. \tag{7}$$

Note the similarity with Eq. (4), making this form even easier to implement into Jimmy.

The Monte Carlo implementation of this procedure is straightforward:

1. Give all events cross section $\sigma_{\rm b}$.
2. In a given event choose $n$ according to Eq. (7).
3. Generate 1 scatter of type b and $n-1$ of type a.

There is one important difference between the cases in which b is distinct from a and b is a subset of a: some of the $n-1$ scatters of type a could also be of type b. Although this is a small fraction of the total, it can be phenomenologically important. As each scatter of type a is generated, a check is made as to whether it is also of type b. The $m$th scatter of type b generated so far is rejected with probability $1/(m+1)$. This ensures that the proposed algorithm is continuous at the boundary of b.

When using JIMMY at the LHC, the tuneable parameters are those described previously [10], with the obvious exception of those parameters which only concern the photon. Those remaining are therefore the minimum transverse momentum of a hard scatter, the proton structure, and the effective radius of the proton. Details on how to adjust these parameters can be found elsewhere [11].

### 2.3 Simulation of Multiple Interactions in `Sherpa`

Given the studies presented in the following sections, and references therein, current multi-purpose event generators rely heavily on the implementation of multiple parton interaction models to describe the final state in hadronic collisions. To allow `Sherpa` to provide a complete description of hadronic events, the module `AMISIC++` has been developed to simulate multiple parton interactions. This module is capable of simulating multiple scatterings according to the formalism initially presented in [3] and in its current implementation acts as a benchmarking tool to cross-check new multiple interaction models [12].

The basic assumption of the multiple interaction formalism according to T. Sjöstrand and M. van Zijl is, that the differential probability $\mathcal{P}(p_\perp^{\rm out})$ to get a (semi-)hard scattering in the underlying event is given by $\mathcal{P}(p_\perp^{\rm out}) = \sigma_{\rm hard}(p_\perp^{\rm out})/\sigma_{\rm ND}$, where $p_\perp^{\rm out}$ is the transverse momentum of the outgoing partons in the scattering. Since $\sigma_{\rm hard}$ is dominated by $2 \to 2$ processes, the definition of $p_\perp^{\rm out}$ is unambiguous. The specific feature of `AMISIC++` is, that it allows for an independent $Q^2$-evolution of initial and final state partons in each (semi-)hard scattering via an interface to `Sherpa`'s parton shower module `APACIC++` [13, 14]. The key point here is, that the parton shower must then respect the initial $p_\perp^{\rm out}$ distribution of each (semi-)hard scattering. In particular, it must not radiate partons with $p_\perp > p_\perp^{\rm out}$. The appropriate way to incorporate this constraint is in fact identical to the realisation of the highest multiplicity treatment in the CKKW approach [15–18]. Our proposed algorithm works as follows:

1. Create a hard scattering process according to the CKKW approach.
   Employ a $K_T$ jet finding algorithm in the $E$-scheme to define final state jets.
   Stop the jet clustering as soon as there remains only one QCD node to be clustered.
   Set the starting scale of the multiple interaction evolution to $p_\perp$ of this node.
2. Select $p_\perp$ of the next (semi-)hard interaction according to [3].
   If done for the first time in the event, select the impact parameter $b$ of the collision.

3. Set the jet veto scale of the parton shower to the transverse momentum $p_\perp$, selected in 2. Start the parton shower at the QCD hard scale $\mu^2_{\rm QCD} = 2\,stu/\left(s^2+t^2+u^2\right)$.
4. Return to step 2.

The above algorithm works for pure QCD hard matrix elements as well as for electroweak processes in the hard scattering. In the QCD case the selected starting scale for the determination of the first additional interaction reduces to $p_\perp^{\rm out}$ and is thus equal to the original ordering parameter. In the case of electroweak core processes, like single $W$- or $Z$-boson production there is no such unique identification. On the other hand the multiple scatterings in the underlying event must not spoil jet topologies described by the hard event through, e.g., using multi-jet matrix elements. However, since the electroweak bosons may be regarded to have been radiated off QCD partons during the parton shower evolution of a hard QCD event, it is appropriate to reinterprete the hard matrix element as such a QCD+EW process, where the simplest is a 1-jet process.

An important question in conjunction with the simulation of underlying events is the assignment of colours to final state particles. In the Sherpa framework, colour connections in any hard $2 \to 2$ QCD process are chosen according to the kinematics of the process. In particular the most probable colour configuration is selected. Additionally, initial state hadrons are considered to be composed from QCD partons in such a way that the colour string lengths in the final state are minimized. In cases, where it is impossible to realise this constraint, the colour configurations of the hard matrix elements are kept but the configuration of the beam remnants is shuffled until a suitable solution is found.

Figures 2–5 show some preliminary results obtained with the above algorithm, implemented in the current Sherpa version, Sherpa-1.0.6. We compare the Sherpa prediction including multiple interactions to the one without multiple interactions and to the result obtained with PYTHIA 6.214, also including multiple interactions and employing the parameters of PYTHIA Tune A [6]. Shown are hadron-level predictions, which are uncorrected for detector acceptance, except for a uniform track finding efficiency as given in [19]. Data were taken at the Fermilab Tevatron during Run I [20]. Good agreement between the simulations and data is observed only if multiple interactions are included. The mean interaction number in Sherpa, including the hard scattering, in this case is $<N_{\rm hard}> = 2.08$, while for PYTHIA 6.214 it is $<N_{\rm hard}> = 7.35$. The lower interaction number in Sherpa can easily be understood, as a decrease of parton multiplicity in the (semi-)hard scatterings due to a rise of the parton multiplicity in the parton showers. PYTHIA 6.214 does not allow for parton showers in the (semi-)hard scatterings in the underlying event. This feature has, however, been added in PYTHIA 6.3 (see Section 2.1), and is also present in JIMMY(Section 2.2).

## 2.4 PHOJET

The physics model used in the MC event generator PHOJET combines the ideas of the DPM [21] with perturbative QCD to give an almost complete picture of high-energy hadron collisions [22].

PHOJET is formulated as a two-component model containing contributions from both soft and hard interactions. The DPM is used to describe the dominant soft processes and perturbative QCD is applied to generate hard interactions.

There has been very little development on PHOJET for the last few years, although it is used quite widely in minimum bias and cosmic ray physics. A major disadvantage for the LHC is that it is not part of a general purpose generator, and therefore cannot be used to generate underlying events to low cross section processes.

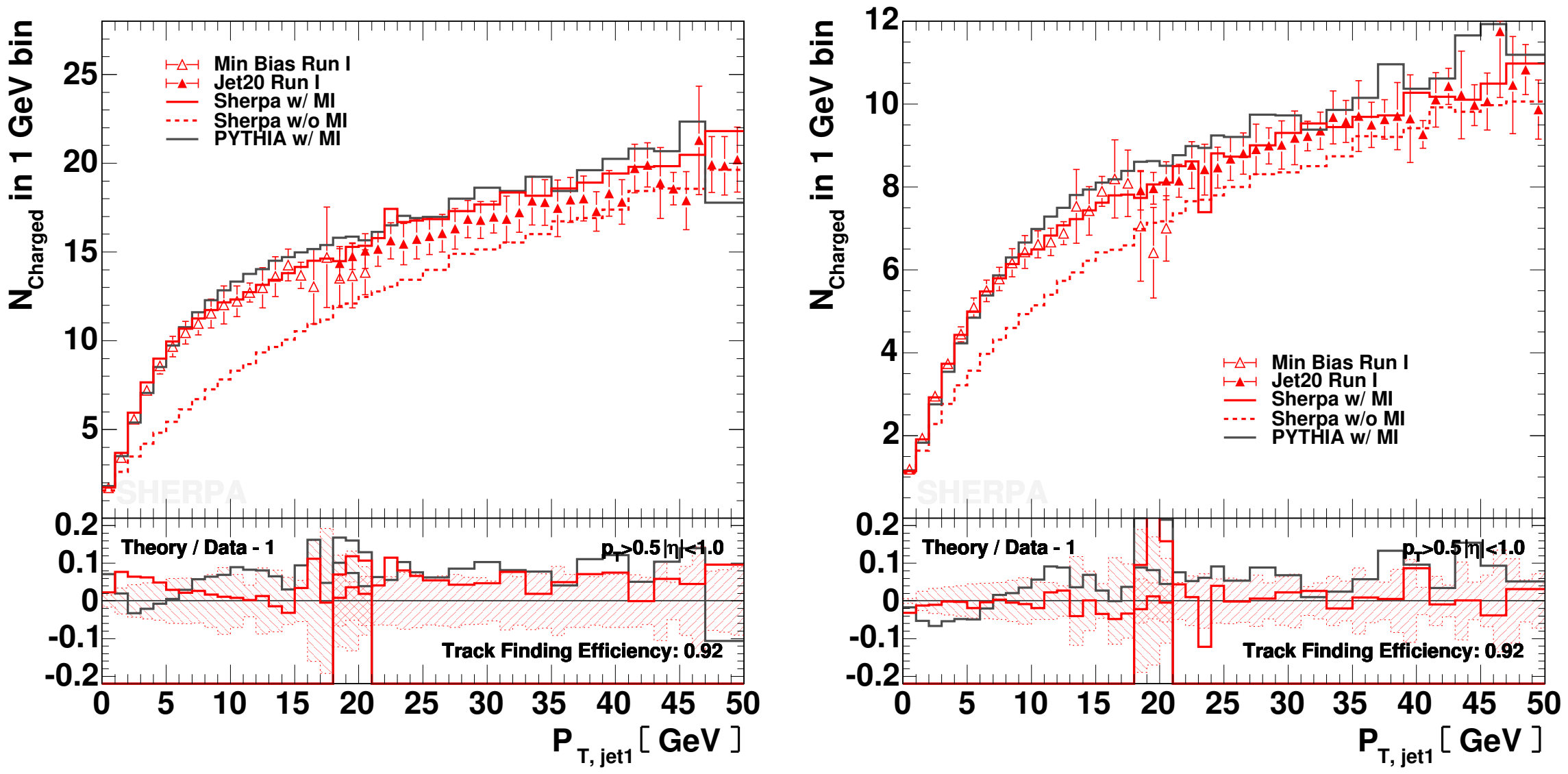

**Fig. 2:** Charged particle multiplicity as a function of $P_T$ of the leading charged particle jet. The left figure shows the total charged particle multiplicity in the selected $p_T$- and $\eta$-range, the right one displays the same in the "Toward" region (for definitions, see Section 3 and [20]).

## 3 Tuning PYTHIA and HERWIG/JIMMY in Run 2 at CDF

The behaviour of the charged particle ($p_T > 0.5\,\mathrm{GeV/c}$, $|\eta| < 1$) and energy ($|\eta| < 1$) components of the UE in hard scattering proton-antiproton collisions at $1.96\,\mathrm{TeV}$ has been studied at CDF. The goal is to produce data on the UE that is corrected to the particle level, so that it can be used to tune the QCD Monte-Carlo models using tools such as those described in the contributions from Group 5 of this workshop without requiring a simulation of the CDF detector. Unlike the previous CDF Run 2 UE analysis which used JetClu to define "jets" and compared uncorrected data with the QCD Monte-Carlo models after detector simulation (i.e., CDFSIM), this analysis uses the midpoint jet algorithm and corrects the observables to the particle level. The corrected observables are then compared with the QCD Monte-Carlo models at the particle level (i.e., generator level). The QCD Monte-Carlo models include PYTHIA Tune A, HERWIG and a tuned version of JIMMY.

One can use the topological structure of hadron-hadron collisions to study the UE [19,23,24]. The direction of the leading calorimeter jet is used to isolate regions of $\eta$-$\phi$ space that are sensitive to the UE. As illustrated in Fig. 6, the direction of the leading jet, jet#1, is used to define correlations in the azimuthal angle, $\Delta\phi$. The angle $\Delta\phi = \phi - \phi_{\mathrm{jet}\#1}$ is the relative azimuthal angle between a charged particle (or a calorimeter tower) and the direction of jet#1. The "transverse" region is perpendicular to the plane of the hard 2-to-2 scattering and is therefore very sensitive to the UE. We restrict ourselves to charged particles in the range $p_T > 0.5\,\mathrm{GeV/c}$ and $|\eta| < 1$ and calorimeter towers with $E_T > 0.1\,\mathrm{GeV}$ and $|\eta| < 1$, but allow the leading jet that is used to define the "transverse" region to have $|\eta(jet\#1)| < 2$. Furthermore, we consider two classes of events. We refer to events in which there are no restrictions placed on the second and third highest $P_T$ jets (jet#2 and jet#3) as "leading jet" events. Events with at least two jets with $P_T > 15\,\mathrm{GeV/c}$ where the leading two jets are nearly "back-to-back" ($|\Delta\phi| > 150°$) with $P_T(jet\#2)/P_T(jet\#1) > 0.8$ and $P_T(jet\#3) < 15\,\mathrm{GeV/c}$ are referred to as "back-to-back" events. "Back-to-back" events are a subset of the "leading jet" events. The idea is to suppress hard initial and final-state radiation thus increasing the sensitivity of the "transverse" region to the "beam-beam remnants" and the multiple parton scattering component of the "underlying event".

As illustrated in Fig. 7, we define a variety of MAX and MIN "transverse" regions which help to separate the "hard component" (initial and final-state radiation) from the "beam-beam remnant" component. MAX (MIN) refer to the "transverse" region containing largest (smallest) number of charged particles or to the region containing the largest (smallest) scalar $PT$sum of charged particles or the region containing the largest (smallest) scalar $ET$sum of particles. Since we will be studying regions in $\eta$-$\phi$ space with different areas, we will construct densities by dividing by the area. For example, the number density, $dN_{chg}/d\phi d\eta$, corresponds to the number of charged particles ($p_T > 0.5\,\text{GeV/c}$) per unit $\eta$-$\phi$ the PTsum density, $dPT_{sum}/d\phi d\eta$, corresponds to the amount of charged particle ($p_T > 0.5\,\text{GeV/c}$) scalar $PT$sum per unit $\eta$-$\phi$, and the transverse energy density, $dET_{sum}/d\phi d\eta$, corresponds the amount of scalar $ET$sum of all particles per unit $\eta$-$\phi$. One expects that the "transMAX" region will pick up the hardest initial or final-state radiation while both the "transMAX" and "transMIN" regions should receive "beam-beam remnant" contributions. Hence one expects the "transMIN" region to be more sensitive to the "beam-beam remnant" component of the "underlying event", while the "transMAX" minus the "transMIN" (i.e., "transDIF") is very sensitive to hard initial and final-state radiation. This idea, was first suggested by Bryan Webber and Pino Marchesini [25], and implemented in a paper by Jon Pumplin [26]. This was also studied by Valeria Tano in her CDF Run 1 analysis of maximum and minimum transverse cones [27].

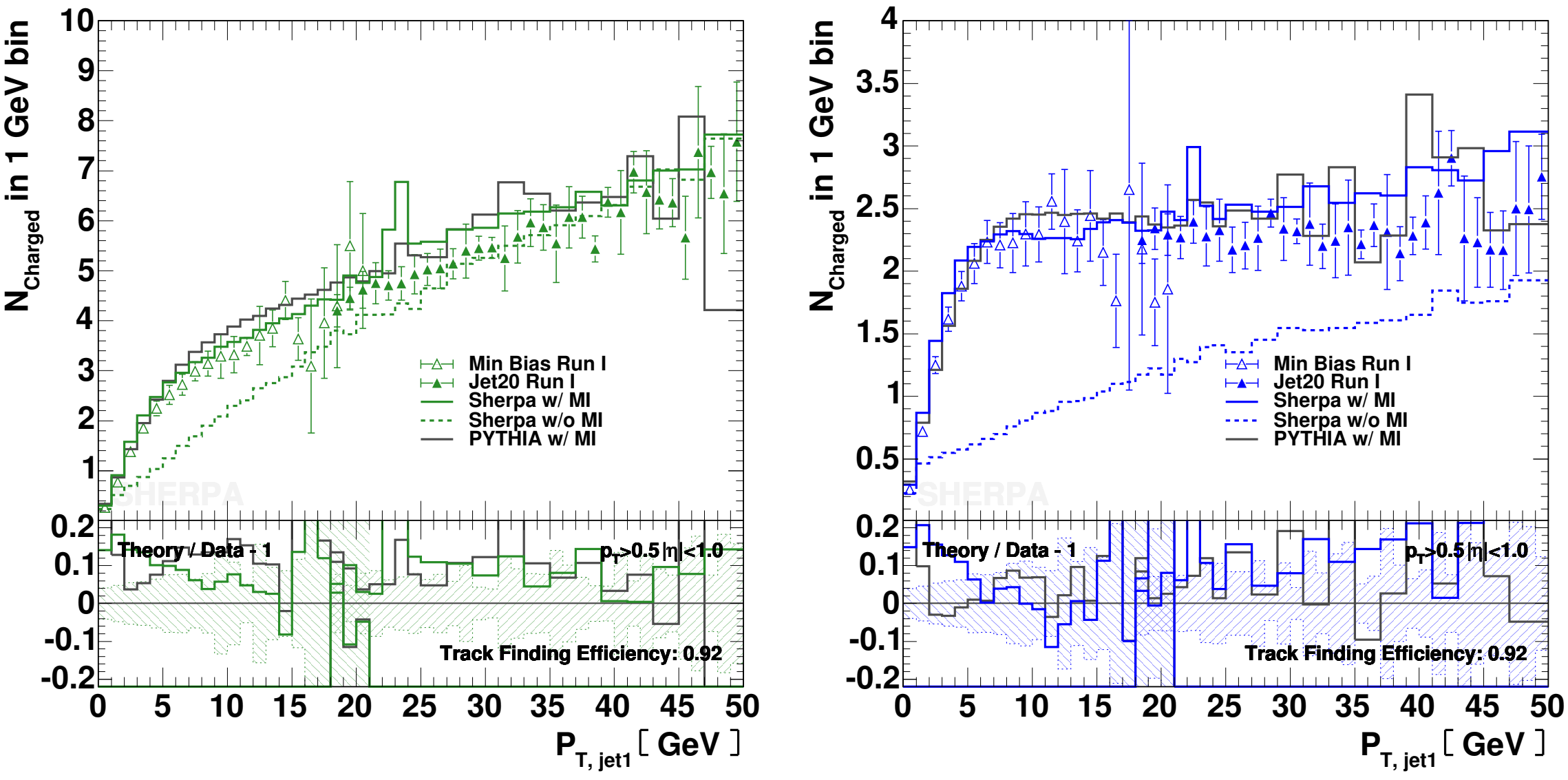

**Fig. 3:** Charged particle multiplicity as a function of $P_T$ of the leading charged particle jet. The left figure shows results for the "Away" side region, the right one displays results for the "Transverse" region.

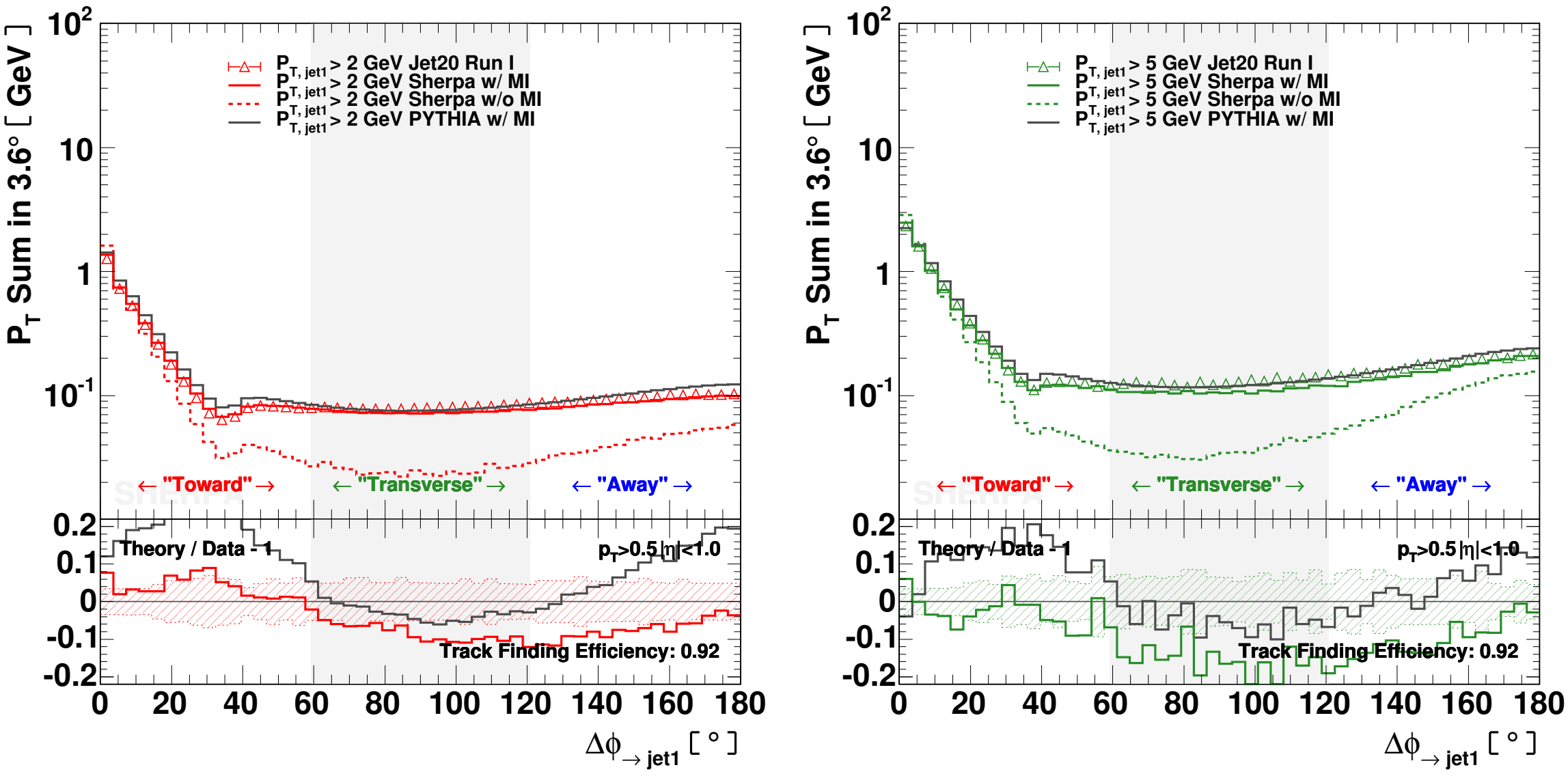

**Fig. 4:** Scalar $P_T$ sum as a function of the azimuthal angle relative to the leading charged particle jet. The left figure shows results for $P_{T,jet1} > 2\,\text{GeV}$, the right one displays results for $P_{T,jet1} > 5\,\text{GeV}$.

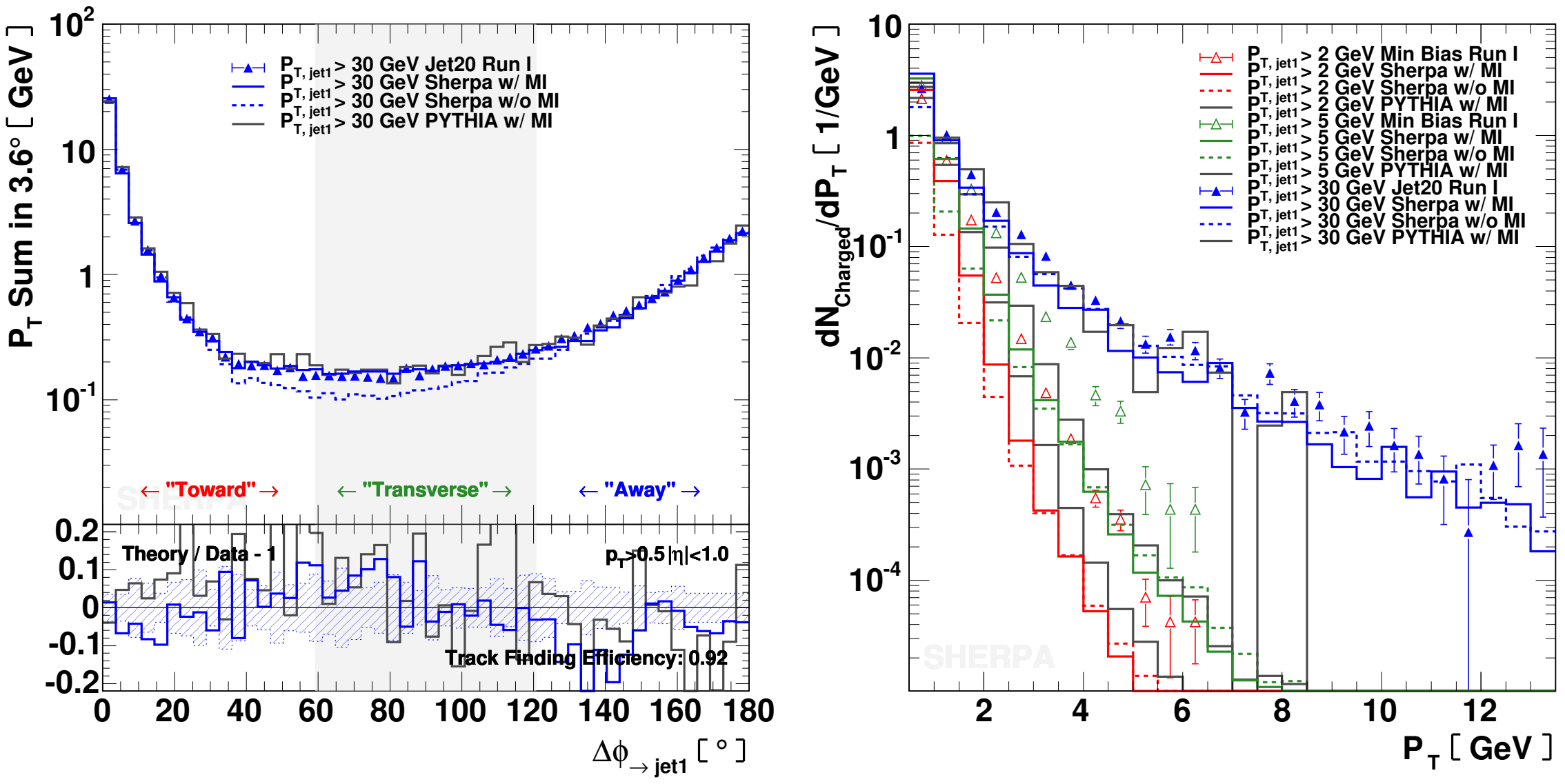

**Fig. 5:** Left: Scalar $P_T$ sum as a function of the azimuthal angle relative to the leading charged particle jet for $P_{T,jet1} > 30\,\text{GeV}$. Right: Charged particle multiplicity as a function of $P_T$ in the "Transverse" region.

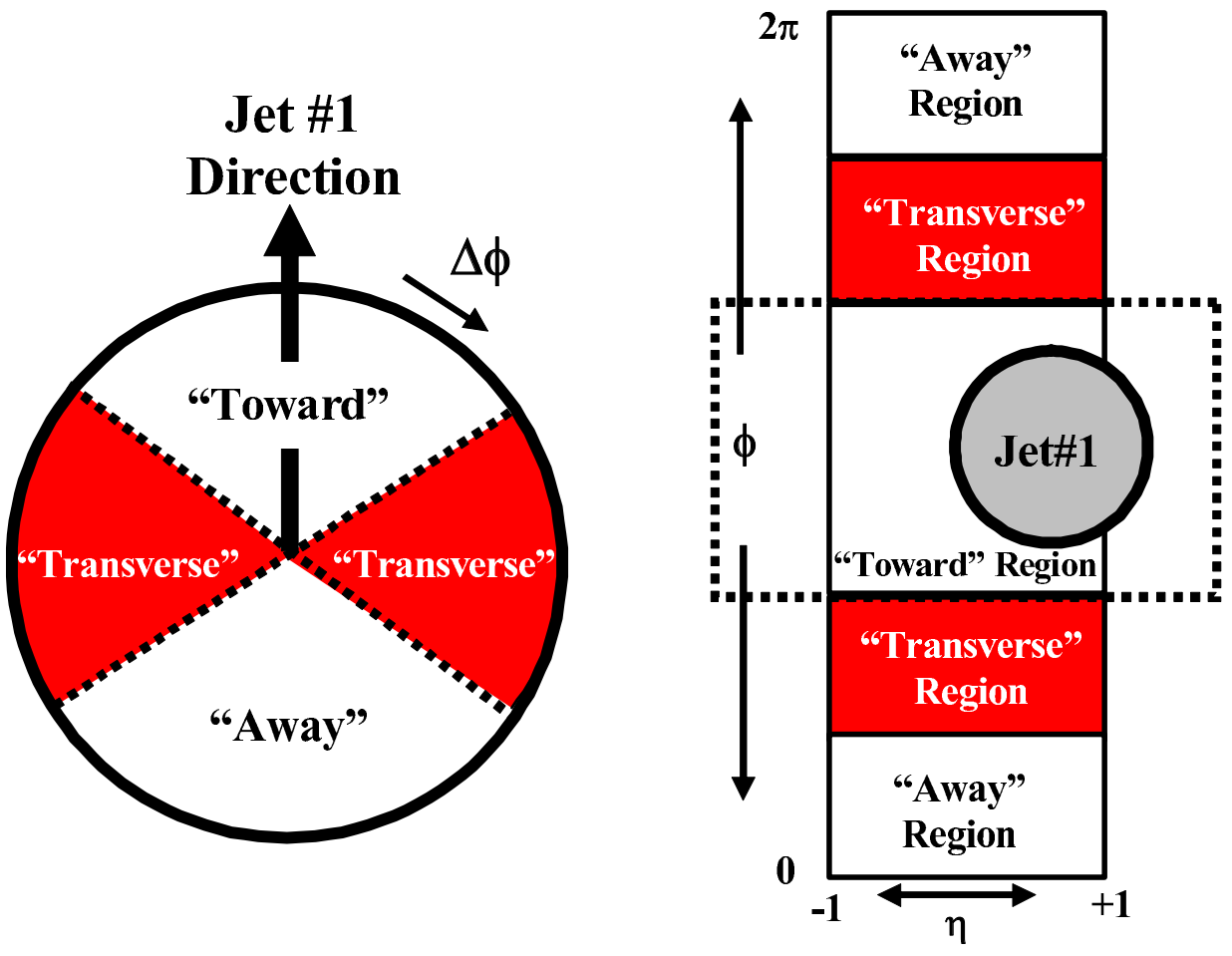

**Fig. 6:** Illustration of correlations in azimuthal angle $\phi$ relative to the direction of the leading jet (MidPoint, $R = 0.7$, $f_{merge} = 0.75$) in the event, jet#1. The angle $\Delta\phi = \phi - \phi_{\text{jet1}}$ is the relative azimuthal angle between charged particles and the direction of jet#1. The "transverse" region is defined by $60° < |\Delta\phi| < 120°$ and $|\eta| < 1$. We examine charged particles in the range $p_T > 0.5\,\text{GeV/c}$ and $|\eta| < 1$ and calorimeter towers with $|\eta| < 1$, but allow the leading jet to be in the region $|\eta(\text{jet\#1})| < 2$.

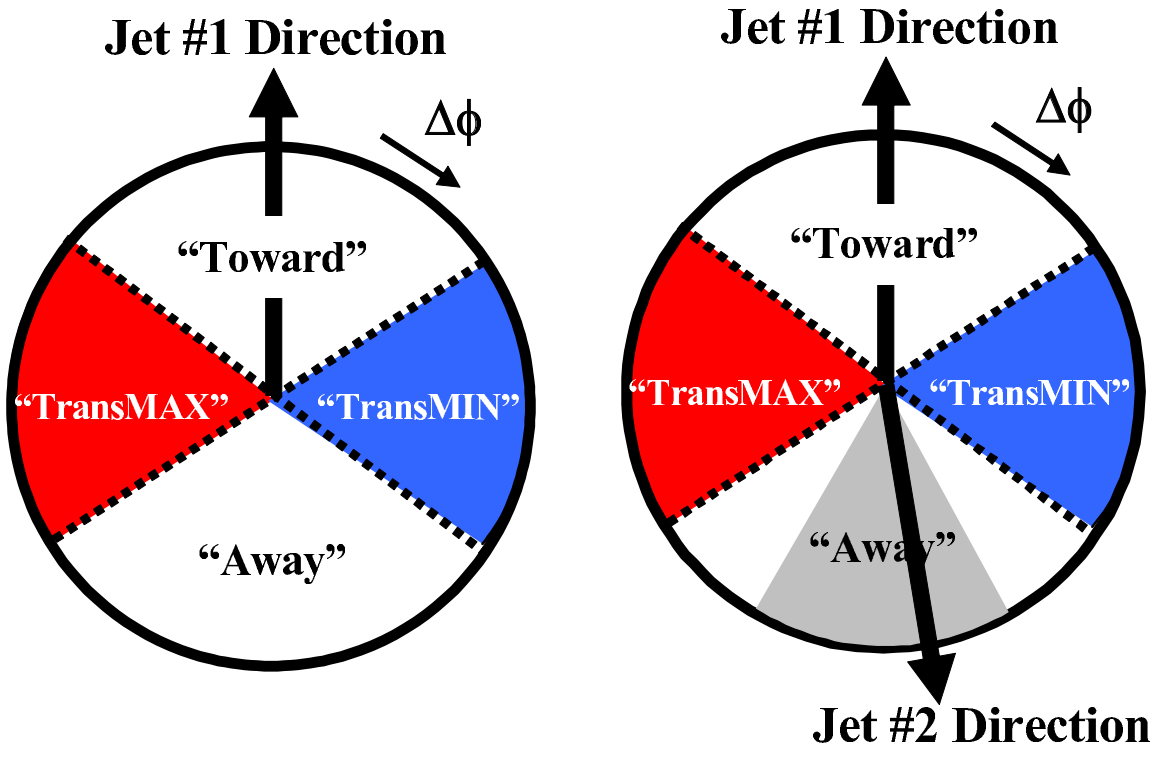

**Fig. 7:** Illustration of correlations in azimuthal angle $\phi$ relative to the direction of the leading jet (highest $P_T$ jet) in the event, jet#1. The angle $\Delta\phi = \phi - \phi_{\text{jet}\#1}$ is the relative azimuthal angle between charged particles and the direction of jet#1. On an event by event basis, we define "transMAX" ("transMIN") to be the maximum (minimum) of the two "transverse" regions, $60° < \Delta\phi < 120°$ and $60° < -\Delta\phi < 120°$. "transMAX" and "transMIN" each have an area in $\eta$-$\phi$ space of $\Delta\eta\Delta\phi = 4\pi/6$. The overall "transverse" region defined in Fig. 6 contains both the "transMAX" and the "transMIN" regions. Events in which there are no restrictions placed on the second and third highest $p_T$ jets (jet#2 and jet#3) are referred to as "leading jet" events (*left*). Events with at least two jets with $p_T > 15$ GeV/c where the leading two jets are nearly "back-to-back" ($|\Delta\phi| > 150°$) with $p_T(\text{jet}\#2)/p_T(\text{jet}\#1) > 0.8$ and $p_T(\text{jet}\#3) < 15$ GeV/c are referred to as "back-to-back" events (*right*).

Our previous Run 2 UE analysis [28] used JetClu to define jets and compared uncorrected data with PYTHIA Tune A [6] and HERWIG after detector simulation (i.e., CDFSIM). This analysis uses the MidPoint jet algorithm ($R = 0.7$, $f_{merge} = 0.75$) and corrects the observables to the particle level. The corrected observables are then compared with the QCD Monte-Carlo models at the particle level (i.e., generator level). The models includes PYTHIA Tune A, HERWIG and HERWIG with a tuned version of JIMMY [10]. In addition, for the first time we study the transverse energy density in the "transverse" region.

Fig. 8 compares the data on the density of charged particles and the charged $PT$sum density in the "transverse" region corrected to the particle level for "leading jet" and "back-to-back" events with PYTHIA Tune A and HERWIG at the particle level. As expected, the "leading jet" and "back-to-back" events behave quite differently. For the "leading jet" case the "transMAX" densities rise with increasing $P_T(\text{jet}\#1)$, while for the "back-to-back" case they fall with increasing $P_T(\text{jet}\#1)$. The rise in the "leading jet" case is, of course, due to hard initial and final-state radiation, which has been suppressed in the "back-to-back" events. The "back-to-back" events allows a closer look at the "beam-beam remnant" and multiple parton scattering component of the UE. PYTHIA Tune A, which includes multiple parton interactions, does a better job of describing the data than HERWIG which does not have multiple parton interactions.

The "transMIN" densities are more sensitive to the "beam-beam remnant" and multiple parton interaction component of the "underlying event". The "back-to-back" data show a decrease in the "transMIN" densities with increasing $P_T(\text{jet}\#1)$ which is described fairly well by PYTHIA Tune A (with multiple parton interactions) but not by HERWIG (without multiple parton interactions). The decrease of the "transMIN" densities with increasing $P_T$(jet#1) for the "back-to-back" events is very interesting and might be due to a "saturation" of the multiple parton interactions at small impact parameter. Such an effect is included in PYTHIA Tune A but not in HERWIG (without multiple parton interactions).

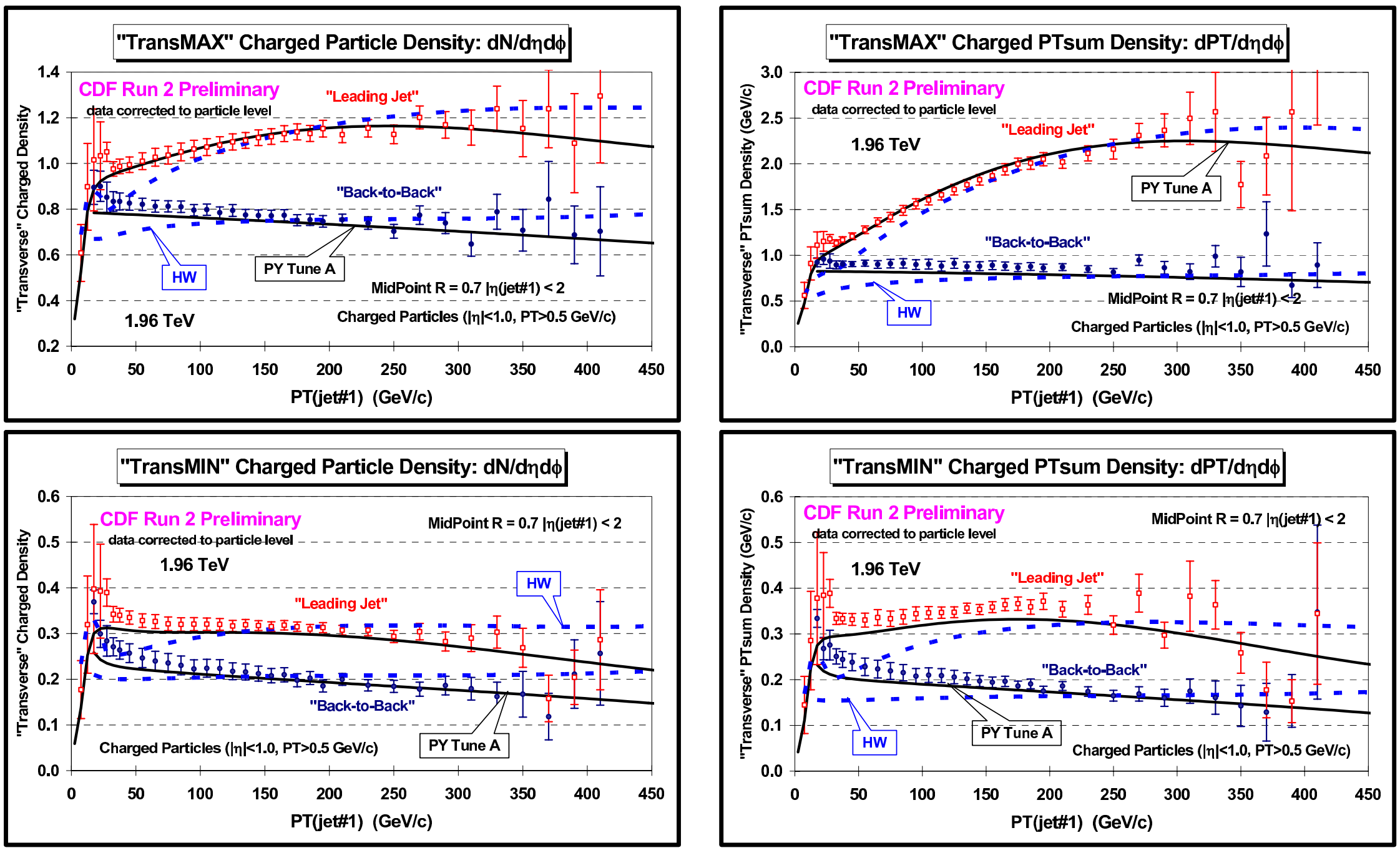

**Fig. 8:** Data at 1.96 TeV on (left) the density of charged particles $dN_{chg}/d\phi d\eta$ and (right) on the scalar $PT$sum density of charged particles, with $p_T > 0.5$ GeV/c and $|\eta| < 1$ in the "transMAX" region (*top*) and the "transMIN" region (*bottom*) for "leading jet" and "back-to-back" events defined in Fig. 7 as a function of the leading jet $P_T$ compared with PYTHIA Tune A and HERWIG. The data are corrected to the particle level (with errors that include both the statistical error and the systematic uncertainty) and compared with the theory at the particle level (i.e., generator level).

Fig. 9(left) compares the data on average $p_T$ of charged particles in the "transverse" region corrected to the particle level for "leading jet" and "back-to-back" events with PYTHIA Tune A and HERWIG at the particle level. Again the "leading jet" and "back-to-back" events behave quite differently.

Fig. 9(right) shows the data corrected to the particle level for the scalar $ET$sum density in the "transverse" region for "leading jet" and "back-to-back" events compared with PYTHIA Tune A and HERWIG. The scalar $ET$sum density has been corrected to correcpond to all particles (all $p_T$, $|\eta| < 1$). Neither PYTHIA Tune A nor HERWIG produce enough energy in the "transverse" region. HERWIG has more "soft" particles than PYTHIA Tune A and does slightly better in describing the energy density in the "transMAX" and "transMIN" regions.

Fig. 10(left) shows the difference of the "transMAX" and "transMIN" regions ("transDIF" = "transMAX" minus "transMIN") for "leading jet" and "back-to-back" events compared with PYTHIA Tune A and HERWIG. "TransDIF" is more sensitive to the hard scattering component of the UE (i.e., initial and final state radiation). Both PYTHIA Tune A and HERWIG underestimate the energy density in the "transMAX" and "transMIN" regions (see Fig. 9). However, they both fit the "transDIF" energy density. This indicates that the excess energy density seen in the data probably arises from the "soft" component of the UE (i.e., beam-beam remnants and/or multiple parton interactions).

JIMMY is a model of multiple parton interaction which can be combined with HERWIG to enhance the UE thereby improving the agreement with data. Fig. 10(right) and Fig. 11(left) show the energy density and charged $PT$sum density, respectively, in the "transMAX" and "transMIN" regions for "lead-ing jet" and "back-to-back" events compared with PYTHIA Tune A and a tuned version of JIMMY˙JIMMY was tuned to fit the "transverse" energy density in "leading jet" events ($PTJIM = 3.25\,\mathrm{GeV/c}$). The default JIMMY ($PTJIM = 2.5\,\mathrm{GeV/c}$) produces too much energy and too much charged $PT$sum in the "transverse" region. Tuned JIMMY does a good job of fitting the energy and charged $PT$sum density in the "transverse" region (although it produces slightly too much charged PTsum at large $P_T$(jet#1)). However, the tuned JIMMY produces too many charged particles with $p_T > 0.5\,\mathrm{GeV/c}$ (see Fig. 11(right)). The particles produced by this tune of JIMMY are too soft. This can be seen clearly in Fig. 12 which shows the average charge particle $p_T$ in the "transverse" region.

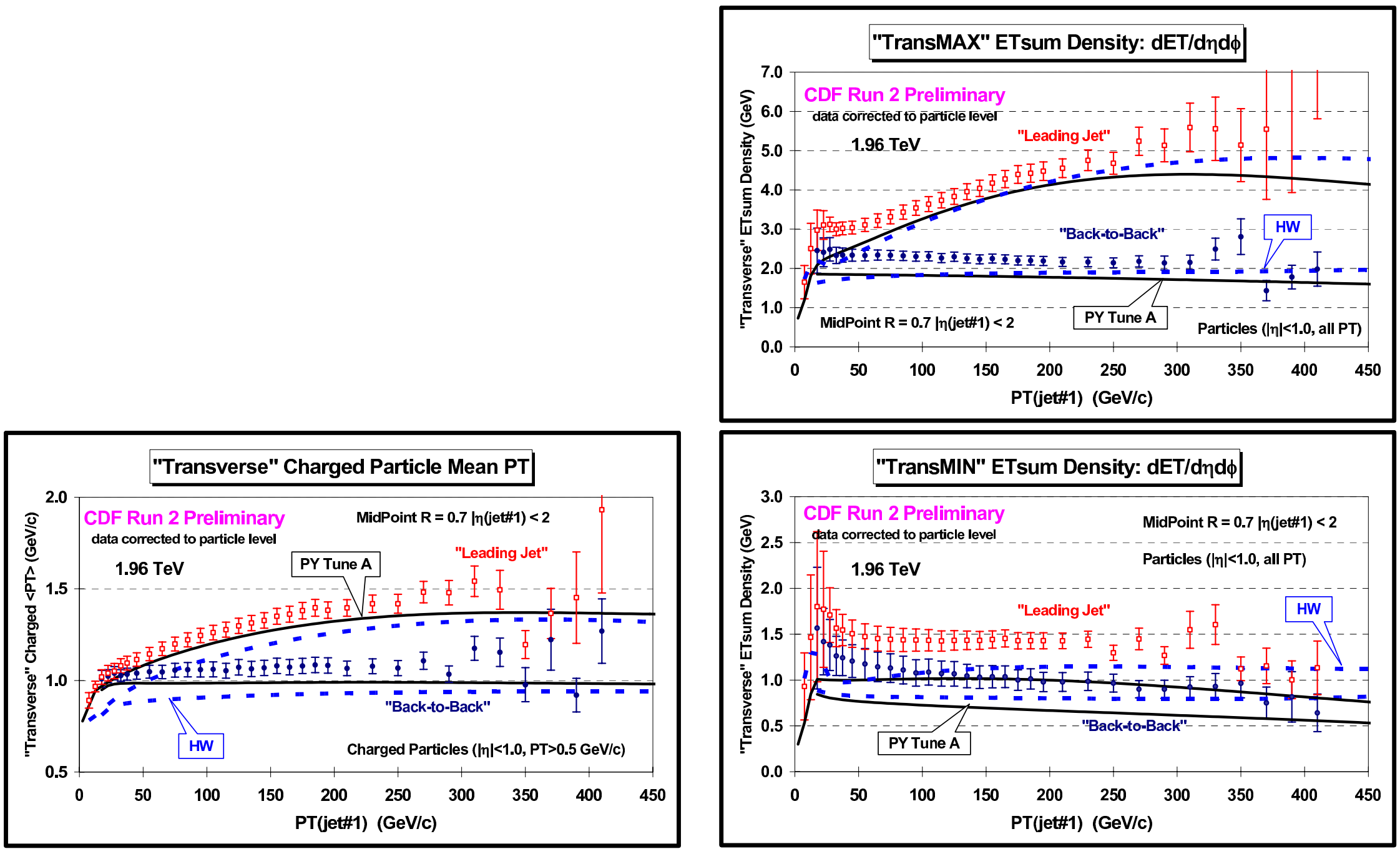

**Fig. 9:** On the left, data at 1.96 TeV on average transverse momentum, $\langle p_T \rangle$, of charged particles $|\eta| < 1$ in the with with $p_T > 0.5\,\mathrm{GeV/c}$ and $|\eta| < 1$ in the "transverse" region. On the right, scalar $ET$sum density, $dET_{sum}/d\phi d\eta$, for particles. with $p_T > 0.5\,\mathrm{GeV/c}$ and $|\eta| < 1$ in the "transMAX" region or the "transMIN" region. The "leading jet" and "back-to-back" events are defined in Fig. 7, and the data are shown as a function of the leading jet $P_T$ and compared with PYTHIA Tune A and HERWIG. The data are corrected to the particle level (with errors that include both the statistical error and the systematic uncertainty) and compared with the theory at the particle level (i.e., generator level).

The goal of this analysis is to produce data on the UE that is corrected to the particle level so that it can be used to tune the QCD Monte-Carlo models without requiring CDF detector simulation. Comparing the corrected observables with PYTHIA Tune A and HERWIG at the particle level (i.e., generator level) leads to the same conclusions as we found when comparing the uncorrected data with the Monte-Carlo models after detector simulation [28]. PYTHIA Tune A (with multiple parton interactions) does a better job in describing the UE (i.e., "transverse" regions) for both "leading jet" and "back-to-back" events than does HERWIG (without multiple parton interactions). HERWIG does not have enough activity in the UE for $P_T$(jet#1) less than about 150 GeV/c, which was also observed in our published Run 1 analysis [19].

This analysis gives our first look at the energy in the UE (i.e., the "transverse" region). Neither PYTHIA Tune A nor HERWIG produce enough transverse energy in the "transverse" region. However, they both fit the "transDIF" energy density ("transMAX" minus "transMIN"). This indicates that the excess energy density seen in the data probably arises from the "soft" component of the UE (i.e., beam-beam remnants and/or multiple parton interactions). HERWIG has more "soft" particles than PYTHIA Tune A and does slightly better in describing the energy density in the "transMAX" and "transMIN" regions. Tuned JIMMY does a good job of fitting the energy and charged $PT$sum density in the "transverse" region (although it produces slightly too much charged $PT$sum at large $P_T$(jet#1)). However, the tuned JIMMY produces too many charged particles with $p_T > 0.5$ GeV/c indicating that the particles produced by this tuned JIMMY are too soft.

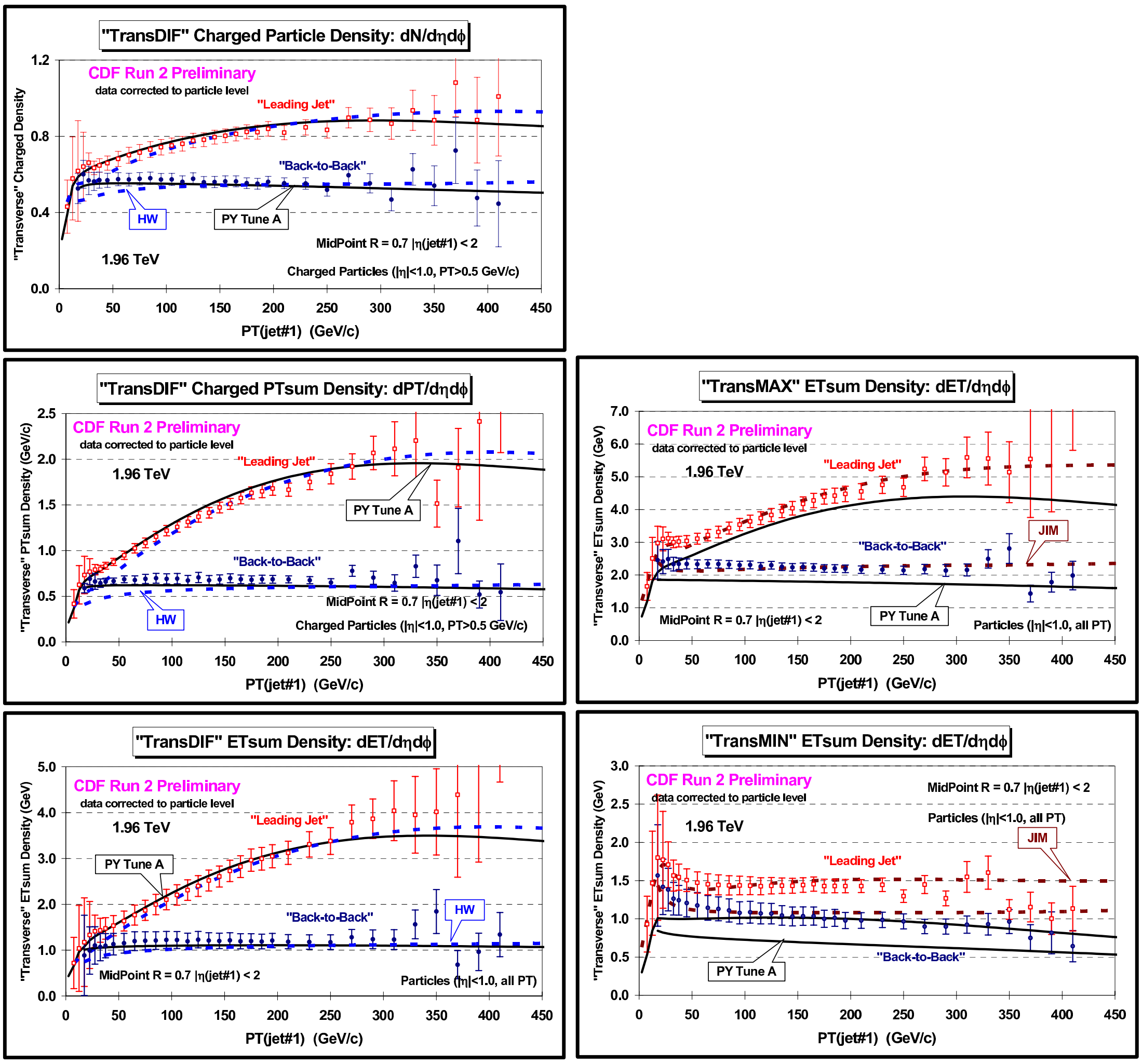

**Fig. 10:** Left: Data at 1.96 TeV on the difference of the "transMAX" and "transMIN" regions ("transDIF" = "transMAX"- "transMIN") for "leading jet" and "back-to-back" events defined in Fig. 7 as a function of the leading jet $P_T$ compared with PYTHIA Tune A and HERWIG.
Right: Data on scalar $ET$sum density, $dET_{sum}/d\phi d\eta$, for particles with $|\eta| < 1$ in the "transMAX" region (*top*) and the "transMIN" region (*bottom*) for "leading jet" and "back-to-back" events defined in Fig. 7 as a function of the leading jet $P_T$ compared with PYTHIA Tune A and tuned JIMMY. JIMMY was tuned to fit the "transverse" energy density in "leading jet" events ($PTJIM = 3.25$ GeV/c). The data are corrected to the particle level (with errors that include both the statistical error and the systematic uncertainty) and compared with the theory at the particle level (i.e., generator level).

In summary, we see an interesting dependence of the UE on the transverse momentum of the leading jet (i.e., the $Q^2$ of the hard scattering). For the "leading jet" case the "transMAX" densities rise with increasing $P_T$(jet#1), while for the "back-to-back" case they fall with increasing $P_T$(jet#1). The rise in the "leading jet" case is due to hard initial and final-state radiation with $p_T > 15$ GeV/c, which has been suppressed in the "back-to-back" events. The "back-to-back" data show a decrease in the "transMIN" densities with increasing $P_T(\text{jet}\#1)$. The decrease of the "transMIN" densities with increasing $P_T(\text{jet}\#1)$ for the "back-to-back" events is very interesting and might be due to a "saturation" of the multiple parton interactions at small impact parameter. Such an effect is included in PYTHIA Tune A (with multiple parton interactions) but not in HERWIG (without multiple parton interactions). PYTHIA Tune A does predict this decrease, while HERWIG shows an increase (due to increasing initial and final state radiation).

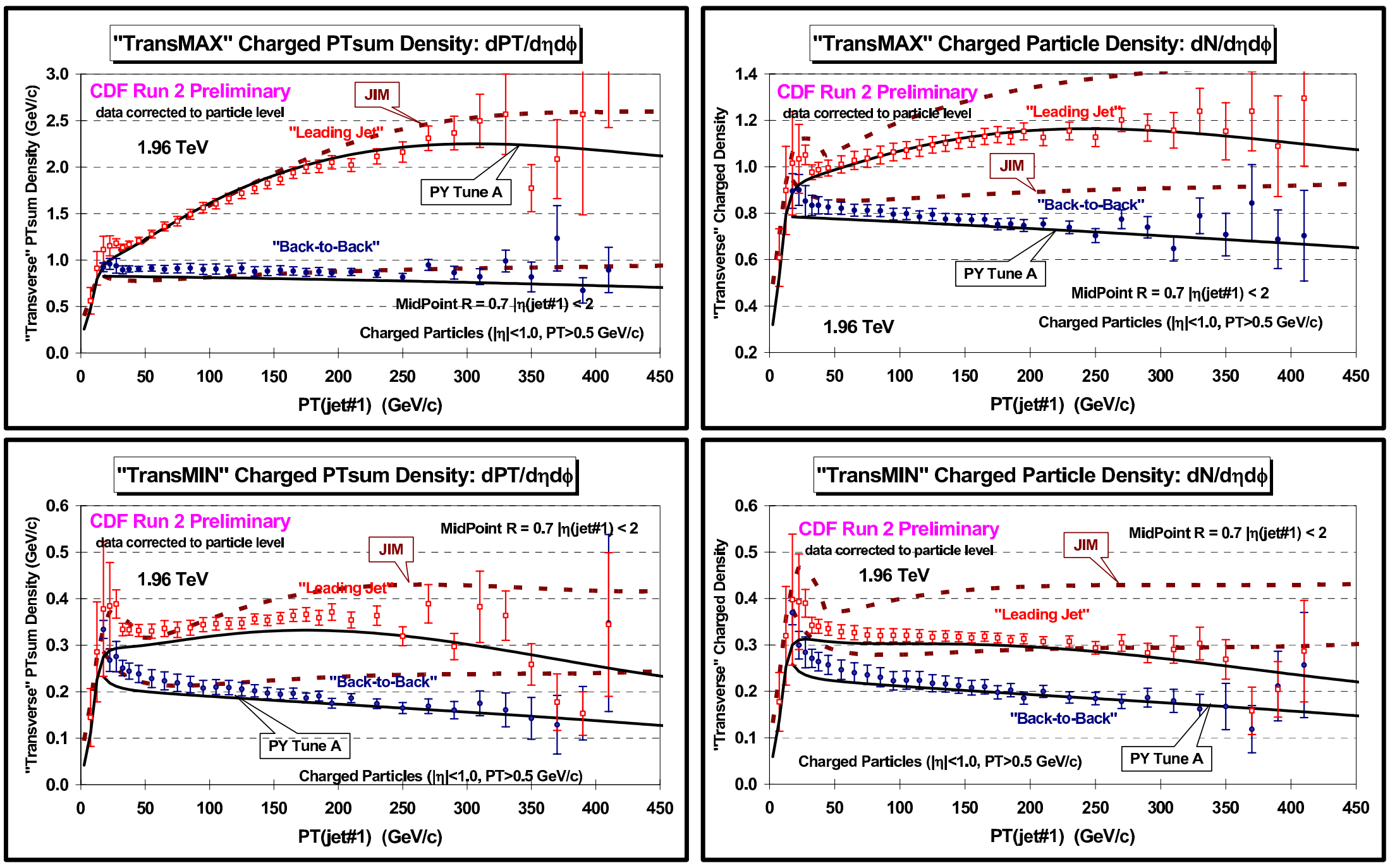

**Fig. 11:** Left: Data at 1.96 TeV on scalar $PT$sum density of charged particles, $dPT_{sum}/d\phi d\eta$, with $p_T > 0.5\,\mathrm{GeV/c}$ and $|\eta| < 1$ in the "transMAX" region (*top*) and the "transMIN" region (*bottom*) for "leading jet" and "back-to-back" events defined in Fig. 7 as a function of the leading jet $P_T$ compared with PYTHIA Tune A and tuned JIMMY. JIMMY was tuned to fit the "transverse" energy density in "leading jet" events ($PTJIM = 3.25$ GeV/c). Right: Data on the density of charged particles, $dN_{chg}/d\phi d\eta$, with $p_T > 0.5\,\mathrm{GeV/c}$ and $|\eta| < 1$ in the "transMAX" region (*top*) and the "transMIN" region (*bottom*) for "leading jet" and "back-to-back" events defined in Fig. 2 as a function of the leading jet $P_T$ compared with PYTHIA Tune A and tuned JIMMY. JIMMY was tuned to fit the "transverse" energy density in "leading jet" events ($PTJIM = 3.25$ GeV/c). The data are corrected to the particle level (with errors that include both the statistical error and the systematic uncertainty) and compared with the theory at the particle level (i.e., generator level).

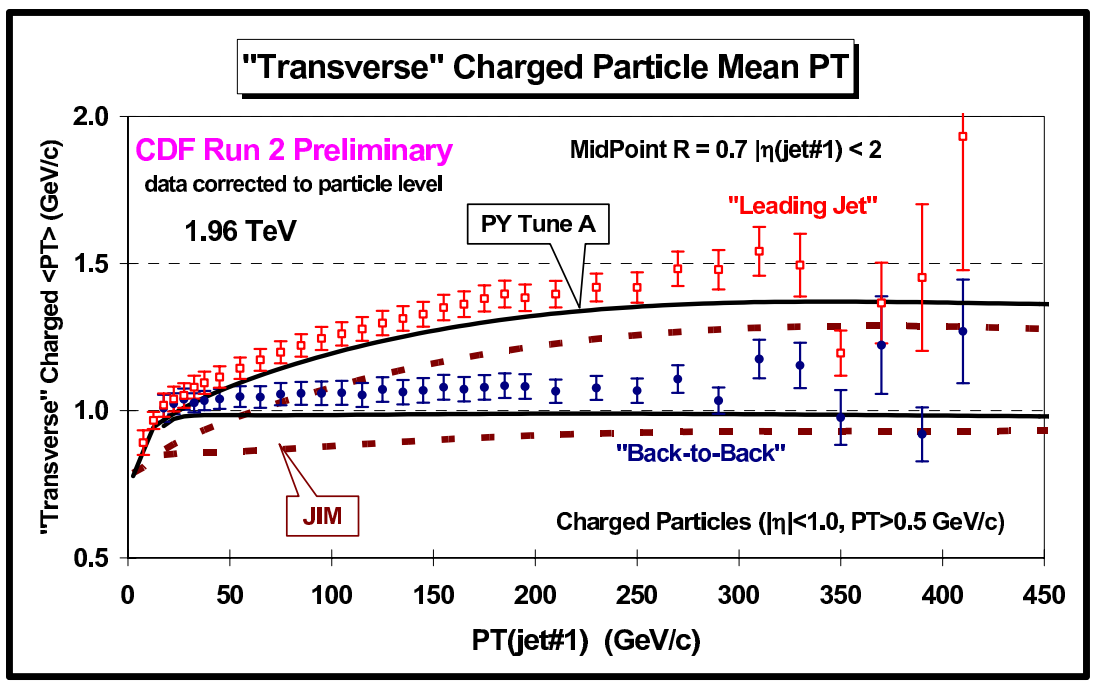

**Fig. 12:** Data at 1.96 TeV on average transverse momentum, $\langle p_T \rangle$, of charged particles with $p_T > 0.5\,\mathrm{GeV/c}$ and $|\eta| < 1$ in the "transverse" region for "leading jet" and "back-to-back" events defined in Fig. 7 as a function of the leading jet $P_T$ compared with PYTHIA Tune A and tuned JIMMY. JIMMY was tuned to fit the "transverse" energy density in "leading jet" events ($PTJIM = 3.25$ GeV/c). The data are corrected to the particle level (with errors that include both the statistical error and the systematic uncertainty) and compared with the theory at the particle level (i.e., generator level).

## 4 Extrapolation to LHC energies

The LHCb experiment [29] is designed to measure CP violation in the B-quark sector at the LHC and expand the current studies underway at the B-factories (BaBar, Belle) and at the Tevatron (CDF, D0). At $\sqrt{s}$=1.8 TeV, 28% of all of the primary produced B-mesons in $p\bar{p}$ collisions are produced in L=1 excited states [30]. These excited states decay via the emission of a charged hadron, allowing the possibility of same-side-tagging (SST) studies. As such, it is important to simulate the production of B mesons as accurately as possible.

The production of primary produced excited meson states are not included in the default PYTHIA [31] settings and including them increases the average multiplicity of an event. An attempt to reproduce the HFAG [32] values whilst retaining the spin counting rule for B** states has been made. This note covers a preliminary re-tuning [33] of PYTHIA v6.224 including these settings.

### 4.1 Method

The main parameter of the multiple-interaction model in PYTHIA v6.224 is the $\hat{p}_T^{\min}$ parameter, which defines the minimum transverse momentum of the parton-parton interactions. This effectively controls the number of parton-parton collisions and hence the average track multiplicity.

The charged particle density measured at $\eta = 0$ in the range of centre-of-mass energies, 52 GeV $< \sqrt{s} <$ 1800 GeV, [34] [35] is used to tune the $\hat{p}_T^{\min}$ parameter of PYTHIA. We define $\rho = \frac{1}{N_{ev}} \frac{dN_{ch}}{d\eta}|_{\eta=0}$ and measure $\rho$ for a range of $\hat{p}_T^{\min}$ values at each $\sqrt{s}$. The quantity $\delta = \rho_{MC} - \rho_{Data}$ is plotted against $\hat{p}_T^{\min}$ and a linear fit performed. In Fig. 13, the re-tuned value of $\hat{p}_T^{\min}$ at $\sqrt{s}$ = 900 GeV is taken to be the point at which the fit crosses the $\hat{p}_T^{\min}$ axis. To extrapolate $\hat{p}_T^{\min}$ to LHC energy, a fit is performed (Figure 14) using the form suggested by PYTHIA:

$$\hat{p}_T^{\min} = \hat{p}_T^{\min}(LHC)\Big(\frac{\sqrt{s}}{14TeV}\Big)^{2\epsilon} \tag{8}$$

### 4.2 Results

Extrapolating to 14 $TeV$ using the tuned values of $\hat{p}_T^{\min}(\sqrt{s})$ and (8), we obtain $\hat{p}_T^{\min}(LHC) = 3.34 \pm 0.13$, with $\epsilon = 0.079 \pm 0.0006$ with a corresponding central multiplicity of $\rho = 6.45 \pm 0.25$. Comparing the output of the re-tuned settings (dashed line) to the old LHCb settings (solid line), *Fig.* 15, 16 and 17, we find that the re-tuned settings produce a slightly lower multiplicity which affects the other distributions accordingly. Note: both the fragmentation parameters and the $\hat{p}_T^{\min}$ parameter affect the multiplicity of a generated event. This re-tuning method varies the $\hat{p}_T^{\min}$ parameter only i.e. it does not alter the fragmentation parameters in any fashion. Further investigations into re-tuning the fragmentation parameters using data from LEP are underway.

### 4.3 Conclusions

The central multiplicity values measured at CDF and UA5 are accurately reproduced using the re-tuned values for $\hat{p}_T^{\min}$ at several $\sqrt{s}$. An extrapolation of $\hat{p}_T^{\min}$ to LHC energies using a form implemented in PYTHIA gives $\hat{p}_T^{\min}(LHC) = 3.34 \pm 0.13$, with $\epsilon = 0.079 \pm 0.0006$ with a corresponding central multiplicity of $\rho_{LHC} = 6.45 \pm 0.25$ in non-diffractive events.

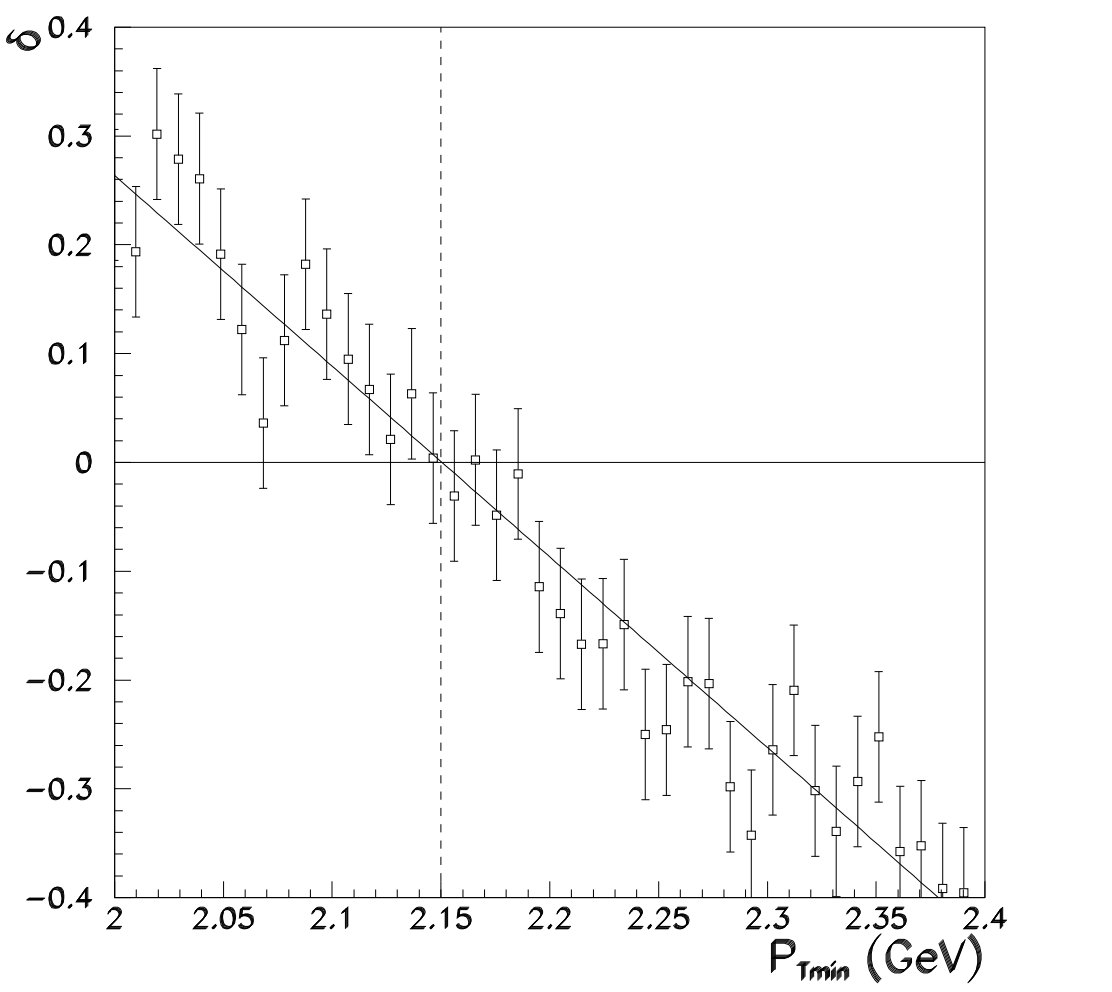

Fig. 13: Determining the value of $\hat{p}_T^{\min}(\sqrt{s} = 900GeV)$, the dashed line shows the point at which $|\delta|$ is minimised.

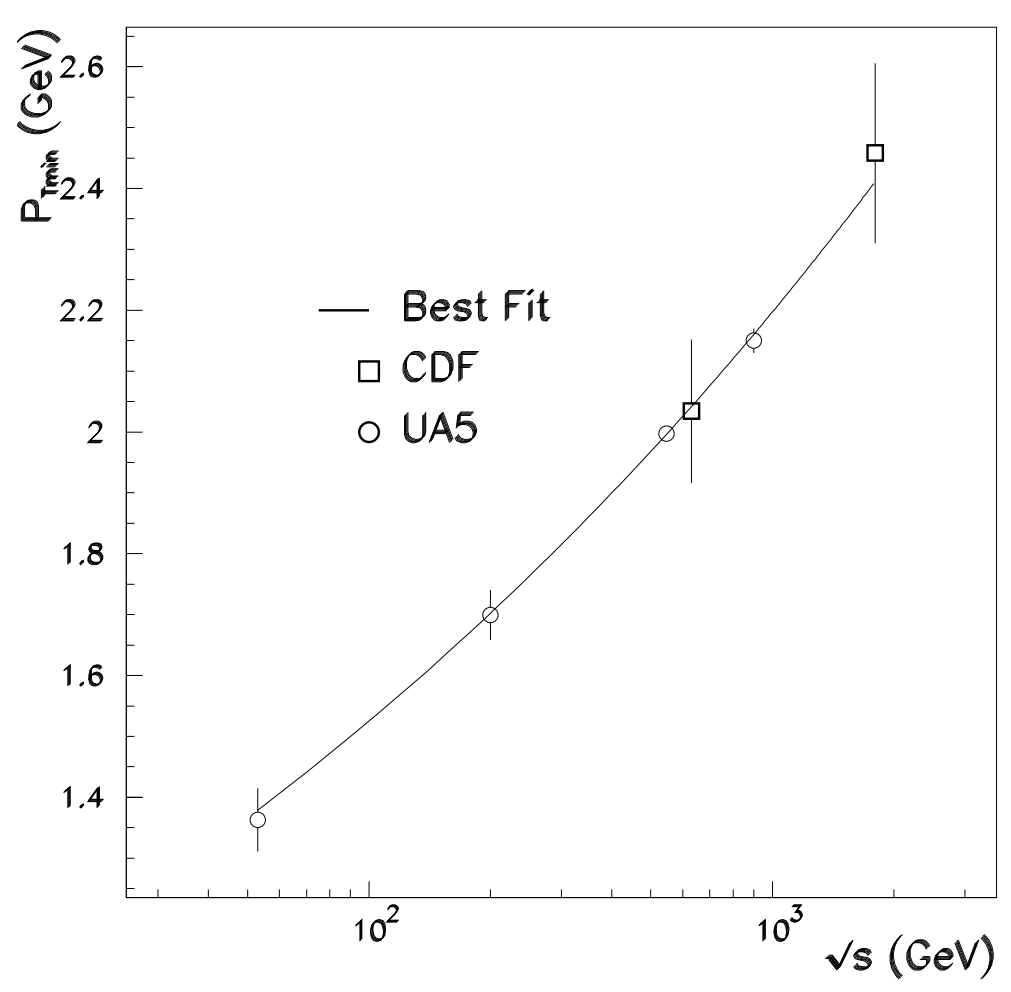

**Fig. 14:** The $\sqrt{s}$ dependance of $\hat{p}_T^{\min}$. The curve is the result of a fit assuming the functional form of (8).

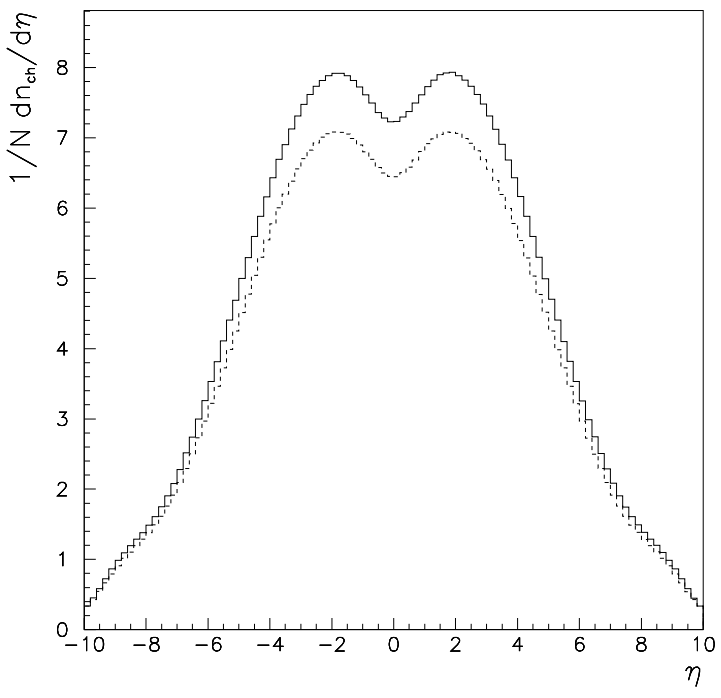

**Fig. 15:** $\eta$ distribution at 14 TeV using the extrapolated value of $P_{T_{\mathrm{Min}}}$

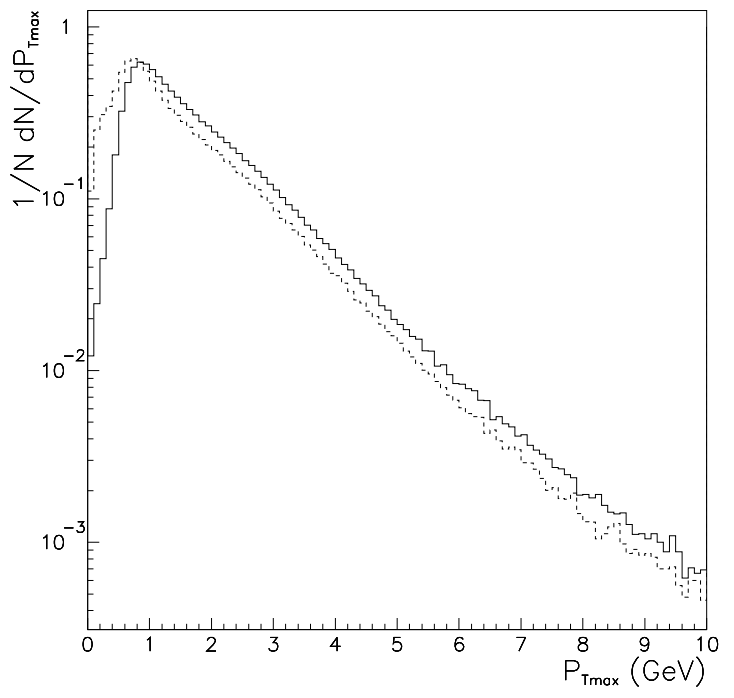

**Fig. 16:** $p_{\perp_{\max}}$ distribution in the LHCb acceptance

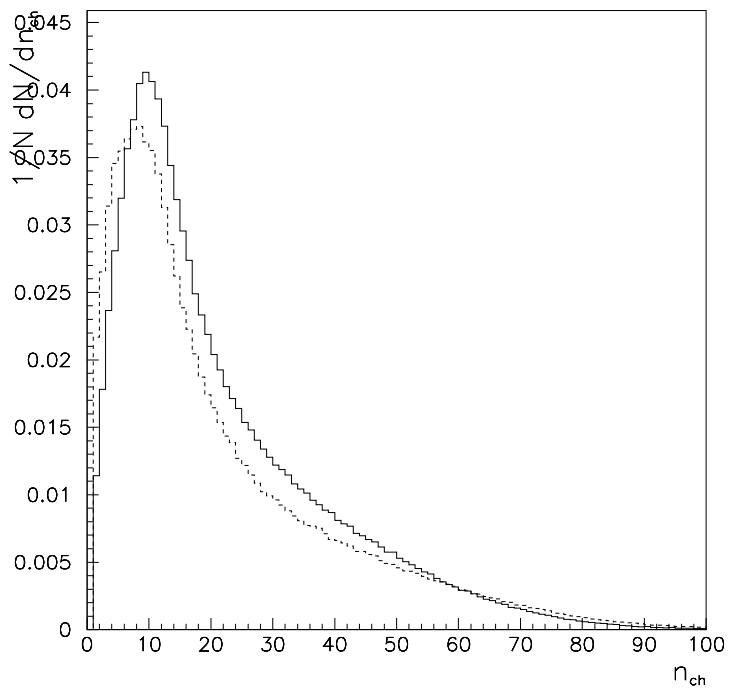

**Fig. 17:** Charged-stable multiplicity distribution in the LHCb acceptance.

## 5 Tuned models for the underlying event and minimum bias interactions

In this section we compare tuned MC generator models for the underlying event and minimum bias interactions. The aim of this study is to predict the event activity of minimum bias and the underlying event at the LHC. The models investigated correspond to tuned versions of PYTHIA, PHOJET and JIMMY.

### 5.1 Tuned models for the underlying event and minimum bias interactions

The starting point for the event generation in PYTHIA and JIMMY is the description of multiple hard interactions in the hadronic collision described in Section 2.1 (for PYTHIA 6.2), Section 2.2 for JIMMY and Section 2.4 for PHOJET.

**Table 1:** PYTHIA 6.214 default, ATLAS and CDF tune A parameters for minimum bias and the underlying event.

| **Default** [31] | **ATLAS** [37] | **CDF tune A** [6] | **Comments** |
|---|---|---|---|
| MSTP(51)=7 | MSTP(51)=7 | MSTP(51)=7 | CTEQ5L - selected p.d.f. |
| MSTP(81)=1 | MSTP(81)=1 | MSTP(81)=1 | multiple interactions |
| MSTP(82)=1 | MSTP(82)=4 | MSTP(82)=4 | complex scenario plus double Gaussian matter distribution |
| PARP(67)=1 | PARP(67)=1 | PARP(67)=4 | parameter regulating initial state radiation |
| PARP(82)=1.9 | PARP(82)=1.8 | PARP(82)=2.0 | $p_{t_{\min}}$ parameter |
| PARP(84)=0.2 | PARP(84)=0.5 | PARP(84)=0.4 | hadronic core radius (only for MSTP(82)=4) |
| PARP(85)=0.33 | PARP(85)=0.33 | PARP(85)=0.9 | probability for gluon production with colour connection to nearest neighbours |
| PARP(86)=0.66 | PARP(86)=0.66 | PARP(86)=0.95 | probability to produce gluons either either as in PARP(85) or as a closed gluon loop |
| PARP(89)=1.0 | PARP(89)=1.0 | PARP(89)=1.8 | energy scale (TeV) used to calculate $p_{t_{\min}}$ |
| PARP(90)=0.16 | PARP(90)=0.16 | PARP(90)=0.25 | power of the energy dependence of $p_{t_{\min}}$ |

PYTHIA and PHOJET have been shown to describe both minimum bias and underlying event data reasonably well when appropriately tuned [3, 6, 36, 37]. JIMMY is limited to the description of the underlying event; again, it has been shown capable of describing this rather well [38].

### 5.2 PYTHIA tunings

Several minimum bias and underlying event (UE) tunings for PYTHIA have been proposed in recent years. Ref. [37] describes how the current ATLAS tuning for PYTHIA was obtained after extensive comparisons to a variety of experimental measurements made at different colliding energies. Similar work has been done by the CDF Collaboration, although their PYTHIA tuning, CDF tune A [6], is primarily based on the description of the underlying event in jet events measured for $p\bar{p}$ at $\sqrt{s}$ = 1.8 TeV.

Table 1 displays the relevant parameters tuned to the data as proposed by the ATLAS [37] and CDF [6] collaborations. For the purpose of comparison, the corresponding default values [31] are also shown in the table.

### 5.3 PHOJET

The parameters used in PHOJET to describe minimum bias and the underlying event can be found in Ref. [22] and are currently set as default in PHOJET1.12, which is used in this study.

### 5.4 JIMMY tunings

We have tuned JIMMY to describe the UE as measured by CDF [19] and the resulting sets of parameters are shown in table 2. Figure 18 shows JIMMY predictions for the UE compared to CDF data for the average charged particle multiplicity (a) and the average $p_t$ sum in the underlying event (b). In Fig.18 we compare JIMMY - default parameters to “Tuning A” and “Tuning B”. Note that for the default parameters JIMMY does not give a correct description of the data. The other two distributions, generated with tuning A and B parameters, agree fairly well with the data.

In this study, JIMMY - tuning A and B will only be used to generate LHC predictions for the underlying event associated to jet events.

**Table 2:** JIMMY 4.1 default, tunings A and B parameters for the underlying event.

| Default | Tuning A | Tuning B | Comments |
|---|---|---|---|
| JMUEO=1 | JMUEO=0 | JMUEO=0 | multiparton interaction model |
| PTMIN=10.0 | PTMIN=3.0 | PTMIN=2.0 | minimum $p_T$ in hadronic jet production |
| PTJIM=3.0 | – | – | minimum $p_T$ of secondary scatters when JMUEO=1 or 2 |
| JMRAD(73)=0.71 | JMRAD(73)=2.13 | JMRAD(73)=0.71 | inverse proton radius squared |
| PRSOF=1.0 | PRSOF=0.0 | PRSOF=0.0 | probability of a soft underlying event |

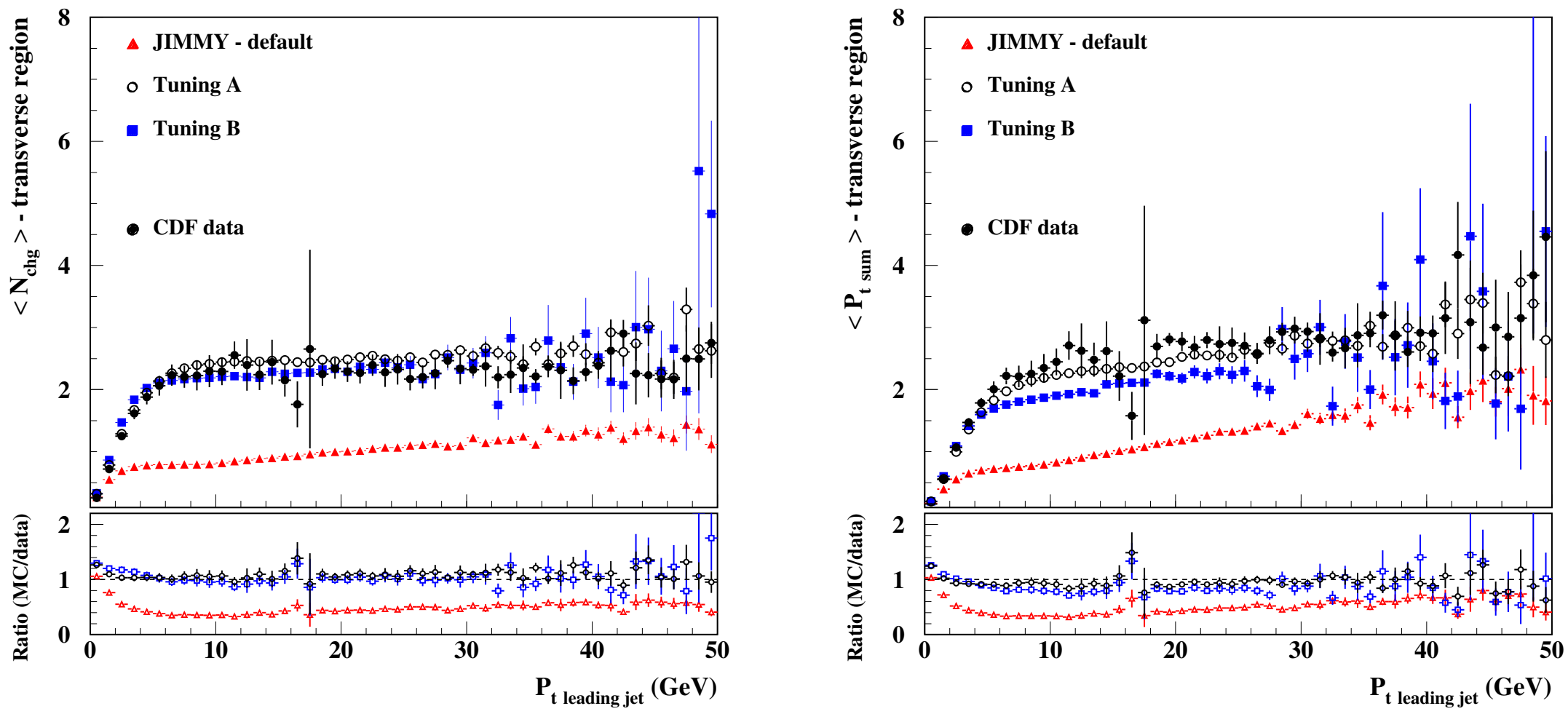

(a) (b)

**Fig. 18:** JIMMY predictions for the UE compared to CDF data. (a) Average charged particles multiplicity in the UE and (b) average $p_t$ sum in the UE.

### 5.5 Minimum bias interactions at the LHC

Throughout this report, minimum bias events will be associated with non-single diffractive inelastic interactions, following the experimental trend (see Ref. [37] and references therein).

For LHC collisions (pp collisions at $\sqrt{s} = 14$ TeV) the minimum bias cross-section estimated by PYTHIA 6.214, regardless of which tuning is used, is $\sigma_{nsd} = 65.7$ mb while PHOJET1.12 predicts $\sigma_{nsd} = 73.8$ mb, 12.3% greater than the former. Hence, for the same luminosity PHOJET1.12 generates more minimum bias pp collisions than PYTHIA 6.214 - tuned. We shall however, focus on the general properties per pp collision not weighted by cross-sections. The results per pp collision can later be easily scaled by the cross-section and luminosity.

Figure 19(a) shows charged particle density distributions in pseudorapidity for minimum bias pp collisions at $\sqrt{s}$ = 14 TeV generated by PHOJET1.12 and PYTHIA 6.214 - ATLAS and CDF tune A. The charged particle density generated by PHOJET1.12 and PYTHIA 6.214 - CDF tune A and ATLAS at $\eta = 0$ is 5.1, 5.3 and 6.8, respectively. Contrasting to the agreement seen in previous studies for $p\bar{p}$ collisions at $\sqrt{s}$ = 200 GeV, 546 GeV, 900 GeV and 1.8 TeV in Ref. [37], at the LHC PYTHIA 6.214 - ATLAS generates $\sim 25\%$ more charged particle density in the central region than PYTHIA 6.214 - CDF tune A and PHOJET1.12.

Compared to the charged particle density $dN_{ch}/d\eta$ measured by the CDF experiment at 1.8 TeV [39], PYTHIA 6.214 - ATLAS indicates a plateau rise of $\sim 70\%$ at the LHC in the central region while PHOJET1.12 and PYTHIA 6.214 - CDF tune A suggest a smaller rise of $\sim 35\%$.

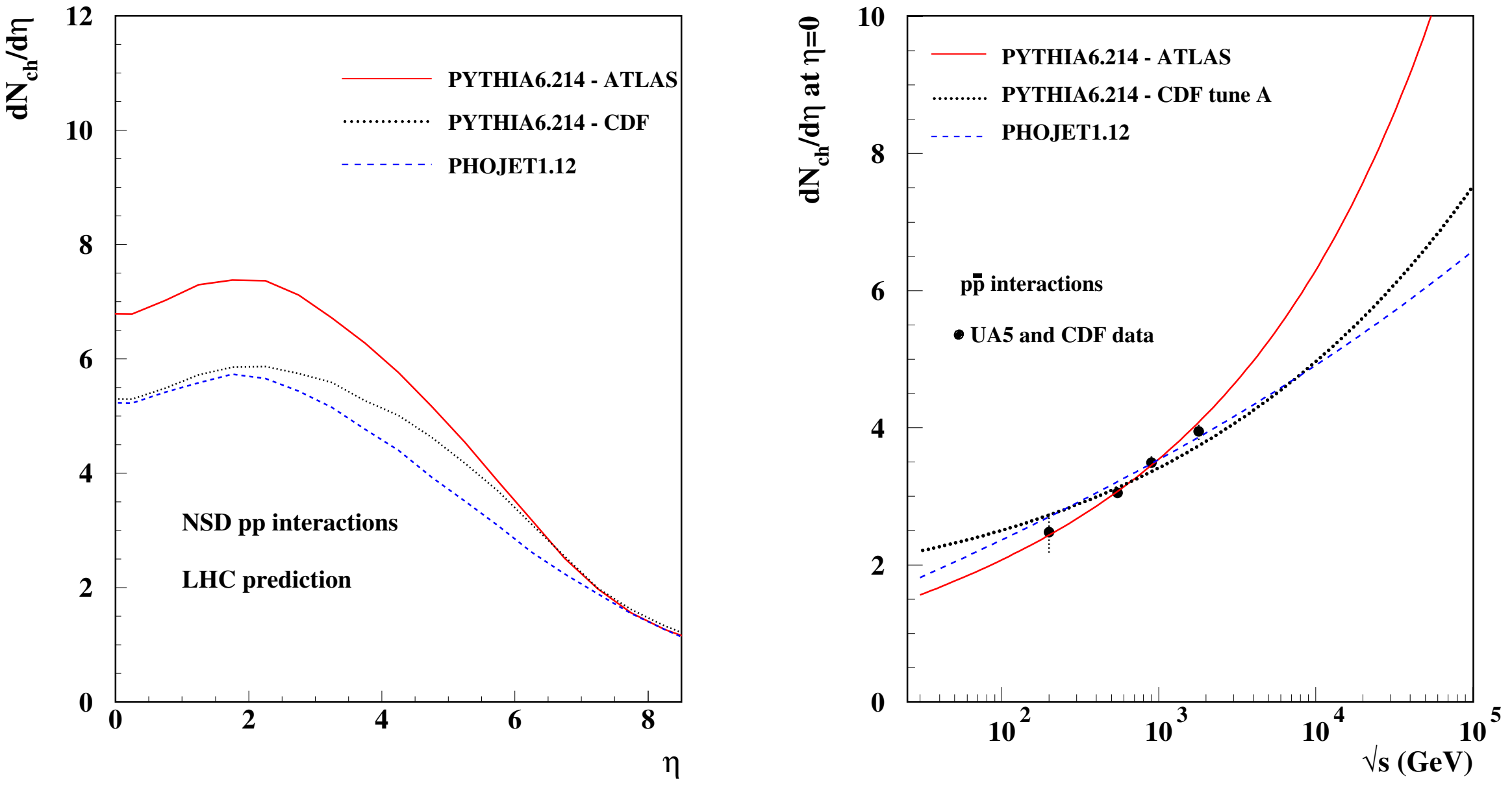

(a) (b)

**Fig. 19:** (a) Charged particle density distributions, d$N_{ch}$/d$\eta$, for NSD pp collisions at $\sqrt{s}$ = 14 TeV. (b) d$N_{ch}$/d$\eta$ at $\eta = 0$ for a wide range of $\sqrt{s}$. Predictions generated by PYTHIA 6.214, ATLAS and CDF tune A and PHOJET1.12.

Figure 19(b) displays d$N_{ch}$/d$\eta$ at $\eta = 0$ plotted as a function of $\sqrt{s}$. For centre-of-mass energies greater than $\sim$ 1 TeV, the multiparton interaction model employed by PYTHIA and the DPM used by PHOJET lead to multiplicity distributions with different rates of increase with the energy. PYTHIA suggests a rise dominated by the $\ln^2$(s) term while PHOJET predicts that the dominant term gives a ln(s) rise for d$N_{ch}$/d$\eta$ at $\eta = 0$. The ATLAS tuning for PYTHIA gives a steeper rise than CDF tune A and PHOJET (Fig. 19(b)) indicating a faster increase in the event activity at the partonic level in the ATLAS tuning when compared to CDF tune A and PHOJET. The average charged particle multiplicity in LHC minimum bias collisions, $< n_{ch} >$, is 69.6, 77.5 and 91.0 charged particles as predicted by PHOJET1.12, PYTHIA 6.214 - CDF tune A and ATLAS, respectively.

The $<$ $p_t$ $>$ at $\eta = 0$ for charged particles in LHC minimum bias collisions predicted by PHOJET1.12 and PYTHIA 6.214 - ATLAS and CDF tune A models is 0.64 GeV, 0.67 GeV and 0.55 GeV, respectively. Generating less particles in an average minimum bias collision at the LHC, PHOJET1.12 predicts that the average $p_t$ per particle at $\eta = 0$ is greater (or harder) than the corresponding prediction from PYTHIA 6.214 - ATLAS. However, amongst the three models, PYTHIA 6.214 - CDF tune A gives the hardest $<$ $p_t$ $>$ at $\eta = 0$. The main reason for this is the increased contribution of harder parton showers used to make the model agree with the $p_t$ spectrum of particles in the UE, and obtained by setting PARP(67)=4 [6].

## 5.6 The underlying event

Based on CDF measurements, we shall use their definition for the UE, i.e., the angular region in $\phi$ which is transverse to the leading charged particle jet as described in Section 3 and shown in Fig. 6. Figure 20(a) displays PYTHIA 6.214 — ATLAS and CDF tune A, and PHOJET1.12 predictions for the average particle multiplicity in the UE for pp collisions at the LHC (charged particles with $p_T > 0.5$ GeV and $|\eta| < 1$). The distributions generated by the three models are fundamentally different. Except for events with $p_{t_{\rm ljet}} \lesssim 3$ GeV, PYTHIA 6.214 — ATLAS generates greater multiplicity in the UE than the other models shown in Fig. 20(a).

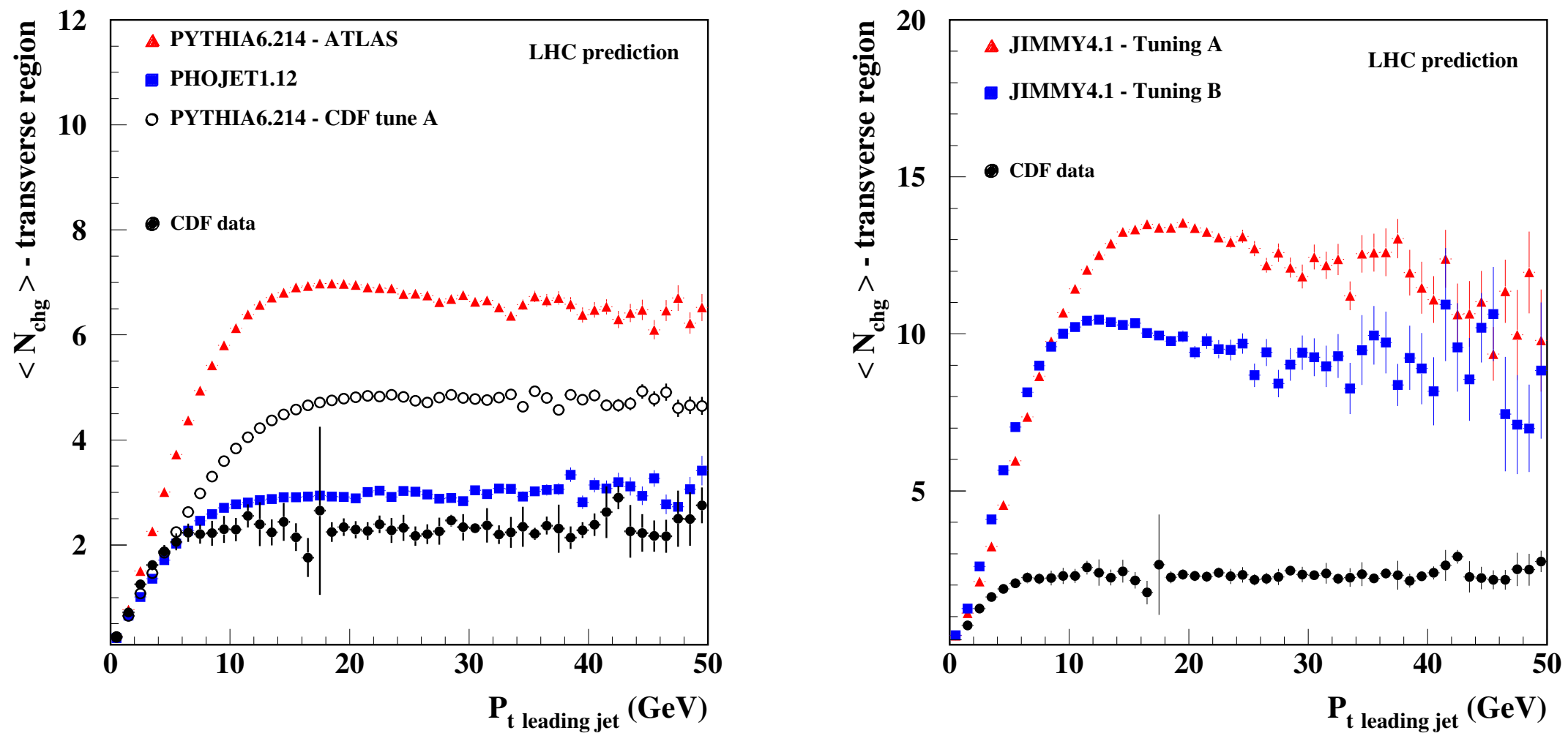

(a) (b)

**Fig. 20:** (a) PYTHIA 6.214 (ATLAS and CDF tune A), PHOJET1.12 and (b) JIMMY 4.1 (tunings A and B) predictions for the average multiplicity in the UE for LHC pp collisions.

A close inspection of predictions for the UE given in Fig. 20(a), shows that the average multiplicity in the UE for $P_{t_{\rm ljet}} > 10$ GeV reaches a plateau at $\sim 6.5$ charged particles according to PYTHIA 6.214 - ATLAS, $\sim 5$ for CDF tune A and $\sim 3.0$ according to PHOJET1.12. Compared to the underlying event distributions measured by CDF at 1.8 TeV, PYTHIA 6.214 - ATLAS indicates a plateau rise of $\sim 200\%$ at the LHC while PYTHIA 6.214 - CDF tune A predicts a rise of $\sim 100\%$ and PHOJET1.12 suggests a much smaller rise of $\sim 40\%$.

In Fig. 20(b) we show JIMMY 4.1 - Tuning A and B predictions for the average particle multiplicity in the UE for LHC collisions. The average multiplicity in the UE for $P_{t_{\rm ljet}} > 10$ GeV reaches a plateau at $\sim 12$ charged particles according to JIMMY 4.1 - Tuning A, and $\sim 9.0$ according to JIMMY 4.1 - Tuning B. Note that, for both JIMMY tunings, the plateau rise for the average multiplicity in the UE is much greater than the ones predicted by any of the PYTHIA tunings or by PHOJET as shown in Figs. 20(a) and (b). Once again, compared to the underlying event distributions measured by CDF at 1.8 TeV, JIMMY 4.1 - Tuning A indicates a five-fold plateau rise at the LHC while JIMMY 4.1 - Tuning B - CDF suggests a four-fold rise.

### 5.7 Conclusion

The minimum bias and underlying event predictions for the LHC generated by models which have been tuned to the available data have been compared. In previous studies, these models have been shown to be able to describe the data distributions for these two classes of interactions. However, in this article, it has been shown that for the models detailed in tables 1 and 2, there can be dramatic disagreements in their predictions at LHC energies. This is especially evident in the distributions for the average multiplicity in the UE (Fig. 20) where, for example, PHOJET1.12 predicts that the distribution's plateau will be at $\sim 3$ charged particles while JIMMY 4.1 - Tuning A predicts for the same distribution, a plateau at $\sim 12$.

Even though models tuned to the data have been used in this study, uncertainties in LHC predictions for minimum bias and the underlying event are still considerable. Improved models for the soft component of hadronic collisions are needed as well as more experimental information which may be used to tune current models. Future studies should focus on tuning the energy dependence for the event activity in both minimum bias and the underlying event, which at the moment seems to be one of the least understood aspects of all the models investigated in this study.

## 6 Can the final state at LHC be determined from ep data at HERA?

### 6.1 Jets and $E_\perp$-flow

A phenomenological fit for a soft-cutoff, $\hat{p}_T^{\min}$, and an extrapolation to LHC energies, was discussed in sections 4.1 and 5.2. However, in the $k_\perp$-factorization formalism the soft divergence is avoided, and it is possible to predict minijets and $E_\perp$-flow from HERA data alone. Thus it is not necessary to rely on a purely phenomenological fit using $\mathrm{p\bar{p}}$ collision data. This gives a better dynamical insight, and avoids the uncertainties associated with the extrapolation to higher energies.

High $p_\perp$ jets are well described by conventional *collinear factorization*, but in this formalism the minijet cross section diverges, $\sigma_{jet} \propto 1/p_\perp^4$. This implies that the total $E_\perp$ also diverges, and therefore a cutoff $\hat{p}_T^{\min}$ is needed. Fits to data give $\hat{p}_T^{\min} \sim 2$ GeV growing with energy [8,9]. There is no theoretical basis for the extrapolation of $\hat{p}_T^{\min}$ from the Tevatron to LHC, which induces an element of uncertainty in the predictions for LHC.

In the $k_\perp$*-factorization* formalism the off shell matrix element for the hard subcollision $k_1 + k_2 \to q_1 + q_2$ does not blow up, when the momentum exchange $k_\perp^2$ is smaller than the incoming virtualities $k_{\perp 1}^2$ and $k_{\perp 2}^2$. The unintegrated structure functions $\mathcal{F}(x, k_\perp^2, Q^2)$ are also suppressed for small $k_\perp$, and as a result the total $E_\perp$ is not divergent but stays finite. An "effective cutoff" increases with energy, but the increase is less steep for larger energies [40].

At high energy $\sigma_{\rm jet}$ is larger than $\sigma_{\rm tot}$, which implies that there usually are *multiple hard subcollisions* in a single event. The experimental evidence for multiple collisions has been discussed in previous sections. It includes multijet events, forward-backward correlations, the pedestal effect, and associated particles in jet events. The data also indicate that the hard subcollisions are not independent. Central collisions contain more, and peripheral collisions fewer, minijets, and the results are well described by a double Gaussian distribution in impact parameter, as suggested in ref. [3].

At high energies the pdfs needed to calculate the minijet cross section have to be evaluated in the BFKL domain of small $x$ and low $k_\perp$. This implies that non-$k_\perp$-ordered parton chains are important. For a $\gamma^*$p collision a single local $k_\perp$-maximum corresponds to a resolved photon interaction. Similarly several local maxima in a single chain correspond to correlated hard subcollisions.

In the BFKL formalism the gluon links in the $t$-channel correspond to reggeized gluons, which means that soft emissions are compensated by virtual corrections. These soft emissions *do not* contribute to the parton distributions or total cross sections, but they *do* contribute to the properties of final states, and should then be added with Sudakov form factors. The CCFM model [41, 42] interpolates between DGLAP and BFKL. Here some soft emissions are included in the initial state radiation, which implies that they must be suppressed by non-eikonal form factors. The Linked Dipole Chain (LDC) model [43] is a reformulation and generalization of CCFM, in which more emissions are treated as final state emissions, in closer agreement with the BFKL picture. In the LDC formalism the chain formed by the initial state radiation is *fully symmetric* with respect to the photon end and the proton end of the ladder. This symmetry implies that the formalism is also directly applicable to hadron-hadron collisions. Thus a fit to DIS data will also give the cross section for a *parton chain* in pp collisions [44].

A potential problem is due to the fact that with a running $\alpha_s$, the enhancement of small $k_\perp$ implies that the result depends on a necessary cutoff $Q_0$. Good fits to DIS data are possible with different $Q_0$, if the input distribution $f_0(x, Q_0^2)$ is adjusted accordingly. However, although a larger cutoff gives fewer hard chains, it also implies a larger number of soft chains, in which no link has a $k_\perp$ larger than $Q_0$. Thus the total number of chains in pp scattering is independent of $Q_0$, and therefore well determined by the fit to DIS data.

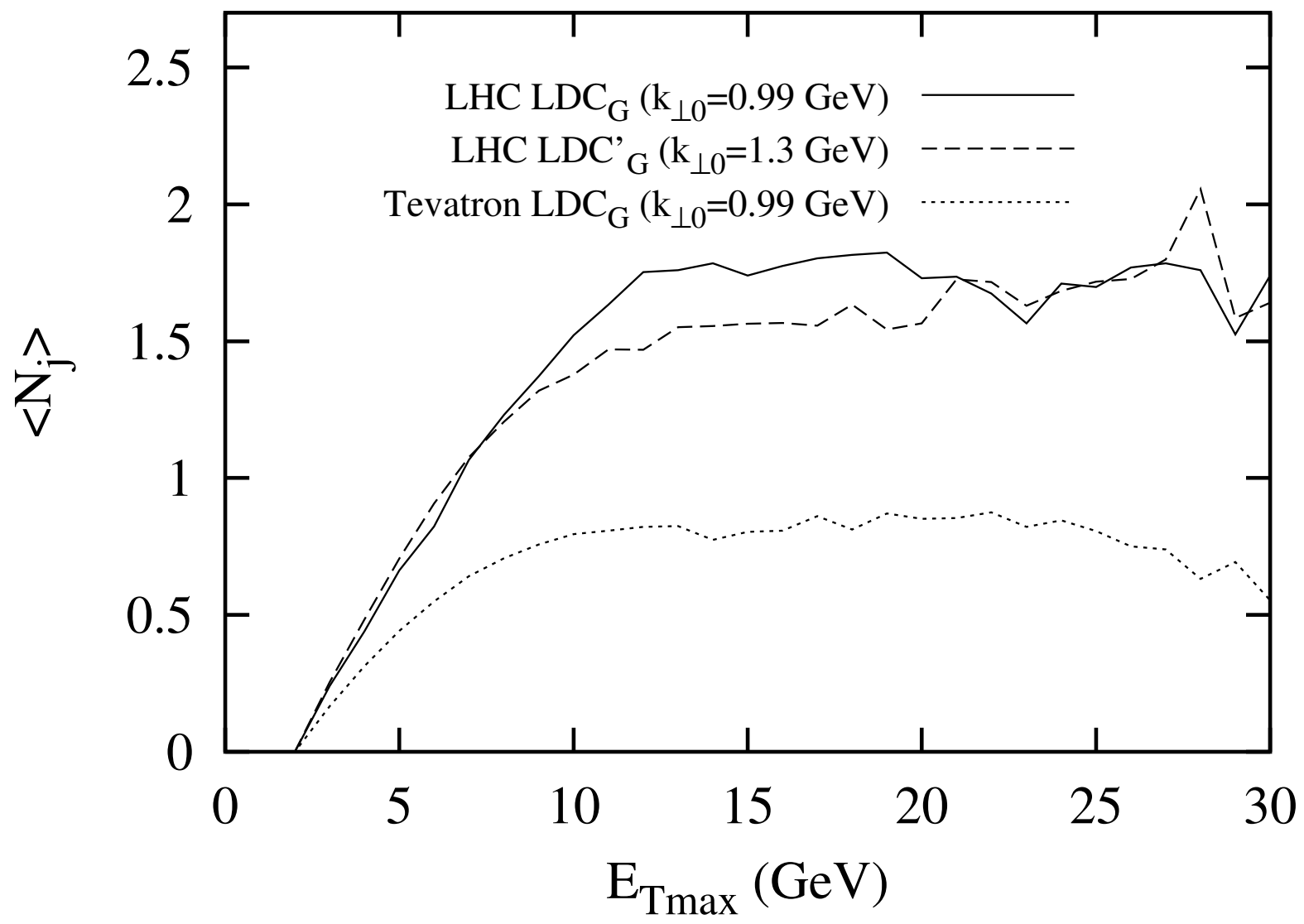

**Fig. 21:** The average number of minijets per event in the "minimum azimuth region", as a function of transverse energy of the trigger jet, $E_{\perp max}$. The figure shows the result for 1.8 TeV and for LHC. The two LHC curves correspond to different values for $Q_0$, showing the stability with respect to the soft cutoff.

When the fit to HERA data in this way is applied to p$\bar{\text{p}}$ scattering at the Tevatron, the predictions for *e.g.* jet multiplicity and the pedestal effect are very close to CDF's tune A, described in Section 3. The result is insensitive to the soft cutoff $Q_0$, which implies that the extrapolation to LHC energies is stable, and does not depend on an uncertain extrapolation of the low-$p_\perp$ cutoff needed in a collinear formalism. As an example fig. 21 shows a prediction for the average number of minijets per event within $60^\circ$ in azimuth perpendicular to a trigger jet, on the side with minimum activity.

As the LDC model is fully symmetric with respect to an interchange of the projectile and the target, the parton chains have to combine at one end at the same rate as they multiply at the other. Therefore the formalism should be suitable for studies of gluon recombination and *saturation*. This work is in progress, and some preliminary results from combining the LDC model with Mueller's dipole formulation in transverse coordinate space [45–47] are presented in ref. [48].

### 6.2 Hadron multiplicities

The hadron multiplicity is much more sensitive to non-perturbative effects. This implies larger uncertainties, and models differ by factors 3-4 in their predictions for LHC (see Section 5). The CDF data also show that the data are best fitted if colours rearrange so that secondary hard scatterings give minimum extra string length, i.e. minimum extra multiplicity. This is very different from the case in $e^+e^-$ annihilation.

In pp collisions the multiplicity of final state hadrons depends very sensitively on the colour connections between the produced partons. This implies that the result depends on soft non-perturbative effects. Multiple interactions are related to multiple pomeron exchange, which is expected to obey the Abramovskyĭ-Gribov-Kancheli cutting rules [49]. These rules are derived for a multiperipheral model, but a multiperipheral chain has important similarities with a gluonic chain. An essential feature is the dominance of small momentum exchanges at each vertex. The colour structure of QCD gives, however, some extra complications as discussed by J. Bartels (see the contribution by Bartels to working group 4).

The pomeron is identified by two gluon exchange, and multiple chains correspond to multi-pomeron exchange. For the example of two pomeron exchange, the AGK rules give the relative weights $1 : -4 : 2$ for cutting 0, 1 or 2 pomerons. These ratios imply that the two-pomeron diagram contributes to the multiplicity *fluctuations*, but has no effect on the number of produced particles, determined by $\sum n\sigma_n$. This result can also be generalized to the exchange of more pomerons.

Similar cutting rules apply to a diagram with two pomerons attached to one proton and one pomeron to the other, connected by a central triple-pomeron coupling. In ref. [49] this and similar diagrams are, however, expected to give smaller contributions.

A hard gg $\rightarrow$ gg subcollision will imply that the two proton remnants carry colour octet charges. This is expected to give two colour triplet strings, or two cluster chains, connecting the two remnants and the two final state gluons. In the string model the strings are stretched between the remnants, with the gluons acting as kinks on the strings. These kinks can either be on different strings or both on the first or both on the second string, with equal probabilities for the three possibilities (see ref. [50]). Including initial state radiation will give extra kinks, which due to colour coherence will be connected so as to result in minimal extra string length.

Multiple collisions with two independent gg $\rightarrow$ gg scatterings would be expected to correspond to two cut pomerons, with four triplet strings stretched between the proton remnants. This would give approximately a doubled multiplicity, in accordance with the AGK cutting rules. However, the CDF data show that this is *far from reality*.

CDF's successful tune A [6] is a fit using an early PYTHIA version. Already in the analysis in ref. [3] it was realized that four strings would give too high multiplicity. Therefore in this early PYTHIA version there are three possible string connections for a secondary hard subcollision. 1) An extra closed string loop between the two final state gluons. 2) A single string between the scattered partons, which are then treated as a $q\bar{q}$ system. 3) The new hard gluons are inserted as extra kinks among the initial state radiations, in a way which corresponds to minimum extra string length. In the successful tune A the last possibility is chosen in 90% of the cases, which corresponds to *minimal extra multiplicity*. The default PYTHIA tune, which contained equal probabilities for the three cases, does not give a good fit. A more advanced treatment of pp collisions [8, 9] is implemented in a new PYTHIA version (6.3) [2] (see Section 2.1). This model does, however, not work as well as Field's tune A of the older model.

Consequently two independent hard collisions do not correspond to two cut pomeron ladders stretched between the proton remnants. It also does not correspond to a cut pomeron loop in the centre. Instead it looks like a single ladder, with a higher density of gluon rungs in the central region.

How can this be understood? It raises a set of important questions: What does it imply for the AGK rules and the diffractive gap survival probability? Do rescattering and unitarity constraints (and AGK) work in the initial perturbative phase? If so, does this correspond to an initial hard collision inside a confining bag, with the final state partons colour connected in a later non-perturbative phase?

We can compare with the situation in $e^+e^-$-annihilation. If two gluons are emitted from the quark or antiquark legs, these gluons form a colour singlet with probability $\sim 1/N_c^2$. They could then hadronize as a separate system. Analyses of data from LEP indicate that such isolated systems are suppressed even more than by a factor $\sim 1/N_c^2$.

In conclusion we have following important questions:

- Why do the strings make the shortest connections in $\approx 100\%$ in pp and almost never in $e^+e^-$?
- How do multiplicity fluctuations and the relation diffraction *diffraction* and *high multiplicity events* reflect features of AGK in ep, $\gamma$p, and pp?
- Do unitarity effects and AGK cutting rules work as expected in an initial perturbative phase, and the colours recombine in a subsequent nonperturbative soft phase?
- Or is the pomeron a much more complicated phenomenon than the simple ladder envisaged by Abramovskyǐ-Gribov-Kancheli?

## 7 Conclusions and the potential for HERA data

This was a very active area of discussion during the workshop. In fact, the area remains so active that firm conclusions are hard to make, and likely to be superceded on a very short timescale. Nevertheless there are some things which do seem clear.

- The underlying event is clearly an topic of substantial importance for the LHC.
- The dominant input data for understanding the underlying event comes at present from the Tevatron, with HERA data primarily featuring indirectly, though importantly, via the parton densities.
- The data strongly indicate that multiple hard scatters are required to adequately describe the final state in high energy hadron collisions.
- The UE depends on the measurement being made as demonstrated by difference between the UE in the CDF leading jet and back-to-back jet analysis.
- The colour structure of the final parton state is an unsolved problem. The CDF data indicate that 'short strings' are strongly favoured.
- There are large uncertainties associated with extrapolating the available models to LHC energies.

As far as the future impact of HERA data on this area goes, some ideas have been discussed in the previous section. In addition, it is worth noting that most of the models discussed here have also been used in high energy photoproduction at HERA [51], where they also improve the description of the data. No study comparable to those carried out at $\mathrm{pp}$ or $\mathrm{p\bar{p}}$ experiments is currently available. The benefits of such a study would be that (a) HERA could add another series of points in energy (around 200 GeV) to help pin down the energy dependence of the underlying event, (b) it is possible to select regions of phase space where resolved (i.e., hadronic) or direct (i.e., pointlike) photons dominate, thus effectively switching on or off the photon PDF (and thus presumably multiparton interactions) and allowing comparison between the two cases, (c) the photon is a new particle with which the physics assumptions of underlying event models can be confronted. The last of these points however also implies that a slew of new parameters will be introduced, and one may learn more about the photon this way than about underlying events themselves. Either way, it is to be hoped that such a study will be carried out before HERA finishes and LHC switches on.

# Forward Jets and Multiple Interactions

*Jacek Turnau*[1], *Leif Lönnblad*[2]
[1]Institute of Nuclear Physics, Polish Academy of Sciences, Cracow, Poland;
[2]Department of Theoretical Physics, Lund University, Sweden.

**Abstract**
HERA provides a unique possibility to investigate the dependence of multiple interactions on transverse interaction sizes through variation of the photon virtuality $Q^2$. In order to observe effects of multiple interactions at $Q^2$ substantially different from zero we have to look into regions of phase space where resolved processes dominate over direct ones. The forward jet production at small values of Bjorken $x$ is one example. PYTHIA and RAPGAP have been employed to estimate contribution of the multiple interactions to forward jet production cross section.

Comparisons of HERA photoproduction data with QCD NLO calculations for high transverse momentum jets revealed that the observed jets are not well described by the calculations. The energy flow adjacent to jets - the underlying event or jet pedestal - was found to be far above QCD expectations [1]. Similar excess of underlying energy was observed in $p\bar{p}$ data, see [2] and [3] for recent studies. It appears that both HERA and TEVATRON data can be described by adding beam remnant interactions, from soft to hard, as first proposed in ref. [4]. The remnant beam-beam interactions can result in multiple hard parton interactions (MI) thus creating additional pairs of jets. Therfore the presence of four high transverse momentum objects in the hadronic final state (e.g. four jets or prompt photon and three jets) allows searches for signatures of multi-parton interactions in a region of phase space where their effects may be maximized. The evidence of MI coming from 4-jet studies is more explicit and is not complicated by initial/final state radiation and soft beam-remnant components of the underlying event. Both ZEUS [5] and CDF [6] observed explicite double parton interactions in rough agreement with PYTHIA [4, 7] simulations.

The very interesting aspect of measurements at HERA is that variation of the photon virtuality $Q^2$ provides information about transverse interaction sizes. Observation of the dependence of MI on $Q^2$ could be important from the phenomenological point of view. In order to see MI at photon virtuality substantially different from zero we have to look into regions of phase space for deep inelastic scattering where the resolved virtual photon processes dominate over direct ones. The forward jet production at small values of Bjorken $x$ is one example. Here one could expect that additional interactions between the remnants of the proton and resolved virtual photon would produce extra hadron multiplicity in an underlying event. Although the transverse momentum of these hadrons would be limited, they could still give a substantial effect on the rate of forward jets which have a steeply falling $p_\perp$ spectrum.

The forward jet cross-section is especially interesting since it has been notoriously difficult to reproduce by standard DGLAP-based parton shower event generators. It has been shown that the description of the forward jet cross section can be improved by adding resolved virtual photon component in eg. the RAPGAP Monte Carlo [8], but the jet rates produced in the simulations are still a bit too low in the small-$x$ region. In order to check if MI can give measurable contribution to this process we have performed a study in which we estimate MI effect using both PYTHIA 6.2 and RAPGAP 3.1. We use PYTHIA since the MI model there has been shown to be able to give a good description of underlying events and jet pedestal effects in hadron-hadron collisions and in photoproduction, and it is fairly easy to apply the same model to the resolved part of the $\gamma^\star - p$ collisions. However, PYTHIA does not describe correctly the transverse energy flow in in DIS at HERA above $Q^2 \approx 5\ \mathrm{GeV}^2$. We can still use PYTHIA to estimate the relative effect of MI and we have generated forward jet cross section with H1 cuts [9]:

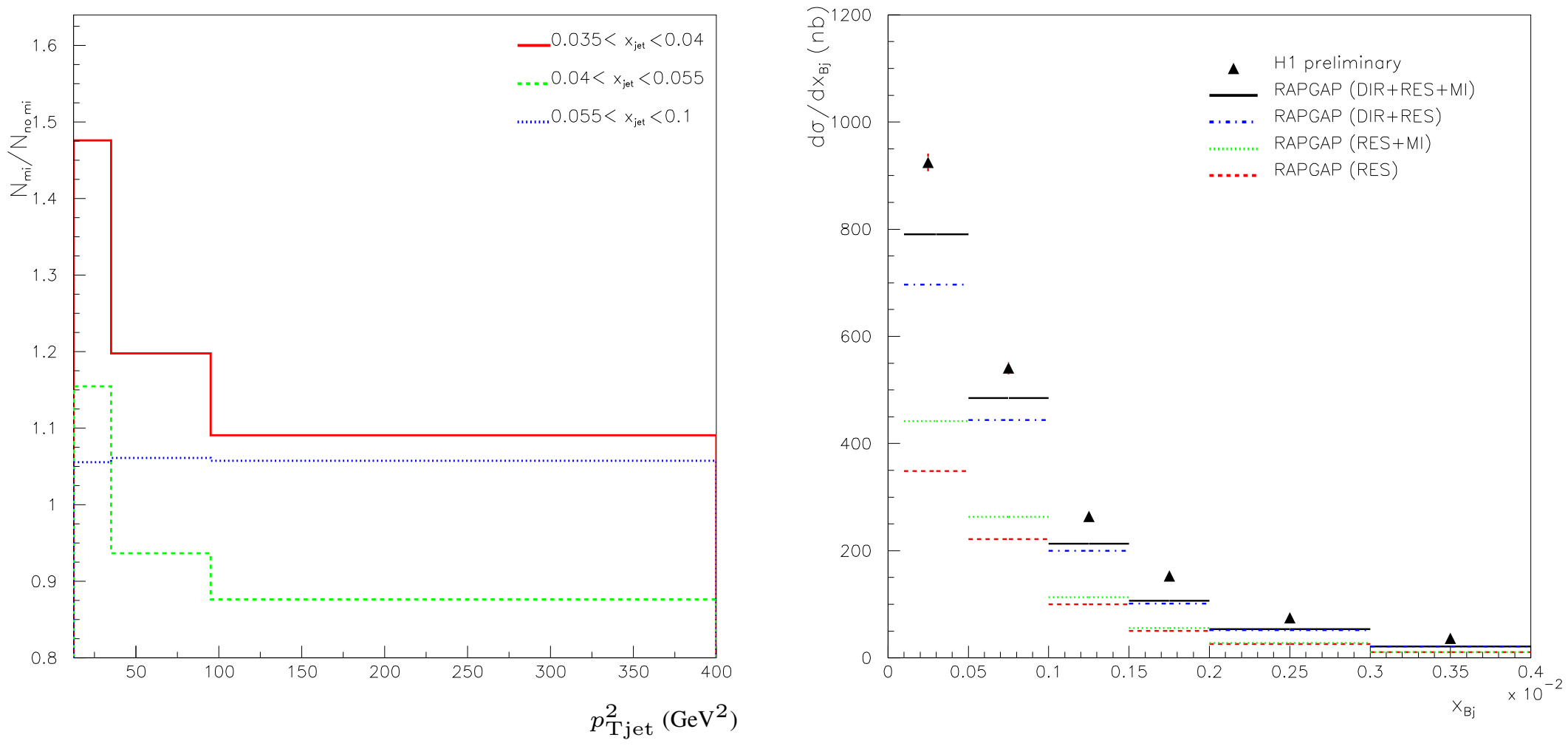

**Fig. 1: Left:** Ratio of forward jets with and without multiple interactions as a function of jet transverse momentum squared for three regions of proton momentum fraction carried by jet **Right:** The H1 forward jet cross section data compared with RAPGAP 3.1 simulation. Multiple interactions are included as $x, Q^2, x_{jet}$ and $p^2_{Tjet}$ dependent weights to resolved component, calculated using PYTHIA 6.2

($p_{Tjet} > 3.5$ GeV,$x_{jet} > 0.035$, $20° > \Theta_{jet} > 7°$ and $0.5 < p^2_{Tjet}/Q^2 < 5$) using PYTHIA 6.2 with default settings in $\gamma p$ mode (MI in mode 2) with $\gamma^\star$ momentum corresponding to several values of $x$ and $Q^2$ within DIS kinematical phase space $0.0001 < x < 0.004$ and $5 < Q^2 < 85$ GeV$^2$.

In Fig. 1 (left) we show example of the ratio of number of the forward jets with and without MI, here for $x = 0.0004$ and $Q^2 = 8$ GeV$^2$, as a function of $p^2_{Tjet}$. It can be seen that effect of MI is quite substantial in the lowest $p^2_{Tjet}$ bin. Treating the above mentioned ratios as weights depending on $x$, $Q^2$, $x_{jet}$ and $p^2_{Tjet}$, we have generated inclusive forward jet cross section using RAPGAP 3.1 within above mentioned H1 cuts. The Fig. 1 (right) shows the result of this calculation. The inclusive forward jet cross section is enhanced by MI for about 15% in the lowest $x$ bin, in fact improving description of the data. The effect of MI diminishes quickly with increasing $x$ as result of decreasing contribution of the resolved photon component.

This very preliminary study suggests that $Q^2$ dependence of multiple interactions can be studied at HERA. This will require large statistics and an improved understanding of the underlying QCD evolution in forward jet production.

# Survival Probability of Large Rapidity Gaps

*E. Gotsman, E. Levin, U. Maor, E. Naftali, A. Prygarin*
HEP Department, School of Physics and Astronomy, Raymond and Beverly Sackler Faculty of Exact Science, Tel Aviv University, Tel Aviv, 69978, ISRAEL.

**Abstract**
Our presentation centers on the consequences of s-channel unitarity, manifested by soft re-scatterings of the spectator partons in a high energy diffractive process, focusing on the calculations of gap survival probabilities. Our emphasis is on recent estimates relevant to exclusive diffractive Higgs production at the LHC. To this end, we critically re-examine the comparison of the theoretical estimates of large rapidity gap hard di-jets with the measured data, and remark on the difficulties in the interpretation of HERA hard di-jet photoproduction.

## 1 Introduction

A large rapidity gap (LRG) in an hadronic, photo or DIS induced final state is experimentally defined as a large gap in the $\eta - \phi$ lego plot devoid of produced hadrons. LRG events were suggested [1–4] as a signature for Higgs production due to a virtual $W - W$ fusion subprocess. An analogous pQCD process, in which a colorless exchange (”hard Pomeron”) replaces the virtual W, has a considerably larger discovery potential as it leads also to an exclusive $p + H + p$ final state. Assuming the Higgs mass to be in the range of $100 - 150\,GeV$, the calculated rates for this channel, utilizing proton tagging are promissing. Indeed, LRG hard di-jets, produced via the same production mechanism, have been observed in the Tevatron [5–17] and HERA [18–29]. The experimental LRG di-jets production rates are much smaller than the pQCD (or Regge) estimates. Following Bjorken [3, 4], the correcting damping factor is called ”LRG survival probability”.

The present summary aims to review and check calculations of the survival probability as applied to the HERA-Tevatron data and explore the consequences for diffractive LRG channels at LHC with a focus on diffractive Higgs production.

We distinguish between three configurations of di-jets (for details see Ref. [13–17]):

1) A LRG separates the di-jets system from the other non diffractive final state particles. On the partonic level this is a single diffraction (SD) Pomeron exchange process denoted GJJ.
2) A LRG separates between the two hard jets. This is a double diffraction (DD) denoted JGJ.
3) Centrally produced di-jets are separated by a LRG on each side of the system. This is a central diffraction (CD) two Pomeron exchange process denoted GJJG. This mechanism also leads to diffractive exclusive Higgs production.

We denote the theoretically calculated rate of a LRG channel by $F_{gap}$. It was noted by Bjorken [3, 4] that we have to distinguish between the theoretically calculated rate and the actual measured rate $f_{gap}$

$$f_{gap} = \langle \mid S \mid^2 \rangle \cdot F_{gap}. \tag{1}$$

The proportionality damping factor [30–33] is the survival probability of a LRG. It is the probability of a given LRG not to be filled by debris (partons and/or hadrons). These debris originate from the soft re-scattering of the spectator partons resulting in a survival probability denoted $\mid S_{spec}(s) \mid^2$, and/or from the gluon radiation emitted by partons taking part in the hard interaction with a corresponding survival probability denoted $\mid S_{brem}(\Delta y) \mid^2$,

$$\langle \mid S(s, \Delta y) \mid^2 \rangle = \langle \mid S_{spec}(s) \mid^2 \rangle \cdot \langle \mid S_{brem}(\Delta y) \mid^2 \rangle. \tag{2}$$

$s$ is the c.m. energy square of the colliding particles and $\Delta y$ is the large rapidity gap. Gluon radiation from the interacting partons is strongly suppressed by the Sudakov factor [34]. However, since this suppression is included in the perturbative calculation (see **4.3**) we can neglect $\langle \mid S_{brem}(\Delta y) \mid^2 \rangle$ in our calculations. In the following we denote $\langle \mid S_{spec} \mid^2 \rangle = S^2$. It is best defined in impact parameter space (see **2.1**)). Following Bjorken [3, 4], the survival probability is determined as the normalized integrated product of two quantities

$$S^2 = \frac{\int d^2b \mid M^H(s,b) \mid^2 P^S(s,b)}{\int d^2b \mid M^H(s,b) \mid^2}. \tag{3}$$

$M^H(s,b)$ is the amplitude for the LRG diffractive process (soft or hard) of interest. $P^S(s,b)$ is the probability that no inelastic soft interaction in the re-scattering eikonal chain results in inelasticity of the final state at $(s,b)$.

The organization of this paper is as follows: In Sec.2 we briefly review the role of s-channel unitarity in high energy soft scattering and the eikonal model. The GLM model [30–33] and its consequent survival probabilities [35–37] are presented in Sec.3, including a generalization to a multi channel re-scattering model [38, 39]. The KKMR model [40–44] and its survival probabilities is presented in Sec.4. A discussion and our conclusions are presented in Sec.5. An added short presentation on Monte Carlo calculations of $S^2$ is given in an Appendix.

## 2 Unitarity

Even though soft high energy scattering has been extensively studied experimentally over the last 50 years, we do not have, as yet, a satisfactory QCD framework to calculate even the gross features of this impressive data base. This is just a reflection of our inability to execute QCD calculations in the non-perturbative regime. High energy soft scattering is, thus, commonly described by the Regge-pole model [45, 46]. The theory, motivated by S matrix approach, was introduced more than 40 years ago and was soon after followed by a very rich phenomenology.

The key ingredient of the Regge pole model is the leading Pomeron, whose linear $t$-dependent trajectory is given by

$$\alpha_{I\!P}(t) = \alpha_{I\!P}(0) + \alpha'_{I\!P} t. \tag{4}$$

A knowledge of $\alpha_{I\!P}(t)$ enables a calculation of $\sigma_{tot}$, $\sigma_{el}$ and $\frac{d\sigma_{el}}{dt}$, whose forward elastic exponential slope is given by

$$B_{el} = 2B_0 + 2\alpha'_{I\!P} ln\left(\frac{s}{s_0}\right). \tag{5}$$

Donnachie and Landshoff (DL) have vigorously promoted [47, 48] an appealing and very simple Regge parametrization for total and forward differential elastic hadron-hadron cross sections in which they offer a global fit to all available hadron-hadron and photon-hadron total and elastic cross section data. This data, above $P_L = 10\,GeV$, is excellently fitted with universal parameters. We shall be interested only in the DL Pomeron with an intercept $\alpha_{I\!P}(0) = 1 + \epsilon$, where $\epsilon = 0.0808$, which accounts for the high energy growing cross sections. Its fitted [49] slope value is $\alpha'_{I\!P} = 0.25\,GeV^{-2}$.

### 2.1 S-channel unitarity

The simple DL parametrization is bound to violate s-channel unitarity at some energy since $\sigma_{el}$ grows with energy as $s^{2\epsilon}$, modulu logarithmic corrections, while $\sigma_{tot}$ grows only as $s^{\epsilon}$. The theoretical problems at stake are easily identified in an impact b-space representation.

The elastic scattering amplitude is normalized so that

$$\frac{d\sigma_{el}}{dt} = \pi \mid f_{el}(s,t) \mid^2, \tag{6}$$

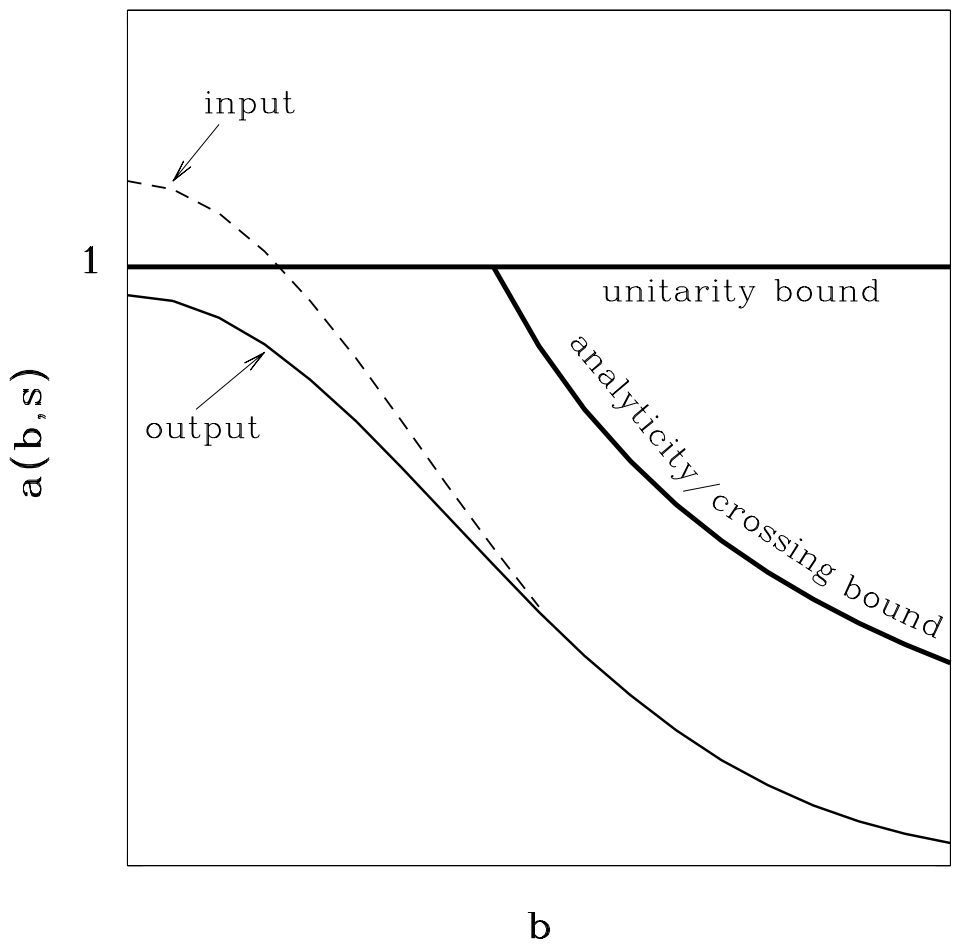

**Fig. 1:** A pictorial illustration of a high energy b-space elastic amplitude bounded by unitarity and analyticity/crossing. In the illustration we have an input amplitude which violates the eikonal unitarity bound and an output amplitude obtained after a unitarization procedure.

$$\sigma_{tot} \; = \; 4\pi Im f_{el}(s,0). \tag{7}$$

The elastic amplitude in b-space is defined as

$$a_{el}(s,b) \; = \; \frac{1}{2\pi}\int d\mathbf{q} e^{-i\mathbf{q}\cdot\mathbf{b}} f_{el}(s,t), \tag{8}$$

where $t = -\mathbf{q}^2$. In this representation

$$\sigma_{tot} \; = \; 2\int d^2b \, Im[a_{el}(s,b)], \tag{9}$$

$$\sigma_{el} \; = \; \int d^2b \mid a_{el}(s,b) \mid^2, \tag{10}$$

$$\sigma_{in} \; = \; \sigma_{tot} \; - \; \sigma_{el}. \tag{11}$$

As noted, a simple Regge pole with $\alpha_{I\!P}(0) > 1$ will eventually violate s-channel unitarity. The question is if this is a future problem to be confronted only at far higher energies than presently available, or is it a phenomena which can be identified through experimental signatures observed within the available high energy data base. It is an easy exercise to check that the DL model [47,48], with its fitted global parameters, will violate the unitarity black bound (see **2.2**) at very small b, just above the present Tevatron energy. Indeed, CDF reports [50] that $a_{el}(b = 0, \sqrt{s} = 1800) = 0.96 \pm 0.04$. A pictorial illustration of the above is presented in Fig.1. Note that the energy dependence of the experimental SD cross section [13–17] in the ISR-Tevatron energy range is much weaker than the power dependences observed for $\sigma_{el}$. Diffractive cross sections are not discussed in the DL model.

### 2.2 The eikonal model

The theoretical difficulties, pointed out in the previous subsection, are eliminated once we take into account the corrections necessitated by unitarity. The problem is that enforcing unitarity is a model dependent procedure. In the following we shall confine ourselves to a Glauber type eikonal model [51]. In this approximation, the scattering matrix is diagonal and only repeated elastic re-scatterings are summed. Accordingly, we write

$$a_{el}(s,b) \;=\; i\left(1-e^{-\Omega(s,b)/2}\right). \tag{12}$$

Since the scattering matrix is diagonal, the unitarity constraint is written as

$$2Im[a_{el}(s,b)] \;=\; \mid a_{el}(s,b)\mid^2 \;+\; G^{in}(s,b), \tag{13}$$

with

$$G^{in} \;=\; 1 \;-\; e^{-\Omega(s,b)}. \tag{14}$$

The eikonal expressions for the soft cross sections of interest are

$$\sigma_{tot} \;=\; 2\int d^2b\left(1 \;-\; e^{-\Omega(s,b)/2}\right), \tag{15}$$

$$\sigma_{el} \;=\; \int d^2b\left(1 \;-\; e^{-\Omega(s,b)/2}\right)^2, \tag{16}$$

$$\sigma_{in} \;=\; \int d^2b\left(1 \;-\; e^{-\Omega(s,b)}\right), \tag{17}$$

and

$$B_{el}(s) \;=\; \frac{\int d^2b\; b^2\;\left(1\;-\;e^{-\Omega(s,b)/2}\right)}{2\int d^2b\;\left(1\;-\;e^{-\Omega(s,b)/2}\right)}. \tag{18}$$

From Eq.(14) it follows that $P^S(s,b) \;=\; e^{-\Omega(s,b)}$ is the probability that the final state of the two initial interacting hadrons is elastic, regardless of the eikonal rescattering chain. It is identified, thus, with $P^S(s,b)$ of Eq.(3).

Following our implicit assumption that, in the high energy limit, hadrons are correct degrees of freedom, i.e. they diagonalize the interaction matrix, Eq.(12) is a general solution of Eq.(13) as long as the input opacity $\Omega$ is arbitrary. In the eikonal model $\Omega$ is real and equals the imaginary part of the iterated input Born amplitude. The eikonalized amplitude is imaginary. Its analyticity and crossing symmetry are easily restored. In a Regge language we substitute, to this end, $s^{\alpha_{I\!P}} \;\rightarrow\; s^{\alpha_{I\!P}}e^{-\frac{1}{2}i\pi\alpha_{I\!P}}$.

In the general case, Eq.(13) implies a general bound, $\mid a_{el}(s,b)\mid\leq 2$, obtained when $G^{in} \;=\; 0$. This is an extreme option in which asymptotically $\sigma_{tot} \;=\; \sigma_{el}$ [52]. This is formally acceptable but not very appealing. Assuming that $a_{el}$ is imaginary, we obtain that the unitarity bound coincides with the black disc bound, $\mid a_{el}(s,b)\mid\leq 1$. Accordingly,

$$\frac{\sigma_{el}}{\sigma_{tot}} \leq \frac{1}{2}. \tag{19}$$

## 3 The GLM Model

The GLM screening correction (SC) model [30–33] is an eikonal model originally conceived so as to explain the exceptionally mild energy dependence of soft diffractive cross sections. It utilized the observation that s-channel unitarization enforced by the eikonal model operates on a diffractive amplitude in a different way than it does on the elastic amplitude. The GLM diffractive damping factor is identical to Bjorken's survival probability.

### 3.1 The GLM SC model

In the GLM model, we take a DL type Pomeron exchange amplitude input in which $\alpha_{I\!P}(0) = 1 + \Delta > 0$. The simplicity of the GLM SC model derives from the observation that the eikonal approximation with a central Gaussian input, corresponding to an exponential slope of $\frac{d\sigma_{el}}{dt}$, can be summed analytically. This is, clearly, an over simplification, but it reproduces the bulk of the data well, i.e. the total and the forward elastic cross sections. Accordingly, the eikonal DL type b-space expression for $\Omega(s, b)$ is:

$$\Omega(s,b) \,=\, \nu(s)\,\Gamma^S(s,b), \tag{20}$$

where,

$$\nu(s) \,=\, \sigma(s_0)\left(\frac{s}{s_0}\right)^{\Delta}, \tag{21}$$

$$R^2(s) \,=\, 4R_0^2 \,+\, 4\alpha'_{I\!P} ln(\frac{s}{s_0}), \tag{22}$$

and the soft profile is defined

$$\Gamma^S(s,b) \,=\, \frac{1}{\pi R^2(s)}\, e^{-\frac{b^2}{R^2(s)}}. \tag{23}$$

It is defined so as to keep the normalization $\int d^2b\,\Gamma^S(s,b) \,=\, 1$.

One has to distinguish between the eikonal model input and output. The key element is that the power $\Delta$, and $\nu$, are input information, not bounded by unitarity, and should not be confused with DL effective power $\epsilon$ and the corresponding total cross section. Since the DL model reproduces the forward elastic amplitude, in the ISR-HERA-Tevatron range, well, we require that the eikonal model output will be compatible with the DL results. Obviously, $\Delta > \epsilon$. In a non screened DL type model with a Gaussian profile the relation $B_{el} = \frac{1}{2}R^2(s)$ is exact. In a screened model, like GLM, $B_{el} > \frac{1}{2}R^2(s)$ due to screening.

With this input we get

$$\sigma_{tot} \,=\, 2\pi R^2(s)\left[ln\left(\frac{\nu(s)}{2}\right) + C - Ei\left(-\frac{\nu(s)}{2}\right)\right] \,\propto\, ln^2(s), \tag{24}$$

$$\sigma_{el} \,=\, \pi R^2(s)\left[ln\left(\frac{\nu(s)}{4}\right) + C - 2Ei\left(-\frac{\nu(s)}{2}\right) + Ei\left(-\nu(s)\right)\right] \,\propto\, \frac{1}{2}ln^2(s), \tag{25}$$

$$\sigma_{in} \,=\, \pi R^2(s)\{ln[\nu(s)] + C - Ei[-\nu(s)]\} \,\propto\, \frac{1}{2}ln^2(s). \tag{26}$$

$Ei(x) = \int_{-\infty}^{x} \frac{e^t}{t} dt$, and $C = 0.5773$ is the Euler constant. An important consequence of the above is that the ratio $\frac{\sigma_{el}}{\sigma_{tot}}$ is a single variable function of $\nu(s)$. In practice it means that given the experimental value of this ratio at a given energy we can obtain an "experimental" value of $\nu$ which does not depend on the adjustment of free parameters.

The formalism presented above is extended to diffractive channels through the observation, traced to Eqs.(3) and (14), that $P^S(s,b) = e^{-\Omega(s,b)}$. Accordingly, a screened non elastic diffractive cross section is obtained by convoluting its b-space amplitude square with the probability $P^S$.

The above has been utilized [30–33] to calculate the soft integrated single diffraction cross section. To this end, we write, in the triple Regge approximation [53], the double differential cross section $\frac{M^2 d\sigma_{sd}}{dM^2 dt}$, where $M$ is the diffracted mass. We, then, transform it to b-space, multiply by $P^S(s,b)$ and integrate. The output $\frac{M^2 d\sigma_{sd}}{dM^2 dt}$, changes its high energy behaviour from $s^{2\Delta}$ modulu $ln(\frac{s}{s_0})$ (which is identical to the behaviour of a DL elastic cross section) to the moderate behaviour of $ln(\frac{s}{s_0})$. Note also a major difference in the diffractive b-space profile which changes from an input central Gaussian to an output peripheral distribution peaking at higher b. Consequently, the GLM model is compatible with the Pumplin bound [54, 55].

$$\frac{\sigma_{el}(s,b)\,+\,\sigma_{diff}(s,b)}{\sigma_{tot}(s,b)}\,\leq\,\frac{1}{2}. \tag{27}$$

### 3.2 Extension to a multi channel model

The most serious deficiency of a single channel eikonal model is inherent, as the model considers only elastic rescatterings. This is incompatible with the relatively large diffractive cross section observed in the ISR-Tevatron energy range. To this we add a specific problematic feature of the GLM model. Whereas, $\sigma_{tot}$, $\sigma_{el}$ and $B_{el}$ are very well fitted, the reproduction of $\sigma_{sd}$, in the available ISR-Tevatron range, is poorer. A possible remedy to these deficiencies is to replace the one channel with a multi channel eikonal model, in which inelastic diffractive intermediate re-scatterings are included as well [38, 39, 56]. However, we have to insure that a multi channel model does improve the diffractive (specifically SD) predictions of the GLM model, while maintaining, simultaneously, its excellent reproductions [30–33] of the forward elastic amplitude, as well as its appealing results on LRG survival probabilities [35–37] to be discussed in **3.3**.

In the simplest approximation we consider diffraction as a single hadronic state. We have, thus, two orthogonal wave functions

$$\langle\Psi_h\mid\Psi_d\rangle\;=\;0. \tag{28}$$

$\Psi_h$ is the wave function of the incoming hadron, and $\Psi_d$ is the wave function of the outgoing diffractive system initiated by the incoming hadron. Denote the interaction operator by **T** and consider two wave functions $\Psi_1$ and $\Psi_2$ which are diagonal with respect to **T**. The amplitude of the interaction is given by

$$A_{i,k}=\langle\Psi_i\Psi_k\mid\mathbf{T}\mid\Psi_{i'}\Psi_{k'}\rangle=a_{i,k}\,\delta_{i,i'}\,\delta_{k,k'}. \tag{29}$$

In a $2\times 2$ model $i,k\;=\;1,2$. The amplitude $a_{i,k}$ satisfies the diagonal unitarity condition (see Eq.(13))

$$2Im\,a_{i,k}(s,b)\;=\mid a_{i,k}(s,b)\mid^2\;+G^{in}_{i,k}(s,b), \tag{30}$$

for which we write the solution

$$a_{i,k}(s,b)\;=\;i\left(1\;-\;e^{-\frac{\Omega_{i,k}(s,b)}{2}}\right), \tag{31}$$

and

$$G^{in}_{i,k}\;=\;1-e^{-\Omega_{i,k}(s,b)}. \tag{32}$$

$\Omega_{i,k}(s,b)$ is the opacity of the $(i,k)$ channel with a wave function $\Psi_i\;\times\;\Psi_k$.

$$\Omega_{i,k}\;=\;\nu_{i,k}(s)\,\Gamma^S_{i,k}(s,b) \tag{33}$$

where

$$\nu_{i,k}\;=\;\sigma^{S0}_{i,k}\left(\frac{s}{s_0}\right)^{\Delta}. \tag{34}$$

The factorizable radii are given by

$$R^2_{i,k}(s)\;=\;2R^2_{i,0}\;+\;2R^2_{0,k}\;+\;4\alpha'_{I\!P}ln(\frac{s}{s_0}). \tag{35}$$

$\Gamma^S_{i,k}(s,b)$ is the soft profile of the (i,k) channel. The probability that the final state of two interacting hadron states, with quantum numbers i and k, will be elastic regardless of the intermediate rescatterings is

$$P^S_{i,k}(s,b)\;=\;e^{-\Omega_{i,k}(s,b)}\;=\;\{1\;-\;a_{i,k}(s,b)\}^2. \tag{36}$$

In the above diagonal representation, $\Psi_h$ and $\Psi_d$ can be written as

$$\Psi_h = \alpha\Psi_1 + \beta\Psi_2, \tag{37}$$

$$\Psi_d = -\beta\Psi_1 + \alpha\Psi_2. \tag{38}$$

$\Psi_1$ and $\Psi_2$ are orthogonal. Since $\mid \Psi_h \mid^2 = 1$, we have

$$\alpha^2 + \beta^2 = 1. \tag{39}$$

The wave function of the final state is

$$\begin{aligned}\Psi_f &= \mid \mathbf{T} \mid \Psi_h \times \Psi_h\rangle = \\ &\alpha^2 a_{1,1}\{\Psi_1 \times \Psi_1\} + \alpha\beta a_{1,2}\{\Psi_1 \times \Psi_2 + \Psi_2 \times \Psi_1\} + \\ &\beta^2 a_{2,2}\{\Psi_2 \times \Psi_2\}.\end{aligned} \tag{40}$$

We have to consider 4 possible re-scattering processes. However, in the case of a $\bar{p}p$ (or $pp$) collision, single diffraction at the proton vertex equals single diffraction at the antiproton vertex. i.e., $a_{1,2} = a_{2,1}$ and we end with three channels whose b-space amplitudes are given by

$$a_{el}(s,b) = \langle\Psi_h \times \Psi_h \mid \Psi_f\rangle = \alpha^4 a_{1,1} + 2\alpha^2\beta^2 a_{1,2} + \beta^4 a_{2,2}, \tag{41}$$

$$a_{sd}(s,b) = \langle\Psi_h \times \Psi_d \mid \Psi_f\rangle = \alpha\beta\{\alpha^2 a_{1,1} + (\alpha^2 - \beta^2)a_{1,2} + \beta^2 a_{2,2}\}, \tag{42}$$

$$a_{dd}(s,b) = \langle\Psi_d \times \Psi_d \mid \Psi_f\rangle = \alpha^2\beta^2\{a_{1,1} - 2a_{1,2} + a_{2,2}\}. \tag{43}$$

In the numeric calculations one may further neglect the double diffraction channel which is exceedingly small in the ISR-Tevatron range. This is obtained by setting $a_{2,2} = 2a_{1,2} - a_{1,1}$. Note that in the limit where $\beta << 1$, we reproduce the single channel model.

As in the single channel, we simplify the calculation assuming a Gaussian b-space distribution of the input opacities soft profiles

$$\Gamma^S_{i,k}(s,b) = \frac{1}{\pi R^2_{i,k}(s)} e^{-\frac{b^2}{R^2_{i,k}(s)}}. \tag{44}$$

The opacity expressions, just presented, allow us to express the physical observables of interest as functions of $\nu_{1,1}$, $\nu_{1,2}$, $R^2_{1,1}$, $R^2_{1,2}$ and $\beta$, which is a constant of the model. The determination of these variables enables us to produce a global fit to the total, elastic and diffractive cross sections as well as the elastic forward slope. This has been done in a two channel model, in which $\sigma_{dd}$ is neglected [38]. The main conclusion of this study is that the extension of the GLM model to a multi channel eikonal results with a very good overall reproduction of the data. The results maintain the b-space peripherality of the diffractive output amplitudes and satisfy the Pumplin bound [54, 55]. Note that since different experimental groups have been using different algorithms to define diffraction, the SD experimental points are too scattered to enable a tight theoretical reproduction of the diffractive data, see Fig.2.

### 3.3 Survival probabilities of LRG in the GLM model

The eikonal model simplifies the calculation of the survival probability, Eq.(3), associated with the soft re-scatterings of the spectator partons. We can, thus, eliminate the nominator and denominator terms in $\mid M^H(s,b) \mid^2$ which depend exclusively on s. In the GLM model we assume a Gaussian b-dependence for $\mid M^H(s,b) \mid^2$ corresponding to a constant hard radius $R^{H^2}$. This choice enables an analytic solution of Eq.(3). More elaborate choices, such as dipole or multi poles distributions, require a numerical evaluation of this equation.

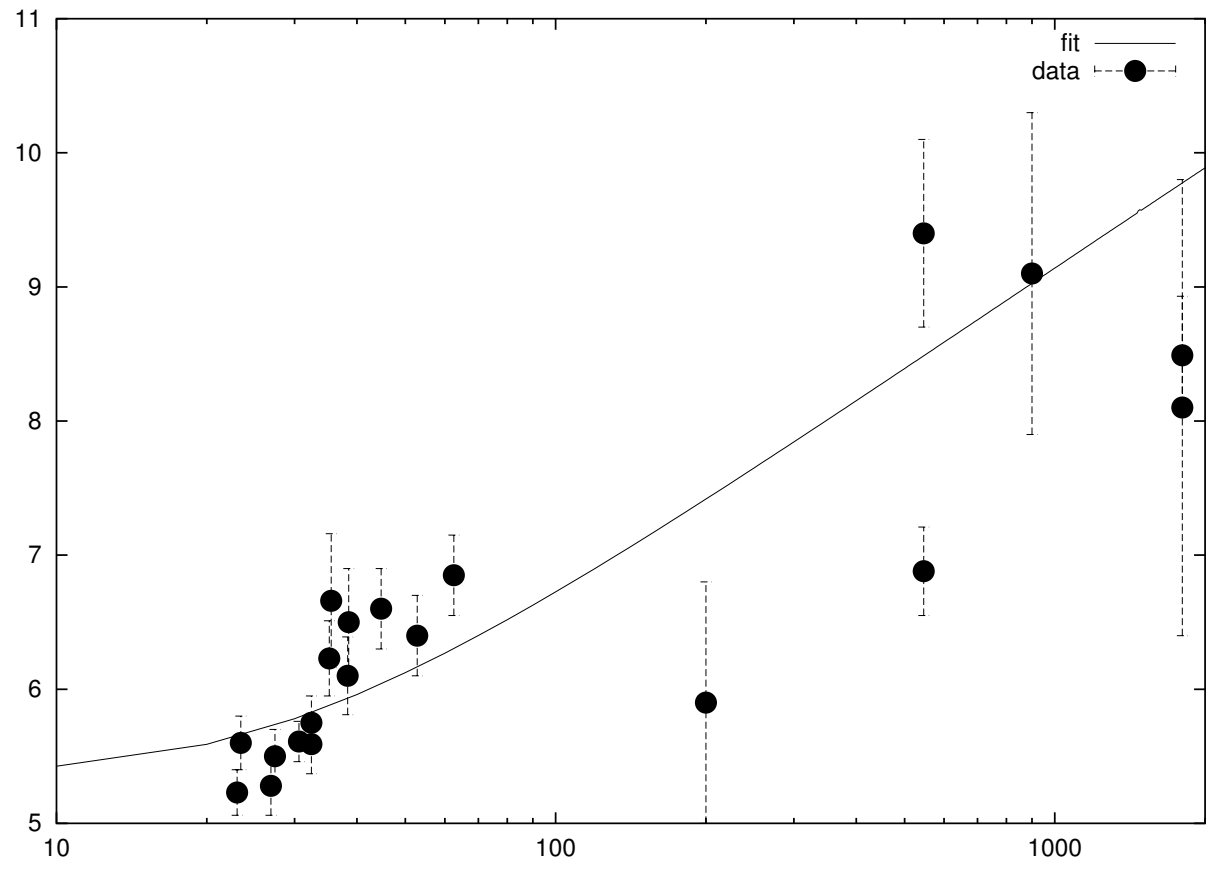

**Fig. 2:** Integrated SD data and a two channel model fit.

Define,

$$a_H(s) = \frac{R^2(s)}{{R^H}^2(s)} > 1. \tag{45}$$

$a_H(s)$ grows logarithmically with $s$. As stated, Eq.(3) can be analytically evaluated with our choice of Gaussian profiles and we get

$$S^2 = \frac{a_H(s)\gamma[a_H(s), \nu(s)]}{[\nu(s)]^{a_H(s)}}, \tag{46}$$

where $\gamma(a, \nu)$ denotes the incomplete Euler gamma function

$$\gamma(a, x) = \int_0^x z^{a-1} e^{-z} dz. \tag{47}$$

The solution of Eq.(46), at a given $s$, depends on the input values of ${R^H}^2$, $R^2$ and $\nu(s)$. In the GLM approach, ${R^H}^2$ is estimated from the excellent HERA data [57–59] on $\gamma + p \to J/\Psi + p$. The values of $\nu(s)$ and $R^2(s)$ are obtained from the experimental $\bar{p}p$ data. This can be attained from a global fit to the soft scattering data [38]. Alternatively, we can obtain $\nu$ from the ratio $\frac{\sigma_{el}}{\sigma_{tot}}$ and then obtain the value of $R^2$ from the explicit expressions given in Eqs.(24,25,26). LHC predictions presently depend on model calculations with which this information can be obtained. Once we have determined $\nu(s)$ and $a_H(s)$, the survival probability is calculated from Eq.(46).

In the GLM three channel model we obtain for central hard diffraction of di-jets or Higgs a survival probability,

$$S^2_{CD}(s) = \frac{\int d^2b \left( \alpha^4 P^S_{1,1} \, {\Omega^H_{1,1}}^2 + 2\alpha^2\beta^2 P^S_{1,2} \, {\Omega^H_{1,2}}^2 + \beta^4 P^S_{2,2} \, {\Omega^H_{2,2}}^2 \right)}{\int d^2b \left( \alpha^4 \, {\Omega^H_{1,1}}^2 + 2\alpha^2\beta^2 \, {\Omega^H_{1,2}}^2 + \beta^4 \, {\Omega^H_{2,2}}^2 \right)}. \tag{48}$$

The hard diffractive cross sections in the (i,k) channel are calculated using the multi particle optical theorem [53]. They are written in the same form as the soft amplitudes

$${\Omega^H_{i,k}}^2 = \nu^H_{i,k}(s)^2 \, \Gamma^H_{i,k}(b), \tag{49}$$

where,

$$\nu^H_{i,k} = \sigma^{H0}_{i,k} \left( \frac{s}{s_0} \right)^{\Delta_H}. \tag{50}$$

As in the single channel calculation we assume that $\Gamma^H_{i,k}(b)$ is Gaussian,

$$\Gamma^H_{i,k}(b) \;=\; \frac{2}{\pi R^2_{i,k}}\, e^{-\frac{2\,b^2}{R^2_{i,k}}}\,. \tag{51}$$

Note, that the hard radii ${R^H_{i,k}}^2$ are constants derived from HERA $J/\Psi$ photo and DIS production [57–59].

As it stands, a three channel calculation is not useful since $\sigma_{dd}$ is very small and the 3'd channel introduces additional parameters which can not be constraint by the meager experimental information on $\sigma_{dd}$ [13–17]. In a two channel model Eq.(48) reduces to

$$S^2_{CD}(s) \;=\; \frac{\int d^2b\,\left(P^S_{1,1}\,{\Omega^H_{1,1}}^2 \;-\; 2\beta^2\,(P^S_{1,1}\,{\Omega^H_{1,1}}^2 \;-\; P^S_{1,2}\,{\Omega^H_{1,2}}^2)\right)}{\int\, d^2b\left({\Omega^H_{1,1}}^2 \;-\; 2\beta^2\,({\Omega^H_{1,1}}^2 \;-\; {\Omega^H_{1,2}}^2)\right)}. \tag{52}$$

A new, unpublished yet, model [60], offers an explicit $S^2$ calculation for the exclusive $NN \to N + LRG + 2J + LRG + N$ final state, both in one and two channel eikonal models. We shall comment on its output in the next subsection.

### 3.4 GLM $S^2$ predictions

Following are a few general comments on the GLM calculations of $S^2$, after which we discuss the input/output features of the single and two channel models. Our objective is to present predictions for LHC.

The only available experimental observable with which we can check the theoretical $S^2$ predictions is the hard LRG di-jets data obtained in the Tevatron and Hera. A comparison between data and our predictions is not immediate as the basic measured observable is $f_{gap}$ and not $S^2$. The application of the GLM models to a calculation of $f_{gap}$ depends on an external input of a hard diffractive LRG cross section which is then corrected by $S^2$ as presented above. Regardless of this deficiency, the introduction of a survival probability is essential so as to understand the huge difference between the pQCD calculated $F_{gap}$ and its experimental value $f_{gap}$. A direct test of the GLM predictions calls for a dedicated experimental determination of $S^2$. The only direct $S^2$ information from the Tevatron is provided by a JGJ ratio measured by D0 [5–7] in which $\frac{S^2(\sqrt{s}=630)}{S^2(\sqrt{s}=1800)} \;=\; 2.2 \pm 0.8$. This is to be compared with a GLM ratio of $1.2 - 1.3 \pm 0.4$ presented below.

The survival probabilities of the CD, SD and DD channels are not identical. The key difference is that each of the above channels has a different hard radius. A measure of the sensitivity of $S^2$ to changes in $\nu$ and $a_H$ is easy to identify in a single channel calculation which is presented in Fig.3. Indeed, preliminary CDF GJJG data [17] suggest that $f_{gap}$ measured for this channel is moderately smaller than the rate measured for the GJJ channel.

GLM soft profile input is a central Gaussian. This is over simplified, and most models assume a power like dipole or multipole b-dependence of $\Gamma^S(s,b)$ and $\Gamma^H(s,b)$. Explicit comparisons [60] of $S^2$ obtained with different input profiles shows a diminishing difference between the survival probability outputs, provided their effective radii are compatible.

Regardless of the attractive simplicity of the single channel model, one should add a cautious reminder that the single channel model does not reproduce $\sigma_{sd}$ well since its survival probabilities are over-estimated. Consequently, we are inclined to suspect that the $S^2$ values presented in Table 1 are over-estimated as well.

As we noted, the soft input can be obtained from either a model fit to the soft scattering data or directly from the measured values of $\sigma_{tot}, \sigma_{el}$ and ${R^H}^2$. The first method is denoted F1C and the second is denoted D1C. Note that having no LHC data, $S^2_{DD}(D1C)$, at this energy, is calculated on the basis of model estimates for the total and elastic cross sections. The constant hard radius $R^{H^2} = 7.2$ is deduced from HERA $J/\Psi$ photoproduction forward exponential slope which shows only diminishing shrinkage [57,58]. This is a conservative choice which may be changed slightly with the improvement of the Tevatron CDF estimates [61] of $R^{H^2}$. The two sets of results obtained are compatible, even though, $S^2(D1C)$ is consistently lower than $S^2(F1C)$. The $S^2$ output presented above depends crucially on the quality of the data base from which we obtain the input parameters. The two sets of Tevatron data at $1800\,GeV$ have a severe $10-15\%$ difference resulting in a non trivial ambiguity of the $S^2$ output.

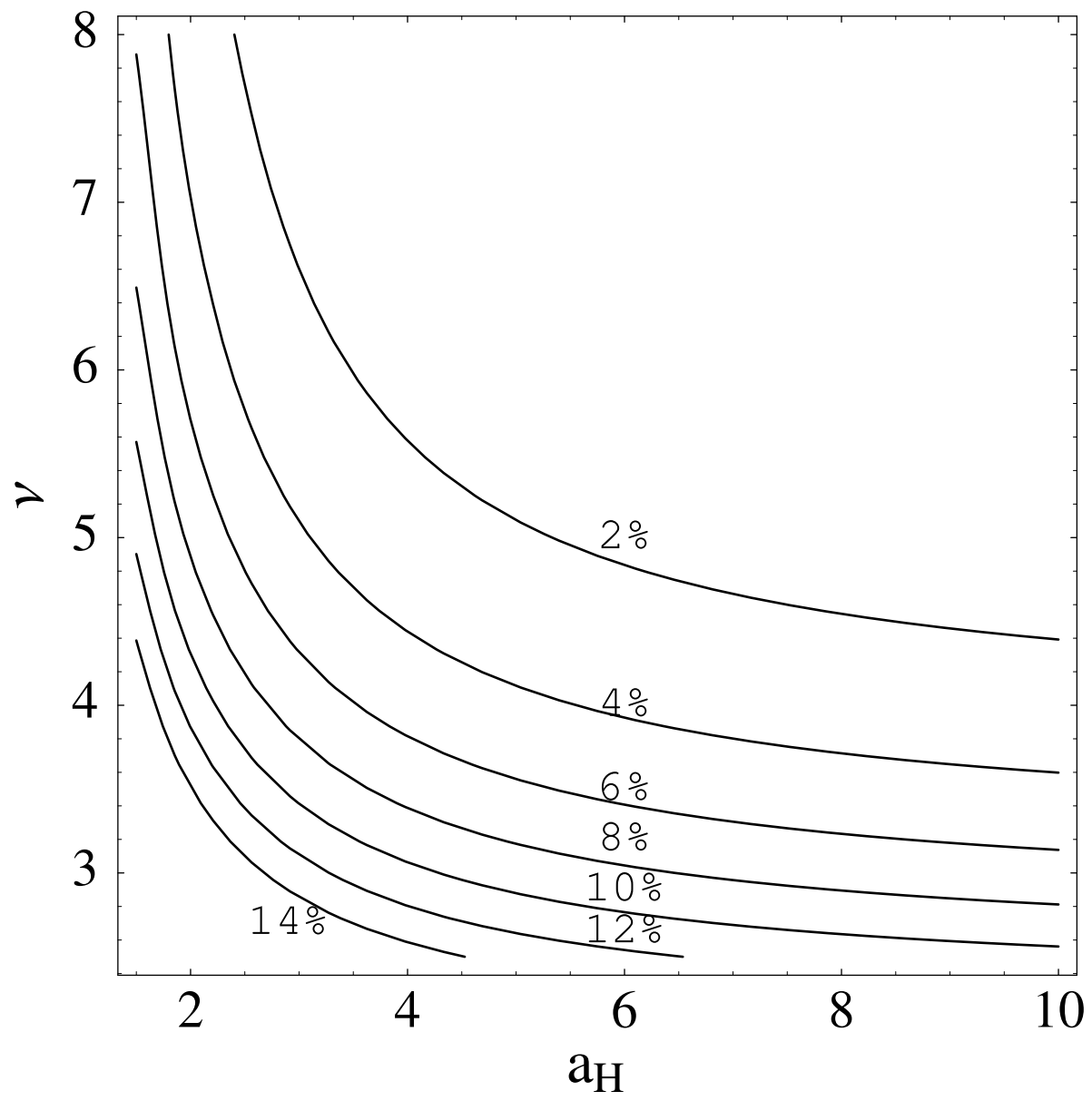

**Fig. 3:** A contour plot of $S^2(1C)$ against $\nu(s)$ and $a_H(s)$.

**Table 1:** Survival probabilities

| $\sqrt{s}$ (GeV) | $S^2_{\mathrm{CD}}$(F1C) | $S^2_{\mathrm{CD}}$(D1C) | $S^2_{\mathrm{SD_{incl}}}$(F1C) | $S^2_{\mathrm{SD_{incl}}}$(D1C) | $S^2_{\mathrm{DD}}$(F1C) | $S^2_{\mathrm{DD}}$(D1C) |
|---|---|---|---|---|---|---|
| 540 | 14.4% | 13.1% | 18.5% | 17.5% | 22.6% | 22.0% |
| 1800 | 10.9% | 8.9% | 14.5% | 12.6% | 18.2% | 16.6% |
| 14000 | 6.0% | 5.2% | 8.6% | 8.1% | 11.5% | 11.2 % |

The global GLM two channel fit [38] reproduces the soft scattering data (including SD) remarkably well with $\beta = 0.464$. Its fitted parameters are used for the soft input required for the $S^2$ calculations. Our cross section predictions for LHC are: $\sigma_{tot} = 103.8\,mb$, $\sigma_{el} = 24.5\,mb$, $\sigma_{sd} = 12\,mb$ and $B_{el} = 20.5\,GeV^{-2}$. The input for the calculation of $S^2$ requires, in addition to the soft parameters, also the values of $\nu^H_{i,k}$ and ${R^H_{i,k}}^2$. The needed hard radii can be estimated, at present, only for the CD channel, where we associate the hard radii $R^H_{1,1}$ with the hard radius obtained in HERA exclusive $J/\Psi$ photoproduction [57,58] and $R^H_{1,2}$ with HERA inclusive $J/\Psi$ DIS production [59]. Accordingly, we have ${R^H_{1,1}}^2 = 7.2\,GeV^{-2}$, and ${R^H_{1,2}}^2 = 2.0\,GeV^{-2}$. We do not have experimental input to determine $\nu^H_{i,k}$. We overcome this difficalty by assuming a Regge-like factorization $\sigma^{H0}_{i,k}/\sigma^{S0}_{i,k} = constant$. Our predictions for the CD survival probabilities are: $6.6\%$ at $540\,GeV$, $5.5\%$ at $1800\,GeV$ and $3.6\%$ at $14000\,GeV$.

These results may be compared with a recent, more elaborate, eikonal formulation [60] aiming to calculate the survival probability of a final exclusive $N + LRG + 2J(orH) + LRG + N$ state. These calculations were done in one and two channel models. The one channel $S^2_{CD}$ predicted values are $14.9\%$ at $540\,GeV$, $10.8\%$ at $1800\,GeV$ and $6.0\%$ at $14000\,GeV$. These values are remarkably similar to the GLM one channel output. In the two channel calculations the corresponding predictions are $5.1\%$, $4.4\%$ and $2.7\%$, which are marginally smaller than the GLM two channel output numbers.

In our assessment, the two channel calculations provide a more reliable estimate of $S^2$ since they reproduce well the soft scattering forward data. Our estimate for the survival probability associated with LHC Higgs production is $2.5\% - 4.0\%$.

## 4 The KKMR Model

The main part of this section (**4.1-4.3**) was written by V.A. Khoze, A.D. Martin and M. Ryskin (KMR) and is presented here without any editing.

The KKMR model calculation [40–44] of the survival probabilities is conceptually quite similar to the GLM model, in as much as unitarization is enforced through an eikonal model whose parameters provide a good reproduction of the high energy soft scattering data. However, the GLM model is confined to a geometrical calculation of $S^2$ for which we need just the value of $R^{H^2}$, without any specification of the hard dynamics. This value is an external input to the model. The KKMR model contains also a detailed pQCD calculation of the hard diffractive proccess, specifically, central diffractive Higgs production. Consequently, it can predict a cross section for the channel under investigation.

### 4.1 KKMR model for soft diffraction

The KMR description [41] of soft diffraction in high energy $pp$ (or $p\bar{p}$) collisions embodies

(i) *pion-loop* insertions in the bare Pomeron pole, which represent the nearest singularity generated by $t$-channel unitarity,

(ii) a *two-channel eikonal* which incorporates the Pomeron cuts generated by elastic and quasi-elastic (with $N^*$ intermediate states) $s$-channel unitarity,

(iii) high-mass *diffractive dissociation*.

The KKMR model gives a good description of the data on the total and differential elastic cross section throughout the ISR-Tevatron energy interval, see [41]. Surprisingly, KMR found the bare Pomeron parameters to be

$$\Delta \;\equiv\; \alpha(0) \,-\, 1 \;\simeq\; 0.10, \qquad \alpha' \;=\; 0. \tag{53}$$

On the other hand it is known that the same data can be described by a simple effective Pomeron pole with [47, 48, 62]

$$\alpha_{I\!P}^{\rm eff}(t) \;=\; 1.08 \,+\, 0.25\,t. \tag{54}$$

In this approach the shrinkage of the diffraction cone comes not from the bare pole ($\alpha' = 0$), but has components from the three ingredients, (i)–(iii), of the model. That is, in the ISR-Tevatron energy range

$$\text{“}\alpha'_{\rm eff}\text{”} \;=\; (0.034 \,+\, 0.15 \,+\, 0.066)\ {\rm GeV}^{-2} \tag{55}$$

from the $\pi$-loop, $s$-channel eikonalisation and diffractive dissociation respectively. Moreover, eikonal rescattering suppresses the growth of the cross section and so $\Delta \simeq 0.10 > \Delta_{\rm eff} \simeq 0.08$.

Since the model [41] embodies all the main features of soft diffraction KMR expect it to give reliable predictions for the *survival probability* $S^2$ of the rapidity gaps against population by secondary hadrons from the underlying event, that is hadrons originating from soft rescattering. In particular, KMR predict $S^2 = 0.10$ $(0.06)$ for single diffractive events and $S^2 = 0.05$ $(0.03)$ for exclusive Higgs boson production, $pp \to p + H + p$, at Tevatron (LHC) energies.

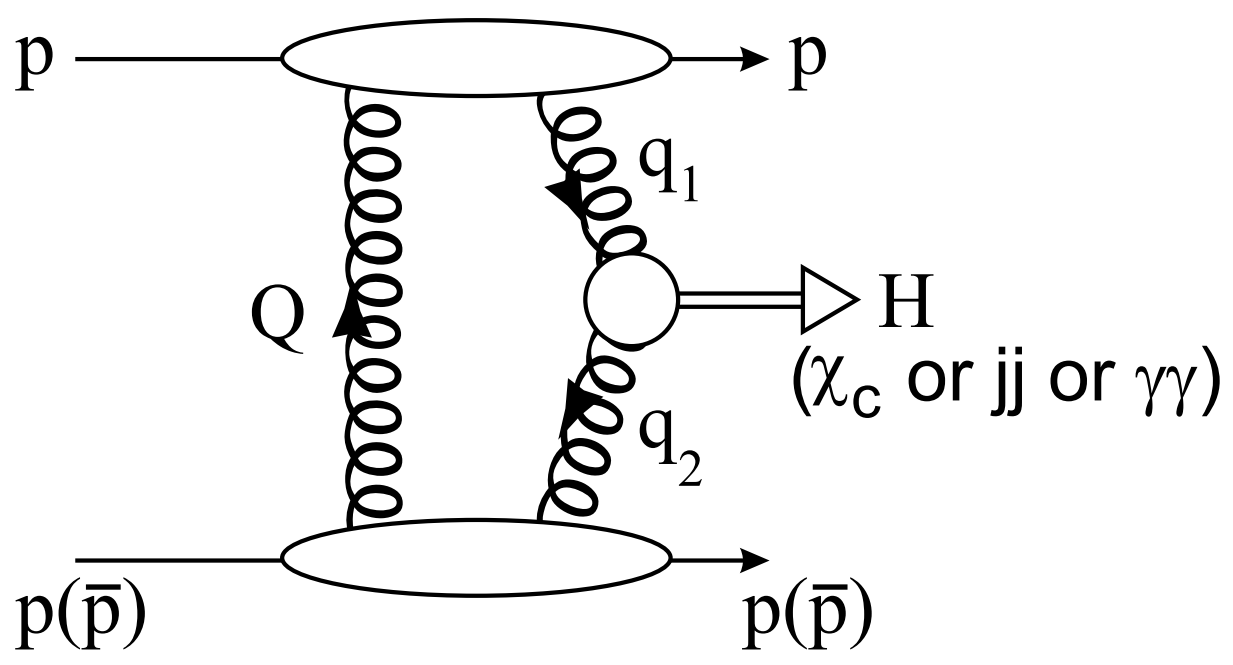

**Fig. 4:** Schematic diagram for central exclusive production, $pp \to p + X + p$. The presence of Sudakov form factors ensures the infrared stability of the $Q_t$ integral over the gluon loop. It is also necessary to compute the probability, $S^2$, that the rapidity gaps survive soft rescattering.

### 4.2 Calculation of the exclusive Higgs signal

The basic mechanism for the exclusive process, $pp \to p + H + p$, is shown in Fig. 4. The left-hand gluon $Q$ is needed to screen the colour flow caused by the active gluons $q_1$ and $q_2$. Since the dominant contribution comes from the region $\Lambda^2_{\rm QCD} \ll Q_t^2 \ll M_H^2$, the amplitude may be calculated using perturbative QCD techniques [40, 63]

$$\mathcal{M}_H \simeq N \int \frac{dQ_t^2}{Q_t^4}\, f_g(x_1, x_1', Q_t^2, \mu^2) f_g(x_2, x_2', Q_t^2, \mu^2), \tag{56}$$

where the overall normalisation constant $N$ can be written in terms of the $H \to gg$ decay width [40, 64]. The probability amplitudes ($f_g$) to find the appropriate pairs of $t$-channel gluons ($Q, q_1$) and ($Q, q_2$) are given by the skewed unintegrated gluon densities at the hard scale $\mu$, taken to be $0.62\ M_H$. Since the momentum fraction $x'$ transfered through the screening gluon $Q$ is much smaller than that ($x$) transfered through the active gluons ($x' \sim Q_t/\sqrt{s} \ll x \sim M_H/\sqrt{s} \ll 1$), it is possible to express $f_g(x, x', Q_t^2, \mu^2)$, to single log accuracy, in terms of the conventional integrated density $g(x)$ [65–68]. The $f_g$'s embody a Sudakov suppression factor $T$, which ensures that the gluon does not radiate in the evolution from $Q_t$ up to the hard scale $\mu \sim M_H/2$, and so preserves the rapidity gaps.

It is often convenient to use the simplified form [40]

$$f_g(x, x', Q_t^2, \mu^2) \ = \ R_g \frac{\partial}{\partial \ln Q_t^2} \left[ \sqrt{T_g(Q_t, \mu)}\, x g(x, Q_t^2) \right], \tag{57}$$

which holds to 10–20% accuracy.[1] The factor $R_g$ accounts for the single $\log Q^2$ skewed effect [67]. It is found to be about 1.4 at the Tevatron energy and about 1.2 at the energy of the LHC.

### 4.3 The Sudakov factor

The Sudakov factor $T_g(Q_t, \mu)$ reads [65, 66, 69]

$$T_g(Q_t, \mu) = \exp\left( - \int_{Q_t^2}^{\mu^2} \frac{\alpha_S(k_t^2)}{2\pi} \frac{dk_t^2}{k_t^2} \left[ \int_{\Delta}^{1-\Delta} z P_{gg}(z) dz \ + \ \int_0^1 \sum_q P_{qg}(z) dz \right] \right), \tag{58}$$

with $\Delta = k_t/(\mu + k_t)$. The square root arises in (57) because the (survival) probability not to emit any additional gluons is only relevant to the hard (active) gluon. It is the presence of this Sudakov factor which makes the integration in (56) infrared stable, and perturbative QCD applicable[2].

[1] In the actual computations a more precise form, as given by Eq. (26) of [68], was used.

[2] Note also that the Sudakov factor inside $t$ integration induces an additional strong decrease (roughly as $M^{-3}$ [44]) of the cross section as the mass $M$ of the centrally produced hard system increases. Therefore, the price to pay for neglecting this suppression effect would be to considerably overestimate the central exclusive cross section at large masses.

**Table 2:** Compilation of $S^2$ values obtained in the KKMR model

| $\sqrt{s}$ (GeV) | $S^2_{2C}$(CD) | $S^2_{2C}$(SD$_{\rm incl}$) | $S^2_{2C}$(DD) |
|---|---|---|---|
| 540 | 6.0% | 13.0% | 20.0% |
| 1800 | 4.5% | 10.0% | 15.0% |
| 14000 | 2.6% | 6.0% | 10.0% |

It should be emphasized that the presence of the double logarithmic $T$-factors is a purely classical effect, which was first discussed in 1956 by Sudakov in QED. There is strong bremsstrahlung when two colour charged gluons 'annihilate' into a heavy neutral object and the probability not to observe such a bremsstrahlung is given by the Sudakov form factor[3]. Therefore, any model (with perturbative or non-perturbative gluons) must account for the Sudakov suppression when producing exclusively a heavy neutral boson via the fusion of two coloured particles.

More details of the role of the Sudakov suppression can be found in J. Forshaw's review in these proceedings [34]. Here KMR would like to recall that the $T$-factors in [44, 70] were calculated to *single* log accuracy. The collinear single logarithms were summed up using the DGLAP equation. To account for the 'soft' logarithms (corresponding to the emission of low energy gluons) the one-loop virtual correction to the $gg \to H$ vertex was calculated explicitly, and then the scale $\mu = 0.62\ M_H$ was chosen in such a way that eq.(58) reproduces the result of this explicit calculation. It is sufficient to calculate just the one-loop correction since it is known that the effect of 'soft' gluon emission exponentiates. Thus (58) gives the $T$-factor to single log accuracy.

In some sense, the $T$-factor may be considered as a 'survival' probability not to produce any hard gluons during the $gg \to H$ fusion subprocess. However, it is not just a number (i.e. a numerical factor) which may be placed in front of the integral (the 'bare amplitude'). Without the $T$-factors hidden in the unintegrated gluon densities $f_g$ the integral (56) diverges. From the formal point of view, the suppression of the amplitude provided by $T$-factors is infinitely strong, and without them the integral depends crucially on an ad hoc infrared cutoff.

### 4.4 Summary of KKMR $S^2$ predictions

Table 2 shows a compilation of $S^2$ values in the KKMR model. A comparison with the corresponding GLM two channel model is possible only for the available GLM CD channel, where, the KKMR output is compatible with GLM. KKMR SD and DD output are compatible with the corresponding GLM single channel numbers. Overall, we consider the two models to be in a reasonable agreement.

A remarkable utilization of the KKMR model is attained when comparing the HERA [18–27] and CDF [8–12, 17] di-jets diffractive structure functions derived for the dynamically similar GJJ channels. To this end, the comparison is made between the kinematically compatible HERA $F^D_{jj}(Q^2 = 75\,GeV^2, \beta)$ and the CDF $F^D_{jj}(< E^2_T >= 75\,GeV^2, \beta)$. The theoretical expectation is that $F^D_{jj}(\beta)$, as measured by the two experiments, should be very similar. As can be seen in Fig.5, the normalizations of the two distributions differ by approximately an order of magnitude and for very small $\beta < 0.15$ there is a suggestive change in the CDF distribution shape. This large discrepancy implies a breaking of QCD and/or Regge factorization. Reconsidering, it is noted, that HERA DIS data is measured at a high $Q^2$ where the partonic interactions induced by the highly virtual photon are point like and, hence, $S^2 = 1$. On the other hand, CDF GJJ measurement is carried out at $1800\,GeV$ and, as we saw, its survival prob-ability is rather small. The convolution between the HERA determined GJJ $F^D_{jj}(\beta)$ and the $\beta$ dependent survival probabilities, as calculated by KKMR, provides the $F^D_{jj}(\beta)$ distribution corrected for the soft rescattering of the spectator partons. This is shown in Fig.5 and provides an impressive reproduction of the experimental distribution. We were informed [71] that this analysis was successfully redone with an updated H1 produced structure function distribution.

[3]It is worth mentioning that the $H \to gg$ width and the normalization factor $N$ in (56) is an 'inclusive' quantity which includes all possible bremsstrahlung processes. To be precise, it is the sum of the $H \to gg + ng$ widths, with $n$=0,1,2,... . The probability of a 'purely exclusive' decay into two gluons is nullified by the same Sudakov suppression.

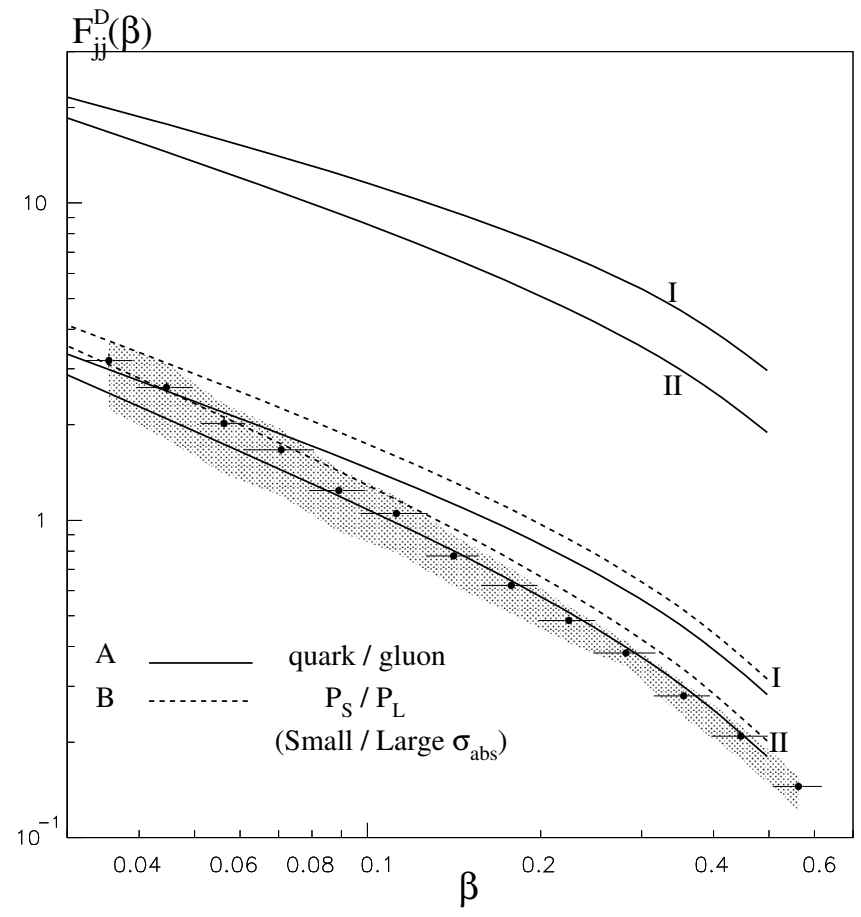

**Fig. 5:** The predictions for the diffractive di-jets production at the Tevatron (lower lines), obtained from two alternative sets of HERA diffractive parton distributions I and II, compared with the CDF data (shaded area). The upper lines correspond to the Tevatron prediction neglecting the survival probability correction.

The weak element in the above analysis is that it is crucially dependent on the H1 determined $F^D_{jj}(\beta)$ distribution. ZEUS has constructed a somewhat different structure function. Clearly, a very different experimental determination of $F^D_{jj}(\beta)$, such as been recently suggested by Arneodo [72], will re-open this analysis for further studies, experimental and theoretical.

### 4.5 A Comparison between KKMR and GLM

The approach of GLM and KKMR to the calculation of forward soft scattering in the ISR-Tevatron range are basically similar. Both models utilize the eikonal model assuming different input soft profiles which have, nevertheless, compatible effective radii. There are, though, a few particular differences between the two sets of calculations:

1) The GLM model, with a Gaussian soft profile, is applicable only in the forward cone ($|t| < 0.3\,GeV^2$), where we have most of the data of interest. KKMR use a multipole power behaviour profile which enables applicability over a, somewhat, wider t range, $|t| < 0.5\,GeV^2$. Note that, the GLM output is not significantly changed with a multipole power behaviour profile provided its radii are compatible with the Gaussian input [60].
2) The GLM input Pomeron trajectory is specified by $\Delta = 0.12$ and $\alpha'_{I\!P} = 0.2$. These evolve due to eikonalisation to an effective output of $\epsilon = 0.08$ and $\alpha'_{I\!P} = 0.25$. Note that, $\Delta$ is obtained in GLM as a fitted output parameter. In KKMR, the relatively high input $\Delta \simeq 0.2$ is theoretically tuned by a pion loop renormalization resulting in an input value of $\Delta \simeq 0.1$. KKMR have a more elaborate treatment of $\alpha_{I\!P}(t)$ than GLM, resulting, nevertheless, with forward cone output predictions similar to GLM. However, KKMR accounts for a somewhat wider t range than GLM and reproduces the t dependence of $B_{el}$ well. Similar results are obtained in a GLM version [39,56] in which the soft profile is given by a dipole distribution. KKMR can predict a few differential properties of $S^2$, which are beyond the scope of GLM.

3) Both models treat the high mass diffraction with the triple Pomeron formalism [53]. In GLM the final SD cross section is obtained by a convolution of the input $\frac{d\sigma_{sd}}{d^2b}$ with $P^S(s,b)$. In KKMR the treatment of the SD amplitude is more elaborate, ending, though, with no detailed SD data reconstruction which is presented in GLM.
4) The LHC predictions of the two models for cross sections and slopes are compatible, with the exception of $\sigma_{dd}$ which is neglected in GLM and acquires a significant KKMR predicted value of $9.5\,mb$.

GLM is a geometrical model where both the input hard LRG non corrected matrix element squared and the soft elastic scattering amplitude, are approximated by central Gaussians in b-space. This property enables us to easily calculate the survival probabilities which depend on $\nu$, $R^2$ and $R^{H^2}$ in a single channel input, and on $\nu_{i,k}$, $R^2_{i,k}$ and ${R^H_{i,k}}^2$ in a two channel input. As we have noted, the GLM model, on its own, cannot provide a calculation of $F_{gap}$ and $f_{gap}$ as it needs the hard radii as an external input. The KKMR model is more sophisticated. This is attributed to the fact that the hard diffractive LRG process is explicitly calculated in pQCD, hence the non corrected $F_{gap}$ and the corrected $f_{gap}$ and $F^D_{jj}$ are model predictions. As we have just noted, given the hard diffractive matrix element, the actual calculation of the diffractive LRG survival probability damping is almost identical to GLM. Keeping this basic observation in mind, it is constructive to compare the features of the two models with a special interest on the input assumptions and output differences of the two models.

The main difference between the two models is reflected in the level of complexity of their inputs. GLM soft input is obtained from a simple eikonal model for the soft forward scattering, to which we add the hard radii which are derived from the HERA data. KKMR calculations of $P^S$ are equally simple. The calculation of the hard sector matrix elements are, naturally, more cumbersome. Given HERA $F^D_{jj}(Q^2,\beta)$, a Tevatron diffractive $F^D_{jj}$ in which $< E_T >$ and $Q^2$ are comparable, can be calculated, parameter free, without the need to calculate the hard amplitude. But this is a particular case and, in general, the KKMR calculation depends on an extended parameter base, such as the the input p.d.f. and pQCD cuts. These input parameters are not constrained tightly enough.

The elaborate structure of the KKMR model provides a rich discovery potential which is reflected in the model being able to define and calculate the dependence of $S^2$ not only on b, but also on other variables, notably $\beta$, and experimental cuts such as the recoil proton transverse momentum. GLM depends on the hard radii external information obtained from HERA data. It lacks the potential richness of KKMR. GLM can serve, though, as a standard through which we can compare different unitarized models. Given such a model, we can extract effective values for $\nu$, $R^2$ and $R^{H^2}$ and proceed to a simple calculation of $S^2$. We shall return to this proposed procedure in the final discussion.

Even though both GLM and KKMR are two channel models, they are dynamically different. GLM two channel formulation relates to the diversity of the intermediate soft re-scatterings, i.e. elastic and diffractive for which we have different soft amplitudes $a_{i,k}$, each of which is convoluted with a different probability $P^S_{i,k}$ which depends on a different interaction radius $R^2_{i,k}$. In the KKMR model the two channels relate to two different dynamical options of the hard process. In model A the separation is between valence and sea interacting partons. In model B the separation is between small and large dipoles. The two models give compatible results. The key point, though, is that the KKMR opacities $\Omega_{i,k}$, in the definition of $P^S_{i,k}$, differ in their normalization, but have the same b-dependence. Regardless of this difference the output of the GLM and KKMR models is reasonably compatible. The compatibility between GLM and KKMR is not surprising since the explicit KKMR calculation of the hard LRG amplitude is approximated relatively well by the GLM simple Gaussian.

Our final conclusion is that the two model output sets are compatible. The richness of the KKMR model has a significant discovery potential lacking in GLM. On the other hand, the GLM simplicity makes it very suitable as a platform to present different models in a uniform way, which enables a transparent comparison.

## 5 Discussion

As we shall see, at the end of this section, there is no significant difference between the values of $\sigma_{tot}$ predicted by DL and GLM up to the top Cosmic Rays energies. This is, even though, DL is a Regge model without unitarity corrections. The explanation for this "paradox" is that the DL amplitude violations of s-unitarity are confined, even at super high energies, to small b which does not contribute significantly to $\sigma_{tot}$. Note, though, that $\frac{\sigma_{el}}{\sigma_{tot}}$ grows in DL like $s^{\epsilon}$ whereas in GLM its growth is continuosly being moderated with increasing s (see table in **5.3**). The DL model predicts that $S^2$ is identical to unity or very close to it in the DL high-$t$ model where a weak $I\!\!P I\!\!P$ cut is added. The need for survival probabilities so as to reproduce the the experimental soft SD cross section values and the hard di-jets rates, is the most compelling evidence in support of unitarization at presently available energies. As such, the study of high energy soft and hard diffraction serves as a unique probe substantiating the importance of s-channel unitarity in the analysis of high energy scattering.

### 5.1 $S^2$ in unitarized models

Most, but not all, of the unitarized models dealing with LHC $S^2$ predictions have roughly the same $S^2$ values. This calls for some clarifications. The first part of our discussion centers on the correlated investigation of two problems:

1) How uniform are the output predictions of different unitarization procedures?
2) How sensitive are the eikonal calculations to the details of the eikonal model they use?

We start with two non eikonal models which have contradictory predictions.

The first is a model suggested by Troshin and Tyurin [52]. In this model the single channel unitarity constraint (Eq.(13)) is enforced with an asymptotic bound where $G_{in} = 0$ and $|a_{el}| = 2$ i.e. asymptotically, $\sigma_{tot} = \sigma_{el}$ and $P^S(s,b) = 1$. The parameters of the model are set so as to obtain a "normal" survival probability monotonically decreasing with energy up to about $2500\,GeV$ where it changes its behavior and rises monotonically to its asymptotic limit of 1. Beside the fact that the model has a legitimate but non appealing asymptotics, its main deficiency is that it suggests a dramatic change in the systematics of $S^2$ without being able to offer any experimental signature to support this claim. Regardless of this criticism, this is a good example of a proper unitarity model whose results are profoundly different from the eikonal model predictions.

Another non eikonal procedure is Goulianos flux renormalization model [17]. This is a phenomenological model which formally does not enforce unitarity, but rather, a bound of unity on the Pomeron flux in diffractive processes. Note that, the Pomeron flux is not uniquely defined so this should be regarded as an ad hoc parametrization. Nevertheless, it has scored an impressive success in reproducing the soft and hard diffractive data in the ISR-Tevatron range. The implied survival probabilities of this procedure are compatible with GLM and KKMR. However, the model predicts suppression factors for the diffractive channels which are $t$-independent and, thus, b-independent. The result is that, even though the output diffractive cross section is properly reduced relative to its input, there is no change of the output profile from its input Gaussian form. Consequently, the Pumplin bound is violated. We are informed that Goulianos plans to improve his model by eikonalizing the output of his present model.

As noted, there are a few eikonal models on the market [73–80], and their predictions are compatible with GLM and KKMR. Reconsidering the procedure of these calculations, their compatibility is not surprising once we translate their input to a GLM format. The GLM eikonal $S^2$ calculation has two input sectors in either a single or a two channel version. They are the soft $\nu$ and $R^2$, and the hard radius $R^{H^2}$. Since the soft input is based on a fit of the soft scattering data base, the potential variance in the soft parameters is relatively small. The input hard radius is obtained from either the HERA data or a theoretical calculation, be it a pQCD diagram or a Regge model. All in all, this is a reasonably stable input. In this context, it is interesting to discuss the eikonal model of Block and Halzen [73], where the calculated survival probabilities for Higgs production through W-W fusion are seemingly too high, $S^2(540) = 27\%$, $S^2(1800) = 21\%$ and $S^2(14000) = 13\%$. Even though, Higgs production is a CD process, the above $S^2$ values are in agreement with the KKMR calculations of $S^2_{DD}$ with a relatively high $R^{H^2} = 11\,GeV^{-2}$. In a proper $S^2_{CD}$ calculation, these high $S^2$ values correspond to an even higher $R^{H^2} \simeq 20\,GeV^{-2}$, which is far too high as an estimate of the hard radius of $WW \rightarrow H$. A possible interpretation of Block-Halzen results is to associate them with a soft, rather than a hard, LRG CD process. This would couple with the non screened interpretation of CD Higgs through the soft CEM model [74, 75], which predicts very high $S^2$ values. Since the CEM model is not screened we may, as well, assign a survival probability to its output result. This translates into $S^2_{CD} = S^2_{BH} S^2_{CEM}$, providing rather reasonable one channel predictions, $S^2_{CD}(540) = 18.9\%$ and $S^2_{CD}(1800) = 7.2\%$.

Obviously, each of the eikonal models, quoted above has its own particular presentation and emphasis. They do, however, have compatible results reflecting the observation that their input translates into similar values of $\nu$, $R^2$ and $R^{H^2}$.

### 5.2 Compatibility between HERA and the Tevatron di-jets data

Much attention has been given recently to the compatibility between the Tevatron and HERA DIS GJJ data. The starting point made by KKMR and CDF is that rather than depend on a p.d.f. input to calculate $F_{gap}$, we may use, the GJJ di-jets diffractive structure function, $F^D_{jj}$, inferred from HERA DIS data [18–27] and associate it with the $F^D_{jj}$ derived from the Tevatron GJJ data. As it stands, this procedure ignores the role of the survival probability. Consequently, $F^D_{jj}$ obtained from the Tevatron is an order of magnitude smaller than the HERA output [8–12, 17, 40–44]. This result led to speculations about a possible breaking of QCD or Regge factorization or both. Once the Tevatron di-jets diffractive structure function is rescaled by the appropriate survival probability, the compatibility between the Tevatron and HERA DIS diffractive data is attained. The conclusion of this analysis is that the breaking of factorization is attributed to the soft re-scatterings of the the colliding projectiles. Additional hard contribution to the factorization breaking due to gluon radiation is suppressed by the Sudakov factor included in the pQCD calculation (see **4.3**).

One should note, though, that the H1 determination [18–27] of $F^D_{jj}$ is not unique. Arneodo [72] has suggested a different $F^D_{jj}$ output based on HERA di-jets data which has a different normalization and $\beta$ dependences. Should this be verified, there might well be a need to revise the KKMR calculations.

The evolution of HERA $F^D_{jj}$ from high $Q^2$ DIS to $Q^2 = 0$ di-jets photoproduction has raised additional concern with regard to the validity of the factorization theorems [28,29]. This is a complicated analysis since one has to be careful on two critical elements of the calculations:

1) The determination of the ratio between direct and resolved exchanged photon (real or virtual). This is a crucial element of the theoretical calculation since survival probability is applicable only to the resolved photon component. For very high $Q^2$ data the hard scattering process with the target partons is direct. At $Q^2 = 0$ there is a significant resolved photon contribution.
2) For di-jets production there is a big difference between the LO and the NLO pQCD calculated cross sections [81–83]. Since the HERA analysis compares the pQCD calculation with the di-jets measured cross section the normalization and shape of the theoretical input is most crucial in the experimental comparison between the high $Q^2$ and $Q^2 = 0$ data.

On the basis of a NLO calculation, Klasen and Kramer [81, 82] conclude that they can reproduce the photoproduction data with $S^2 = 0.34$, applied to the resolved sector. This survival probability is in agreement with KKMR and GLM calculations.

Regardless of the above, preliminary photoproduction GJJ HERA data [28, 29] suggest that both the direct and resolved photon sectors are suppressed at $Q^2 = 0$. A verification of this observation has severe consequences for our understanding of the evolution of the diffractive structure function from DIS to photoproduction. It does not directly relate, though, to the issue of soft survival probability which apply, per definition, only to the resolved photon sector. The suggested effect in the direct photon sector should, obviously be subject to a good measure of caution before being substantiated by further independent analysis.

**Table 3:** GLM two-channel predictions at a few energies

| $\sqrt{s}$ [GeV] | $\sigma_{\rm tot}^{\rm DL}$ [mb] | $\sigma_{\rm tot}^{\rm GLM}$ [mb] | $\sigma_{\rm el}^{\rm GLM}$ [mb] | $\sigma_{\rm sd}^{\rm GLM}$ [mb] | $B_{\rm el}^{\rm GLM}$ [GeV$^{-2}$] | $S_{\rm CD}^{\rm GLM 2}$ |
|---|---|---|---|---|---|---|
| 540 | 60.1 | 62.0 | 12.3 | 8.7 | 14.9 | 0.066 |
| 1800 | 72.9 | 74.9 | 15.9 | 10.0 | 16.8 | 0.055 |
| 14000 | 101.5 | 103.8 | 24.5 | 12.0 | 20.5 | 0.036 |
| 30000 | 114.8 | 116.3 | 28.6 | 12.7 | 22.0 | 0.029 |
| 60000 | 128.4 | 128.7 | 32.8 | 13.2 | 23.4 | 0.023 |
| 90000 | 137.2 | 136.5 | 35.6 | 13.5 | 24.3 | 0.019 |
| 120000 | 143.6 | 142.2 | 37.6 | 13.7 | 24.9 | 0.017 |

### 5.3 Diffraction at energies above the LHC

We end with Table 3, which shows the GLM two channel predictions for energies including the LHC, and up to the top Cosmic Rays energies. The, somewhat, surprizing observation is that the GLM calculated total cross sections are compatible with the DL simple Regge predictions all over the above energy range. This is a reflection of the fact that even at exceedingly high energies unitarization reduces the elastic amplitude at small enough b values to be relatively insensitive to the calculation of $\sigma_{tot}$. On the other hand, we see that $\sigma_{el}$ becomes more moderate in its energy dependence and $\sigma_{el}/\sigma_{tot}$ which is $23.6\%$ at the LHC is no more than $26.4\%$ at the highest Cosmic Rays energy, $120\,TeV$. The implication of this observation is that the nucleon profile becomes darker at a very slow rate and is grey (well below the black disc bound) even at the highest energy at which we can hope for a measurment. A check of our results at the Planck scale shows $\sigma_{tot} = 1010\,mb$ and the profile to be entirely black. i.e., $\frac{\sigma_{el}}{\sigma_{tot}} = \frac{1}{2}$. $\sigma_{sd}$ is even more moderate in its very slow rise with energy. The diminishing rates for soft and hard diffraction at exceedingly high energies are a consequence of the monotonic reduction in the values of $S^2$ with a Planck scale limit of $S^2 = 0$. This picture is bound to have its effect on Cosmic Rays studies.

Our LHC predictions are compatible with KMR. Note, though, that: i) $\sigma_{sd}^{GLM}$ is rising slowly with $s$ gaining $20\%$ from the Tevatron to LHC. KMR has a much faster rise with energy, where, $\sigma_{sd}^{KMR}$ is gaining $77\% - 92\%$ over the same energy interval. ii) At the LHC $B_{el}^{GLM} = 20.5\,GeV^{-2}$, to be compared with a DL slope of $19\,GeV^{-2}$ and a KMR slope of $22\,GeV^{-2}$. The GLM $30\,TeV$ cross sections are compatible with Block-Halzen.

## 6 Acknowledgements

We are very thankful to our colleagues Valery Khoze, Alan Martin, Misha Ryskin and Leif Lönnblad, who generously contributed to **Section 4** and the **Appendix**. Needless to say, they bear no responsibility for the rest of this review.

## Appendix: Monte Carlo modeling of gap survival

The following was contributed by Leif Lönnblad and is presented without any editing.

An alternative approach to gap survival and factorization breaking is to implement multiple interactions in Monte Carlo event generators. These models are typically based on the eikonalization of the partonic cross section in hadronic collisions and can be combined with any hard sub process to describe the additional production of hadrons due to secondary partonic scatterings. Some of these programs, such as PYTHIA [84, 85] and HERWIG/JIMMY [86–88], are described in some detail elsewhere in these proceedings [89]. Common for all these models is that they include exact kinematics and flavour conservation, which introduces some non-trivial effects and makes the multiple scatterings process-dependent. Also, the predictions of the models are very sensitive to the cutoff used to regularize the partonic cross section and to the assumptions made about the distribution of partons in impact parameter space. Nevertheless, the models are quite successful in describing sensitive final-state observables such as multiplicity distributions and jet-pedestal effects [89]. In particular this is true for the model in PYTHIA which has been successfully tuned to Tevatron data[4] by Rick Field [90], the so-called *CDF tune A*.

The PYTHIA model does not make any prediction for the energy dependence of the total cross section - rather this is an input to the model used to obtain the distribution in the number of multiple interactions. PYTHIA can, however, make predictions for gap survival probabilities. This was first done for Higgs production via W-fusion [2], and amounts to simply counting the fraction of events which do not have any additional scatterings besides the W-fusion process. The basic assumption is that any additional partonic scattering would involve a colour exchange which would destroy any rapidity gap introduced by W-fusion process. Since PYTHIA produces complete events, these can also be directly analyzed with the proper experimental cuts. A similar estimate was obtained for the gaps between jets process, both for the Tevatron and HERA case [91].

Recently, PYTHIA was used to estimate gap survival probabilities also for the case of central exclusive Higgs production [92]. As in the case of gaps between jets, the actual signal process is not implemented in PYTHIA, so direct analysis with proper experimental cuts was not possible. Instead a similar hard sub process was used (standard inclusive Higgs production via gluon fusion in this case) and the fraction of events without additional secondary partonic scatterings was identified with the gap survival probability. Using the *CDF tune A* the gap survival probability was estimated to be 0.040 for the Tevatron and 0.026 for the LHC. This is remarkably close both to the values used in [64] obtained in the KKMR model [43], and to the GLM values presented in section 3.4 especially the two-channel ones obtained in [60].

[4]Note that the model in PYTHIA has recently been revised [89]. However, the reproduction of Tevatron data is not as good for the revised model.

# Multi-Jet Production and Multi-Scale QCD

*Z. Czyczula*[1,2], *G. Davatz*[3], *A. Nikitenko*[4], *E. Richter-Was*[1,5,*] , *E. Rodrigues*[6], *N. Tuning*[6]
[1] Institute of Physics, Jagiellonian University, Krakow, Poland
[2] Niels Bohr Institute, University of Copenhagen, Copenhagen, Denmark
[3] Institute for Particle Physics, ETH Zürich, Switzerland
[4] Imperial College, London, UK
[5] Institute of Nuclear Physics PAN, Krakow, Poland
[6] NIKHEF, Amsterdam, The Netherlands

**Abstract**

We summarize the contributions in Working Group II on "Multi-jet final states and energy flows" related to the topic of jet production, multi-jet topologies and multi-scale QCD. Different parton shower models will lead to systematic differences in the event topology. This may have a significant impact on predictions for the LHC. Here we will look at a few examples, such as the acceptance of $H \to \tau\tau$ events and in applying a jet veto in the non-hadronic $H \to WW \to l\nu l\nu$ decay channel. We also study the effect of CCFM evolution on the jet veto and on the event topology at the LHC in the forward region. Finally, we show that the choice of the QCD scale leads to large uncertainties in e.g. the $H \to \tau\tau$ analysis.

## 1 Introduction

In simulating high-energy interactions, the sequence of branchings such as $q \to qg$, can be modelled by calculating the exact amplitude of the Feynman diagrams, known as the matrix-element method, or, alternatively, can be modelled using the parton-shower approach. Matrix elements are in principle the exact approach but lead to increasingly complicated calculations in higher orders, and are therefore only used for specific exclusive physics applications, such as background estimates with multiple hard jets (see also [1]).

Since no exact way of treating partonic cascades exist, various Monte Carlo programs model the parton showers in different ways. In HERWIG [2] the parton showers are performed in the soft or collinear approximation, treating the soft gluon emission correctly. The shower is strictly angular ordered, where the angle between emitted partons is smaller at each branching. The hardest gluon emission is then matched to the first order matrix-element. This matrix-element correction has recently been implemented for $gg \to H$, leading to harder jets, and thus a more stringent jet veto in e.g. the non-hadronic decay $H \to WW \to l\nu l\nu$, where the jet veto is crucial to reduce the top background. PYTHIA [3] applies the collinear algorithm with the cascade ordered according to the virtuality $Q^2$. Corrections to the leading-log picture using an angular veto, lead to an angular ordering of subsequent emissions. The initial parton branchings are weighted to agree with matrix-elements. ARIADNE [4] on the other hand, does not emit gluons from single partons, but rather from the colour dipoles between two dipoles, thus automatically including the coherence effects approximated by angular ordering in HERWIG. From the resulting two dipoles softer emission occurs, resulting in a $p_T$ ordering of subsequent emissions. ARIADNE has proven to predict the event shapes at HERA accurately [5], and could be explored more widely for simulation studies for the LHC.

The way parton showers are implemented affects the emission of soft gluons, and therefore affect both the transverse momentum of the produced Higgs, as well as the $p_T$ of the balancing jets. In the following we will discuss the effect of the different parton showers on the selection of $H \rightarrow \tau\tau$ by applying angular cuts on the jets and on the selection of $H \rightarrow WW \rightarrow l\nu l\nu$ by rejecting events with jets with large $p_T$.

* Supported in part by the Polish Government grant KBN 1 P03 091 27 (years 2004-2006) and by the EU grant MTKD-CT-2004-510126, in partnership with the CERN Physics Department.

Both PYTHIA and HERWIG are general purpose leading order (LO) parton shower Monte Carlo programs, based on LO matrix elements. MC@NLO [6] on the other hand, uses exact next-to-leading order (NLO) calculations and is matched to the HERWIG parton shower Monte Carlo. Its total cross section is normalized to NLO predictions. The different predictions of these programs for the high part of the transverse momentum spectrum of the Higgs will be described in detail.

In the parton cascade as implemented in e.g. PYTHIA, the parton emissions are calculated using the DGLAP approach [7], with the partons ordered in virtuality. DGLAP accurately describes high-energy collisions of particles at moderate values of the Bjorken-$x$ by resummation of the leading log terms of transverse momenta ($(\alpha_s \ln Q^2)^n$). However, to fixed order, the QCD scale used in the ladder is not uniquely defined. Different choices of the scale lead to large differences in the average transverse momentum of the Higgs in e.g. the processes $gb \rightarrow bH$ and $gg \rightarrow bbH$.

In the CCFM formalism [8] there is no strict ordering along the parton ladder in transverse energy, contrary to the DGLAP formalism. The CASCADE Monte Carlo program [9] has implemented the CCFM formalism, inspired by the low-$x$ $F_2$ data and forward jet data from HERA, and became recently available for $pp$ scattering processes. Until now, CASCADE only includes gluon chains in the initial state cascade. Different sets of unintegrated gluon densities are available, which all describe HERA data equally well [9]. Note, however, that it is questionable if these densities are constrained enough for Higgs production, as discussed elsewhere in these proceedings [10].

CCFM is expected to provide a better description of the gluon evolution at very low values of $x$ compared to DGLAP, as it also takes leading-logs of longitudinal momenta ($(\alpha_s \ln x)^n$) into account. Since the partons at the bottom of the ladder (furthest away from the hard scatter) are closest in rapidity to the outgoing proton, effects might be expected in the forward region. The event topology in terms of jets and charged multiplicity is investigated at rapidities $2<\eta<5$, corresponding to the acceptance of the LHCb detector.

## 2 MSSM Higgs production with the Yukawa bbH coupling induced mechanisms

In the MSSM, the Yukawa coupling of the heavy neutral Higgs bosons to the bottom quarks is strongly enhanced for large tan($\beta$) with respect its SM value, which makes the Higgs boson production in association with bottom quarks the primary production mechanism in LHC $pp$ collisions. Currently, the inclusive cross section for this process is under good control up to NNLO, both in the so called fixed-flavour-scheme (FFS) and varying-flavour-scheme (VFS). The impressive level of theoretical uncertainty in the order of 15% is achieved on the predictions for the total cross-section for $m_H$=120 GeV [11, 12].

The observability potential for the $H \rightarrow \tau\tau$ channel [13] is, however, very sensitive to the topology of the events, due to the reconstruction of the invariant mass of the tau-pair, using the collinear approximation of $\tau$-leptons decay, in order to account for the neutrino momenta. The impact of the event topology on the final acceptance of the signal has been discussed elsewhere [14]. Here, we pursue the subject further and we study more quantitatively the systematic effects from the parton shower model and the choice of the QCD scale selected in the event generation.

Currently available Monte Carlo generators for the Higgs boson production are based on the LO matrix elements, with the QCD part of physics event simulated with a parton shower approach. Clearly, the kinematics of the Higgs boson (and therefore the final acceptance for the signal) depends strongly on the algorithm used to simulate the QCD cascade. At tree level, the following exclusive processes have been studied, combining the observability of events with and without spectator b-tagged jets accompanying the reconstructed tau-pair: $gb \rightarrow bH$ (VFS), $gg \rightarrow b\bar{b}H$ (FFS), $b\bar{b} \rightarrow H$ (VFS) and $gg \rightarrow H$.

For the purpose of the discussion presented here we have studied the SM Higgs boson production with a mass of 120 GeV, decaying into a tau pair, where one tau decays hadronically and one leptonically. The reconstruction of the Higgs boson mass and the selection criteria were performed at the level of generated particles (leptons, hadrons) or, where necessary (missing energy, b-jets), on objects reconstructed from simplified simulation of the detector response [15].

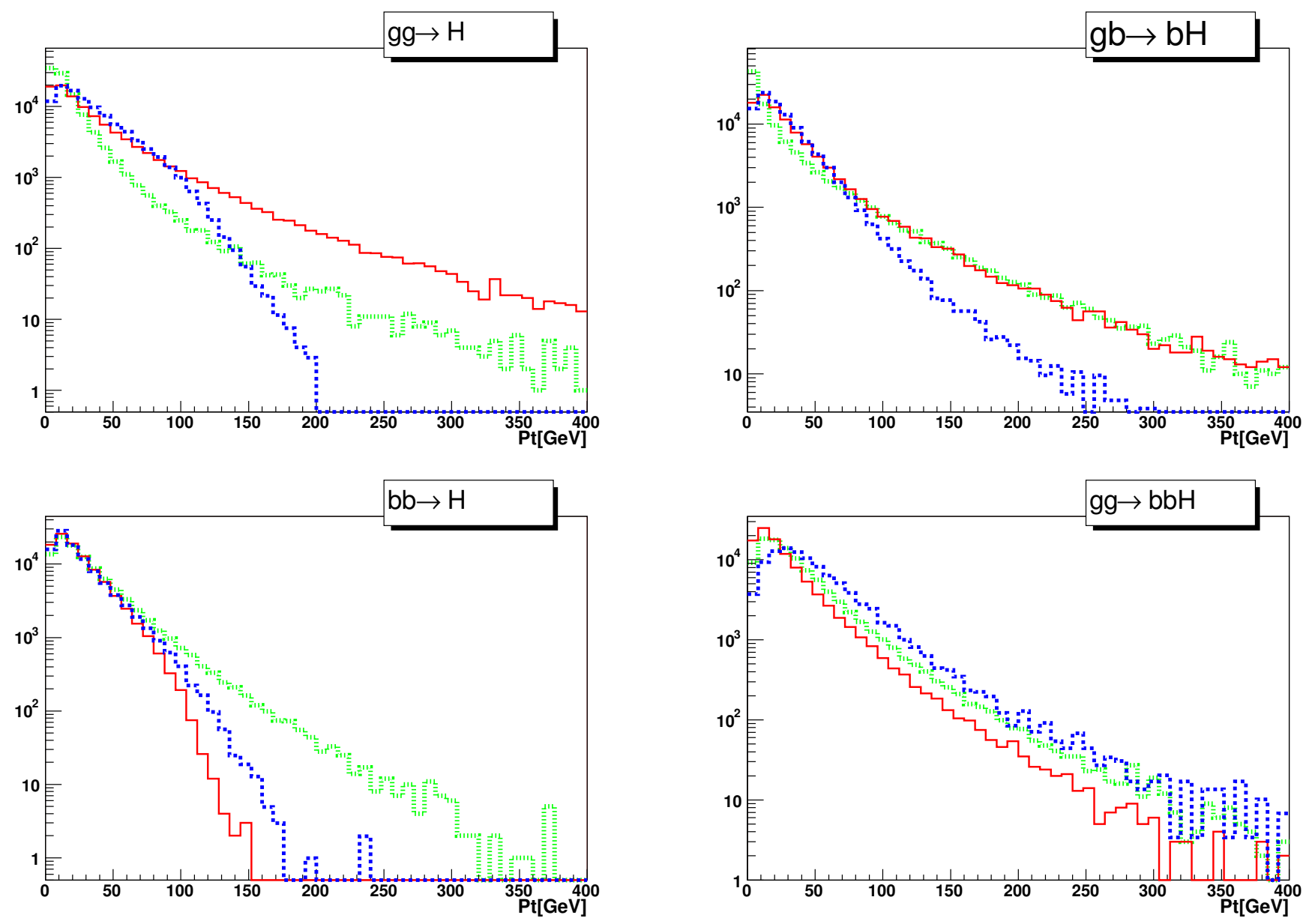

**Fig. 1:** The transverse momenta of the Higgs boson, $p_T^{\mathrm{Higgs}}$ for 3 different shower models for each production mechanism. The red solid line represents PYTHIA, the dashed green line ARIADNE and the dotted blue line HERWIG events. The vertical scale gives the number of events per bin, and a total of $10^5$ events have been generated with each program.

### 2.1 Systematics from the choice of parton shower model

As discussed in the introduction, the various parton shower models predict different spectra of the transverse momentum, $p_T^{\mathrm{Higgs}}$, of the produced Higgs boson. This leads to a large variation in the prediction for the fraction of accepted events. The obvious starting point for the discussion is the Higgs boson transverse momentum spectra in complete physics events [1]. In case of the 2→2 and 2→3 processes, the $p_T$ of the Higgs boson arrises predominantly from matrix elements, whereas in the 2→1 events $p_T^{\mathrm{Higgs}}$ purely comes from the parton shower. Therefore, the Higgs transverse momentum spectra differ significantly for different models of the QCD cascade. Figure 1 shows these spectra for each production mechanism [2].

Clearly, the spectra of the Higgs boson transverse momenta show substantial dependence not only on the topology of the hard process, but also on the shower model used in the simulation of the event. The shower model as implemented in PYTHIA includes hard matrix element corrections for inclusive gluon-gluon fusion, $gg \to H$, hence leading to a harder spectrum compared to the one obtained from the standard HERWIG shower. In this production mode the shower model from ARIADNE fails because of the missing splitting kernel for $g \to q\bar{q}$. On the other hand, the ARIADNE model predicts the hardest spectra for the process $b\bar{b} \to H$. In this production channel, predictions from PYTHIA and HERWIG are in quite good agreement. However, almost a factor of two difference for the prediction of the mean transverse momenta can be reported between PYTHIA and HERWIG in $gg \to b\bar{b}H$ process.

[1] The AcerMC 2.4 framework [16] with interfaces to PYTHIA 6.2, ARIADNE 4.12 and HERWIG 6.5 was used to generate events and AcerDET [15] was used to simulate the detector performance.

[2] The CTEQ5L parton density functions were used in all simulations. It has been checked that both final acceptance of the signal and the mean Higgs boson transverse momentum is almost independent of the pdf parametrization. Uncertainties below 10% are observed by using CTEQ5L, CTEQ6L, MRST2001 interfaced with LHAPDF [17]).

**Table 1:** The average transverse momenta of the Higgs boson and acceptance of selection criteria for different hard processes and parton shower models. Events were generated with default initialization of these generators. Columns marked PY, AR and HW denote results from PYTHIA, ARIADNE and HERWIG shower model respectively.

| Hard process | $\mathbf{gg \to H}$ | | | $\mathbf{b\bar{b} \to H}$ | | |
|---|---|---|---|---|---|---|
| Shower model | PY | AR | HW | PY | AR | HW |
| $< p_T^{\rm Higgs}$ (generated)> (GeV) | 37.2 | X | 32.2 | 23.1 | 29.9 | 24.6 |
| $< p_T^{\rm Higgs}$ (accepted)> (GeV) | 129.4 | X | 75.27 | 58.6 | 91.64 | 68.4 |
| basic selection | 14.2% | X | 12.7% | 12.8% | 13.8% | 11.8% |
| +( $cos(\phi)$ >-0.9 , $\|sin(\phi)\|$ >0.2 ) | 5.5% | X | 4.5% | 2.9% | 4.3% | 2.7% |
| +($p_T^{miss} >$ 30 GeV, $m_T^{lep-miss}$ <50GeV) | 3.8% | X | 2.3% | 1.4% | 2.3% | 1.5% |
| +( mass window: 120 $\pm$ 20 GeV ) | 2.4% | X | 1.3% | 0.6% | 1.3% | 0.6% |
| +( 1 tagged b-jet) | | | | 0.4% | 1.0% | 0.4% |
| Hard process | $\mathbf{gb \to bH}$ | | | $\mathbf{gg \to b\bar{b}H}$ | | |
| Shower model | PY | AR | HW | PY | AR | HW |
| $< p_T^{\rm Higgs}$ (generated)> [GeV] | 32.5 | 26.0 | 26.9 | 27.2 | 35.8 | 47.4 |
| $< p_T^{\rm Higgs}$ (accepted)> [GeV] | 125.1 | 133.9 | 82.1 | 95.0 | 99.6 | 105.3 |
| basic selection | 13.3% | 12.6% | 11.7% | 13.0% | 13.6% | 12.1% |
| +( $cos(\phi) > -0.9, \|sin(\phi)\|$ >0.2 ) | 4.4% | 3.4% | 3.2% | 3.5% | 5.1% | 6.7% |
| +($p_T^{miss} >$ 30 GeV,$m_T^{lep-miss}$ <50GeV) | 2.7% | 2.4% | 1.7% | 2.0% | 2.9% | 3.8% |
| +( mass window: 120 $\pm$ 20 GeV ) | 1.7% | 1.5% | 0.9% | 1.1% | 1.8% | 2.6% |
| +( 1 tagged b-jet) | 1.3% | 1.4% | 0.6% | 0.9% | 1.2% | 2.1% |

Numerical values for the average Higgs boson transverse momentum in different production processes and parton shower models are given in Table 1. It is important to stress that these results were obtained with default settings of the parameters for each parton shower model.

The steps of the analysis that lead to the reconstruction of the tau-pair invariant mass are indicated in Table 1, including the acceptances for all the discussed production processes and parton shower models. They consist of the basic selection (including the trigger and $p_T$ and $|\eta|$ cuts on the lepton and jet), and the additional selection that is needed to improve the mass resolution of the accepted tau-pair. The acceptance of the signal after the basic selection is rather stable, at the level of 12%-14% depending on the production mechanism. The significant differences start to appear when a cut on the angle between the lepton and hadron is applied. A difference of almost a factor two is observed for the $b\bar{b} \to H$ production process with the parton shower from the HERWIG or ARIADNE model, respectively.

For the final acceptance values, the uncertainty from the parton shower model varies between 85% for inclusive gluon fusion to 135% for $gg \to bbH$ (between HERWIG and PYTHIA models). In the case of the Higgs production through $bb \to H$, predictions from HERWIG and PYTHIA models are in excellent agreement. However, the prediction of the acceptance in this production channel differs by 115% if the parton shower from ARIADNE is used. For the $gb \to bH$ production mechanism, the uncertainty due to the shower model from either PYTHIA or HERWIG is about 90%.

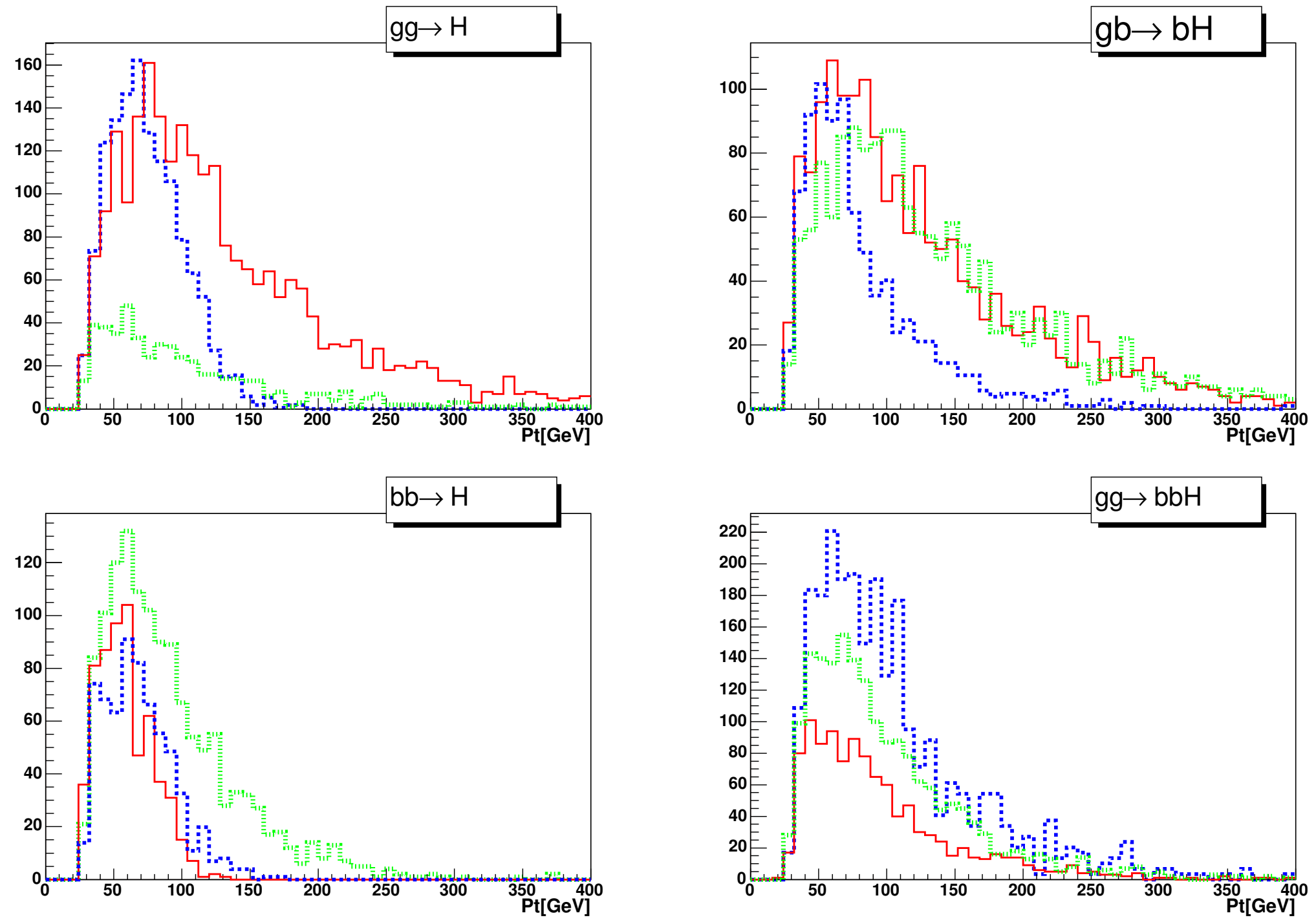

**Fig. 2:** Same as Fig. 1 but after selection presented in Table 1. The vertical scale in in number of events entering given bin after selection procedure, in each case $10^5$ events were initially generated.

The systematic theoretical uncertainty on the predictions for the final acceptance ranges from 85% to 135% for the three different shower models studied here. The uncertainty is even larger, when the requirement of an additional tagging b-jet is introduced, up to 170% for $bb \to H$ [3]. Figure 2 shows the Higgs boson transverse momentum for those events that passed all selection criteria. As can be observed, the selection criteria rejected most of events with $p_T^{\rm Higgs} < 40$ GeV.

### 2.2 Systematics from the choice of QCD scale

Having considered here the available Monte Carlo generators with the overall precision of the leading order only, large uncertainties are expected for the predictions coming from different scale choices. Here we concentrate only on the effects on the event topology, neglecting the effects from the choice of the QCD scale on the total cross-section. Table 2 shows the Higgs boson mean transverse momentum and final acceptance of the signal for 2→2 and 2→3 processes for some possible choices in PYTHIA and ARIADNE. The $Q^2$ value sets the scale not only for the hard scattering process, but also for the initial state parton shower. For the 2→1 production, the $Q^2$ scale is naturally set to be the mass of the Higgs boson mass. The uncertainty in the acceptance due to scale choice for the $gg \to b\bar{b}H$ production mechanism is about 60% in the case of PYTHIA and 25% in the case of ARIADNE parton shower model. For the exclusive process $gb \to bH$, the uncertainties are 75% and 100%, respectively.

## 3 gg → H at the LHC: Uncertainty due to a Jet Veto

In the Higgs mass range between 155 and 180 GeV, $H \to W^+W^- \to \ell\nu\ell\nu$ is considered to be the main Higgs discovery channel [18, 19]. The signal consists of two isolated leptons with large missing $E_T$ and with a small opening angle in the plane transverse to the beam, due to spin correlations of the $W$-pair. In order to reduce the top background, a jet veto has to be applied. The signal over background ratio is found to be around 2:1 for Higgs masses around 165 GeV. For lower and higher Higgs masses, the signal over background ratio decreases slightly [19]. The experimental cross section $\sigma_{meas}$ of the Higgs signal and other final states is given by:

$$\sigma_{meas} = N_s/(\epsilon_{sel} \times L_{pp}), \tag{1}$$

with $N_s$ being the number of signal events, $\epsilon_{sel}$ the efficiency after all signal selection cuts are applied and $L_{pp}$ the proton-proton luminosity. In order to get an estimate of the cross section uncertainty, the statistical and systematic uncertainties have to be determined. The systematic uncertainties come from the experimental selection, background and luminosity uncertainties. As the signal over background ratio is small in the channel under study, the systematic uncertainties should be known precisely. This study concentrates on the uncertainty of the signal efficiency due to the jet veto, by studying the systematics using different Monte Carlo simulations. To do so, four different parton-shower Monte Carlo programs were used, as described in the introduction. The effect of different parton shower models are discussed by comparing PYTHIA 6.225 [3] and HERWIG 6.505 [2], whereas the comparison to MC@NLO 2.31 [6] leads to an uncertainty estimate of higher-order effects [4]. Then, also CASCADE 1.2009 [9] is studied to compare the DGLAP approach to the CCFM formalism.

[3] It should be stressed, that the problem of the efficiency of b-jet tagging was not touched upon, nor was the problem of the efficiency for the reconstruction of the $\tau$-jet. Discussing these effects, very important for complete experimental analysis, would complicate the problem and dilute the aim of the phenomenological studies presented here.

**Table 2:** The average transverse momenta of the Higgs boson and acceptance of selection criteria for different scale choices. Events were generated with default initialization of these generators. Events marked `PY` and `AR` denote results from PYTHIA and ARIADNE shower model respectively.

| Hard process | $\mathbf{gb} \to \mathbf{bH}$ | | | $\mathbf{gg} \to \mathbf{b\bar{b}H}$ | | | |
|---|---|---|---|---|---|---|---|
| $Q^2$ scale | `default` | $\hat{s}$ | $\frac{2\hat{s}\hat{t}\hat{u}}{\hat{s}^2+\hat{t}^2+\hat{u}^2}$ | `default` | $m_b^2$ | $m_b^2$ | $\hat{s}$ |
| <Q> (GeV) | 94 | 257 | 49 | 27 | 4.8 | 120 | 255 |
| < $p_T^{\text{Higgs}}$ (generated)> (GeV)[`PY`] | 32.5 | 42.7 | 43.2 | 27.2 | 29.8 | 32.1 | 36.2 |
| Acceptance (%) [`PY`] | 1.7 | 2.6 | 2.96 | 1.1 | 1.3 | 1.4 | 1.8 |
| < $p_T^{\text{Higgs}}$ (generated)> (GeV)[`AR`] | 26.0 | 25.5 | 44.9 | 35.8 | 38. | 35.3 | 34.5 |
| Acceptance (%) [`AR`] | 1.5 | 1.6 | 3.1 | 1.8 | 2.1 | 1.7 | 1.7 |

Jets are reconstructed using an iterative cone algorithm with a cone size of 0.5. The leading particle (seed) of the jet is required to have a $p_T$ larger than 1 GeV. The pseudo-rapidity $|\eta|$ of the jet should be smaller than 4.5, corresponding to the CMS detector acceptance [20]. The event is rejected if it contains a jet with a $p_T$ higher than 30 GeV. The Higgs mass for this study was chosen to be 165 GeV, corresponding to the region of phase space with the highest signal over background ratio. First, all events are studied without considering the underlying event. Finally, PYTHIA is also studied including different underlying event schemes.

### 3.1 Matrix Element Corrections

At leading order, the transverse momentum of the Higgs boson, $p_T^{\text{Higgs}}$, is zero. However, parton shower Monte Carlos emit soft gluons which balance the Higgs and introduce a transverse momentum in LO parton shower Monte Carlos. As the Higgs is balanced by jets, the transverse momentum is very sensitive to the jet veto and therefore also the efficiency of a jet veto dependends stongly on $p_T^{\text{Higgs}}$.

In Fig. 3, the normalized $p_T^{\text{Higgs}}$ spectra are shown for PYTHIA, HERWIG and MC@NLO. HERWIG and MC@NLO are very similar at low $p_T$, as can be seen on the linear scale, which is to be expected as the soft and collinear emissions of MC@NLO are treated by HERWIG. Figure 4 shows that PYTHIA predicts a softer leading jet spectrum than HERWIG and therefore also a softer $p_T^{\mathrm{Higgs}}$ spectrum. HERWIG implements angular ordering exactly and thus correctly sums the $LL$ (Leading Log) and part of the $N^kLL$ (Next-to..Leading Log) contributions. However, the current version of HERWIG available does not treat hard radiations in a consistent way. Hence the spectrum drops quickly at high $p_T$, see Fig. 3b). PYTHIA on the other hand does not treat angular ordering in an exact way, but includes hard matrix element corrections. Therefore PYTHIA looks more similar to MC@NLO at high $p_T$. MC@NLO correctly treats the hard radiation up to NLO, combining the high $p_T$ spectrum with the soft radiation of HERWIG.

[4] In the following, HERWIG and PYTHIA use the pdf-set CTEQ5L, whereas MC@NLO uses CTEQ5M.

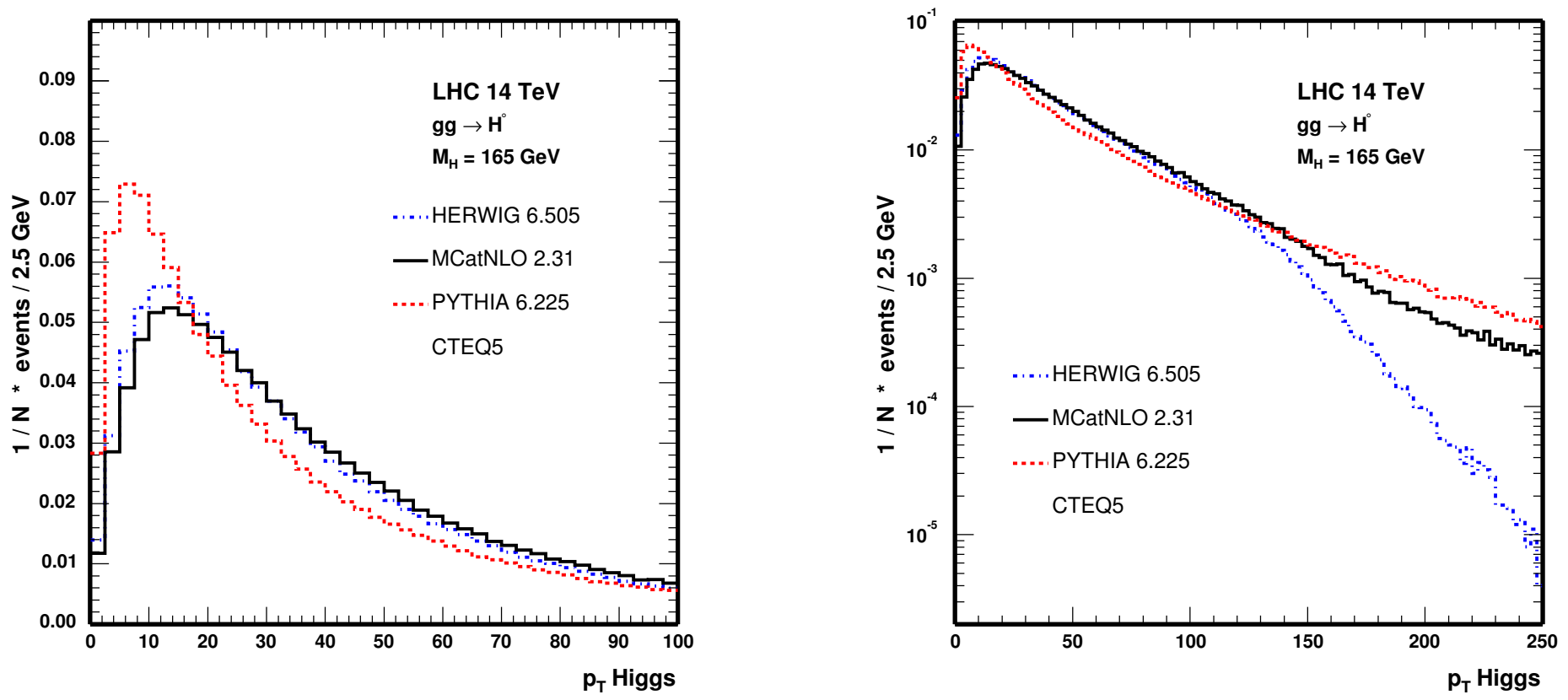

**Fig. 3:** $p_T^{\mathrm{Higgs}}$ spectra for PYTHIA, HERWIG and MC@NLO in linear and logarithmic scale.

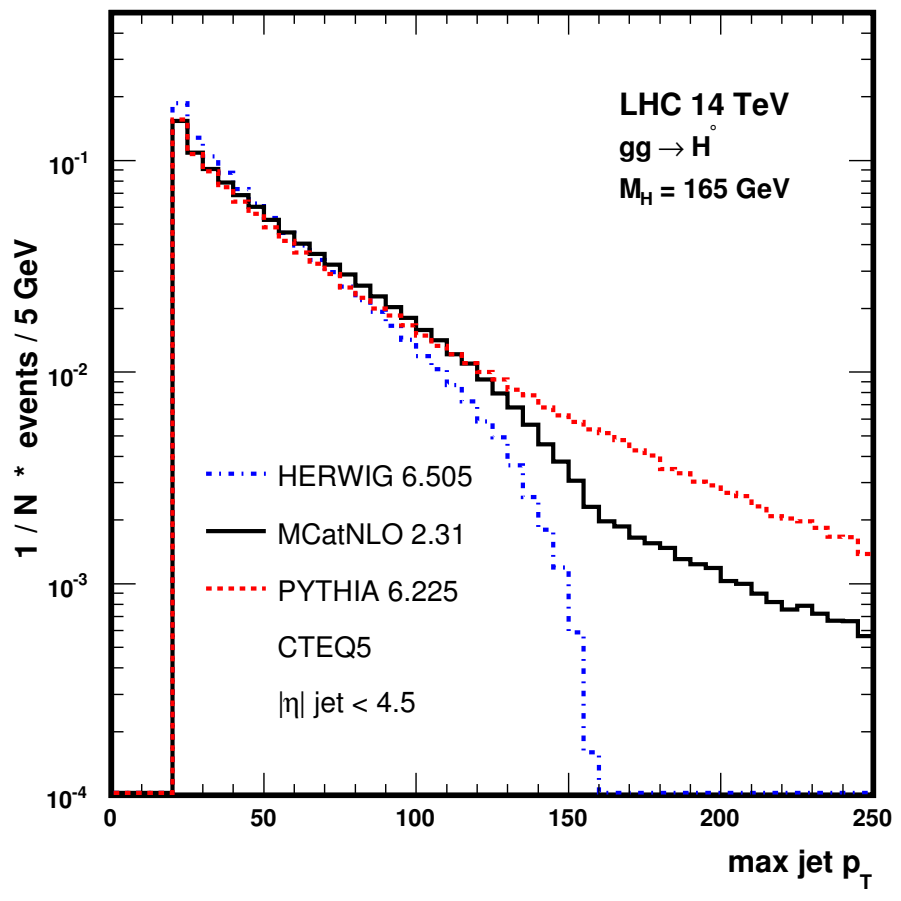

**Fig. 4:** $p_T$ of the leading jet for PYTHIA, HERWIG and MC@NLO

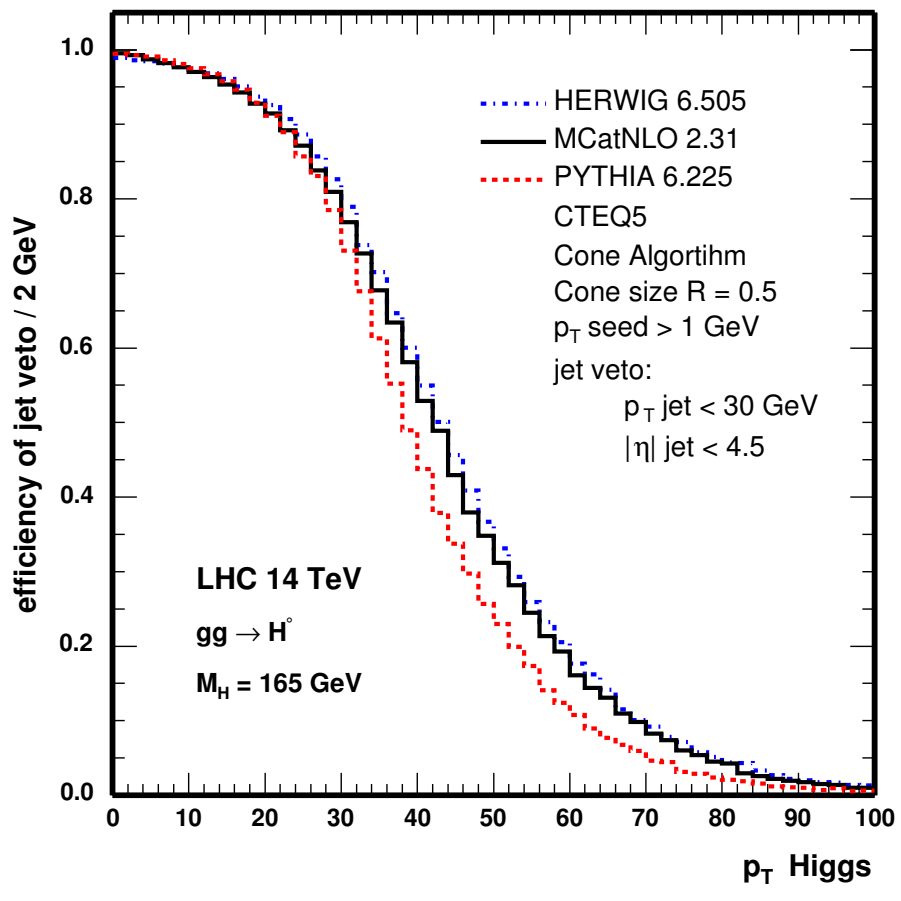

**Fig. 5:** Efficiency of the jet veto of 30 GeV as a function of $p_T^{\mathrm{Higgs}}$.

In Fig. 5, the efficiency of the jet veto is shown for the three different Monte Carlos as a function of $p_T^{\mathrm{Higgs}}$. One observes a strong dependency of $p_T^{\mathrm{Higgs}}$ on the jet veto. Once a jet veto is defined, the efficiency starts to drop quickly as soon as $p_T^{\mathrm{Higgs}}$ is close to the $p_T$ used to define a jet veto. However, as the transverse momentum of the Higgs can be balanced by more than one jet, the efficiency is not zero above this value.

G. Corcella provided a preliminary version of HERWIG including hard matrix element corrections for $gg \to H$ [21]. The hard matrix element corrections lead to harder jets, see Fig. 6, and therefore the jet veto is more effective. At high $p_T$, PYTHIA and HERWIG now show very similar predictions. Table 3 shows the efficiencies of the jet veto of 30 GeV for MC@NLO, PYTHIA and HERWIG with and without matrix element corrections. In addition, the numbers for CASCADE are shown, which will be discussed in more detail later. In the first row, the number of the efficiency for $p_T^{\text{Higgs}}$ between 0 and 80 GeV is shown. The second column shows the inclusive efficiency for all events. One has to keep in mind that after all selection cuts, only the low $p_T$ region is important [19].

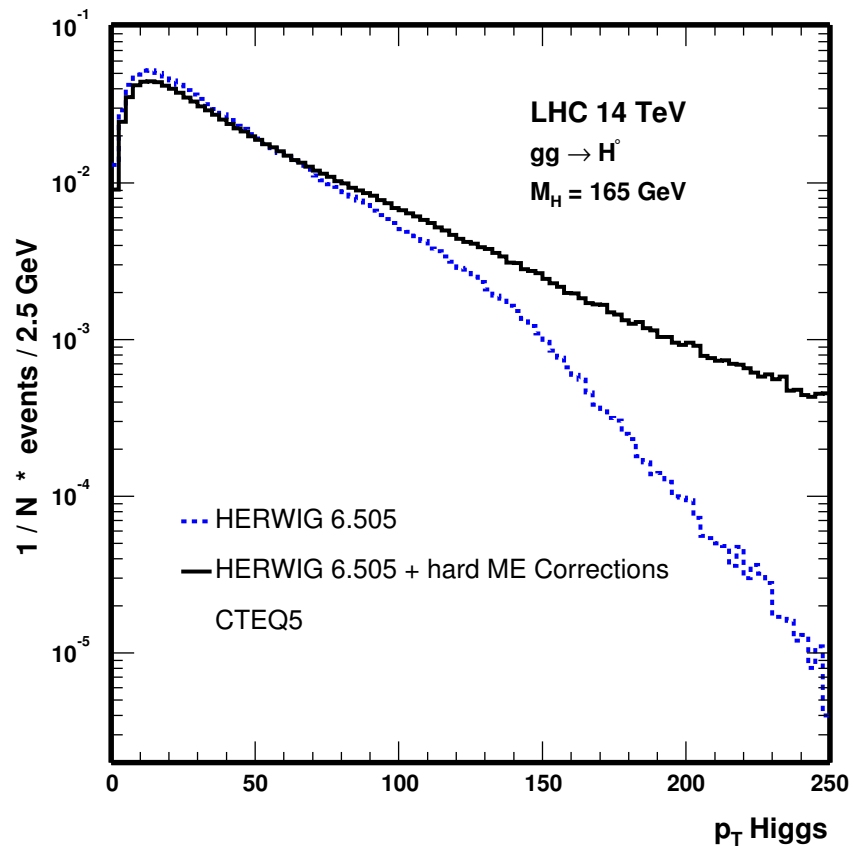

**Fig. 6:** HERWIG with and without hard Matrix Element Corrections, logarithmic scale.

**Table 3:** Efficiency of jet veto for MC@NLO, PYTHIA, HERWIG, HERWIG + ME Corrections and CASCADE

| | Efficiency for events with a $p_T$ Higgs between 0 and 80 GeV | Inclusive efficiency (all events) |
|---|---|---|
| MC@NLO 2.31 | 0.69 | 0.58 |
| PYTHIA 6.225 | 0.73 | 0.62 |
| HERWIG 6.505 | 0.70 | 0.63 |
| HERWIG 6.505 + ME Corrections | 0.68 | 0.54 |
| CASCADE 1.2009 | 0.65 | 0.55 |

In order to estimate the effect from the detector resolution on the jet veto, the $E_T$ of the jet is smeared with the jet resolution of e.g. CMS, as given by [20]:

$$\Delta E_T/E_T = 118\%/\sqrt{E_T} + 7\%. \tag{2}$$

More jets at initially low $p_T$ are shifted to higher $p_T$ than vice versa, as the jets are generally soft. However, the effect of the smearing is limited and the difference between the smeared and unsmeared case is smaller than $1\%$.

In the last years, a lot of progress has been made in understanding the Higgs boson production and decays on a theoretical basis. The gluon fusion cross section has been calculated up to NNLO [22]. Such corrections are known to increase the LO cross section by a factor of more than two. In order to include these higher order corrections in a parton shower Monte Carlo, each event is reweighted with its corresponding $p_T$-dependent effective K-factor (which includes all selection cuts) [19]. This technique can be applied to other processes which are sensitive to jet activity, e.g. the $WW$ background for this channel. The result is an overall effective K-factor of 2.04 for a Higgs mass of 165 GeV, which is only about 15% lower than the inclusive K-factor (without any cuts) for the same mass. This reweighting method allows to optimize the selection cuts and thus also helps to improve the discovery potential. We observe that the uncertaintiy of the jet veto efficiency does not change significantly including those higher order corrections.

**Table 4:** Efficiency numbers for different underlying event tunings in PYTHIA.

| | Efficiency for events with a $p_T$ Higgs between 0 and 80 GeV | Inclusive efficiency (all events) |
|---|---|---|
| PYTHIA no UE | 0.730 | 0.620 |
| PYTHIA default | 0.723 | 0.613 |
| ATLAS tune | 0.706 | 0.600 |
| CDF tune | 0.709 | 0.596 |

### 3.2 Underlying Event

So far all events were generated without considering the underlying event. However, to study a jet veto, it is important to consider also the effect of the underlying event. Therefore, PYTHIA was studied with different underlying event tuning schemes, which are the ATLAS Tune [23], CDF Tune A [24] and PYTHIA default (MSTP(81)=1, MSTP(82)=3 [3]). The different tunings lead to approximately the same efficiency, and also the difference in the efficiency with and without underlying event is smaller than 1%, see Table 4.

### 3.3 Comparing to CCFM evolution

Finally, we compared the PYTHIA, HERWIG and MC@NLO predictions with the ones obtained using CASCADE. One has to keep in mind that this Monte Carlo is dedicated to low-$x$ physics, and is about to be released for LHC physics applications. There were many improvements implemented during this workshop. In Fig. 7, the $p_T^{\rm Higgs}$ spectra for PYTHIA, HERWIG+ME Corrections, MC@NLO and CASCADE are shown. The prediction from CASCADE lies within the ones from PYTHIA and HERWIG. When looking at different $p_T$ regions, one generally observes that CASCADE produces more jets compared to the other Monte Carlos, and the jets are harder. The jet veto efficiency as a function of the $p_T$ of the Higgs is shown in Fig. 8, indicating that the main differences are in the low $p_T$ range and that the efficiency for CASCADE is slightly smaller than unity at a $p_T^{\rm Higgs}$ of zero. A reason for this is that the Higgs boson is balanced by more than one jet, with at least one of the jets with a $p_T$ higher than 30 GeV and thus vetoed. For the same reason, the efficiency in general is lower than for the other Monte Carlo programs at low $p_T^{\rm Higgs}$. Results in the high $p_T$ region have to be studied carefully.

## 4 Forward Studies with CASCADE at LHC Energies

The applicability of DGLAP evolution [7] is known to be limited in the very forward region, that is at small values of Bjorken-$x$, where $ln(x)$ terms are expected to become large [25]. Since the partons at the bottom of the ladder (furthest away from the hard scatter) are closest in rapidity to the outgoing proton, effects might be expected in the forward region. The CCFM evolution [8] takes these BFKL-like terms into account, and is implemented in the CASCADE Monte Carlo program [9].

We have studied the topology of forward particle and jet production in the LHCb detector at the LHC. LHCb is a forward spectrometer covering roughly the forward region 1.8<$\eta$<4.9 [26]. Its main goal is the study of CP violation in the $B$-meson sector and the measurement of rare $B$-decays. But its very nature makes LHCb a suitable environment for QCD forward studies.

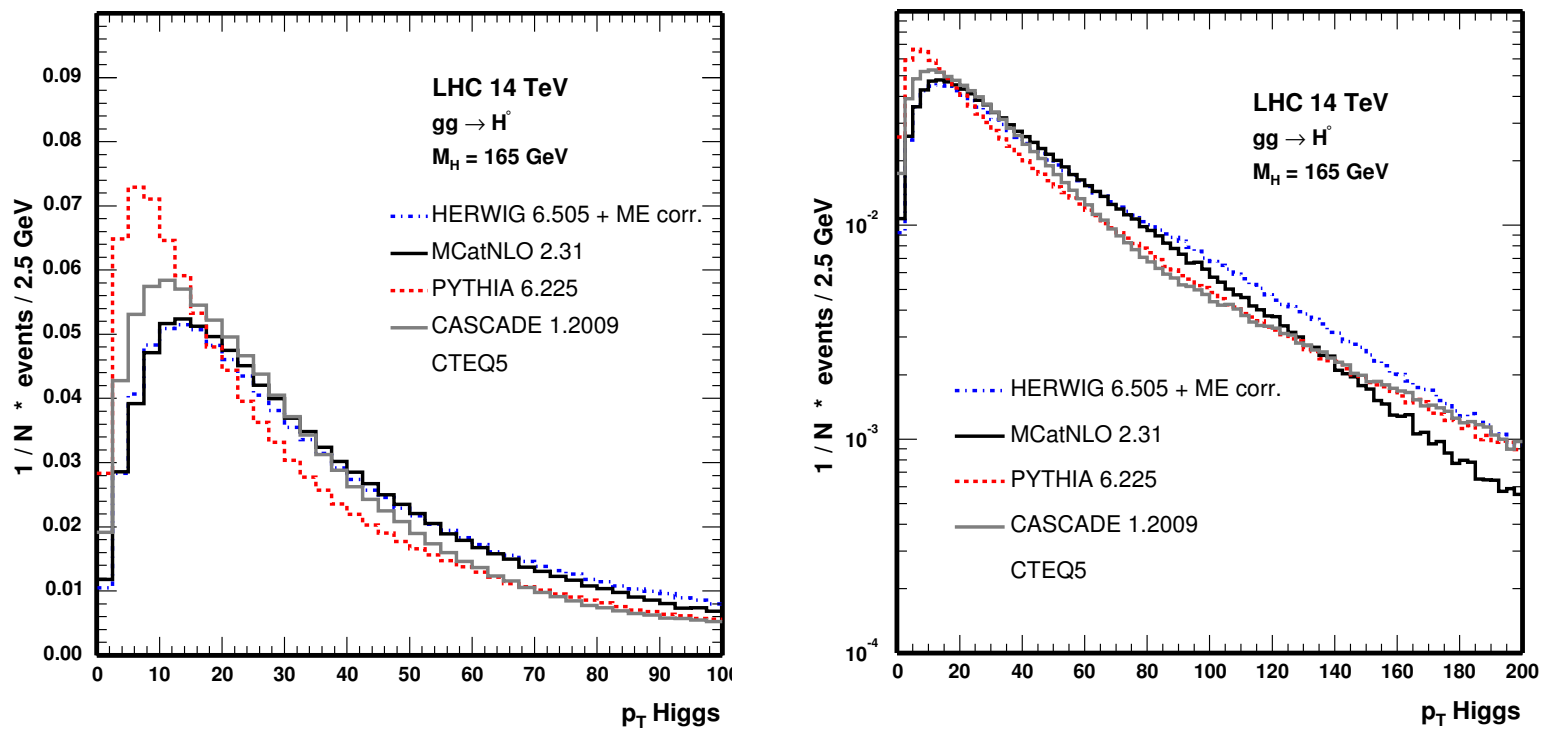

**Fig. 7:** $p_T^{\mathrm{Higgs}}$ Higgs of PYTHIA, HERWIG + ME Corrections, MC@NLO and CASCADE, linear and logarithmic scale.

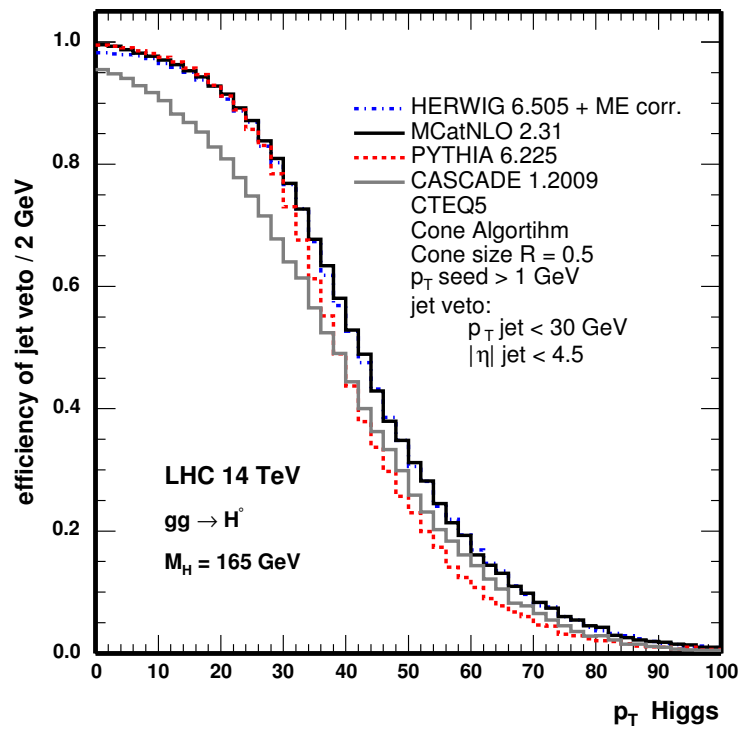

**Fig. 8:** Efficiency as a function of $p_T$ for PYTHIA, HERWIG+ME Corrections, MC@NLO and CASCADE.

The usage of another Monte Carlo program in LHCb is important in order to estimate the uncertainty on the PYTHIA [3] predictions. In particular, the track multiplicity seen in the detector is an important factor to take into account, as it affects the performance of the trigger, the tracking and the $B$-tagging. But here we will concentrate on another aspect: the study of the QCD evolution itself, proving that LHCb has the potential to be a natural test bed of QCD in the forward region, complementing the studies done at present at the Tevatron and the future studies to be made with the central detectors – ATLAS and CMS – at the LHC. The predictions in the forward region as given by CASCADE are here compared with that of PYTHIA, the default Monte Carlo generator used in LHCb. This is a "natural" way to test CCFM versus DGLAP QCD evolution in the region of the phase space where differences are most likely to show.

In what follows we will compare both predictions for the event kinematics and topology, and the particle and jet production. We used CASCADE version 1.2009 "out of the box" and PYTHIA 6.227 with the LHCb tune. We used for the comparisons a sub-sample of the QCD processes of PYTHIA, as CASCADE only includes (unintegrated) gluons. PYTHIA was run with the only sub-processes $fg \to fg$, $gg \to f\bar{f}$ and $gg \to gg$, and multiple interactions (MI) were also switched off, since they are as yet not implemented in CASCADE; this version is denoted *"PYTHIA gluon"* in the plots. Another configuration named *"PYTHIA gluon incl MI"* has the multiple interactions switched on, for a cross-check of the influence of such inclusion. All the plots refer to minimum bias events.

### 4.1 Event Kinematics

Figure 9 shows the kinematic variables $Q^2$ and Bjorken-$x$ variables $x_1$ and $x_2$ (referring to both LHC proton beams of energy $E_p$), using the definitions given below. For PYTHIA the standard definitions from the PYPARS common block were used:

$$x_1 = \text{PARI(33)} \quad x_2 = \text{PARI(34)};$$
$$Q^2 = \text{PARI(18)},$$

whereas for CASCADE we set [5]:

$$x_{1,2} = \frac{(E + |p_z|)_{in.\ parton\ 1,2}}{2E_p};$$

[5] The two incoming partons in the hard interaction are obtained from the variables NIA1 and NIA2, corresponding to the positions 4 and 6 in the CASCADE event record, whereas the outgoing partons are at positions 7 and 8.

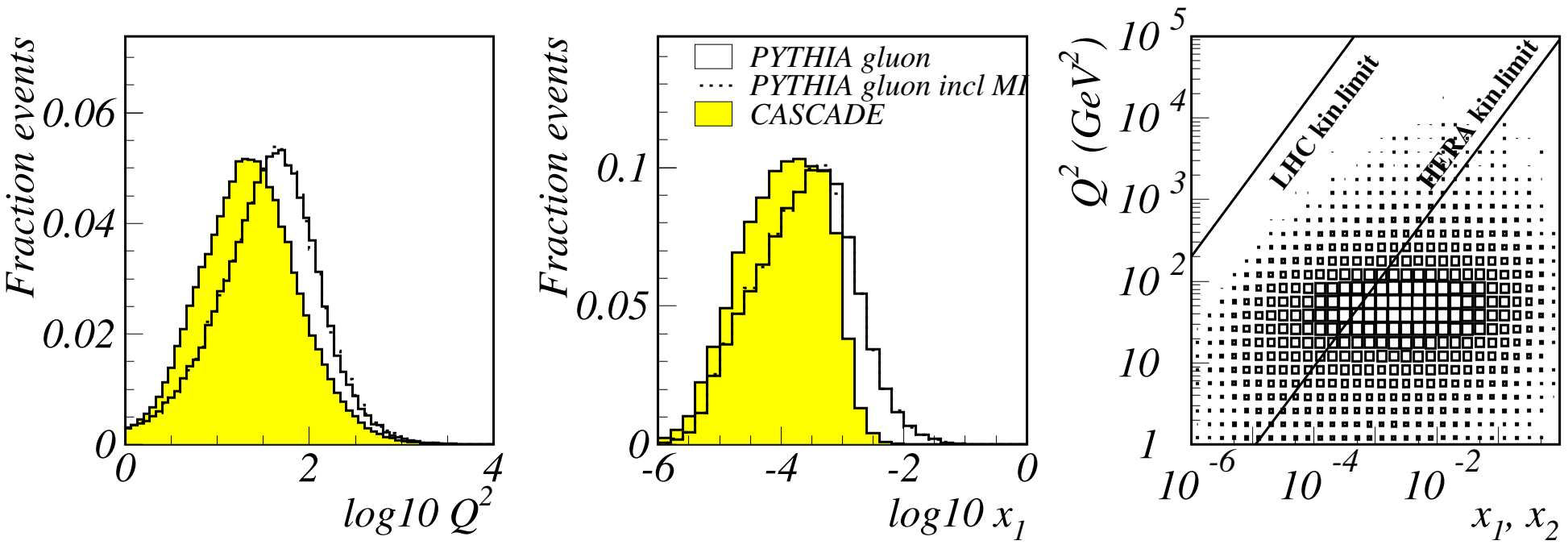

**Fig. 9:** Comparison between CASCADE and PYTHIA for the general event kinematics variables (refer to the text for the definitions). Note that $x_1 < x_2$ by construction.

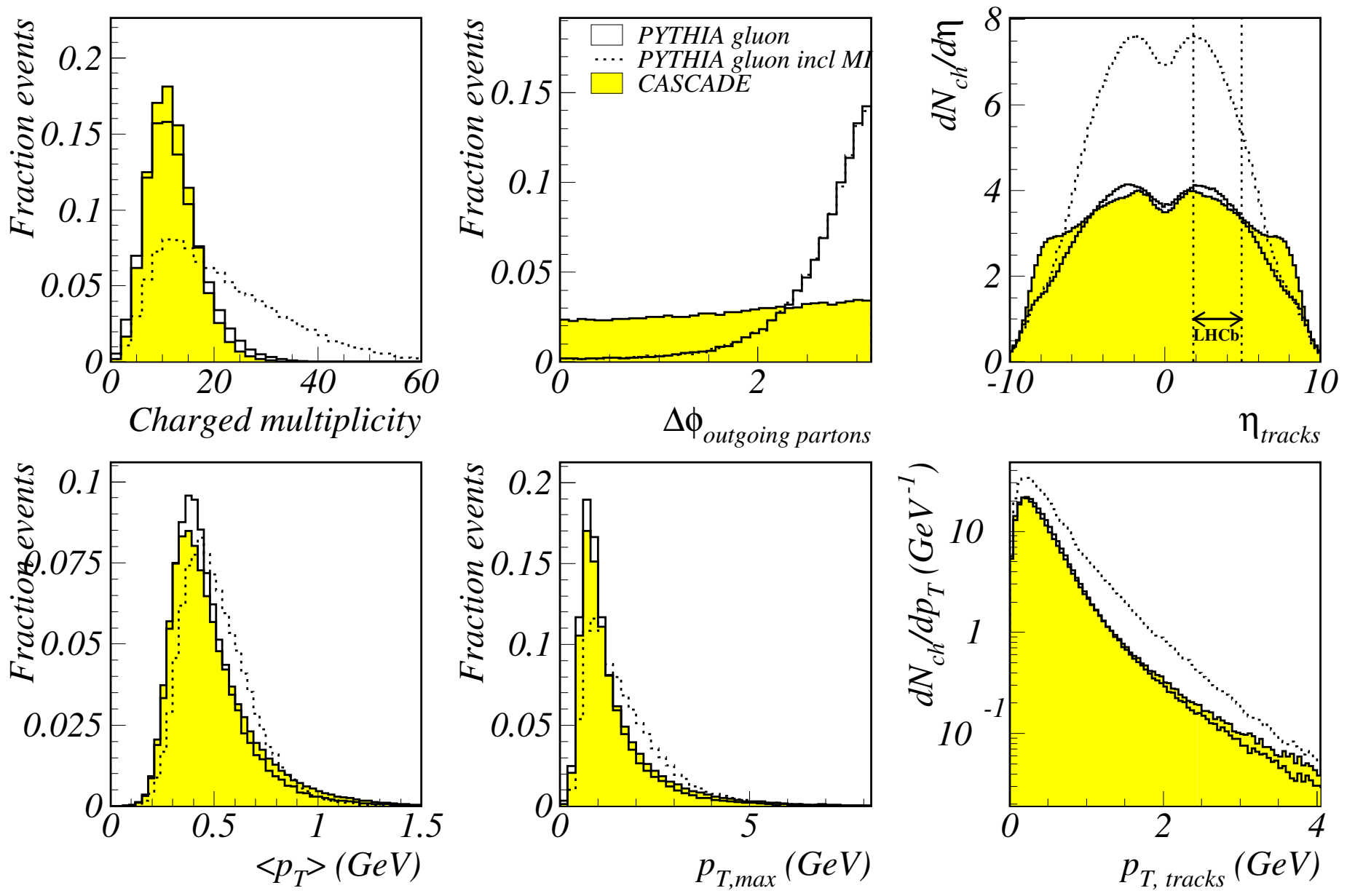

**Fig. 10:** Comparison between CASCADE and PYTHIA for general event variables, and between charged tracks variables in the region of the LHCb acceptance defined as $1.8 < \eta < 4.9$. No acceptance cuts are applied on the $\Delta\phi$ and $\eta_{tracks}$ distributions.

$$Q^2 = p^2_{T\ out.\ parton}.$$

There is a reasonable agreement between both Monte Carlo programs, although a direct comparison seems difficult and unnatural given the definitions above. The phase space spanned by the kinematic variables $x_{1,2}$ and $Q^2$ is shown also in Fig. 9 for PYTHIA.

### 4.2 Forward Particle Production

Some general event variables are compared in Fig. 10 in the region of the LHCb acceptance, $1.8<\eta<4.9$, including the charged track multiplicity, the acoplanarity ($\Delta\phi$) of the outgoing partons, the average track transverse momentum in the event $<p_T>$ and the maximum track transverse momentum $p_{T,max}$. The predictions from both Monte Carlo programs agree well – neglecting the multiple interactions in PYTHIA – likely because the same final state parton showering is performed. The effect of including the multiple interactions is seen mainly in the event multiplicity, as expected. Interesting is the distribution of the acoplanarity of the two outgoing partons: PYTHIA predicts a strong (anti-)correlation whereas CASCADE exibits a distribution that is nearly flat.

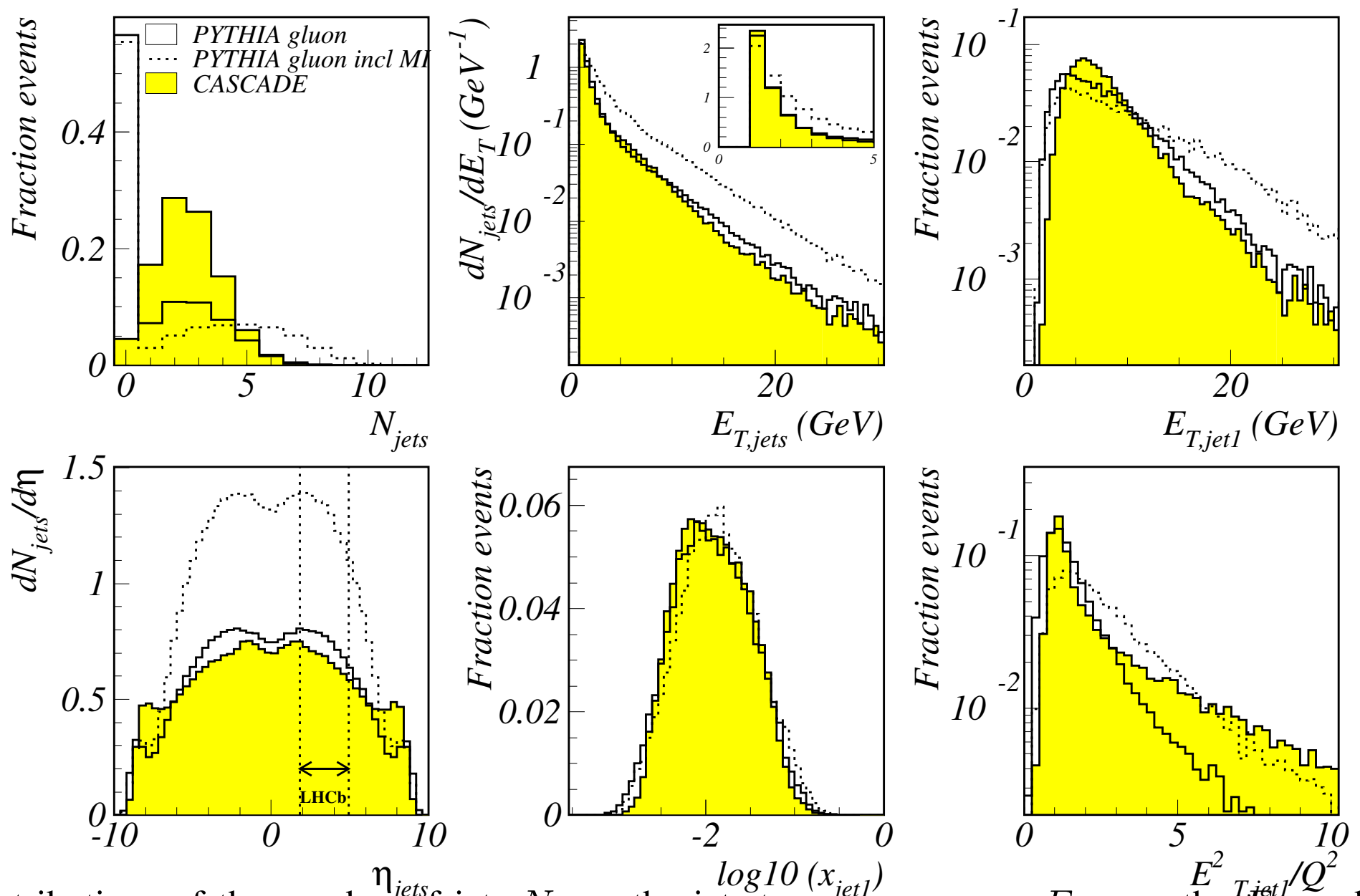

**Fig. 11:** Distributions of the number of jets $N_{jets}$, the jets transverse energy $E_{T,jets}$, the $E_T$ and $x_{jet}$ of the highest-$E_T$ jet, $jet1$, all in the LHCb acceptance. Also the number of jets per unit pseudorapidity is shown. The distribution of the ratio of $E^2_{T,jet}/Q^2$ in the LHCb acceptance shows a comparison of the two scales. Jets were selected with $E_{T,jet} > 1$ GeV.

The number of charged tracks per unit rapidity, $dN/d\eta_{tracks}$, and the differential distribution of the number of charged tracks (in the LHCb acceptance) as a function of the transverse momentum $p_{T,tracks}$ are also included in Fig. 10. Note that these 2 distributions were normalized to the mean track multiplicity in the full and LHCb acceptance, respectively. The $p_T$ distributions compare very well, leading us to conclude that the general hard dynamics of the event is predicted in a rather similar way by both programs. CASCADE however, produces more forward tracks than PYTHIA, as the $\eta$-distribution is clearly flatter than the rather steep distribution of PYTHIA. This is particularly true in the region $5<\eta<8$, just beyond the acceptance of the LHCb spectrometer – shown between the 2 vertical dashed lines – , but could make LHCb a candidate environment to discriminate between the two predicted forward behaviours.

### 4.3 Forward Jet Production

We have also looked at jet production. Jets were found in the laboratory frame with the KTCLUS algorithm on all stable hadrons, in the longitudinally invariant inclusive mode. We looked at the jet production in the LHCb acceptance with a rather loose selection of $E_{T,jets} > 1$ GeV. The number of jets found in PYTHIA or CASCADE is shown in Fig. 11. The number of events with no jets satisfying $E_{T,jet} > 1$ GeV inside $1.8<\eta<4.9$ is much larger for PYTHIA. In other words, CASCADE predicts a jet cross-section larger than PYTHIA, a fact already shown by the HERA experiments in low-$x$ jet analyses. This difference leads us to believe that strong angular ordering in CASCADE favours a "clustered production" of particles and therefore the production of jets, whereas PYTHIA tends to give a more spreaded transverse energy flow. Furthermore, though the effect is small, we already saw from Fig. 10 that the highest-$p_T$ track is somewhat softer in PYTHIA compared to CASCADE.

The rapidity distribution and the transverse energy distribution of the jets is also shown in Fig. 11; they have been normalized to the average number of jets per event in the full acceptance and LHCb acceptance, respectively. PYTHIA and CASCADE predict similar jets in the LHCb acceptance, but the inclusion of multiple interactions gives a harder spectrum.

Also shown are the event distributions in the LHCb acceptance of the highest-$E_T$ jet in the event, $E_{T,jet1}$, and the energy fraction of the proton carried by the highest-$E_T$ jet, $x_{jet1} = E_{jet1}/E_p$. The hardest jet in the event is on average harder in CASCADE compared to PYTHIA. The distributions of $x_{jet}$ and $E^2_{T,jet}/Q^2$ are interesting in that they correspond to variables now in standard use within the HERA experiments as a means of selecting samples where forward effects are expected. Indeed both experiments have published a series of "forward QCD" analyses [25] applying cuts of the kind $E^2_{T,jet} \sim Q^2$ and $x_{jet} \gg x_{Bjorken}$. The phase space is selected such that it suppresses jet production via DGLAP evolution and enhances production from BFKL dynamics:

– DGLAP evolution is suppressed in the small phase space for $Q^2$ evolution requiring $E^2_{T,jet} \sim Q^2$;
– CCFM evolution enhanced when large phase space for $x$ evolution requiring $x_{jet} \gg x_{Bjorken}$.

At the LHC such a selection becomes rather delicate, since there are two proton beams and the comparison of $x_{jet}$ with $x_{Bjorken}$ gets an ambiguity between the choice of $x_1$ or $x_2$. A way out – though it lowers significantly the statistics – would be to make the selection based on $x_{jet} \gg max(x_1, x_2)$. From the distributions presented in this paper we are lead to believe that such a forward selection is indeed possible. But we leave this issue open for further investigation.

## 5 Summary

Various ways of treating parton showers have been compared, as implemented by the HERWIG, PYTHIA and ARIADNE Monte Carlo programs. We have studied the uncertainties that arrise from these different models to the $p_T$-spectrum of the jets, and the $p_T$-spectrum of the Higgs boson.

The theoretical systematic uncertainty on predictions for inclusive cross section at NNLO for Higgs production with $bbH$ Yukawa coupling is under good theoretical control with an uncertainty of about 15% for a Higgs mass around 120 GeV. However, the predictions for the exclusive cross section determined by the event selection of a simplified experimental analysis indicates at present an order by magnitude larger uncertainty in e.g. $H \to \tau\tau$ events. Uncertainties due to the shower model can reach 170% and depend strongly on the production mechanism. Another factor of two arises from the choice of the QCD scale. Higher order Monte Carlo generators will therefore be mandatory to achieve better precision on the theoretical predictions.

On the other hand, the uncertainty of the jet veto efficiency in the $H \to WW \to l\nu l\nu$ decay channel by using different Monte Carlo generators in the $gg \to H$ process is estimated to be around 10%. Including higher order QCD corrections does not enhance the uncertainty significantly. Also the effect of including a realistic jet-$E_T$ resolution is very small. The effect of including an underlying event in the simulation is smaller than 1%, and does not vary significantly for various tuning models.

Furthermore we have studied the predictions at the LHC using the CCFM formalism as implemented in the full hadron level Monte Carlo generator. We conclude that CASCADE produces more and harder jets compared to the other Monte Carlo programs, leading to a bigger uncertainty of the jet veto efficiency in the small $p_T^{\rm Higgs}$ range. In the forward region larger differences are expected between the DGLAP and CCFM approach, but in the moderate forward rapidity range $2<\eta<5$, as covered by the LHCb detector, a fairly good agreement between CASCADE and PYTHIA is observed for most of the distributions looked at, and despite their different philosophies. However, this result has to be treated with care, as the program is only recently developped for proton physics at such high energies as produced in the future LHC. It also comes out of this simple study that CASCADE is indeed a potential Monte Carlo tool to use for QCD studies at the LHC in the forward region. In the future one should further investigate regions of phase space where large differences in behaviour are expected at the LHC from DGLAP and BFKL dynamics. LHCb seems a natural experimental environment in which to study such differences.

Finally, we would like to encourage the community by stating that it is very interesting and instruc-tive to study the predictions at the LHC by using tools developed and tuned at HERA, such as the CCFM Monte Carlo CASCADE, and by using parton shower models such as ARIADNE, that have proven their validity at HERA.

# Unintegrated parton density functions

*John Collins*[1]*, Markus Diehl*[2]*, Hannes Jung*[2]*, Leif Lönnblad*[3]*, Michael Lublinsky*[4]*, Thomas Teubner*[5]
[1] Physics Department,Penn State University, U.S.A.
[2] Deutsches Elektronen-Synchroton Hamburg, FRG
[3] Department of Theoretical Physics, Lund University, Sweden
[4] University of Connecticut, U.S.A.
[5] University of Liverpool, U.K

**Abstract**
An overview on activities to determine unintegrated parton density functions is given and the concept and need for unintegrated PDFs is discussed. It is also argued that it is important to reformulate perturbative QCD results in terms of fully unintegrated parton densities, differential in all components of the parton momentum. Also the need for non-linear BFKL evolution is discussed and results using the BK equation supplemented by DGLAP corrections at short distances is reviewed. Finally the use unintegrated generalized parton distributions for hard diffractive processes is discussed.

## 1 Unintegrated parton density functions[1]

The parton distributions of hadrons still cannot be calculated from first principles, but have to be determined experimentally. However, once the initial distributions $f_i^0(x, \mu_0^2)$ at the hadronic scale ($\mu^2 \sim$ 1 GeV$^2$) are determined, different approximations allow to calculate the parton density functions (PDFs) for different kinematic regions:

- DGLAP [1–4] describes the evolution with the scale $\mu^2$
- BFKL [5–7] describes the evolution in the longitudinal momenta $x$
- CCFM [8–11] describes the evolution in an angular ordered region of phase space while reproducing DGLAP and BFKL in the appropriate asymptotic limits

The different evolution equations attempt to describe different regions of phase space on the basis of in perturbative QCD (pQCD).

### 1.1 Introduction to uPDFs and $k_\perp$ factorization

In the collinear factorization ansatz the cross sections are described by $x$-dependent density functions $f_i(x, \mu^2)$ of parton $i$ at a given factorization scale $\mu$ convoluted with an (on-shell) coefficient function (matrix element):

$$\sigma(a + b \to X) = \int dx_1 dx_2 f_i(x_1, \mu^2) f_j(x_2, \mu^2) \hat{\sigma}_{ij}(x_1, x_2, \mu^2) \tag{1}$$

with $\hat{\sigma}_{ij}(x_1, x_2, \mu^2)$ being the hard scattering process for the partons $i+j \to X$. In this equation we have left implicit all external kinematic variables, keeping only the variables used in the parton densities. This ansatz is very successful in describing inclusive cross sections, such as the structure function $F_2(x, Q^2)$ at HERA or the inclusive production of vector bosons or Drell-Yan in proton proton collisions. The free parameters of the starting distributions $f_i^0(x, \mu_0^2)$ are determined such that after a DGLAP evolution to the scale $\mu^2 = Q^2$ and convolution with the coefficient functions the measured structure function

[1]Authors: Hannes Jung and Leif Lönnblad.

$F_2(x, Q^2)$ at HERA (and, usually, some other cross sections, e.g., in hadron-hadron and neutrino-hadron scattering) are best described.

However, as soon as, for example, final-state processes are considered, the collinear factorization ansatz becomes more and more unreliable, because neglecting the transverse momenta of the partons during the (DGLAP) evolution leads to inconsistencies, as will be discussed in more detail in section 2. Collinear factorization is only appropriate when (a) the transverse momentum (and virtuality) of the struck parton(s) can be neglected with respect to $Q$, and (b) the integrals over these variables can be treated as independent and unrestricted up to the scale $Q$. (Certain complications concerning high transverse momentum partons are correctly treated by NLO and higher corrections to the hard scattering.) When these requirements are not met, a more general treatment using unintegrated parton densities (uPDFs) is better.

For example, in the small $x$ regime, when the transverse momenta of the partons are of the same order as their longitudinal momenta, the collinear approximation is no longer appropriate and high energy or $k_\perp$ - factorization has to applied, with the appropriate BFKL or CCFM evolution equations. Cross sections are then $k_\perp$- factorized [12–15] into an off-mass-shell ($k_\perp$- dependent) partonic cross section $\hat{\sigma}(x_1, x_2, k_{\perp 1}, k_{\perp 2})$ and a $k_\perp$- unintegrated parton density function (uPDF) $\mathcal{F}(z, k_\perp)$:

$$\sigma = \int dx_1 dx_2 d^2k_{\perp 1} d^2k_{\perp 2} \hat{\sigma}_{ij}(x_1, x_2, k_{\perp 1}, k_{\perp 2}) \mathcal{F}(x_1, k_{\perp 1}) \mathcal{F}(x_2, k_{\perp 2}) \quad (2)$$

The unintegrated gluon density $\mathcal{F}(z, k_\perp)$ is described by the BFKL evolution equation in the region of asymptotically large energies (small $x$). It is important to note that only when the $k_\perp$dependence of the hard scattering process $\hat{\sigma}$ can be neglected, i.e. if $\hat{\sigma}(x_1, x_2, k_{\perp 1}, k_{\perp 2}) \sim \hat{\sigma}(x_1, x_2, 0, 0)$, then the $k_\perp$integration can be factorized and an expression formally similar to eq.(1) is obtained.

An appropriate description valid for both small and large $x$, is given by the CCFM evolution equation, resulting in an unintegrated gluon density $\mathcal{A}(x, k_\perp, \mu)$, which is a function also of the additional evolution scale $\mu$. This scale is connected to the factorization scale in the collinear approach.

Further examples where uPDFs are needed are the Drell-Yan and related processes at low transverse momentum, as in the CSS formalism [16]. However, the relation between CSS method (which does not need small $x$) and $k_\perp$-factorization of the BFKL/CCFM kind (for small $x$) has not yet been properly worked out.

### 1.2 Extraction and determination of uPDFs

In this section we will review how measurements of uPDFs have been extracted from DIS data at small $x$, mostly from the inclusive structure function $F_2$. For measurements of the uPDFs in Drell-Yan processes using the CSS formalism, see [17].

From the DIS data, the uPDF can be obtained by adjusting the non-perturbative input distribution $f_i^0(x, \mu_0^2)$ and the free parameters of the perturbative evolution such that after convolution with the appropriate off-shell matrix element (according to eq.(2)) a measured cross section is best described.

Applying $k_\perp$-factorization to determine the uPDF from DIS data until now mainly the inclusive structure function measurements of $F_2(x, Q^2)$ at HERA have been used. The exceptions are those which are simply derivatives of integrated PDFs, which then neglects fully the transverse momentum dependence of the matrix element. Extracting a uPDF from the integrated PDF is appropriate only if the $k_\perp$-dependence of the hard scattering process $\hat{\sigma}$ in eq.(2) can be neglected. In addition, contributions from $k_\perp > \mu$, which are present in a full calculation, are entirely neglected. It thus can only provide an estimate of the proper kinematics in the collinear approach, which is otherwise fully neglected when using integrated PDFs.

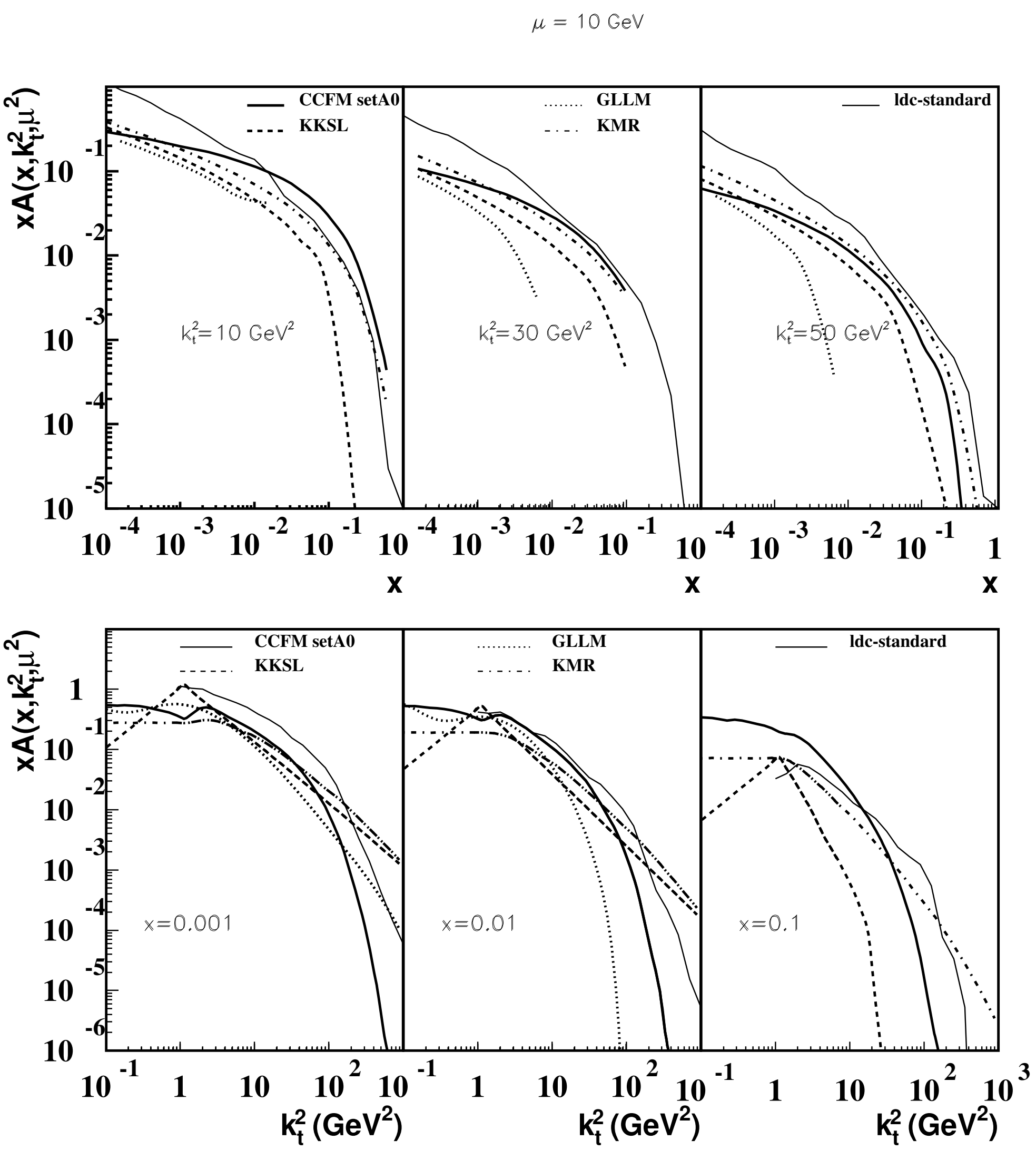

**Fig. 1:** Comparison of different uPDFs at $\mu = 10$ GeV.

Here we compare some of these parameterizations which have been obtained in different ways:

- **CCFM set A0** was obtained using CCFM evolution in [18, 19].
- **LDC standard** was similarly obtained in [20] using LDC evolution [21], which is a reformulation and generalization of CCFM.
- **KKSL** [22] was obtained from a combined BFKL and DGLAP evolution following [23].
- **GLLM** [24] was obtained applying the BK equation to HERA $F_2$ measurements, as described in Section 3.
- **KMR** is one of the more advanced derivatives of integrated PDFs, using Sudakov form factors [25].

In Fig. 1 we show a comparison of the different uPDFs as a function of $x$ and $k_\perp$ at a factorization scale $\mu = 10$ GeV. All the parameterizations are able to describe the measured $F_2(x, Q^2)$ in the small $x$ range reasonably well, with a $\chi^2/ndf \sim 1$. In Fig. 2 the same uPDFs are compared at a factorization scale which is relevant at LHC energies, e.g. for inclusive Higgs production ($\mu = 120$ GeV). One should note that the uPDFs from KKSL and GLLM have no explicit factorization scale dependence, therefore they are the same as in Fig 1.

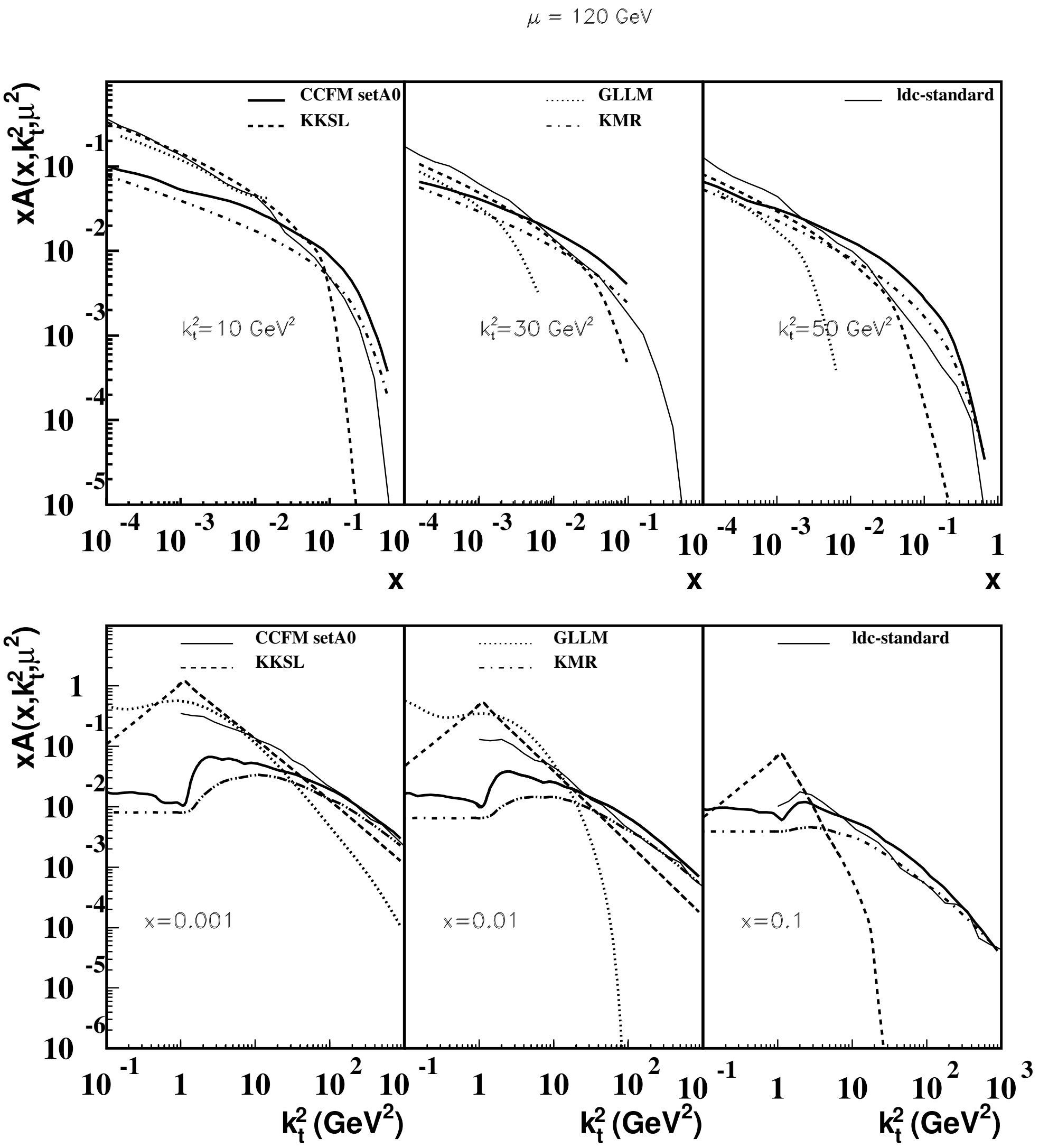

**Fig. 2:** Comparison of different uPDFs at $\mu = 120$ GeV.

### 1.3 Extrapolation to LHC energies

All the parameterizations of uPDFs considered in this report give a fairly good fit to HERA $F_2$ data. This means that they are well constrained mainly in the region of small $x$ and relatively small $Q^2$, where the bulk of the HERA data is concentrated. For higher $x$ and $Q^2$, a fit to HERA data is less constraining, and indeed some of the parameterizations based on the CCFM and LDC evolution of the gluon alone are only fitted in the small-$x$ region (typically $x < 0.01$, $Q^2 < 100$ GeV$^2$).

When evolving the uPDFs to apply them to the processes of main interest at the LHC, such as Higgs production, this is a serious concern. Although the $x$-values in such processes are typically below $0.01$, the scales involved are of the order of $10^4$ GeV$^2$ or more. Through the evolution one then becomes sensitive to larger $x$-values at lower scales where the current parameterizations are less constrained.

A notable exception is the KMR [25] densities which are obtained from a global fit of integrated PDFs, which should give reliable prediction at LHC at least for integrated observables such as the inclusive Higgs cross section. In contrast, it was shown in [20] that the CCFM [8–11] and LDC [21] evolved uPDFs have unreasonably large uncertainties for such cross sections. On the other hand it was also shown in [20] that there are some questions about the constraint of the actual $k_\perp$ distribution of the KMR uPDFs resulting eg. in a too soft $p_\perp$ spectrum of W or Z production at the Tevatron for small transverse momenta. Hence, although the KMR prediction for inclusive quantities may be reliable at the LHC, the predictions of eg. the detailed distribution of low-$p_\perp$ Higgs may be questionable.

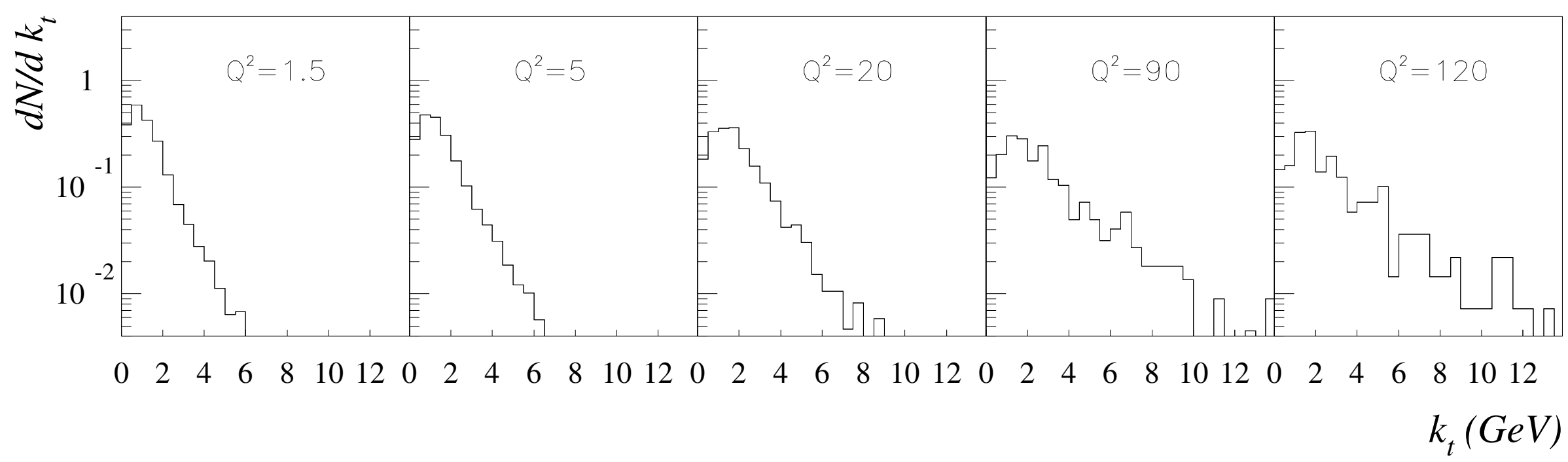

**Fig. 3:** $k_\perp$distribution in different $Q^2$ bins used in $F_2(x, Q^2)$ at HERA.

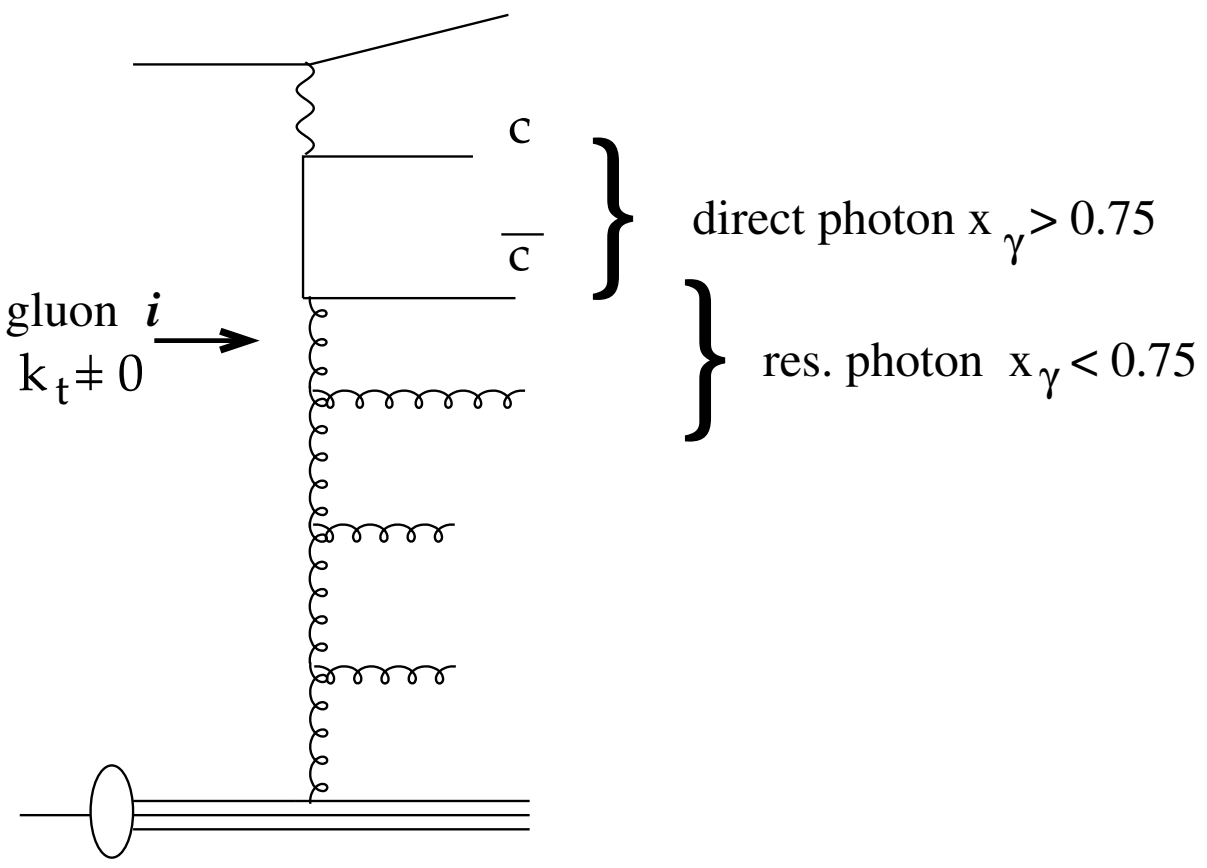

**Fig. 4:** Diagram of charm photoproduction, showing the sensitivity to the gluon transverse momentum

What is needed is clearly to obtain fits of the uPDFs, not only to HERA $F_2$ data, but also to observables more sensitive to higher $x$ and $Q^2$ values, as well as to observables directly sensitive to the $k_\perp$ distribution. To obtain such global fits there is a need for both theoretical and phenomenological developments. Examples of the former is the inclusion of quarks in the CCFM evolution, while the latter involves the development of $k_\perp$-sensitive observables, where HERA data at small $x$, such as forward jet or heavy quark production, will play an important role, as discussed in the following.

### 1.4 Global uPDF fits

Until now the uPDFs obtained from DIS were only determined and constrained by the inclusive structure function $F_2(x, Q^2)$. It is clear that the inclusive measurements are not very sensitive to the details of the $k_\perp$dependence. In Fig. 3 we show the $k_\perp$distribution of the gluon in $\gamma^* g^* \to q\bar{q}$ which is the relevant process for $F_2$ at small $x$. The $k_\perp$-distributions in Fig. 3 are obtained with CASCADE [26, 27] using the CCFM uPDFs. The bins in $Q^2$ are typical for HERA $F_2$ measurements. It is interesting to observe that even at large $Q^2$ essentially only the small $k_\perp$region is probed by $F_2$.

A larger lever arm for the $k_\perp$distribution can be obtained with photoproduction of $D^*$ + jet events at HERA. In Fig. 4 the relevant diagram is shown. The quantity $x_\gamma$, normally designed to separate direct from resolved photon processes, can be also used to distinguish small and large $k_\perp$- regions. The region of large $x_\gamma$ corresponds to measuring jets coming from the quark-box. The region of small $x_\gamma$ corresponds to the situation where one of the jets originates from a gluon, as indicated in Fig. 4. Thus, the transverse momentum of the gluon $i$ can be probed, as shown in Fig. 5 for two different regions of $x_\gamma$ using CASCADE . It is interesting to note that the average $k_\perp$distribution for bottom production at the Tevatron is similar to what it shown in Fig. 5.

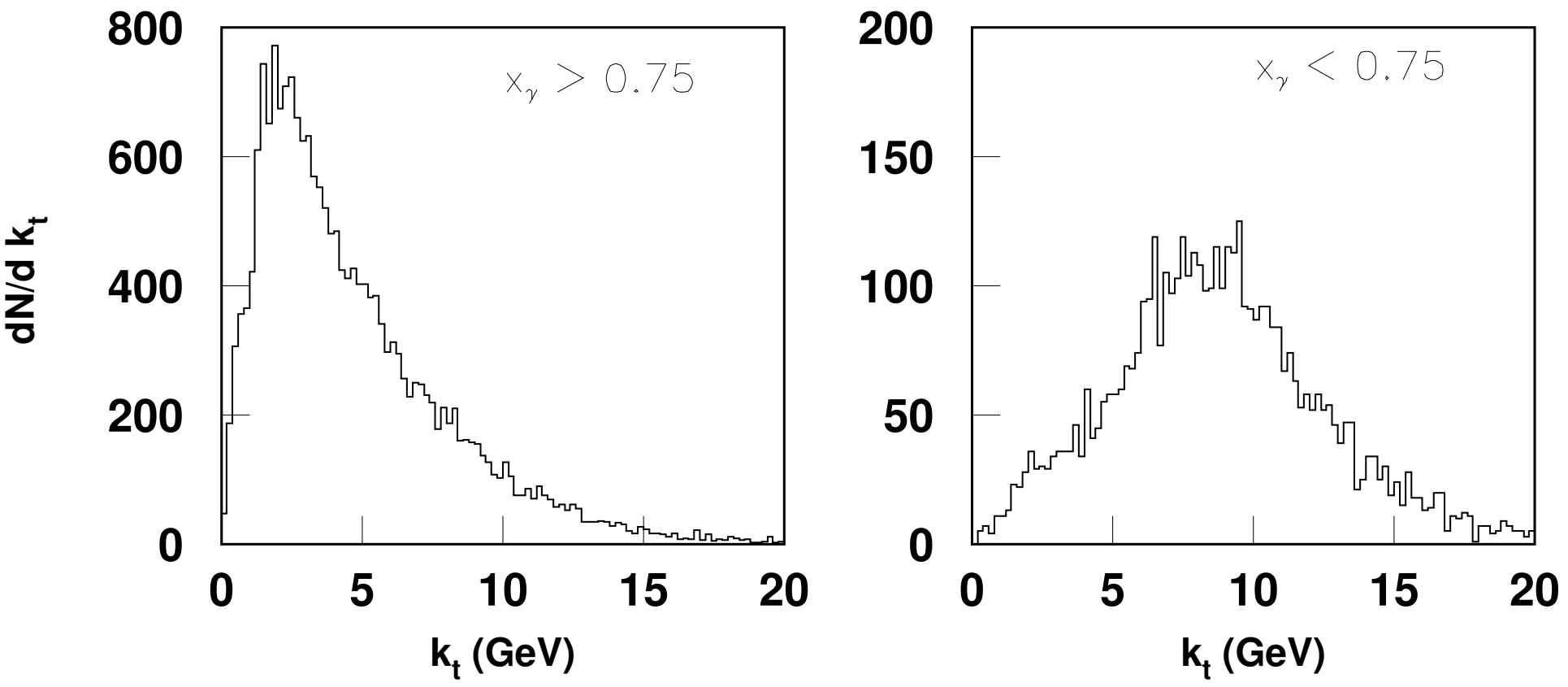

**Fig. 5:** $k_\perp$distribution in different $x_\gamma$ bins obtained from $D^*$+jet photo-production at HERA.

To further constrain the uPDF it would be desirable to perform a common fit to inclusive measurements like $F_2$ and simultaneously to final state measurements.

Once the data sets and the sensitivity to the uPDFs have been identified, a systematic error treatment of the data used in the uPDF fits can be performed. Until now, the uPDFs are not really the result of a fit but rather a proof that the uPDF is consistent with various measurements.

A uPDF fit would require a systematic variation of the parameters used to specify the non-perturbative input gluon distribution as well as a systematic treatment of the experimental systematic uncertainties. Only then an uncertainty band of the uPDFs can be given. To consider the uncertainty of the uPDF given from the spread of different available parameterizations is a very rough estimate.

### 1.5 Outlook and Summary

Clearly, the extraction of uPDFs from data is still in its infancy, especially if compared to the well developed industry of fitting integrated PDFs. The uPDFs are only leading order parameterizations, they have mainly been fitted to $F_2$ data at small $x$, and besides the KMR and LDC parameterizations, no attempts have been made to obtain unintegrated quark densities. Taken together, this means that the applicability to LHC processes are uncertain. However, the field is maturing and we hope to soon be able to do more global uPDF fits which will greatly enhance the reliability of the predictions for the LHC. In doing so the small-$x$ data from HERA will be very important, but also eg. Tevatron data will be able to provide important constraints.

## 2 Need for fully unintegrated parton densities[2]

### 2.1 Introduction

Conventional parton densities are defined in terms of an integral over all transverse momentum and virtuality for a parton that initiates a hard scattering. While such a definition of an integrated parton density is appropriate for very inclusive quantities, such as the ordinary structure functions $F_1$ and $F_2$ in DIS, the definition becomes increasingly unsuitable as one studies less inclusive cross sections. Associated with the use of integrated parton densities are approximations on parton kinematics that can readily lead to unphysical cross sections when enough details of the final state are investigated.

[2]Authors: John Collins and Hannes Jung.

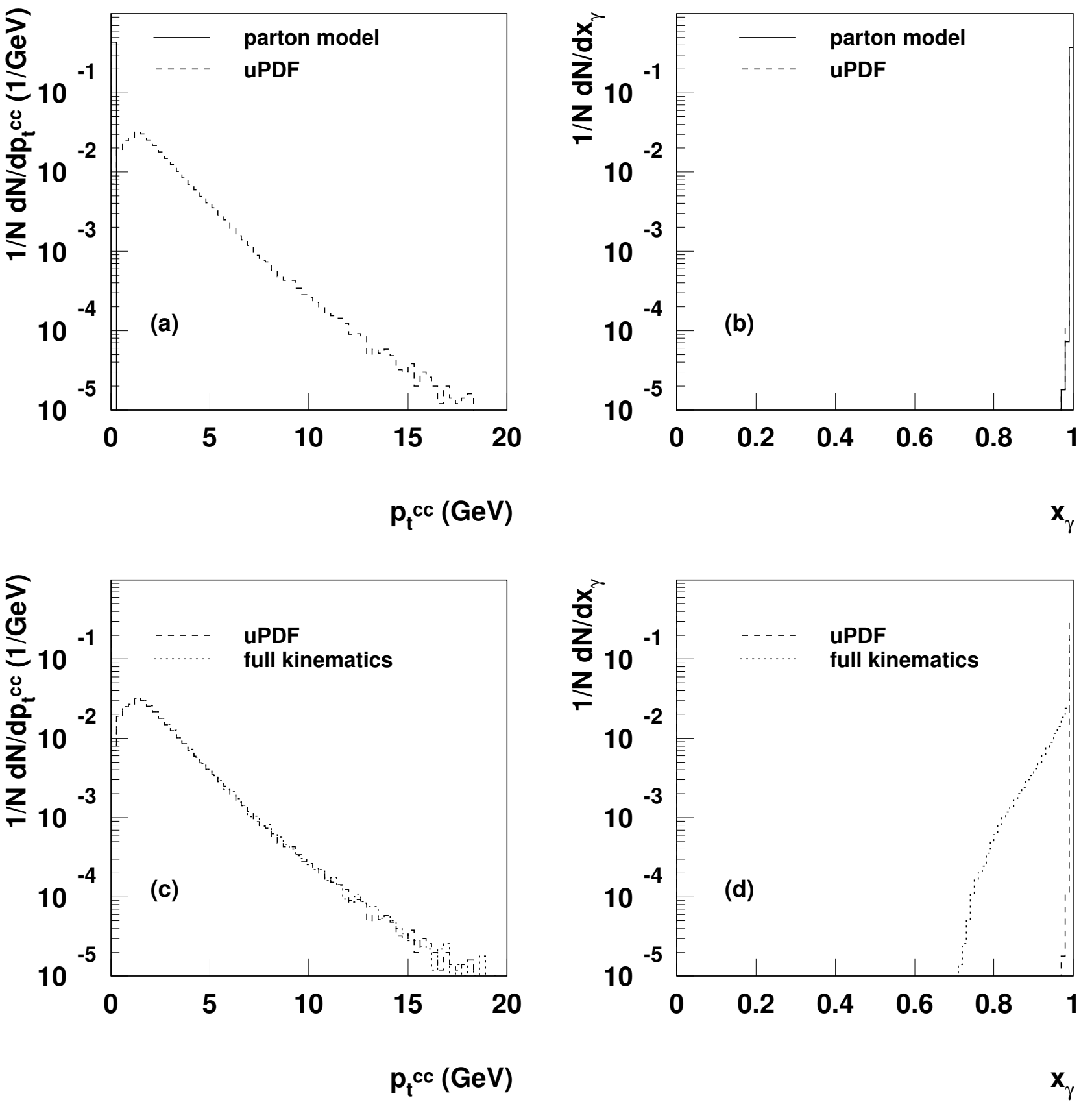

**Fig. 6:** (a) and (b): comparison between use of simple LO parton model approximation and of the use of $k_\perp$ densities for the $p_T$ of $c\bar{c}$ pairs in photoproduction, and for the $x_\gamma$. (c) and (d): comparison of use of $k_\perp$ densities and full simulation.

We propose that it is important to the future use of pQCD that a systematic program be undertaken to reformulate factorization results in terms of fully unintegrated densities, which are differential in both transverse momentum and virtuality. These densities are called "doubly unintegrated parton densities" by Watt, Martin and Ryskin [28, 29], and "parton correlation functions" by Collins and Zu [30]; these authors have presented the reasoning for the inadequacy, in different contexts, of the more conventional approach. The new methods have their motivation in contexts such as Monte-Carlo event generators where final-state kinematics are studied in detail. Even so, a systematic reformulation for other processes to use unintegrated densities would present a unified methodology.

These methods form an extension of $k_\perp$-factorization. See Sec. 1 for a review of $k_\perp$-factorization, which currently involves two different formalisms, the BFKL/CCFM methods [5–11] and the CSS method [16].

## 2.2 Inadequacy of conventional PDFs

The problem that is addressed is nicely illustrated by considering photoproduction of $c\bar{c}$ pairs. In Figs. 6, we compare three methods of calculation carried out within the CASCADE event generator [26, 27]:

- Use of a conventional gluon density that is a function of parton $x$ alone.
- Use of a $k_\perp$ density that is a function of parton $x$ and $k_\perp$. These are the "unintegrated parton densities" (uPDFs) that are discussed in Sec. 1

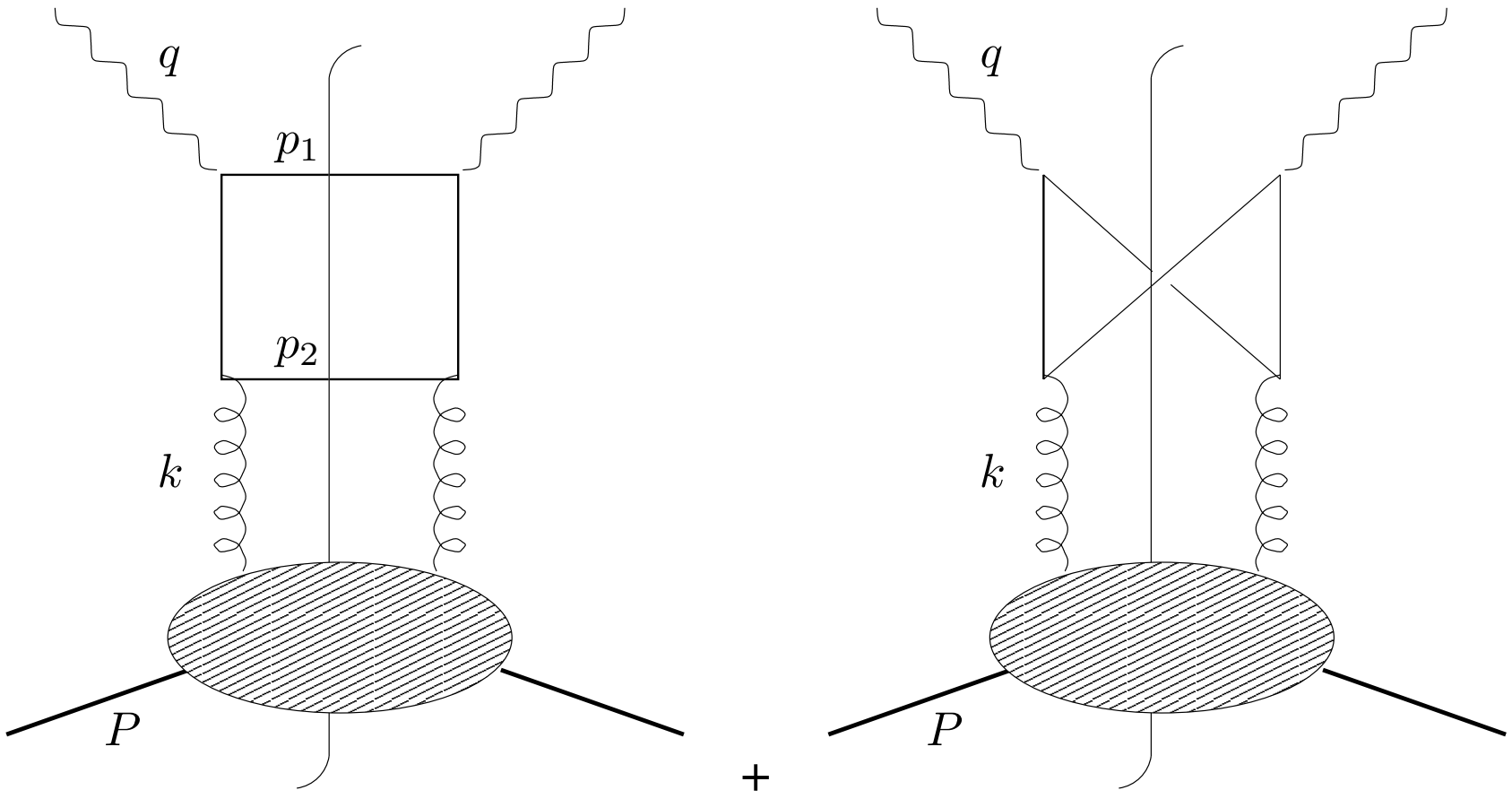

**Fig. 7:** Photon-gluon fusion.

– Use of a doubly unintegrated density that is a function of parton $x$, $k_\perp$ and virtuality, that is, of the complete parton 4-momentum.

The partonic subprocess in all cases is the lowest order photon-gluon-fusion process $\gamma + g \longrightarrow c + \bar{c}$ (Fig. 7). Two differential cross sections are plotted: one as a function of the transverse momentum of the $c\bar{c}$ pair, and the other as a function of the $x_\gamma$ of the pair. By $x_\gamma$ is meant the fractional momentum of the photon carried by the $c\bar{c}$ pair, calculated in the light-front sense as

$$x_\gamma = \frac{\sum_{i=c,\bar{c}}(E_i - p_{z\,i})}{2yE_e} = \frac{p^-_{c\bar{c}}}{q^-}.$$

Here $E_e$ is the electron beam energy and the coordinates are oriented so that the electron and proton beams are in the $-z$ and $+z$ directions respectively.

In the normal parton model approximation for the hard scattering, the gluon is assigned zero transverse momentum and virtuality, so that the cross section is restricted to $p_{Tc\bar{c}} = 0$ and $x_\gamma = 1$, as shown by the solid lines in Fig. 6(a,b). When a $k_\perp$ dependent gluon density is used, quite large gluonic $k_\perp$ can be generated, so that the $p_{Tc\bar{c}}$ distribution is spread out in a much more physical way, as given by the dashed line in Fig. 6(a). But as shown in plot (b), $x_\gamma$ stays close to unity. Neglecting the full recoil mass $m_{\rm rem}$ (produced in the shaded subgraph in Fig 7) is equivalent of taking $k^2 = -k_\perp^2/(1-x)$ with $k^2$ being the virtuality of the gluon in Fig. 7, $k_\perp$ its transverse momentum and $x$ its light cone energy fraction. This gives a particular value to the gluon's $k^-$. When we also take into account the correct virtuality of gluon, there is no noticeable change in the $p_{Tc\bar{c}}$ distribution — see Fig. 6(c) (dashed line) — since that is already made broad by the transverse momentum of the gluon. But the gluon's $k^-$ is able to spread out the $x_\gamma$ distribution, as in Fig. 6(d) with the dashed line. This is equivalent with a proper treatment of the kinematics and results in $k^2 = -(k_\perp^2 + xm_{\rm rem}^2)/(1-x)$, where $m_{\rm rem}$ is the invariant mass of the beam remnant, the part of the final state in the shaded blob in Fig. 7. This change can be particularly significant if $x$ is not very small.

Note that if partons are assigned approximated 4-momenta during generation of an event in a MC event generator, the momenta need to be reassigned later, to produce an event that conserves total 4-momentum. The prescription for the reassignment is somewhat arbitrary, and it is far from obvious what constitutes a correct prescription, especially when the partons are far from a collinear limit. A treatment with fully unintegrated PDFs should solve these problems.

If, as we claim, an incorrect treatment of parton kinematics changes certain measurable cross sections by large amounts, then we should verify directly that there are large discrepancies in the distributions in partonic variables themselves. We see this in Fig. 8. Graph (a) plots the gluonic transverse momentum divided by the charm-pair mass. As is to be expected, the typical values are less than one, but there is a long tail to high values. But the use of full parton kinematics does not have much of an effect, the unintegrated parton distributions already providing realistic distributions in transverse momentum.

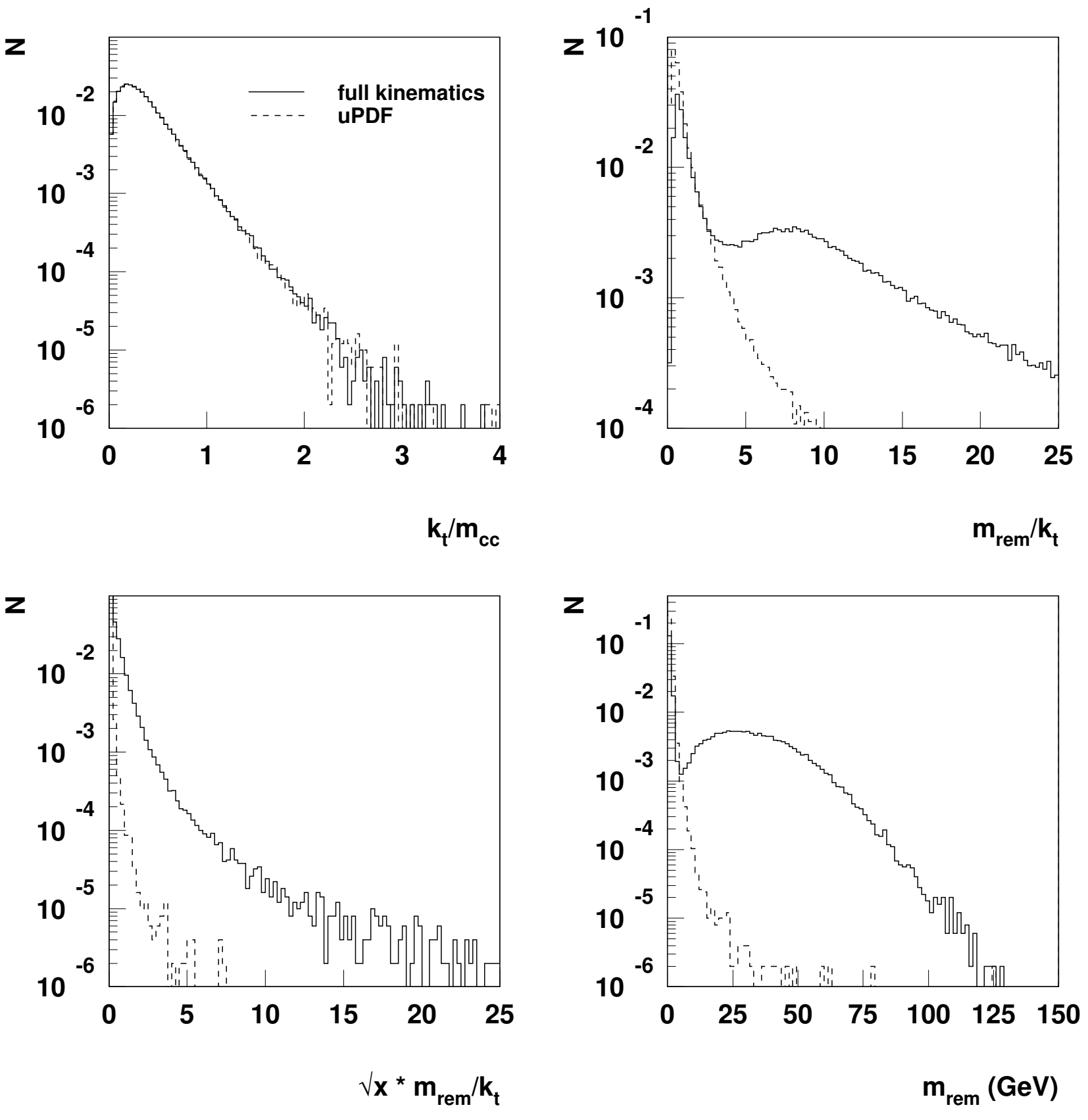

**Fig. 8:** Comparison of distributions in partonic variables between calculations with full parton kinematics and with ordinary unintegrated PDFs.

On the other hand, a simple collinear approximation for showering sets the remnant mass, $m_{\rm rem}$, to zero. As can be seen from the formulae for the gluon virtuality, this only provides a good approximation to the gluon kinematics if $m_{\rm rem}$ is much less than $k_\perp$. In reality, as we see from graph (b), there is a long tail to large values of $m_{\rm rem}/k_\perp$, and the tail is much bigger when correct kinematics are used. A more correct comparison uses $xm^2_{\rm rem}$, with an extra factor of $x$. Even then, there is a large effect, shown in graph (c). The vertical scale is logarithmic, so the absolute numbers of events are relatively small, but the tail is broad. Finally, graph (d) shows that the distribution in $m_{\rm rem}$ itself is very broad, extending to many tens of GeV. This again supports the argument that unless a correct treatment of parton kinematics is made, very incorrect results are easily obtained.

It is important to note that, for the cross sections themselves, the kinematic variables used in Fig. 6 are normal ones that are in common use. Many other examples are easily constructed. Clearly, the use of the simple parton-model kinematic approximation gives unphysically narrow distributions. The correct physical situation is that the gluon surely has a distribution in transverse momentum and virtuality, and for the considered cross sections neglect of parton transverse momentum and virtuality leads to wrong results. It is clearly better to have a correct starting point even at LO, for differential cross sections such as we have plotted.

### 2.3 Kinematic approximations

The standard treatment of parton kinematics involves replacing the incoming parton momentum $k$ by its plus component only: $k^\mu \mapsto \hat{k}^\mu \equiv (k^+, 0, 0_T)$. There are actually two parts to this. The first is to neglect the $^-$ and transverse components of $k$ with respect to the large transverse momenta in the calculation of the numerical value of the hard-scattering amplitude; this is a legitimate approximation, readily corrected by higher order terms in the hard scattering. The second part is to change the kinematics of the final-state particles, $p_1$ and $p_2$, so that their sum is $q$ plus the approximated gluon momentum. It is this second part that is problematic, for it amounts to the replacement of the momentum conservation delta function $\delta^{(4)}(k + q - p_1 - p_2)$ by $\delta^{(4)}(\hat{k} + q - p_1 - p_2)$. These delta-functions are infinitely different, point-by-point. Only when integrated with a sufficiently smooth test function can they be regarded as being approximately the same, as in a fully inclusive cross section.

In an event generator, the effect is to break momentum conservation, which is restored by an ad hoc correction of the parton kinematics. Note that the change of parton kinematics is only in the hard scattering, i.e., in the upper parts of the graphs. Parton kinematics are left unaltered within the parton density part, and the integrals over $k_\perp$ and virtuality are part of the standard definition of integrated PDFs.

The situation is ameliorated by inclusion of NLO terms, and perhaps also by some kind of resummation. But these do not correct the initial errors in the approximation, and lead to a very restricted sense in which the derivation of the cross section can be regarded as valid. Furthermore, when much of the effect of NLO terms is to correct the kinematic approximations made in LO, this is an inefficient use of the enormous time and effort going into NLO calculations. A case in point is the BFKL equation, where $70\%$ of the (large) NLO corrections are accounted for [31] by the correction of kinematic constraints in the LO calculation.

### 2.4 Conclusions

The physical reasoning for the absolute necessity of fully unintegrated densities is, we believe, unquestionable. Therefore it is highly desirable to reformulate perturbative QCD methods in terms of doubly unintegrated parton densities from the beginning. A full implementation will be able to use the full power of calculations at NLO and beyond.

Among other things, a full implementation, as in [30], will provide extra factorization formulae for obtaining the values of the unintegrated densities at large parton transverse momentum and virtuality. This will incorporate all possible perturbatively calculable information, so that the irreducible nonperturbative information, that must be obtained from data, will be at low transverse momentum and virtuality. In addition, the implementation will quantify the relations to conventional parton densities. With the most obvious definitions, the integrated PDFs are simple integrals of the unintegrated densities. However, in full QCD a number of modifications are required [30,32], so that the relations between integrated and unintegrated PDFs are distorted.

The fact that we propose new and improved methods does not invalidate old results in their domain of applicability. The work of Watt, Martin and Ryskin, and of Collins and Zu provides a start on this project; but much remains to be done to provide a complete implementation in QCD; for example, there is as yet no precise, valid, and complete gauge-invariant operator definition of the doubly unintegrated densities in a gauge theory.

The outcome of such a program should have the following results:

1. Lowest order calculations will give a kinematically much more realistic description of cross sections. This may well lead to NLO and higher corrections being much smaller numerically than they typically are at present, since the LO description will be better.

2. It will also obviate the need for separate methods (resummation or the CSS technique), which are currently applied to certain individual cross sections like the transverse-momentum distribution for the Drell-Yan process. All these and others will be subsumed and be given a unified treatment.
3. A unified treatment will be possible for both inclusive cross sections using fixed order matrix element calculations and for Monte-Carlo event generators.
4. For a long-term theoretical perspective, the doubly unintegrated distributions will interface to methods of conventional quantum many-body physics much more easily than regular parton densities, whose definitions are tuned to their use in ultra-relativistic situations.

This program is, of course, technically highly nontrivial if it is to be used in place of conventional methods with no loss of predictive power. A start is made in the cited work.

Among the main symptoms of the difficulties are that the most obvious definition of a fully unintegrated density is a matrix element of two parton fields at different space-time points, which is not gauge-invariant. It is often said that the solution is to use a light-like axial gauge $A^+ = 0$. However, in unintegrated densities, this leads to divergences — see [32] for a review — and the definitions need important modification, in such a way that a valid factorization theorem can be derived.

We also have to ask to what extent factorization can remain true in a generalized sense. Hadron-hadron collisions pose a particular problem here, because factorization needs a quite nontrivial cancellation arising from a sum over final-state interactions. This is not compatible with simple factorization for the exclusive components of the cross section, and makes a distinction between these processes and exclusive components of DIS, for example.

## 3 PDF extrapolation to LHC energies based on combined BK/DGLAP equations [3]

### 3.1 Introduction

In recent years it became clear that the DGLAP evolution is likely to fail in certain kinematics associated with the *low* $x$ domain. This might be a dangerous problem for certain DGLAP based predictions made for the LHC. The reasons for the failure are well known.

– DGLAP predicts a very steep rise of gluon densities with energy. If not suppressed this rise will eventually violate unitarity.
– The leading twist evolution breaks down when higher twists become of the same order as the leading one. We have to recall here that higher twists are estimated to rise with energy much faster than the leading one [33].
– The DGLAP evolution is totally unable to describe physics of low photon virtualities.

It is most important to stress that NLO corrections are in principal unable to solve any of the above problems, though they can potentially help to delay their onset.

Fortunately, a solution to the low $x$ problem does exist. We have to rely on a nonlinear evolution based on the BFKL dynamics. So far the best candidate on the market is the Balitsky-Kovchegov (BK) equation [34, 35], which is a nonlinear version of the LO BFKL equation. Compared to the DGLAP equation it has the following advantages:

– it accounts for saturation effects due to high parton densities.
– it sums higher twist contributions.
– it allows an extrapolation to large distances.

Though the BK evolution takes care of the low $x$ domain, it misses the essential part of the short distance physics correctly accounted for by the DGLAP evolution. The reason is that the BFKL kernel involves the $1/z$ part only of the full gluon-gluon splitting function $P_{gg}(z)$. Thus we have to develop a scheme which in a consistent manner would use elements of both the equations. Such scheme was proposed in Ref. [36] and realized in a successful fit to $F_2$ data in Ref. [37].

[3] Author: Michael Lublinsky.

One of the main problems of the DGLAP evolution is a necessity to specify the $x$ dependence of the distributions in the initial conditions of the evolution. The scheme which we propose generally avoids this problem and thus can be used for future more elaborated analysis including NLO corrections and the quark sector.

At low $x$ it is very convenient to use the dipole picture. In this approach the structure function $F_2$ can be expressed through the universal dipole cross section $\sigma^{dipole}$:

$$F_2(y,Q^2) \;=\; \frac{Q^2}{4\pi^2}\int d^2 r \int d\, z \; P^{\gamma^*}(Q^2;r,z)\; \sigma^{dipole}(r,y)\;. \qquad (3)$$

with the probability to find a dipole of the transverse size $r$ in the photon's wavefunction given by

$$P^{\gamma^*}(Q^2;r,z)^2 \;=\; \frac{N_c}{2\,\pi^2}\sum_{f=1}^{3} Z_f^2\,\left\{(z^2+(1-z)^2)\,a^2\,K_1^2(a\,r) \;+\; 4\,Q^2\,z^2\,(1-z)^2\,K_0^2(a\,r)\right\}\;,$$

where $a^2 = Q^2 z(1-z)$, $Z_f$ are the quark charges, and $K_i$ the standard modified Bessel functions.

The dipole cross section is determined through the evolution of the imaginary part of the dipole target elastic amplitude $N$ subsequently integrated over the impact parameter $b$ (in the analysis of Ref. [37] the dependence on $b$ was modeled):

$$\sigma^{dipole}(r,y) \;=\; 2\int d^2\, b\; N(r,y;b)\,.$$

In our approach, the amplitude $\tilde{N}$ is given by a sum of two terms

$$N \;=\; \tilde{N} \,+\, \Delta N$$

The first term $\tilde{N}$ follows from the solution of the BK equation whereas $\Delta N$ is a DGLAP correction to it (Fig. 9). The strategy of the fit is the following. We trust the DGLAP evolution for $x$ above $x_0 = 10^{-2}$. The gluon density obtained as a result of this evolution is then used as a initial condition for the low $x$ evolution based on the BK equation. In practice the CTEQ6 gluon was used as an input. The large distance behavior was extrapolated using the method proposed in Ref. [38]. The extrapolation is based on the geometrical scaling [39], a phenomenon experimentally observed by HERA. The BK evolved function $N$ is fitted to the low $Q^2$ data, with the effective proton size being the only fitting parameter entering the $b$ dependence ansatz. As the last step, the DGLAP correction $\Delta N$ is switched on and computed by solving a DGLAP-type equation. An inhomogeneous $N$-dependent term in the equation acts as a source term for $\Delta N$. This allows to have zero initial condition for the DGLAP correction. [4]

### 3.2 Results

We skip most of the technical details reported in Ref. [37] and present a result of the fit with $\chi^2/d.o.f. \simeq 1$. Fig. 10 displays the results vs. a combined set of experimental data for $x$ below $10^{-2}$. The solid line is the final parameterization. The dashed line on plot (b) is the result without DGLAP corrections added. Figure 11, a presents our results for the logarithmic derivative of $F_2$ with respect to $lnx$. This graph illustrates the hard-soft pomeron transition as a result of multiple rescattering of the BFKL pomeron. The intercept decreases from the LO BFKL intercept of the order 0.3 to the hadronic value of the order 0.1. As clearly observed from Fig. 11a, the intercept depends strongly on the photon virtuality $Q^2$ and decreases towards hadronic value when the virtuality decreases. If we further increase the energy, the intercept would eventually vanish in accord with the unitarity requirements. The band of our estimates for the value of saturation scale at LHC is displayed on Fig. 11b together with the most popular Golec-Biernat Wüsthoff saturation model [40]. Based on our analysis we predict much stronger saturation effects compared to the ones which could be anticipated from the GBW model. Though the power growth of the saturation scale in both cases is given by the very same exponent of the order $\lambda \simeq 0.3$, we had to take a much stronger saturation input at the beginning of the evolution.

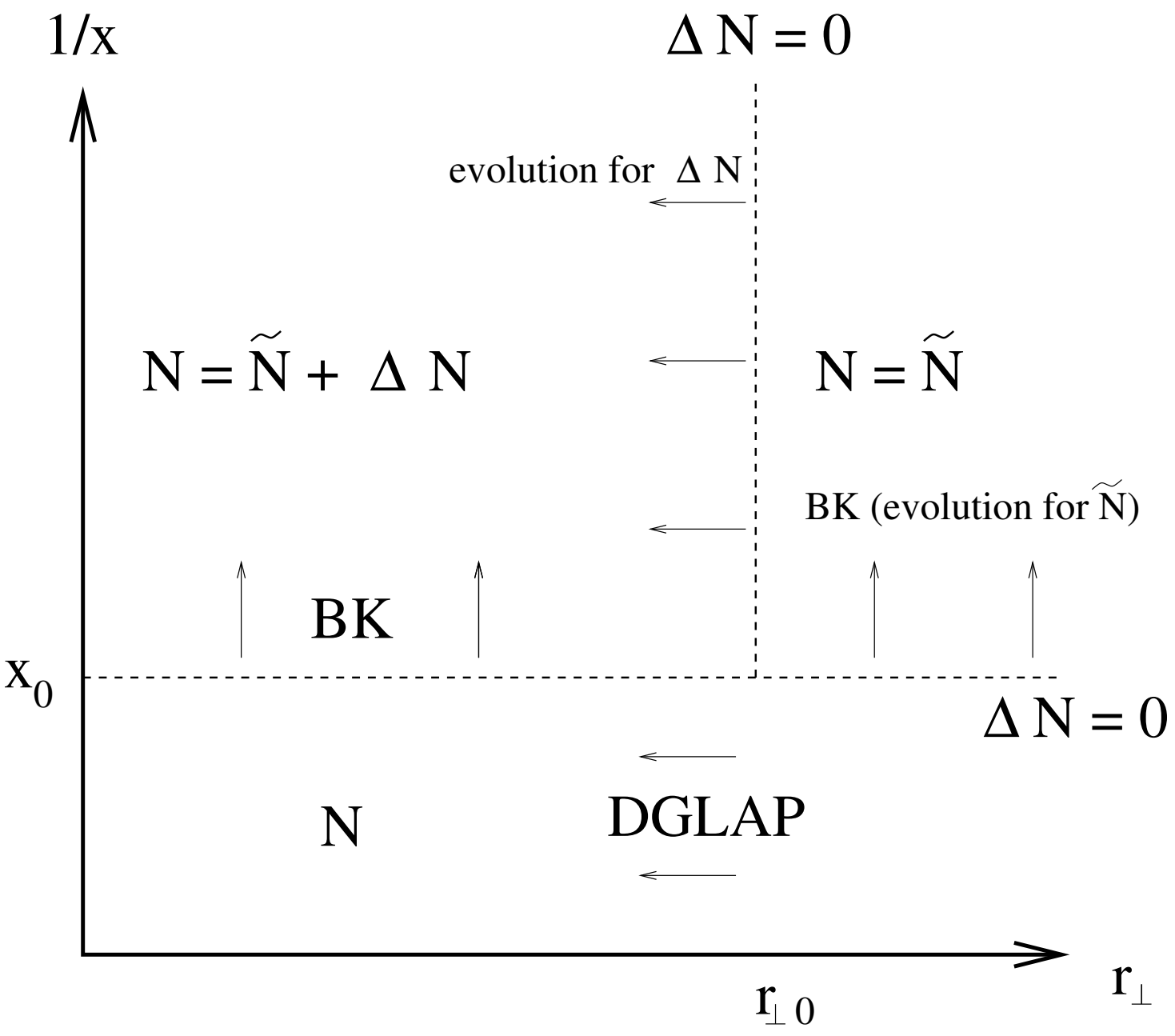

**Fig. 9:** The kinematic map for the solutions.

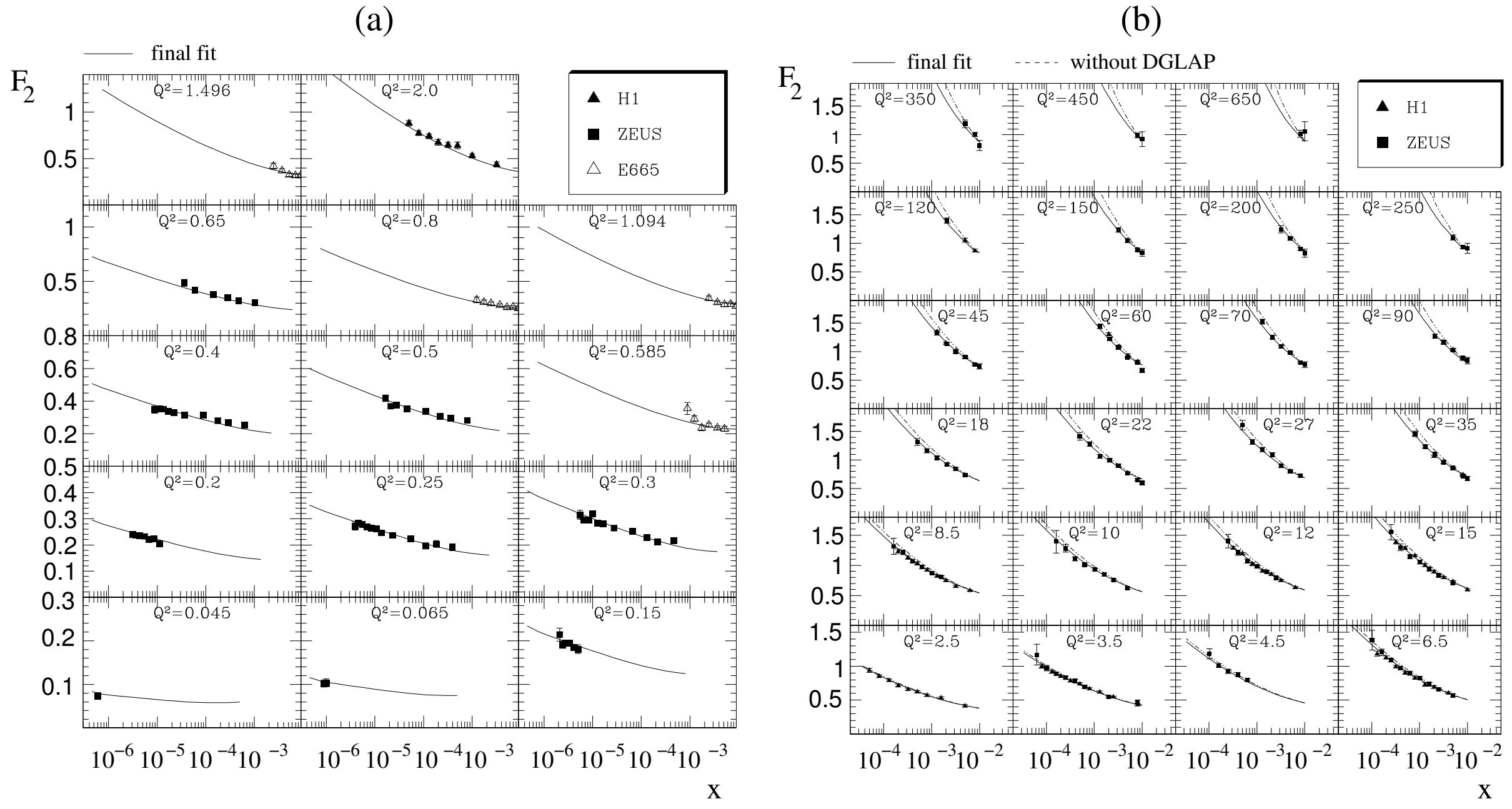

**Fig. 10:** Fit to the $F_2$ structure function.

[4]The initial condition for the BK equation is CTEQ gluon distribution. In the DGLAP-type equation for $\Delta N$ an initial condition at $r = r_0$ is required, which is set to zero and no modelling of the small $x$ behavior is needed.

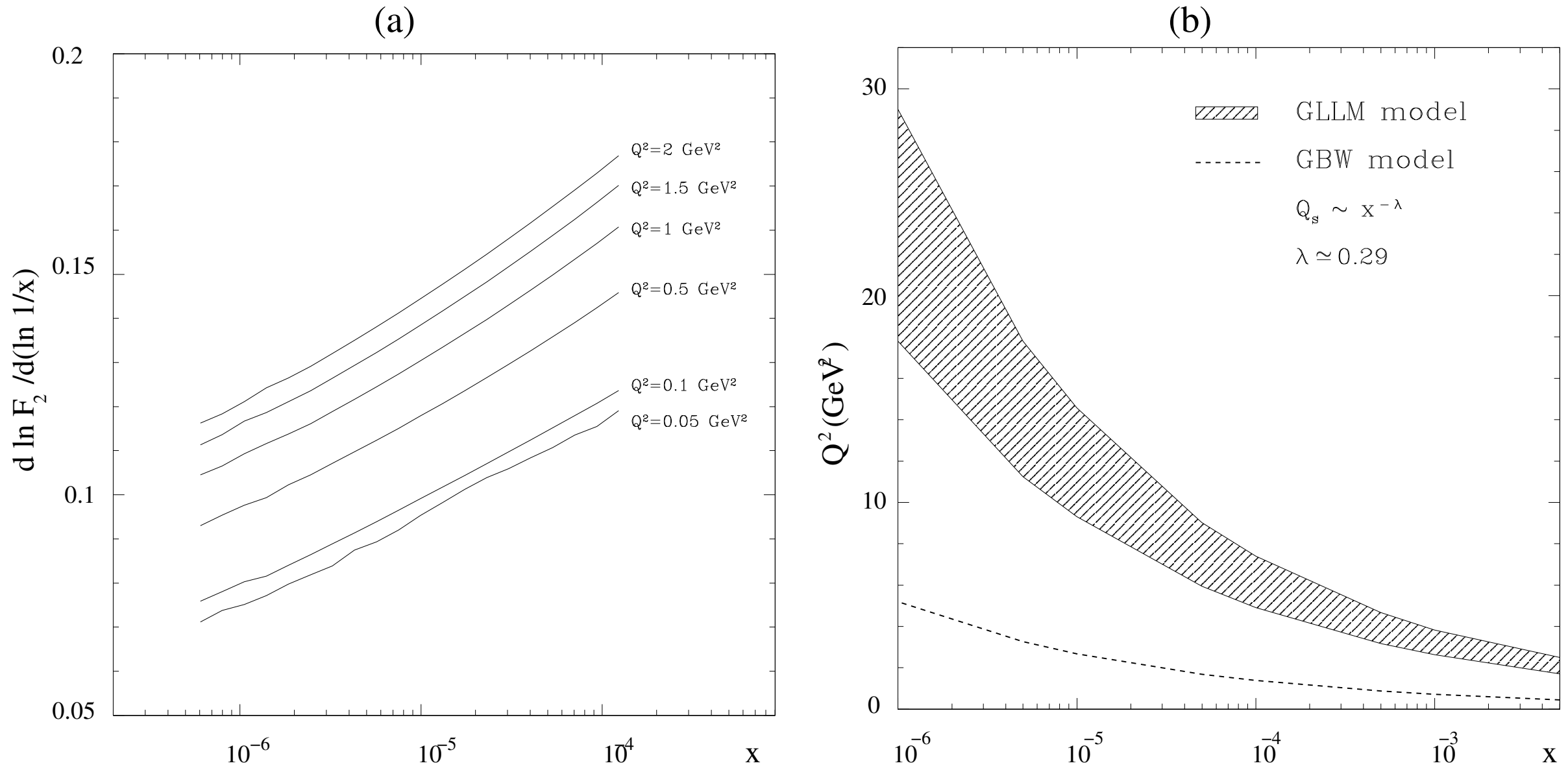

**Fig. 11:** (a) The logarithmic derivative $\lambda = \partial \ln F_2 / \partial \ln 1/x$ plotted at low $Q^2$ and very low $x$. (b) Saturation scale. the hatched area defines a prediction band of Ref. [37]; dashed line is the GBW model.

Model predictions for $F_L$ at HERA and $F_2$ at LHC can be found in Ref. [37]. Having determined the dipole cross section we can relate it to the unintegrated gluon distribution $f(k, y)$:

$$\sigma^{dipole}(r, y) = \frac{4\pi^2}{N_c} \int \frac{d k^2}{k^4} \left[1 - J_0(k\, r)\right] \alpha_s(k^2)\, f(k, y)\,. \tag{4}$$

The relation (4) can be inverted for $f$ which can be then used as an input for any computation based on the $k_t$ factorization scheme. The data set for the dipole cross section $\sigma^{dipole}$ as well as for the unintegrated gluon $f$ can be found in [24]. The uPDF is compared to other parameterizations in Fig.1.

### 3.3 Outlook

We have reported on, so far, the most advanced analysis of the $F_2$ data based on combined BK/DGLAP evolution equations. Though our approach incorporates most of the knowledge accumulated in saturation physics, it is not yet fully developed. The next essential steps would be to include NLO corrections both to BFKL and DGLAP. The quark sector should be also added into a unique scheme.

## 4 Generalized parton distributions[5]

The theoretical description of hard diffractive processes involves the gluon distribution in the proton. Such processes have a proton in the final state which carries almost the same momentum as the incident proton. Due to the small but finite momentum transfer, it is not the usual gluon distribution which appears, but its generalization to nonforward kinematics. Prominent example processes are the exclusive production of mesons from real or virtual photons (Figure 12a) when either the photon virtuality or the meson mass provides a hard scale, virtual Compton scattering $\gamma^* p \to \gamma p$, and the diffractive production of a quark-antiquark pair (Figure 12b) in suitable kinematics. The generalized gluon distribution depends on the longitudinal momentum fractions $x$ and $x'$ of the emitted and reabsorbed gluon (which differ because of the longitudinal momentum transfer to the proton) and on the invariant momentum transfer $t = -(p - p')^2$. In its "unintegrated" form it depends in addition on the transverse momentum $k_t$ of the emitted gluon. Another important process involving this distribution is exclusive diffractive production of a Higgs in $pp$ scattering (Figure 13), discussed in detail in [41]. Note that the description of this process requires the gluon distribution to be unintegrated with respect to $k_t$, whereas the processes in $\gamma^{(*)}p$ collisions mentioned above can be treated either in $k_t$-factorization or in the collinear factorization framework, where $k_t$-integrated generalized parton distributions occur. Note also that Figures 12 and 13 show graphs for the process *amplitudes*: the cross section depends hence on the square of the gluon distribution for Figure 12, and on its fourth power for Figure 13.

[5]Authors: Markus Diehl and Thomas Teubner.

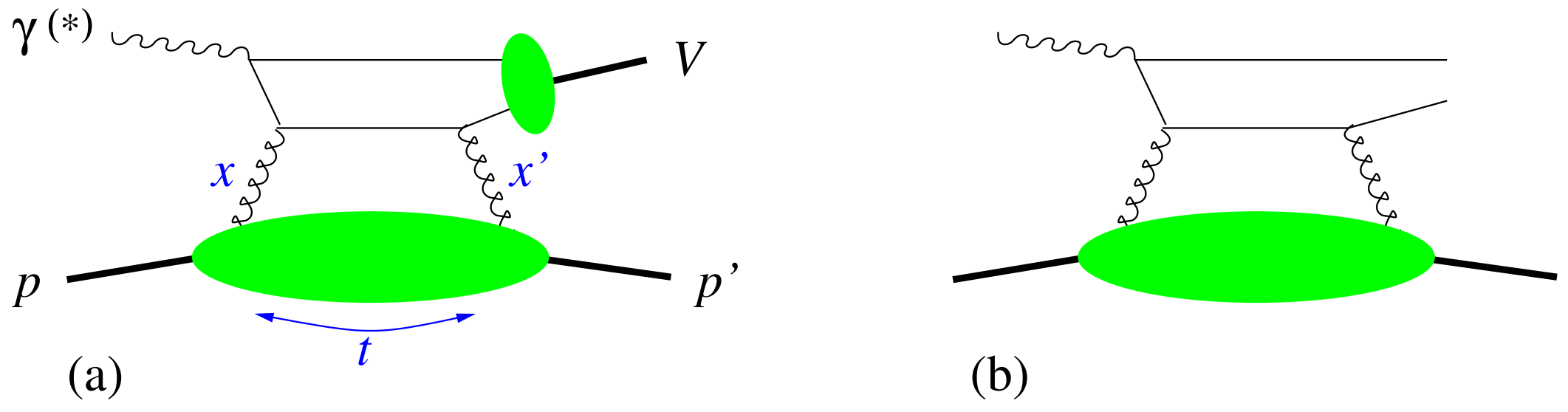

**Fig. 12:** Example graphs for the diffractive production of (a) a vector meson $V$ or (b) a quark-antiquark pair. The large blob denotes the generalized gluon distribution of the proton and the small one the vector meson wave function.

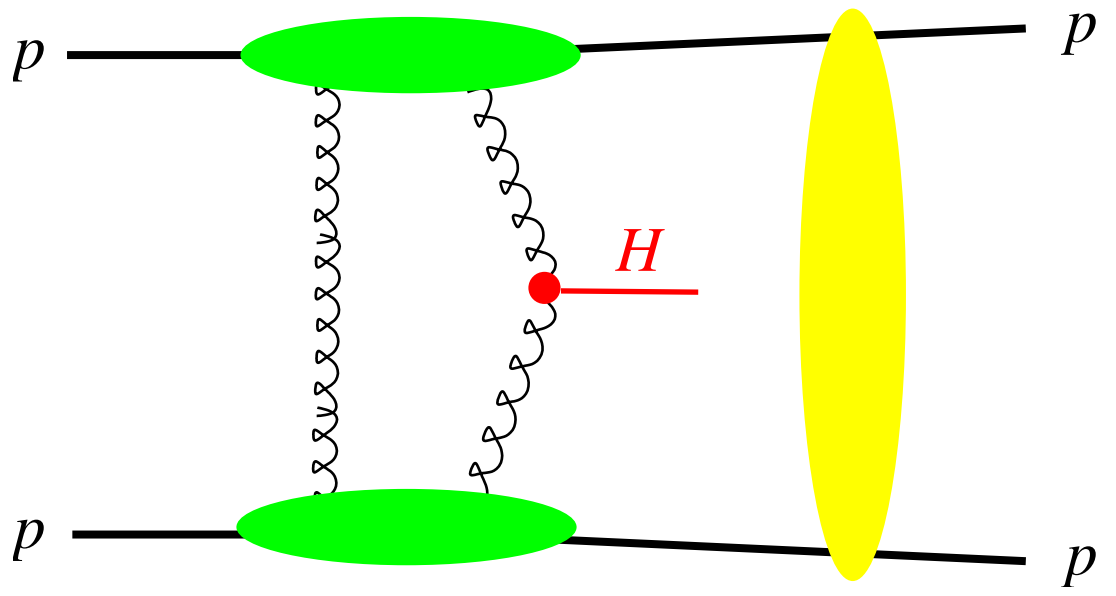

**Fig. 13:** Graph for the exclusive diffractive production of a Higgs boson, $p + p \to p + H + p$. The horizontal blobs indicate generalized gluon distributions, and the vertical blob represents secondary interactions between the projectiles.

To extract the generalized gluon distribution from vector meson production data requires knowledge of the meson wave function, which is an important source of uncertainty for the $\rho^0$ and $\phi$ and, to a lesser extent, for the $J/\Psi$. In this respect $\Upsilon$ production is by far the cleanest channel but experimentally challenging because of its relatively low production rate. An approach due to Martin, Ryskin and Teubner (MRT) [42] circumvents the use of the meson wave function by appealing to local parton-hadron duality, where the meson production cross section is obtained from the one for open quark-antiquark production, integrated over an interval of the invariant $q\bar{q}$ mass around the meson mass. The choice of that interval is then mainly reflected in an uncertainty in the overall normalization of the cross section. Virtual Compton scattering $\gamma^* p \to \gamma p$ does not involve any meson wave function and for sufficiently large $Q^2$ is again theoretically very clean.

By a series of steps one can relate the generalized gluon distribution to the usual gluon density, obtained for instance in global parton distribution fits.

1. The $t$ dependence is typically parameterized by multiplying the distribution at $t = 0$ with an exponential $\exp(-b|t|)$, whose slope $b$ has to be determined from measurement. In more refined models this slope parameter may be taken to depend on the other kinematic variables of the process.

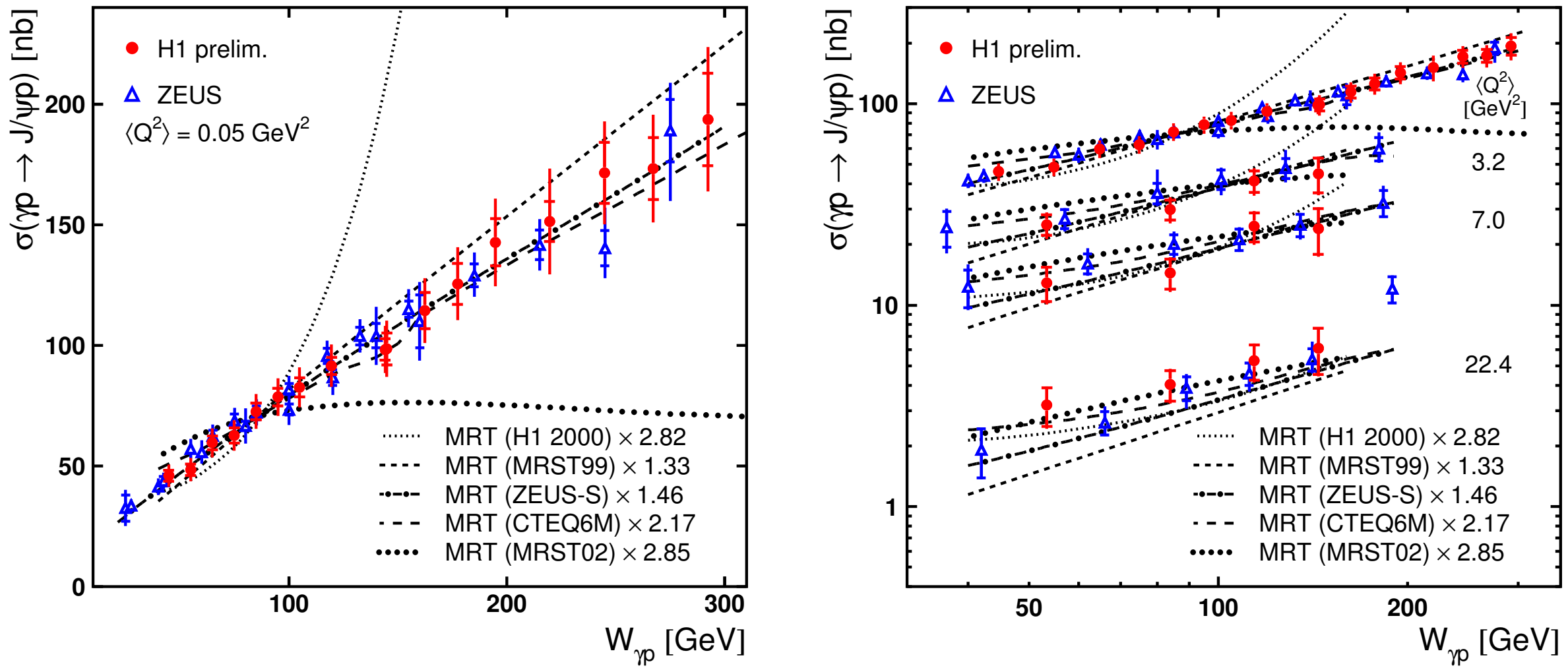

**Fig. 14:** Data for the $\gamma^* p \to J/\Psi\, p$ cross section from H1 [47] and ZEUS [48, 49] compared to calculations in the MRT approach [42, 46] with different gluon densities. The upper data points in the right panel correspond to those in the left one. The ZEUS data has been shifted to the $Q^2$ values of the H1 analysis using the $Q^2$ dependence measured by ZEUS, as described in [47]. Figure courtesy of Philipp Fleischmann (H1 Collaboration).

2. To leading logarithmic accuracy in $\log(1/x)$ one can neglect the difference between the longitudinal momentum fractions of the two gluons. The amplitude for meson production is then proportional to the usual gluon density evaluated at $x_g = (M_V^2 + Q^2)/W^2$, where $M_V$ is the meson mass, $Q^2$ the photon virtuality, and $W$ the $\gamma^* p$ c.m. energy. For phenomenology this leading logarithmic approximation is however insufficient. A weaker approximation allows one to express the amplitude in terms of the gluon density at $x_g$ times a correction factor for the kinematic asymmetry ("skewing") between the two momentum fractions [43].
3. The problem to relate the $k_t$ unintegrated gluon distribution to the $k_t$ integrated one is quite analogous to the case of the usual forward gluon density (see Sect. 1.1), with some specifics concerning Sudakov form factors in the nonforward case [44].

An overview and discussion of theoretical aspects and uncertainties in describing vector meson production in this framework can be found in [45].

To illustrate the sensitivity of such processes to the gluon distribution we show in Figure 14 data for photo- and electroproduction of $J/\Psi$ compared to calculations in the MRT approach [46], with different gluon densities taken as input to construct the generalized gluon distribution as just described. The potential of such processes to constrain the gluon distribution is evident from this plot.

We finally note that the theoretical description of diffractive Higgs production in $pp$ collisions is very similar to the description of diffractive processes in $ep$ scattering using $k_t$ factorization (much more than to the description of, say, inclusive DIS in collinear factorization, which provides the main input to the determination of conventional gluon densities at small $x$), see [41, 50] for further discussion. The analysis of diffractive $ep$ scattering is hence well suited to provide input to estimate the diffractive Higgs cross section at the LHC.

## Acknowledgments

This work is supported in part (JC) by the U.S. DOE.

# Resummation

*A. Banfi*[1], *G. Corcella*[2], *M. Dasgupta*[3], *Y. Delenda*[3], *G.P. Salam*[4] *and G. Zanderighi*[5].
[1] University of Cambridge, Madingley Road, Cambridge CB3 0HE ,U.K
[2] Department of Physics, Theory Division, CH-1211 Geneva 23, Switzerland.
[3] University of Manchester, Oxford Road, Manchester, M13 9PL, U.K
[4] LPTHE, Universities of Paris VI and VII and CNRS, 75005, Paris, France.
[5] Fermilab, Batavia, IL 60510, USA.

**Abstract**

We review the work discussed and developed under the topic "Resummation" at Working Group 2 "Multijet final states and energy flow" , of the HERA-LHC Workshop. We emphasise the role played by HERA observables in the development of resummation tools via, for instance, the discovery and resummation of non-global logarithms. We describe the event-shapes subsequently developed for hadron colliders and present resummed predictions for the same using the automated resummation program CAESAR. We also point to ongoing studies at HERA which can be of benefit for future measurements at hadron colliders such as the LHC, specifically dijet $E_t$ and angular spectra and the transverse momentum of the Breit current hemisphere.

## 1 Introduction

Resummed calculations are an invaluable tool, both for the understanding of perturbative QCD dynamics at all orders as well as for extracting, as accurately as possible, QCD parameters such as the strong coupling, quark masses and parton distribution functions. These parameters, which cannot be directly computed from QCD perturbation theory itself, will be vital inputs in new physics searches at the LHC. Moreover, resummed expressions are also an important stepping stone to probing observable distributions in regions where non-perturbative power corrections make a significant contribution. In this region one may expect a smearing of the resummed perturbative result with a non-perturbative function (for which one can adopt, for example, a renormalon-inspired model), and the resulting spectrum can be confronted with data to test our understanding of non-perturbative dynamics. In all these aspects, HERA data and observables have played an important role (sometimes significantly underrated in the literature) in furthering our knowledge, without which accurate studies of several observables at the LHC would simply not be possible.

A concrete example of HERA's important role in this regard is the case of event shape distributions [1, 2], theoretical studies of which led to the finding of non-global single-logarithmic [3] effects (discussed in more detail below). Prior to these studies it was widely believed that the HERA distributions, measured in the current hemisphere Breit frame, were trivially related to their $e^+e^-$ counterparts. Had such ideas, based on independent soft gluon emission by the hard partons, been applied directly to similar variables at the LHC, such as energy flows away from jets, the accuracy of theoretical predictions would have been severely compromised leading almost certainly to erroneous claims and conclusions.

Another area where HERA has played a vital role is in the testing of renormalon inspired models for power corrections, most significantly the dispersive approach [4] to $1/Q$ power corrections, tested against HERA event-shape distributions and mean-values [5]. The fact that HERA data seem to confirm such models , where one can think of the power corrections as arising from the emission of a gluon with transverse momentum $\mathcal{O}(\Lambda_{\mathrm{QCD}})$, is significant for the LHC. This is because the agreement of the renormalon model with data demonstrates that the presence of initial state protons does not affect significantly the form of $1/Q$ corrections. It thus sets limits on the additional non-perturbative contribution that may potentially be generated by the flight of struck partons through the proton cloud, which therefore does not appear to be significant. Once again it is accurate resummed predictions [6] which have allowed us access to the non-perturbative domain hence strengthening our understanding of power corrections.

One important aspect of resummed studies, till date, is that stringent comparisons of next-to–leading logarithmic resummed predictions with data have only been carried out in cases involving observables that vanish in the limit of two hard partons. Prominent examples reflecting the success of this program are provided by $e^+e^- \to 2$ jet event shapes and DIS (1+1) jet event shapes as well as Drell-Yan vector boson transverse momentum spectra at hadron colliders. At the LHC (and hadron colliders in general) one already has two hard incoming partons and any observable dealing with final state jet production would take us beyond the tested two hard parton situation. Thus dijet event shapes at hadron colliders (discussed in detail later), which involve much more complicated considerations as far as the resummation goes, represent a situation where NLL resummations and power corrections are as yet untested. Bearing in mind the hadronic activity due to the underlying event at hadron colliders, it is important to test the picture of resummations and power corrections for these multiparton event shapes in cleaner environments. Thus LEP three-jet event shapes and similar $2+1$ jet event shapes at HERA become important to study in conjunction with looking at resummation of event shapes at hadron colliders.

Predictions for several LEP and HERA three-jet event shapes already exist (see e.g [7] and for a full list of variables studied Ref. [8]) and at this workshop a prominent development presented was the proposal of several dijet event-shapes in hadron-hadron collisions and the resummed predictions for their distributions [9].

Existing HERA data can also be usefully employed to study soft gluon radiation dynamics from multi-hard–parton ensembles, in the study of dijet $E_t$ and angular spectra. These quantities are somewhat different from event shapes since one defines observables based on aggregate jet-momenta and angles rather than directly constructing them from final-state hadron momenta. Examples are the transverse energy, $E_t$, mismatch between the leading $E_t$ jets in dijet production and the azimuthal correlation between jets $\phi_{jj}$, once again refering to the highest $E_t$ jets in dijet production. For the former quantity there are no direct experimental data as yet, but it is simply related to the dijet total rate in the region of symmetric $E_t$ cuts for which data does exist . For the latter quantity similarly there are direct experimental data [10]. These observables have smaller hadronisation corrections scaling as $1/Q^2$ rather than $1/Q$ as for most event shapes. They thus offer a good opportunity to test the NLL perturbative predictions alone without necessarily probing non-perturbative effects at the same time [1].

At this workshop developments were reported on extending existing calculations [11] for cone dijets, to different jet algorithms, such as the $k_t$ algorithm, comparing to fixed order estimates and performing the leading order matching. Once the HERA data has been well described similar studies can be carried out for hadron–hadron dijets. In fact predictions already exist for hadron-hadron dijet masses near threshold [12] but are not in a form conducive to direct comparisons with data containing neither the jet algorithms in the form actually employed in experiment, nor the matching to fixed order. However these calculations provided a useful starting point for the calculations presented here, which should eventually lead to direct comparisons with data.

Another area where HERA may play an important role is to establish whether unaccounted for small $x$ effects may be significant in comparing theoretical resummations for e.g. vector boson $p_t$ spectra with experimental data. It has been suggested that a non-perturbative intrinsic $k_t$, growing steeply with $x$, is required to accomodate HERA data for semi-inclusive DIS processes [13]. When this observation is extrapolated to the LHC kinematical region there is apparently significant small $x$ broadening in the vector boson $p_t$ distribution. Similar effects may well arise in the case of the Higgs boson too. However DIS event shape studies in the Breit current hemisphere [6] apparently do not acquire such corrections since they are well described by conventional NLL resummations supported by dispersive

[1]Although effects to do with intrinsic $k_t$ will eventually have to be accounted for similar to the case of Drell-Yan vector boson $p_t$ spectra.

power corrections [5], which are $x$ independent [2]. However there are some important caveats:

- Unlike vector boson $p_t$ spectra, event shapes receive $1/Q$ hadronisation corrections unrelated to intrinsic $k_t$. These could mask $1/Q^2$ terms originating from intrinsic $k_t$ which may yet contain the $x$ dependence in question.
- It has already been observed that including H1 data for $Q < 30$ GeV does spoil somewhat the agreement with the dispersive prediction of universal power corrections to event shapes [6]. The origin of this effect could well be extra non-perturbative $k_t$ broadening related to the effects described above for vector boson $p_t$.

To get to the heart of this matter a useful variable that has been suggested (see plenary talk by G. Salam at the first meeting of this workshop) is the modulus of the vector transverse momentum $\sum_{i\in H_c} \vec{k}_{t,i}$ of the current hemisphere in the DIS Breit frame. This quantity is simply related to the Drell-Yan $p_t$ spectra and comparing theoretical predictions, presented here, with data from HERA should help to finalise whether additional small-$x$ enhanced non-perturbative terms are needed to accomodate the data. We begin by first describing the results for hadron-hadron event shape variables, discussed by G. Salam at this workshop. Then we describe the progress in studying dijet $E_t$ and angular spectra (presented by M. Dasgupta and G. Corcella at the working group meetings). Finally we mention the results obtained thus far, for the $Q_t$ distribution of the current hemisphere and end with a look at prospects for continuing phenomenology at HERA, that would be of direct relevance to the LHC.

## 2 Event shapes for hadron colliders

Event shape distributions at hadron colliders, as has been the case at LEP and HERA, are important collinear and infrared safe quantities, that can be used as tools for the extraction of QCD parameters, for instance $\alpha_s$, by comparing theory and data. In contrast however to more inclusive sources of the same information (e.g the ratio of 3 jet to 2 jet rates), event shape distributions provide a wealth of other information, some of which ought to be crucial in disentangling and further understanding the different physics effects, relevant at hadron colliders. These range from fixed-order predictions to resummations, hadronisation corrections and, in conjunction with more detailed studies assesing the structure of, and role played by, the underlying event (beam fragmentation).

Until recently there have only been limited experimental studies of jet-shapes at hadron colliders [15] and no resummed theoretical predictions for dijet shape variables at hadron colliders. Rapid recent developments (see Ref. [9] and references therin) in the field of perturbative resummations have now made theoretical estimates possible for a number of such distributions, introduced in [9] which we report on below.

The three main theoretical developments that have led to the studies of Ref. [9] are:

- Resummation for hadron-hadron dijet observables depends on describing multiple soft gluon emission from a system of four hard partons. The colour structure of the resulting soft anomalous dimensions is highly non-trivial and was explicitly computed by the Stony Brook group in a series of papers (see e.g [12] and references therin).
- The discovery of non-global observables [3]. The realisation that standard resummation techniques based on angular ordering/independent-emission of soft gluons by the hard-parton ensemble, are not valid for observables that are sensitive to emissions in a limited angular range, has led to the introduction of observables that are made global by construction. This means that one can apply the technology developed by the Stony-Brook group to obtain accurate NLL predictions for these observables, without having to resort to large $N_c$ approximations.

[2] An exception is the jet broadening [14] but the $x$ dependence there is of an entirely different origin and nature.

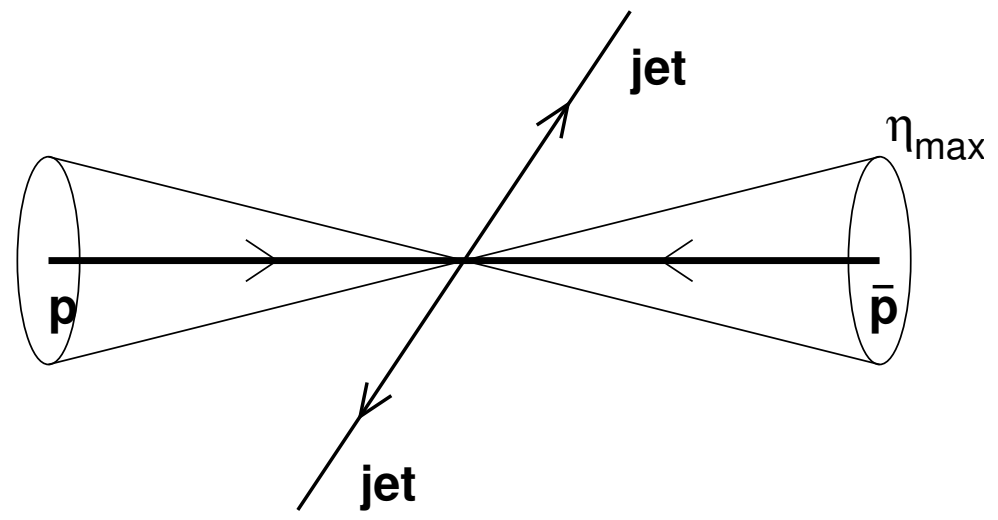

**Fig. 1:** Cut around the beam direction beyond rapidity $\eta_{\max}$ corresponding to the maximum rapidity reach of the detectors.

– The advent of automated resummation [16]. The development of generalised resummation formulae and powerful numerical methods to determine the parameters and compute the functions thereof, has made it possible to study several variables at once rather than having to perform copious, and in some cases previously unfeasible, calculations for each separate observable.

We now discuss the different types of variables defined and resummed in [9]. The first issue one has to deal with is the fact that experimental detectors have a limited rapidity range, which can be modeled by a cut around the beam direction.

This cut would then correspond to a position in rapidity of the edge of the most forward detector with momentum or energy resolution and the relevant values of the maximum rapidity for measurements is 3.5 units at the Tevatron and 5 units at the LHC. One may then worry about gluon emissions beyond this rapidity (i.e. inside the beam cut, see Fig. 1) that emit softer gluons into the allowed rapidity range, outside the cones depicted in Fig. 1. Such a configuration would of course render the observable non-global.

To get around this potential problem, one can employ an idea suggested for 3-jet observables such as out-of–plane momentum flows in hadron-hadron collisions [17], which helps side-step the issue of non-globalness. We note that all the observables studied here have the following functional dependence on a soft emission, $k$, collinear to a given hard leg [3] (common to all event shapes studied here and in other processes)

$$V(\tilde{p}, k) = d\left(\frac{k_t}{Q}\right)^a e^{-b\eta} g(\phi), \tag{1}$$

where $k_t$ , $\eta$ and $\phi$ are measured wrt a given hard leg and $\tilde{p}$ represent the set of hard parton momenta including recoil against $k$ while $Q$ is the hard-scale of the process. We are particularly interested in emissions soft and collinear to the beam (incoming) partons. Then an emission beyond the maximum detector rapidity $\eta \geq \eta_{\max}$ corresponds to at most a contribution to the observable $V \sim e^{-(a+b_{\min})\eta_{\max}}$ with $b_{\min} = \min(b_1, b_2)$ and $b_1$ and $b_2$ are the values of $b$ associated with collinear emission near beam-partons 1 and 2.

If one then choses to study the observable over a range of values such that

$$L \leq (a + b_{\min})\eta_{\max}, \;\; L \equiv \ln 1/V, \tag{2}$$

then emissions more forward than $\eta_{\max}$ do not affect the observable in the measured range of values. One can thus include the negligible contribution from this region and do the calculation *as if the observable were global*, ignoring the cut around the beam. Including the region beyond $\eta_{\max}$ does not alter the NLL resummed result in the suitably selected range Eq. 2.

[3] In general the values of parameters $d, a, b$ and the function $g$ depend on the observable considered. For more details and constraints on the various parameters that ensure globalness and infrared and collinear safety etc., see Ref. [16].

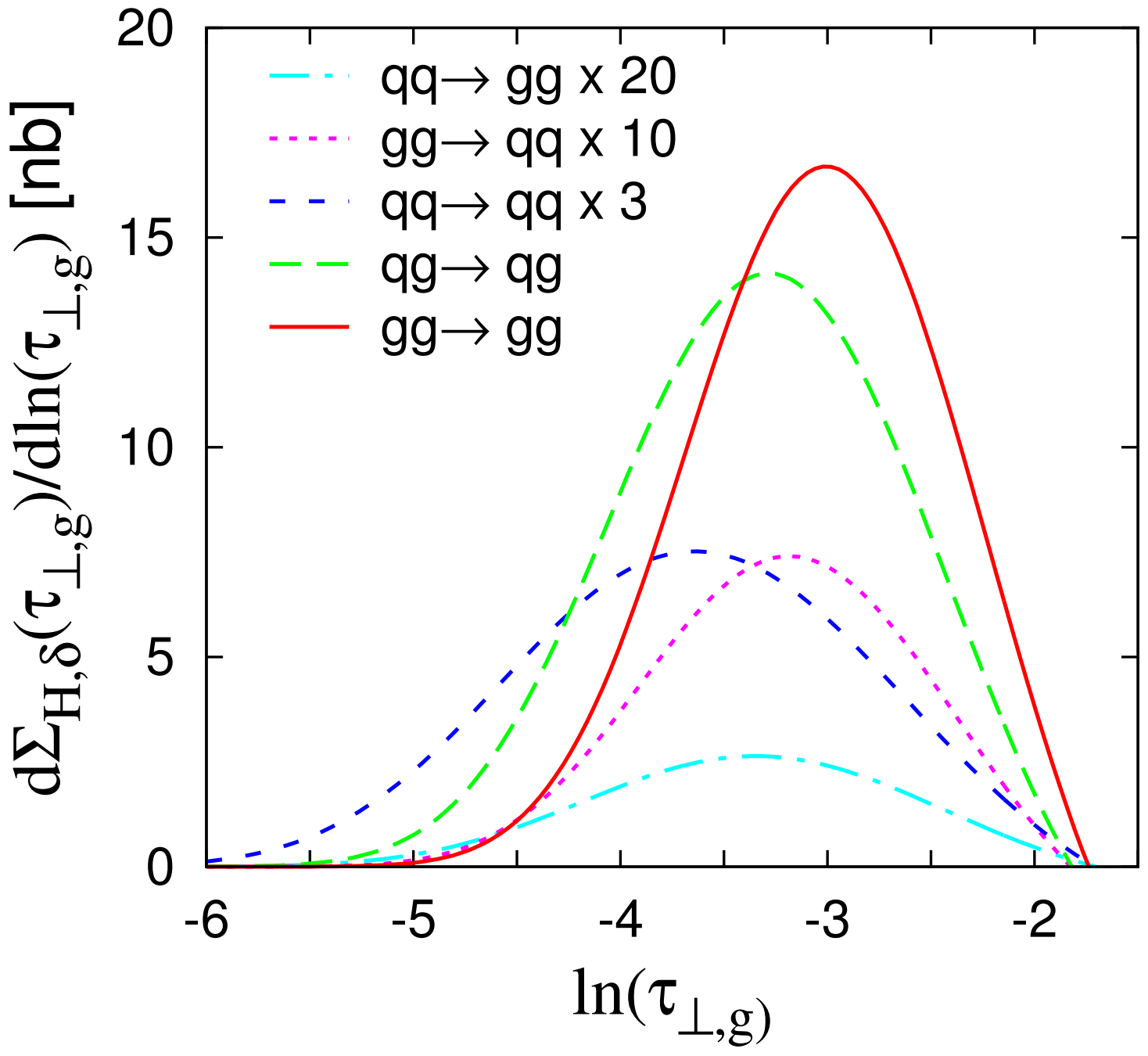

**Fig. 2:** The global transverse thrust distribution with the contribution from different partonic channels explicitly displayed.

The price one has to pay is to limit the range of the study of the observable $V$, such that emissions beyond $\eta_{\max}$ make a negligible contribution. As we will mention later this is a more significant restriction for some variables compared to others (depending on the parameters $a$ and $b$) but a range of study can always be found over which the observable can be treated as global.

### 2.1 Global event shapes

With the above caveat in place several variables can be safely studied (treated as global) over a wide range of values. An explicit example is the *global transverse thrust* defined as:

$$T_{\perp,g} \equiv \max_{\vec{n}_T} \frac{\sum_i |\vec{q}_{\perp i} \cdot \vec{n}_T|}{\sum_i q_{\perp i}}\,, \qquad \tau_{\perp,g} = 1 - T_{\perp,g}\,, \tag{3}$$

where the thrust axis $\vec{n}_T$ is defined in the plane transverse to the beam axis. The probability $P(v)$, that the event shape is smaller than some value $v$ behaves as:

$$P(v) = \exp\left[-G_{12}\frac{\alpha_s}{2\pi}L^2 + \cdots\right],\; L = \ln 1/v, \tag{4}$$

with $G_{12} = 2C_B + C_J$, where $C_B$ and $C_J$ represent the total colour charges of the beam and jet (outgoing) partons. The above represents just the double-logarithmic contribution. The full result with control of up to next-to–leading single-logarithms in the exponent is considerably more complicated. It contains both the Stony-Brook colour evolution matrices as well as multiple emission effects (generated by phase-space factorisation). The automated resummation program CAESAR [16] is used to generate the NLL resummed result shown in Fig. 2. In this particular case the effect of the cut around the beam direction can be ignored for values $\tau_{\perp,g} \geq 0.15e^{-\eta_{\max}}$. We note that it is advisable to leave a safety margin between this value and the values included in measurement.

Other global variables studied include the *global thrust minor* and the three jet-resolution threshold parameter $y_{23}$. For detailed definitions and studies of these variables, the reader is refered to [9].

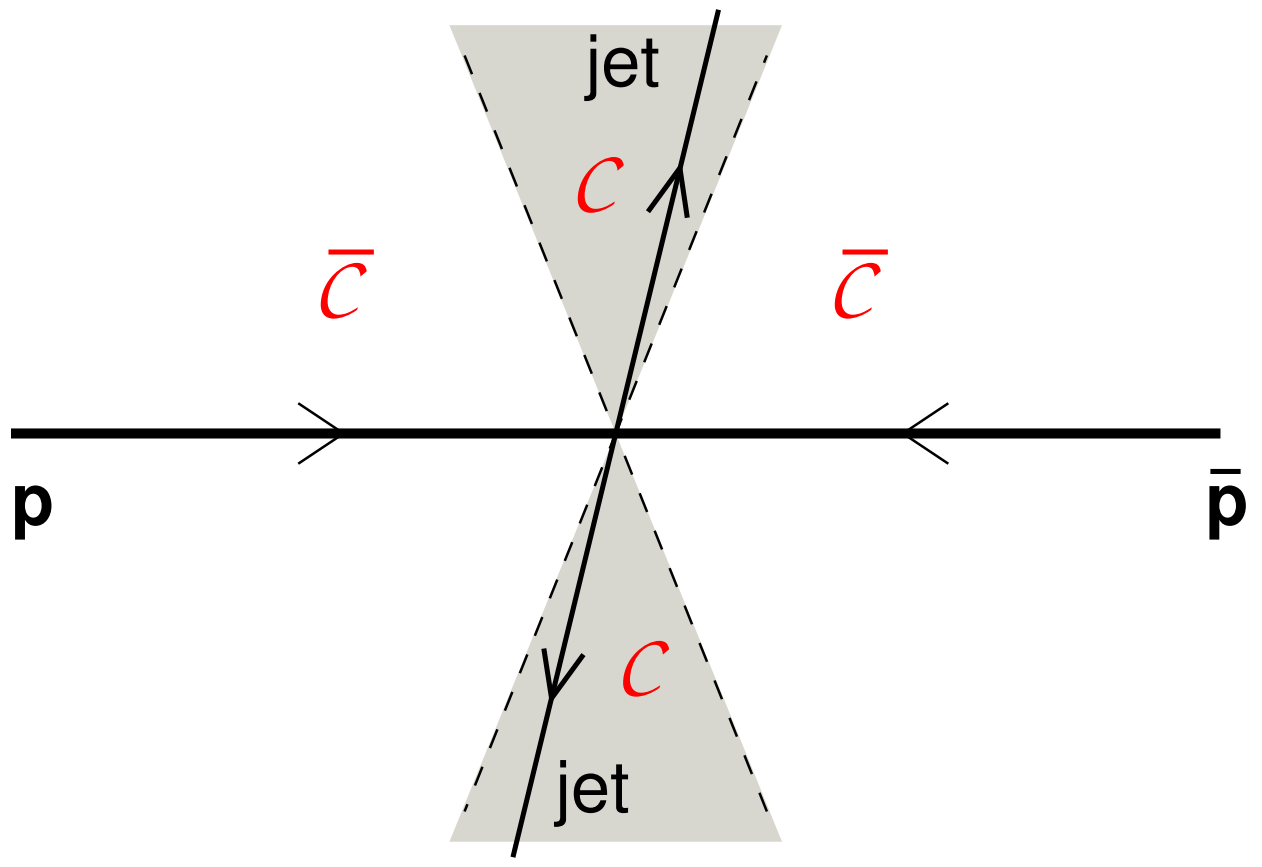

**Fig. 3:** Figure depicting the central region marked $\mathcal{C}$ , containing the two hard jets.

We shall now proceed to look at two different ways of defining event shapes in a given central region, which on its own would lead to non-globalness, and then adding terms that render them global.

## 2.2 Forward suppressed observables

Here we shall examine event shapes defined in a chosen central region $\mathcal{C}$ well away from the forward detector edges.

First we define central $\perp$ momentum, and rapidity:

$$Q_{\perp,\mathcal{C}} = \sum_{i\in\mathcal{C}} q_{\perp i}\,, \quad \eta_{\mathcal{C}} = \frac{1}{Q_{\perp,\mathcal{C}}}\sum_{i\in\mathcal{C}} \eta_i\, q_{\perp i} \tag{5}$$

and an *exponentially suppressed forward term,*

$$\mathcal{E}_{\bar{\mathcal{C}}} = \frac{1}{Q_{\perp,\mathcal{C}}}\sum_{i\notin\mathcal{C}} q_{\perp i}\, e^{-|\eta_i-\eta_{\mathcal{C}}|}\,. \tag{6}$$

Then we can define an event shape in the central region $\mathcal{C}$[4] which on its own would be non-global since we measure emissions just in $\mathcal{C}$. The addition of $\mathcal{E}_{\bar{\mathcal{C}}}$ to the event-shape renders the observable global as this term includes suitably the effect of emissions in the remaining region $\bar{\mathcal{C}}$. The exponential suppression of the added term reduces sensitivity to emissions in the forward region which in turn reduces the effect of the beam cut $\eta_{\max}$ considerably, pushing its impact to values of the observable where the shape cross-section is highly suppressed and thus too small to be of interest.

The event shapes are constructed as described stepwise below:

– Split $\mathcal{C}$ into two pieces: *Up, Down*
– Define *jet masses* for each

$$\rho_{X,\mathcal{C}} \equiv \frac{1}{Q^2_{\perp,\mathcal{C}}}\left(\sum_{i\in\mathcal{C}_X} q_i\right)^2, \qquad X = U, D\,. \tag{7}$$

Define sum and heavy-jet masses

$$\rho_{S,\mathcal{C}} \equiv \rho_{U,\mathcal{C}} + \rho_{D,\mathcal{C}}\,, \qquad \rho_{H,\mathcal{C}} \equiv \max\{\rho_{U,\mathcal{C}}, \rho_{D,\mathcal{C}}\}\,. \tag{8}$$

[4]There is considerable freedom on the choice of the central region. For instance this could be a region explicitly delimited in rapidity or the two hard jets themselves.

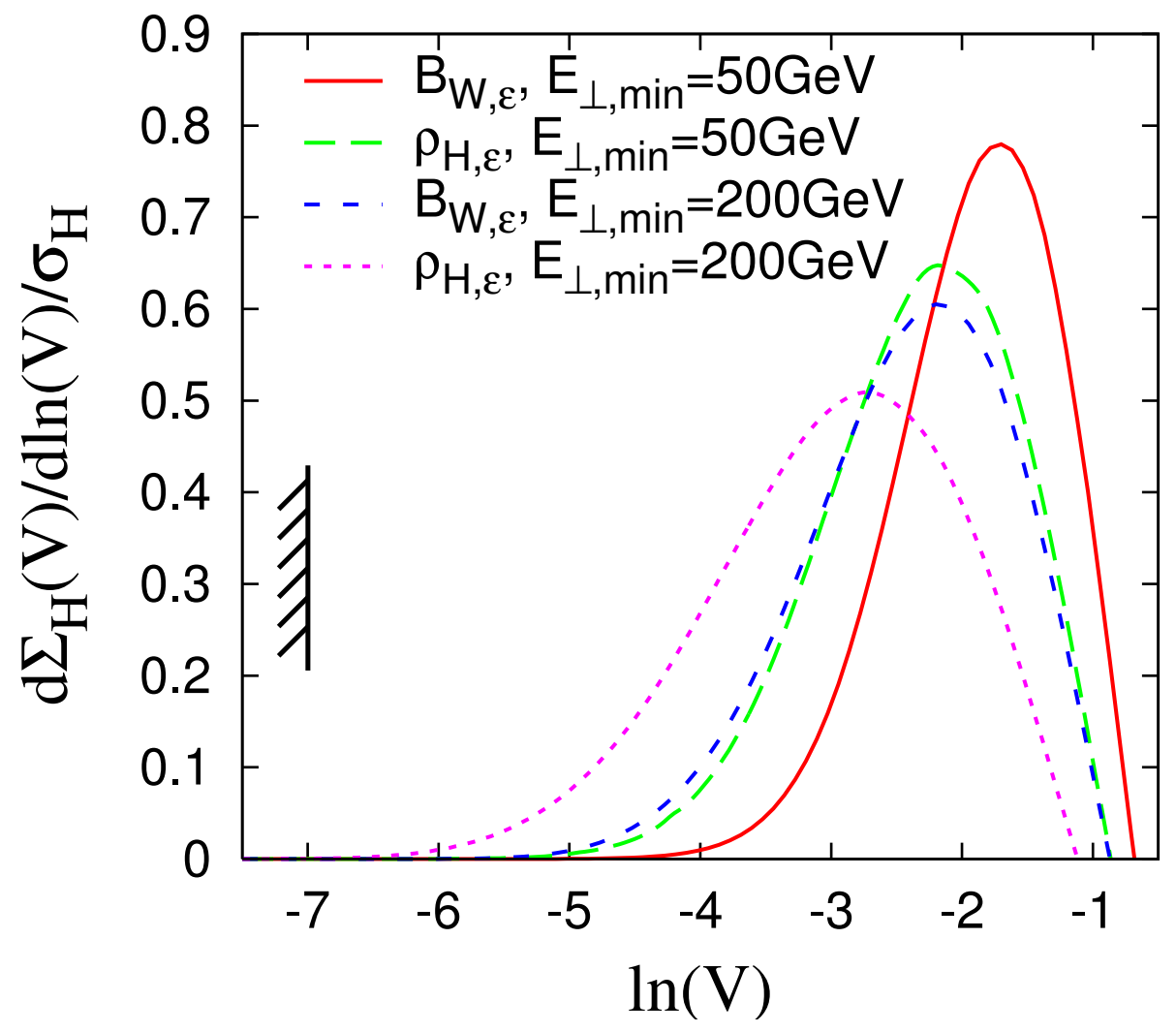

**Fig. 4:** NLL resummed predictions from CAESAR for the heavy jet-mass and the wide jet-broadening with the minimum jet transverse energy $E_{\perp,\min}$ values of 50 and 200 GeV as shown.

Define global extension, with extra forward-suppressed term

$$\rho_{S,\mathcal{E}} \equiv \rho_{S,\mathcal{C}} + \mathcal{E}_{\bar{\mathcal{C}}}\,, \qquad \rho_{H,\mathcal{E}} \equiv \rho_{H,\mathcal{C}} + \mathcal{E}_{\bar{\mathcal{C}}}\,. \tag{9}$$

– Similarly: *total and wide jet-broadenings*

$$B_{T,\mathcal{E}} \equiv B_{T,\mathcal{C}} + \mathcal{E}_{\bar{\mathcal{C}}}\,, \qquad B_{W,\mathcal{E}} \equiv B_{W,\mathcal{C}} + \mathcal{E}_{\bar{\mathcal{C}}}\,. \tag{10}$$

At the double-log level the results assume an identical form to Eq. 4 with $G_{12}$ representing a combination of total incoming (beam) and outgoing (jet) parton colour charges [9]. The full NLL resummed results have a substantially more complex form and results from CAESAR [16] are plotted in Fig. 4.

### 2.3 Indirectly global recoil observables

Here we study observables that are defined exclusively in terms of particles in the central region but are global. Such observables are already familiar from HERA studies. As an example, although the current-jet broadening wrt the photon axis of the DIS Breit frame involves only particles that enter the current hemisphere, the current quark acquires transverse momentum by *recoil* against remnant hemisphere particles. This recoil means that the observable is indirectly sensitive to emissions in the remnant hemisphere which makes the observables global.

To construct similar observables in the hadron-hadron case we observe that by momentum conservation, the following relation holds :

$$\sum_{i\in\mathcal{C}} \vec{q}_{\perp i} = -\sum_{i\notin\mathcal{C}} \vec{q}_{\perp i} \tag{11}$$

which relates the sum of transverse momenta in $\mathcal{C}$ to that in the complementary region. Then the central particles can be used to define a recoil term:

$$\mathcal{R}_{\perp,\mathcal{C}} \equiv \frac{1}{Q_{\perp,\mathcal{C}}} \left| \sum_{i\in\mathcal{C}} \vec{q}_{\perp i} \right| \,, \tag{12}$$

which contains an indirect dependence on non-central emissions.

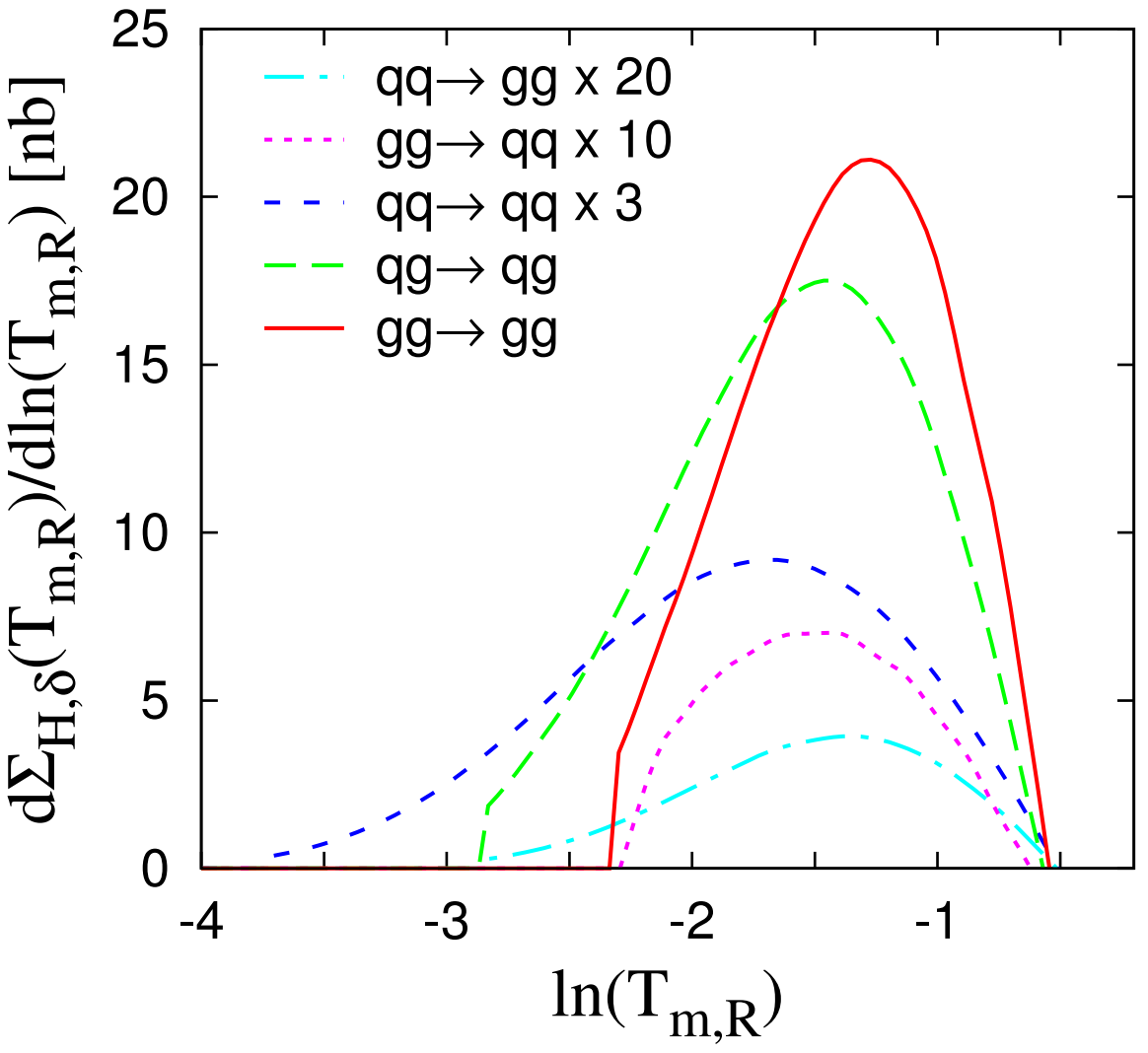

**Fig. 5:** The recoil thrust minor as predicted by CAESAR, with a cutoff before the divergence. Only a small fraction of the cross-section is beyond the cutoff.

Now we can define event shapes explicitly in terms of central particle momenta in $\mathcal{C}$. Examples are the recoil jet-masses and broadenings

$$\rho_{X,\mathcal{R}} \equiv \rho_{X,\mathcal{C}} + \mathcal{R}_{\perp,\mathcal{C}}\,, \qquad B_{X,\mathcal{R}} \equiv B_{X,\mathcal{C}} + \mathcal{R}_{\perp,\mathcal{C}}\,, \ldots \tag{13}$$

It is clear that since these observables are defined in terms of central particles alone, the cut around the beam direction is not an issue here. There is however another potential problem. Due to the addition of the recoil term we lose direct exponentiation of the result in variable space. Exponentiation to NLL accuracy only holds in impact-parameter or $b$ space .

The physical effect in question here is similar to Drell-Yan $Q_T$ spectra where there are two competing mechanisms that lead to a given small $Q_T$, Sudakov suppression of soft emissions and vectorial cancellation between harder emissions. Where the latter effect takes over (typically in the region where single-logs are large $\alpha_s L \sim 1$) we get a breakdown of the Sudakov result generated by CAESAR. This result is of the general form:

$$P(V) = e^{Lg_1(\alpha_s L)+g_2(\alpha_s L)+\cdots}. \tag{14}$$

The result for recoil observables produced by CAESAR will contain a divergence in the single-log function $g_2$ and is cut before the divergence. Again for some variables this cut is at a position that significantly reduces the range of possible phenomenological studies. For other variables the divergence is at values of the observable that are sufficiently small so that only a few percent of the cross-section is beyond the cutoff. An example of the former is the recoil transverse thrust where 15% of the cross-section lies beyond the cut-off. For the recoil thrust minor, in contrast, the cutoff has only a moderate effect and much less of the cross-section is cutoff, due to the divergence in $g_2$.

Table 1 contains the different event shapes mentioned here and the impact of the two main limitations we discussed, the beam-cut $\eta_{\max}$ and the breakdown of resummation due to divergences of $g_2$. Additionally we mention the expected impact of hadronisation corrections (not yet computed in full) on the different observables as well as the form of the estimated contribution from the underlying event. The entries marked * are subject to uncertainty at present.

Further work is needed before the resummed expressions presented here can be compared with data including the matching to fixed order and computation of the power corrections for the various observables. This is currently in progress.

**Table 1:** Event shapes and their characteristics

| Event-shape | Impact of $\eta_{\max}$ | Resummation breakdown | Underlying Event | Jet hadronisation |
|---|---|---|---|---|
| $\tau_{\perp,g}$ | tolerable* | none | $\sim \eta_{\max}/Q$ | $\sim 1/Q$ |
| $T_{m,g}$ | tolerable | none | $\sim \eta_{\max}/Q$ | $\sim 1/(\sqrt{\alpha_s}Q)$ |
| $y_{23}$ | tolerable | none | $\sim \sqrt{y_{23}}/Q^*$ | $\sim \sqrt{y_{23}}/Q^*$ |
| $\tau_{\perp,\mathcal{E}}$, $\rho_{X,\mathcal{E}}$ | negligible | none | $\sim 1/Q$ | $\sim 1/Q$ |
| $B_{X,\mathcal{E}}$ | negligible | none | $\sim 1/Q$ | $\sim 1/(\sqrt{\alpha_s}Q)$ |
| $T_{m,\mathcal{E}}$ | negligible | serious | $\sim 1/Q$ | $\sim 1/(\sqrt{\alpha_s}Q)$ |
| $y_{23,\mathcal{E}}$ | negligible | none | $\sim 1/Q$ | $\sim \sqrt{y_{23}}/Q^*$ |
| $\tau_{\perp,\mathcal{R}}$, $\rho_{X,\mathcal{R}}$ | none | serious | $\sim 1/Q$ | $\sim 1/Q$ |
| $T_{m,\mathcal{R}}$, $B_{X,\mathcal{R}}$ | none | tolerable | $\sim 1/Q$ | $\sim 1/(\sqrt{\alpha_s}Q)$ |
| $y_{23,\mathcal{R}}$ | none | intermediate* | $\sim \sqrt{y_{23}}/Q^*$ | $\sim \sqrt{y_{23}}/Q^*$ |

Having discussed the hadron-hadron event shapes we now move on to describe resummed studies concerning dijet production at HERA which can also be straightforwardly extended to hadron-hadron collisons.

## 3 Dijet $p_t$ and angular spectra

It has been known for some time that dijet total rates cannot be predicted within fixed-order QCD if symmetric cuts are applied to the two highest $p_t$ dijets [18]. While it was understood that the problems are to do with constraints on soft gluon emission, the exact nature of this constraint was only made clear in Ref. [11]. There it was pointed out that there are large double logarithms (aside from single logarithms and less singular pieces) in the slope $\sigma'(\Delta)$ of the total rate, as a function of $\Delta$ the difference in minimum $p_t$ values of the two highest $p_t$ jets. These logarithms were resummed and it was shown that the slope of the total rate $\sigma' \to 0$ as $\Delta \to 0$. This leads to a physical behaviour of the total rate as reflected by the data [10].

To perform the comparison to data accurately however, requires two improvements to be made to the calculations of Ref. [11]. Firstly the exact same jet algorithm has to be employed in the theoretical calculations and experimental measurements. The current algorithm used by H1 and ZEUS experiments is the inclusive $k_t$ algorithm. At hadron colliders variants of the cone algorithm are used and it is in fact a cone algorithm that was employed in Ref. [11]. However the details of the calculation need to be ammended to define the cones in $\eta, \phi$ space as is done experimentally and calculations concerning this were presented at the working group meeting. The second important step is matching to fixed order estimates. We report below on the leading order matching to DISENT [19] while a full NLO matching is still awaited.

We also introduce and study two variables of related interest, the first is the difference in $p_t$, between the highest $p_t$ jets $\Delta p_{t,jj} = p_{t1} - p_{t2}$ (note that here we talk about the $p_t$ difference rather than the difference in the minimum $E_{\text{cut}}$, that we mentioned earlier. The resummation of this distribution $\frac{d\sigma}{d\Delta p_{t,jj}}$ is essentially identical to that carried out in Ref. [11], except that here we compute the next-to–leading logarithms in different versions of the jet algorithm, which should help with direct experimental comparsions. We also perform the leading-order matching to DISENT.

Having developed the calculational techniques for $d\sigma/d\Delta p_{t,jj}$ it is then straightforward to generate the results for the distribution in azimuthal angle between jets $d\sigma/d\phi_{jj}$ which requires resummation in the region $\phi_{jj} = \pi$. These distributions have been measured at HERA and the Tevatron (most recently by the D0 collaboration). Comparing the resummation with data would represent an interesting challenge for the theory insofar as the status of resummation tools is concerned, and is potentially very instructive.

### 3.1 The $\Delta p_{t,jj}$ and $\phi_{jj}$ distributions

We shall consider dijet production in the DIS Breit frame. For the jet definition we can consider either an $\eta, \phi$ cone algorithm (such as the infrared and collinear safe midpoint cone algorithm) or the inclusive $k_t$ algorithm. We shall point out to what level the two algorithms would give the same result and where they can be expected to differ. We shall use a four-vector recombination scheme where the jet four-momentum is the sum of individual constituent hadron four-momenta. We also impose cuts on the highest $p_t$ jets such that $|\eta_{1,2}| \leq 1$ and $p_{t1,t2} \geq E_{\min}$.

We then consider the quantity $\Delta p_{t,jj} = p_{t1} - p_{t2}$ which vanishes at Born order and hence the distribution at this order is just $\frac{d\sigma}{dp_{t,jj}} \propto \delta(p_{t,jj})$.

Beyond leading order the kinematical situation in the plane normal to the Breit axis is represented as before [11]:

$$\begin{aligned} \vec{p_{t1}} &= p_{t1}(1,0) && (15)\\ \vec{p_{t2}} &= p_{t2}\left(\cos(\pi \pm \epsilon), \sin(\pi \pm \epsilon)\right) && (16)\\ \vec{k_t} &= k_t\left(\cos\phi, \sin\phi\right) && (17) \end{aligned}$$

Thus we are considering a small deviation from the Born configuration of jets back-to–back in azimuth, induced by the presence of a soft gluon with transverse momentum $k_t \ll p_{t1,t2}$ (which is not recombined by the algorithm with either hard parton) and with azimuthal angle $\phi$. In the above $\epsilon$ represents the recoil angle due to soft emission. We then have

$$\Delta p_{t,jj} = |p_{t1} - p_{t2}| \approx |k_t \cos\phi|, \tag{18}$$

which accounts for the recoil $\epsilon$ to first order and hence is correct to NLL accuracy. Thus for the emission of several soft gluons we have the $p_t$ mismatch given by

$$\Delta p_{t,jj} = |\sum_{i \notin j} k_{xi}|, \tag{19}$$

where $k_x$ denotes the single component of gluon transverse momentum, along the direction of the hard jets, which are nearly back-to–back in the transverse plane. The sum includes only partons not merged by the algorithm into the highest $E_t$ jets.

Similarly for the dijet azimuthal angle distribution[5], we have :

$$\pi - \phi_{jj} \approx \frac{1}{p_t}|\sum_{i \notin j} k_{yi}|. \tag{20}$$

where $\phi_{jj}$ is the azimuthal angle between the two highest $p_t$ jets. Note that in the above we have set $p_{t1} = p_{t2} = p_t$ since we are considering a small deviation from the Born configuration and this approximation is correct to NLL accuracy. We also introduced $k_y$, the component of soft gluon momentum normal to the jet axis in the transverse plane.

In either of the above two cases, i.e the $\Delta p_{t,jj}$ or $\phi_{jj}$ distributions, an identical resummation is involved , due to the similar role of soft partons not recombined into jets. Henceforth we shall proceed with just the $\Delta p_{t,jj}$ resummation results, it being understood that similar considerations apply to $\phi_{jj}$ in the region $\phi_{jj} \sim \pi$.

Assuming independent emission of soft gluons by the hard three-parton system (the incoming parton and the two outgoing partons that initiate the dijets) and factorising the phase-space Eq. 19 as

[5]Note that the kinematical relations we derive here would be equally valid for dijets produced in hadron-hadron collisions at the Tevatron or LHC and just the dynamics of multisoft gluon emission would be more complex.

below[6]:

$$\Theta\left(\Delta p_{t,jj} - |\sum_{i\notin j} k_{x,i}|\right) = \frac{1}{\pi}\int_{-\infty}^{\infty}\frac{db}{b}\sin(b\Delta p_{t,jj})\prod_{i\notin j} e^{ibk_{xi}}, \tag{21}$$

the resummed result for the $\Delta p_{t,jj}$ distribution can be expressed as

$$\frac{d^3\sigma}{dxdQ^2d\Delta p_{t,jj}}(E_{\min},\Delta p_{t,jj}) = \sum_{\delta=q,g}\int_x^1\frac{d\xi}{\xi}\int_0^1 dz\sum_{a=T,L}F_a(y)C_\delta^a(\xi,z,E_{\min})w_\delta(Q,\Delta p_{t,jj}). \tag{22}$$

In the above $\xi$ and $z$ are phase-space variables that parametrise the Born dijet configuration, $F_{a=T,L}$ denotes the $y = Q^2/xs$ dependence associated to the transverse or longitudinal structure function while $C^a$ is the Born matrix-element squared. The function $w$ represents the result of resummation.

The resummed expression $w$ requires some explanation. Its form is as follows

$$w_\delta(p_{t,jj}) = \int_0^\infty\frac{db}{b}\sin(b\Delta p_{t,jj})\exp[-R_\delta(b)]\mathcal{S}(b)q_\delta\left(x/\xi,1/b^2\right). \tag{23}$$

Note the fact that the exponentiation holds only in $b$ space where $b$ is the impact parameter. The function $R(b)$ (we ignore the subscript $\delta$ which describes either incoming quarks or gluons) is the Sudakov exponent which can be computed up to NLL accuracy,

$$R(b) = Lg_1(\alpha_s L) + g_2(\alpha_s L),\ L \sim \ln(bQ). \tag{24}$$

while $S(b)$ is the non-global contribution that arises from soft partons inside the jet emitting outside it. $q_\delta$ is the incoming quark or gluon density and its scale depends on the variable $b$. The functions $g_1$ and $g_2$ are the leading-logarithmic and next-to–leading logarithmic resummed quantities.

For the leading logarithms $g_1$ and a subset of next-to–leading logarithms $g_2$, generated essentially by exponentiation of the single-log result in $b$ space, the cone and inclusive $k_t$ algorithms would give the same result, which we have computed. Starting from terms that begin with $\alpha_s^2\ln^2 b$ in $g_2$ (specifically two soft wide-angle gluons), the following two effects become important:

– For cone algorithms the implementation of the split/merge stage affects the $g_2$ piece. Present calculations [11] are valid to NLL accuracy if *all* the energy shared by overlapping jets is given to the jet that would have highest $p_t$. Note that this is different from merging the overlapping jets themselves. If other merging procedures are used the calculation becomes more complex but is still tractable.
– For the $k_t$ algorithm it is just being realised that running the algorithm generates terms that start at $\alpha_s^2\ln^2 b$ in the exponent, which are not correctly treated by naive Sudakov exponentiation. These terms, which are generated by the clustering procedure, can also be numerically accounted for in our case, but this is work in progress.

The effects that we mention above cause a similar impact on the final result as the non-global term $S(b)$ which was shown to be at around the 10% level in Ref. [11]. Hence the current results for the $k_t$ algorithm that do not account for the recently found additional terms and only approximately for the non-global logs, can be expected to change by around 10% when these effects will be included correctly.

We present in Fig. 6 preliminary results for the $\Delta p_{t,jj}$ distribution matched to the leading order DISENT prediction, using the $k_t$ algorithm. The matching at present combines quark and gluon channels wheras ideally one would like to separate the incoming quark and gluon channels with the right weights ($\mathcal{O}(\alpha_s)$ coefficient functions). This would be possible if, for instance, there was parton flavour information explicit in the fixed order codes, a limitation of the fixed-order codes that needs to be addressed also for hadron-hadron event shapes to be matched to NLO predictions.

[6]We compute here the cross-section for the observable to be less than $\Delta p_{t,jj}$ from which we can easily obtain the corresponding distribution.

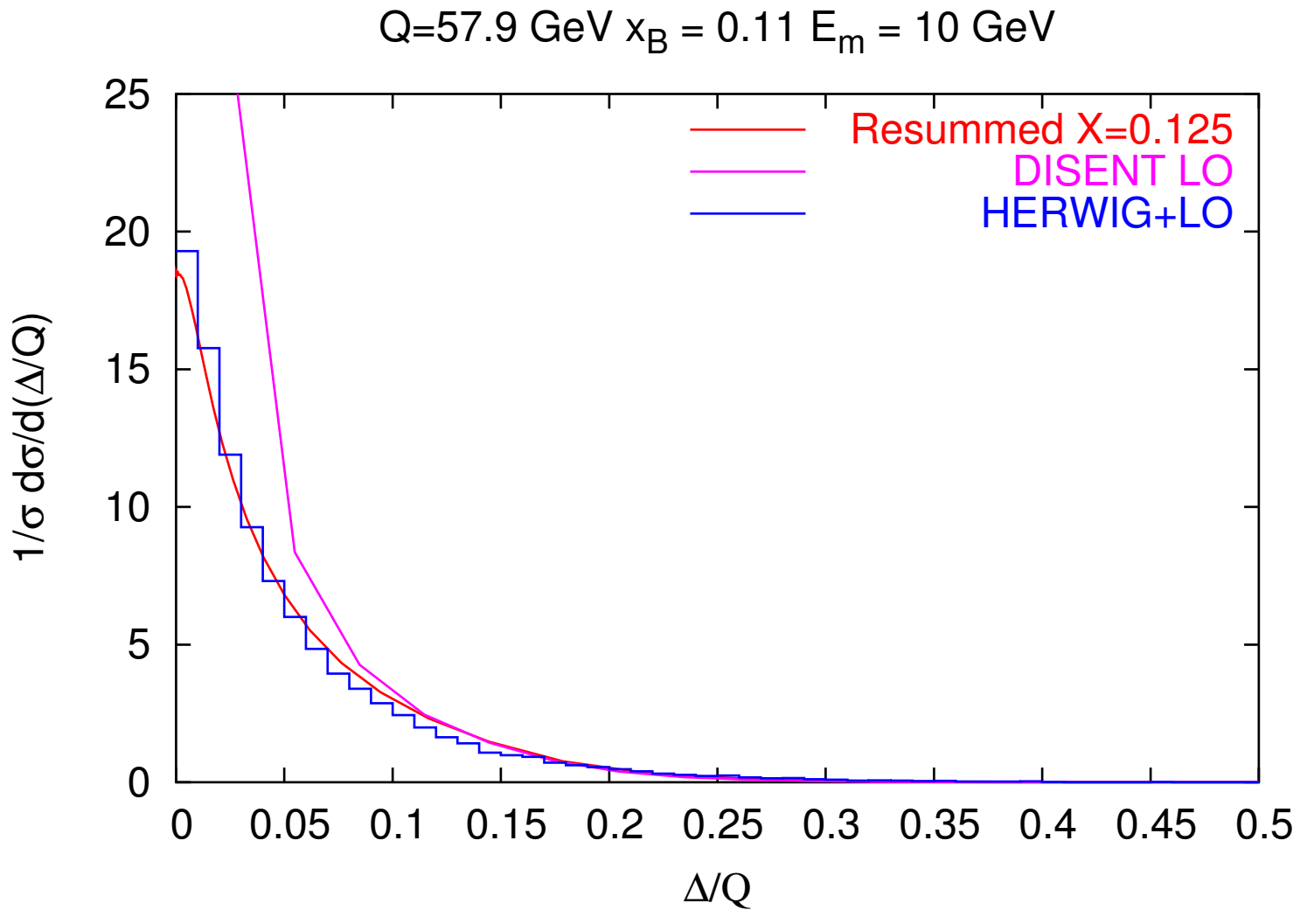

**Fig. 6:** Figure showing the resummed result matched to fixed-order DISENT results for the variable $\Delta = Q\Delta p_{t,jj}$. Also shown, for comparison, are HERWIG results with matrix-element corrections and the DISENT result alone.

We also present a comparison with HERWIG [20] results on the same quantity. The variable $X$ in the figure merely refers to the effect of using the jet $p_t$ as the hard scale rather than the photon virtuality $Q^2$, formally a NNLL effect. It is amusing to note the very good agreement of the resummation with HERWIG but not too much can be read into it at this stage. Given the minor role of non-global effects we would expect HERWIG and our predictions to indeed have a broad resemblence. However we should mention that the resummed result in Fig. 6 is at present subject to change pending proper inclusion of non-global logs and the effect of independent soft emission at large angles. The latter is partly included in the results shown, through exponentiation of the one-gluon result as we pointed out before, but the clustering procedure changes this result at about the same level as the non-global logs ($\mathcal{O}(\alpha_s^2 \ln^2 b)$ in the exponent), and this feature needs to be accounted for still. Secondly the matching to LO DISENT combines channels and this spoils control over the $\alpha_s^2 \ln^2 Q/\Delta p_{t,jj}$ term in the expansion of the resummation to NLO. A full NLO matching with proper separation of the channels is awaited. The HERWIG curve also includes an intrinsic $k_t$ component that lowers the height of the result at small $p_{t,jj}$, which can be easily included in the theoretical resummation but at present is excluded. Given these differences the very good agreement one sees with HERWIG is expected to change to some extent although broadly speaking the shapes of the two curves are expected to be similar. Similar conclusions apply for the $\phi_{jj}$ observable.

## 4 The vector $Q_t$ of the current hemisphere

Next we examine a quantity that, as mentioned in the introduction, makes a very good analogy with Drell-Yan transverse momentum, $Q_t$, distributions. Comparison of the resummation of this observable with data could help to understand whether extra broadening of conventionally resummed $Q_t$ spectra, is generated at small $x$. If so this will be a significant factor at the LHC. The observable in question is the (modulus of) the vectorially summed transverse momenta of all particles in the Breit frame current hemisphere:

$$Q_t = |\sum_{i\in\mathcal{H}_c} \vec{k}_{t,i}|. \tag{25}$$

Using momentum conservation this quantity is simply equal to the modulus of the transverse momenta of emissions in the remnant hemisphere. These emissions can all be ascribed to the incoming quark to NLL accuracy, *apart* from the soft wide-angle component where large-angle emissions in the current hemisphere can emit softer gluons into the remnant hemisphere (the by now familiar non-global logarithms).

The resummed result for this observable can be expressed as :

$$\frac{d\sigma}{dQ_T^2} \sim \sigma_0 \int_0^\infty bdbJ_0(bQ_t)\exp[-R(b)]S(b)q(x,1/b^2) \tag{26}$$

where $J_0$ is the zeroth order Bessel function, $R(b)$ is the Sudakov exponent (the "radiator") , $S(b)$ the non-global contribution and $q$ denotes the quark distribution summed over quark flavours with appropriate weights (charges).

The result for the radiator to NLL accuracy can be expressed, as before, in terms of a leading-log and next-to–leading log function:

$$R(b) = Lg_1(\alpha_s L) + g_2(\alpha_s L),\, L = \ln(bQ). \tag{27}$$

We have

$$g_1 = \frac{C_F}{2\pi\beta_0\lambda}[-\lambda - \ln(1-\lambda)], \tag{28}$$

$$\begin{aligned} g_2 &= \frac{3C_F}{4\pi\beta_0}\ln(1-\lambda) + \frac{KC_F}{4\pi^2\beta_0^2}\left[\frac{\lambda}{1-\lambda} + \ln(1-\lambda)\right] \\ &+ \frac{C_F}{2\pi}\left(\frac{\beta_1}{\beta_0^3}\right)\left[-\frac{1}{2}\ln^2(1-\lambda) - \frac{\lambda + (1-\lambda)}{1-\lambda}\right], \end{aligned} \tag{29}$$

where we have $\lambda = \beta_0\alpha_s \ln[Q^2(\bar{b})^2]$, $\bar{b} = be^{\gamma_E}/2$ and $K = (67/18 - \pi^2/6)C_A - 5/9\, n_f$.

It is straightforward to express the result directly in $Q_t$ space and one has for the pure NLL resummed terms:

$$\frac{d\sigma}{dQ_T^2} \sim \frac{d}{dQ_T^2}\left[e^{-R(Q/Q_t) - \gamma_E R'(Q/Q_t)}\frac{\Gamma(1-R'/2)}{\Gamma(1+R'/2)}q(x,Q_T^2)S(Q/Q_t)\right] \tag{30}$$

where $R' = dR/d\ln(Q/Q_t)$. The result has a divergence at $R' = 2$ which is due to retaining just NLL terms and is of the same nature as that discussed before for certain hadron-hadron event shapes and the Drell-Yan $Q_t$ distribution. However in the present case the divergence is at quite low values of $Q_t$, e.g for $Q = 100$ GeV, the divergence is at around 0.5 GeV (depending on the exact choice for $\Lambda_{\rm QCD}$). Thus it is possible to safely study the distribution down to $Q_t$ values of a few GeV using the simple form Eq. 30. We note that is is also possible to eliminate the divergence if one defines the radiator such that $R(b) \to R(b)\theta(\bar{b}Q - 1)$, which is a restriction that follows from leading-order kinematics (that one assumes to hold at all orders). The resultant modification has only a negligible impact in the $Q_t$ range that we expect to study phenomenologically.

After the matching to fixed-order is performed, we can probe the non-perturbative smearing $e^{-gb^2}$ that one can apply to the $b$ space resummed result. Comparisons with data should hopefully reveal whether the NLL resummed result + 'intrinsic $k_t$' smearing, mentioned above, is sufficient at smaller values of $x$ or whether extra broadening is generated in the small $x$ region, that has a significant effect on the result. Data from H1 are already available for this distribution [21] and this should enable rapid developments concerning the above issue.

## 5 Conclusions

In this article we have provided a summary of the developments discussed at the HERA-LHC workshop working group 2, concerning the topic of all-order QCD resummations. Specifically we have mentioned recent work carried out for hadronic dijet event shapes, dijet $E_t$ and angular spectra and resummation of the current-hemisphere transverse momentum distribution in the DIS Breit frame.

We have stressed the important role of HERA studies in the development of the subject from the LEP era and the fact that, in this regard, HERA has acted as a bridge between LEP studies of the past (although LEP analysis of data continues and is an important source of information) and future studies at both the Tevatron and the LHC.

We have particularly tried to stress the continuing crucial role of HERA in testing all-order QCD dynamics, especially in the context of multi-hard parton observables where studies are currently ongoing. Careful experimental and theoretical collaborative effort is needed here in order to confirm the picture developed for NLL resummations and power corrections. If this program is successful it will greatly ease the way for accurate QCD studies at more complex hadronic environments, such as the LHC.

## Acknowledgements

We would like to thank the convenors and organisers of the series of meetings which were a part of the HERA-LHC workshop, for their skillful organisation and for providing us all the necessary facilities needed to present and develop our respective contributions here.

# Matching Parton Showers and Matrix Elements

*Stefan Höche*[1], *Frank Krauss*[1], *Nils Lavesson*[2], *Leif Lönnblad*[2], *Michelangelo Mangano*[3], *Andreas Schälicke*[1], *Steffen Schumann*[1]
[1]Institut für Theoretische Physik, TU Dresden, Germany; [2]Department of Theoretical Physics, Lund University, Sweden; [3]CERN, Geneva, Switzerland.

**Abstract**
We compare different procedures for combining fixed-order tree-level matrix element generators with parton showers. We use the case of W-production at the Tevatron and the LHC to compare different implementations of the so-called CKKW scheme and one based on the so-called MLM scheme using different matrix element generators and different parton cascades. We find that although similar results are obtained in all cases, there are important differences.

## 1 Introduction

One of the most striking features of LHC final states will be the large number of events with several hard jets. Final states with 6 jets from $t\bar{t}$ decays will have a rate of almost 1Hz, with 10-100 times more coming from prompt QCD processes. The immense amount of available phase-space, and the large acceptance of the detectors, with calorimeters covering a region of almost 10 units of pseudorapidity ($\eta$), will lead to production and identification of final states with 10 or more jets. These events will hide or strongly modify all possible signals of new physics which involve the chain decay of heavy coloured particles, such as squarks, gluinos or the heavier partners of the top which appear in little-Higgs models. Being able to predict their features is therefore essential.

To achieve this, our calculations need to describe as accurately as possible both the full matrix elements for the underlying hard processes, as well as the subsequent development of the hard partons into jets of hadrons. For the complex final-state topologies we are interested in, no factorization theorem exists however to rigorously separate these two components, providing a constructive algorithm for the implementation of such separation. The main obstacle is the existence of several hard scales, like the jet transverse energies and dijet invariant masses, which for a generic multijet event will span a wide range. This makes it difficult to unambiguously separate the components of the event which belong to the "hard process" (to be calculated using a multiparton amplitude) from those developing during its evolution (described by the parton shower). A given $(N+1)$-jet event can be obtained in two ways: from the collinear/soft-radiation evolution of an appropriate $(N+1)$-parton final state, or from an $N$-parton configuration where hard, large-angle emission during its evolution leads to the extra jet. A factorization prescription (in this context this is often called a "matching scheme") defines, on an event-by-event basis, which of the two paths should be followed. The primary goal of a matching scheme is therefore to avoid double counting (by preventing some events to appear twice, once for each path), as well as dead regions (by ensuring that each configuration is generated by at least one of the allowed paths). Furthermore, a good matching scheme will optimize the choice of the path, using the one which guarantees the best possible approximation to a given kinematics. It is possible to consider therefore different matching schemes, all avoiding the double counting and dead regions, but leading to different results in view of the different ways the calculation is distributed between the matrix element and the shower evolution. As in any factorization scheme, the physics is independent of the separation between phases only if we have complete control over the perturbative expansion. Otherwise a residual scheme-dependence is left. Exploring different matching schemes is therefore crucial to assess the systematic uncertainties of multijet calculations.

In this work we present a first comparison of the three approaches which have been proposed so far, the so-called CKKW scheme, the Lönnblad scheme, and the MLM scheme. After shortly reviewing them, we present predictions for a set of $W$+multijet distributions at the Tevatron collider and at the LHC.

## 2 Matching procedures

In general, the different merging procedures all follow a similar strategy:

1. A jet measure is defined and all relevant cross sections including jets are calculated for the process under consideration. I.e. for the production of a final state $X$ in $pp$-collisions, the cross sections for the processes $pp \to X + n$jets with $n = 0,\, 1,\, \ldots n_{\max}$ are evaluated.
2. Hard parton samples are produced with a probability proportional to the respective total cross section, in a corresponding kinematic configuration following the matrix element.
3. The individual configurations are accepted or rejected with a dynamical, kinematics-dependent probability that includes both effects of running coupling constants and of Sudakov effects. In case the event is rejected, step 2 is repeated, i.e. a new parton sample is selected, possibly with a new number of jets.
4. The parton shower is invoked with suitable initial conditions for each of the legs. In some cases, like, e.g. in the MLM procedure described below, this step is performed together with the step before, i.e. the acceptance/rejection of the jet configuration. In all cases the parton shower is constrained not to produce any extra jet; stated in other words: Configurations that would fall into the realm of matrix elements with a higher jet multiplicity are vetoed in the parton shower step.

From the description above it is clear that the merging procedures discussed in this contribution differ mainly

– in the jet definition used in the matrix elements;
– in the way the acceptance/rejection of jet configurations stemming from the matrix element is performed;
– and in details concerning the starting conditions of and the jet vetoing inside the parton showering.

### 2.1 CKKW

In the original merging description according to [1, 2], which has been implemented [3] in SHERPA [4] in full generality, the acceptance/rejection of jet configurations from the matrix elements and the parton showering step are well-separated.

In this realisation of what is known as the CKKW-prescription the phase space separation for the different multijet processes is achieved through a $k_\perp$-measure [5–7]. For the case of hadron–hadron collisions, two final-state particles belong to two different jets, if their relative transverse momentum

$$k_\perp^{(ij)2} = 2\min\left\{p_\perp^{(i)},\, p_\perp^{(j)}\right\}^2 \left[\cosh(\eta^{(i)} - \eta^{(j)}) - \cos(\phi^{(i)} - \phi^{(j)})\right] \tag{1}$$

is larger than a critical value, $k^2_{\perp,0}$. In addition, the transverse momentum of each jet has to be larger than $k_{\perp,0}$. The matrix elements are then reweighted by appropriate Sudakov and coupling weights. The task of the weight attached to a matrix element is to take into account terms that would appear in a corresponding parton shower evolution. Therefore, a "shower history" is reconstructed by clustering the initial and final state partons according to the $k_\perp$-algorithm. The resulting chain of nodal $k_\perp$-measures is interpreted as the sequence of relative transverse momenta of multiple jet production. The first ingredient of the weight are the strong coupling constants taken at the respective nodal values, divided by the value of $\alpha_S$ used during the matrix element evaluation. The other part of the correction weight is provided by NLL-Sudakov form factors defined by

$$\Delta_{q,g}(Q,Q_0) := \exp\left[-\int\limits_{Q_0}^{Q} \mathrm{d}q\Gamma_{q,g}(Q,q)\right] , \tag{2}$$

where the integrated splitting functions $\Gamma_{q,g}$ are given by

$$\Gamma_{q,g}(Q,q) := \begin{cases} \frac{2C_F\alpha_s(q)}{\pi q}\left[\log\frac{Q}{q} - \frac{3}{4}\right] \\ \frac{2C_A\alpha_s(q)}{\pi q}\left[\log\frac{Q}{q} - \frac{11}{12}\right] \end{cases} \tag{3}$$

and contain the running coupling constant and the two leading, logarithmically enhanced terms in the limit when $Q_0 \ll Q$. The two finite, non-logarithmic terms $-3/4$ and $-11/12$, respectively emerge when integrating the non-singular part of the corresponding splitting function in the limits [0, 1]. Potentially, when $q/Q$ is not going to zero, these finite terms are larger than the logarithmic terms and thus spoil an interpretation of the emerging NLL-Sudakov form factor as a non-branching probability. Therefore, without affecting the logarithmic order of the Sudakov form factors, these finite terms are integrated over the interval $[q/Q,\, 1 - q/Q]$ rather than over $[q,\, Q]$. This way a Sudakov form factor determines the probability for having no emission resolvable at scale $Q_0$ during the evolution from a higher scale $Q$ to a lower scale $Q_0$. A ratio of two Sudakov form factors $\Delta(Q,Q_0)/\Delta(q,Q_0)$ then gives the probability for having no emission resolvable at scale $Q_0$ during the evolution from $Q$ to $q$. Having reweighted the matrix element, a smooth transition between this and the parton shower region is achieved by choosing suitable starting conditions for the shower evolution of the parton ensemble and vetoing any parton shower emission that is harder than the separation cut $k_{\perp,0}$.

Within SHERPA the required matrix elements are provided by its internal matrix element generator AMEGIC++ [8] and the parton shower phase is handled by APACIC++ [9, 10]. Beyond the comparisons presented here the SHERPA predictions for $W$+multijets have already been validated and studied for Tevatron and LHC energies in [11, 12]. Results for the production of pairs of $W$-bosons have been presented in [13].

### 2.2 The Dipole Cascade and CKKW

The dipole model [14,15] as implemented in the ARIADNE program [16] is based around iterating $2 \to 3$ partonic splitting instead of the usual $1 \to 2$ partonic splittings in a conventional parton shower. Gluon radiation is modeled as being radiated coherently from a color–anticolor charged parton pair. This has the advantage of eg. including first order correction to the matrix elements for $e^+e^- \to q\bar{q}$ in a natural way and it also automatically includes the coherence effects modeled by angular ordering in conventional showers. The process of quark antiquark production does not come in as naturally, but can be added [17]. The emissions in the dipole cascade is ordered according to invariant transverse momentum defined as

$$p_\perp^2 = \frac{s_{12}s_{23}}{s_{123}}, \tag{4}$$

where $s_{ij}$ is the squared invariant mass of parton $i$ and $j$, with the emitted parton having index 2.

When applied to hadronic collisions, the dipole model does not separate between initial and final state radiation. Instead all emissions are treated as coming from final state dipoles [18, 19]. To be able to extend the dipole model to hadron collisions, extended colored objects are introduced to model the hadron remnants. Dipoles involving hadron remnants are treated in a similar manner to the normal final-state dipoles. However, since the hadron remnant is considered to be an extended object, emissions with small wavelength are suppressed. This is modeled by only letting a fraction of the remnant take part in the emission. The fraction that is resolved during the emission is given by

$$a(p_\perp) = \left(\frac{\mu}{p_\perp}\right)^\alpha, \tag{5}$$

where $\mu$ is the inverse size of the remnant and $\alpha$ is the dimensionality.

There are two additional forms of emissions which need to be included in the case of hadronic collisions. One corresponds to an initial state $g \to q\bar{q}$ [20]. This does not come in naturally in the dipole model, but is added by hand in a way similar to that of a conventional initial-state parton shower [20]. The other corresponds to the initial-state $q \to gq$ (with the gluon entering into the hard sub-process) which could be added in a similar way, but this has not been implemented in ARIADNE yet.

When implementing CKKW for the dipole cascade, the procedure is slightly different from what has been described above [21, 22]. First, rather than just reconstructing emission scales using the $k_\perp$-algorithm, a complete dipole shower history is constructed for each state produced by the Matrix Element generator, basically answering the question *how would ARIADNE have generated this state*. This will produce a complete set of intermediate partonic states, $S_i$, and the corresponding emission scales, $p_{\perp i}$.

The Sudakov form factors are then introduced using the Sudakov veto algorithm. The idea is that we want to reproduce the Sudakov form factors used in Ariadne. This is done by performing a trial emission starting from each intermediate state $S_i$ with $p_{\perp i}$ as a starting scale. If the emitted parton has a $p_\perp$ higher than $p_{\perp i+1}$ the state is rejected. This correspond to keeping the state according to the no emission probability in Ariadne, which is exactly the Sudakov form factor.

It should be noted that for initial-state showers, there are two alternative ways of defining the Sudakov form factor. The definition in eq. (2) is used in eg. HERWIG [23], while eg. PYTHIA [24, 25] uses a form which includes ratios of parton densities. Although formally equivalent to leading logarithmic accuracy, only the latter corresponds exactly to a no-emission probability, and this is the one generated by the Sudakov-veto algorithm. This, however, also means that the reconstructed emissions need not only be reweighted by the running $\alpha_S$ as in the standard CKKW procedure above, but also with ratios of parton densities, which in the case of gluon emissions correspond to the suppression due to the extended remnants in eq. (5) as explained in more detail in [22], where the complete algorithm is presented.

### 2.3 The MLM proceedure

In this approach we match the partons from the ME calculation to the jets reconstructed after the perturbative shower. Parton-level events are defined by a minimum $E_T$ threshold $E_T^{min}$ for the partons, and a minimum separation among them, $\Delta R_{jj} > R_{min}$. A tree structure is defined in analogy with the CKKW algorithm, starting however from the colour-flow extracted from the matrix-element calculation [26], thus defining the scales at which the various powers of $\alpha_s$ are calculated. However, no Sudakov reweighting is applied. Rather, events are showered, without any hard-emission veto during the shower. After evolution, a jet cone algorithm with cone size $R_{min}$ and minimum transverse energy $E_T^{min}$ is applied to the final state. Starting from the hardest parton, the jet which is closest to it in $(\eta, \phi)$ is selected. If the distance between the parton and the jet centroid is smaller than $R_{min}$, the parton and the jet match. The matched jet is removed from the list of jets, and matching for subsequent partons is performed. The event is fully matched if each parton has a matched jet. Events which do not match are rejected. A typical example is when two partons are so close that they cannot generate independent jets, and therefore cannot match. Rejection removes double counting of the leading double logarithms associated to the collinear behaviour of the amplitude when two partons get close. Another example is when a parton is too soft to generate its own jet, again failing matching. This removes double counting of some single logarithms. For events which satisfy matching, it is furthermore required that no extra jet, in addition to those matching the partons, be present. Events with extra jets are rejected, a suppression replacing the Sudakov reweighting used in the CKKW approach. Events obtained by applying this procedure to the parton level with increasing multiplicity can then be combined to obtain fully inclusive samples spanning a large multiplicity range. Events with extra jets are not rejected in the case of the sample with highest partonic multiplicity. The distributions of observables measured on this inclusive data set should not depend on the value of the parameters $E_T^{min}$ and $R_{min}$, similar to the $k_{\perp,0}$ independence of the CKKW approach. This algorithm is encoded in the ALPGEN generator [27, 28], where evolution with both HERWIG and PYTHIA are enabled. In the following studies, the results quoted as "ALPGEN" employ the MLM matching scheme, and use ALPGEN for the generation of the parton-level matrix elements and HERWIG for the shower evolution and hadronisation.

## 3 Examples and comparisons

We present in this Section some concrete examples. We concentrate on the case of $W$+multijet production, which is one of the most studied final states because of its important role as a background to top quark studies at the Tevatron. At the LHC, $W$+jets, as well as the similar $Z$+jets processes, will provide the main irreducible backgrounds to signals such as multijet plus missing transverse energy, typical of Supersymmetry and of other manifestations of new physics. The understanding of $W$+multijet production at the Tevatron is therefore an essential step towards the validation and tuning of the tools presented here, prior to their utilization at the LHC.

For each of the three codes we calculated a large set of observables, addressing inclusive properties of the events ($p_T$ spectrum of the $W$ and of leading jets), geometric correlations between the jets, and intrinsic properties of the jets themselves, such as energy shapes. In view of the limited space available here we present only a subset of our studies, which will be documented in more detail in a future publication. An independent study of the systematics in the implementation of the CKKW prescription in HERWIG and PYTHIA was documented in [29].

The comparison between the respective results shows a reasonable agreement among the three approaches, but points also to differences, in absolute rates as well as in the shape of individual distributions, which underscore the existence of an underlying systematic uncertainty. The differences are nevertheless by and large consistent with the intrinsic systematic uncertainties of each of the codes, such as the dependence on the generation cuts or on the choice of renormalization scale. There are also differences due to the choice of parton cascade. In particular the ARIADNE cascade is quite different from a conventional parton shower, and it has been shown in this workshop [30] that ARIADNE eg. gives a much harder $p_{\perp \mathrm{W}}$ spectrum than does HERWIG or PYTHIA. Now, although the hard emissions in the matching proceedures should be described by the exact matrix element, the Sudakov formfactors in the ARIADNE matching (and indirectly in the MLM scheme) are generated by the cascade. In addition, the events in the ARIADNE matching are reweighted by PDF ratios in the same way as is done in the plain cascade. This means that some properties of the cascade may affect also the hard emissions in the matching procedure in these cases.

The existence in each of the codes of parameters specifying the details of the matching algorithms presents therefore an opportunity to tune each code so as to best describe the data. This tuning should be seen as a prerequisite for a quantitative study of the overall theoretical systematics: after the tuning is performed on a given set of final states (e.g. the $W$+jets considered here), the systematics for other observables or for the extrapolation to the LHC can be obtained by comparing the difference in extrapolation between the various codes. It is therefore auspicable that future analysis of Tevatron data will provide us with spectra corrected for detector effects in a fashion suitable to a direct comparison against theoretical predictions.

The following two sections present results for the Tevatron ($p\bar{p}$ collisions at 1.96 TeV) and for the LHC ($pp$ at 14 TeV), considering events with a positively charged $W$. Jets are defined by Paige's GETJET cone-clustering algorithm, with a calorimeter segmentation of ($\Delta\eta$, $\Delta\phi$) = (0.1,6°) and a cone size of 0.7 and 0.4 for Tevatron and LHC, respectively. At the Tevatron (LHC) we consider jets with $E_T > 10(20)$ GeV, within $|\eta| < 2(4.5)$. We use the PDF set CTEQ6L, with $\alpha_S(M_Z) = 0.118$.

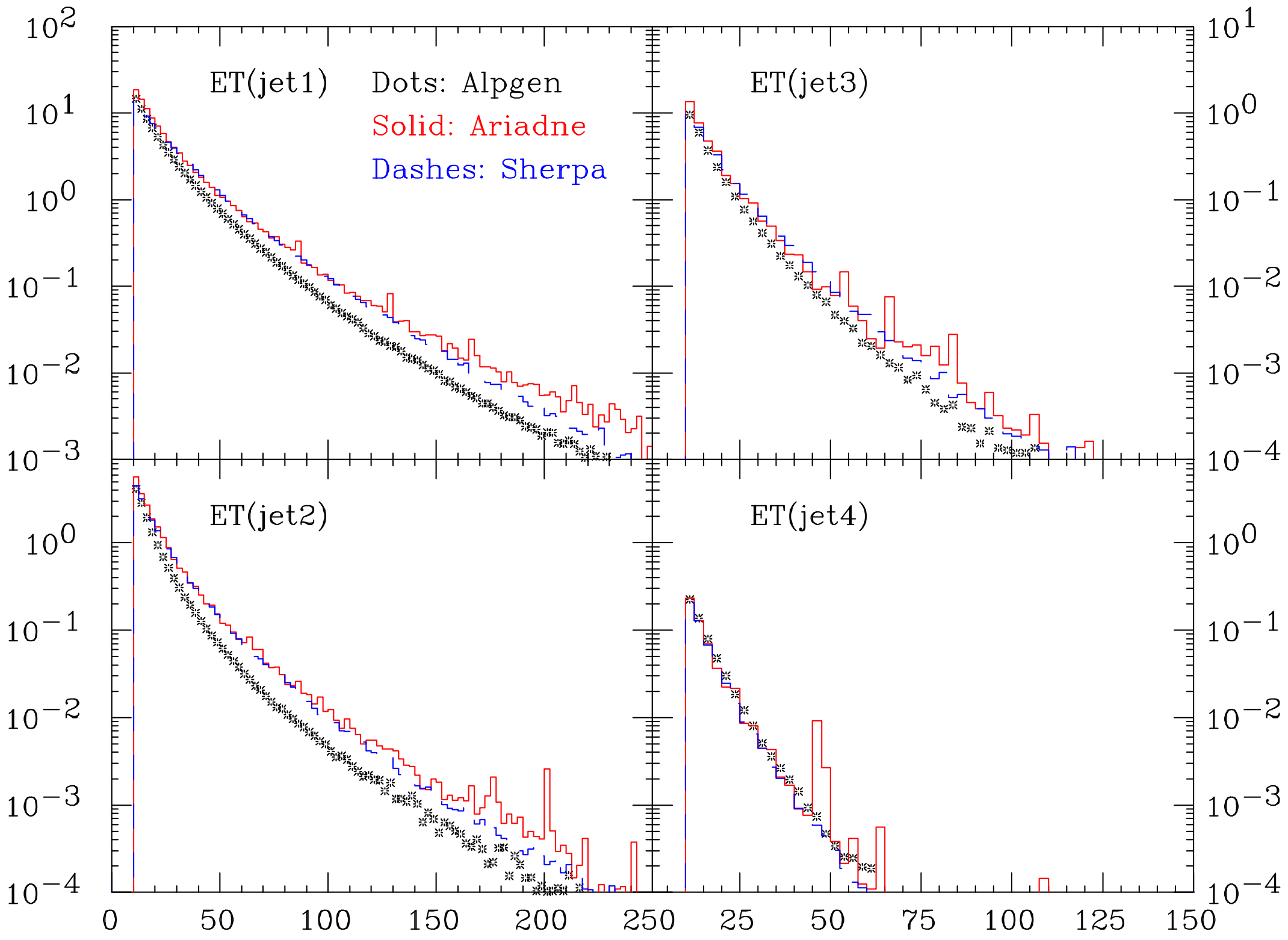

**Fig. 1:** Inclusive $E_T$ spectra of the leading 4 jets at the Tevatron (pb/GeV).

For our default distributions, the ALPGEN results for the Tevatron (LHC) were obtained using parton level cuts of $p_{T,min} = 10(20)$ GeV, $|\eta| < 2.5(5)$, $Rjj < 0.7(0.4)$ and matching defined by $E_{Tmin} = 10$ GeV and $R = 0.7$. The SHERPA samples have been generated using matrix elements with up to four extra jets and the value of the merging scale has been chosen to $k_{\perp,0} = 10(20)$ GeV, respectively. Finally, for ARIADNE, the parton level cuts were $p_{T,min} = 10(20)$, $Rjj < 0.5(0.35)$ and, in addition, a cut on the maximum pseudorapidity of jets, $\eta_{j\,\max} = 2.5(5.0)$.

In all cases, the analysis is done at the hadron level, but without including the underlying event.

### 3.1 Tevatron Studies

We start by showing in fig. 1 the inclusive $E_T$ spectra of the leading 4 jets. The absolute rate predicted by each code is used, in units of pb/GeV. We notice that the ALPGEN spectrum for the first two jets is softer than both SHERPA and ARIADNE, with the latter having even harder tails. The spectra for the third and fourth jet are instead in very good agreement, both in shape and normalization. As an indication of possible sources of systematics in these calculations, we rescaled the renormalization scale used in ALPGEN by a factor of 1/2. As seen in fig. 2 the distributions for the leading jets is now in perfect agreement with SHERPA, with an increase in rate for the third and fourth jet. These plots give us an idea of the level of flexibility which is intrinsic in the calculation of higher-order jet production. One should not forget that the rate for production of $N$ jets is proportional to the $N$th power of $\alpha_s$, and the absence of the full set of virtual corrections unavoidably leads to a large scale uncertainty.

Figure 3 shows the inclusive $\eta$ spectra of the leading 4 jets, all normalized to unit area. The asymmetry for the first two jets is due to the $W+$, which preferentially moves in the direction of the proton (positive $\eta$). This is partially washed out in the case of the third and fourth jet. There is a good agreement between the spectra of ALPGEN and SHERPA, while ARIADNE spectra appear to be broader, in particular for the subleading jets. This broadening is expected since the gluon emissions in ARIADNE are essentially unordered in rapidity, which means that the Sudakov form factors applied to the ME-generated states include also a $\log 1/x$ resummation absent in the other programs.

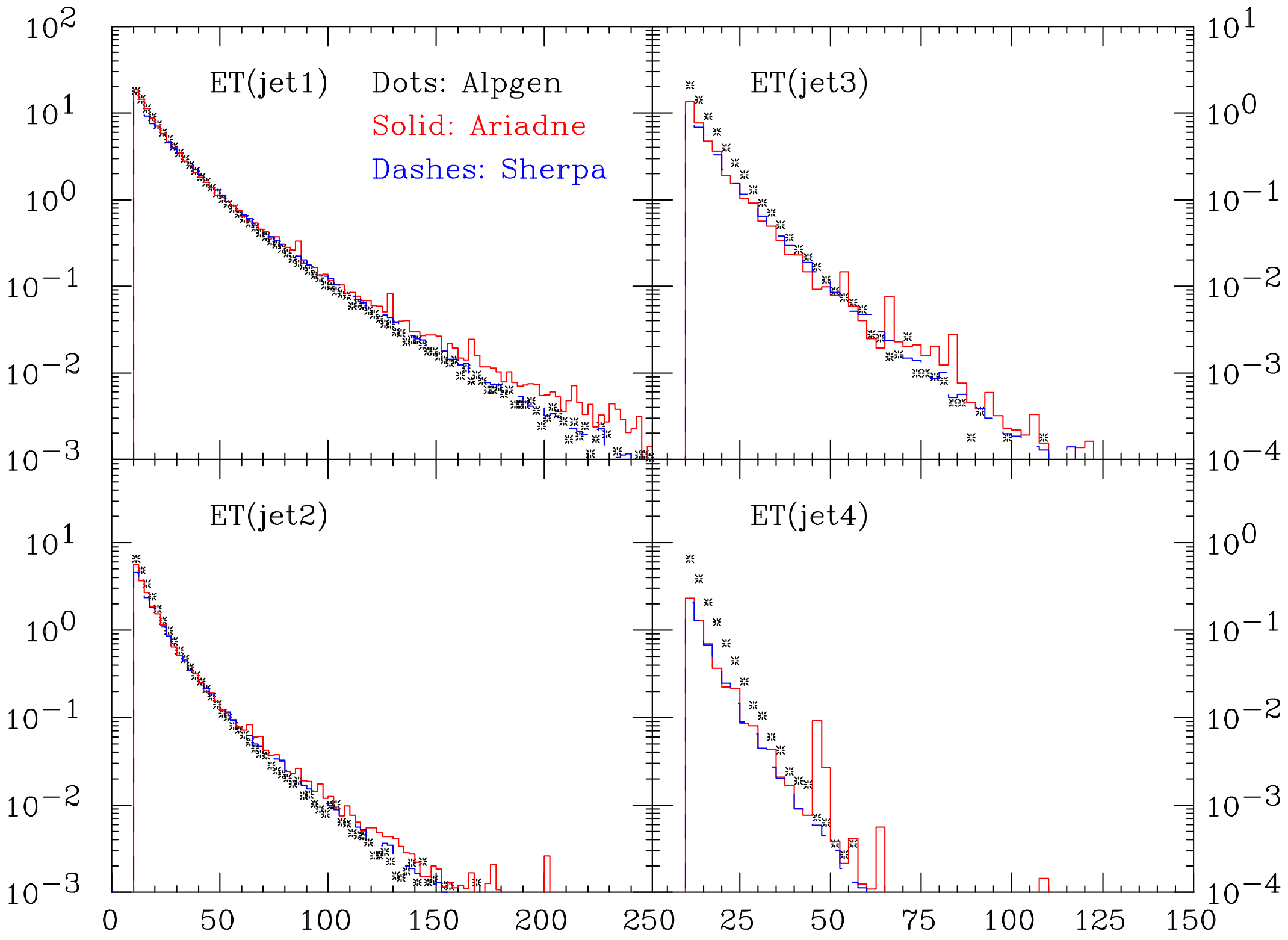

**Fig. 2:** Same as Fig. 1, but with the ALPGEN renormalization scale reduced by a factor 2.

The top-left plot of fig. 4 shows the inclusive $p_T$ distribution of the $W^+$ boson, with absolute normalization in pb/GeV. This distribution reflects in part the behaviour observed for the spectrum of the leading jet, with ALPGEN slightly softer, and ARIADNE slightly harder than SHERPA. The $|\eta|$ separation between the $W$ and the leading jet of the event is shown in the top-right plot. The two lower plots show instead the distributions of $|\eta(\text{jet}_1) - \eta(\text{jet}_2)|$ and $|\eta(\text{jet}_2) - \eta(\text{jet}_3)|$. These last three plots are normalized to unit area. In all these cases, we observe once more a reflection of the behaviour observed in the inclusive $\eta$ distributions of the jets: ALPGEN is slightly narrower than SHERPA, and ARIADNE is slightly broader.

### 3.2 LHC Predictions

In this section we confine ourselves to ALPGEN and SHERPA. It turns out that ARIADNE has a problem in the reweighting related to the fact that initial-state $g \to q\bar{q}$ emissions, contrary to the gluon emissions, are ordered both in $p_\perp$ and rapidity. With the extra phase space available at the LHC this leads to unnatural reconstructions which, in turn, gives rise to a systematically too high reweighting. A solution for this problem is under investigation and a fuller comparison including ARIADNE will be documented in a future publication.

Following the same sequence of the Tevatron study, we start by showing in fig. 5 the inclusive $E_T$ spectra of the leading 4 jets. The absolute rate predicted by each code is used, in units of pb/GeV. The relative behaviour of the predictions by ALPGEN and SHERPA follows the pattern observed in the

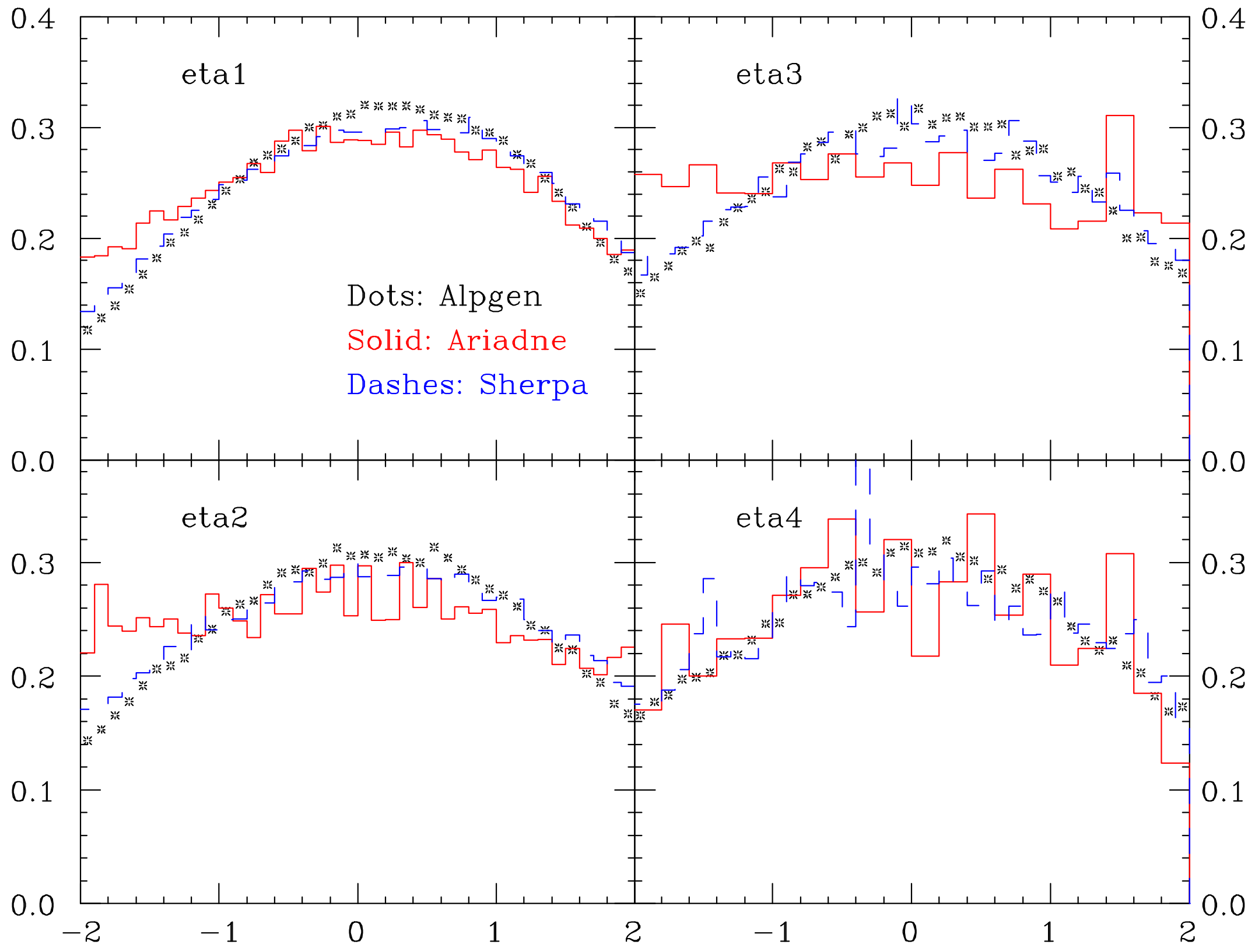

**Fig. 3:** Inclusive $\eta$ spectra of the leading 4 jets at the Tevatron, normalized to unit area.

Tevatron case, with ALPGEN being softer in the case of the leading two jets. We do not notice however a deterioration of the discrepancy going from the Tevatron to the LHC, suggesting that once a proper tuning is achieved at lower energy the predictions of two codes for the LHC should be comparable.

Figure 6 shows the inclusive $\eta$ spectra of the leading 4 jets, all normalized to unit area. The asymmetry now is not present, because of the symmetric rapidity distribution of the $W^+$ in $pp$ collisions. As in the case of the Tevatorn, jet production in ALPGEN is slightly more central than in SHERPA.

The top-left plot of fig. 7 shows the inclusive $p_T$ distribution of the $W^+$ boson, with absolute normalization in pb/GeV. The $|\eta|$ separation between the $W$ and the leading jet of the event is shown in the top-right plot. The two lower plots show instead the distributions of $|\eta(\text{jet}_1) - \eta(\text{jet}_2)|$ and $|\eta(\text{jet}_2) - \eta(\text{jet}_3)|$. These last three plots are normalized to unit area. As before, the features of these comparisons reflect what observed in the inclusive jet properties.

## 4 Conclusions

This document summarizes our study of a preliminary comparison of three independent approaches to the problems of merging matrix element and parton shower evolution for multijet final states. Overall, the picture shows a general consistency between the three approaches, although there are occasional differences. The origin of these differences is under study. It could be based on intrinsic differences between the matching schemes, as well as to differences between the different shower algorithms used in the three cases. We expect nevertheless that these differences be reconciled with appropriate changes in the default parameter settings for the matching schemes, as partly supported by the few systematic studies presented here. Validation and tuning on current Tevatron data is essential, and will allow to reduce the systematics.

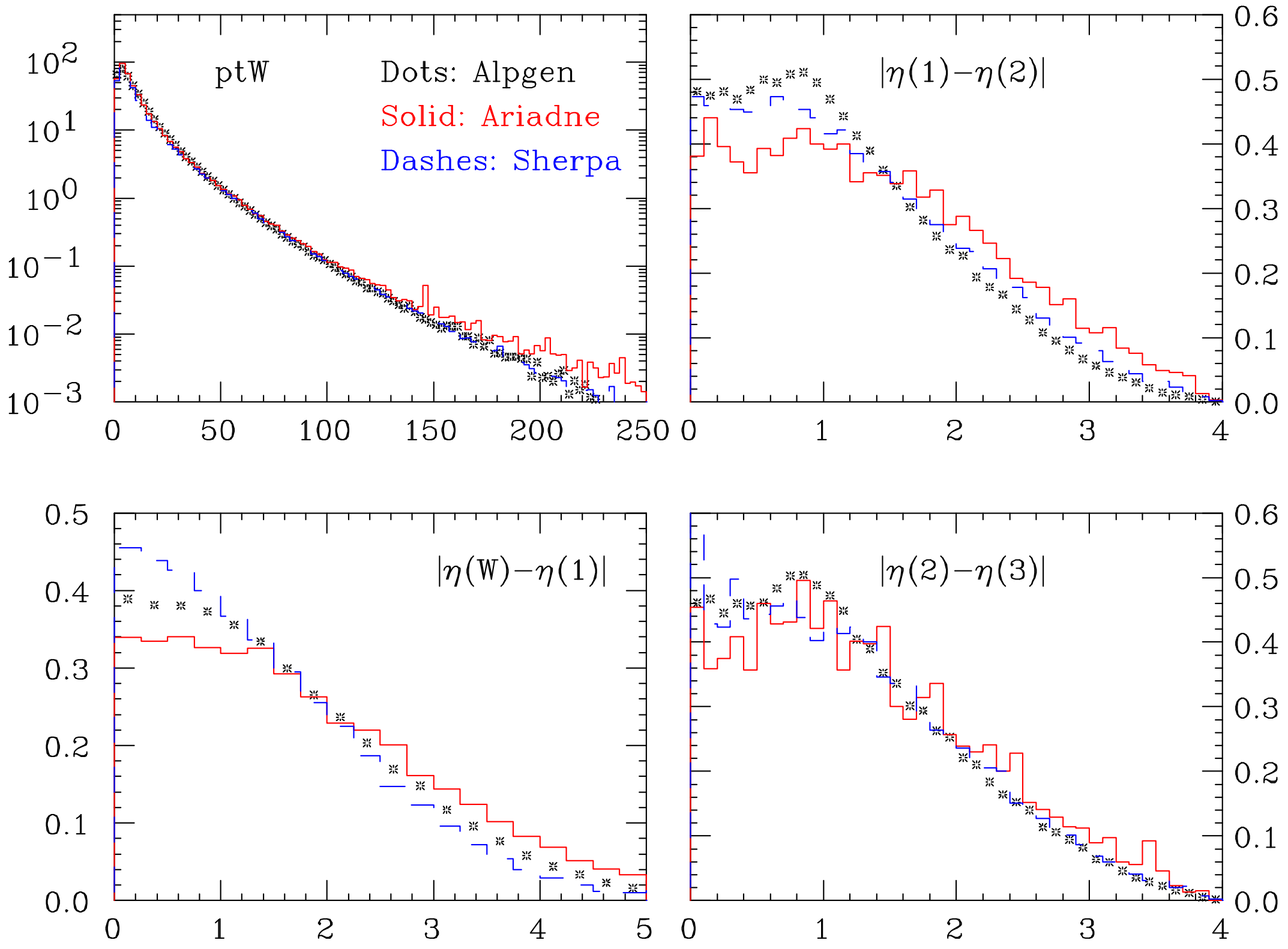

**Fig. 4:** Top left: inclusive $p_\perp(W^+)$ spectrum, pb/GeV. Bottom left: $|\eta(W^+) - \eta(\mathrm{jet}_1)|$ (unit area). Top right: $|\eta(\mathrm{jet}_1) - \eta(\mathrm{jet}_2)|$ and bottom right: $|\eta(\mathrm{jet}_2) - \eta(\mathrm{jet}_3)|$ (unit area).

It is also important to compare these models to HERA data. However, besides some preliminary investigations for ARIADNE [31], there is no program which properly implement a CKKW or MLM matching scheme for DIS. The energy of HERA is, of course, lower, as are the jet multiplicities and jet energies, but HERA has the advantage of providing a large phase space for jet production which is not mainly determined by the hard scale, $Q^2$, but rather by the total energy, giving rise to large logarithms of $x \approx Q^2/W^2$ which need to be resummed to all orders. This is in contrast to the Tevatron, where the phase space for additional jets in W-production mainly are determined by $m_W$. However, when going to the LHC there may also be important effects of the increased energy, and there will be large logarithms of $x \propto m_W/\sqrt{S}$ present, which may need to be resummed. The peculiar treatment of the available phase space in the plain ARIADNE cascade means that some logarithms of $x$ are resummed in contrast to conventional initial-state parton cascades. This feature survives the matching procedure and is the reason for the broader rapidity spectra presented in the figures above. In DIS this is reflected by the increased rate of forward jets, and such measurements are known to be well reproduced by ARIADNE while conventional parton showers fail. It would be very interesting if the matching of these conventional showers with higher order matrix elements would improve the description of forward jets. In that case the extrapolation of the Tevatron results to the LHC would be on much safer grounds.

As our study of the LHC distributions suggests, the increase in energy exhibits the same pattern of discrepancies observed at the Tevatron. We therefore expect that if different algorithms are tuned on the same set of data, say Tevatron $W$+jets, they will extrapolate in the same way to the LHC or to different final states, for example multijet configurations without $W$ bosons. While these systematics studies can be performed directly at the Monte Carlo level, only the availability of real measurements from the Tevatron can inject the necessary level or realism in these exploration. We look forward to the availability of such data.

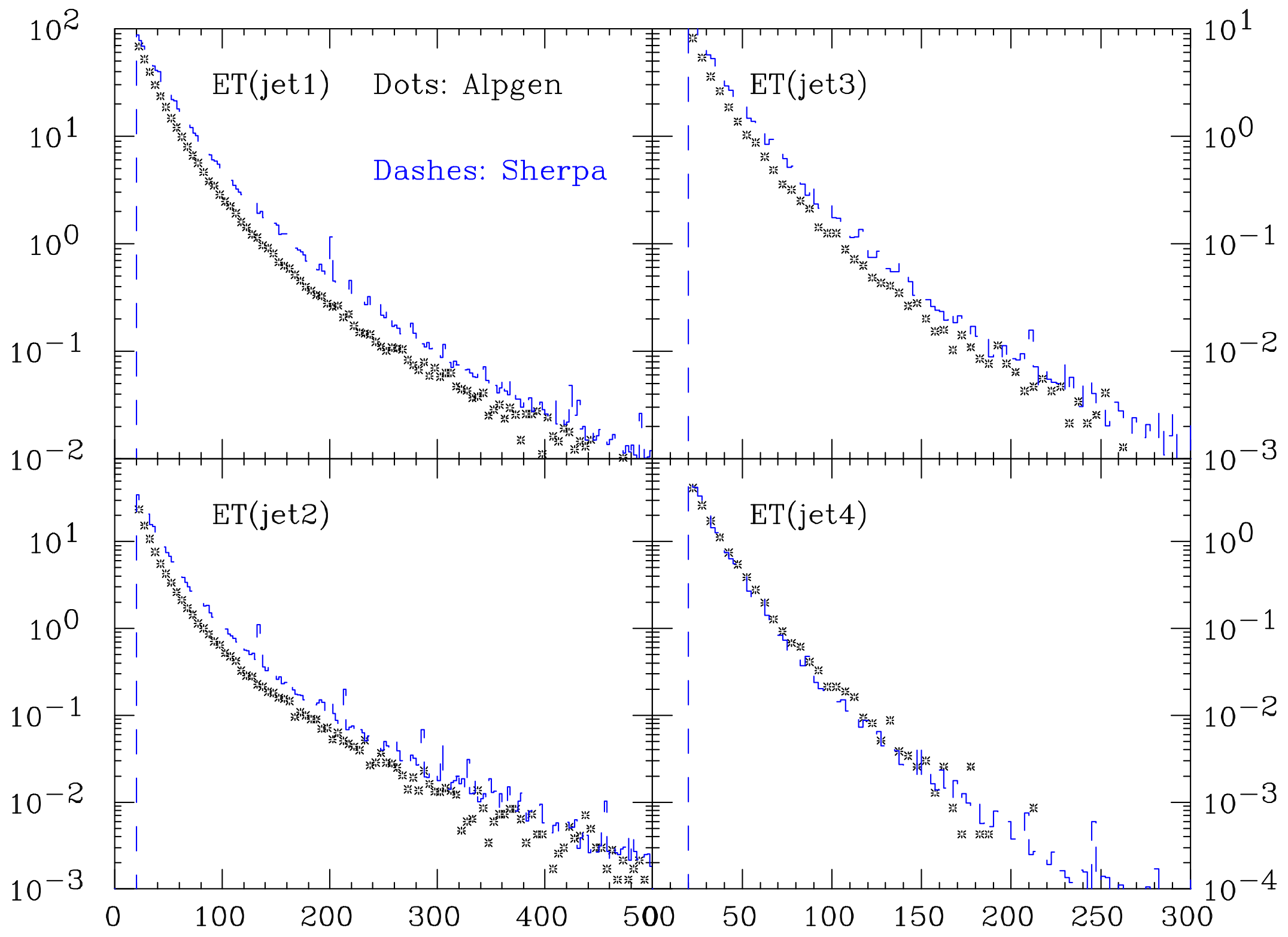

**Fig. 5:** Inclusive $E_T$ spectra of the leading 4 jets at the LHC (pb/GeV).

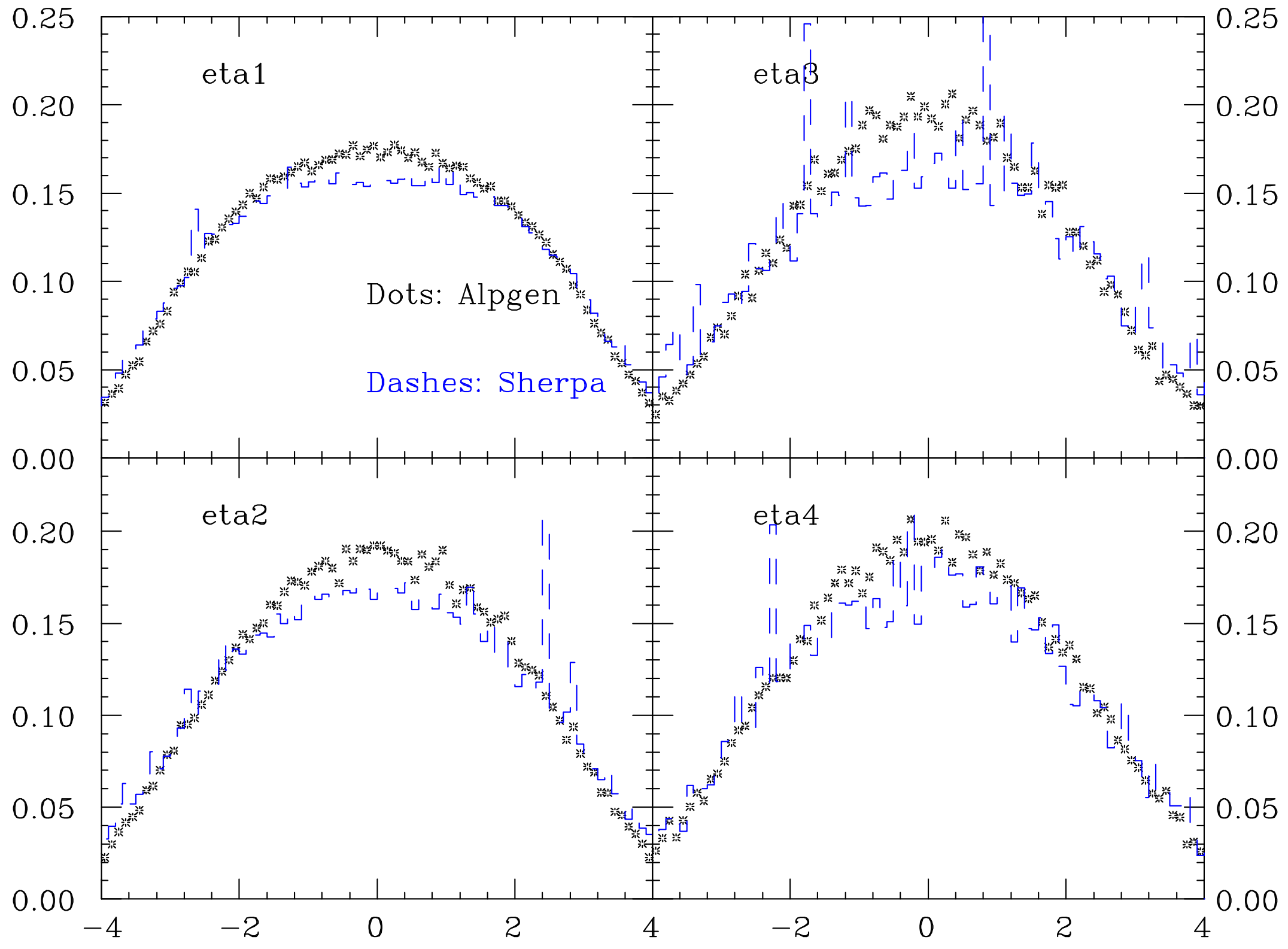

**Fig. 6:** Inclusive $\eta$ spectra of the leading 4 jets at the LHC, normalized to unit area.

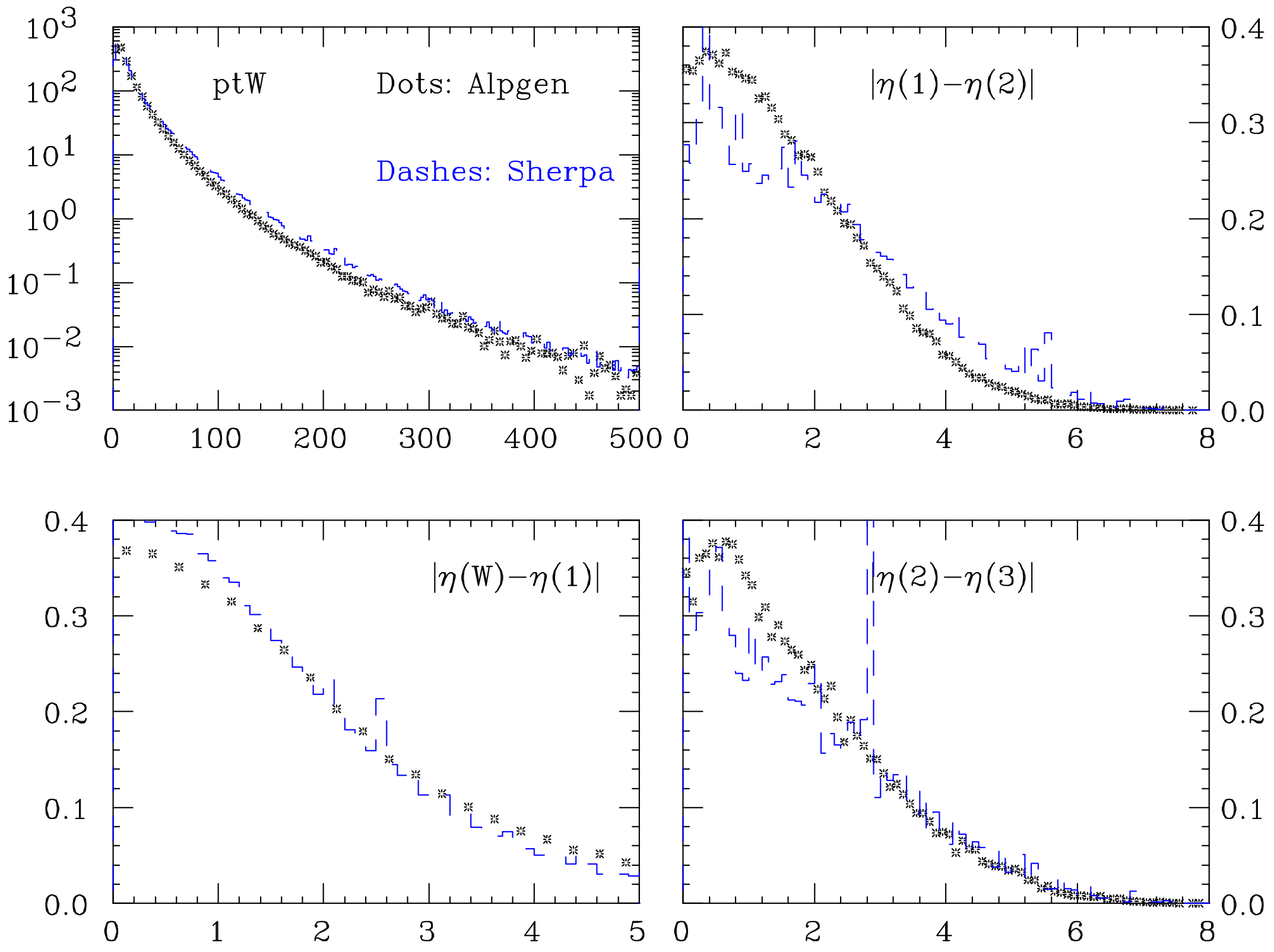

**Fig. 7:** Top left: inclusive $pt_T(W^+)$ spectrum, pb/GeV. Bottom left: $|\eta(W^+) - \eta(\mathrm{jet}_1)|$ (unit area). Top right: $|\eta(\mathrm{jet}_1) - \eta(\mathrm{jet}_2)|$ and bottom right: $|\eta(\mathrm{jet}_2) - \eta(\mathrm{jet}_3)|$ (unit area).

# Constrained non-Markovian Monte Carlo modeling of the evolution equation in QCD*

*S. Jadach and M. Skrzypek*
Institute of Nuclear Physics, Academy of Sciences, ul. Radzikowskiego 152, 31-342 Cracow, Poland and CERN Department of Physics, Theory Division, CH-1211 Geneva 23, Switzerland

**Abstract**
A new class of the constrained Monte Carlo (CMC) algorithms for the QCD evolution equation was recently discovered. The constraint is imposed on the type and the total longitudinal energy of the parton exiting QCD evolution and entering a hard process. The efficiency of the new CMCs is found to be reasonable.

This brief report summarizes the recent developments in the area of the Monte Carlo (MC) techniques for the perturbative QCD calculations. Most of it was done at the time of the present HERA–LHC workshop, partial results being presented at several of its meetings. At present, two papers, [1] and [2], demonstrating the principal results are already available. Generally, these MC techniques concern the QCD evolution of the parton distribution functions (PDFs) $D_k(x,Q)$, where $k$ denotes the type of the parton (quark, gluon), $x$ the fraction of longitudinal momentum of the initial hadron carried by the parton, and the size of the available real/virtual emission phase space is $Q$. The evolution equation describes the response of the PDF to an increase of $Q$; $D_k(x,Q)$ is an inclusive distribution and can be measured almost directly in hadron — lepton scattering. On the other hand, it was always known that there exists in QCD an *exclusive* picture of the PDF, the so-called parton-shower process, in which $D_k(x,Q)$ is the distribution of the parton exiting the emission chain and entering the hard process (lepton–quark for example). The kernel functions $P_{kj}(Q,z)$, that govern the differential evolution equations of PDFs are closely related to distributions governing a single emission process $(i-1) \to i$ in the parton shower: $P_{k_i k_{i-1}}(Q_i, x_i/x_{i-1})$.

In other words, the evolution ($Q$-dependence) of PDFs and the parton shower represent two faces of the same QCD reality. The first one (inclusive) is well suited for basic precision tests of QCD at hadron–lepton colliders, while the second one (exclusive) provides realistic exclusive Monte Carlo modeling, vitally needed for experiments at high-energy particle colliders.

At this point, it is worth stressing that, so far, we were referring to DGLAP-type PDFs [3] and their evolution, and to constructing a parton-shower MC starting from them, as was done two decades ago and is still done today. This involves a certain amount of "backward engineering" and educated guesses, because the classical inclusive PDFs integrate over the $p_T$ of the exiting parton. The so-called unintegrated PDFs (UPDFs) $D_k(x,p_T,Q)$ would be more suitable for the purpose, leading to higher-quality QCD calculations. UPDFs are, however, more complicated to handle, both numerically and theoretically. (It is still a challenge to construct a parton-shower MC based consistently on the theoretically well defined UPDFs.)

Another interesting "entanglement" of the evolution of PDFs on one side and of the parton shower (PS) MC on the other side is also present in the modeling of the showering of the incoming hadron — mostly for technical reasons and convenience. The Markovian nature of the QCD evolution can be exploited directly in the PS MC, where partons split/decay as long as there is enough energy to dissipate (final state) or the upper boundary $Q$ of the phase space is hit (initial state). The multiparton distribution in such a MC is a product of the evolution kernels. However, such a direct Markovian MC simulation of a shower is hopelessly inefficient in the initial state, because the hard process accepts only certain types and momenta of the incoming partons — most of the *shower histories* are rejected (zero MC weight) by the hard process, in particular when forming narrow resonances such as electroweak bosons or Higgs boson at the LHC. A well-known "workaround" is Sjöstrand's backward evolution MC algorithm, used currently in all PS MCs, e.g., HERWIG [4] and PYTHIA [5]. Contrary to the forward Markovian MC, where the physics inputs are PDFs at low $Q_0 \sim 1$ GeV and the evolution kernels, in the backward evolution MC one has to know PDFs in the entire range $(Q_0, Q)$ from a separate non-MC numerical program solving the evolution equation to provide look-up tables (or numerical parametrization) for them[1].

*Supported in part by the EU grant MTKD-CT-2004-510126, in partnership with the CERN Physics Department.

The following question has been pending in the parton-shower MC methodology for a long time: Could one invent an efficient "monolithic" MC algorithm for the parton shower from the incoming hadron, in which no external PDFs are needed and the only input are PDFs at $Q_0$ and the evolution kernel (the QCD evolution being a built-in feature of the parton shower MC)? Another question rises immediately: Why bother? Especially since this is a tough technical problem. This cannot still be fully answered before the above technique is applied in the full-scale (four-momentum level) PS MC. Generally, we hope that this technique will open new avenues in the development of the PS MC at the next-to-leading-logarithmic (NLL) level. In particular, it may help in constructing PS MCs closely related to unintegrated structure functions and, secondly, it may provide a better integration of the NLL parton shower (yet to be implemented!) with the NLL calculation for the hard process.

The first solution of the above problem of finding an efficient "constrained MC" (CMC) algorithm for the QCD evolution was presented in refs. [1, 6]. This solution belongs to what we call a CMC class II, and it relies on the observation that all initial PDFs at $Q_0$ can be approximated by const $\cdot\, x_0^{\eta-1}$; this is to be corrected by the MC weight at a later stage. This allows elimination of the constraint $x = \prod_i z_i$, at the expense of $x_0$, keeping the factorized form of the products of the kernels. Simplifying phase-space boundaries in the space of $z_i$ is the next ingredient of the algorithm. Finally, in order to reach a reasonable MC efficiency for the pure bremsstrahlung case out of the gluon emission line, one has to generate a $1/z$ singularity in the $G \to G$ kernel in a separate branch of the MC. The overall efficiency of the MC is satisfactory, as is demonstrated in Ref. [1] for the case of the pure bremsstrahlung out of the gluon and quark colour charge. Generalization to the quark–gluon transition is outlined, but not yet implemented. The main drawback of this method is its algebraic complexity. Further improvement of its relatively low MC efficiency is possible (even though it could lead to even more algebraic complexity).

The second, more efficient, CMC algorithm was presented in Ref. [2] (as well as during the October 2004 meeting of the workshop). It belongs to what we call a CMC class I. The main idea is to project/map points from the hyperspace defined by the energy constraint $x = \prod_i z_i$, into a simpler hyperspace, defined by the hardest emission, $x = \min z_i$. This mapping is accompanied by the appropriate MC weight, which compensates exactly for the deformation of the distributions involved, and the bookkeeping of the hyperspace boundaries is rigorous. The above describes a CMC for the pure bremsstrahlung segment of the gluon emission out of a quark or gluon chain. Many such segments are interconnected by the quark–gluon transitions. The algebraic hierarchic reorganization of the emission chain into a super-level of the quark–gluon transitions and sub-level of the pure bremsstrahlung is an important ingredient in all CMC algorithms and will be published separately [7]. The basic observation made in Ref. [8] is that the average number of super-level transitions is low, $\sim 1$; hence for precision of a $10^{-4}$ it is sufficient to limit it to three or four transitions. The integration/simulation of the super-level variables is done efficiently using the general-purpose MC tool FOAM [9, 10]. The above proof of the correctness of the CMC class I algorithm concept was given in Ref. [2] for the full DGLAP-type QCD evolution with the LL kernels (including quark–gluon transitions).

[1] Backward evolution is basically a change in the order of the generation of the variables: Consider generating $\rho(x, y)$, where one generates first $x$ according to $\rho(x) = \int dy\ \rho(x, y)$, and next $y$ according to $\rho(x, y)$, by means of *analytical* mappings of $x$ and $y$ into uniform random numbers. However, such analytical mappings may not exist, if we insist on generating first $x$ and next $y$! Nevertheless, we may still proceed with the same method by "brute force", if we pretabulate and invert numerically the functions $R(x) = \int^x \int dx'dy'\ \rho(x', y')$ and $R_x(y) = \int^y dy'\ \rho(x, y')$. This is what is done in a more dimensional case of the backward-evolution MC; it also explains why pretabulated PDFs are needed in these methods.

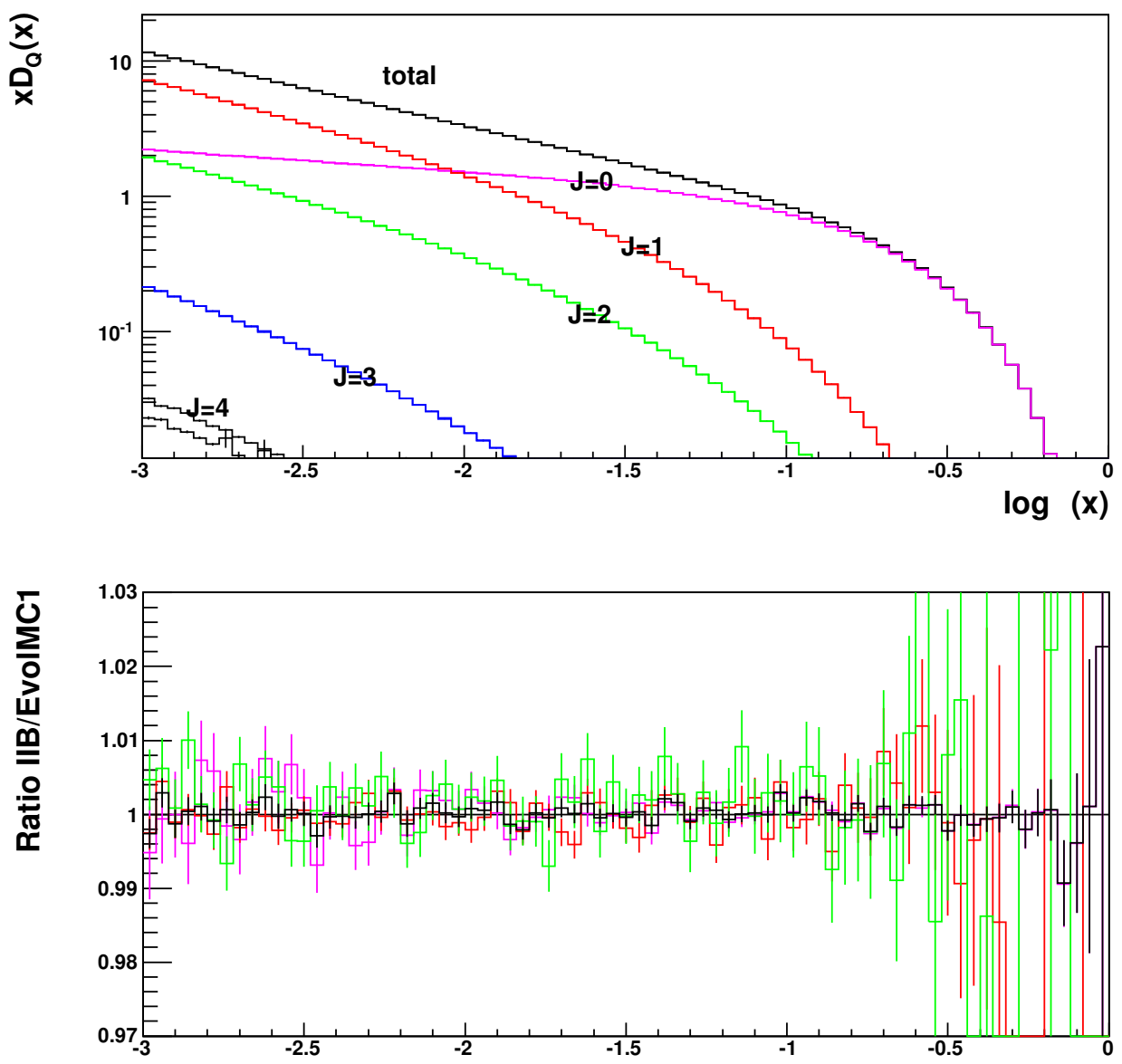

**Fig. 1:** CMC of the one-loop CCFM versus the corresponding MMC for quarks; number of quark–gluon transitions $J = 0, 1, 2, 3, 4$, and the total. The ratios in the lower plot are for $n = 0, 1$ and the total (blue).

Although our main aim is to construct the non-Markovian CMC class of algorithms, we have developed in parallel the family of Markovian MC (MMC) algorithms/programs, which provide numerical solutions of the QCD evolution equations with high precision, $\sim 10^{-3}$. We use them at each step of the CMC development as numerical benchmarks for the precision tests of the algorithms and their software implementations. The first example of MMC for DGLAP at LL was defined/examined in Ref. [8] and tested using the non-MC program QCDnum16 [11][2]. In some cases our MMC programs stand ahead of their CMC brothers; for instance, they already include NLL DGLAP kernels. A systematic description of the MMC family of our MC toolbox is still under preparation [13].

The last development at the time of the workshop was an extension of the CMC type-I algorithm from DGLAP to CCFM one-loop evolution [14] (also referred to as HERWIG evolution [15]), in which the strong coupling constant gains $z$-dependence, $\alpha_s(Q) \to \alpha_s(Q(1-z))$, as advocated in Ref. [16], confirmed by NLL calculations [17]. The above ansatz also compels introduction of a $Q$-dependent IR cutoff, $\varepsilon = Q_\varepsilon/Q$: another departure from DGLAP. This version of the CMC is still unpublished. Its version for the pure bremsstrahlung was presented at the March 2005 meeting of the workshop; in particular a perfect numerical agreement with the couterpartner MMC was demonstrated. Recently both CMC and MMC for the one-loop CCFM were extended to quark–gluon transitions, and again perfect agreement was found.

For the detailed description of the new CMC algorithm, we refer the reader to the corresponding papers [1] and [2] and workshop presentations[3]. Here, let us only show one essential step in the development of the CMC for the one-loop CCFM model — the mapping of the Sudakov variables for the pure bremsstrahlung:

$$
\begin{aligned}
I &= \int_{t_0}^{t_1} dt \int_0^{z_1} dz\; \alpha(Q(1-z))\; zP^{\Theta}_{GG}(z,t) \\
&= \frac{2}{\beta_0}\int_0^{z_1} dz \int_{t_0}^{t_1} dt\; \frac{1}{\hat{t}+\ln(1-z)}\, \frac{\theta_{\ln(1-z)>\hat{t}_\varepsilon-\hat{t}}}{1-z} = \frac{2}{\beta_0}\int_0^{y_{\max}} dy(z) \int_0^1 ds(t).
\end{aligned}
\tag{1}
$$

[2]It was also compared with the non-MC program APCheb [12].

[3]To be found at `http://jadach.home.cern.ch/jadach/`.

The short-hand notation $\hat{t} = \hat{t}(t) \equiv t - t_\Lambda$ and $v = \ln(1-z)$ supplements that of Ref. [2] in use, and the mapping reads

$$y(z) = \rho(v_1; \hat{t}_1, \hat{t}_0) = \rho(v_1 + \hat{t}_1) - \theta_{v_1 > t_\varepsilon - t_0} \rho(v_1 + \hat{t}_0), \quad s(t) = \frac{\ln(\hat{t} + v)}{\rho'(v; \hat{t}_1, \hat{t}_0)},$$
$$\rho'(v; \hat{t}_1, \hat{t}_0) = \theta_{v < t_\varepsilon - t_0} \rho'(v + \hat{t}_1) + \theta_{v > t_\varepsilon - t_0} [\rho'(v + \hat{t}_1) - \rho'(v + \hat{t}_0)], \tag{2}$$

where $\rho(t) \equiv \hat{t}(\ln \hat{t} - \ln \hat{t}_\varepsilon) + \hat{t}_\varepsilon - \hat{t}$. Once the above mapping is set, the same algorithm, with the parallel shift $y_i \to y_i + Y$, can be used in this case. The super-level of quark–gluon transitions is again implemented using FOAM[4]. A numerical comparison of the corresponding CMC and MMC programs is shown in fig. 1. The MC efficiency is comparable with that of the DGLAP case.

*Summary:* We have constructed and tested new, efficient, constrained MC algorithms for the initial-state parton-emission process in QCD.

[4]The $z$-independent $\alpha_s(t)$ is set in front of the relevant flavour-changing kernels to simplify the program.

# QED⊗QCD Exponentiation and Shower/ME Matching at the LHC*

*B.F.L. Ward and S.A. Yost*
Department of Physics, Baylor University, Waco, TX, USA

**Abstract**
We present the elements of QED⊗QCD exponentiation and its interplay with shower/ME matching in precision LHC physics scenarios. Applications to single heavy gauge boson production at hadron colliders are illustrated.

In the LHC environment, precision predictions for the effects of multiple gluon and multiple photon radiative processes will be needed to realize the true potential of the attendant physics program. For example, while the current precision tag for the luminosity at FNAL is at the $\sim 7\%$ level [1], the high precision requirements for the LHC dictate an experimental precision tag for the luminosity at the $2\%$ level [2]. This means that the theoretical precision tag requirement for the corresponding luminosity processes, such as single W,Z production with the subsequent decay into light lepton pairs, must be at the $1\%$ level in order not to spoil the over-all precision of the respective luminosity determinations at the LHC. This theoretical precision tag means that multiple gluon and multiple photon radiative effects in the latter processes must be controlled to the stated precision. With this objective in mind, we have developed the theory of $QED \otimes QCD$ exponentiation to allow the simultaneous resummation of the multiple gluon and multiple photon radiative effects in LHC physics processes, to be realized ultimately by MC methods on an event-by-event basis in the presence of parton showers in a framework which allows us to systematically improve the accuracy of the calculations without double-counting of effects in principle to all orders in both $\alpha_s$ and $\alpha$.

Specifically, the new $QED \otimes QCD$ exponentiation theory is an extension of the $QCD$ exponentiation theory presented in Refs. [3][1]. We recall that in the latter references it has been established that the following result holds for a process such as $q + \bar{q}' \to V + n(G) + X \to \bar{\ell}\ell' + n(g) + X$:

$$\begin{aligned} d\hat{\sigma}_{\rm exp} = \sum_n d\hat{\sigma}^n = e^{\rm SUM_{IR}(QCD)} \sum_{n=0}^{\infty} \int \prod_{j=1}^{n} \frac{d^3k_j}{k_j} \\ \int \frac{d^4y}{(2\pi)^4} e^{iy\cdot(P_1+P_2-Q_1-Q_2-\sum k_j)+D_{\rm QCD}} \\ * \tilde{\bar{\beta}}_n(k_1,\ldots,k_n) \frac{d^3P_2}{P_2^0}\frac{d^3Q_2}{Q_2^0} \end{aligned} \tag{1}$$

where gluon residuals $\tilde{\bar{\beta}}_n(k_1,\ldots,k_n)$ , defined by Ref. [3], are free of all infrared divergences to all orders in $\alpha_s(Q)$. The functions $SUM_{IR}(QCD), D_{\rm QCD}$, together with the basic infrared functions $B^{\rm nls}_{\rm QCD}, \tilde{B}^{\rm nls}_{\rm QCD}, \tilde{S}^{\rm nls}_{\rm QCD}$ are specified in Ref. [3]. Here $V = W^{\pm}, Z$,and $\ell = e, \mu,\ \ell' = \nu_e, \nu_\mu(e,\mu)$ respectively for $V = W^+(Z)$, and $\ell = \nu_e, \nu_\mu,\ \ell' = e, \mu$ respectively for $V = W^-$. We call attention to the essential compensation between the left over genuine non-Abelian IR virtual and real singularities between $\int dPh\bar{\beta}_n$ and $\int dPh\bar{\beta}_{n+1}$ respectively that really allows us to isolate $\tilde{\bar{\beta}}_j$ and distinguishes QCD from QED, where no such compensation occurs. The result in (1) has been realized by Monte Carlo methods [3]. See also Refs. [5–7] for exact $\mathcal{O}(\alpha_s^2)$ and Refs. [8–10] for exact $\mathcal{O}(\alpha)$ results on the W,Z production processes which we discuss here.

*Work partly supported by US DOE grant DE-FG02-05ER41399 and by NATO grant PST.CLG.980342.

[1]We stress that the formal proof of exponentiation in non-Abelian gauge theories *in the eikonal approximation* is given in Ref. [4]. The results in Ref. [3] are in contrast *exact* but have an exponent that only contains the leading contribution of the exponent in Ref. [4].

The new $QED \otimes QCD$ theory is obtained by simultaneously exponentiating the large IR terms in QCD and the exact IR divergent terms in QED, so that we arrive at the new result

$$
\begin{aligned}
d\hat{\sigma}_{\rm exp} &= e^{\rm SUM_{IR}(QCED)} \\
&\sum_{n,m=0}^{\infty} \int \prod_{j_1=1}^{n} \frac{d^3 k_{j_1}}{k_{j_1}} \prod_{j_2=1}^{m} \frac{d^3 {k'}_{j_2}}{{k'}_{j_2}} \int \frac{d^4 y}{(2\pi)^4} \\
&e^{iy\cdot(p_1+q_1-p_2-q_2-\sum k_{j_1}-\sum {k'}_{j_2})+D_{\rm QCED}} \\
&\tilde{\bar{\beta}}_{n,m}(k_1,\ldots,k_n;k'_1,\ldots,k'_m)\frac{d^3p_2}{p_2^0}\frac{d^3q_2}{q_2^0},
\end{aligned}
\tag{2}
$$

where the new YFS [11, 12] residuals, defined in Ref. [13], $\tilde{\bar{\beta}}_{n,m}(k_1,\ldots,k_n;k'_1,\ldots,k'_m)$, with $n$ hard gluons and $m$ hard photons, represent the successive application of the YFS expansion first for QCD and subsequently for QED. The functions $\rm SUM_{IR}(QCED), D_{QCED}$ are determined from their analogs $\rm SUM_{IR}(QCD), D_{QCD}$ via the substitutions

$$
\begin{aligned}
B^{\rm nls}_{\rm QCD} &\rightarrow B^{\rm nls}_{\rm QCD} + B^{\rm nls}_{\rm QED} \equiv B^{\rm nls}_{\rm QCED}\,, \\
\tilde{B}^{\rm nls}_{\rm QCD} &\rightarrow \tilde{B}^{\rm nls}_{\rm QCD} + \tilde{B}^{\rm nls}_{\rm QED} \equiv \tilde{B}^{\rm nls}_{\rm QCED}\,, \\
\tilde{S}^{\rm nls}_{\rm QCD} &\rightarrow \tilde{S}^{\rm nls}_{\rm QCD} + \tilde{S}^{\rm nls}_{\rm QED} \equiv \tilde{S}^{\rm nls}_{\rm QCED}
\end{aligned}
\tag{3}
$$

everywhere in expressions for the latter functions given in Refs. [3]. The residuals $\tilde{\bar{\beta}}_{n,m}$ are free of all infrared singularities and the result in (2) is a representation that is exact and that can therefore be used to make contact with parton shower MC's without double counting or the unnecessary averaging of effects such as the gluon azimuthal angular distribution relative to its parent's momentum direction.

In the respective infrared algebra (QCED) in (2), the average Bjorken $x$ values

$$
\begin{aligned}
x_{avg}({\rm QED}) &\cong \gamma({\rm QED})/(1+\gamma({\rm QED})) \\
x_{avg}({\rm QCD}) &\cong \gamma({\rm QCD})/(1+\gamma({\rm QCD}))
\end{aligned}
$$

where $\gamma(A) = \frac{2\alpha_A \mathcal{C}_A}{\pi}(L_s-1)$, $A =$QED, QCD, with $\mathcal{C}_A = Q_f^2, C_F$, respectively, for $A =$QED, QCD and the big log $L_s$, imply that QCD dominant corrections happen an order of magnitude earlier than those for QED. This means that the leading $\tilde{\bar{\beta}}_{0,0}$-level gives already a good estimate of the size of the interplay between the higher order QED and QCD effects which we will use to illustrate (2) here.

More precisely, for the processes $pp \rightarrow V + n(\gamma) + m(g) + X \rightarrow \bar{\ell}\ell' + n'(\gamma) + m(g) + X$, where $V = W^{\pm}, Z$,and $\ell = e, \mu,\ \ell' = \nu_e, \nu_\mu(e,\mu)$ respectively for $V = W^+(Z)$, and $\ell = \nu_e, \nu_\mu,\ \ell' = e, \mu$ respectively for $V = W^-$, we have the usual formula (we use the standard notation here [13])

$$
\begin{aligned}
&d\sigma_{\rm exp}(pp \rightarrow V + X \rightarrow \bar{\ell}\ell' + X') = \\
&\sum_{i,j}\int dx_i dx_j F_i(x_i) F_j(x_j) d\hat{\sigma}_{\rm exp}(x_i x_j s),
\end{aligned}
\tag{4}
$$

and we use the result in (2) here with semi-analytical methods and structure functions from Ref. [14]. A Monte Carlo realization will appear elsewhere [15].

We *do not* attempt in the *present* discussion to replace HERWIG [16] and/or PYTHIA [17] – we intend *here* to combine our exact YFS calculus with HERWIG and/or PYTHIA *by using the latter to generate a parton shower starting from the initial* $(x_1, x_2)$ *point at factorization scale* $\mu$ *after this point is provided by the* $\{F_i\}$. This combination of theoretical constructs can be systematically improved with exact results order-by-order in $\alpha_s$, where currently the state of the art in such a calculation is the work in Refs. [18] which accomplishes the combination of an exact $\mathcal{O}(\alpha_s)$ correction with HERWIG. We note that, even in this latter result, the gluon azimuthal angle is averaged in the combination. We note that the recent alternative parton distribution function evolution MC algorithm in Refs. [19] can also be used in our theoretical construction here. Due to its lack of the appropriate color coherence [20], we do not consider ISAJET [21] here.

To illustrate how the combination with Pythia/Herwig can proceed, we note that, for example, if we use a quark mass $m_q$ as our collinear limit regulator, DGLAP [22] evolution of the structure functions allows us to factorize all the terms that involve powers of the big log $L_c = \ln \mu^2/m_q^2 - 1$ in such a way that the evolved structure function contains the effects of summing the leading big logs $L = \ln \mu^2/\mu_0^2$ where we have in mind that the evolution involves initial data at the scale $\mu_0$. The result is therefore independent of $m_q$ for $m_q \downarrow 0$. In the context of the DGLAP theory, the factorization scale $\mu$ represents the largest $p_\perp$ of the gluon emission included in the structure function. In practice, when we use these structure functions with an exact result for the residuals in (2), it means that we must in the residuals omit the contributions from gluon radiation at scales below $\mu$. This can be shown to amount in most cases to replacing $L_s = \ln \hat{s}/m_q^2 - 1 \rightarrow L_{\rm nls} = \ln \hat{s}/\mu^2$ but in any case it is immediate how to limit the $p_T$ in the gluon emission [2] so that we do not double count effects. In other words, we apply the standard QCD factorization of mass singularities to the cross section in (2) in the standard way. We may do it with either the mass regulator for the collinear singularities or with dimensional regularization of such singularities – the final result should be independent of this regulator. This would in practice mean the following: We first make an event with the formula in (4) which would produce an initial beam state at $(x_1, x_2)$ for the two hard interacting partons at the factorization scale $\mu$ from the structure functions $\{F_j\}$ and a corresponding final state X from the exponentiated cross section in $d\hat{\sigma}_{\rm exp}(x_i x_j s)$ ; the standard Les Houches procedure [23] of showering this event $(x_1, x_2, X)$ would then be used, employing backward evolution of the initial partons. If we restrict the $p_T$ as we have indicated above, there would be no double counting of effects. Let us call this $p_T$ matching of the shower from the backward evolution and the matrix elements in the QCED exponentiated cross section.

However, one could ask if it is possible to be more accurate in the use of the exact result in (2)? Indeed, it is. Just as the residuals $\tilde{\bar{\beta}}_{n,m}(k_1, \ldots, k_n; k'_1, \ldots, k'_m)$are computed order by order in perturbation theory from the corresponding exact perturbative results by expanding the exponents in (2) and comparing the appropriate corresponding coefficients of the respective powers of $\alpha^n \alpha_s^m$, so too can the shower formula which is used to generate the backward evolution be expanded so that the product of the shower formula's perturbative expansion, the perturbative expansion of the exponents in (2), and the perturbative expansions of the residuals can be written as an over-all expansion in powers of $\alpha^n \alpha_s^m$ and required to match the respective calculated exact result for given order. In this way, new shower subtracted residuals, $\{\hat{\tilde{\bar{\beta}}}_{n,m}(k_1, \ldots, k_n; k'_1, \ldots, k'_m)\}$, are calculated that can be used for the entire gluon $p_T$ phase space with an accuracy of the cross section that should in principle be improved compared with the first procedure for shower matching presented above. Both approaches are under investigation.

Returning to the general discussion, we compute, with and without QED, $r_{\rm exp} = \sigma_{\rm exp}/\sigma_{\rm Born}$. For this ratio we do not use the narrow resonance approximation; for, we wish to set a paradigm for precision heavy vector boson studies. The formula which we use for $\sigma_{\rm Born}$ is obtained from that in (4) by substituting $d\hat{\sigma}_{\rm Born}$ for $d\hat{\sigma}_{\rm exp}$ therein, where $d\hat{\sigma}_{\rm Born}$ is the respective parton-level Born cross section. Specifically, we have from (1) the $\tilde{\bar{\beta}}_{0,0}$-level result

$$\hat{\sigma}_{\rm exp}(x_1 x_2 s) = \int_0^{v_{max}} dv \gamma_{\rm QCED} v^{\gamma_{\rm QCED}-1} F_{\rm YFS}(\gamma_{\rm QCED}) e^{\delta_{\rm YFS}} \hat{\sigma}_{\rm Born}((1-v)x_1 x_2 s) \tag{5}$$

where we intend the well-known results for the respective parton-level Born cross sections and the value of $v_{max}$ implied by the experimental cuts under study. What is new here is the value for the QED$\otimes$QCD exponent

$$\gamma_{\rm QCED} = \left\{ 2Q_f^2 \frac{\alpha}{\pi} + 2C_F \frac{\alpha_s}{\pi} \right\} L_{\rm nls} \tag{6}$$

where $L_{\rm nls} = \ln x_1 x_2 s/\mu^2$ when $\mu$ is the factorization scale.

[2]Here, we refer to both on-shell and off-shell emitted gluons.

The functions $F_{\rm YFS}(\gamma_{\rm QCED})$ and $\delta_{\rm YFS}(\gamma_{\rm QCED})$ are well-known [12] as well:

$$\begin{aligned} F_{\rm YFS}(\gamma_{\rm QCED}) &= \frac{e^{-\gamma_{\rm QCED}\gamma_E}}{\Gamma(1+\gamma_{\rm QCED})}, \\ \delta_{\rm YFS}(\gamma_{\rm QCED}) &= \frac{1}{4}\gamma_{\rm QCED} + (Q_f^2\frac{\alpha}{\pi} + C_F\frac{\alpha_s}{\pi})(2\zeta(2) - \frac{1}{2}), \end{aligned} \tag{7}$$

where $\zeta(2)$ is Riemann's zeta function of argument 2, i.e., $\pi^2/6$, and $\gamma_E$ is Euler's constant, i.e., $0.5772\ldots$ Using these formulas in (4) allows us to get the results

$$r_{\rm exp} = \begin{cases} 1.1901 & ,\text{QCED} \equiv \text{QCD+QED, LHC} \\ 1.1872 & ,\text{QCD, LHC} \\ 1.1911 & ,\text{QCED} \equiv \text{QCD+QED, Tevatron} \\ 1.1879 & ,\text{QCD, Tevatron.} \end{cases} \tag{8}$$

We see that QED is at the level of .3% at both LHC and FNAL. This is stable under scale variations [13]. We agree with the results in Refs. [5, 6, 8–10] on both of the respective sizes of the QED and QCD effects. The QED effect is similar in size to structure function results found in Refs. [24–28], for further reference.

We have shown that YFS theory (EEX and CEEX) extends to non-Abelian gauge theory and allows simultaneous exponentiation of QED and QCD, QED⊗QCD exponentiation. For QED⊗QCD we find that full MC event generator realization is possible in a way that combines our calculus with Herwig and Pythia in principle. Semi-analytical results for QED (and QCD) threshold effects agree with literature on Z production. As QED is at the .3% level, it is needed for 1% LHC theory predictions. We have demonstrated a firm basis for the complete $\mathcal{O}(\alpha_s^2, \alpha\alpha_s, \alpha^2)$ results needed for the FNAL/LHC/RHIC/ILC physics and all of the latter are in progress.

## Acknowledgments

One of us ( B.F.L.W.) thanks Prof. W. Hollik for the support and kind hospitality of the MPI, Munich, while a part of this work was completed.

# PHOTOS as a pocket parton shower: flexibility tests for the algorithm*

*Piotr Golonka and Zbigniew Was*
CERN, 1211 Geneva 23, Switzerland, and Institute of Nuclear Physics, ul. Radzikowskiego 152, 31-342 Kraków, Poland

**Abstract**
PHOTOS is widely used for generation of bremsstrahlung in decays of particles and resonances in LHC applications. We document here its recent tests and variants. Special emphasis is on those aspects which may be useful for new applications in QED or QCD.

Recently version 2.14 of the PHOTOS Monte Carlo algorithm, written for bremsstrahlung generation in decays became available. In Ref. [1] detailed instructions on how to use the program are given. With respect to older versions [2,3] of PHOTOS, it now features: improved implementation of QED interference and multiple-photon radiation. The numerical stability of the code was significantly improved as well. Thanks to these changes, PHOTOS generates bremsstrahlung corrections in $Z$ and $W$ decays with a precision of 0.1%. This precision was established in [4] with the help of a multitude of distributions and of a specially designed numerical test (SDP), see Ref. [1], section 5 for the definition. The tests for other channels, such as semileptonic $K$ decays and leptonic decays of the Higgs boson and the $\tau$-lepton, are presented in [4] as well. In those cases the level of theoretical sophistication for the reference distribution was lower though.

In this note we will not repeat a discussion of the design properties, but we will recall the main tests that document robustness and flexibility of the PHOTOS design. The results of the comparisons of PHOTOS running with different options of separation of its physical content into functional parts of the algorithm will be shown. The design of the program, i.e. the relation between the parts of the algorithm remained unchanged for these tests. This aspect may be of broader use and may find extensions in future applications, also outside the simple case of purely QED bremsstrahlung in decays.

In the calculations that led to the construction of PHOTOS we had to deal with the diagrams generated by photon couplings to the charged fermions, scalars or vectors. They were definitely simpler than the ones required for the QCD, nonetheless they offered a place to develop solutions which may be of some use there as well. Having such possibility in mind, yet not having any extension to QCD at hand, we have called PHOTOS a *pocket parton shower*. We hope that the methods we developed would be useful for QCD at least as pedagogical examples.

We begin with a presentation of the components of the PHOTOS algorithm using operator language. The consecutive approximations used in the construction of the crude distribution for photon generation, and the correcting weights used to construct the physically complete distributions are listed, but can not be defined in detail here. Instead, we present the variations of the algorithm. Comparisons between different options of the algorithm provide an important class of technical tests, and also help to explore the limits of the universality of the PHOTOS solution. The results of some of these tests will be listed later in the contribution (for the remaining ones and the details we address the reader to refs. [1,4]). In the comparisons we use the SDP universal test based on MC-TESTER [5] as in Ref. [1]. We skip its definition here as well.

The starting point for the development of PHOTOS was the observation that, at first order, the bremsstrahlung corrections in the $Z \to \mu^+\mu^-$ process can be written as a convolution of the Born-level distribution with the single-photon emission kernels for the emission from $\mu^+$ and $\mu^-$.

*Supported in part by the EU grant MTKD-CT-2004-510126, in partnership with the CERN Physics Department, and the Polish State Committee for Scientific Research (KBN) grant 2 P03B 091 27 for the years 2004–2006.

The formulae for the emission kernels are 3-dimensional and can be parametrized using the angles and the invariant mass, which are the same variables as those used in the parametrization of the three-body phase space (the kernels use only a subset of the complete set of phase-space parametrization variables). The remaining two angular variables, not used in the kernels, can be identified as the angles defining the orientation of the $\mu^+$ and/or $\mu^-$ directions (for a detailed definition, see e.g. [2]).

The principle of the single-photon algorithm working on $n$-body decay is to replace a point in the $n$-body phase space $\Omega_2$, with either the point in the original $\Omega_2$, or the point in the $(n+1)$-body phase space $\Omega_3$ (with generated photon). The overall normalization of the decay rate has to change as well and, for example, in the case of $Z \to \mu^+\mu^-$, due to the action of the single-photon algorithm, it needs to be multiplied by a factor of $1 + \frac{3}{4}\frac{\alpha}{\pi}$.

Subsequent steps of the PHOTOS algorithm are described in terms of the evolution operators. Let us stress the relations of these operators to the matrix elements and phase-space parametrizations. We will present the decomposition of the operators in the top–down order, starting with the definition of $R_\alpha$, the operator describing the complete PHOTOS algorithm for single emission (which at least in the case of $Z$ and leptonic $\tau$ decays originates from field theory calculations without any approximation). Then, we will gradually decompose the operators (they differ from decay channel to decay channel) so that we will end up with the single well-defined, elementary operator for the emission from a single charged particle in the final state. By aggregation of these elementary operators, the $R_\alpha$ may be reconstructed for any decay channel. Let us point out that the expression of theoretical calculations in the form of operators is particularly suitable in computer programs implementation.

We skip here a separate discussion of the factorization properties, in particular to define/optimize the way the iteration of $R$'s is performed in PHOTOS. Not only the first-order calculations are needed, but also higher-order ones, including mixed virtual–real corrections. For practical reasons, the $R_\alpha$ operator needs to be regularized with the minimum energy for the explicitly generated photons: the part of the real-photon phase space, under threshold, is integrated, and the resulting factor is summed with the virtual correction.

**• 1**

Let us define the five steps in $R_\alpha$ separation. In the first one, the $R_\alpha$ is replaced by (we use two-body decay as an example) $R_\alpha = R_I(R_S(\mu^+) + R_S(\mu^-))$, where $R_I$ is a generalized interference operator and $R_S$ is a generalized operator responsible for photon generation from a single, charged decay-product.

Let us point out here, that we use the word *interference* here having in mind its usual quantum-mechanical sense. The interference is introduced simultaneously for the real and the virtual photon correction. As a consequence, it changes, for instance, the hard-photon energy spectrum, and the action of $R_I$ looks like kinematic reshuffling of events around the phase space. This interpretation of the interference was particularly clear in the case of the $Z$ decays where the $R_I$ operator can introduce *exact and complete* first-order radiative corrections.

It is important to firstly define the amplitudes, the sum of which is squared, in physically meaningful way, that is in gauge-invariant way, to produce interference. Our approach has changed with time, and we relaxed this requirement; at present we simply request that the action of $R_I$ properly introduces interference effects. We also require that the generalized interference operator respects energy–momentum conservation, and also overall normalization of the distribution under construction. The freedom of choice in the separation of $R_\alpha$ into $R_I$ and $R_S$ we obtained this way is used to create different variants of the PHOTOS algorithm.

The $R_S$ operator acts on the points from the $\Omega_2$ phase space, and the results of its action belong either to $\Omega_2$ or to $\Omega_3$. The domain of the $R_I$ operator has to be $\Omega_2 + \Omega_3$, and the results are also in $\Omega_2 + \Omega_3$. In our solution we required that $R_I$ acts as a unit operator on the $\Omega_2$-part of its domain and, with some probability, returns the points from $\Omega_3$ back to the original points in $\Omega_2$, thus reverting the action of the $R_S$.

Let us stress that in practical applications, to ease the extension of the algorithm to "any" decay mode, we used in PHOTOS a simplification for $R_I$. Obviously, the exact representation of the first-order result would require $R_I$ to be decay-channel-dependent. Instead, we used an approximation that ensures the proper behaviour of the photon distribution in the soft limit. Certain deficiencies at the hard-photon limit of the phase space appear as a consequence, and are the subject of studies that need to be performed individually for every decay channel of interest. The comparisons with matrix-element formulae, as in [6], or experimental data, have to be performed for the sake of precision; they may result in dedicated weights to be incorporated into PHOTOS. In principle, there is no problem to install a particular decay-channel matrix element, but there has not been much need for this yet. So far, the precision of the PHOTOS algorithm could always be raised to a satisfactory level by implementing some excluded parts of formulae, being the case of $W$ decay [6] an exception.

The density generated by the $R_S$ operator is normally twice that of real photons at the end of generation and all over the phase space; it can also overpopulate only those regions of phase space where it is necessary for $R_I$. The excess of these photons is then reduced by Monte Carlo with the action of $R_I$.

**• 2**

In the next step of the algorithm construction, we have separated $R_S = R_B R_A$, where $R_B$ was responsible for the implementation of the spin-dependent part of the emission, and the $R_A$ part was independent of the spin of the emitting final-state particle. Note that this step of the algorithm can be performed at the earlier stage of generation as well, that is before the full angular construction of the event. $R_B$ is again, as $R_I$, it moves the hard bremsstrahlung events in excess back to the original no-bremsstrahlung ones. $R_B$ operates on the internal variables of PHOTOS rather than on the fully constructed events.

**• 3**

The definition of the $R_I$, $R_B$, $R_A$ operators was initially based on the inspection of the first-order matrix elements for the two-body decays. In the general solution for $R_A$, the process of multiple-body decay of particle $X$ is temporarily replaced by the two-body decay $X \to CY$, in which particle $X$ decays to the charged particle $C$, which "emits" the photon, and the "spectator system" $Y$. The action of the operator is repeated for each charged decay product: the subsequent charged particle takes the role of the photon emitter $C$; all the others, including the photons generated in the previous steps, become a part of the spectator system $Y$. The independence of the emissions from each charged product then has to be ensured. This organization works well and can be understood with the help of the exact parametrization of multibody phase space. It is helpful for iteration in multiple-photon emission. It also helps to implement some genuine second-order matrix elements. This conclusion can be drawn from an inspection of the second-order matrix elements, as in [7].

**• 4**

In the next step, we decompose the $R_A$ operator, splitting it in two parts: $R_A = R_a R_x$. The $R_x$ operator generates the energy of the (to be generated) photon, and $R_a$ generates its explicit kinematic configuration.

The $R_x$ operator acts on points from the $\Omega_2$ phase space, and generates a single real number $x$; the $R_a$ operator transforms this point from $\Omega_2$ and the number $x$ to a point in $\Omega_3$, or leaves the original point in $\Omega_2$. Note that again, as $R_I$, the $R_a$ operator has to be unitary and has to conserve energy–momentum[1].

An analogy between $R_x$ and the kernel for structure-function evolution should be mentioned. However, there are notable differences: the $x$ variable is associated more with the ratio of the invariant mass of decay products of $X$, photon excluded, and the mass of $X$, than with the fraction of energy taken away by the photons from the outgoing charged product $C$. Also, $R_x$ can be simplified by moving its parts to $R_a$, $R_S$ or even $R_I$. Note that in $R_x$ the contributions of radiation from all charged final states are summed.

**• 5**

The $R_x$ operator is iterated, in the solutions for double, triple, and quartic photon emission. The iterated $R_x$ can also be shifted and grouped at the beginning of the generation, because they are free from the phase-space constraints. The iterated $R_x$ takes a form similar to a formal solution for structure-function evolution, but with exceptionally simple kernels. The phase-space constraints are introduced later, with the action of the $R_a$ operators. Because of this, the iteration of $R_x$ can go up to fixed or infinite order. The algorithm is then organized in two steps. At first, a crude distribution for the number of photon candidates is generated; then, their energies are defined. For that purpose we can perform a further separation: $R_x = R_f R_0 R_N$, where the $R_0$ operator determines whether a photon candidate has to be generated at all, and $R_f$ defines the fraction of its energy (without energy–momentum-conservation constraint). From the iteration of $R_0$, we obtain a Poisson distribution, but any other analytically solvable distribution would be equally good.

The overall factor, such as $1 + \frac{3}{4}\frac{\alpha}{\pi}$ in $Z$ leptonic partial width, does not need to be lost. It finds its way to the $R_N$, which is a trivial overall normalization constant in the case of the final-state radiation discussed here. In the cases where precision requirements are particularly high, the users of PHOTOS should include this (process-dependent) factor into the decay tables in their main generator for decays. However, until now, the effects on the normalization due to $R_N$ are too small and were usually neglected. We rise the attention to this point, because it may be important for generalizations, when different organization of $R_f$, $R_0$ and $R_N$ may be enforced by the properties of the matrix elements.

---

The input data for the algorithm are taken from the event record, the kinematic configurations of all particles, and the mother–daughter relations between particles in the decay process (which could be a part of the decay cascade) should be available in a coherent way.

This wraps up, a basic, presentation of the steps performed by the PHOTOS algorithm. For more details see [1, 8].

**Tests performed on the algorithm:**

1. The comparison of PHOTOS running in the quartic-photon emission mode and the exponentiated mode for the leptonic $Z$ and $W$ decays may be found on our web page which documents the results of the tests [4]. The agreement in branching ratios and shapes of the distributions is better than

[1] On the contrary, the $R_x$ operator can not, in general, fulfill the unitarity requirement. For example, the part of $R_\alpha$ leading to $1 + \frac{3}{4}\frac{\alpha}{\pi}$ for the $Z$ decay can not be placed elsewhere but in $R_x$. The energy–momentum conservation does not apply directly to $R_x$, as it does not change the kinematic configuration, but only supplements it with $x$, the energy of the photon to be generated. However, for multiple-photon generation, the limits for generated $x$ for subsequent generated photons are the same as for the first photon, which may be in potential conflict with energy–momentum conservation constraint.

0.07% for all the cases that were tested. It can be concluded that changing the relative order for the iterated $R_0$ and the rest of $R_\alpha$ operators does not lead to significant differences. This test, if understood as a technical test, is slightly biased by the uncontrolled higher-than-fourth-order terms which are missing in the quartic-emission option of `PHOTOS`. Also, the technical bias, due to the minimal photon energy in generation, present in the fixed-order options of `PHOTOS` may contribute to the residual difference.

2. The comparison of `PHOTOS` with different options for the relative separation between $R_I$ and $R_S$. The tests performed for the fixed-order and exponentiated modes indicated that the differences in results produced by the two variants of the algorithm are below the level of statistical error for the runs of $10^8$ events. In the code these two options are marked respectively as `VARIANT-A` and `VARIANT-B`.
3. The comparisons of `PHOTOS` with different algorithms for the implementation of the $R_I$ operator. In `PHOTOS` up to version 2.12, the calculations were performed using internal variables in the angular parametrization. This algorithm was limited to the cases of decays of a neutral particle into two charged particles. In later versions, the calculations are performed using the 4-momenta of particles, hence for any decay mode. The tests performed for leptonic $Z$ decays indicated that the differences are below the statistical error of the runs of $10^8$ events.
4. The comparisons of `PHOTOS` with different options for the relative separation between $R_0$ and $R_x$, consisting of an increase in the crude probability of hard emission at $R_0$. The tests performed for the exponentiated mode of `PHOTOS` indicated that the differences are below the statistical error of the runs of up to $10^8$ events.
5. The remaining tests, including new tests for the effects of the interference weights in cascade decays, are more about the physics content of the program than on the technical or algorithmic aspects. They are presented in Ref. [1] and the results are collected on the web page [4].

Multiple options for `PHOTOS` running and technical compatibility of results even for $10^8$ event samples generated in a short CPU cycle time are encouraging. They indicate the potential for algorithm extensions. Note that `PHOTOS` was found to work for decays of up to 10 charged particles in the final state.

**Acknowledgements:** The authors are indebted to members of BELLE, BaBar, NA48, KTeV, ATLAS, CMS, D0, CDF collaborations for useful comments and suggestions.